\documentclass[final,3p,times,authoryear]{elsarticle}

\newcommand{\bra}[1]{\ensuremath{\langle\, \mathrm{#1} \,|}}

\newcommand{\ket}[1]{\ensuremath{|\, \mathrm{#1}\,\rangle}}

\newcommand{\ME}[3]{ \mbox{$\langle #1\,|\,#2\,|\,#3\rangle $} }

\newcommand{\Skalar}[2]{ \mbox{$\langle\,#1\,|\,#2\,\rangle $} }

\usepackage[T1]{fontenc} 
\usepackage{lineno,longtable}

\modulolinenumbers[5]

\journal{Physics Reports}

\biboptions{authoryear}
\begin{document}

\begin{frontmatter}

\title{Atomic, molecular and optical physics applications of longitudinally coherent and narrow bandwidth Free-Electron Lasers}

\author[elettra]{Carlo Callegari\corref{ca}}
\ead{carlo.callegari@elettra.eu}
\author[moscow]{Alexei N. Grum-Grzhimailo}
\ead{grum@sinp.msu.ru}
\author[tokyo]{Kenichi L. Ishikawa}
\ead{ishiken@n.t.u-tokyo.ac.jp}
\author[elettra]{Kevin C. Prince\corref{ca}}
\ead{kevin.prince@elettra.eu}
\author[freiburg]{Giuseppe Sansone}
\ead{giuseppe.sansone@physik.uni-freiburg.de}
\author[tohoku]{Kiyoshi Ueda}
\ead{kiyoshi.ueda@tohoku.ac.jp}
\address[elettra]{Elettra Sincrotrone Trieste, 34039 Basovizza, Italy}
\address[moscow]{Lomonosov Moscow State University, Moscow 119991, Russia}
\address[freiburg]{University of Freiburg, Physikalisches Institut, Albert-Ludwigs-Universit\"at Freiburg, 79106 Freiburg, Germany}
\address[tohoku]{Institute of Multidisciplinary Research for Advanced Materials,
Tohoku University, Sendai 980-8577, Japan}
\address[tokyo]{Graduate School of Engineering,
The University of Tokyo, 7-3-1 Hongo, and Bunkyo-ku, Tokyo 113-8656, and Research Institute for Photon Science and Laser Technology,
The University of Tokyo, 7-3-1 Hongo,
Bunkyo-ku, Tokyo 113-0033, Japan}
\cortext[ca]{Corresponding author}

\begin{abstract}
Short wavelength Free-Electron Lasers (FELs) are the newest light sources available to scientists to probe a wide range of phenomena, with chemical, physical and biological applications, using soft and hard X-rays. These sources include the currently most powerful light sources in the world (hard X-ray sources) and are characterised by extremely high powers and high transverse coherence, but the first FELs had reduced longitudinal coherence. Now it is possible to achieve good longitudinal coherence (narrow bandwidth in the frequency domain) and here we discuss and illustrate a range of experiments  utilising this property, and their underlying physics. 
The primary applications are those which require high resolution (for example resonant experiments), or temporal coherence (for example coherent control experiments). The currently available light sources extend the vast range of laboratory laser techniques to short wavelengths.

\end{abstract}

\end{frontmatter}

%\linenumbers
\tableofcontents
\section{Introduction}

Free-Electron Lasers are modern light sources which use high energy beams of electrons moving in vacuum to generate radiation. Although there are many such sources producing infrared light, in this paper we focus on those producing ionizing radiation, that is, wavelengths from the Extreme Ultraviolet (XUV; 124--10 nm) to soft and hard X-rays ($<$10 nm).

\subsection{Historical background of FEL sources}
The interaction of light with matter is governed by the Einstein coefficients, which describe the spontaneous emission, absorption and stimulated emission of light. Einstein's recognition of the importance of stimulated emission was the basis for the invention (much later) of the maser~\citep{Gordon1954,Gordon1955} and laser~\citep{Schawlow1958,Maiman1960}. In atoms, molecules, and solids the electrons involved in the lasing process are all bound in specific orbitals or bands, that is, they reside in a strong electromagnetic field. The wavelengths which such lasers can emit is limited by the quantum transitions which are possible, so that it is not possible to tune over wide ranges. 
Long ago it was recognised that stimulated emission may occur in free electrons, that is, electrons moving in vacuum. The word ``free'' is perhaps slightly misleading, as an electron moving in field-free space cannot be stimulated to emit radiation. Rather, the electron is free in the sense that it is not confined to an atom or molecule but moves in an external field. Microwave devices, such as klystrons, are mostly based on electric fields, whereas short-wavelength devices are generally based on relativistic particle accelerator technology, where high-energy electrons are deflected, focused, and guided by magnetic fields. Since the electron energy and magnetic fields can be tuned continuously, the idea of using light from free electrons travelling through a periodic magnetic structure~\citep{Ginzburg1947,Motz1951} offered the possibility of a light source with continuously tunable wavelength~\citep{Motz1953}. 
Let us note that because of the methods used to produce and accelerate the electrons, the latter are not uniformly distributed, but rather are grouped in packets (``bunches''). 
A strong distinction must be made between microwave emission at one end (where the bunch size can be externally controlled, and be shorter than the radiation wavelength) and X-rays at the other end (where the bunch size is much longer than the radiation wavelength). In the latter case one generally ignores the finite length of the bunch, and analyzes the periodic grouping of the electrons \emph{within} the bunch. This grouping, which grows as part of the amplification process, is termed ``microbunching''~\citep{Yu1984}, and the short form ``bunching'', is a confusing but accepted terminology. \citet{Motz1951} recognized the importance of pre-bunching to make the radiation coherent, and even analyzed the case of an electron moving through an electromagnetic wave traveling in the opposite direction, but did not consider the process of stimulated emission and the associated bunching growth. 
This was later done by~\citet{Madey1971}, who is generally credited as the father of the Free-Electron Laser; an analysis of the formal equivalence of the spontaneous and stimulated emission from electrons in atoms and that from quasifree electrons was later done by \citet{Friedman1988}.

There exist several reviews of the history of Free-Electron Lasers: \citet{Colson1985a} trace the evolution of free-electron sources back to the 1940s, from non-relativistic electron tubes, and list a number of schemes suitable for long-wavelength operation (down to $\sim 5000$~{\AA}).
\citet{Madey2014} gives a personal account of how short-wavelength free-electron lasers are the result of a cross fertilization between the fields of electron tubes, masers/lasers, and particle accelerators, and borrow fundamental concepts from all these fields. 
The considered electron energies varied over a broad range, hundreds of keV to hundreds of MeV; modern FELs all utilise electron beams of GeV energies, and are thus highly relativistic devices (we recall that the rest energy of the electron is 0.511~MeV); a non-relativistic precursor -- the ``ubitron'' -- was built by \citet{Phillips1960,Phillips1988}. 

Just as for optical lasers, long-wavelength FELs can  be classified as oscillators or amplifiers, depending on the presence or absence of a resonant cavity~\citep{Colson1985a}. Early FELs were conceived as low-gain amplifiers~\citep{Madey1971,Madey1973} and later as oscillators, for wavelengths where mirrors were available. The efficiency and tunability were seen as promising characteristics, and applied research on single-pass FELs was carried out, for example by the United States Office of Naval Research~\citep{Roberson1989Book}.

The coherence properties of FELs were immediately recognised as important, and at long wavelength longitudinal coherence was not difficult to attain. There was clearly great interest in FELs as sources for short wavelength ranges, where no convenient alternatives were available; high harmonic generation from FELs operating as single-pass amplifiers was recognized as a means of overcoming the limitations of mirrors in the UV region and beyond~\citep{Colson1981,Colson1985}.
\citet{Kondratenko1979,Kondratenko1980} proposed a single-pass amplifier (in the infrared) starting from noise, and \citet{Murphy1985} extended the concept to soft X-rays. The process is called Self-Amplified Spontaneous Emission (SASE), and implies exponential growth up to saturation; this was observed at 530 nm by \citet{Milton2000,Milton2001}. The first successful demonstration of production of XUV radiation was the TESLA Test Facility (TTF) VUV-FEL~\citep{Ayvazyan2002}, which later evolved into FLASH~\citep{Ayvazyan2006}. Abandoning the use of a cavity and/or pre-bunching of the electrons, made the property of longitudinal coherence secondary: as remarked in the review work of \citet{Bonifacio1990a} the initial intensity is linearly proportional to the electron current, i.e., the radiation is incoherent. We will elaborate this aspect below; for now we note that for this reason, longitudinal coherence has not been a distinctive feature of short-wavelength (SASE) FELs, and of the experiments performed with them, until the advent of seeded sources, which are the subject of this review.

In this context, what we call ``FEL spontaneous emission'' is exactly the same as the synchrotron radiation produced by an undulator, with the difference that in the latter case the back-action of the radiation on the electrons is negligible, i.e., there is no amplification: the spectro-temporal properties of synchrotron radiation are determined by the incoherent sum of the emission of the single electons in the bunch. In the rest frame of the electrons, one can see synchrotron radiation as the pseudo-radiation field of the moving magnetic structure (``undulator'') being reflected by the electron, through Compton back-scattering. 

For a discussion of the most important formulas, it is useful to introduce some physical quantities: $c$ is the speed of light, $m_\mathrm{e}$ the rest mass of the electron, $\lambda_\mathrm{u}$ and $B_0$ the undulator period and the peak magnetic field on the undulator axis in the laboratory frame, $\mathcal{E}$ the electron beam energy, $\gamma = \mathcal{E}/(m_\mathrm{e} c^2)$ the associated Lorentz factor, $\gamma_{\parallel}$ the Lorentz factor in the forward direction.
In the laboratory frame the radiation peaks at a wavelength $\lambda_\mathrm{s} = \lambda_\mathrm{u}/(2\gamma_{\parallel}^2)$; the relation can be interpreted as the Lorentz-FitzGerald contraction of the undulator period in the rest frame of the electrons \textit{and} of the radiation emitted in the forward direction in the laboratory frame.

Alternatively, one can observe that the electrons have a forward velocity $v_\parallel < c$, and see the resonance condition as that for which they lag the radiation by one unit of \(\lambda_\mathrm{s}\) per undulator period,  i.e., \(\lambda_\mathrm{u}(c - v_\parallel)/v_\parallel =\lambda_\mathrm{s}\), noting that $(c - v_\parallel)/v_\parallel \approx (c - v_\parallel)/c \approx 1/(2\gamma_{\parallel}^2)$. This condition is technically known as ``slippage'', and the total slippage length ($N_\mathrm{u} \lambda_\mathrm{s}$, with $N_\mathrm{u}$ the number of undulator periods) defines the maximum separation between two electrons such that their respective emission can still overlap in time (i.e., interfere); thus the slippage length is important in discussing the longitudinal coherence of the pulse. In the presence of gain, it is replaced by the shorter cooperation length. 

The resonance condition can be tuned either by changing the electron beam energy, or by changing the amount of energy transferred into the transverse direction, via the strength of the undulator field. 
Both methods have their advantages and disadvantages. The former is more complicated in that the electron accelerating cavities and transport optics have to be tuned; the latter in that it requires variable-gap undulators, which are mechanically demanding. Nevertheless, the latter solution is generally more flexible, also in terms of controlling the polarization of the radiation~\citep{Sasaki1994}. 
It is convenient to introduce an \textit{undulator parameter} $K = e B_0 \lambda_\mathrm{u}/(2\pi m_\mathrm{e} c)$; the resonance condition becomes $\lambda_\mathrm{s} = \lambda_\mathrm{u}(1+K^2/2)/(2\gamma^2)$ \citep[see, e.g.,][]{Attwood2016}.
The fractional bandwidth of this radiation is of order $1/N_{\mathrm{u}}$; the rms angular divergence and source size are respectively of order $\sqrt{\lambda_\mathrm{s}/(2\lambda_\mathrm{u} N_\mathrm{u})} \approx \frac{1}{2\gamma}\sqrt{\frac{1+K^2/2}{N_\mathrm{u}}}$ and $\sqrt{\lambda_\mathrm{s}\lambda_\mathrm{u} N_\mathrm{u}/2)}$, i.e., their product is diffraction-limited~\citep[$\sim \lambda_\mathrm{s}$;][]{Kim2017}.

Let us now discuss the back-action of the radiation on the electrons, i.e., the amplification: \citet{Bonifacio1990a} explain that in the presence of a radiation field of wavelength $\lambda \approx \lambda_{\mathrm{s}}$, electrons at the resonant energy remain in phase with the radiation, with which they interact undergoing either absorption or stimulated emission according to the value of their phase. 
This implies that for a monoenergetic electron beam, with uniformly distributed (random) phase, the net gain will be zero; this is however a condition of unstable equilibrium, and random fluctuations that fall within the gain bandwidth will be amplified. 
In the low-gain condition, the resulting FEL equations are the same as those describing a simple pendulum~\citep{Colson1977,Bonifacio1990a,Seddon2017,Kim2017}. The exact treatment of FEL gain is instructive but lengthy and  complex~\citep{Kim2017}, and we will mention here some simplified results of interest for longitudinal coherence; in a 1-dimensional model one studies the evolution of the phase-space of an initially uniform (in space and in a narrow energy band) distribution of electrons; we note that the phase relative to the radiation field is customarily used as the spatial coordinate. 
In the low-gain initial phase, the first departure from uniformity is an energy modulation at the resonant wavelength, whereas the spatial density remains initially constant (to first perturbation order), but is then driven to grow by the energy modulation; seeded FELs are based on the concept that an initial energy modulation much stronger than that driven by spontaneous emission can be imprinted onto the electron bunch. 
It is important to remark that the emitted intensity contains a term that scales with the square of the number of electrons in the bunch, $N_\mathrm{e}^2$, times the absolute square of the ``bunching factor'', defined as the Fourier coefficient of the electron density modulation at the resonant frequency of the amplifier.   

The cooperation length (defined as the slippage length accrued over one gain length) versus the length of the electron bunch directly determines the coherence properties of the radiation. Note that (as for undulator radiation in synchrotrons) relativistic effects confine the emitted light to a forward cone of aperture $1/\gamma_{\parallel}$, and that leading electrons that are spatially separated by more than the slippage length cannot be reached by the light emitted by trailing electrons, so they emit as independent sources. Thus, the shape in both time and wavelength of a SASE pulse from a long electron bunch is the incoherent sum of a set of spikes~\citep{Bonifacio1994,Yu2003,Penco2015}; the coherence properties of SASE FELs have been discussed by \citet{Saldin2008}, and references therein. In the hard X-ray region, short, intense pulses are obtained by passively restricting below the cooperation length the portion of the emitting electrons~\citep{Coffee2019}.

\begin{figure}[ht]
    \centering
    \includegraphics[width = 0.9\textwidth]{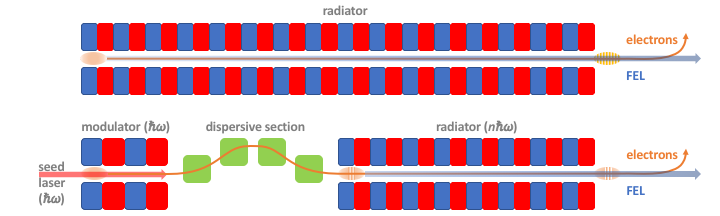}
    \caption{Schematic of a SASE FEL (top) and a seeded FEL (bottom). Although not to scale, the figure reflects the fact that a seeded FEL radiator is generally much shorter, and that only the central portion of the electron bunch, that has interacted with the seed laser, develops significant microbunching.}
    \label{fig:seedingschemes}
\end{figure}

External seeding is an alternative approach to manipulating the electron bunch, and we will discuss in Section~\ref{ssec:selfseeded} the concept of self-seeding, i.e., amplifying a monochromated (or cleaned-up) pulse from a SASE FEL in a second stage of undulators. Historically, the concept of seeding has been associated with that of High Gain Harmonic Generation  (HGHG, Section~\ref{ssec:seeded}) This scheme~\citep{Yu1991,Bonifacio1990,Yu2000,Yu2003} is based on the concept of introducing, by interaction with a seed laser in a first, dedicated, undulator (the modulator) a small energy modulation of the electron beam, see Fig. \ref{fig:seedingschemes}. 
This is then converted into a spatial modulation, via a magnetic chicane (dispersive section), and the modulation contains the fundamental seed frequency and -- with progressively smaller coefficients -- its higher harmonics. 
A second undulator (the radiator) tuned to the desired harmonic causes emission and exponential amplification of that harmonic, with ``a single phase determined by the seed laser, and [a] spectral bandwidth [which] is Fourier transform limited.''~\citep{Yu2000}.
Yu et al.\ explicitly mentioned the excellent temporal coherence of the resulting pulses, and the fact that ``the output properties at the harmonic wavelength are a map of the characteristics of the high-quality fundamental seed laser. This results in a high degree of stability and control of the central wavelength, bandwidth, energy, and duration of the output pulse.''~\citep{Yu2000}. 
The first demonstrations of the HGHG scheme were done at far- or near-IR seed wavelengths~\citep{Yu2000,Yu2003};
\citet{Feldhaus2005} reviewed externally seeded and self-seeded FELs, concentrating on the machine physics aspects. When that paper was written, seeding had been demonstrated at optical and UV wavelengths, with few if any applications. Then, in 2012 the FERMI FEL1~\citep{Allaria2012} became the first short-wavelength, externally seeded facility open to users.

Several recent reviews have covered Free-Electron Lasers and their role as instruments for research: \citet{Seddon2017} have provided a wide-ranging overview of the physics of FELs, beamline and diagnostics design, and a very wide range of applications, including crystallography and high energy density matter. \citet{Bostedt2009} summarised experiments at the FLASH FEL in Hamburg, while \citet{Coffee2019} discussed ultrafast X-ray techniques at the Linac Coherent Light Source, concentrating on instrumental aspects.

In the present work, we concentrate on applications of longitudinally coherent and narrow bandwidth light, to atomic, molecular and cluster systems. The narrow resonances and line widths of atomic and molecular systems find many uses for narrow bandwidth light. We will not discuss results in the fields of coherent diffractive imaging, warm dense matter, and will say little about condensed matter studies with FELs.

\subsection{Introduction to Atomic, Molecular and Optical science with FELs}

Atomic, molecular and optical (AMO) physics is a branch of science in which it is possible to study phenomena at their most fundamental level and with very high precision. Einstein developed his interpretation of Planck`s constant and quantization by recognising the similarity of radiation in a box to a gas, and analysing the energy and entropy \citep{Einstein1905}. Quantum mechanics evolved further based on the study of atomic spectra; for example, the concept of the fine-structure constant, developed by \citet{Sommerfeld1916,Sommerfeld1916a}, arose from the interpretation of atomic spectra, which are still used today for high precision measurements.

It is therefore no surprise that AMO physics was a leading field of research in the early development of laboratory lasers, for example to study dynamical phenomena in molecules, for which Ahmed Zewail was awarded the Nobel Prize in Chemistry in 1999~\citep{Zewail2000}. Now AMO is a field of intense activity for the newest light sources, i.e., short-wavelength Free-Electron Lasers, as seen in a series of recent review articles~\citep{Berrah_JModOpt_2010, Ullrich2012, Yabashi2013, Feldhaus2013, Bostedt2016, Seddon2017, Meister2020, Fukuzawa2020}. 
Some of the phenomena studied by optical lasers can now be studied at XUV and X-ray wavelengths, but it is also possible to observe new phenomena. One factor leading to new physics at short wavelengths is that when core levels are excited, they have an additional decay channel not available for low energy excitation, namely Auger decay. 
A second reason is that the ponderomotive shift for high fields and long wavelengths is very large and dominates the physics (see  Section \ref{ssec:SFA}, Strong Field Approximations). At short wavelengths, the ponderomotive shift may become negligible and a different description is required.

The field of attosecond physics is growing rapidly, and many experiments are carried out on atoms and molecules, thereby avoiding various complications associated with condensed matter. The Fourier limit implies that an attosecond pulse of long wavelength radiation has a very broad frequency distribution on the scale of molecular excitations. However a Fourier transform limited attosecond X-ray pulse can have a bandwidth comparable, for example, to the (Auger decay) lifetimes of core holes, permitting high resolution spectroscopy.

Multiphoton processes are an important class of non-linear optical experiments, and only became possible with the invention of the laser. They have been explored for many years at long wavelengths, for example to ionize molecules by the absorption of many photons of the same colour, or of photons of different colour. The process may be non-resonant, occurring via virtual intermediate states, or resonant, with a much higher cross section due to the resonances; this is a standard technique in laboratory laser optics, known as REMPI [Resonance Enhanced Multi Photon Ionization;~\citet{Zandee1979} and references therein]. 
The process has been studied and widely used in FEL experiments, e.g., \citet{YoungNature2010, Rudek2012}, but there is an important experimental difference with respect to long wavelengths. At short wavelengths, it is often possible to ionize a target more than once by consecutive single photon processes, and when this situation occurs, the process is described as sequential multiphoton ionization, to distinguish it from processes requiring the absorption of two or more photons simultaneously. Generally, when the two processes are in competition, the sequential process dominates, as single-photon processes generally have higher cross sections than multi-photon processes.

The narrow bandwidth and high intensity of laboratory lasers has often been used to control the outcome of a photochemical or photophysical event, for example the momentum of the ejected products, or ratio between the products. Both light and matter are described by waves which have an amplitude and a phase, and the light properties can be imprinted on atoms, to manipulate the outcome of an interaction. 
Generally these methods require fully coherent light in both the transverse and temporal domains. While optical lasers have long demonstrated these characteristics, only recently has fully coherent FEL light become available, and in this review we describe some of the ongoing experiments. 
The techniques may use temporally overlapping, mutually coherent wavelengths; a single wavelength split into temporally separated pulses with controlled phase; or other optical arrangements.

Because of the short time structure of pulsed lasers, they have been instrumental in studying the dynamics of matter after absorption of a photon. The advantage of FELs with respect to laboratory sources  for valence ionization studies is a very large increase in signal-to-noise, and a wider spectral range. FELs can also photoionize core levels for femtosecond dynamics studies, and have enough intensity to perform core level photoelectron spectroscopy, which has not been demonstrated with other sources. In this case, coherence is not the primary concern, but rather a sufficiently narrow bandwidth to provide good resolution.

\subsection{Role of transverse and longitudinal coherence.}
Radiation is transverse coherent when the electric field at two points across a wavefront have a fixed and well-defined phase relationship. The concept arose from Young's slit diffraction experiments and is now widely applied in other fields. Mathematically, the degree of coherence is described by the correlation function $g^{(1)}$, and measures amplitude-amplitude correlations. Most Free-Electron Lasers produce radiation with a high degree of transverse coherence, which is important for applications such as Coherent Diffractive Imaging~\citep{Miao2015}.

Longitudinal or temporal coherence relates to the phase relation of a wave at a given point and at different times. If the phase is perfectly periodic at that point, then the wave is temporally coherent. 
Thus in the frequency domain, temporally coherent light has a single frequency (or narrow band of frequencies), i.e. it is monochromatic. Temporally incoherent light consists of a large range of frequencies, and so has a large bandwidth. The degree of coherence of a beam is not a constant, but can be varied, for example by spatial or frequency filtering. 
Free-Electron Lasers based on SASE have rather low degrees of temporal coherence, which hinders high resolution spectroscopy. In this review, we will concentrate on experiments and applications where the degree of longitudinal coherence (or monochromaticity) of the light played a significant role. 
In the following sections, we give a brief description of the types of applications where this property is important, and what other properties of FELs are important.

\section{\label{sec:methods}Methods}

Most FELs operating in the world today are based on the Self Amplified Spontaneous Emission (SASE) process, Table~\ref{tab:facilities}, and all of the FELs in this Table are based on SASE except FERMI and DCLS. As described above, in the SASE process, radiation is initially generated incoherently as the electron pulse passes through an undulator.
For seeded FELs (FERMI and DCLS), the accelerator functions as an amplifier and wavelength shifter, so that the light has a naturally narrow bandwidth, and inherits the properties of the seed laser.

\begin{table}[!ht]
    \centering
    \begin{tabular}{c|c|c|c|c|c}
Name & starting year & wavelength (nm)& pulse duration (fs) &rep rate & References\\
\hline
\href{https://flash.desy.de}{FLASH} & 2005 &4.2--52 & 30--200 &1 Hz--1 MHz & \citep{Ayvazyan2006} \\
\href{https://flash.desy.de/flash_upgrades/}{FLASH2} & 2016 & 4--90 & $<$10--200 &1 Hz--1MHz& \citep{Faatz2017} \\
\href{https://lcls.slac.stanford.edu}{LCLS} & 2009 & 0.11--6.2& $<$10--250& 120 Hz & \citep{Emma2010}\\
\href{https://lcls.slac.stanford.edu/lcls-ii}{LCLS-II} &[late 2020]& [0.05--6.2] & [30] & [120 Hz--1 MHz] & \citep{Raubenheimer2018}\\
SCSS$^*$ & 2008 & 50--60 & 30--100 & 10--60 Hz & \citep{Shintake2008}\\
\href{http://xfel.riken.jp/eng/index.html}{SACLA (HX)} & 2012 & 0.062--0.31 & 2--10 & 30--60 Hz& \citep{Ishikawa2012}\\
\href{http://xfel.riken.jp/eng/index.html}{SACLA (SX)} & 2016 & 8.3--31 & 70 & 30--60 Hz& \citep{Owada2018}\\
\href{https://www.xfel.eu}{EuXFEL} & 2017 & 0.05--4.7 & $<$ 100  & 27 kHz & \citep{Tschentscher2017}\\%\citep{Abela2006}
&&&&&\citep{Decking2020}\\
\href{https://www.psi.ch/en/swissfel}{SwissFEL} & 2017 & 0.1--5 & 1--20 & 100 Hz & \citep{Milne2017} \\
\href{http://pal.postech.ac.kr/paleng/}{PAL} & 2017 &0.083--0.56; 1--6.2 & 25; 80 & 10, 30, 60 Hz & \citep{Kang2017} \\
\href{https://www.elettra.eu/lightsources/fermi.html}{FERMI, FEL1} & 2012 & 20--120 & 35--80 & 50 Hz & \citep{Allaria2012}\\
\href{https://www.elettra.eu/lightsources/fermi.html}{FERMI, FEL2} & 2016 & 4--20 & 5--35 & 10; 50 Hz & \citep{Allaria2013}\\
DCLS & 2018 & 50--150 & 100--1000 & 20 Hz & \citep{Yong2019}\\
Shanghai SXFEL & [late 2020] & [2--24] & [50--200] & [10--50 Hz] & \citep{Zhao2017,Zhao2017a,Zhao2019}\\
\hline
\multicolumn{6}{p{\textwidth}}{$^*$Now upgraded and incorporated in SACLA (SX) as BL1 \citep{Owada2018}.}\\
\hline
    \end{tabular}
    \caption{\label{tab:facilities}List of short wavelength FELs open to users. ``Starting year'' is the actual (or expected) year of opening to users. In the electronic version, click on the name of the FEL to follow the associated hyperlink.}
\end{table}

\subsection{\label{ssec:seeded}Seeded soft X-ray and XUV FELs.}

The FERMI FEL, or more accurately FELs \citep{Allaria2012,Allaria2013}, are seeded and have shown remarkable flexibility and potential for development of novel operating modes \citep{Giannessi2018,Gauthier2016a,Roussel2015}. 
The first report of lasing was published in 2012~\citep{Allaria2012}, and was followed shortly after by the announcement of cascaded lasing, which produced soft X-ray emission at 10.8 nm~\citep{Allaria2013}. 
Later developments pushed the wavelength down to 4 nm and below, with significant pulse energies. 
The predictability of the seed pulses supported by established theory allows the reliable prediction of the temporal duration and saturation behaviour of the pulses~\citep{Finetti2017}.

\begin{figure}[ht]
    \centering
    \includegraphics[width = 0.99\textwidth]{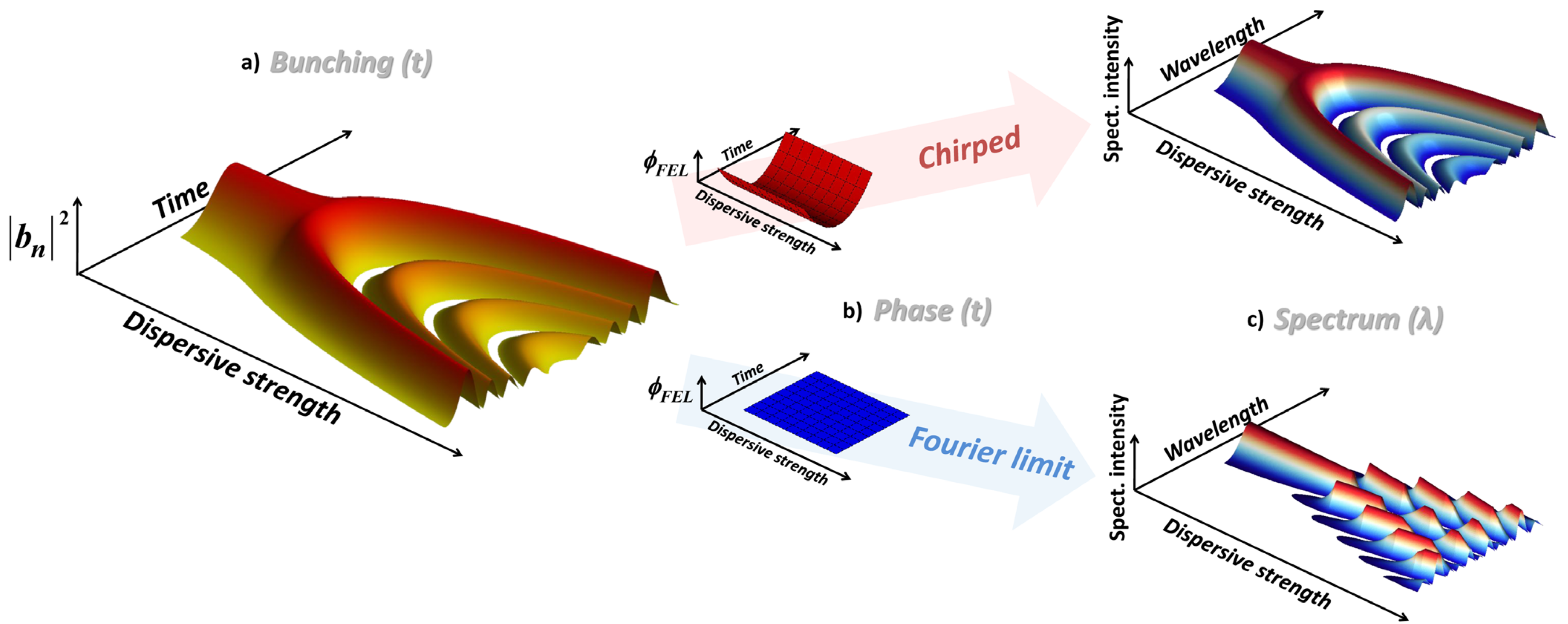}
    \caption{Simulated temporal FEL pulse shape (left) and FEL spectrum (right) as a function of the strength of the seeding process for strong linear chirp (middle, top) versus flat phase (middle, bottom). Figure reprinted from \citet{Gauthier2015}, with permission. Copyright (2015) by the American Physical Society.}
    \label{fig:chirp}
\end{figure}

 Due to the longitudinal coherence provided by the seed pulse, FEL pulses can be analysed shot-by-shot using the technique of SPIDER, which has been applied in the optical wavelength range~\citep{DeNinno2015}. Note that the possibility of deducing from the optical seed pulse the phase of the FEL pulse is invaluable when characterizing the latter, and can be exploited to fine-tune its properties. \citet{Gauthier2015} compared the experimental measurements of the FEL spectrum as a function of the strength of the initial modulation with existing theory, and were able: to produce the first direct evidence of the full temporal coherence of the pulses; to retrieve the phase of the FEL pulse, showing the presence of a linear chirp introduced by the electron bunch; to use the linear frequency chirp of the seed laser to precisely cancel that from the electron bunch, thus attaining Fourier-transform-limited pulses (Fig.~\ref{fig:chirp}). 
 The Figure is easily understood in the two limit-case scenarios of strongly-chirped pulse versus transform-limited (zero-chirp) pulse: the time behaviour of the pulse envelope only depends on the strength of the seed process (the instantaneous intensity of the seed laser scaled by the dispersive strength of the subsequent magnetic chicane) and develops into a set of maxima and minima corresponding to an instantaneous condition of optimal seeding or overseeding. For a dominating linear chirp, the time envelope is streaked onto the wavelength coordinate, and the pulse structure has the same appearance in wavelength as in time; for a Fourier-limited pulse, the spectrum corresponds to the interference of two or more pulses of the same carrier frequency, separated in time.

Importantly, since FERMI's inception, a number of developments of the sources, directly related to various seeding schemes, have permitted operation with: double pulses of the same wavelength~\citep{Gauthier2016}, or different wavelengths~\citep{Allaria2013a, Ferrari2016}; single-seed multiple wavelengths~\citep{Roussel2015}, and other variations.
The machine physics results include the realisation of a proof-of-principle demonstration of chirped pulse amplification~\citep{Gauthier2016a}, which is the state-of-the-art method for producing short intense pulses in optical and infrared lasers, as recognised by the award of the 2018 Nobel prize for physics~\citep{Mourou2019,Strickland2019}.
Novel ways of generating light have been demonstrated, in particular Echo Enabled Harmonic Generation~\citep{RebernikRibic_NatPhot_2019} promises to provide shorter wavelengths with higher pulse energies and greater pulse-to-pulse stability. Another mode of operation is superradiance~\citep{Giannessi_PRL_2013,Xi2020}, which can provide pulses with durations below 10 fs, and higher peak power than normal operation.

Conceptually, the radiator section of a FEL consists of a single long undulator; in reality, the practical length of an undulator is typically much shorter than the gain length, and the radiator consists of a sequence of identical modules. There are two technical reasons for using undulators shorter than the gain length. The electron beam is divergent, and has to be periodically refocused by magnets. Also there is a mechanical limitation due to the strong magnetic forces: it is difficult to attain sufficient rigidity over more than a few meters with forces of $10^4$ Newton or more, corresponding to a few metres of magnets. 
For example, FEL-1 of FERMI consists of six identical 2.4 m modules separated by 1.3 m breaks~\citep{Allaria2012,Diviacco2011}. The need then arises to preserve the phase relationship between electron bunching and the emitted light between two modules. 
This is accomplished by the introduction within the gap of a short magnetic chicane \citep[``phase shifter'';][]{Diviacco2011} to slightly delay the electrons. This lengthens the path of the relativistic electrons with respect to the light by distances on the nm scale.
The trimming of the delay is easily accomplished by setting the phase to maximise the FEL intensity, and it is reproducible, so that a lookup table can be created~\citep{Diviacco2011}. 
Indeed, the fields produced in different modules need not be of the same polarization, and one can use the phase shifter to control the \emph{vector sum} of the two fields, producing for example circularly polarized pulses from crossed linearly-polarized undulators, or vice versa \citep{Ferrari2019}.
Note that this scheme remains valid for a coherence length shorter than the length of the pulse, and was first implemented at a synchrotron source~\citep{Bahrdt1992}; interestingly, it has also been applied for the production of polarized gamma rays~\citep{Yan2019}. By further extension, one need not consider a single wavelength for all undulators: we will discuss in detail in sections~\ref{sec:cc_Ne} to \ref{sec:atto} specific experiments which use a combination of two or more phase-locked harmonics generated in distinct undulators. 

\citet{Feng2018} provide a recent overview of the development of fully coherent FEL facilities in China. The Dalian light source is in operation at long wavelengths, and there are ambitious plans to build SXFEL, a seeded source in the soft x-ray region.

\subsection{\label{ssec:selfseeded}Self-seeded soft and hard X-ray FELs.}

The considerable advantages of seeded FELs with respect to SASE suggest that this technique would be useful in the hard X-ray range as well as the soft X-ray range. However there is a lack of sufficiently intense and monochromatic X-ray sources to use as a seed, although FELs can be used indirectly, see Section \ref{ssec:SASE_pumps}.
For now, external seeding cannot be applied, but an attractive alternative is to use self-seeding. The potential of self-seeding was recognised quite early \citep{Feldhaus1997}, but it took many years before it was implemented. 
The subject has been reviewed recently by \citet{Geloni2016}, and will be briefly summarised here. In self-seeding, SASE lasing is initiated, and then the electron and photon beams are separated: the electrons are deviated in a magnetic chicane while the photon beam is filtered. The photon beam is then recombined with the electron beam and acts as a seed.

This was first demonstrated in the hard X-ray region at the LCLS \citep{Amann2012}, based on a previously published scheme \citep{Geloni2011}. The resulting X-rays were near Fourier-transform limited, had a bandwidth of 0.5~eV at 9~keV, corresponding to a bandwidth reduction by a factor of 40--50 with respect to SASE. Soon after, self-seeding was demonstrated for soft X-rays \citep{Ratner2015}, using a monochromator.

The implementation of hard X-ray self-seeding is more complicated than this simple description suggests. If a conventional X-ray monochromator is used, the path length of the X-rays destined to act as seed is increased by centimetres, so the electron beam path must be increased by the same amount to ensure overlap on recombination. High energy electrons cannot be deflected easily, so the chicane needs to be tens of metres long, which is a severe technical limitation.

A solution to this problem is to substitute the monochromator, which is a bandpass filter, with a notch filter \citep{Geloni2011}. This counter-intuitive idea works because a broad pulse, from which a narrow band has been removed in the frequency domain, consists of a pulse with a monochromatic tail in the time domain. 
The wavelength of the tail or ``wake'' corresponds to the notch in the frequency domain. 
This structure arises primarily because of dynamical scattering in the diffraction process; the thickness of the crystal and the extinction length must be carefully optimised. In these calculations, the monochromatic wake appears at about 20 fs (or 6 microns in path length) after the main, multi-spike pulse. By using a very short electron pulse, with a duration of about 6 fs, it is possible to overlap this seed with the electron bunch, while the broad-band part of the seed pulse does not overlap, because it precedes the electron bunch.

 Technically, this scheme is easy to implement. Thin diamond crystals (about 0.1 mm) diffract well and are resistant to the high thermal load of the FEL pulse. The crystal is set at the Bragg angle for the central energy of the SASE pulse, and reflects most of the X-rays in a narrow band out of the beam, but also creates the monochromatic tail. 
 The requirements on the chicane, which deviates the electrons so that they do not strike the diamond crystal, are much reduced, and it has three functions. It separates and recombines the electron and photon beams. It delays the electrons so that they now overlap the monochromatic wake of the photon beam; now the delay is of the order fs in the time domain or microns in path length. 
 The third function of the chicane is to smooth out the micro-bunching in the electron beam, which was created by lasing during the creation of the seed pulse. If this is not done, the microbunched electrons continue to lase in SASE mode in the second stage.

In the soft X-ray region, this scheme is more difficult to implement, because of a lack of tunable notch filters. For this reason the first implementation of self-seeding used a conventional monochromator, and the beam was seeded with monochromatic light \citep{Ratner2015}.
Soon after, an application was demonstrated in a near edge x-ray absorption fine structure experiment \citep{kroll2016}, where the bandwidth of SASE pulses is generally too large to allow accurate measurements.
This method of self-seeding has the drawback that a ``pedestal'' is created, that is, there is a broad background of other wavelengths. For experiments where only the monochromatic beam produces signal this is not problematic, but for experiments where the other wavelengths produce background signal, this is an issue \citep{Marcus2019, Hemsing2019, Hemsing2020}.

Since the development of self-seeding, the method has been applied at other light sources and
there have been various other schemes reported to further improve self-seeding \citep{Min2019,Emma2017,Inoue2019,Prat2018,Halavanau2019}.
These rapid developments mean that X-ray beams of narrower bandwidth, shorter duration and higher peak power are becoming available at an increasing number of SASE FELs. 

\subsection{Alternative light sources.}

\begin{figure}[ht]
    \centering
    \includegraphics{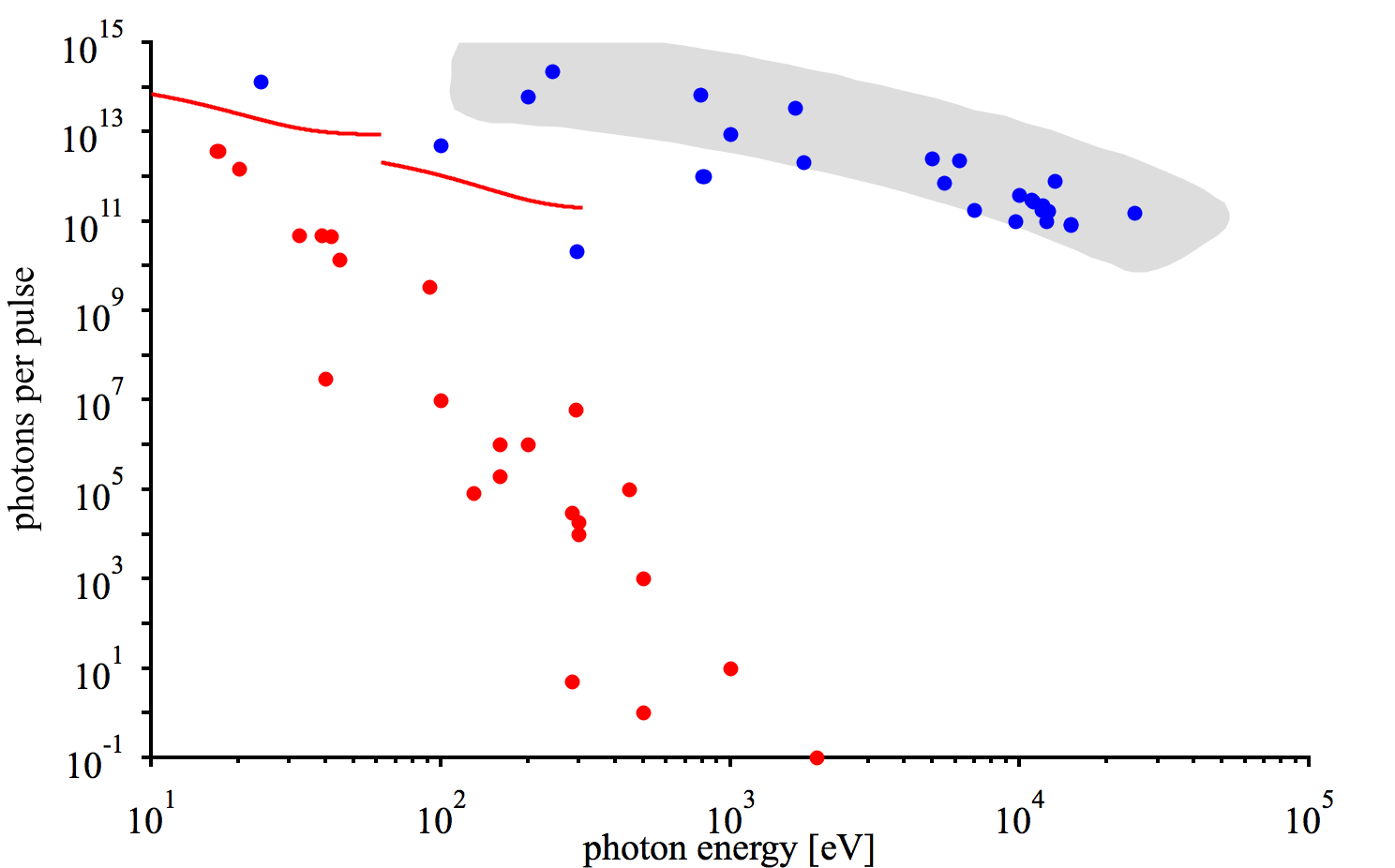}
    \caption{Typical numbers of photons per pulse for FERMI FELs 1 and 2 (red lines), soft and hard x-ray FELs (blue markers and laser driven HHG sources (red markers). The shaded area is derived from the data of \citet{Seddon2017}. The FEL data are from FLASH, LCLS, PAL-FEL, SACLA, SwissFEL, EuXFEL: references in Table~\ref{tab:facilities} or the respective webpages. Even for a fixed photon energy and machine, the actual pulse energies will vary over a range depending on machine parameters, pulse duration, and other factors. The HHG data are from~\cite{Chen2010,Cousin2014,Ding2014,Feng2020,Hong2012,Hong2014,Nayak2018,Rudawski2013,Seres2006,Takahashi2002,Teichmann2016,Wang2018,Zhang2018}, and aim to show the state-of-the-art for various ranges of wavelengths. Note that some of the vertical scatter in the data is due to the range of bandwidths used to report the data in the original work, and to the varying repetition rates. As a rule of thumb, for a given wavelength the energy per pulse scales inversely with the repetition rate: see for example \cite{Heyl2016}.}
    \label{fig:Seddon2017_f19}
\end{figure}

Other light sources which could be used for some of the experiments described here are HHG and plasma sources. Comparisons have been made recently by \citet{Seddon2017} and 
\citet{Schoenlein2019} of X-ray FELs and other short wavelength sources. At long wavelength (ca.~100 nm, 12 eV), laboratory sources can provide good numbers of photons per pulse, but this declines rapidly for shorter wavelengths (ca.~10 nm, 120 eV; Fig.~\ref{fig:Seddon2017_f19})

Optical lasers are based on the principle of creating a population inversion by pumping the lasing medium, and the upper state generally has a long lifetime. Thus the population can be built up over a period of time before lasing is initiated. It has long been a goal in photon science to create a coherent X-ray laser, but the technical difficulty is that all excited states at very short wavelength, which may potentially be used for lasing, have extremely short lifetimes, of the order of femtoseconds or less. Thus the population inversion cannot be built up over time by a weak source, but must be created in an extremely short time, i.e., a very intense source is required. 
This difficulty has now been overcome, adopting the approach of the old proverb ``set a thief to catch a thief'', that is, use a laser to pump a laser.

At a wavelength of 20 nm (62 eV) \citet{Matthews1985} used a laser pulse to excite a thin film of selenium, and \citet{Wang2008} used similar methods at 13.9 (89 eV) and 18.9 nm (65.6 eV). At a wavelength of 18.2 nm (68 eV), a laser-induced plasma column was used to create the necessary population inversion to sustain stimulated emission \citep{Suckewer1985}. 
These papers and others demonstrated the principles of the techniques, and have been reviewed \citep{Suckewer2009}. The light produced is significantly less bright than that of FELs, and being based on atomic resonances, it is not tunable. However these methods have the considerable advantages of being laboratory based and cheaper than large facilities. In some cases, for example when the signal is a linear function of the pulse energy, experiments are possible even with pJ pulse energies~\citep{Kleine2019,Bhattacherjee2017}

\section{\label{sec:optical}Optical experiments with FELs}

\subsection{\label{ssec:SASE_pumps}SASE FELs as pumps for 
fully coherent X-ray FELs}

The XUV laboratory lasers mentioned in the last section used optical lasers to pump a medium and create a population inversion, which then lased. It is also possible to use a SASE FEL to pump a medium which then lases in the soft or hard X-ray regions.
\citet{Rohringer2012} used intense soft X-ray pulses from the Linac Coherent Light Source to produce X-ray lasing at the wavelength of 1.46 nm (849 eV). The FEL pulses ionized Ne 1s core levels, which have a lifetime of 2.4 fs and normally undergo Auger decay as the main de-excitation mechanism, with fluorescence a much weaker channel. However the FEL pulses had sufficiently high intensity and short duration to create a  high enough population of core ionized states, providing the necessary population inversion and leading to lasing. 
The efficiency is far below 100\% so the reader may ask why one would create a much weaker laser pulse than the FEL pump pulse. The point is that the FEL is very intense, but is created by the SASE process; it is a laser if one defines the term to mean a device with a very large number of photons per unit phase space and time. It has poor longitudinal coherence, and the statistical properties of a chaotic source \citep{Gorobtsov2018}, whereas a true laser pulse is fully coherent, according to the definition of \citet{Glauber1963}, and among other things, has different statistical properties. The pulse created in the experiment of \citet{Rohringer2012} has narrow bandwidth, and is a laser pulse according to this definition.

Lasing has also been achieved at 0.15 nm (8 keV), using the SACLA FEL as the pump \citep{Yoneda2015}. In this case, FEL radiation was focused to a 120 nm spot on a 20 micron thick copper target. The pump radiation was bichromatic, consisting of a pump wavelength of 0.14 nm which ionized the Cu 1s shell, and a seed pulse tuned to Cu K$\alpha_1$ or Cu K$\alpha_2$. Thus the setup functioned as an amplifier rather than an oscillator. The radiation produced had a narrower line width, and higher density of photons in phase space than the FEL radiation used as pump.

\subsection{\label{ssec:Superfluorescence}Superfluorescence and related phenomena}

There have been optical experiments at FELs which involve coherence and population inversion, but do not result in lasing. \citet{Nagasono2011} and \citet{Harries2018} investigated  superfluorescence and other phenomena 
by performing experiments in which He was irradiated by strong SASE radiation.
In the first experiment, ground state He was excited to the 1s3p Rydberg state. This can decay spontaneously by fluorescence to the ground state, or to the 1s2s state by emitting a photon of wavelength 501.6~nm. If however the number density of excited states (denoted $\rho$) is sufficiently high, they interact and a new decay channel opens. This is termed superradiance, superfluorescence or collective spontaneous emission, and the condition for the process to occur is that the mean distance between atoms is less than the wavelength of the emitted radiation.
Superfluorescence of an excited state population is manifested with a characteristic delay between excitation and emission, is strongly directional, has an intensity proportional to $\rho^2$, and the emitted pulse has a temporal duration proportional to $\rho^{-1}$. All of these signatures were observed, confirming that the superfluorescence mechanism was operative.

\citet{Harries2018} irradiated He gas with light of central wavelength 24.3~nm (51.0~eV), again with SASE radiation. 
All photons ionize the gas, and those photons with a wavelength of exactly 24.302~nm excite the He$^{+}$~1s ions to He$^{+}$~4p states. These may decay, as above, by superfluorescence, but the authors observed other phenomena in addition. They assigned emission at 469 and 164~nm to cascade superfluorescence, at 30.4~nm to yoked superfluorescence, and at 25.6~nm to free induction decay.

These experiments were carried out using SASE light, which is of course not narrow bandwidth radiation. However the sample itself acted as a filter, as the resonances were very narrow; clearly narrow bandwidth radiation would make the experiments easier.

\section{\label{sec:Few_photon_single}Few photon, single ionization}
\subsection{\label{ssec:photo_intro}Brief introduction to photoionization}

In the previous Section, optical experiments with FEL light were described. In this and the following Sections, we describe experiments based on photoionization, one of the most important methods used in soft X-ray experiments. 
Photoionization (the photoeffect) dominates photon interaction with matter in the XUV and soft X-ray domain, because the cross sections are much greater than those of elastic and inelastic photon scattering.
The advantage of photoelectron spectroscopy, compared with other techniques, is that it is fast, and ``photographs'' the target on a time scale of a few attoseconds or femtoseconds. 
Other methods, such as ion spectroscopy or optical fluorescence, usually have time scales of picoseconds or nanoseconds, and so are much slower. The ultrafast nature of photoelectron spectroscopy is therefore very compatible with the femtosecond time structure of FELs.

In photoionization, one or more photons are absorbed by a target, and one or more electrons are ejected. The simplest case is of a single photon ejecting a single electron into the continuum, and 
Section \ref{sec:pump-probe}
describes pump-probe experiments, in which an excited target is singly photoionized by one photon. 
In the present Section we will discuss several examples of experiments on targets in their ground state, with emission of a single electron by one or a few photons.

In other experiments, absorption of two or more photons can lead to the emission of two or more electrons. Short wavelength FELs produce radiation which is very often above the first ionization potential of the sample being studied, and is intense. This leads to the situation that sequential ionization dominates ionization by simultaneous absorption of two photons. Some cases of sequential ionization are examined in Section~\ref{sec:Few_photon_sequential}.

An important phenomenon in photoionization is resonant photoemission. If an electron or electrons are excited to states in the continuum, and there is no resonance, the outgoing electron waves are described simply by angular momentum selection rules, and the cross section varies smoothly. If there is a resonant, or quasi-bound, state embedded in the continuum, the cross section may vary rapidly with photon energy and the state may be subject to autoionization. 
Resonances have been studied for their own intrinsic interest, and also to gain access to other phenomena: we discuss some examples in Section~\ref{sec:multi-photon-resonant}. 
Here we give a brief introduction to autoionization, the emission of an electron following resonant excitation. For historical reasons, the term autoionization has been mostly used to describe
phenomena at low energy, such as the decay of neutral, valence excited states. 
However autoionization includes Auger or Resonant Auger processes, in which an ion or neutral with a core hole decays; Interatomic Coulombic Decay, Sections~\ref{ssec:two_photon_ICD} and \ref{sec:multi-photon-resonant}, is also in this class.

Autoionization is among the most prominent manifestations of electronic correlation in atomic and molecular physics and indeed a seminal paper on autoionization by~\citet{Fano1961} has been cited more than eight thousand times.  The interference between direct photoionization and time-delayed decay of a quasi-bound state in the continuum strongly affects the cross section and angular distributions of photoelectrons. These states have been widely investigated in the spectral domain \citep{Beutler1935, Madden1963, Madden1969, Codling1971, Maeda1993, Sorensen1994, Domke1995, Domke1996}  
and temporal domain \citep{PRL-Chang-2010, Ott2013, Ott2014, Kotur2016, Gruson2016, Kaldun2016, Cirelli2018}. 

The excited state embedded in the continuum may have two electrons in excited orbitals, i.e., it is a doubly excited state. 
In the autoionization process, one of these electrons is emitted while the other is de-excited. In this case, initial state correlation plays an important role in the excitation of the system from the ground state to the autoionizing state by single photon absorption. 
Employing two-photon absorption, however, one can access such doubly excited states rather straightforwardly. Autoionizing doubly excited states in atoms, dimers and clusters accessed by two-photon absorption will be further discussed in Section~\ref{sec:Few_photon_sequential}.

Autoionization may also take place when an electron is promoted from an inner orbital (i.e., not from the highest occupied orbital) to an unoccupied Rydberg orbital by one-photon absorption. If the energy of the state is high enough, this may decay leaving a residual ion with a single hole.  
Here the initial state correlation in the ground state does not play an important role in the process of excitation to the autoionizing state. Window resonances of the type $n\mathrm{s} \rightarrow m \mathrm{p}$ in rare gas atoms are typical examples of this class of autoionization~\citep{Madden1969,Codling1971}. For example, in Ar these authors reported the $3\mathrm{s}^2 3\mathrm{p}^6 \rightarrow 3 \mathrm{s} 3\mathrm{p}^6 m \mathrm{p}$ resonances. 
In Section~\ref{two-dim-spec}, we will describe a novel technique to study such autoionization resonances, or dephasing time in general, employing a unique feature of FERMI, i.e, temporal coherence of multiple light pulses.

\subsection{\label{ssec:one colour two photon}One-colour, two-photon ionization}

Two-photon absorption leading to single ionization
of hydrogen and helium atoms may be the simplest single-color
multiphoton process, and 
such two photon processes have been extensively studied theoretically \citep {Ishikawa2010,Ishikawa2012a,Ishikawa2013}.
\citet{Ma_JPhysB_2013} investigated two-photon single ionization of He in the photon
energy range between 20 and 24 eV by
velocity map imaging electron spectroscopy at SCSS,
and discussed the coexistence of resonant versus direct
two-photon ionization pathways, following theoretical
predictions described by \citet{Ishikawa2012a}.

Using XUV radiation, two-photon above threshold ionization of the inner shell
has been realized at FLASH in Xe for ionization of the 4d shell in the region of the giant $4\mathrm{d} \rightarrow \varepsilon \mathrm{f}$ resonance~\citep{Richardson2010}. This process proceeds mainly
 along the path $4\mathrm{d} \rightarrow \varepsilon \mathrm{f} \rightarrow
\varepsilon \mathrm{g}$. Later this process was used to reveal a doublet structure of the
giant resonance in Xe~\citep{Mazza2015a,Chen2015}.

\begin{figure}[ht]
    \centering
    \includegraphics[width = 0.5\textwidth]{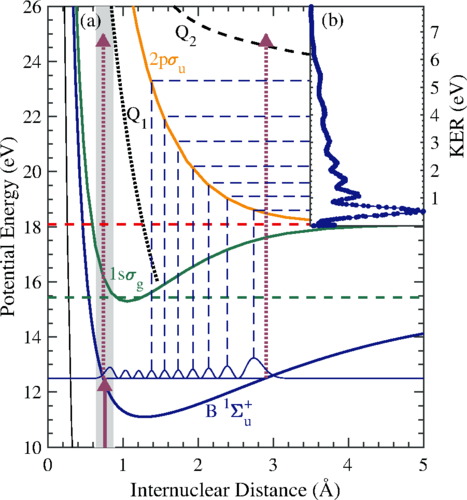}
    \caption{Potential energy curves of the singly and doubly excited states of the hydrogen molecule. The first photon excites the molecule to the  B$^1\Sigma_\mathrm{u}^+ (v =9)$ intermediate state. Grey shaded area: Franck-Condon region for the ground state geometry. The Q$_1$ and Q$_2$ states are the lowest doubly excited states of $^1\Sigma_\mathrm{g}^+$ symmetry. Inset (b) shows the measured Kinetic Energy Release spectrum for dissociative ionization via the $\mathrm{B}(v=9)$ state. Figure reprinted from \cite{Holzmeier2018}, with permission. Copyright (2018) by the American Physical Society.}
    \label{fig:Holzmeier}
\end{figure}

\citet{Holzmeier2018} employed resonant two-photon spectroscopy of H$_2$ to investigate nuclear dynamics in the ionization process.  
Molecular ionization is much more complicated than the atomic case, even for a system as simple as hydrogen, because of the additional nuclear degrees of freedom, especially the vibrational degrees. For this experiment, resolution and wavelength stability  sufficient to select single vibronic states were required, about 150 meV.
When hydrogen is ionized by a single photon, for example at about $\hbar\omega = 25.5$~eV, the main outcome is a stable molecular ion, and the probability of dissociative ionization (to $\mathrm{H}^{+} + \mathrm{H}$) is very low.  Fig.~\ref{fig:Holzmeier} shows that at the ground state internuclear distance, only continuum ionic states can be reached, see grey shaded region.
Absorption of two photons of energy $\hbar\omega/2$ implies the molecule absorbs the same amount of energy, but there is an important difference. If the photon energy $\hbar\omega/2$ excites a resonance, the two photons may be absorbed sequentially, rather than simultaneously.
The vibrational wavefunction of the intermediate state is much more extended than that of the ground state, as seen in Fig.~~\ref{fig:Holzmeier}.
The Franck-Condon region for transitions from the intermediate state to the final state is also at larger internuclear distance, so that different potential energy curves are probed.

This was what was predicted in the calculations and observed in the experiment  by \citet{Holzmeier2018}. Resonant excitation to the bound B$^1\Sigma_\mathrm{u}^+$ state causes an expansion of the vibrational wavefunction, and the expectation value of the internuclear distance is larger, Fig.~\ref{fig:Holzmeier}. The full and dotted red curves on the left indicate the situation for simultaneous absorption of two photons; the 2p$\pi_\mathrm{u}$ final state is not accessible. For sequential absorption, the transitions lying between the two dotted red arrows are possible.
The ionized 2p$\pi_\mathrm{u}$ final state then becomes accessible, Fig.~\ref{fig:Holzmeier}: this is a strongly repulsive state, and so the molecular ion can dissociate into a proton and a hydrogen atom.  Between the ground and the excited state, the internuclear distance changes, but this is not to be seen as classical vibrational motion. Rather, the wavefunction evolves to yield a larger expectation value of the bond length.

The photoelectrons and kinetic-energy resolved ions were detected and were interpreted with detailed calculations. By accessing this state, the ratio between dissociative and non-dissociative ionization was increased to values above~1, i.e., the two channels have comparable probability.

The time scale of this experiment was determined by the pulse duration, about 100 fs. The dynamics observed were thus intermediate between what is expected from attosecond/few femtosecond pulses~\citep{Palacios2006}, where the absorption of two photons is almost simultaneous, and high resolution experiments, where 
the coherence time is nearly infinite.

Detailed theoretical calculations by \cite{Holzmeier2018} revealed that it is important to take account of the doubly excited autoionizing states in order for theory to reproduce the experimental results quantitatively. 
In the experiment, the bound B$^1\Sigma^+_\mathrm{u}$ state was resonantly excited by a first photon. The second photon 
accesses not only the 2p$\pi_\mathrm{u}$ final state, but also the doubly excited states Q$_1$ and Q$_2$ in Fig.~\ref{fig:Holzmeier}, at certain internuclear distances. 
The doubly excited states thus populated dissociate along the repulsive potential curves and in parallel, autoionize mainly to a continuum state associated with the 1s$\sigma_\mathrm{g}$ ionic ground state above the dissociation limit, contributing to dissociative ionization. 
As a result of competition between dissociation and autoionization, which varies the partition between kinetic energy of the electron and the kinetic energy release of the ion, the kinetic energy of the electron is spread over a wide band of energy, see Fig.~4 of \citet{Holzmeier2018}. 
It is worth noting that the autoionization is the main contribution to dissociative ionization when the pulse duration is short ($\lesssim 10$~fs) and the range of nuclear distances is limited close to the Frank-Condon region of the neutral ground state~\citep{Palacios2006}.

\subsection{\label{ssec:XUV+NIR}XUV/X-ray + Near Infrared ionization and Circular Dichroism}

The photoionization of atomic and molecular systems in a field consisting of a (usually weak) XUV or X-ray pulse and a (usually stronger) infrared (IR)/visible pulse, is a widely studied topic.
One of the simplest examples is the use of an XUV pulse to promote an electron from an occupied orbital to an unoccupied Rydberg orbital and then use a near IR (800 nm) pulse to ionize it. The XUV pulse can be a laboratory laser based HHG source~\citep{Haber2009, Haber2010} synchrotron radiation~\citep{OKeeffe2013}, or FEL~\citep{Mondal2013, Mondal2014}. 
Interestingly, significant pulse delay effects in the photoelectron angular distribution of near threshold XUV + NIR two-photon ionization were found for Ne~\citep{Mondal2014} while there were no significant effects in He~\citep{Haber2009, Haber2010}. Later, these observations were reproduced by theoretical simulations~\citep{Ishikawa2014}.

Another example is the study of sidebands. Here an XUV pulse promotes a bound electron to the continuum, and simultaneous absorption or emission of an IR or visible photon by the outgoing photoelectron gives rise to sidebands in the electron spectrum.  Two-colour photoionization of He in XUV free-electron and visible laser fields was studied immediately after FLASH started two-colour experiments, and the absolute amplitude ratio of the s and d partial waves was extracted~\citep{Meyer2006, Meyer2010}. 
Sidebands can also be observed in the Auger spectrum for the case of overlapping IR/visible and X-ray pulses, known as laser-assisted Auger decay~\citep{Schins1994}. \citet{Meyer2012} investigated the fine details of laser-assisted KLL Auger decay following 1s photoionization of atomic Ne with few-femtosecond X-ray pulses from the LCLS.

Optical pulses can be combined with FEL pulses to provide a useful diagnostic to characterise the time structure of the FEL pulses. The sideband intensity as a function of delay between the pulses is the cross-correlation curve, and its width is equal to the convolution of the temporal widths of the two pulses. If the duration of the optical laser pulse is known, the FEL pulse duration can be extracted. This method is usually applied to FEL pulses of tens of femtoseconds or longer.

For shorter FEL pulses, the methods of attosecond science can be applied~\citep{Krausz2009}, which also involve the use of temporally overlapping optical and short wavelength pulses; see also Sections \ref{sec:ews} and \ref{sec:atto}. The usual experimental approach is to overlap an IR pulse with an XUV attosecond pulse or an attosecond pulse train, in order to characterize the XUV pulse~\citep{Paul2001, Goulielmakis2004} or to perform time-resolved investigations of electron dynamics, such as photoionization delay~\citep{Dahlstrom2012, Pazourek2015}.
In this case, attosecond synchronisation of the pulses is required, that is, phase-sensitive control.  Frequently employed approaches are photoelectron streaking~\citep{Paul2001,Pazourek2015} for an isolated attosecond pulse and RABBITT (Reconstruction of Attosecond Beating By Interference of Two-photon Transitions)~\citep{Muller2002} for attosecond pulse trains, see also the review of \citet{Dahlstrom2012}, and Section \ref{sec:atto}. 
Streaking has been employed for characterizing XFEL pulses~\citep{Helml2014}. See also \cite{Helml2017} for characterization of XFEL pulses by various methods, and Sections \ref{sec:ews} and \ref{sec:atto} for a more detailed description of these methods.

Circular dichroism (CD), i.e. the different response of a system to left and right circularly polarized radiation, has been widely investigated in the visible and infrared range~\citep{Barron2004,Starke2000} and also in the XUV and X-ray spectral region, exploiting the availability of circularly polarized pulses by synchrotron sources~\citep{GrumGrzhimailo2009}. The recent availability of FELs delivering femtosecond pulses with variable polarization offers the opportunity to extend these investigations to time-resolved studies. In photoionization CD, the measurement of the photoelectron spectra and the photoelectron angular distributions gives access to additional information on the photoionization pathways with the possibility to reconstruct the amplitude and (relative) phases of the single photoionization channels.

Circular dichroism may arise because the target is chiral (natural circular dichroism), or because the conditions of the experiment break the reflection symmetry of the target. Thus a system containing a totally symmetric atom in a monochromatic, circularly polarized light field is not dichroic, because reflection symmetry is preserved. However if a second circularly polarized field at a different wavelength is applied, the symmetry is broken, and dichroism may be detected. These are the often-used conditions for two-colour photoionization.

\citet{Mazza2014} exploited the effect to measure the degree of circular polarization of light from FERMI. Although the FEL produces light with a high degree of helicity, the transport optics may depolarize it to a certain extent, and it is challenging to measure the degree of polarization at the end-station. 
In this experiment, helium was irradiated simultaneously with circularly polarized light of photon energy 48.4 eV (25.6 nm) and IR light of energy 1.58 eV (784 nm), and two-photon emission was observed as sidebands.
The helicity of the IR was reversed, and the spectra re-measured; the dichroism was defined as the difference in intensity divided by the sum of intensities of the two situations. 
The dichroism was found to be $0.04 \pm 0.004$, corresponding to a degree of polarization of $0.95\pm 0.05$, in agreement with the estimated value, $0.92$, calculated from the optical constants of the transport mirrors. The importance of this work is not so much the physics, but the metrological application, providing a determination of the polarization at short wavelength.

Later, a detailed analysis of the photoelectron 
angular distribution and CD
in the two-colour XUV+IR above-threshold ionization of helium, both
experimental and theoretical, was presented by \citet{Mazza2016}.
In particular, the first ``complete
experiment'' of two-colour, two-photon above-threshold ionization
was realised, providing the partial s- and d-amplitudes, together
with their relative phase, of transitions between the continuum states. See also Section \ref{sec:complete experiments} for a detailed discussion of a single-colour, two-photon complete experiment.
The analysis covered also the second and third sidebands for higher
intensities of the IR light. 
A short overview of early studies of the 
dichroism of the sidebands, generated by two-color XUV FEL + IR ionization is given in \citet{Mazza2015}.

\subsection{\label{ssec:OAM}Orbital Angular Momentum}

Photons have spin angular momentum of $\pm\hbar$ depending on the polarization of the light. They may also possess orbital angular momentum, which depends on the spatial distribution of the phase of the wavefront. For example, ``twisted light'' can be created by passing a plane wave through a phase mask which transforms the wavefront into a vortex, where the phase varies in a spiral manner around the axis. 
While absorption of twisted light by a trapped atom had been previously observed~\citep{Schmiegelow2016,Afanasev2018}, an effect in the photoelectron angular distribution of an extended sample was not observed, and it was argued that both a well defined phase singularity \emph{and} the localization of sample atoms near the singularity are required to selectively observe nondipole transitions~\citep{Kaneyasu2017}. 
In a recent experiment \citep{DeNinno2020} the interaction of twisted light with atoms was investigated. He atoms were two-photon ionized by a Near Infrared photon and an XUV photon to generate sidebands (see Section \ref{ssec:XUV+NIR}). The IR field was circularly polarized and carried orbital angular momentum, while the XUV photon was circularly polarized. Dichroic effects were observed on changing the sign of the orbital angular momentum, confirming that effects are in fact observable. Theoretical analysis showed that the effect can be traced back to non-dipole transitions that would be optically forbidden for a gaussian beam and become allowed when the orbital angular momentum is taken into account.

\subsection{\label{two-dim-spec}Towards two-dimensional XUV spectroscopy}

A large number of experiments have been performed with optical lasers using phase-coherent, single-colour pulses, where the phase between two pulses separated in time is varied and the signal measured. 
It is a form of coherent control, with various names such as pump-dump spectroscopy or ``wave packet dancing'' \citep{Kosloff1989}. The name pump-dump arises from the simple interpretation that the first pulse pumps the target to an excited state, and if the second pulse is in antiphase, it pumps the target back down to the ground state. 
If it is in phase, it pumps the system further, that is, it increases the excited state population. The method is however much more sophisticated than this simple view suggests, and was used initially to control photochemical processes and yield a desired end-product \citep{Rice1992, Gordon1997}. Furthermore, adding one more pulse as a probe or read-out, one can read off the phase of the wavepacket created by a sequence of the two pulses that are phase-controlled~\citep{Ohmori2006}.  
This branch of optical laser spectroscopy has evolved into multi-dimensional spectroscopy, with sequences of pulses, and control of parameters such as wavelength, time delay, chirp, etc.~\citep{Jonas2003, Goswami2003, Tan2008}, The methods were originally developed in nuclear magnetic resonance~\citep{Aue1976,Ernst1990}, and adapted to lasers.

As a first step towards multi-dimensional spectroscopy with FELs,
phase-coherent double pulses have been generated at FERMI~\citep{Gauthier2016a}. The method is in principle simple: FERMI is a seeded FEL, so it is sufficient to duplicate the optical seed pulse, and seed with two identical pulses separated in time. 
In reality, it is not easy to maintain the required attosecond-scale phase stability between the two pulses. \citet{Gauthier2016a} used a birefringent crystal to split the seed pulse. 
The thickness of the crystal determined the coarse delay, and the phase was tuned by rotating the crystal, producing small path length differences between the two pulses on the nm scale. 
%\red{
Phase-coherent double-pulse generation has also been 
demonstrated at the XUV FEL facility FLASH~\citep{Usenko2017, Ding2019} using a split-and-delay unit based on diffractive optics. In contrast to the above scheme, single pulses are generated, and then split and delayed, rather than generating two phase-coherent pulses in the accelerator.
\cite{Hikosaka2019} and \cite{Kaneyasu2019} claim to have demonstrated double pulse, coherent control using the synchrotron radiation facility UV-SOR. They used two undulators with a phase shifter between them. 

At FERMI, \citet{Wituschek2020} built on the approach of~\cite{Gauthier2016a}, by introducing
a monolithic Mach-Zehnder beam splitter. This showed improved temporal resolution and much improved flexibility, as the phase could be scanned over a range of 1 ps (limited only by the length and uniformity of the electron bunch), which was not possible using the birefringent crystals described above. 
Most importantly, phase-locked detection was implemented, with modulation provided by acousto-optic modulators, acting on the seed pulses. Since FERMI acts as a wavelength shifter and amplifier, the phase information was imprinted on all harmonics of the seed.

\begin{figure}[ht]
    \centering
    \includegraphics[width = 0.9\textwidth]{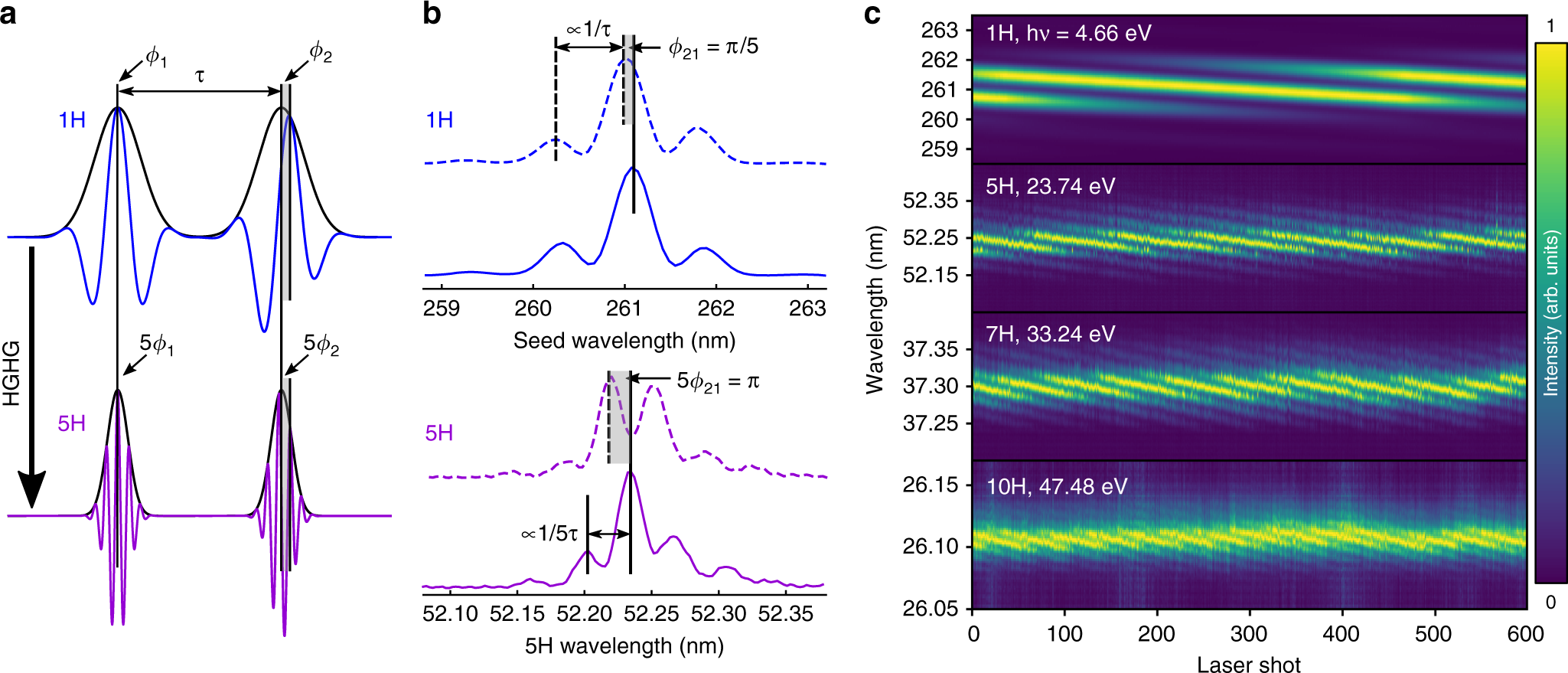}
    \caption{(a) Scheme of the timing and phase control of the XUV pulse pair (violet) by manipulation of the seed pulse (blue). Control of the pulse delay $\tau$ and the phase difference $\phi_{21} = \phi_2-\phi_1$ are decoupled. (b) Example  of interference fringes in the first harmonic and fifth harmonic spectra for fixed delay $\tau = 250$~fs, and two values of phase $\phi_{21}$. The fringe spacing is inversely proportional to the delay $\tau$ and the fringe phase is proportional to $\phi_{21}$. (c) Ramsey fringes for the first, fifth, seventh and tenth harmonics. $\phi_{21}$ was incremented by 15 mrad steps between each laser shot. The data are intensity-normalised single-shot spectra with no additional processing. Reproduced from \cite{Wituschek2020}; copyright 2020 by the Authors. The original figure has been published under a Creative Commons Attribution 4.0 license (CC BY) \url{http://creativecommons.org/licenses/by/4.0/}}
    \label{fig:Wituscheck_2}
\end{figure}

When two coherent and collinear pulses of the same wavelength are separated in time, they interfere and this is visible in their spectrum (Figure~\ref{fig:Wituscheck_2}). If they are in phase, there is a maximum at the central wavelength, and the fringe spacing is inversely proportional to the delay between the pulses. 
If they are in anti-phase, the spacing is the same, but there is a minimum at the central wavelength. This is another way of looking at the pump-dump interpretation: in-phase pulses have a maximum at the central wavelength, and so may excite a resonance at that wavelength. 
In antiphase, there is no net intensity, so the resonance is not excited. 

The frequency domain picture gives an immediate and intuitive view of why at the end of a double pulse sequence, there is no excited state population (pump-dump): the excitation frequency was not present in the Fourier transform of the two light pulses. However it does not explain intuitively the effect of dynamics, or transient population of states, in which case the time domain picture may give a more intuitive description. 
For a closed quantum system (a few quantum states without a bath), such as the above example, the ``pump-dump'' time domain description or the ``no absorption'' frequency domain description are both correct and their relation is just a Fourier transform. 

\citet{Wituschek2020} demonstrated the method for the case of the argon $3\mathrm{s} \rightarrow 6 \mathrm{p}$ resonance, which is embedded in the continuum and decays by autoionization on a femtosecond time scale. 
The lifetime was already known from high resolution absorption spectra~\citep{Sorensen1994}. 
However the new detection scheme measures both the real and imaginary parts of the susceptibility, whereas absorption provides only the imaginary part. 
This proof of principle experiment paves the way for full development of multi-dimensional spectroscopy, where a sequence of pulses is used~\citep{Jonas2003}. .  
It promises to reveal phenomena, such as corelation between different states (peaks in the absorption spectrum), which cannot be observed in conventional one-dimensional spectroscopy (absorption spectroscopy).

\subsection{\label{sec:cc_Ne}Coherent control of one-colour ionization via biharmonic ionization} 

Historically the first experiment which demonstrated coherent control in the XUV domain was performed in 2016~\citep{Prince2016} at the seeded FEL FERMI
exploiting the longitudinal coherence of the radiation, which is not present in radiation from SASE FELs.
The theoretically predicted phase correlation between two different XUV beams was demonstrated for the fundamental frequency and its second harmonic. By varying the relative phase of the two fields, the angular distribution of photoelectrons ejected from the neon atomic target was controlled. The relative phase was, in turn, controlled using an accelerator physics method, so that the femtosecond pulse of one of the harmonics relative to the other was delayed with attosecond precision. 
For pulses of $\sim$100~fs duration, the time shift of the envelopes on the scale of a few attoseconds is not important, while the relative phase of the two XUV fields was controlled with a resolution of 3 as time delay. As mentioned at the end of Section~\ref{ssec:seeded}, the delay was provided by an electron delay line, which shifts in time the electron bunch generating the second colour. 
It is very problematic to control phase at short 
wavelengths, especially that of different harmonics, using laboratory optics methods, due to a lack of the required optical elements and strong absorption of the XUV radiation in matter. The need for optical elements to control the light is avoided by controlling the electron beam, and therefore the light. 
This is in contrast to physically equivalent experiments in the optical domain where variable pressure gas cells~\citep{Baranova1992,Yin1992,Wang2001} or a rotating transparent plate~\citep{Yamazaki2007} were used as phase shifters. 

\begin{figure}[ht]
    \centering
    \includegraphics[width=0.8\textwidth] {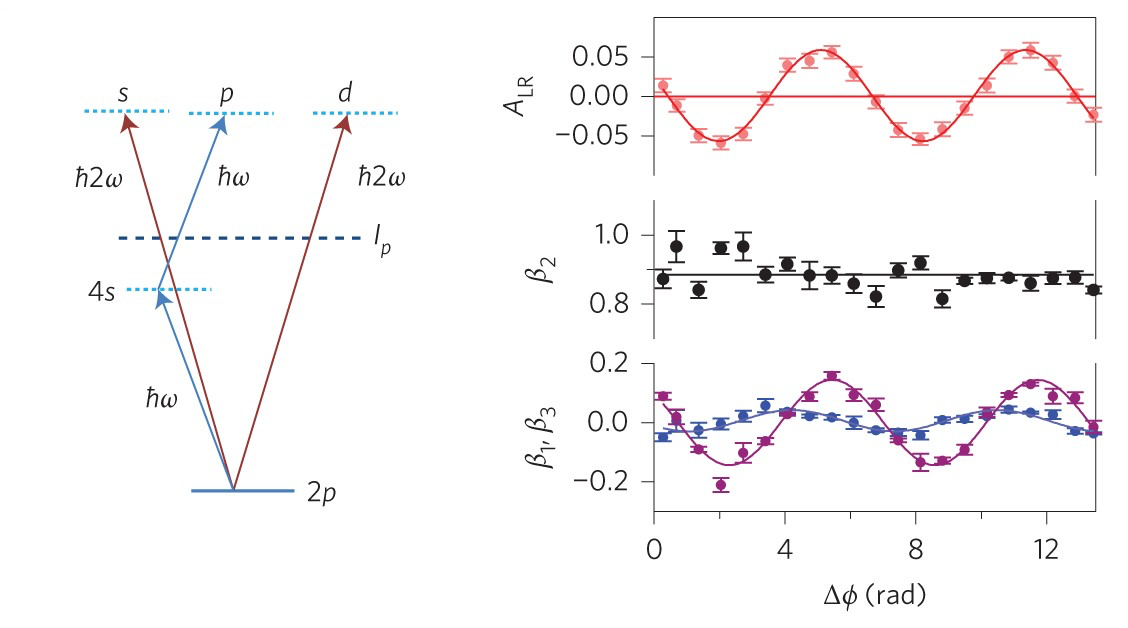}
    \caption{Left: Level diagram for coherent control of neon with short wavelength radiation. Single-photon (second harmonic) ionization leads to emission of s and d partial waves. Two-photon (fundamental) ionization via the 4s resonance leads to emission of p partial waves. The amplitude of f partial waves is negligible compared with the resonant path. Right: Asymmetry parameter $A_{\mathrm{LR}}$ and asymmetry parameters $\beta_1$ (blue), $\beta_3$ (magenta) and $\beta_2$ (black) as a function of phase. Figure reprinted from~\citet{Prince2016}, with permission. Copyright (2016) by the Authors.}
    \label{fig:CControl1}
\end{figure}

The fundamental frequency in the FERMI experiment corresponded to the $2\mathrm{p}^6 \rightarrow 2\mathrm{p}^5(^2\mathrm{P}_{3/2}^\circ)4\mathrm{s}$ resonance of Ne at 62.97~nm, selected to enhance the two-photon ionization branch and thus to provide stronger interference between one- and two-photon ionization, Fig.~\ref{fig:CControl1}. 
The collinear beams of both harmonics were linearly polarized along the same direction. The photoelectron angular distributions were measured by a VMI (velocity map imaging) spectrometer. 

To describe theoretically the effect of varying the relative phase of the harmonics~\citep{Douguet2017,Gryzlova2018}, the electric field may be written in the form
\begin{equation} \label{eq:nefield}
    E(t) = F(t) \, \left[ \cos \omega t + \eta \cos (2\omega t + \phi) \right]
\end{equation}
where $F(t)$ is the pulse envelope, $\eta$ is the amplitude ratio of the fundamental and second harmonic.
In the lowest non-vanishing order of the perturbation theory (see Section~\ref{ssec:PT}), the differential ionization probability for an initial atom with zero angular momentum ($J_\mathrm{i}=0$) can be expressed as
\begin{equation} \label{eq:neon1}
    \frac{d W}{d \Omega} \equiv W(\vartheta, \varphi) = \sum_{M_f} \Big| \, \eta \, \ME{f}{T}{i}_1 + \ME{f}{T}{i}_2 \Big|^2
\end{equation}
where the first and the second order amplitudes are given (in the dipole approximation) by Eqs.~(\ref{eq:pt1t}, \ref{eq:pt2t}), Section \ref{ssec:PT}, $M_\mathrm{f}$ is the magnetic quantum number of the residual ion (with total electronic angular momentum $J_\mathrm{f}$). Eq.~(\ref{eq:neon1}) can be written as a sum of three terms,
\begin{equation} \label{eq:neon2}
W(\vartheta, \varphi)  =  W^{(1)}(\vartheta, \varphi) + W^{(2)}(\vartheta, \varphi) + W^{(12)}(\vartheta, \varphi)  \,,
\end{equation}
with the axial symmetric angular dependence described by the sum of Legendre polynomials $P_k(\cos \vartheta)$:
\begin{equation} \label{eq:neon3}
W(\vartheta, \varphi) =  \frac{W_0}{4 \pi} \left( 1+ \sum_{k=1}^4 \beta_k P_k(\cos \vartheta) \right)
\end{equation}
Here the first and the second terms in Eq.~(\ref{eq:neon2}) are the ionization rates due to the (two-photon) first and (single-photon) second harmonics, respectively, 
while the third term is due to the interference between the two paths; the angle $\vartheta$ is measured from the direction of polarization.
The interference term in Eq.~(\ref{eq:neon2}) gives rise to odd Legendre polynomials ($k=1,\, 3$) in Eq.~(\ref{eq:neon3}), due to the opposite parities of the final states in one- and two-photon ionization. As a result, a ``forward-backward'' asymmetry of the angular distribution Eq.~(\ref{eq:neon3}) along the direction of the polarization vector appears, Fig.~\ref{fig:CControl1}. 

This asymmetry has an oscillatory dependence on the relative phase of the harmonics $\phi$ and is quantified by the value 
\begin{equation} \label{eq:neon4}
A_\mathrm{LR} = \frac{W(\vartheta=0) - W(\vartheta = \pi)}{W(\vartheta=0) + W(\vartheta = \pi)} = 
\frac{\sum_{k=1,3} \beta_k}{1+ \sum_{k=2,4} \beta_k} = A_m \cos \, (\phi - \phi_m)
\end{equation}
where $A_m$ = the amplitude of the oscillations  and  $\phi_m$ = the phase at which the asymmetry reaches its maximum value. 
The anisotropy parameters $\beta_1$ and $\beta_3$ are described by similar oscillations with their own amplitudes and phases analogous respectively to $A_m$ and $\phi_m$. 
Such oscillations were indeed observed in the experiment~\citep{Prince2016} and were confirmation of the mutual coherence of the two XUV harmonics and coherent control of the angular distribution. 
Theoretical calculations~\citep{Gryzlova2018} were performed in the lowest non-vanishing order perturbation theory, taking into account the neighbouring excited states $2\mathrm{p}^5 4\mathrm{s}, \, 3\mathrm{d}$ of Ne, including their fine structure, and by solving the TDSE on the space-time grid (see Section~\ref{ssec:Grid}) with a local potential and the perturbation theory, in the limit of infinite pulse duration. 
The resonance behaviour of the $\beta_k$ parameters and the parameters $A_m$ and $\phi_m$ of the asymmetry, Eq.~(\ref{eq:neon4}), as a function of the XUV photon frequency was revealed. A similar theoretical study of the coherent control for the region of the Ne$(2\mathrm{p}^5 3\mathrm{s}\,^1\mathrm{P})$ state~\citep{Douguet2017} and before that, for the  H(2p) state~\citep{Grum2015}, were performed.

Some theoretical predictions for the same experiment on neon with two circularly polarized harmonics are available~\citep{Gryzlova2019}. In this case, the 
coherent control of the photoelectron angular distribution manifests itself through a change from a one-lobe to a three-lobe shape, for co- and counter-rotating harmonics respectively, a variation of the polar asymmetry with the light frequency, and a rotation of the distributions around the direction of the beam depending on the relative phase between the harmonics. The controlled parameters change sharply when the fundamental frequency passes through an intermediate resonance. These effects were first predicted for the hydrogen atom~\citep{Douguet2016}.

\subsection{\label{sec:DiFraia}Experimental determination of optical phase}

The experiment described in Section~\ref{sec:cc_Ne} was carried out at a single fundamental wavelength and its second harmonic, and while their $\it{relative}$ phase was controlled, the $\it{absolute}$ optical phase difference was unknown. Contrary to optical experiments, the characterization of this phase is not straightforward, because of the lack of suitable optical materials. 
To extend bichromatic coherent control to other applications, it would be useful to know the absolute optical phase at any chosen wavelength. This challenge was met by \citet{DiFraia2019} who used He atoms as a nonlinear mixer, and measured the absolute phase relationships of fundamental wavelengths of FERMI FEL-1 and their second harmonics. 
They used bichromatic light, and measured the interference for single ionization of He in the photon energy range 14.3 to 19.1 eV (i.e., below the first 1s to 2p excitation) as a function of relative optical phase.
The measurement exploits the asymmetry induced in the angle-resolved photoelectron distribution, and results in a system of equations that becomes solvable if the value of one of the scattering phase shifts is known. 
For He, scattering phase shifts over a wide range of photoelectron energies are available in the literature. The absolute phase value was measured at three photon energies (14.3, 15.9, 19.1 eV) with an error of 0.01 -- 0.04 rad. 
An advantage of the method is that it also provides the effective relative photon intensities relevant for a bichromatic process, which is useful for the case of imperfect experimental conditions.  The method is also of interest because it can potentially be extended to other s shell electrons (e.g., C$_\mathrm{1s}$) at much shorter wavelengths.

\subsection{\label{sec:ews}Photoemission time delay}
In Section~\ref{ssec:Amplitudes} below, we give an expression for the Eisenbud-Wigner-Smith (EWS) time delay~\citep{Eisenbud1948,Wigner1955,Smith1960}, Eq. (\ref{eq:ews}).
This is the delay (with respect to a freely propagating electron) that a photoelectron experiences when exiting the potential of an atom or molecule, and it may depend on both the kinetic energy and emission direction of the electron. 
It can be shown that under appropriate circumstances, the derivative with respect to energy of the phase of the photoionization amplitude, $\partial \eta(\epsilon) / \partial \epsilon$, is equal to the EWS time delay.
Currently, the main methods applied to measure these time delays are based on laboratory ultrafast lasers, and the application of two techniques, attosecond streaking and RABBITT (Reconstruction of Attosecond Beating By Interference of Two-photon Transitions; see also Section~\ref{sec:atto}).
Usually two photoelectron signals are measured simultaneously, for example from two different atoms, or from two different levels of the same atom, so that the difference in time delay $\Delta \tau$, or the corresponding difference in phase $\Delta\eta$, is determined. 
Thus in attosecond streaking and RABBITT, $\partial \Delta\eta(\epsilon) / \partial \epsilon$ is measured.
Both of these methods depend on the presence of an infrared pulse which is synchronised to an attosecond XUV pulse (streaking) or to an attosecond pulse train (RABBITT) with attosecond precision, and require correction due to effects of the infrared pulse.

In Section~\ref{sec:cc_Ne}, we discussed the application of fully coherent FEL light to control the emission of photoelectrons. Extremely high phase resolution corresponding to a few attoseconds was demonstrated, and the question immediately arises as to whether this exquisite control can be used to measure phenomena on this time scale. The first experiment aiming to do so has been carried out recently~\citep{You2020b} on a Ne target.

The method is based on the technique described above, namely the control of the phase difference between a fundamental and its second harmonic, and measurement of the PADs (photoelectron angular distributions) as a function of the phase difference between the two wavelengths at two or more kinetic energies of the photoelectron. From these measurements, intensities as a function of phase for a series of angles were derived; the relative phase between the two-photon and the single-photon ionization channels was extracted; and $\partial \Delta\eta(\epsilon) / \partial \epsilon$ was calculated in the finite-difference approximation.
For the ionization of an $n$s state, such as He 1s, this quantity is identified with the group delay of the photoelectron wave packet, as in the case of the EWS delay. Ionization of an $n$s state by linearly-polarized bichromatic light is a relatively simple case, as there are only three outgoing partial waves, a p wave due to single-photon ionization, and $\mathrm{s} + \mathrm{d}$ waves, due to two-photon ionization; these interfering partial waves all have magnetic quantum number $m=0$.

The situation is more complicated for the ionization of an $n$p orbital, as in the experiment on Ne 2p~\citep{You2020b}. There are four outgoing partial waves: $\mathrm{s} + \mathrm{d}$ from single-photon ionization, and $\mathrm{p} + \mathrm{f}$ from two-photon ionization. As well, the outgoing waves may have magnetic quantum number $m=-1, 0, 1$. Waves with the same value of $m$ interfere, and the oscillations as a function of the relative phase between the fundamental and second harmonic then add incoherently, i.e., in terms of intensity rather than amplitude, for the three values of $m$.
Nevertheless, clear oscillations were observed, but the interpretation is more complicated than for the case of s initial states. What is clear is that an average phase of photoionization amplitudes can be measured, and this is an important quantum mechanical quantity.

The derivative of the phase with respect to energy was taken, but the interpretation in terms of a photoemission time delay is more subtle than the case of He. A {\it generalized delay} for multiple wave packets was proposed, defined as
\begin{equation}
	\Delta \tilde \tau = \frac{\partial \Delta \tilde \eta \left( \bar{\epsilon} \right)}{\partial \tilde{\epsilon}},
	\label{eq:tilde-tau}
\end{equation}
where the phase difference of photoionization amplitudes was averaged (in the sense of sum of phasors) over $m$ and then the resulting $\Delta \tilde \eta$ was differentiated with respect to the average photoelectron energy, $\tilde{\epsilon}$ (where ${\tilde \epsilon}$ rather than $\epsilon$ takes account of the finite bandwidth of the electron wave packet.) This is in contrast to the definition of~\cite{Smith1960}, where the scattering phase of each wave packet is first 

\subsection{\label{sec:atto}Generation and characterization of attosecond pulse trains}

The demonstration of coherence between two different harmonics of the FEL described in the previous Sections and also in \cite{Prince2016}, naturally raises the question whether multiple harmonics of the seed wavelength are expected to be coherent. For the typical configuration of FERMI with an ultraviolet  seed laser
of angular frequency $\omega_\mathrm{UV}$, the coherent superposition of two harmonics ($q_1$ and $q_2$) would lead to beating in the temporal domain, and to an intensity profile $I(t)$ characterized by an attosecond time structure, which repeats itself periodically with the period between zeroes $T=1/[(q_2-q_1)\omega_\mathrm{UV}]$. For consecutive (even and odd) harmonics of the seed laser this period is $\simeq 880$ as.

\begin{figure}[ht]
    \centering
    \includegraphics[width=100mm]{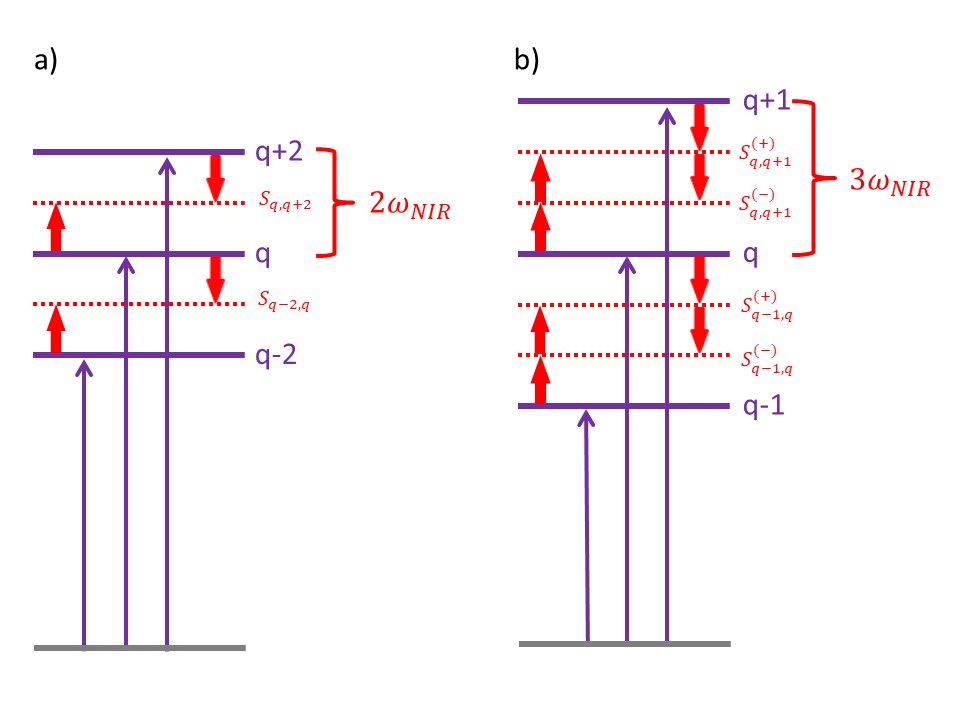}
    \caption{a) RABBITT method implemented for the temporal characterization of the relative phase between the consecutive odd harmonics $q-2$, $q$, and $q+2$. The energy difference between consecutive main photoelectron peaks is $2\omega_\mathrm{NIR}$. One-NIR-photon transitions give rise to sidebands of the main peaks $S_{q-2,q}$ and $S_{q,q+2}$, whose intensity depends on the interference between two different pathways.
    b) Two-colour photoionization scheme for the characterization of the phase difference between consecutive harmonics ($q-1$, $q$, and $q+1$) of the seeded FEL FERMI. The energy difference between consecutive main photoelectron peaks is $3\omega_\mathrm{NIR}$. One- and two-NIR-photon transitions give rise to the sidebands of the main peaks 
    $S^{(\pm)}_{q-1,q}$ and $S^{(\pm)}_{q,q+1}$,
    whose intensity depends on the interference between two different pathways.
    }
    \label{Figure_sideband}
\end{figure}

Attosecond pulse trains generated by high-order harmonic generation usually consist of the odd harmonics of the fundamental near-infrared driving field with angular frequency $\omega_\mathrm{NIR}$. The characterization of the relative phase between the harmonics of the fundamental radiation is accomplished through a cross-correlation between the pulse train and a synchronized infrared pulse. 
The signal is the photoelectron spectrum generated by the two-colour fields, which, besides the photoelectrons corresponding to the absorption of a single XUV photon, contains additional peaks due to the absorption (emission) of one IR photon, indicated as $S_{q-2,q}$ and $S_{q,q+2}$ in Fig. \ref{Figure_sideband}a. These additional peaks are usually named sidebands and their intensity depends on the relative phase between the harmonics and on the relative delay $\tau$ between the XUV and IR fields according to the relation:

\begin{equation}\label{Eq_sidebands}
S_{q,q+2}(\tau)\propto\cos(\varphi_{q+2}-\varphi_{q}+\Delta\varphi_\mathrm{at}+2\omega_\mathrm{NIR}\tau),
\end{equation}
where $\varphi_q$ and $\varphi_{q+2}$ indicate the phases of the two consecutive, odd harmonics $q$ and $q+2$, respectively, and $\Delta\varphi_{at}$ indicates the difference of the additional phases accumulated along the two ionization pathways by the photoelectron wave packet due to the bound-continuum  and continuum-continuum transitions.
By measuring the variation of the sideband intensities as a function of the relative delay (which can be controlled with sub-fs accuracy in HHG-based experimental setups), the phase difference between consecutive odd harmonics can be extracted and the attosecond waveform can be reconstructed \citep{Paul2001,Muller2002}. The unknown phase term $\Delta\varphi_\mathrm{at}$ is derived from theoretical calculations. This reconstruction method is usually known as Reconstruction of Attosecond Bursts By Interference of Two-photon Transitions or RABBITT \citep{Muller2002}, see also Section \ref{sec:ews}.

The harmonics of the seeded FEL FERMI are the even and odd harmonics of the seed laser pulse, obtained by frequency tripling of the output of either a Ti:Sa system or of an optical parametric amplifier (OPA)~\citep{Danailov2011,Allaria2012a}.
In contrast to the HHG-based case, one-NIR-photon transitions are not sufficient to create two interfering pathways contributing to the same sideband.
The interference between two different pathways encoding the relative phase between the harmonics needs (at least) two-NIR-photon transitions, as can be seen in Fig.~\ref{Figure_sideband}b. For example, the sideband $S^{(-)}_{q,q+1}$ is populated either by the absorption of a photon of the harmonic $q$ and subsequent absorption of one NIR photon, or by the absorption of a photon of the harmonic $q+1$ and emission of two NIR photons.

Multiple sidebands of the fundamental photoelectron peaks were observed in the interaction of a single harmonic and a NIR field \citep{Mazza2014}, thus indicating the possibility to use the fundamental laser ($\omega_\mathrm{NIR}$) for cross-correlation measurements.
Simulations based on the strong-field-approximation and solution of the TDSE in the single-active electron approximation indicate that, also in the case of multiple NIR photon transitions, the sideband intensity oscillates depending on the relative phase between the harmonics($\varphi_{q+1}-\varphi_{q}$) and the temporal delay $\tau$ between the fields according to the relation:
\begin{equation}\label{eq:sidebdand_3omega}
S^{(\pm)}_{q,q+1}=a^{(\pm)}_{q,q+1}\pm b^{(\pm)}_{q,q+1}\cos(\varphi_{q+1}-\varphi_{q}+\Delta\varphi_{at}+3\omega_\mathrm{NIR}\tau),
\end{equation}
where the coefficient $a^{(\pm)}_{q,q+1}$ and $b^{(\pm)}_{q,q+1}$ on the photoelectron energy and on the intensity of the NIR pulse.
Also in this case, the relative phase between the harmonics can be extracted by the oscillations of the sidebands as a function of the relative delay.

This approach, however, cannot be directly applied to the measurement of the relative phase between consecutive harmonics of FERMI, due to the lack of sub-femtosecond synchronization between the XUV waveform and the NIR lasers. For SASE FELs, the timing jitter can be as large as ten to a hundred femtoseconds~\citep{Coffee2019}. For FERMI the timing jitter is of the order of a few femtoseconds~\citep{Finetti2017} and is, therefore, still too large for observing the sub-cycle intensity dependence of the sidebands.

Information about the relative synchronization between the harmonics can still be retrieved by correlating the intensities of different sidebands measured shot-by-shot. Indeed, using Eq. (\ref{eq:sidebdand_3omega}) we observe that the curve representing the variation of the intensity of the sideband $S^{(+)}_{q-1,q}$ ($S^{(-)}_{q-1,q}$) with respect to $S^{(+)}_{q,q+1}$ ($S^{(-)}_{q,q+1}$) is an ellipse, whose shape depends on the phase difference:
\begin{equation}
\label{eq:phase_diff}
    \Delta\varphi_{q-1,q,q+1}=\varphi_{q+1}+\varphi_{q-1}-2\varphi_{q}
\end{equation}\\
which can be regarded as the difference between the differences of the phases $\varphi_{q-1},\varphi_{q},\varphi_{q+1}$ of three consecutive harmonics $q-1,q,q+1$. The phase difference $\Delta\varphi_{q-1,q,q+1}$ is proportional to the group-delay dispersion of the phase $\varphi(\omega)$ of the harmonic spectrum:
\begin{equation}\label{Eq_GDD}
    \mathit{GDD}(\omega)=\frac{d^2\varphi}{d\omega^2}\simeq\frac{\Delta\varphi_{q-1,q,q+1}}{\omega_\mathrm{UV}^2},
\end{equation}
where $\omega_\mathrm{UV}$ indicates the angular frequency of the ultraviolet seed laser. 
In particular, if the three harmonics are in phase (i.e. $\varphi_{q+1}-\varphi_{q}=\varphi_{q}-\varphi_{q-1}$), the phase difference $\Delta\varphi_{q+1,q,q-1}$ is zero and the ellipse reduces to a line.
Therefore, the shape of the correlation plots reveals the relative phase between the group of three harmonics.
By using this approach, the total electric field, which depends only on these phase differences (apart from an overall time shift), can then be reconstructed.

This approach was recently demonstrated at FERMI for the temporal characterization of attosecond pulse trains composed of three and four harmonics \citep{Maroju2020}. In the three-harmonics scheme, each harmonic was generated by two undulators.
The relative phase between the harmonics was changed by delaying the electron bunch using the phase shifters between each pair of undulators.
The characterization of the phases between the harmonics was based on the single-shot correlation analysis of the sideband variations. In order to minimize the effect of intensity fluctuations of the single harmonics, the oscillating component of the sideband was isolated by considering the quantity:
\begin{equation}
\label{eq:sideband_parameter}
    P_{q,q,+1}=\frac{S^{(+)}_{q,q+1}-S^{(-)}_{q,q+1}}{S^{(+)}_{q,q+1}+S^{(-)}_{q,q+1}}
\end{equation}
It can be easily observed from Eq.~\ref{eq:sidebdand_3omega}, that, under the approximation $a^{(+)}_{q,q+1}\simeq a^{(-)}_{q,q+1}$) and $b^{(+)}_{q,q+1}\simeq b^{(-)}_{q,q+1}$ also the correlation plots $(P_{q-1,q},P_{q,q+1})$ are described by ellipses, whose shape depends on $\Delta\varphi_{q-1,q,q+1}$.

\begin{figure}[ht]
    \centering
    \includegraphics[width=150mm]{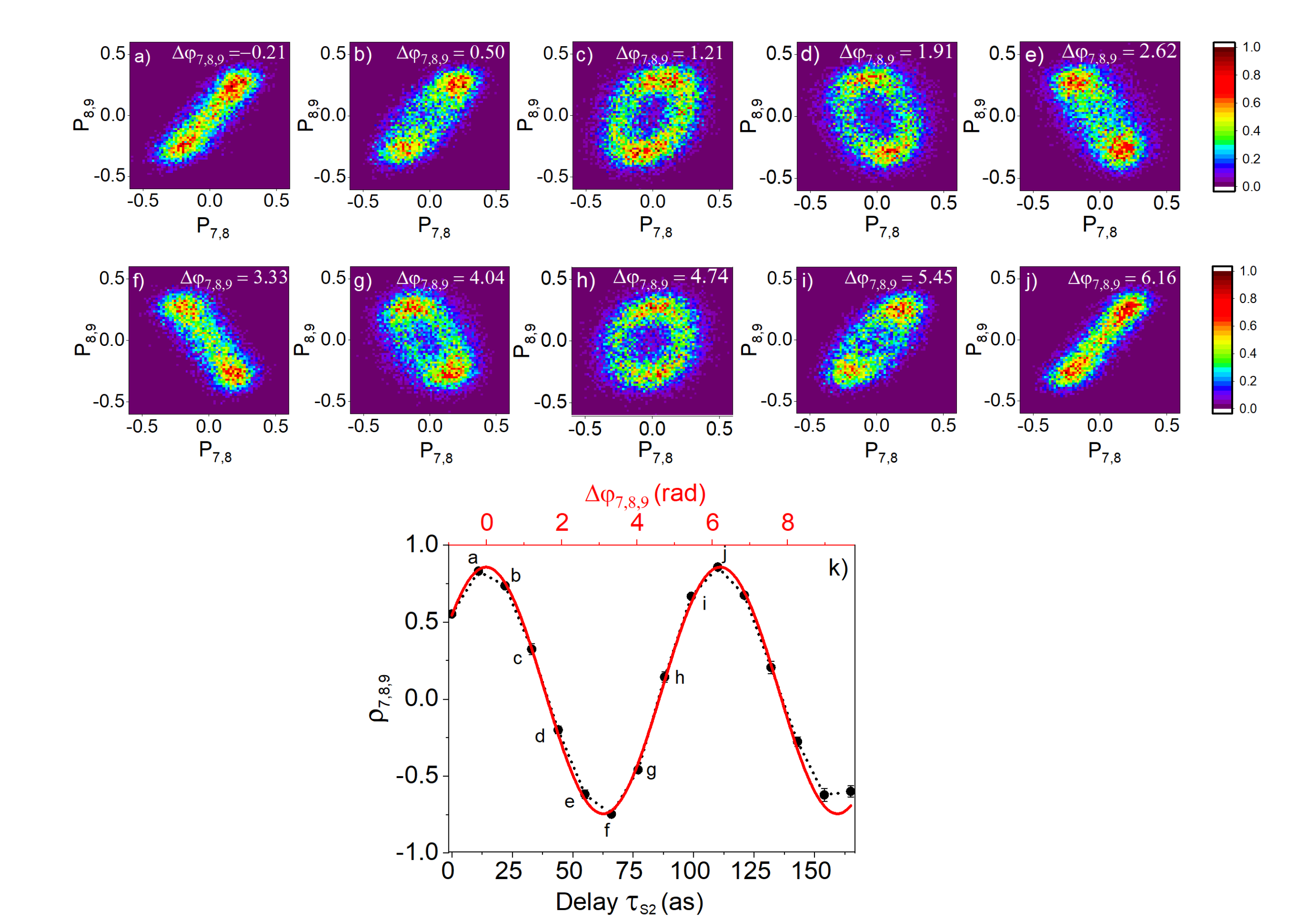}
    \caption{a-j) Correlation plots $P_{8,9}$ vs $P_{7,8}$ for different delays $\tau_{s2}$ introduced by the phase shifter PS2. The corresponding phase difference $\Delta\varphi_{7,8,9}$ is shown in the right-upper corner of each panel. k) Evolution of the correlation parameter $\rho_{7,8,9}$ as a function of the delay $\tau_{s2}$ (black points and dashed line). The letters a-j indicate the experimental points corresponding to the correlation plots shown in panels a-j. The red curve is a sinusoidal fit of the experimental points. The error bars are the standard deviation of the average correlation coefficients evaluated over ten sets of experimental data, with 1200 single shots per set. The red upper axis of panel k) was obtained by fixing the maxima of the fit equal to $\delta\varphi_{7,8,9}=2m\pi$, where $m$ is an integer. Figure reprinted from~\citet{Maroju2020},  with permission. Copyright (2020) by the Authors.
    }
    \label{FigMaroju2020_f2}
\end{figure}

Figure \ref{FigMaroju2020_f2}a-j reports the correlation plots of the oscillating components of the sidebands $P_{7,8}$ and $P_{8,9}$ for different delays $\tau_{s2}$ introduced by phase shifter PS2, which controls the phase of the ninth harmonic relative to the eighth and seventh harmonics. The plot evolves from a linear, to an elliptical, to a circular, to a linear plot with negative slope, then back to a linear plot with positive slope. For each delay $\tau_{s2}$ the relative group delay dispersion between the three harmonics can be retrieved by estimating the correlation parameter $\rho_{789}$ of the corresponding correlation plot. The values of the correlation parameter as a function of the delay are shown in Fig. \ref{FigMaroju2020_f2}k and clearly show a periodic evolution, which is well approximated by a sinusoidal.  By assigning phase difference $\Delta\varphi_{789}=2m\pi$ (with m integer) to the maxima of the sinusoidal fit, a correspondence between phase difference and delay is determined, giving access to the group delay dispersion of the harmonics for each position of the delay shifter. Using this information the pulse, the temporal structure of the attosecond train can be determined. 

The experiment was also performed with four harmonics and, in principle, extension up to six harmonics is technically feasible, since six undulators are available at FERMI.
Thus FERMI offers great flexibility, giving access to attosecond pulse trains in which the electric field is reproduced in each pulse (by using odd and even harmonics) or with a $\pi$-phase jump between consecutive pulses (only odd harmonics) and with a variable time-spacing (around 1 fs).

Attosecond pulse trains produced at a seeded FEL offer major advantages compared to HHG-based attosecond sources. Indeed, due to selective control of the relative phases and amplitudes of the different harmonic components, complete pulse-shaping control of the attosecond waveform was demonstrated. This capability goes beyond the shaping demonstrated so far for HHG-based sources \citep{Martens2005,Hofstetter2011,Bartels2000}. Moreover the very high energies per pulse (in the $\mu$J range) allow investigation of nonlinear optical processes induced by the absorption of multiple XUV photons.

The coherence between the different harmonics of the FEL is a fundamental prerequisite for the application of this approach. For SASE FELs, which can generate sub-femtosecond pulses, a single-shot technique \citep{Li_OE_2018} is mandatory for the characterisation of the temporal structure of the attosecond waveform \citep{Duris2020}. For future experiments, the angular information contained in the photoelectron spectra could provide additional information about the different phases accumulated by the wave packets during the two-colour photoionization process \citep{Dahlstrom2012}. Velocity Map Imaging (VMI) spectrometers can provide this data \citep{OKeeffe2012}.

The development of  attosecond pulses from a fully coherent FEL echoes the development of attosecond pulses from laboratory lasers, which first produced short pulses, then attosecond pulse trains, and now isolated attosecond pulses. Apart from the technical differences of FELs, the wavelengths are shorter, and the pulse energies much higher. More importantly, the method can be applied to even shorter wavelengths, in principle reaching core level energies.

\section{\label{sec:Few_photon_sequential}Few-photon sequential multiple ionization}

\subsection{\label{sec:Early_studies}Early studies} 
In this Section, we mainly focus on few-photon sequential multiple ionization of atoms, as \citet{Seddon2017} have recently reviewed FEL studies of multiphoton multiple ionization of molecules.
Early studies of multiphoton ionization using intense XUV laser sources were performed at FLASH by measuring the different ionic charges states of Ne and He. These preliminary investigations indicated that sequential ionization through ionic and resonance states dominates the direct multiphoton ionization through virtual states \citep{Sorokin2007}. 

At SCSS, with a photon energy of 24 eV and a FEL power density $\sim 10^{14}$ W/cm$^2$, formation of Ar$^{7+}$ and Kr$^{8+}$ was found~\citep{Berrah_JModOpt_2010}. The total energies required to remove seven electrons from the Ar atom and eight electrons from the Kr atom are $\sim 434$ eV and $\sim 508$ eV, respectively, i.e., more than 18 and 21 times the photon energy of 24 eV. Such multiple ionization is likely to be due to sequential stripping of the outermost electrons~\citep{Motomura2009}. With this assumption, the
total numbers of photons absorbed by single Ar and Kr atoms should be 22 and 26, respectively. Details of these multi-photon multiple-ionization pathways, however, have not been fully analysed.
Later, such ion charge distribution measurements have been extended to X-ray ranges~\citep{YoungNature2010,Rudek2012,FukuzawaPRL2013}. Modern \emph{ab initio} theoretical calculations can reproduce these charge distributions accurately~\citep{HoPRL2014,Rudek2018}. See also Section~\ref{ssec:photo_intro} for the role of resonant excitation in these studies.

Information about the momentum of the ejected electron or recoil ion provides additional insight into multiple ionization processes. \citet{Rudenko2008} investigated the recoil ion momentum distribution of singly and doubly ionized helium and neon after irradiation with intense XUV pulses. 
In the photon energy reange 20--24 eV available at SCSS, single photon ionization of Ar first takes place followed by two-photon ionization of Ar$^+$~\citep{Fukuzawa2010,Hikosaka2010,Miyauchi2011}. In particular, \citet{Hikosaka2010} elucidated the role of an
intermediate resonance in the two-photon ionization of
Ar$^+$, using a magnetic bottle electron spectrometer. 
At the higher photon energy of 93 eV at FLASH, three-photon triple ionization of Ne was investigated using a velocity-map imaging electron spectrometer~\citep{Rouzee2011}. 

In more refined experiments, both the photoion and the photoelectron momenta were measured to retrieve additional information on the photoionization process. At FLASH, with a photon energy of 44 eV, \citet{Kurka2009} performed a kinematically complete experiment for sequential two-photon double ionization
of neon atoms, by measuring the momenta of both electrons in coincidence with Ne$^{2+}$ ions. A
reaction microscope was used and the process fully characterized. 

Another type of refined experiment is to measure energy correlations of two ejected electrons. At LCLS, \citet{Frasinski2013} investigated competition between hollow atom formation and sequential ionization by intense X-ray pulses using covariance mapping~\citep{Frasinski1989,Frasinski2016}. 

It is worth noting that several theoretical works have  investigated electronic correlation effects exploiting the correlated momentum measurement of two-photon double ionization of helium \citep{Ishikawa2005,Burgdorfer2009}.
Extensive theoretical studies of angular distributions of the electron emission in the sequential two-photon double ionization of
the valence shell of the noble gases started simultaneously
with the corresponding experiments at FLASH (see references in Section~\ref{ssec:PT}).

\subsection{\label{sec:complete experiments}Complete experiments on Ne}
Upon ionization by absorption of an XUV photon, an electronic wave packet is released into the continuum. Depending on the initial electronic state and on the light polarization, several outgoing partial waves characterized by different angular momenta can be emitted. 
Their coherent superposition results in the three-dimensional photoelectron angular distribution (PAD) observed in experiments. For a complete characterization of the photoionization process, the complex amplitudes (that is, the modulus of the amplitudes and relative phases) of these partial waves must be measured. 
Such information cannot be obtained, in general, by measuring only the photoelectron angular distribution and the total cross-section, but additional information is required.
The high intensity available with FELs offers the opportunity to investigate the characteristics of the sequential multiple photoionization of atomic and molecular targets. In particular, it is known that photoionization leaves the ion in a polarized state, in which the angular momentum is preferentially aligned along a certain direction \citep{Greene_ARPC_1982}. This characteristic of the ionic electronic distribution influences the angular distribution of the photoelectron emitted by the absorption of a second XUV photon. 
By measuring the PADs of the two photoelectrons for different polarization states of the incident XUV light, and with suitable assumptions, information about the polarization state of the ion and the complete charaterization of the two photoionization processes is possible. 
Such a scheme imposes significant constraints on the XUV source, which need to have high intensities and a controllable polarization state. These two conditions can be achieved by the seeded FEL FERMI. 
According to perturbation theory, for the absorption of two photons the PADs are described by the relation:
\begin{equation} \label{eq:ang}
I^{\nu}(\vartheta) = \frac{I^{\nu}_0}{4 \pi} \left[ 1+\beta^{\nu}_2 P_2(\cos\vartheta)+
\beta^{\nu}_4 P_4(\cos\vartheta) \right] \,,
\end{equation}
where $P_k(x)$ is the $k^{\mathrm{th}}$ Legendre polynomial, $\vartheta$ is the angle between the axis of
symmetry $z$ and the emission direction of the photoelectron, $I^{\nu}_0$ is the
angle-integrated intensity, the superscript $\nu$ denotes linearly or circularly polarized pulses. In the cases of linearly and of circularly polarized pulses,
the axis of symmetry $z$ corresponds to the polarization direction and to the propagation direction of the
XUV pulses, respectively. The polarization of the ion is linked to the coefficient $\beta^{\nu}_4$ and it can be derived by measuring the three-dimensional PADs. 

\begin{figure}[ht]
    \centering
    \includegraphics[width=150mm]{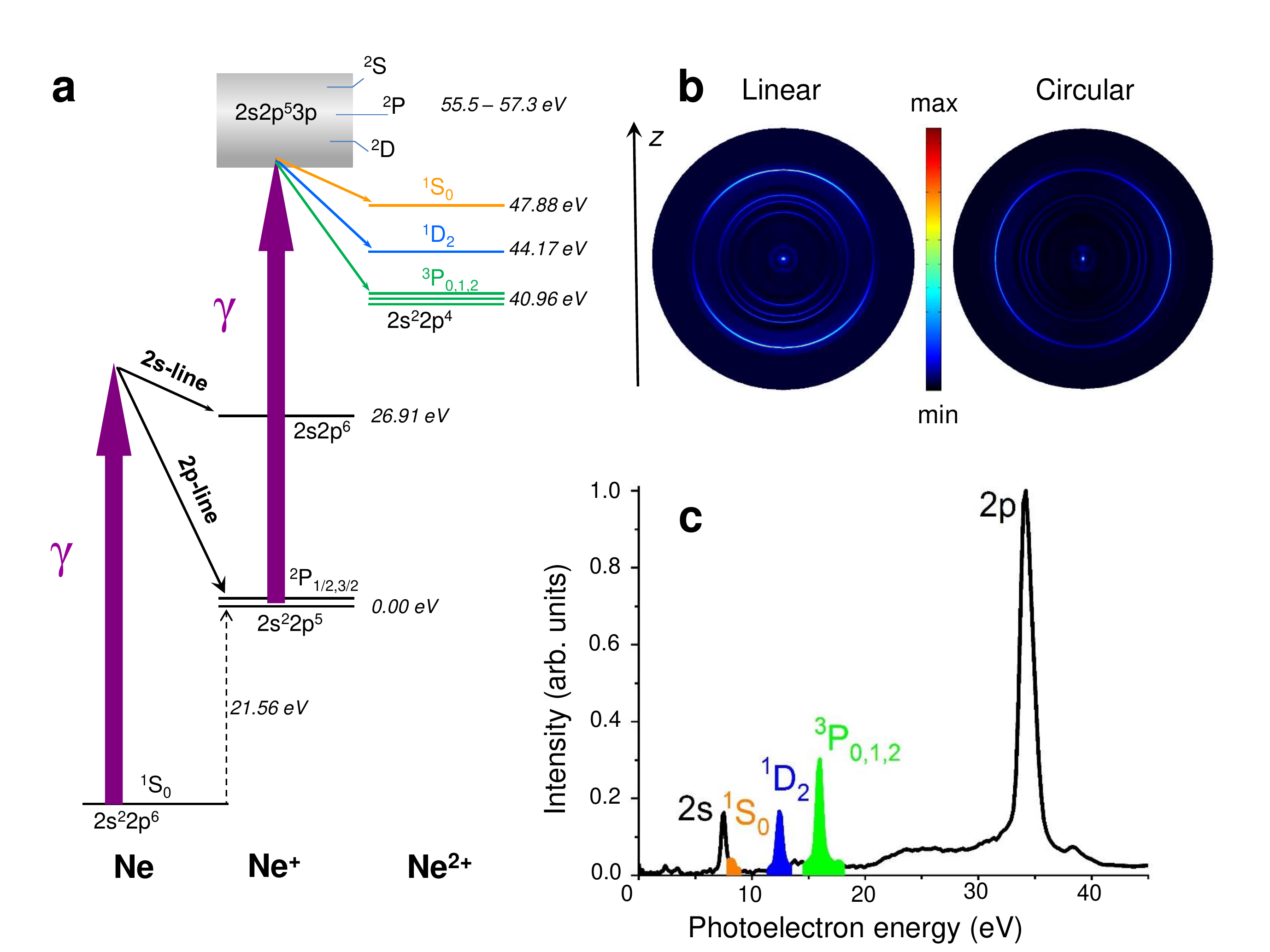}
    \caption{a) Scheme of the energy levels of Ne, Ne$^{+}$ and Ne$^{2+}$. The three final ionic states of Ne$^{2+}$, $^1$S$_0$, $^1$D$_2$, $^3$P$_{0,1,2}$ are indicated by yellow, blue, and green lines, respectively.  b) Inverted images of the photoelectron angular distributions obtained for linearly and circularly polarized XUV pulses using a velocity map imaging spectrometer. c) Photoelectron spectra obtained from Neon using linearly polarized intense XUV pulses. The three final ionic states of Ne$^{2+}$ ($^1$S$_0$, $^1$D$_2$, $^3$P$_{0,1,2}$) as well as the photoelectrons corresponding to ionization from the 2s and 2p levels of Neon are indicated. Figure reprinted from~\cite{Carpeggiani2019}, with permission. Copyright (2019) by the Authors.}
    \label{fig:Carpeggiani}
\end{figure}

In the experiment performed at FERMI by \citet{Carpeggiani2019}, Fig. \ref{fig:Carpeggiani}, the PADs were measured using a velocity map imaging spectrometer, which gives access to the three-dimensional PADs, provided that the interaction is characterized by an axis of symmetry. As stated, this is the polarization direction (linear polarization) or propagation direction (circular polarization). 
The experimental results were compared with  predictions based on the perturbation theory, within the single-configuration non-relativistic approximation, and a fitting procedure was implemented to determine the values of the ratio between the amplitude of the s and d-partial waves and their relative phase for the two photoionization steps. The photoelectron angular distribution of the residual ion and those of the scattering states of the corresponding photoelectron(s) were retrieved, providing a complete characterization of the two-photon double ionization of neon \citep{Carpeggiani2019}.\\

\subsection{\label{ssec:autoionizing2I}The role of autoionizing resonances in double ionization}
As mentioned in Section~\ref{ssec:photo_intro}, autoionization is an important phenomenon giving access to a number of physical quantities.
Studies of this effect with synchrotron~\citep{Madden1963, Maeda1993, Sorensen1994, Domke1995, Domke1996} and attosecond XUV 
pulses~\citep{Ott2013, Ott2014, Kotur2016, Gruson2016, Cirelli2018} have been limited to the autoinizing states of neutral systems, which are embedded in the continuum state associated with the cation in the ground state and thus fragment into a cation and a free electron.  With the intense pulses provided by FELs, one can ionize the sample first and then excite autoionizing resonances of cations, which are embedded in the continuum of the ground state dication, and therefore fragment into the dication and a free electron. Thus the pulse has two functions and the experiment  elucidates the role of the autoionizing resonances in double ionization. 
The importance of autoionization resonances in both steps of
sequential double ionization was confirmed by experiments on Ar at FLASH
in the region of the Rydberg autoionizing states of Ar$^+$~\citep{Augustin2018}. A clearly measurable value of $\beta_4$ in the 
PAD [Eq. (\ref{eq:ang})] of the first electron in the double ionization process confirmed
a theoretically predicted correlation between both electrons, resulting from the polarization of the intermediate state, the Ar$^+$ ion.

It is worth noting that similar electronic decay pathways that enhance the degree of ionization can also be seen using X-rays. \citet{Rudek2012} reported an unprecedentedly high degree of ionization of xenon atoms by 1.5 keV free-electron laser pulses to charge states with ionization energies far exceeding the photon energy. Comparing ion charge-state distributions and fluorescence spectra with ab initio calculations, they speculated that these surprisingly high charge states are created via resonant excitation in highly charged ions, which are subject to autoionization, or resonant Auger decay. Later, this resonance-enabled X-ray multiple ionization (REXMI) pathway was fully confirmed experimentally and theoretically, at higher photon energies and with higher X-ray fluxes~\citep{Rudek2018}.

\subsection{\label{sec:spLEAD}Single-photon Laser Enabled Auger Decay}

The phenomenon of Auger decay is well-known as a relaxation mechanism after core ionization. In this process, a singly charged core hole state undergoes a valence-core or core-core transition, and simultaneously a second electron (the Auger electron) is emitted
to form a doubly charged final state; the excess energy is carried away from the atom by the kinetic energy of the Auger electron. 
The field of Auger spectroscopy includes a number of  other cases with their own terminology, for example, core-excited neutral states decay by resonant Auger Raman processes; when the process involves one electron from the same shell as the core hole, and one electron from another shell, it is said to be a Coster-Kronig transition. 
When both electrons  are in the same shell as the hole, it is a super Coster-Kronig transition, and is usually a very fast processes. Auger electrons from a doubly core ionized atom are known as Auger hypersatellites.

The usual Auger process is equivalent to autoionization of the atomic ion discussed in the previous subsection. Historically, autoionization refers to inner valence and low energy excited states, whereas Auger processes refer to core ionized or core excited states.  
For shallow holes, the energy of the doubly charged state may be higher than the energy of the singly charged state, in which case Auger decay is energetically forbidden. However, if additional energy is available, the Auger decay becomes allowed. 
When the energy is provided by a laser, the process is known as Laser Enabled Auger Decay, and this has been observed in experiments where multiple laser photons provided the ``enabling'' energy \citep{Ranitovic2011, Tong2011, Hogle2015}. 
Thus, while normal Auger can be a single-photon, double ionization process, Laser Enabled Auger Decay is a few-photon, double ionization process.

If however the missing energy is provided by a single photon, then selection rules apply \citep{Cooper2013}. For a pure electronic state consisting of a single configuration, single-photon Laser Enabled Auger Decay (spLEAD) is forbidden. Due to the phenomenon of correlation, many electronic states are not pure, but are mixed with other configurations, so that spLEAD then becomes allowed. Thus this process represents a probe of the extent of correlation in a given ionic state.

\citet{Iablonskyi2017} reported the first observation of spLEAD, using the Free-Electron Laser FERMI. The sample was neon and the method employed was rather sophisticated, and took advantage of the fact that FERMI is longitudinally coherent. The sample was irradiated with phase-locked fundamental and second harmonic radiation and then the phase was tuned. The fundamental wavelength was chosen to be at the ionic resonance
\begin{equation}
\label{eq:eq_spLEAD}
2\mathrm{s}^2 2\mathrm{p}^5 + \omega \rightleftharpoons 2\mathrm{s} 2\mathrm{p}^6
\end{equation}
This arrangement ensured that the final, doubly ionized Auger state could be
reached by two different paths, one of which involved the spLEAD process. Interference occurred as the optical phase was scanned, and was manifested as an oscillation of the yield of photoelectrons corresponding to the final states. In the absence of the spLEAD process, the phase would have no effect on the yield. The final states are represented by the terms $^1$S, $^1$D and $^3$P, and of these, $^1$S was not resolved from another spectral feature, so the work concentrated on $^1$D and $^3$P. 

In the paper of \citet{Iablonskyi2017}, it appeared that the triplet $^3$P term did not oscillate and this was interpreted as an indication that $L-S$ coupling was a good approximation. However later calculations, and more accurate analysis, indicated that also this peak oscillated \citep{You2019}. 

To understand the process, consider the main configurations which mix into the state which undergoes spLEAD, in this case 2s ionized Ne. The main configuration is clearly 2s2p$^5$ (80\%), but there are admixtures of $2\mathrm{s}^2 2\mathrm{p}^4 (^1\mathrm{S})n\mathrm{s}$, $2\mathrm{s}^2 2\mathrm{p}^4 (^1\mathrm{D})n\mathrm{d}$ and $2\mathrm{s}2\mathrm{p}^5(^3\mathrm{P})n\mathrm{p}$, as well as $2\mathrm{s}2\mathrm{p}^5(^1\mathrm{P})n\mathrm{p}$. 
For $2\mathrm{s}^2 2\mathrm{p}^4nl$ configurations, the spLEAD process can be visualised as the ionization of the outer electron of these configurations by the ``extra'' energy supplied by the laser, leaving the ionic core of these states as the final doubly ionized state. 
The first three configurations lead to doubly charged ions with both holes in the 2p shell. The calculations show that the $2\mathrm{s}2\mathrm{p}^5(^3\mathrm{P})n\mathrm{p}$ configuration also has a significant electron yield, and this is due to a two-electron process, where one electron fills the 2s hole and another is ejected. 
\cite{You2019} pointed out that \citet{Iablonskyi2017} had considered only configurations built on $2\mathrm{s}^2 2\mathrm{p}^4 (^3\mathrm{P})$ cores for the spLEAD process, but it is the configurations with $2\mathrm{s}2\mathrm{p}^5 (^3\mathrm{P})$ cores that make a significant contribution to final triplet states.

The newer theoretical analysis predicted all phenomena observed, and reproduced qualitatively or quantitatively the branching ratios of $^3\mathrm{P}$ to $^1\mathrm{D}$ final states, the ratios of oscillation amplitudes and the phase lags between the oscillations of different ionic states. An important aspect of this work is that it represents the observation and study of a new de-excitation process, and it is rare to discover such new processes. To perform this work, a key experimental requirement was the narrow bandwidth of the photons, which was tuned to the resonance described by Eq.~(\ref{eq:eq_spLEAD}).

\subsection{\label{ssec:two_photon_ICD}Two-photon excitation leading to interatomic Coulombic decay}

Interatomic Coulombic Decay (ICD) is a decay process which occurs in weakly bound systems, and was predicted more than 20 years ago \citep{Cederbaum1997}. Since then it has been observed in many different systems in various forms; see \citet{Jahnke2015} for a review.  If an inner valence level of an atom or molecule is ionized, and the energy is below the double ionization threshold, then it cannot decay by emission of an electron to form a doubly charged state.
However, if the excited atom is weakly bound to another atom, and the internal energy is greater than the sum of the ionization potentials of the two, then it may decay to two singly ionized atoms, with the emission of an electron. 
A neutral atom weakly bound to a doubly charged ion clearly has a higher energy than two weakly bound singly charged ions, because the Coulomb repulsion is reduced. The ICD process may be viewed as autoionization occurring in a weakly bound system, and 
is generally a very fast process. It often competes with fluorescence decay, which is usually much slower.

\begin{figure}[!ht]
    \centering
    \includegraphics [width=0.9\textwidth] {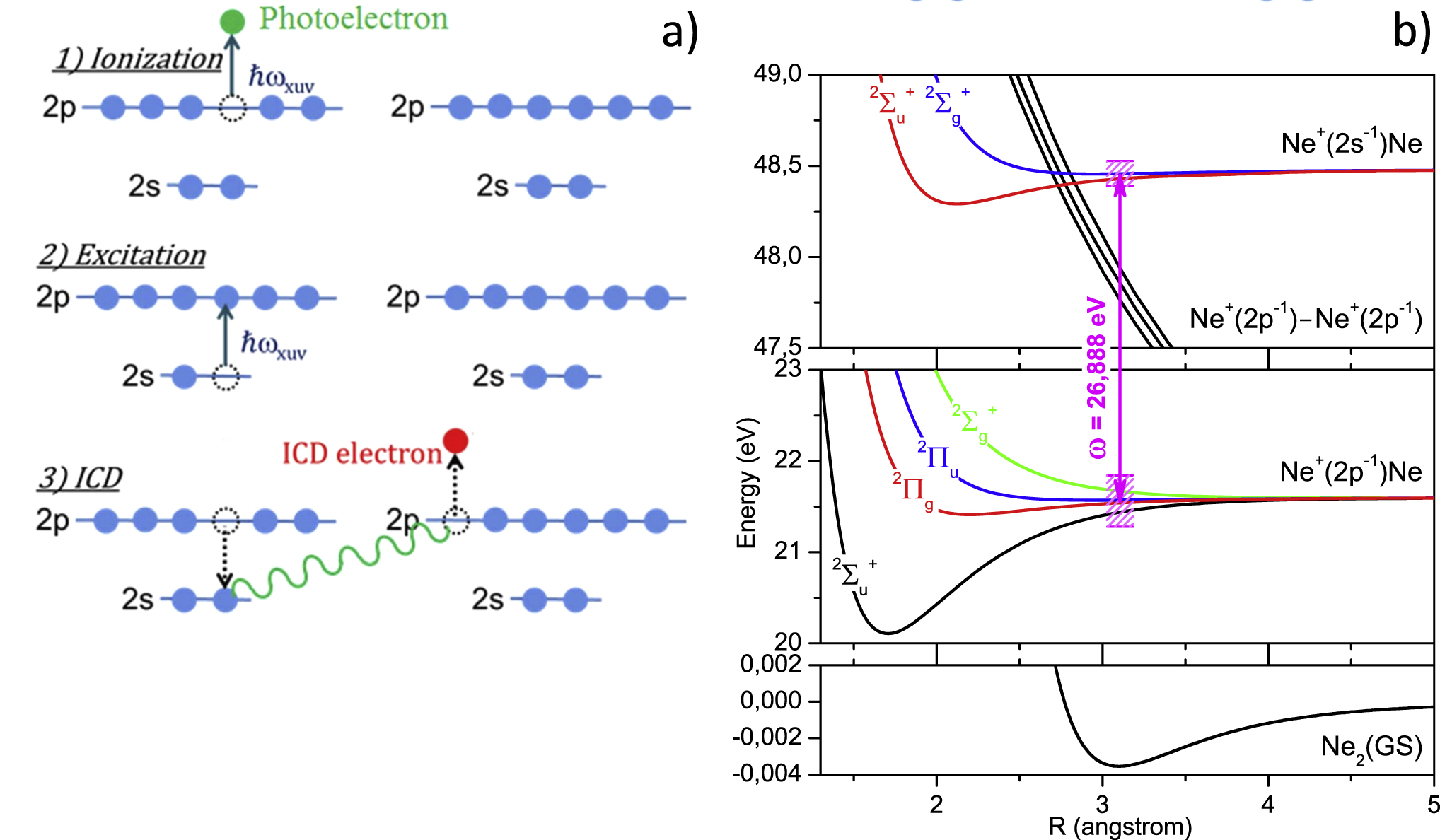}
    \caption{(a) Schematic representation of the two-photon
excitation of ICD. In the Ionization step, the 2p electron is
ionized by a photon from the XUV pulse with the emission
of a photoelectron and population of the Ne$_2$(2p$^{-1}$) states of the dimer. 
In the Excitation step, a second photon is resonantly absorbed leading to the population of the inner valence ionized Ne$_2$(2s$^{-1}$) states. In the ICD step, the 2s hole of the
initially ionized atom is filled by a 2p electron, and the excess energy is transferred to the neighboring atom, leading to the
emission of the ICD electron. As a result, the two-site outer valence
doubly-ionized dimer Ne$_2$(2p$^{-1}$2p$^{-1}$) is populated, and undergoes Coulomb
explosion. (b) Potential energy curves of the relevant neutral
ground state, the outer and inner valence singly ionized states, as well as
outer valence doubly ionized states of the neon dimer. Panel (a) is reprinted from \citet{Dubrouil2015},  \copyright{} IOP Publishing. Reproduced with permission. All rights reserved. Panel (b) is reprinted from \citet{Demekhin2011}, with permission. Copyright (2011) by the American Physical Society.
}
    \label{fig:Dubrouil}
\end{figure}

In neon atoms, an inner-valence 2s hole cannot decay via an Auger process, because the energy 
of an ion with two 2p holes is greater than the energy of a 2s hole.
In a neon dimer, however, the final double positive charge can be shared between the two atoms, thus lowering the energy required for the second ionization step and the ICD decay channel may open up. ICD has been widely investigated using single-photon absorption of photons from synchrotron and FEL sources \citep{Jahnke2015,Schnorr2013, Nagaya2016}. In particular, taking advantage of the development of tunable, femtosecond, intense XUV sources, \citet{Demekhin2011} investigated theoretically the nonlinear excitation of the ICD mechanism through the absorption of two XUV photons.
The energy scheme corresponding to the nonlinear excitation of the ICD process is illustrated in Fig.~\ref{fig:Dubrouil} (a): the dimer is first ionized, with the ejection of a 2p electron and then excited by the absorption of a second photon to a state characterized by a 2s hole in one atom (assuming a description of separated electronic levels for the atoms composing the dimer). 
The 2s hole is filled by a 2p electron and the excess energy is transferred to the second atom, inducing the ejection of a low energy electron (ICD electron).
This work indicated that the preparation of the 2s hole in a dimer by sequential two-photon excitation should be more efficient than the single-photon ionization.

This excitation mechanism requires a resonant step corresponding to the excitation of a 2s electron to the 2p at 26.9~eV. At  different energies, the resonant excitation step cannot take place, and also the ICD process cannot occur.

Figure~\ref{fig:Dubrouil} (b) presents the molecular potential energy curves for the neutral neon dimer Ne$_2$ (lower panel), for the lowest energy states of the singly ionized dimer (corresponding to the configuration Ne$^+$(2p$^{-1}$)Ne, central panel), and for the excited states of the ionized dimer corresponding to the configuration Ne$^+$(2s$^{-1}$)Ne, upper panel. The Coulomb-repulsive energy curves of the double ionized neon dimer are also reported [Ne$^+$(2p$^{-1}$)-Ne$^+$(2p$^{-1}$)]. 
An experiment on neon dimers was performed by \citep{Dubrouil2015} to test this theoretical prediction, and it faced a number of challenges. Neon dimers were produced by a cold supersonic expansion of neon gas, resulting in a concentration of a few percent of dimers in atomic neon. The photons ionize both the dimer and the monomer, resulting in a very strong monomer 2p photoelectron signal, with the risk of saturating the atomic signal so that the weak dimer signal (with only a slightly different kinetic energy) could not be detected. This risk was avoided by measuring the dimer and monomer ion signals, which have a large dynamic range, and are well separated in the spectrum.

The photon energy was scanned across the predicted resonance energy \citep{Demekhin2011}, and the ratio of dimer ions to monomer ions showed a minimum at the resonant energy. The potential energy curves of the dimer are shown in Fig.~\ref{fig:Dubrouil} (b), where it can be seen that the Franck-Condon regions of the 2p and 2s ionized ions correspond to weakly bound ionic states. However the doubly ionized states, with one hole on each atom, are very strongly repulsive.  

In the resonance region, the ionization probability of both dimers and monomers is constant, so the observation of a reduction of dimer ion intensity indicated that they were decaying faster at the resonant energy. This occurs because the 2p ionized dimer resonantly absorbs a second photon to become a 2s ionized dimer, which then decays via ICD, depleting the dimer ion signal.

\subsection{Multiphoton XUV + Near Infrared ionization and Circular Dichoism}

In Section~\ref{ssec:XUV+NIR}, we discussed two-photon ionization by a single XUV and a single NIR photon. 
The availability of intense XUV pulses opens new frontiers for the investigation of the dynamical response of electronic systems to intense XUV and visible/near-infrared pulses. In this context, it is interesting to investigate the interplay between the dynamics induced by both fields.

\citet{Ilchen2017} investigated the photoelectron spectra of an ionic Rydberg state in a two-colour field for different relative helicities of the fields. The atom investigated was helium, which was first ionized to the ground state He$^+$(1s) by single photon absorption and then further excited to He$^+$(3p) by absorption of a second XUV photon. This situation is related to that  discussed in Section \ref{ssec:autoionizing2I}, where ionization followed by resonant excitation was used to access autoioinizing states; here the final state is long-lived, because it can decay only by fluorescence. 
The intense near-infrared pulse was co- or counter-rotating with respect to the circular polarization direction of the XUV pulse (as above), and photoionized the ion, thereby ejecting a second electron. Depending on the relative helicity between the two fields, different angular momentum states dominate the photoelectron angular distribution of the second electron. 
For low photoelectron energies (around 150 meV), in the co-rotating case, only the final continuum state characterized by quantum numbers $l=5$ and $m=5$ contribute to the photoelectron angular distribution. On the other hand, in the case of counter-rotating fields, two states with $l=3,m=-3$ and $l=5,m=-5$ contributed to the photoionization at low kinetic energies, and the photoelectron angular distribution depends, therefore, on the relative phase between these two contributions.

The CD is defined as:
\begin{equation}
\label{eq:CD}
CD=\frac{P_{+}-P_{-}}{P_{+}+P_{-}}
\end{equation}
where $P_{+}$ and $P_{-}$ indicate the probability of ionization for fields with the same or opposite helicity, respectively. In the experiment a strong dependence of the CD as a function of intensity of the IR field was observed. In particular, at the intensity of $I=6\times10^{11}$ W/cm$^2$, the CD reached almost the maximum value of unity, indicating the photoionization for the co-rotating case is more favoured with respect to the counter-rotating case. This difference can be interpreted by observing that four dipole transitions are required to reach the state $l=5, m=5$ from the initial ionic state He$^{+}(n=3,l=1,m=1)$. These transitions require in each step a change of one unit of the angular and magnetic quantum number: $l\rightarrow l+1$ and $m\rightarrow m+1$, which are the most favourable dipole-allowed transitions for increasing $l$. For the counter-rotating case, the probability to reach the two final states $l=5, m=-5$ and $l=3,m=-3$ is much smaller (a factor 50) due to angular-momentum factors. This conclusion was confirmed also by TDSE simulation, which predicted an ionization probability two orders of magnitude smaller.

\begin{figure}[tb]
    \centering
    \includegraphics[width = 3.375 in]{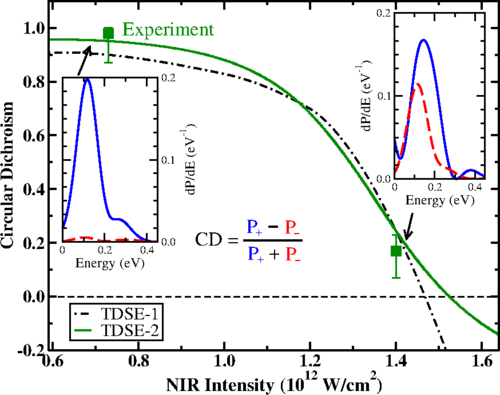}
    \caption{Experimental (green squares) and theoretical (green full line and black dashed line) circular dichroism CD for the electron peak at 200 meV as a function of the NIR intensity. The two theoretical curves were obtained using two TDSE models: black dash-dotted curve \citep{Kazansky2007}; and green curve \citep{Douguet2016}. The insets present the electron energy spectra obtained from the TDSE calculations in correspondence to the two experimental points. Figure reprinted from \cite{Ilchen2017}, with permission. Copyright (2017) by the American Physical Society.}
    \label{fig5}
\end{figure}

At higher IR intensities, the CD reduces and TDSE even predicts a change of sign for intensities exceeding $1.5\times10^{12}$ W/cm$^2$, see Fig.~\ref{fig5}. This variation can be interpreted by investigating the role of the intense IR field on the population of the oriented ionic state He$^+$(3p). In the case of co-rotating pulses, the IR induces a shift of the energy of the $(3\mathrm{p}, m=+1$) level, which strongly reduces the population of this level. This shift is almost absent in the counter-rotating case. Therefore, the initial population of the ionic states compensates for the much less favourable photoionization pathway of the counter-rotating case. The results of \citet{Ilchen2017} were analysed in greater detail in a follow-up paper by~\citet{Grum2019}.

\subsection{\label{sec:XTPPS}Two-photon, double core ionization}
For a sufficiently intense X-ray pulse, it is possible to ionize two core levels with the same pulse. 
Core level photoelectron spectra are sensitive to the chemical surrounding of the ionized atom, observed as binding energy differences in molecules and solids~\citep{Siegbahn1971}.  
\citet{Santra2009} demonstrated theoretically that the advent of XFELs
enabled a novel core spectroscopy: X-ray Two-Photon Photoelectron Spectroscopy
(XTPPS). In this method, two core levels are ionized within the core hole lifetime, usually a few fs.  Twenty four years earlier,~\citet{Cederbaum1986} had demonstrated theoretically that the creation of double core vacancies in molecular systems probes the chemical environment more sensitively than the creation of single core vacancies, and that double ionization potentials (DIP) of two-site double-core-hole (tsDCH) states are particularly sensitive to the chemical environment. 
These tsDCH states, however, were not observed until 2011~\citep{Lablanquie2011} because of extremely small cross sections for single-photon, double core ionization.
Intense and short XFEL pulses made it possible to observe these tsDCH states via XTPPS where the two photons are absorbed sequentially by two atoms at different sites. Following the proposal of XTPPS~\citep{Santra2009}, a series of of calculations were performed to predict DIPs of tsDCH states for a number of molecules as a guide for experiments, for example~\citep{Tashiro2010, Tashiro2010a, Takahashi2011}. Experiments at LCLS~\citep{Berrah2011, Salen2012} demonstrated the feasibility of XTPPS and at the same time its limitation; the photon bandwidth of the SASE beam was broad, hampering precise measurements. Use of narrow band and short (less than 5 fs) soft X-ray pulses, e.g., generated by seeded FELs (see Sections \ref{ssec:seeded} and \ref{ssec:selfseeded}), is expected to improve significantly the performance of this method.

\section{\label{sec:multi-photon-resonant}Multi-photon and multiple resonant excitation}
\subsection{Two-photon excitation of two electrons in atoms: Early studies}

The simultaneous excitation of both electrons in the helium atom by a single photon was first observed by \citet{Madden1963} using synchrotron radiation.  
The theoretical basis for describing this experiment had already been laid by \citet{Fano1961}, who had developed his earlier theory \citep{Fano1935} to describe the electron scattering data of Lassettre and Silverman. The double excitation of He has since been  extensively studied, see \cite{Domke1995, Domke1996}, and references to these papers. 
The importance of this system is that double excitation does not occur in an independent electron model, and becomes allowed by initial state correlation.
Although these doubly excited states are subject to autoionization, they are qualitatively different from the autoionizing states discussed in Section~\ref{ssec:autoionizing2I}, where one-photon excitation to the autoionizing states does not rely on the initial state correlation.
For most of the cases discussed in that Section, two electrons were excited in two separate and sequential steps, for example ionization followed by resonant excitation, and this led to double ionization. In this Section we discuss cases where the excitation occurs by simultaneous absorption of more than one photon.

The helium atom is the simplest three-body system for investigating correlation, and the narrow resonances can be measured and calculated with high precision. Double excitation is therefore a benchmark phenomenon for testing theoretical approaches and approximations.
Indeed, for this reason, intensive state-of-the-art attosecond time-resolved studies have been carried out on helium doubly excited states~\citep{Ott2013, Ott2014,Gruson2016,Kaldun2016}.

With intense FEL pulses, one can promote both electrons to these excited states with two or three photon absorption, without relying on initial state correlation: each photon can interact with each electron separately.  At SCSS, \citet{Hishikawa2011} investigated three-photon single ionization of He employing a magnetic bottle electron spectrometer. 
The photon energy was $\approx24$ eV and the FEL power density was 10$^{13}$ W/cm$^2$. Under these conditions one electron was promoted to the Rydberg manifold by single photon absorption and at the same time another electron was promoted to the Rydberg manifold by two-photon absorption. As a result, three-photon single ionization via autoionization of the doubly excited states was anomalously enhanced.

The role of doubly excited autoionizing states in Ar$^+$ populated via two-photon absorption of the ionic ground state of Ar$^+$ was studied by \citet{Gryzlova2011} using a velocity map imaging electron spectrometer and ab initio theory calculations.

\subsection{Two-photon excitation of two electrons in atoms: Narrow-band studies}

Single-photon excitation of He can access only $^1\mathrm{P}_\mathrm{o}$ final states, but the helium atom has many other resonances of different angular momenta. In particular, states with term values $^1\mathrm{S}_\mathrm{e}$ and $^1\mathrm{D}_\mathrm{e}$ can be excited by two photons. 
At FERMI, \citet{Zitnik_PRL_2014} investigated
two-photon excitation of these even-parity, doubly
excited, autoionizing states, taking advantage of the very
narrow photon bandwidth of fully coherent FEL pulses. 

The experiment is challenging because the ion or electron signal has a very large background due to non-resonant ionization. To overcome this difficulty, \citet{Zitnik_PRL_2014} used an alternative detection scheme: instead of charged particles, they detected neutral metastable atoms. The resonantly excited atoms decay by autoionization to ionic states, and by fluorescence to neutral states. Some of these neutral states undergo cascades by fluorescent decay to the ground state, but some reach metastable states, which are long lived, as they cannot decay by fluorescence. The detector was very simple and consisted of a microchannel plate placed in front of the atomic He beam, with suitable meshes to repel charged particles. The yield of metastable atoms was low, but crucially it was background-free: the metastable atoms were created only at the resonance energies.
This example shows how important narrow bandwidth is for the study of resonant phenomena, as well as the fundamental importance of good design of experiments.\\

\subsection{Two-photon excitation of two electrons in dimers}

In the previous subsection section we discussed the case where two-photon absorption of an atom promotes two electrons to two unoccupied Rydberg orbitals. In this section we discuss the case where a dimer absorbs two photons and both atoms are excited.    
\citet{Takanashi2017} investigated such a doubly excited state in Ne dimers. 
In the separated-atom picture, each atom in the dimer is resonantly excited to the 2p$^{-1}$3s state: overall the order of the process is two-photon, the resulting molecular state is a neutral doubly-excited autoionizing one, and its energy is close to, but lower than, that needed for the double ionization of the system. The system decays by ICD to a  singly-ionized dimer Ne$_2^+$ and a photoelectron e$_\mathrm{ICD}$ as the final products. 

\begin{figure}[ht]
    \centering
    \includegraphics[width=0.45\textwidth]{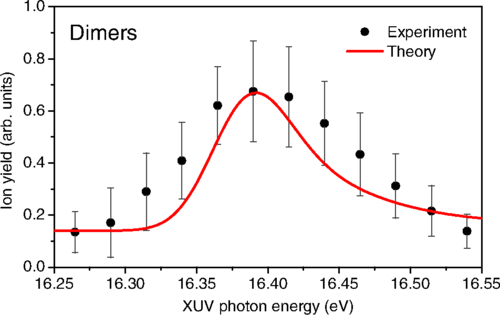}
    \includegraphics[width=0.45\textwidth]{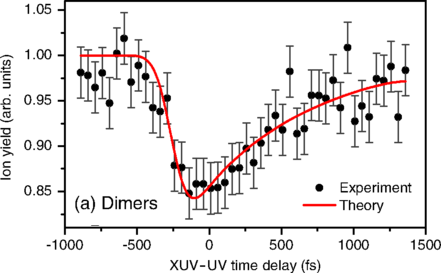}
    \caption{Left: Ne$_2^+$ ion yield as a function of FEL wavelength. Right: normalized Ne$_2^+$ ion yield as a function of pump-probe delay. Figures reprinted from \cite{Takanashi2017}, with permission. Copyright (2017) by the American Physical Society.}
    \label{fig:Takanashi}
\end{figure}

The goal of this experiment was to measure the lifetime (inverse rate) of the ICD process. This approach, which depends critically on the rapid tunability and on the spectral purity of the exciting pulse, has two advantages: first, the single-photon energy is insufficient to ionize bare Ne atoms (the majority component of the gas jet) which then do not contribute a background signal; second, the resonant condition (at a photon energy of 16.39~eV, established by monitoring the Ne$_2^+$ ion yield while scanning the FEL wavelength, Fig.~\ref{fig:Takanashi}) allows a precise selection of the target excited state. 
If, before the system decays, one of the excited electrons is ionized by an external laser pulse (in this case: a UV laser; photon energy 4.75~eV), then the remaining Ne$_2^+$ ion is left in a repulsive state and dissociates into two atoms; the Ne$_2^+$ yield decreases, and a measurement of this as a function of the separation between the two pulses (XUV-UV delay, Fig~\ref{fig:Takanashi}) lets one extract the lifetime of the ICD process. The extracted lifetimes of the long-lived doubly excited states, 390 $^{-130}_{+450}$~fs, and of the short-lived ones, $<150$~fs, are in good agreement with \emph{ab initio} quantum mechanical calculations~\citep{Demekhin2013}.

\subsection{Multi-photon, site-specific, multiple excitation of clusters and droplets}

Multiple excitation ICD, in which more than one atom in a loosely bound system is excited~\citep{Kuleff2010}, may occur in more extended systems than dimers. Indeed much work has been done on large aggregates of atoms, namely clusters, using FEL light~\citep{Ovcharenko2014, LaForge2014, Nagaya2016} as well as intense NIR laser pulses~\citep{Schuette2015}. Here we discuss experiments on neon clusters and helium droplets
for the case of multiple atoms excited by single photons.

\citet{Iablonskyi2016} studied clusters of about 5000 Ne atoms, with resonant excitation of the 2p electrons to 3s states. The resonance energy of surface atoms, 17.12 eV, is slightly shifted with respect to the bulk, 17.65 eV. Thus by tuning the photon energy precisely, it was possible to excite  surface or bulk atoms, and study them independently. In this case, ICD occurs as a de-excitation of adjacent excited atoms, neither of which has sufficient energy to emit an electron.
They observed the evolution of the resulting electron spectrum as a function of the FEL intensity; the latter determines the number of excited atoms per cluster, which can easily exceed 2. The presence of many excited atoms within a cluster sets an upper limit to their abundance, by shifting the energy of the transition out of resonance (Coulomb blockade).
\citet{Iablonskyi2016} found that the relaxation of excited surface atoms proceeded via a sequence of ICD processes, and as well, inelastic electron scattering in the surrounding cluster affected ICD of bulk atoms. 
Comparison with the Ne$_2$ data described above \citep{Takanashi2017} provided an estimate of the maximum excitation limit, found to be 8\%. Although the experiment does not include pump-probe measurements, a lower limit of $\sim 1$~ps was deduced for the ICD lifetime in a cluster, based on the Ne$_2$ data and on the known inverse-sixth-power dependence of the ICD rate on interatomic distance. As a consequence of this long lifetime, cluster relaxation processes can occur before ICD takes place.

More recently, \citet{Ovcharenko2020} have investigated the dynamics of resonant, multiple excitation of He droplets for a wide range of droplet sizes and pulse intensity. 
Depending on conditions, a range of processes occurred: multi-step ionization, electronic relaxation, ICD, secondary inelastic collisions, desorption of electronically excited atoms, collective autoionization, and nanoplasma formation. The spectra were successfully described by numerical calculations based on rate equations.

\section{\label{sec:pump-probe}Pump-probe studies}

Ultrafast, short wavelength FEL pulses are a very useful tool for studying dynamics in molecules. The nuclei typically move on the fs time scale, while electrons move on the attosecond or few femtosecond scale, so that experiments are challenging with sources that range in duration from tens of femtoseconds to longer times. In pump-probe experiments, a sample is typically pumped by a laser pulse at one wavelength, and probed by a pulse at another wavelength. Most commonly the pump pulse is provided by an optical laser, and the FEL provides the probe pulse~\citep{Erk2014, Gorkhover2016, Wolf2017a, Wolf2017, Leitner2018, Nishiyama2019} but other arrangements are possible, e.g., FEL-pump, laser-probe \citep{Takanashi2017, Kumagai2018, Fukuzawa2019}, or FEL-pump, FEL-probe \citep{Jiang2010, Liekhus-Schmaltz2015, Ferguson2016, Berrah2019}. If the signal measured is, for example, total ion yield or Auger spectra \citep{Wolf2017a}, then high resolution of the probe may not be required. In this section, we give some examples of pump-probe studies which required narrow bandwidth in order to achieve sufficient spectroscopic resolution.

\subsection{The dynamics of photo-excited thymine} \label{ssec:pump-probe thymine}

\citet{Wolf2017} have used transient core level absorption to investigate the dynamics of photo-excited thymine. Thymine, see Fig.~\ref{fig:three_molecules}, is one of the nucleobases of DNA, and like the others, is more resistant to damage by UV radiation than many other organic molecules. This suggests that its inclusion in DNA conferred an important characteristic for survival in a world before the ozone layer existed. The electronic energy deposited by UV absorption is converted to thermal energy in a very short time, but the exact mechanism of this process is still debated.

In their experiment, \citet{Wolf2017} used 267 nm pulses to excite free thymine molecules to the $\pi\pi^*$ state, that is, a state with a hole in a bonding $\pi$ orbital, and an electron in a previously unoccupied $\pi^*$ orbital. They probed the ensuing dynamics using Near Edge X-ray Absorption Fine Structure (NEXAFS) spectroscopy at the oxygen K edge, which requires a reasonably small bandwidth for the radiation. They obtained this from the LCLS SASE source by using a monochromator, which provided a bandwidth of 0.1$\%$; for a SASE source this has the effect of increasing the intensity fluctuations, which can be as large as 0 to 100$\%$. This was managed by post-sorting the data into bins of similar pulse energy. 

The selection rules for NEXAFS permit transitions from the O 1s core level to unoccupied states of $\pi$ symmetry. The UV excitation creates a hole in a valence orbital, thus allowing a new transition from the core to this hole; this occurs at lower photon energy than transitions to the unoccupied states of the neutral molecule. Such a feature was observed in the experiment, but theoretical calculations showed that it was not due to the initially excited $\pi\pi^*$ state, but rather to an $n\pi^*$ state, in which the hole is located in a non-bonding valence orbital. From this it was concluded that the excited state converts within 60 fs to the $n\pi^*$ state, via a conical intersection. This state then undergoes further conversion with bi-exponential decay to a lower state, and with time constants of 1.9 and 10.5 ps.

We note in passing that this method, also known as transient absorption, has been used in experiments with laboratory sources \citep{Attar2017}. The details are different -- a super-continuum soft X-ray light source was used with a spectrograph detector -- but similar information was obtained. Both laboratory lasers and FELs are making rapid advances, and for now the techniques are complementary.

\subsection{Photo-excited acetylacetone} \label{ssec:pump-probe acetylacetone}
\begin{figure}[ht]
    \centering
    \includegraphics[width = 0.3\textwidth]{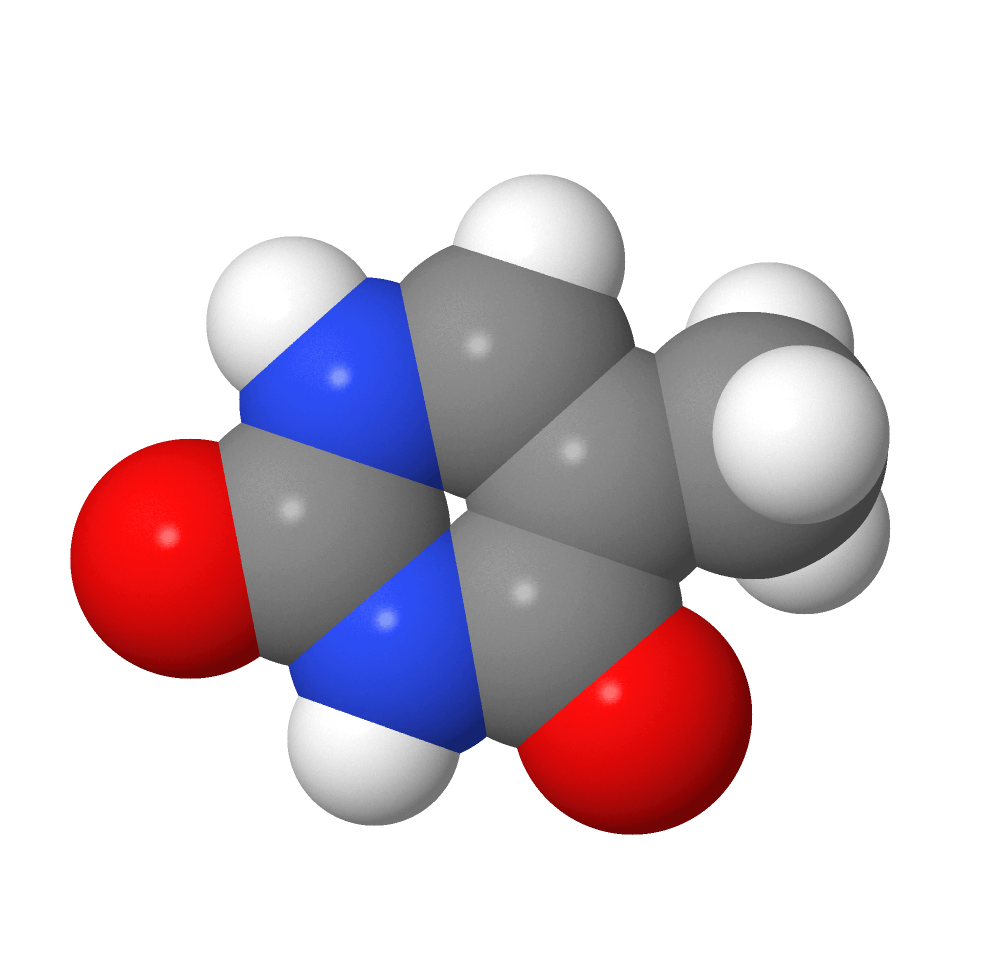}
    \includegraphics[width = 0.3\textwidth]{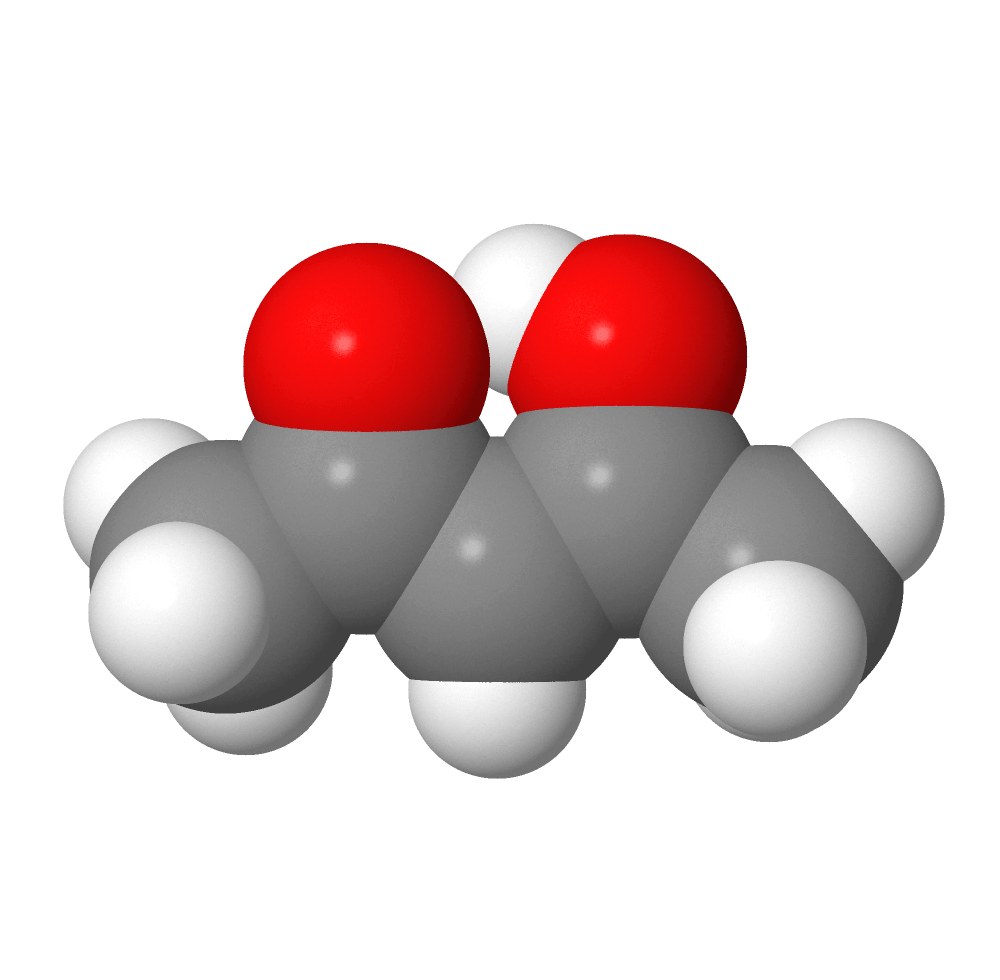}
    \includegraphics[width = 0.3\textwidth]{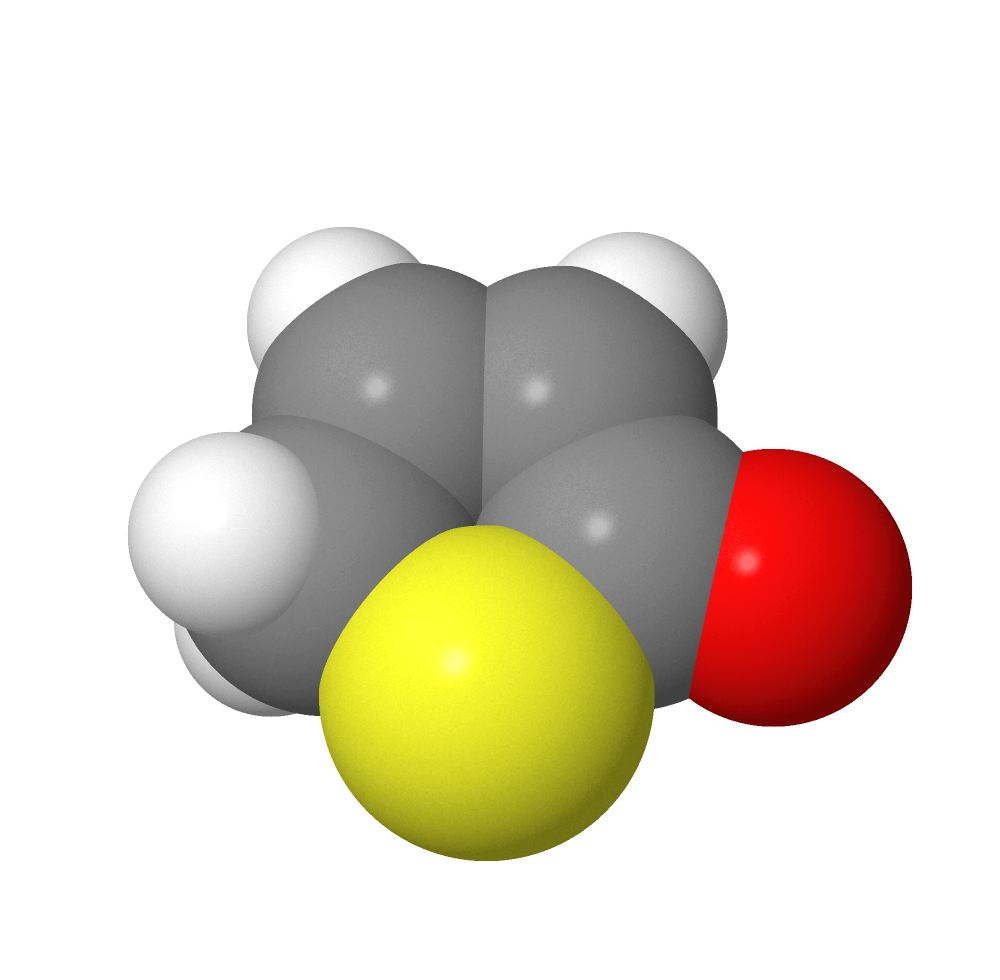}
    \caption{Schematic structure of thymine (left), acetylacetone (middle), 2(5H)-Thiophenone (right). Colours represent atoms of: grey, carbon; blue, nitrogen; red, oxygen; yellow, sulfur; white, hydrogen. Generated with Jmol: an open-source Java viewer for chemical structures in 3D. \url{http://www.jmol.org}}
    \label{fig:three_molecules}
\end{figure}

XUV FEL light can also be used to probe the valence band, and here we discuss the example of photodynamics of acetylacetone, see Fig.~\ref{fig:three_molecules}. This representation portrays the position of the hydrogen between the two carbon atoms as asymmetric, but note that it can also be considered as effectively symmetric \citep{Feyer2018}.
\citet{Squibb2018} pumped this molecule with 261 nm radiation to the S$_2$ state, which has  $\pi\pi^*$ character, and probed the valence band spectrum using XUV radiation (24.0 eV, 51.57 nm) as a function of time after the pump. In core level spectroscopy, the identity of the emitting atom is evident (carbon, oxygen, etc) and its chemical state can be determined. For valence spectroscopy, theoretical calculations are necessary to determine the identity of a given feature in a spectrum.

%\begin{figure}[ht]
%    \centering
%    \includegraphics[width = 0.3\textwidth]{figures/AcAc_fat.png}
%    \caption{Schematic structure of acetylacetone. Colours represent atoms of: grey, carbon; red, oxygen; white, hydrogen. Generated with Jmol: an open-source Java viewer for chemical structures in 3D. \url{http://www.jmol.org}}
%    \label{fig:AcAc}
%\end{figure}

\begin{figure}[ht]
    \centering
    \includegraphics[width=0.45\textwidth]{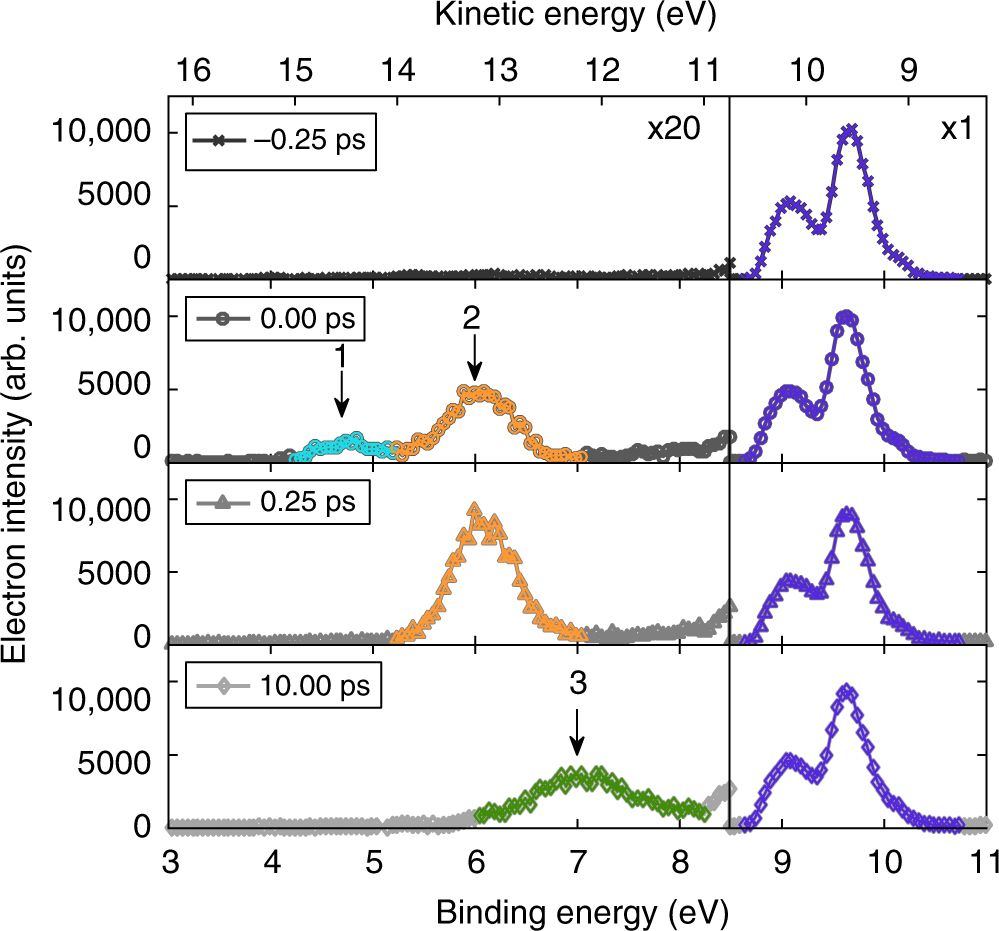}
    \hspace{0.08\textwidth}
    \includegraphics[width=0.45\textwidth]{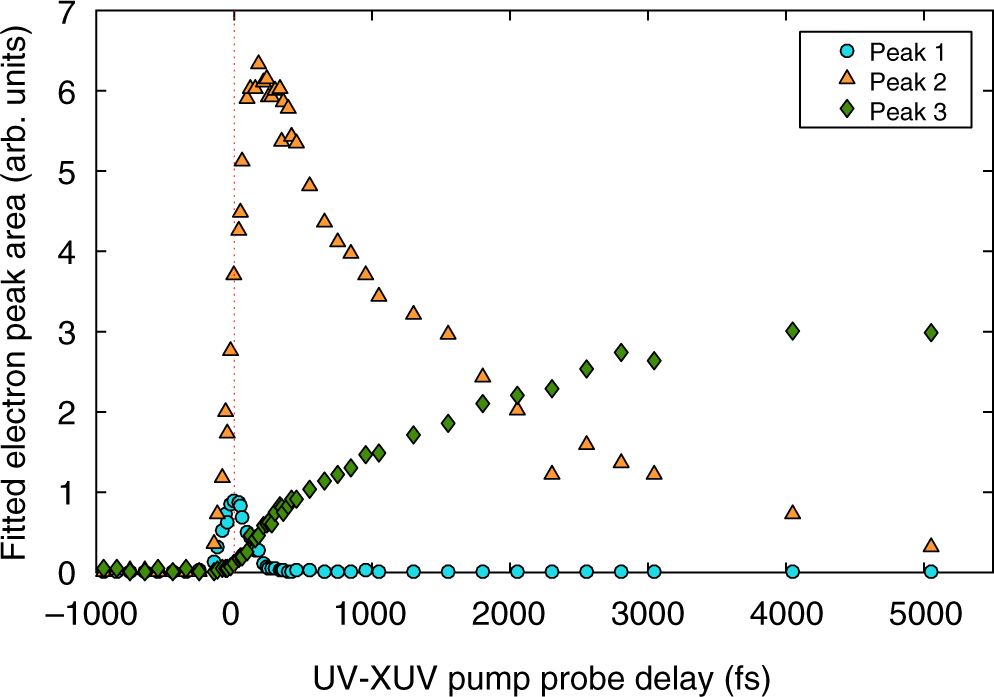}
    \caption{Photoelectron spectra and intensity versus time. Reproduced from~\cite{Squibb2018}; copyright 2018 by the Authors. The original figures have been published under a Creative Commons Attribution 4.0 license (CC BY) \url{http://creativecommons.org/licenses/by/4.0/}}
    \label{fig:Squibb}
\end{figure}

After excitation, a number of peaks appear in the spectrum at lower binding energy than the ground state, so they are background free. Extra features may appear and overlap the ground state spectrum, however the excited state population is usually only a few $\%$ and the spectral features may be masked by the noise of the ground state.
In Fig.~\ref{fig:Squibb}, the valence spectrum shows three clear features at lower binding energy than the ground state (purple features), and these are plotted in colours light blue (peak 1), orange (peak 2) and green (peak 3). 
Their integrated intensities are plotted on the right as a function of time. Peak 1 appears immediately on excitation, then vanishes; it is identified from calculations as the S$_2$ $\pi\pi^*$ state, which converts rapidly to an S$_1$ state, peak 2. 
This has $n\pi^*$ character, and then decays more slowly to a triplet T$_1$, peak 3, from which further decay occurs over long time scales, for example by fragmentation to yield methyl and hydroxyl radicals.

The theoretical description of this process is obviously complicated, and requires the calculation of the electronic structure of the ground and excited states, their ionization spectra, and the potential energy surfaces. 
These were compared with the experimental values of electron binding energies, time constants and ion yields to produce a detailed picture of the dynamics. A feature of this work was the high signal levels and high signal-to-noise ratio achieved, compared to laboratory HHG studies. 
However as laboratory sources continuously improve, we may expect that they will become competitive.

\subsection{Isomerization of thiophenone} \label{sssec:thiophene}

Besides fragmentation, there are many other important photo-induced chemical reactions, such as isomerization, and one of these is ring-opening, in which one of the bonds in a cyclic compound is broken. A well-known example is the synthesis of vitamin D in humans, one step of which involves exposure of the skin to sunlight, which in turn induces a ring-opening reaction. Many model systems of ring-opening reactions have been studied by a wide variety of techniques, and here we discuss a valence band photoemission study of the dynamics of thiophenone, Fig. \ref{fig:three_molecules}, using FEL light \citep{Pathak2020b}.

%\begin{figure}[ht]
%    \centering
%    \includegraphics[width = 0.3\textwidth]{figures/thiophenone_fat.png}
%    \caption{Schematic structure of thiophenone. Colours represent atoms of: grey, carbon; red, oxygen; yellow, sulfur; white, hydrogen. Generated with Jmol: an open-source Java viewer for chemical structures in 3D. \url{http://www.jmol.org}}
%    \label{fig:thiophenone}
%\end{figure}

The methods and apparatus used were similar to those described in the previous section, namely a magnetic bottle spectrometer at the Low Density Matter beamline of FERMI. High level calculations of electronic structure and molecular dynamics were carried out to understand the numerous possible pathways to final photoproducts, via a complex multi-dimensional potential energy landscape. 
The ring opening occurs via the decay of the initially excited S$_2$ state, via the S$_1$ state, to the electronic ground state, accompanied by the lengthening and later scission of the S-CO bond. The initial electronic energy due to photoexcitation is converted to vibrational energy, and at least 3 open-ring structures were identified, some of which may re-form the initial thiophenone molecule. 
In this work, a key point was that it was possible to identify a range of possible photo products by correlating experimental and theoretical results.

\subsection{Dynamics of photo-excited He droplets} \label{ssec:pump-probe droplets}
\citet{Mudrich2020} studied He nanodroplets and directly probed the excited state with a UV pulse (266 nm) and measured the outgoing photoelectrons. 
By working at much lower FEL intensity than the maximum available, they produced a low population of excited atoms in each droplet.  They concentrated on the case of a single excitation per droplet and were able to follow in detail the rapid ($<0.25 $~ps) 1s2p$\rightarrow$1s2s interband relaxation, the slow ($\sim 1$~ps) rearrangement of the local environment (bubble formation), and the ultimate bursting of the bubble (several ps). 
The experimental results were supported by detailed time-dependent density functional theory calculations; note that in this case, the density in question is that of the helium, not of the electrons, and the density functional is a bosonic one, specialized for $^4$He~\citep{Ancilotto2017}.
\section{Theoretical approaches} \label{sec:Theory}

Many reviews and books are devoted to the theory of atoms and molecules in a radiation field~\citep[for example:][]{Manakov1986,Faisal1987,Fedorov1998,Lambropoulos1998,Delone2000,Joachain2012}.
However describing dynamics of many-electron systems interacting with atto- and femtosecond laser pulses is a challenge for theory.
Some recent reviews of theoretical approaches for multielectron dynamics in pulsed electromagnetic fields are given 
in~\citet{Ishikawa2015,Palacios2015,Palacios2020}, as well as in the
book by~\citet{Vrakking2018}. Here we briefly outline modern theoretical methods in treating photoionization with FELs in the XUV domain.
Inevitably, many excellent works on the interaction of radiation pulses in the visible and infrared regions with atoms and molecules will 
not be covered here. Atomic units are used throughout this section.

\subsection{Amplitudes and description of physical quantities} \label{ssec:Amplitudes}

In quantum theory a physical quantity (`observable') after a reaction is obtained from the (complex) amplitudes of the corresponding process.  Thus, evaluation of the amplitudes is a key element of theory and the  approximations used in these evaluations are the most important characteristics of the theoretical approach. Hereafter we limit ourselves to processes where the field may be described classically, and matter is described by quantum mechanics. Such an approach is widely used and justified when the number of photons in a mode of the radiation field is high.

We denote by $\ME{f}{\hat{T}}{i}$ the transition amplitude from an initial state described at time $t_0$ by a state vector (wave function) $\ket{i}$ with a set of quantum numbers \{i\}, to a final state $\ket{f}$ with a set of quantum numbers \{f\}. The $\hat{T}$-operator is the infinite time limit of the evolution operator $\hat{U}(t,t_0)$, which may be formally written as
\begin{equation} \label{eq:unit}
\hat{T} = \lim_{t \rightarrow \infty} \hat{U}(t,t_0) = \lim_{t \rightarrow \infty} {\cal T} \exp \, \left[- i \int_{t_0}^t dt^\prime \, \hat{H}(t^\prime) \right] \,. 
\end{equation}
Here the exponential of an operator is defined in terms of its Taylor series
expansion, while ${\cal T}$ is a time-ordering operator, which
by definition orders operators in the Taylor series 
chronologically~\citep{Tannor2007}.

The choice of the sets of quantum numbers \{i\} and $\{\mathrm{f}\}$ is largely a matter of convenience. Their selection is normally dictated either by conservation laws or by features of the experimental setup. Generally, the system, both in the initial and the final state is in a mixed state, which cannot be described by a single state vector, but rather by the density matrices $\ME{i}{\hat{\rho}}{i'}$ and $\ME{f}{\hat{\rho}}{f'}$, respectively. The density matrix of the final state is found from the density matrix of the initial state in terms of the amplitudes:
\begin{equation} \label{eq:den}
\ME{f}{\hat{\rho}}{f'} = \sum_\mathrm{ii'} \ME{f}{\hat{T}}{i} \, \ME{f'}{\hat{T}}{i'}^{\ast} \, \ME{i}{\hat{\rho}}{i'}. 
\end{equation}

Standard quantum mechanical prescriptions are used to obtain a physical quantity from the known density matrix of the final state, for example, by taking a trace over unobserved quantum numbers and taking diagonal elements for observed characteristics. However this recipe is not always appropriate for mixed states. The concept of efficiency matrix of a detector or system of detectors $\ME{f}{\hat{\epsilon}}{f'}$ is more 
general~\citep{Devons1957,Ferguson1965}. The probability to detect a mixed state is then expressed as $W = {\rm Tr} \, [\hat{\rho} \, \hat{\epsilon}]$. Irreducible components of the density matrix of the angular momentum (equivalent terms such as ``statistical tensors'', ``state multipoles'' and some others are used) and the corresponding efficiency tensors of the detector, in combination with the partial-wave amplitudes are especially convenient in treating different vector correlations. Many expressions for the observable quantities in terms of statistical tensors and the amplitudes can be found in~\citet{Balashov2000}.

The amplitudes $\ME{f}{\hat{T}}{i}$ define an important quantity, the time delay (sojourn time).
A scattered particle is `delayed' in time relative to the free propagating particle. This effect is extensively discussed in the literature~\citep[see, e.g., the reviews by][]{SassolideBianchi2012,Maquet2014,Pazourek2015}. In the formulation of \citet{Eisenbud1948}, \citet{Wigner1955} and \citet{Smith1960}, this EWS time delay in the channel $\ket{f}$ is expressed in terms of the derivative
\begin{equation} \label{eq:ews}
\tau_\mathrm{f} = \frac{d}{d \, E} \arg \, \ME{f}{\hat{T}}{i} \, ,
\end{equation}
where $E$ is the energy of the particle. The concept of the EWS time delay has been extended to photoionization, as a half collision process~\citep{Schultze2010,Dahlstrom2012,Maquet2014,Pazourek2015}. It is widely used currently in studies of photoprocesses on the attosecond time scale. Section~\ref{sec:ews} of the present review gives more details and further references on time delays in photoionization.

\subsection{Time-dependent Schr\"odinger equation (TDSE)} \label{ssec:TDSE}

The time evolution of pure states is described, in the nonrelativistic case, by the time-dependent Schr\"odinger equation (TDSE). In the coordinate representation
\begin{equation} \label{eq:tdse}
i \frac{\partial}{\partial t} \Psi(\xi, t) = \hat{H}(\xi,t) \, \Psi(\xi,t)
\end{equation}
with the initial condition $\Psi(\xi,t=t_0) = \Phi_i(\xi)$. In the present case of an external classical electromagnetic field, 
\begin{equation} \label{eq:ham}
\hat{H}(\xi,t) = \hat{H}_0(\xi) + \hat{V}(\xi,t), 
\end{equation}
where $\hat{H}_0(\xi)$ is the Hamiltonian of the system, $\hat{V}(\xi,t)$ is its interaction with the field, and $\xi$ is a set of particle coordinates of the target. In molecular problems, $\hat{H}_0(\xi)$ is further subdivided into nuclear and electronic parts and electron-nuclei interaction.
The Born-Oppenheimer approximation is often applied, which assumes that the electron motion is much faster than the nuclear motion.
Moreover, for pulses as short as a few femtoseconds, the nuclei can be treated as fixed in space during the pulse.

In the vast
majority of studies the interaction is taken in the long-wave electric dipole (E1) approximation 
$\hat{V}(\xi,t) = \mathbf{E}(t)  \! \cdot  \! \mathbf{D}(\xi)$, where $\mathbf{E}(t)$ is the electric field, $\mathbf{D}(\xi) = \sum_j e_j \mathbf{r}_j$
is the electric dipole operator; $e_j$ and $\mathbf{r}_j$ are the electric charge and radius vector of the particle $j$. In the fixed-nuclei approximation the nuclei are not affected by the external electromagnetic field, the summation is over electrons, and  $e_j = e$. Another frequently used and physically equivalent formulation of the interaction in terms of the vector potential $\mathbf{A}(t)$ uses
$\hat{V}(\xi,t) = - \frac{1}{c} \mathbf{A}(t) \! \cdot \! \sum_j \mathbf{p}_j$, where $\mathbf{p}_j$ is the linear momentum of the particle $j$.
The TDSE~(\ref{eq:tdse}) is equivalent to 
\begin{equation} \label{eq:eq}
\Psi(\xi,t) = {\cal T} \exp \left[- i \, \int_{t_0}^t
d t' \, \hat{H}(\xi,t')  \right] \Phi_i(\xi)
\end{equation}
and the amplitude is expressed (see Eq.~(\ref{eq:unit}))
as the projection $\ME{f}{\hat{T}}{i} = \Skalar{f}{\Psi(\mathit{t \rightarrow \infty})}$. 
Approximating the evolution operator in Eq. (\ref{eq:eq}), i.e., propagating the wave function in time, is a key problem for solving the TDSE. This problem is on the border of theoretical physics and theory of numerical methods and has been discussed for decades~\citep[e.g.,][]{Moler2003,Bandrauk2013,Han2019}. 
Not all methods aim to find the wave function $\Psi(t \rightarrow \infty)$, for example, the time-dependent density functional theory (DFT). 
The DFT is applied to calculations of wave packets in continua, in combination
with expansion methods (see Section~\ref{ssec:Expansion}).

\subsection{Solution of the TDSE on the space-time grid} \label{ssec:Grid}

Modern computing capabilities make it easy enough to cope with solving the TDSE (\ref{eq:tdse}) on the space-time grid for one or two active electrons. 
Hundreds of papers using this approach have been published since the first such calculations for one-~\citep{Kulander1987} and two-electron~\citep{Smyth1998,Parker2001} systems. 
The three active electron problem has also been solved~\citep{Colgan2012} and attempts to solve the TDSE on the grid for four electrons have  begun~\citep{Pindzola2019}. 
In fact the multidimensional TDSE is solved on the grid, especially for two or more active electrons, not directly, but after expansion of the solution in a complete basis in part of the variables. For example, in atomic problems expansions in spherical harmonics are used,
\begin{equation} \label{eq:one}
\Psi(\mathbf{r},t) = \sum_{lm} P_{lm}(r,t) \, Y_{lm}(\Omega) \,,
\end{equation}
where $Y_{lm}(\Omega)$ is a spherical harmonic, for a single active electron, or
\begin{equation} \label{eq:two}
\Psi(\mathbf{r}_1, \mathbf{r}_2,t) = \sum_{l_1 l_2 LM} P_{l_1 l_2}^{LM}(r_1,r_2,t) \, Y_{l_1l_2}^{LM}(\Omega_1,\Omega_2) \,,
\end{equation}
where  $Y_{l_1l_2}^{LM}(\Omega_1,\Omega_2)$ is a bipolar spherical harmonic, for two active electrons, etc. Then Eq.~(\ref{eq:tdse}) reduces to the set of coupled partial differential equations (close-coupling equations) for the radial functions $P_{lm}(r,t)$
and $P_{l_1 l_2}^{LM}(r_1,r_2,t)$, respectively, which are then solved on the one or two dimensional space grids with time propagation. Basis sets were applied to separate part of the variables in prolate spheroidal coordinates for the H$_2$ molecule~\citep{Awasthi2005,Guan2011}.
Two- and three dimensional grids in cylindrical coordinates without basis expansions were used in solving the TDSE for single-electron problems~\citep{Kulander1987,Yuan2019}.

The solution of the TDSE on the grid can be obtained in a wide range of the field frequencies and intensities for pulses with various envelopes and relative carrier-envelope phases. However, direct numerical solution on a space-time grid is a computationally laborious method with rapidly increasing computational cost not only with increasing number of spatial dimensions, but also with the pulse length and field intensity. Furthermore, essential properties of many-electron atoms and molecules are neglected. Only part of them can be taken into account by effective local potentials for the target description. Nevertheless, the method may be useful in simple cases for many-electron atoms too, when the process does not involve the inner shells
and the continuum is flat. Appropriate examples are~\citet{Douguet2017} and~\citet{Gryzlova2018,Gryzlova2019}, 
where photoelectron angular distributions in ionization by XUV pulses in the region of intermediate excited states were described.

\subsection{Expansion methods} \label{ssec:Expansion}

In this class of methods, instead of solving Eq.~(\ref{eq:tdse}) on the space-time grid, the series expansions of the solution in basis functions in the entire space of the variables are applied:
\begin{equation} \label{eq:expa}
\Psi(\xi, t) = \sum_n c_n(t) \, \phi_n(\xi) \,,
\end{equation}
where $\phi_n(\xi)$ is a basis set and $c_n(t)$ are unknown coefficients. Substitution of Eq.~(\ref{eq:expa}) into Eq.~(\ref{eq:tdse}) reduces the latter to a set of ordinary differential equations (equations of motion) depending on the coefficients $c_n(t)$. 
In practice different basis sets $\phi_n(\xi)$ are used, giving rise to a variety of methods, which differ by the basis set chosen and technical realization applied. 
A natural choice is the set of eigenfunctions of the field-free Hamiltonian $\hat{H}_0(\xi)$: $\hat{H}_0(\xi) \, \phi_n(\xi)=\epsilon_n \phi_n(\xi)$ \citep{Mercouris1994,Mercouris2010,Mercouris2016}. 
Further decomposition of the functions $\phi_n(\xi)$ may be performed, e.g. in B-splines~\citep{Martin1999,Bachau2001}, Lagrange interpolating 
polynomials~\citep{Artemyev2015,Muller2018}, or  Coulomb-Sturmian functions~\citep{Foumouo2006}. 
The critical points are the choice of states to include in the basis, and the estimate of the number of basis functions necessary to reproduce the essential physics of the problem. 

In the time-dependent Feshbach close-coupling
method, the wave function is split by the projection operators
$\hat{\cal P}$ and $\hat{\cal Q}$ ($\hat{\cal P} + \hat{\cal Q} = \hat{1}$,
$\hat{\cal P}^2 = \hat{\cal P}$, $\hat{\cal Q}^2 = \hat{\cal Q}$, 
$\hat{\cal P} \hat{\cal Q} = \hat{\cal Q} \hat{\cal P} = 0$) as
\begin{equation} \label{eq:fesh}
\Psi(\xi, t) =  \hat{\cal P} \, \Psi(\xi, t) + \hat{\cal Q} \, \Psi(\xi, t)
\end{equation}
with subsequent series expansion of both terms in Eq.~(\ref{eq:fesh}), similar
to Eq.~(\ref{eq:expa}).
The operators $\hat{\cal P}$ and $\hat{\cal Q}$ project on two subspaces,
coupled by a residual interaction. This approach is successfully used 
in calculations of molecular photoionization by short pulses beyond 
the Born-Oppenheimer 
approximation~\citep{Palacios2015,Palacios2020}.

In the ``time-dependent configuration-interaction singles method''~\citep{Greenman2010,Hochstuhl2012,Karamatskou2014,Sato2018a} only the Hartree-Fock ground state and one-particle-one-hole excitations with respect to this state are included in the basis. 
The corresponding basis functions are built as time-independent Slater determinants. Thus, it includes an essential part of many-electron effects, which is a big step in comparison with the models which use the effective potential. 
The method was applied to studies of energy- and angle-resolved photoelectron spectra in one-photon and above-threshold ionization of argon following strong XUV
irradiation~\citep{Karamatskou2014,Goetz2016}, and the multi-photon process of above-threshold ionization for the light elements in the hard X-ray regime~\citep{Tilley2015}. The method can be further improved by inclusion of two-particle-two-hole states in the expansion Eq.~(\ref{eq:expa})~\citep{Bauch2014}. 
The methods based on the expansion Eq.~(\ref{eq:expa}) may be characterized as configuration interaction methods with a fixed number of certain stable configurations mixed in a time-dependent way.

The multicofigutaion time-dependent Hartree-Fock (MCTDHF) method is a further extension, where the basis functions in Eq.~(\ref{eq:expa}), expressed as Slater determinants, depend on time, $\phi_n(\xi) = \phi_n(\xi,t)$ through the time-dependent electron 
orbitals \citep{Beck2000,Zanghellini2003,Caillat2005,Kato2004,Nest2009,Alon2009,Haxton2011,Haxton2015}.
In this case, equations for $c_n(t)$ and $\phi_n(\xi,t)$ are obtained not directly from Eq.~(\ref{eq:tdse}) but by using the time-dependent variational principle
(TDVP) leading to the relation:
\begin{equation} \label{eq:vari}
\delta \, \ME{\Psi(t)}{\hat{H}(t)}{\Psi(t)} - i \left( \Skalar{\delta \Psi(t)}{\frac{\partial \Psi(t)}{\partial t}} \,
- \Skalar{\frac{\partial \Psi(t)}{\partial t}}{\delta \Psi(t)} \right) = 0 \,.
\end{equation}
The MCTDHF method considers all the possible Slater determinants for a given number of orbitals. Thus, though powerful, the computational time for solving the corresponding set of equations for the time-dependent coefficients and orbitals scales factorially with the number of electrons. Therefore this approach is hardly tractable for systems with a large number of electrons. A more flexible method was introduced by \citet{Sato2013} and by \citet{Miyagi2013}, where orbital sub-spaces are selected, and are treated in different approximations, for example, time-independent tightly bound core electrons and active electrons to describe the ionization. Various options of this method may be applied~\citep{Sato2015}. For an extended list of references, see \citet{Omiste2019}, while more details of this method are given in
Section~\ref{ssec:ab initio}.

In the time-dependent R-matrix approach~\citep{Burke1997}, an expansion of the following type is used 
\begin{equation} \label{eq:rmat}
\Psi(x_1,x_2,...,x_{N+1},t) =  \hat{A} \sum_n \phi_n(x_1,x_2,...,\Omega_N, \sigma_N) \, r_{N+1}^{-1} \, \psi_n(r_{N+1},t) \,,
\end{equation}
where $\hat{A}$ is the antisymmetrization operator, $x = \{ \mathbf{r}, \sigma \}$ denote space $(\mathbf{r} = \{r,\Omega \})$ and spin ($\sigma$) coordinates, $\phi_n$ is a basis (channel) function formed by coupling of the $N$-electron residual state with the angular and spin parts of the outer electron wave function, and $\psi_n$ is the radial function that describes the electron motion in the $n$-th channel. 
Substitution of Eq. (\ref{eq:rmat}) into Eq. (\ref{eq:tdse}) gives a set of integro-differential equations for the radial time-dependent functions $\psi_n$. A characteristic of the method is partitioning of the configuration space into internal and external (asymptotic) regions. 
Account is taken of electron correlations and channel coupling in the internal region, and a local long-range atomic potential, together with the laser field, are assumed in the external region. The solutions in the two regions are matched.

The details of the method are presented in the literature, where versions of the time-dependent R-matrix approach are being developed~\citep{vanderHart2007,Guan2007,Nikolopoulos2008,Lysaght2009,Moore2011,Hutchinson2013,Feist2014,Wragg2015}. A simple option for one-electron systems is discussed in~\citet{Broin2017}. 
Despite the computational awkwardness, this method uses numerical codes developed over years for stationary R-matrix theory and can be successful in the domain of its applicability~\citep{Zatsarinny2013}, although the boundaries of this domain are still to be established. For many-electron atoms examples of the application of the time-dependent R-matrix method so far include ionization of Ne and 
Ar~\citep{vanderHart2007,vanderHart2008,Guan2007,Guan2008,Lysaght2008,Moore2011,Hutchinson2013,Chen2019}. The method is under development for molecules~\citep{Broin2015}. 

\subsection{Perturbation theory} \label{ssec:PT}

It is often assumed that for ionization in the XUV region, perturbation theory (PT)  is sufficiently accurate to describe the interaction $\hat{V}(\xi,t)$. This assumption is justified in many cases of interest, since the ratio of the ponderomotive electron energy to the photon energy decreases as the inverse cube of the photon frequency. 
Therefore, while in the IR and visible range for intensities of the order 10$^{13}$-10$^{14}$ W/cm$^{2}$, the field is considered as strong, and processes take place which cannot be described within the perturbation theory (such as tunneling and high harmonic generation), the lowest possible order PT gives an adequate description of photoionization for these intensities in the XUV. 
However, at a given intensity, the perturbation theory may become unreliable near resonances.

For pulses with finite time duration, the first and second order PT amplitudes are, respectively, of the form
\begin{equation} \label{eq:pt1t}
\ME{f}{\hat{T}}{i}_1 = -i \int_{-\infty}^{\infty} dt \exp \left[i \, (\epsilon_f - \epsilon_i)\, t \right]
\ME{f}{\mathbf{E}(t) \! \cdot \! \mathbf{D}}{i} \,,
\end{equation}
and
\begin{eqnarray} \label{eq:pt2t}
\ME{f}{\hat{T}}{i}_2 & = & i^2 \sum_p \int_{-\infty}^{\infty} dt \exp \left[i \, (\epsilon_f - \epsilon_p) \, t \right]
\ME{f}{\mathbf{E}(t) \! \cdot \! \mathbf{D}}{p}   \nonumber \\
 & & ~~~~~~~~ \times
 \int_{-\infty}^{t} dt' \exp \left[i \, (\epsilon_p - \epsilon_i) \, t' \right]   \ME{p}{\mathbf{E}(t') \!\cdot \! \mathbf{D}}{i} \,,
\end{eqnarray}
where the summation is over eigenstates of the field-free Hamiltonian, including integration over the continuum. In practice, most
essential near-resonant intermediate states are included in Eq.~(\ref{eq:pt2t}). The advantages of the second order PT are its flexibility in selecting and analysing individual reaction channels, straightforward procedure for improving the target description and low computational cost. 
The PT amplitudes explicitly show simple tensorial structure, which is very convenient in analytical transformations, especially for treating vector correlations. Comparison with non-perturbative calculations, for example, with the solution of the TDSE, clarifies the domain of applicability of the PT~\citep{Grum2015,Douguet2016}.

In the limit of infinite pulses with constant amplitudes, the time integrals in Eq. (\ref{eq:pt1t}) and Eq. (\ref{eq:pt2t}) are taken and become the relationships representing energy conservation. 
For the second and higher order amplitude in this limit, the summation over the intermediate states can be performed using additional approaches, for example by using Green's function~\citep{Manakov1986,Maquet1998}
or by variational procedures~\citep{Gao1989,Machado2004,
Douguet2017}. In this limit, developed program packages for the first-order PT amplitudes and characteristics of the atomic electron emission are available, based on an R-matrix approach [RMATRX~\citep{Berrington1995}; BSR~\citep{Zatsarinny2006}], multiconfiguration Dirac-Fock [RATIP~\citep{Fritzsche2001,Fritzsche2012}], Hartree-Fock [MCHF~\citep{Froese-Fischer1997}] approximations, and the random phase approximation with exchange [ATOM,~\citep{Amusia2016}]. Also for molecules, various methods are used for calculating the first-order amplitudes, for example: approaches based on the DFT with B-splines~\citep{Toffoli2002,Stener2007},
including vibrationally resolved 
photoionization~\citep{Plesiat2012,Engin2019}; single center 
methods~\citep{Demekhin2011a}; the random phase approximation~\citep{Semenov2000}; and the multichannel Schwinger configuration interaction method~\citep{Lucchese1983,Stratmann1995}.

The first order PT is extensively used in the calculation of few-photon, sequential ionization by FELs, especially the photoelectron angular distributions and angular correlations~\citep{Kheifets2007,Fritzsche2008,Gryzlova2010,Gryzlova2012,Gryzlova2014,Gryzlova2015}. When combined with the statistical tensor formalism, it conveniently takes into account polarization of intermediate ionic states. This approach was used in the description of sequential ionization of Ne~\citep{Kurka2009,Rouzee2011}, 
Ar~\citep{Fukuzawa2010,Gryzlova2011,Augustin2018,Gryzlova2019a}, Kr~\citep{Fritzsche2009}, 
and Xe~\citep{Fritzsche2011} in the XUV. The field was reviewed in~\citet{Grum2016}. Non-dipole effects in sequential two-photon double ionization were predicted in the framework of the PT~\citep{Grum2012,Grum2015a} and measured in argon at FERMI~\citep{Ilchen2018}.

The photoelectron angular distribution and various types of dichroism for the sidebands in two-photon, two-colour above-threshold atomic ionization were studied within the second order perturbation theory by \citet{Grum2014} taking account of the full multipole expansion of the radiation field and independently variable polarizations of the XUV and IR radiation beams. The calculations predicted nondipole effects in the sidebands for Ne 1s ionization, in some cases much larger than in the main photoelectron line.

\subsection{Strong-field-type approximations} \label{ssec:SFA}

The strong-field approximation (SFA)~\citep{Keldysh1965_en,Faisal1973,Reiss1980} has a large number of variations~\citep{Lewenstein1994,Quere2005,Maquet2007,Kazansky2006,Galan2013,Boll2014}; see also the reviews by~\citet{Ivanov2005}, and by~\citet{Karnakov2015}. This approximation assumes that the photoelectron moves as a particle in the electromagnetic field, while the atomic or molecular potential is considered as a perturbation. In most applications, the latter is neglected and the electron is described by the non-relativistic Volkov wave function~\citep{Volkov1935}
\begin{equation} \label{eq:volk1}
\psi_{\mathbf p}(\mathbf{r},t) = \exp \{ i [{\mathbf p} - {\mathbf A}(t)] \mathbf{r} - i \Phi( \mathbf{p},t) \} \,,
\end{equation}
where $\mathbf{p}$ is the electron momentum, $\mathbf{A}(t)$ is vector potential of the strong field and
\begin{equation} \label{eq:volk2}
\Phi(\mathbf{p},t) = \frac{1}{2} \int_t^{\infty} dt' \left[\mathbf{p} - \mathbf{A}(t') \right]^2 \,.
\end{equation}
The SFA is used when the XUV radiation from the FEL ionizing the target is combined with a strong IR field. Interaction of the XUV field with the target is treated in the first-order PT. Therefore, to obtain the amplitude in the SFA, the wave function (Eq.~\ref{eq:volk2}) combined with the wave function of the residual ion is substituted in Eq.~(\ref{eq:pt1t}) as the final state $\ket{f}$. The domain of applicability of this approach is usually limited to high energies of the outgoing electrons in comparison with the energy of the IR photon. The Coulomb interaction between the photoelectron and the residual ion, which modifies the final state wave function (Coulomb-Volkov approximation) is generally important and may be taken into account. An extensive literature on the effects of the Coulomb field on the ionization process within the strong-field approach is available \citep[see, e.g.,][for further references]{Goreslavski2004,Karnakov2015}. The SFA is regularly used in description of the sidebands of the main photoelectron line in XUV + IR ionization by atto- and femtosecond pulses~\citep{Kazansky2010,Kazansky2011,Galan2013,Picard2014}.
It allows significant advances in developing analytical derivations, in particular, closed-form expressions for the angular distributions of photoelectrons in  two-colour XUV + IR ionization were obtained~\citep{Boll2016}.

\subsection{Density matrix approach} \label{ssec:Density}

The time evolution of the density matrix is described by the Liouville equation~\citep{Blum2012}:
\begin{equation} \label{eq:liou}
i \frac{\partial \hat{\rho}(t)}{\partial t} = \left[ \hat{H}(t), \, \hat{\rho}(t) \right] \,,
\end{equation}
which is written here in the operator form with $\hat{\rho}(t)$ being the density operator. The density matrix formalism consistently takes into account coupling to an external reservoir and relaxation processes. Relaxation is especially important in studies of the inner-shell excitation/ionization by XFELs on the femtosecond time scale. This scale
 is usually determined by the lifetime of the hole decaying by Auger decay, and competing with photoionization by the XFEL pulse. Furthermore, considering the reservoir with short correlation time, i.e., one which quickly ``forgets'' its interaction with the system,
 allows the introduction of the Markov approximation~\citep{Blum2012,CohenTannoudji2004,Lambropoulos2007}.
 Together with the rotating wave approximation, it simplifies
 drastically Eq.~(\ref{eq:liou}) up to the master equations
 for the elements of the density matrix~\citep{Zwanzig1964,Haake1973} 
 and, moreover, eliminates the density matrix elements coupling the 
 continuum states~\citep{Hanson1997}.
 Solutions of the master equations are repeatedly used in treating problems related to the processes generated by 
 FELs~\citep{Sun2010,Nikolopoulos2011,Middleton2012,Rohringer2012a}. 
 For example,~\citet{Nikolopoulos2013} calculated the sequential two-photon
 double ionization of neon by a FEL pulse and successfully reproduced the angular correlation function of the two emitted electrons, 
 while~\citet{Nikolopoulos2015} revealed a role of a small admixture of 
 FEL harmonics on photoionization.

After making some additional approximations~\citep{Blum2012}, Eq.~(\ref{eq:liou}) transforms into a set of rate equations for population of the levels of the system
\begin{equation} \label{eq:rate}
\frac{d N_n(t)}{d t} = \sum_{n' \neq n} W_{n' \rightarrow n} N_{n'}(t) - N_n(t) \sum_{n' \neq n} W_{n \rightarrow n'} \,,
\end{equation}
where $N_n$ is the population of level $n$ and $W_{n \rightarrow n'}$ is the transition rate from level $n'$ to level $n$. Generally, 
$W_{n \rightarrow n'}$ is also time dependent, as in the case of photoionization when the rate varies with the field intensity during the pulse.
The rate equations is a convenient tool for calculating ionic charge
distributions and state populations within and after the pulse. 
They are extensively used in the description of sequential multiple ionization by XFEL pulses of atoms~\citep[for example,][]{Makris2009,Nakajima2002, Son2011,Son2012,Lorenz2012,Lambropoulos2013,Lunin2015,Serkez2018,Buth2018} and molecules~\citep{Liu2016,Inhester2016}.

\subsection{Real-time {\it ab initio} simulations: Time-dependent multiconfiguration self-consistent-field method} 

\label{ssec:ab initio}

Here we give a brief description of the time-dependent complete-active-space self-consistent field (TD-CASSCF) method
\citep{Sato2013,Sato2016,Sato2018} and the time-dependent occupation-restricted multiple-active-space (TD-ORMAS) method \citep{Sato2015}. 
The dynamics of the laser-driven $N$-electron system is described by the 
TDSE~(\ref{eq:tdse})
\begin{equation}
\label{eq:tdse2}
i\frac{\partial\Psi (t)}{\partial t} = \hat{H}(t)\Psi (t),
\end{equation}
where the time-dependent Hamiltonian is
\begin{equation}
\label{eq:TDSE1}
\hat{H}(t)=\hat{H}_1(t)+\hat{H}_2,
\end{equation}
with the one-electron part
\begin{equation}
\label{eq:H1}
\hat{H}_1(t) = \sum_{i=1}^{N} \hat{h}({\bf r}_i,t) 
\end{equation}
and the two-electron part
\begin{equation}
\label{eq:H2}
\hat{H}_2 = \sum_{i=1}^{N} \sum_{j < i} \frac{1}{|{\bf r}_i - {\bf r}_j|}.
\end{equation}
The one-body Hamiltonian $\hat{h}({\bf r},t)$ for an atomic system 
within the dipole approximation is given in the length gauge by,
\begin{equation}
\label{eq:length-gauge}
\hat{h}({\bf r},t) = \frac{{\bf p}^2}{2} + {\bf E}(t) \cdot {\bf r} -\frac{Z}{|{\bf r}|},
\end{equation}
and in the velocity gauge by,
\begin{equation}
\label{eq:velocity-gauge}
\hat{h}({\bf r},t) = \frac{{\bf p}^2}{2} + {\bf A}(t) \cdot {\bf p} -\frac{Z}{|{\bf r}|},
\end{equation}
where ${\bf A}(t) = -\int {\bf E}(t)dt$ is the vector potential, ${\bf E}(t)$ is the laser electric field, and $Z$ the atomic number.

The TD-CASSCF and TD-ORMAS methods belong to a class of {\it ab initio} approach called the time-dependent multiconfiguration self-consistent field (TD-MCSCF) methods, which express the total electronic wave function in the multiconfiguration expansion:
%%%%%%
\begin{equation}
\label{eq:MC_expansion}
	\Psi (x_1, x_2, \cdots, x_N, t) = \sum_{I} C_I(t) \, 
	\Phi_I(x_1, x_2, \cdots, x_N, t) \,,
\end{equation}
where $x=\{\bf r, \sigma \}$.
The electronic configuration $\Phi_I(x_1, x_2, \cdots, x_N, t)$ is a Slater determinant composed of spin orbital functions $\{\psi_p({\bf r},t) \times s(\sigma)\}$, where $\psi_p({\bf r},t)$ and $s(\sigma)$ denote spatial orbitals and spin functions, respectively. 
The coefficients $C_I$ and orbitals are both time dependent.

In the TD-CASSCF and TD-ORMAS methods the spatial orbitals are classified into three groups: doubly occupied and time-independent frozen core (FC), doubly occupied and time-dependent dynamical core (DC), and correlated active orbitals.
The total wave function is given by:
\begin{equation}
\label{eq:TD-CASSCF}
	\Psi(t) = \hat{A}\left[\Phi_{\mathrm{fc}}\Phi_{\mathrm{dc}}(t)
	\sum_I C_I(t) \Phi_I(t) \right],
\end{equation}
where $\hat{A}$ is the antisymmetrization operator, $\Phi_{\mathrm{fc}}$ and $\Phi_{\mathrm{dc}}$ the closed-shell determinants formed with numbers $n_{\mathrm{fc}}$ FC orbitals and $n_{\mathrm{dc}}$ DC orbitals, respectively, and $\Phi_I$ the determinants constructed from $n_{\mathrm{a}}$ active orbitals.
Whereas all the possible distributions of active electrons among active orbitals are considered in the TD-CASSCF method, the TD-ORMAS method further subdivides the active orbitals into an arbitrary number of subgroups, specifying the minimum and maximum number of electrons accommodated in each subgroup.
This decomposition permits us to significantly reduce the computational cost without sacrificing accuracy in the description of correlated multielectron dynamics.

The equations of motion describing the temporal evolution of the coefficients $C_I$ and the orbitals $\psi_p$ are derived using the 
TDVP Eq.~(\ref{eq:vari}) and read, suppressing the argument $t$ for notational convenience,
\begin{eqnarray}
    \label{eomci}
    i \frac{d}{dt} C_I &=& \sum_J \langle \Phi_I| \hat{H} - \hat{R}| \Phi_J \rangle  \,, \\
    \label{eomorb}
	i \frac{d}{dt} \ket{\psi_p} &=& \hat{h} \ket{\psi_p} + \hat{Q} \hat{F}  \ket{\psi_p} + \sum_{q} \ket{\psi_q}  R^q_p,
\end{eqnarray}
where $\hat{Q} = 1 - \sum_{q} \ket{\psi_q} \bra{\psi_q}$ the projector onto the orthogonal complement of the occupied orbital space.
$\hat{F}$ is a non-local operator describing the contribution from the interelectronic Coulomb interaction, defined as, 
\begin{equation}
	\hat{F} \ket{\psi_p} = \sum_{oqsr} (D^{-1})^o_p \, P^{qs}_{or} \, \hat{W}^r_s \, \ket{\psi_q},
\end{equation}
where $D$ and $P$ are the one- and two-electron reduced density matrices, and $\hat{W}^r_s$ is given, in the coordinate space, by,
\begin{equation}
	W^{r}_{s} \left({\bf r} \right) = \int d {\bf r}^\prime \frac{\psi_{r}^{*} ({\bf r}^\prime) \psi_{s} ( {\bf r}^\prime )}{| {\bf r} - {\bf r}^\prime | } .
	\label{eq:W}
\end{equation}
The matrix element $R^q_p$ is given by,
\begin{equation}
\label{eq:orbital-time-derivative}
	R^q_p = i \langle \psi_q | \dot{\psi_p} \rangle - h^q_p,
\end{equation}
with $h^q_p = \langle \psi_q|\hat{h}|\psi_p \rangle $. 
The elements $R^q_p$ within one orbital subspace (frozen core, dynamical core and each subdivided active space) can be arbitrary Hermitian matrix elements and set to zero. 
On the other hand, the elements between different orbital subspaces are determined by the TDVP. Their concrete expressions are given in~\cite{Sato2015}, where $iX^q_p = R^p_q + h^q_p$ is used for working variables. The numerical implementation for atoms with the use of infinite-range exterior complex scaling 
as an efficient absorbing boundary, and the time-dependent surface flux method 
for extracting angle-resolved photoelectron energy spectra, are desscribed in detail in \citet{Sato2016} and \citet{Orimo2018,Orimo2019}.

TD-CASSCF and TD-ORMAS can systematically control the accuracy, until numerical convergence, through the number of orbitals and flexible orbital subspace decomposition. 
With TD-CASSCF, especially, we can take account of as many electron excitations as we want. 
Both methods are gauge invariant by virtue of the time-dependent variationally optimized orbitals.
Moreover, TD-CASSCF is size extensive while TD-ORMAS is not, which is not relevant to the atomic case. 
Whereas the computational cost of TD-CASSCF scales factorially with the number of active electrons, 
TD-ORMAS has the advantage of polynomial cost scaling, enabling even more efficient simulations.

The methods described above and in Section~\ref{ssec:Expansion} are based on the idea of expressing the total wave function as a linear combination.
As an alternative, approaches based on the coupled-cluster expansion,
\begin{equation}
\label{eq:ccwfn}
\Psi = \exp(\tau^a_i\hat{E}^a_i+\tau^{ab}_{ij}\hat{E}^{ab}_{ij}+\tau^{abc}_{ijk}\hat{E}^{abc}_{ijk}+\cdots) \, \Phi,
\end{equation}
of the time-dependent wave function $\Psi (t)$ using time-dependent orbitals for electron dynamics have recently been developed.
Here, the operator
$\hat{E}^{abc\cdots}_{ijk\cdots}$ excites electron(s) from orbital(s) $i,j,k\cdots$, occupied in the
reference determinant $\Phi$ (hole), to those $a,b,c\cdots$, not occupied
in the reference (particle).
Einstein's summation convention is used for the orbital indices.
Both amplitudes $\{\tau^{abc\cdots}_{ijk\cdots}(t)\}$ and orbitals are propagated in time.
The methods are gauge invariant and extensive and scale polynomially with respect to the number of active electrons. Their details are presented, e.g., in \citet{Kvaal2012,Sato2018b,Pathak2020}.

\subsection{Theoretical proposals for XFEL experiments}  \label{ssec:Predicted_phe}
Above we have reviewed experiments using light from 
narrow bandwidth FELs, and the powerful theoretical frameworks available to describe them. In addition, theoreticians have made numerous proposals for specific experiments,
some of which have been tried, with or without success so far, and some of which are waiting to be attempted. Here we describe  a selection of these proposed experiments.

X-Ray Photoelectron Spectroscopy (XPS) probes the chemical environment of individdual atoms by measuring the core-level chemical shift, and has been widely used for chemical analysis of various kinds of samples.
An obvious new direction for XPS with FELs is time-resolved XPS for probing the temporal evolution of the chemical environment in photoexcited systems. Time-resolved XPS has been extensively employed for solids at FLASH, see, for example, \citet{Hellmann2012}, where a monochromatized FEL beam was employed. 
This approach has, however, not been used widely for gas-phase samples because of the rather low fluence of monochromatized FEL light. Recent developments of XFEL technologies, such as seeded FELs in the soft X-ray regime, and special accelerator modes (e.g. \cite{Prat2018}) are changing the situation. Also, the generation of soft X-ray two-colour double FEL pulses, especially attosecond/few-femtosecond double pulses~\citep{Duris2020}, are atractrive and stimulated renewed interest in time-resolved XPS.
\cite{Oberli2019}, for example, 
showed that the N 1s $\rightarrow \pi^*$ transition can induce an isomerization reaction in formamide and that it is possible to observe in real time hydrogen migration by time-resolved XPS, measuring the chemical shifts with the X-ray probe pulse. 

Regarding X-ray double pulses, experiments on some other systems have been suggested with other detection schemes. For example, for the glycine molecule, \citet{Cooper2014} 
showed that the dynamics of a hole arising from ionization in
the inner valence region evolves with a timescale in the range of current XFEL capabilities. The proposed 
pump-probe scheme uses X-rays with photon energy below the K edge of carbon (275--280
eV) that can ionize the inner valence region. A second probe X-ray at the same
energy can excite an electron from the core to fill the vacancy in the inner-valence
region. The dynamics of the inner valence hole can be tracked by measuring the Auger
electrons produced by the subsequent refilling of the core hole as a function of pump-probe delay. \citet{Li2014} studied the dynamics of an electron hole created by
photoionization of the valence shell of protonated water clusters H$^+$(H$_2$O)$_n$ and demonstrated 
that the electron hole is strongly correlated with the protons forming the
hydrogen bond network. They showed that it is possible to probe key aspects of the
valence electron hole dynamics and the coupled nuclear motion with femtosecond
time resolution by resonantly exciting 1$s$ electrons to fill the valence hole. They proposed the use of transient X-ray absorption as a probe.  \citet{Kuleff2016} found that when a hole is created in the core, it is stationary, but in response to this hole, interesting electron dynamics takes place,
which can lead to intense charge migration in the valence shell. This migration is typically faster than that after the ionization of a valence electron and occurs on a shorter time scale than the natural decay of the
core hole by the Auger process, making the subject challenging to experimentalists in attosecond science.

XPS and XTPPS (see Section~\ref{sec:XTPPS}) measure energies of single and double core hole states, and Auger and X-ray absorption spectroscopy also measure energetics. Today,
more advanced experimental tools, which go beyond measuring photoelectron energies, are in our toolbox.
Measurements of electron angular distributions in the molecular frame of reference (molecular-frame photoelectron angular distributions, MFPADs) are performed routinely using synchrotron radiation. Such MFPADs are highly sensitive probes of the electronic and molecular structures. Time-resolved MFPAD measurements are, however, still limited (see, e.g., the review article by \citet{Suzuki2006} for valence photoemission, and \citet{Rouzee2013, Boll2013, Minemoto2016} for inner-shell photoemission). There are some instructive theoretical predictions, for example, \citet{Arasaki2011} demonstrated that time-resolved valence-band MPFADs can monitor electron-ion coupling via a conical intersection, whereas~\citet{Krasniqi2010} demonstrated time-resolved core-level MFPADs can monitor structural changes of photoexcited molecules. 
High-repetition rate soft X-ray FELs will certainly make time-resolved core-level MFPAD measurements a tangible reality.  
Very recently, on the SQS beam line of SASE3 at the European XFEL, \citet{Kastirke2020} succeeded in measuring O 1s MFPADs of dissociating O$_2^{2+}$ molecular dications created by the Auger decay that follows O 1$s$ photoionization of neutral O$_2$ molecules, demonstrating that time-resolved core-level MFPAD measurements are indeed possible when  two-colour X-ray double pulses become available~\citep{Serkez2020}.  

The above described predictions are based on linear spectroscopy or sequential two-photon absorption. For non-linear spectroscopic experiments, perhaps the most prolific theoretical  group for devising new ideas for exploiting coherent, intense and short duration  X-ray 
pulses is that of Mukamel and co-authors. In an early review, \citet{Mukamel2009} showed how coherent techniques developed for nuclear magnetic resonance and optical spectroscopy could be applied at shorter wavelengths, in particular with femtosecond and attosecond X-rays. This group 
has performed theoretical calculations investigating circular dichroism in achiral molecules, where the circular polarization of the light induces dichroism in the excited state \citep{Rouxel2017}, four-wave mixing in molecules \citep{Tanaka2002}, coherent control of electron transfer \citep{Dorfman2016}, etc. Some of the theoretical approaches and proposed experimental measurements for studying conical intersections in molecules have been summarised in \cite{Kowalewski2017}. 
These include X-ray Hybrid Stimulated Raman Detection: Attosecond Stimulated X-ray Raman Spectroscopy (ASRS), in which a combination of narrow and broadband (attosecond) pulses is used; and the TRUECARS technique \citep[Transient Redistribution of Ultrafast Electronic Coherences in Attosecond Raman Signals:][]{Kowalewski2015}, which also requires narrow and broadband X-ray pulses.
In another investigation, \citet{Le2019} suggested the use of X-rays carrying orbital angular momentum (twisted light, see also Section \ref{ssec:OAM}) to measure the dichroism of chiral molecules; since the method is based on resonant absorption, narrow bandwidth is required. 
The four-wave-mixing signals for the CO molecule in the gas phase have been calculated for XUV and core (carbon K edge) excitation \citep{Cho2018}. 
Many of these experiments are at present beyond the capabilities of present FELs: while phase resolution of a few attoseconds and pulse durations of less than a femtosecond have been demonstrated, they have not been demonstrated for the same beam. Longer pulses with phase control, or shorter pulses without phase control are available.

\citet{Ho2015} studied theoretically the effect of extremely high intensities on resonance-mediated atomic ionization by tender X-rays (few keV). 
Noble gas targets were considered, and broad band SASE pulses were compared with seeded (narrow band) pulses. At high energies, Kr and Xe absorb many photons and undergo cascade decay by Auger and other processes to produce highly charged states. 
SASE leads to higher ionization states, basically because the broad bandwidth increases the probability that an intermediate ionic state underwent resonant excitation. Shorter pulses, approaching the lifetime of some excited states, generally leads to less highly charged states because less decay occurs, and channels involving sequential ionization are reduced. 
However under some circumstances the opposite may occur: if the pulse has a duration shorter than the lifetime of a deep core hole, the atom may undergo excitation of this hole state, followed by ionization of valence and shallower core holes, after which the deep core hole decays. This leads to higher charge states.

\citet{Zhang2016} have suggested using a soft x-ray pump-control scheme with double pulses to investigate molecular wave packets. A core level resonance is excited by the first pulse, and then the second, strong control pulse couples this excited state to a valence excited state. 
The system is then interrogated by transient X-ray absorption, or by measuring the vibrationally resolved Auger spectrum as a function of the delay between the two pulses. This experiment would be very demanding. Apart from the necessity to create two pulses with well-defined photon energies and durations and relative delays of about 1 fs or less, very high spectroscopic resolution is required.

\citet{Khokhlova2019} have suggested molecular Auger interferometry, which is based on the coherent phase control of Auger dynamics in a two-color $\omega-2\omega$ laser field.
They showed that, in contrast to atoms, in oriented molecules of certain point groups, the relative $\omega-2\omega$ phase modulates the total ionization yield. They derived a simple analytical formula for the extraction of the lifetimes of Auger-active states from a molecular Auger interferogram,
circumventing the need for either high-resolution or attosecond spectroscopy. Another innovative idea was put forward by \cite{Liu2019}, who have proposed second-harmonic generation using noble gas atoms: this process is normally forbidden in systems containing inversion symmetry.  

Proposals have also been put forward to utilise the polarization of the light to gain new insights. \citet{Agueny2020} has proposed using orthogonally polarized beams to investigate attosecond dynamics. \citet{Hofbrucker2018} have proposed an investigation of two-photon, elliptical dichroism. In this experiment, the dichroism is zero if the light polarization is linear or circular, but non-zero for elliptical polarization. 

The availability of short-wavelength coherent sources has also prompted interest in coherent \emph{nuclear} excitation \citep{Adams2013,Adams2019}. Indeed Rabi oscillations, one of the hallmarks of coherent excitation, have been observed \citep{Haber2017} with a synchrotron source in combination with a resonant cavity scheme. Nuclear excitations are also of interest for the next-generation of scientific clocks \citep{Masuda2019,nuclock}.\\

\section{Perspectives}

Short wavelength Free-Electron Lasers have provided scientists with new light sources for experiments requiring intense light, and have expanded the range of available wavelengths from the UV region to soft and hard X-rays. At hard X-ray wavelengths, a major application  so far of the transverse coherence has been the measurement of coherent diffraction images of single molecules and nanocrystals, providing  fundamental information for structural biology. The absence of longitudinal coherence for SASE sources has meant that there has been little or no exploration of the possibilities of full coherence. 
Several ideas have been published for controlling the bandwidth of SASE FELs, for example modified self-seeding to produce narrower line width radiation 
\citep{Prat2018}, or machine methods to enhance the brightness \citep{Prat2019}. For some experiments, a \emph{larger} bandwidth is advantageous and this can also be provided \citep{Prat2020}. For crystallography, large bandwidth increases the probability of finding Bragg reflections, and may have applications in spectroscopy. For example, a spectrographic setup provides a multiplexing advantage, as in transient absorption \citep{Bhattacherjee2017}, and ghost spectroscopy \citep{Driver2020} exploits correlation techniques to obtain clean data from fluctuating light pulses. We are now at the stage where accelerator physicists can control more and more parameters, not only bandwidth but also pulse duration and polarization.

Experiments in the XUV and soft X-ray range have made the greatest use of the longitudinal coherence properties available at short wavelengths, and we foresee a wide range of interesting developments. In the first instance, many of the techniques developed by optical laser scientists are now available at much shorter wavelengths. Quite a few of these have been tried, or are being tried with short wavelengths: bichromatic experiments, multi-dimensional spectroscopy, STIRAP (STImulated Raman Adiabatic Passage), second harmonic generation \citep[at solid surfaces;][]{Lam2018}, four wave mixing \citep{Bencivenga2015}, Hanbury Brown and Twiss interferometry \citep{Gorobtsov2018}, etc. 
To date many of these methods have been restricted to bound states by the low photon energy of the laboratory sources used, but FEL light permits the extension into the ionizing region, with all the advantages and disadvantages that this implies. 
Whenever a non-linear process is involved, then FELs have a significant advantage compared to most laboratory sources, due to their high pulse energy. These optical techniques allow the discovery of molecular coherences not accessible by other methods, and the exploration of non-linear phenomena \citep{Wituschek2020}.
The present level of phase control, a few attoseconds, permits the measurement of physical quantities on this time scale by interferometric methods. An immediate goal is to use this capability for the measurement of physical quantities of interest, such as photoemission delays \citep{Cirelli2018}.

An exciting future development will be based on the recently demonstrated capability of a seeded FEL to produce a ``sculpted'' pulse train, that is, one in which the phase and amplitude of the contributing wavelengths is controlled \citep{Maroju2020}. The reader will recall that attosecond science with laboratory lasers began with the demonstration of attosecond pulse trains, which are still widely used, and was followed by the production of single attosecond pulses: we may expect, or at least hope, for a similar evolution for longitudinally coherent FELs.
Regarding this experiment, we note that a crucial breakthrough in permitting the ``sculpture'' of pulses was the use of intensity correlation methods. This is a form of covariance analysis, which was implemented some time ago for optical lasers \citep{Frasinski1989}, as well as more recently for FELs \citep{Frasinski2013, Kornilov2013, You2020}: we expect to see increasing applications of correlation techniques. A recent example is ghost spectroscopy, or ``spooktroscopy'' \citep{Driver2020}.

In the long term is is possible that FELs will be applied to investigate quantum optical phenomena. The field was inspired by the experiments of \citet{HBT1956,HBT1956a}, and \citet{Glauber1963} won a Nobel prize for his insights. A challenge here is to discover and control exotic states of light fields at short wavelengths.

Other novel experiments may become possible due to machine developments. Echo Enabled Harmonic Generation has been demonstrated as a method for providing shorter wavelengths, more stable intensity and a larger choice of wavelengths \citep{RebernikRibic_NatPhot_2019}, and signifies a marked improvement in seeded FEL performance. There may also be demand for narrowband pulses for precision spectroscopy; transform limited pulses are now possible with a duration of up to a ps.
At the other extreme, hard X-ray pulses with a duration of less than 1 fs, and peak power of over 100 GW can now be generated \citep{Duris2020}, and coherent soft x-ray FEL pulses with a duration of less than 5 fs have been produced.

As discussed in the previous section, theoreticians continue to provide new insights and ideas for new experiments. Accelerator physicists have made rapid progress in improving light sources - shorter pulses, higher peak powers, better stability, etc. This ensures that much new physics will be discovered at short wavelength, coherent FELs in the future. 

\clearpage
\section[Appendix: List of symbols]{List of symbols}
\begin{longtable}{c|p{0.6\textwidth}|c}
\caption[]{\label{tab:symbols}List of symbols used in this work}\\
    name & definition & equations\\
    \hline
    \endfirsthead
    \caption[]{List of symbols used in this work (continued)}\\
    name & definition & equations\\
    \hline
    \endhead
    $c$ &  speed of light\\
    $e$ & electron charge \\
    $h$; ($\hbar$) & Planck's constant; (reduced) \\
    $i$ & imaginary unit \\
    $A$ & asymmetry of angular distribution $W(\vartheta,\varphi)$ & \ref{eq:neon4}\\
    $A_m$ & amplitude of the asymmetry oscillation & \ref{eq:neon4}\\
    $\hat{A}$ & antisymmetrization operator & \ref{eq:rmat}, \ref{eq:TD-CASSCF}\\
    $\mathbf{A}(t)$ & vector potential & \ref{eq:eq}, \ref{eq:volk1}, \ref{eq:volk2}, \ref{eq:velocity-gauge}\\
    $\beta_{k}$ &coefficients of Legendre expansion (``beta parameters'') & \ref{eq:neon3}, \ref{eq:neon4}\\ 
    $\beta_{k}^{\nu}$ &coefficients of Legendre expansion (``beta parameters''). \par
    ${\nu} = \{\mathrm{lin}, \mathrm{circ}\}$: light polarization & \ref{eq:ang}\\
    $c_n(t)$ & wavefunction expansion coefficients & \ref{eq:expa}\\
    $C_I(t)$ & CI expansion coefficients & \ref{eq:MC_expansion}, \ref{eq:TD-CASSCF}\\
    $\mathbf{D}(t)$ & electric dipole operator & \ref{eq:eq}, \ref{eq:pt1t}, \ref{eq:pt2t}\\
    $e_j$ & electric charge of particle $j$ & \ref{eq:eq}\\
    $E$ & particle energy & \ref{eq:ews}\\
    $E(t)$ & electric field  & \ref{eq:nefield}\\
    $\mathbf{E}(t)$ & electric field vector & \ref{eq:eq}, \ref{eq:pt1t}, \ref{eq:pt2t}\\
    $\epsilon_i, \epsilon_p, \epsilon_f$ & energies of initial, intermediate, and final state & \ref{eq:pt1t}, \ref{eq:pt2t}\\
    $\eta$ & amplitude ratio between FEL harmonics & \ref{eq:nefield}\\
    $\eta$ & spectral phase & Sec.~\ref{sec:ews}\\
    $F(t)$ & pulse envelope & \ref{eq:nefield}\\
    $\phi$ & relative phase between FEL harmonics & \ref{eq:nefield}, \ref{eq:neon4}\\
    $\phi_m$ & phase at the value of maximum asymmetry & \ref{eq:neon4}\\
    $\varphi_q$ & absolute phase of FEL $q^\mathrm{th}$ harmonic & \ref{Eq_sidebands}, \ref{eq:sidebdand_3omega}, \ref{eq:phase_diff}\\
    $\Delta\varphi_{q-1,q,q+1}$ & phase difference between three consecutive  FEL harmonics & \ref{eq:phase_diff}, \ref{Eq_GDD}\\
     $\Delta\varphi_{at}$ & difference of photoionisation phases & \ref{Eq_sidebands}, \ref{eq:sidebdand_3omega}\\
    $\phi_n(\xi)$ & wavefunction expansion basis function & \ref{eq:expa}\\
    $\phi_n(x_1,x_2,...,\Omega_N, \sigma_N)$ & wavefunction expansion basis function & \ref{eq:rmat}\\
    $\Phi_i(\xi)$ & initial-condition wavefunction & \ref{eq:ham}\\
    $\Phi(\mathbf{p},t)$ & wavefunction written as a function of time and electron momentum & \ref{eq:volk2} \\
    $\Phi_I(\vec{x}_1, \vec{x}_2, \cdots, \vec{x}_N, t)$ & electronic configuration & \ref{eq:MC_expansion}, \ref{eq:TD-CASSCF}\\
    $\Phi_{\mathrm{fc}}$, $\Phi_{\mathrm{dc}}$ & frozen core and dynamic core closed-shell determinants & \ref{eq:TD-CASSCF}\\
    $\hat{h}({\bf r},t)$ & one-body Hamiltonian & \ref{eq:H1}, \ref{eq:velocity-gauge}\\
    $\hat{H}(t)$ & Hamiltonian & \ref{eq:unit}, \ref{eq:vari}, \ref{eq:liou}, \ref{eq:tdse2}, \ref{eq:TDSE1}\\
    $\hat{H}_1(t)$ & one-electron Hamiltonian & \ref{eq:TDSE1}, \ref{eq:H1}\\
    $\hat{H}_2(t)$ & two-electron Hamiltonian & \ref{eq:TDSE1}, \ref{eq:H2}\\
    $\hat{H}_0(\xi)$ & unperturbed Hamiltonian & \ref{eq:ham}\\
    $H(\xi,t)$ & full Hamiltonian & \ref{eq:tdse}, \ref{eq:ham}\\
    $I^{\nu}(\theta)$ & photoelectron angular distribution. ${\nu} \in \{\mathrm{lin}, \mathrm{circ}\}$: light polarization & \ref{eq:ang}\\
    $J_i$, $J_f$ & total electronic angular momentum of initial atom, and of residual ion & \ref{eq:neon1}\\
    $\mathbf{p}$ & electron momentum & \ref{eq:volk1}, \ref{eq:volk2}, \ref{eq:velocity-gauge} \\
    $M_f$ & magnetic quantum number of residual ion & \ref{eq:neon1}\\
    $N_n(t)$ & population of level $n$ & \ref{eq:rate}\\
    $\omega$ & fundamental FEL frequency & \ref{eq:nefield}, \ref{eq:eq_spLEAD}\\
    $\omega_\mathrm{NIR}$ & NIR laser frequency & \ref{Eq_sidebands}, \ref{eq:sidebdand_3omega}\\
    $\omega_\mathrm{UV}$ & seed laser frequency & \ref{Eq_GDD}\\
    $\Omega \equiv (\vartheta, \varphi)$ & spherical coordinates & \ref{eq:neon1}, \ref{eq:neon2}, \ref{eq:one}, \ref{eq:two}\\
    $\mathbf{p}_j$ & linear momentum of particle $j$ & \ref{eq:eq}\\
    $P_k(\cos\vartheta)$ & Legendre polynomial of order $k$ & \ref{eq:neon3}, \ref{eq:ang}\\
    $P_{+}$, $P_{-}$ & probability of ionization in bichromatic fields of same or opposite helicities & \ref{eq:CD}\\
    $P_{q,q+1}$ & oscillating component of the sidebands & \ref{eq:sideband_parameter}\\
    $P_{lm}(r,t)$ & one-active-electron radial function & \ref{eq:one}\\
    $P_{l_1 l_2}^{LM}(r_1,r_2,t)$ & two-active-electron radial function & \ref{eq:two}\\
    $\psi_n(r_{N+1},t)$ & $n$-th channel radial function & \ref{eq:rmat}\\
    $\psi_{\mathbf p}$ & Volkov wavefunction & \ref{eq:volk1}\\
%    $\Psi$ & wavefunction & \ref{eq:vari}\\
    $\Psi (t) $ & wavefunction & \ref{eq:vari}, \ref{eq:tdse2}, \ref{eq:TD-CASSCF} \\
    $\Psi(\xi,t)$ & wavefunction & \ref{eq:tdse}, \ref{eq:expa} \\
    $\Psi(x_1,x_2,\dots,x_{N+1},t)$ & wavefunction & \ref{eq:rmat}\\
    $\Phi_I(\vec{x}_1, \vec{x}_2, \cdots, \vec{x}_N, t)$ & electronic wavefunction & \ref{eq:MC_expansion}, \ref{eq:TD-CASSCF}\\
    $\Psi(\mathbf{r},t)$ & one-active-electron wavefunction & \ref{eq:one}\\
    $\Psi(\mathbf{r}_1,\mathbf{r}_2,t)$ & two-active-electron wavefunction & \ref{eq:two}\\
    %$\Psi(\xi,t)$ & wavefunction & \ref{eq:ham} \\
    $\mathbf{r}$ & photoelectron position & \ref{eq:volk1}, \ref{eq:volk2}, \ref{eq:velocity-gauge}\\
    $\mathbf{r}_j$ & position of particle $j$ & \ref{eq:eq}, \ref{eq:H1}, \ref{eq:H2}\\
    $r_1,r_2,\dots,r_{N+1}$ & radial coordinates & \ref{eq:rmat}\\
    $\hat{\rho}, \, \hat{\rho}(t)$ & density operator & \ref{eq:den}, \ref{eq:liou} \\
    $\rho$ & number density of excited states & Sec.~\ref{ssec:Superfluorescence}\\
    $S_{q,q+2}$ & intensity of the sideband between the $q^\mathrm{th}$ and $(q+2)^\mathrm{th}$ harmonics (HHG)&
    \ref{Eq_sidebands}\\%https://www.overleaf.com/project/5bae779f58464c4c2da31982
    $S^{(\pm)}_{q,q+1}$ & intensity of the sidebands between the  $q^\mathrm{th}$ and $(q+1)^\mathrm{th}$ harmonics (FEL) & \ref{eq:sidebdand_3omega}, \ref{eq:sideband_parameter}\\
    %$\sigma_1,\sigma_2,\dots,\sigma_{N+1}$ & spin coordinates & \ref{eq:rmat}\\
    $\sigma$ & spin coordinate & \ref{eq:rmat}, \ref{eq:MC_expansion}\\
    $\tau$ & delay between NIR and FEL harmonics &  \ref{Eq_sidebands}, \ref{eq:sidebdand_3omega}\\
    $\tau_f$ & EWS time delay in the channel $\ket{f}$ & \ref{eq:ews}\\
    $T$ & temporal evolution operator, infinite time limit & \ref{eq:neon1}, \ref{eq:unit}\\
    $\ME{f}{T}{i}_n$ & $n^\mathrm{th}$ order transition amplitudes $(n= 1,2)$ & \ref{eq:neon1}, \ref{eq:pt1t}, \ref{eq:pt2t}\\
    $U(t,t_0)$ & temporal evolution operator & \ref{eq:unit}\\
    $V(\xi,t)$ & external-field perturbation Hamiltonian & \ref{eq:tdse}, \ref{eq:ham}\\
    $W(\vartheta,\varphi)$ & differential ionization probability & \ref{eq:neon1}\\
    $W^{(i)}(\vartheta,\varphi)$ & first-harmonic $(i=1)$, second-harmonic $(i=2)$, and cross-term $(i=12)$ contributions to $W(\vartheta,\varphi)$ & \ref{eq:neon1}\\
    $W_0$ & isotropic component of $W(\vartheta,\varphi)$ & \ref{eq:neon3}\\
    $W_{n \rightarrow n'}$ & transition rate from level $n^{\prime}$ to level $n$ & \ref{eq:rate}\\
    $x_1,x_2,\dots,x_N$ & particle coordinates (space + spin)& \ref{eq:rmat}, \ref{eq:MC_expansion} \\
    $\xi$ & particle coordinates & \ref{eq:ham}, \ref{eq:expa}\\
    $Y_{lm}(\Omega)$ & spherical harmonic & \ref{eq:one}\\
    $Y_{l_1l_2}^{LM}(\Omega_1,\Omega_2)$ & bipolar spherical harmonic & \ref{eq:two}\\
    $Z$ & atomic number & \ref{eq:velocity-gauge}\\
\end{longtable}

\section{Acknowledgements} 

KU acknowledges the Ministry of
Education, Culture, Sports, Science, and Technology of Japan (MEXT) for funding via the X-ray Free Electron Laser Utilization Research Project, the X-ray Free Electron Laser Priority Strategy Program and the Dynamic Alliance for Open Innovation Bridging Human, Environment and Materials program, and the IMRAM program of Tohoku University.
GS acknowledges funding from the  Deutsche Forschungsgemeinschaft (DFG, German Research Foundation) Projekt: 429805582. ANG acknowledges
funding from the Russian Foundation for Basic Research under the Project No. 20-52-12023.

%\section*{References}

\bibliography{PhysRep}

\providecommand{\noopsort}[1]{}
\begin{thebibliography}{488}
\expandafter\ifx\csname natexlab\endcsname\relax\def\natexlab#1{#1}\fi
\providecommand{\url}[1]{\texttt{#1}}
\providecommand{\href}[2]{#2}
\providecommand{\path}[1]{#1}
\providecommand{\DOIprefix}{doi:}
\providecommand{\ArXivprefix}{arXiv:}
\providecommand{\URLprefix}{URL: }
\providecommand{\Pubmedprefix}{pmid:}
\providecommand{\doi}[1]{\href{http://dx.doi.org/#1}{\path{#1}}}
\providecommand{\Pubmed}[1]{\href{pmid:#1}{\path{#1}}}
\providecommand{\bibinfo}[2]{#2}
\ifx\xfnm\relax \def\xfnm[#1]{\unskip,\space#1}\fi
%Type = Misc
\bibitem[{Adams et~al.(2019)Adams, Aeppli, Allison, Baron, Bucksbaum, Chumakov,
  Corder, Cramer, DeBeer, Ding, Evers, Frisch, Fuchs, Gr\"ubel, Hastings, Heyl,
  Holberg, Huang, Ishikawa, Kaldun, Kim, Kolodziej, Krzywinski, Li, Liao,
  Lindberg, Madsen, Maxwell, Monaco, Nelson, Palffy, Porat, Qin, Raubenheimer,
  Reis, R\"{o}hlsberger, Santra, Schoenlein, Sch\"unemann, Shpyrko, Shvyd'ko,
  Shwartz, Singer, Sinha, Sutton, Tamasaku, Wille, Yabashi, Ye and
  Zhu}]{Adams2019}
\bibinfo{author}{Adams, B.}, \bibinfo{author}{Aeppli, G.},
  \bibinfo{author}{Allison, T.}, \bibinfo{author}{Baron, A.Q.R.},
  \bibinfo{author}{Bucksbaum, P.}, \bibinfo{author}{Chumakov, A.I.},
  \bibinfo{author}{Corder, C.}, \bibinfo{author}{Cramer, S.P.},
  \bibinfo{author}{DeBeer, S.}, \bibinfo{author}{Ding, Y.},
  \bibinfo{author}{Evers, J.}, \bibinfo{author}{Frisch, J.},
  \bibinfo{author}{Fuchs, M.}, \bibinfo{author}{Gr\"ubel, G.},
  \bibinfo{author}{Hastings, J.B.}, \bibinfo{author}{Heyl, C.M.},
  \bibinfo{author}{Holberg, L.}, \bibinfo{author}{Huang, Z.},
  \bibinfo{author}{Ishikawa, T.}, \bibinfo{author}{Kaldun, A.},
  \bibinfo{author}{Kim, K.J.}, \bibinfo{author}{Kolodziej, T.},
  \bibinfo{author}{Krzywinski, J.}, \bibinfo{author}{Li, Z.},
  \bibinfo{author}{Liao, W.T.}, \bibinfo{author}{Lindberg, R.},
  \bibinfo{author}{Madsen, A.}, \bibinfo{author}{Maxwell, T.},
  \bibinfo{author}{Monaco, G.}, \bibinfo{author}{Nelson, K.},
  \bibinfo{author}{Palffy, A.}, \bibinfo{author}{Porat, G.},
  \bibinfo{author}{Qin, W.}, \bibinfo{author}{Raubenheimer, T.},
  \bibinfo{author}{Reis, D.A.}, \bibinfo{author}{R\"{o}hlsberger, R.},
  \bibinfo{author}{Santra, R.}, \bibinfo{author}{Schoenlein, R.},
  \bibinfo{author}{Sch\"unemann, V.}, \bibinfo{author}{Shpyrko, O.},
  \bibinfo{author}{Shvyd'ko, Y.}, \bibinfo{author}{Shwartz, S.},
  \bibinfo{author}{Singer, A.}, \bibinfo{author}{Sinha, S.K.},
  \bibinfo{author}{Sutton, M.}, \bibinfo{author}{Tamasaku, K.},
  \bibinfo{author}{Wille, H.C.}, \bibinfo{author}{Yabashi, M.},
  \bibinfo{author}{Ye, J.}, \bibinfo{author}{Zhu, D.}, \bibinfo{year}{2019}.
\newblock \bibinfo{title}{Scientific opportunities with an {X}-ray
  free-electron laser oscillator}.
\newblock \href{http://arxiv.org/abs/1903.09317}{\tt arXiv:1903.09317}.
%Type = Article
\bibitem[{Adams et~al.(2013)Adams, Buth, Cavaletto, Evers, Harman, Keitel,
  P{\'{a}}lffy, Pic{\'{o}}n, R\"ohlsberger, Rostovtsev and
  Tamasaku}]{Adams2013}
\bibinfo{author}{Adams, B.W.}, \bibinfo{author}{Buth, C.},
  \bibinfo{author}{Cavaletto, S.M.}, \bibinfo{author}{Evers, J.},
  \bibinfo{author}{Harman, Z.}, \bibinfo{author}{Keitel, C.H.},
  \bibinfo{author}{P{\'{a}}lffy, A.}, \bibinfo{author}{Pic{\'{o}}n, A.},
  \bibinfo{author}{R\"ohlsberger, R.}, \bibinfo{author}{Rostovtsev, Y.},
  \bibinfo{author}{Tamasaku, K.}, \bibinfo{year}{2013}.
\newblock \bibinfo{title}{{X}-ray quantum optics}.
\newblock \bibinfo{journal}{J. Mod. Opt.} \bibinfo{volume}{60},
  \bibinfo{pages}{2--21}.
\newblock \DOIprefix\doi{10.1080/09500340.2012.752113}.
%Type = Article
\bibitem[{Afanasev et~al.(2018)Afanasev, Carlson, Schmiegelow, Schulz,
  Schmidt-Kaler and Solyanik}]{Afanasev2018}
\bibinfo{author}{Afanasev, A.}, \bibinfo{author}{Carlson, C.E.},
  \bibinfo{author}{Schmiegelow, C.T.}, \bibinfo{author}{Schulz, J.},
  \bibinfo{author}{Schmidt-Kaler, F.}, \bibinfo{author}{Solyanik, M.},
  \bibinfo{year}{2018}.
\newblock \bibinfo{title}{Experimental verification of position-dependent
  angular-momentum selection rules for absorption of twisted light by a bound
  electron}.
\newblock \bibinfo{journal}{New J. Phys.} \bibinfo{volume}{20},
  \bibinfo{pages}{023032}.
\newblock \DOIprefix\doi{10.1088/1367-2630/aaa63d}.
%Type = Article
\bibitem[{Agueny(2020)}]{Agueny2020}
\bibinfo{author}{Agueny, H.}, \bibinfo{year}{2020}.
\newblock \bibinfo{title}{Quantum control and characterization of ultrafast
  ionization with orthogonal two-color laser pulses}.
\newblock \bibinfo{journal}{Sci. Rep.} \bibinfo{volume}{10},
  \bibinfo{pages}{239}.
\newblock \DOIprefix\doi{10.1038/s41598-019-57125-z}.
%Type = Article
\bibitem[{Allaria et~al.(2012a)Allaria, Appio, Badano, Barletta, Bassanese,
  Biedron, Borga, Busetto, Castronovo, Cinquegrana, Cleva, Cocco, Cornacchia,
  Craievich, Cudin, D'Auria, Dal~Forno, Danailov, De~Monte, De~Ninno,
  Delgiusto, Demidovich, Di~Mitri, Diviacco, Fabris, Fabris, Fawley, Ferianis,
  Ferrari, Ferry, Froehlich, Furlan, Gaio, Gelmetti, Giannessi, Giannini,
  Gobessi, Ivanov, Karantzoulis, Lonza, Lutman, Mahieu, Milloch, Milton,
  Musardo, Nikolov, Noe, Parmigiani, Penco, Petronio, Pivetta, Predonzani,
  Rossi, Rumiz, Salom, Scafuri, Serpico, Sigalotti, Spampinati, Spezzani,
  Svandrlik, Svetina, Tazzari, Trovo, Umer, Vascotto, Veronese, Visintini,
  Zaccaria, Zangrando and Zangrando}]{Allaria2012}
\bibinfo{author}{Allaria, E.}, \bibinfo{author}{Appio, R.},
  \bibinfo{author}{Badano, L.}, \bibinfo{author}{Barletta, W.A.},
  \bibinfo{author}{Bassanese, S.}, \bibinfo{author}{Biedron, S.G.},
  \bibinfo{author}{Borga, A.}, \bibinfo{author}{Busetto, E.},
  \bibinfo{author}{Castronovo, D.}, \bibinfo{author}{Cinquegrana, P.},
  \bibinfo{author}{Cleva, S.}, \bibinfo{author}{Cocco, D.},
  \bibinfo{author}{Cornacchia, M.}, \bibinfo{author}{Craievich, P.},
  \bibinfo{author}{Cudin, I.}, \bibinfo{author}{D'Auria, G.},
  \bibinfo{author}{Dal~Forno, M.}, \bibinfo{author}{Danailov, M.B.},
  \bibinfo{author}{De~Monte, R.}, \bibinfo{author}{De~Ninno, G.},
  \bibinfo{author}{Delgiusto, P.}, \bibinfo{author}{Demidovich, A.},
  \bibinfo{author}{Di~Mitri, S.}, \bibinfo{author}{Diviacco, B.},
  \bibinfo{author}{Fabris, A.}, \bibinfo{author}{Fabris, R.},
  \bibinfo{author}{Fawley, W.}, \bibinfo{author}{Ferianis, M.},
  \bibinfo{author}{Ferrari, E.}, \bibinfo{author}{Ferry, S.},
  \bibinfo{author}{Froehlich, L.}, \bibinfo{author}{Furlan, P.},
  \bibinfo{author}{Gaio, G.}, \bibinfo{author}{Gelmetti, F.},
  \bibinfo{author}{Giannessi, L.}, \bibinfo{author}{Giannini, M.},
  \bibinfo{author}{Gobessi, R.}, \bibinfo{author}{Ivanov, R.},
  \bibinfo{author}{Karantzoulis, E.}, \bibinfo{author}{Lonza, M.},
  \bibinfo{author}{Lutman, A.}, \bibinfo{author}{Mahieu, B.},
  \bibinfo{author}{Milloch, M.}, \bibinfo{author}{Milton, S.V.},
  \bibinfo{author}{Musardo, M.}, \bibinfo{author}{Nikolov, I.},
  \bibinfo{author}{Noe, S.}, \bibinfo{author}{Parmigiani, F.},
  \bibinfo{author}{Penco, G.}, \bibinfo{author}{Petronio, M.},
  \bibinfo{author}{Pivetta, L.}, \bibinfo{author}{Predonzani, M.},
  \bibinfo{author}{Rossi, F.}, \bibinfo{author}{Rumiz, L.},
  \bibinfo{author}{Salom, A.}, \bibinfo{author}{Scafuri, C.},
  \bibinfo{author}{Serpico, C.}, \bibinfo{author}{Sigalotti, P.},
  \bibinfo{author}{Spampinati, S.}, \bibinfo{author}{Spezzani, C.},
  \bibinfo{author}{Svandrlik, M.}, \bibinfo{author}{Svetina, C.},
  \bibinfo{author}{Tazzari, S.}, \bibinfo{author}{Trovo, M.},
  \bibinfo{author}{Umer, R.}, \bibinfo{author}{Vascotto, A.},
  \bibinfo{author}{Veronese, M.}, \bibinfo{author}{Visintini, R.},
  \bibinfo{author}{Zaccaria, M.}, \bibinfo{author}{Zangrando, D.},
  \bibinfo{author}{Zangrando, M.}, \bibinfo{year}{2012}a.
\newblock \bibinfo{title}{Highly coherent and stable pulses from the {FERMI}
  seeded free-electron laser in the extreme ultraviolet}.
\newblock \bibinfo{journal}{Nat. Photonics} \bibinfo{volume}{6},
  \bibinfo{pages}{699--704}.
\newblock \DOIprefix\doi{10.1038/nphoton.2012.233}.
%Type = Article
\bibitem[{Allaria et~al.(2012b)Allaria, Battistoni, Bencivenga, Borghes,
  Callegari, Capotondi, Castronovo, Cinquegrana, Cocco, Coreno, Craievich,
  Cucini, D'Amico, Danailov, Demidovich, De~Ninno, Di~Cicco, Di~Fonzo,
  Di~Fraia, Di~Mitri, Diviacco, Fawley, Ferrari, Filipponi, Froehlich, Gessini,
  Giangrisostomi, Giannessi, Giuressi, Grazioli, Gunnella, Ivanov, Mahieu,
  Mahne, Masciovecchio, Nikolov, Passos, Pedersoli, Penco, Principi, Raimondi,
  Sergo, Sigalotti, Spezzani, Svetina, Trov{\`{o}} and
  Zangrando}]{Allaria2012a}
\bibinfo{author}{Allaria, E.}, \bibinfo{author}{Battistoni, A.},
  \bibinfo{author}{Bencivenga, F.}, \bibinfo{author}{Borghes, R.},
  \bibinfo{author}{Callegari, C.}, \bibinfo{author}{Capotondi, F.},
  \bibinfo{author}{Castronovo, D.}, \bibinfo{author}{Cinquegrana, P.},
  \bibinfo{author}{Cocco, D.}, \bibinfo{author}{Coreno, M.},
  \bibinfo{author}{Craievich, P.}, \bibinfo{author}{Cucini, R.},
  \bibinfo{author}{D'Amico, F.}, \bibinfo{author}{Danailov, M.B.},
  \bibinfo{author}{Demidovich, A.}, \bibinfo{author}{De~Ninno, G.},
  \bibinfo{author}{Di~Cicco, A.}, \bibinfo{author}{Di~Fonzo, S.},
  \bibinfo{author}{Di~Fraia, M.}, \bibinfo{author}{Di~Mitri, S.},
  \bibinfo{author}{Diviacco, B.}, \bibinfo{author}{Fawley, W.M.},
  \bibinfo{author}{Ferrari, E.}, \bibinfo{author}{Filipponi, A.},
  \bibinfo{author}{Froehlich, L.}, \bibinfo{author}{Gessini, A.},
  \bibinfo{author}{Giangrisostomi, E.}, \bibinfo{author}{Giannessi, L.},
  \bibinfo{author}{Giuressi, D.}, \bibinfo{author}{Grazioli, C.},
  \bibinfo{author}{Gunnella, R.}, \bibinfo{author}{Ivanov, R.},
  \bibinfo{author}{Mahieu, B.}, \bibinfo{author}{Mahne, N.},
  \bibinfo{author}{Masciovecchio, C.}, \bibinfo{author}{Nikolov, I.P.},
  \bibinfo{author}{Passos, G.}, \bibinfo{author}{Pedersoli, E.},
  \bibinfo{author}{Penco, G.}, \bibinfo{author}{Principi, E.},
  \bibinfo{author}{Raimondi, L.}, \bibinfo{author}{Sergo, R.},
  \bibinfo{author}{Sigalotti, P.}, \bibinfo{author}{Spezzani, C.},
  \bibinfo{author}{Svetina, C.}, \bibinfo{author}{Trov{\`{o}}, M.},
  \bibinfo{author}{Zangrando, M.}, \bibinfo{year}{2012}b.
\newblock \bibinfo{title}{Tunability experiments at {FERMI@E}lettra free
  electron laser}.
\newblock \bibinfo{journal}{New J. Phys.} \bibinfo{volume}{14},
  \bibinfo{pages}{113009}.
\newblock \DOIprefix\doi{10.1088/1367-2630/14/11/113009}.
%Type = Article
\bibitem[{Allaria et~al.(2013a)Allaria, Bencivenga, Borghes, Capotondi,
  Castronovo, Charalambous, Cinquegrana, Danailov, De~Ninno, Demidovich,
  Di~Mitri, Diviacco, Fausti, Fawley, Ferrari, Froehlich, Gauthier, Gessini,
  Giannessi, Ivanov, Kiskinova, Kurdi, Mahieu, Mahne, Nikolov, Masciovecchio,
  Pedersoli, Penco, Raimondi, Serpico, Sigalotti, Spampinati, Spezzani,
  Svetina, Trov{\`{o}} and Zangrando}]{Allaria2013a}
\bibinfo{author}{Allaria, E.}, \bibinfo{author}{Bencivenga, F.},
  \bibinfo{author}{Borghes, R.}, \bibinfo{author}{Capotondi, F.},
  \bibinfo{author}{Castronovo, D.}, \bibinfo{author}{Charalambous, P.},
  \bibinfo{author}{Cinquegrana, P.}, \bibinfo{author}{Danailov, M.B.},
  \bibinfo{author}{De~Ninno, G.}, \bibinfo{author}{Demidovich, A.},
  \bibinfo{author}{Di~Mitri, S.}, \bibinfo{author}{Diviacco, B.},
  \bibinfo{author}{Fausti, D.}, \bibinfo{author}{Fawley, W.M.},
  \bibinfo{author}{Ferrari, E.}, \bibinfo{author}{Froehlich, L.},
  \bibinfo{author}{Gauthier, D.}, \bibinfo{author}{Gessini, A.},
  \bibinfo{author}{Giannessi, L.}, \bibinfo{author}{Ivanov, R.},
  \bibinfo{author}{Kiskinova, M.}, \bibinfo{author}{Kurdi, G.},
  \bibinfo{author}{Mahieu, B.}, \bibinfo{author}{Mahne, N.},
  \bibinfo{author}{Nikolov, I.}, \bibinfo{author}{Masciovecchio, C.},
  \bibinfo{author}{Pedersoli, E.}, \bibinfo{author}{Penco, G.},
  \bibinfo{author}{Raimondi, L.}, \bibinfo{author}{Serpico, C.},
  \bibinfo{author}{Sigalotti, P.}, \bibinfo{author}{Spampinati, S.},
  \bibinfo{author}{Spezzani, C.}, \bibinfo{author}{Svetina, C.},
  \bibinfo{author}{Trov{\`{o}}, M.}, \bibinfo{author}{Zangrando, M.},
  \bibinfo{year}{2013}a.
\newblock \bibinfo{title}{Two-colour pump--probe experiments with a
  twin-pulse-seed extreme ultraviolet free-electron laser}.
\newblock \bibinfo{journal}{Nat. Commun.} \bibinfo{volume}{4},
  \bibinfo{pages}{2476}.
\newblock \DOIprefix\doi{10.1038/ncomms3476}.
%Type = Article
\bibitem[{Allaria et~al.(2013b)Allaria, Castronovo, Cinquegrana, Craievich,
  Dal~Forno, Danailov, D'Auria, Demidovich, De~Ninno, Di~Mitri, Diviacco,
  Fawley, Ferianis, Ferrari, Froehlich, Gaio, Gauthier, Giannessi, Ivanov,
  Mahieu, Mahne, Nikolov, Parmigiani, Penco, Raimondi, Scafuri, Serpico,
  Sigalotti, Spampinati, Spezzani, Svandrlik, Svetina, Trovo, Veronese,
  Zangrando and Zangrando}]{Allaria2013}
\bibinfo{author}{Allaria, E.}, \bibinfo{author}{Castronovo, D.},
  \bibinfo{author}{Cinquegrana, P.}, \bibinfo{author}{Craievich, P.},
  \bibinfo{author}{Dal~Forno, M.}, \bibinfo{author}{Danailov, M.B.},
  \bibinfo{author}{D'Auria, G.}, \bibinfo{author}{Demidovich, A.},
  \bibinfo{author}{De~Ninno, G.}, \bibinfo{author}{Di~Mitri, S.},
  \bibinfo{author}{Diviacco, B.}, \bibinfo{author}{Fawley, W.M.},
  \bibinfo{author}{Ferianis, M.}, \bibinfo{author}{Ferrari, E.},
  \bibinfo{author}{Froehlich, L.}, \bibinfo{author}{Gaio, G.},
  \bibinfo{author}{Gauthier, D.}, \bibinfo{author}{Giannessi, L.},
  \bibinfo{author}{Ivanov, R.}, \bibinfo{author}{Mahieu, B.},
  \bibinfo{author}{Mahne, N.}, \bibinfo{author}{Nikolov, I.},
  \bibinfo{author}{Parmigiani, F.}, \bibinfo{author}{Penco, G.},
  \bibinfo{author}{Raimondi, L.}, \bibinfo{author}{Scafuri, C.},
  \bibinfo{author}{Serpico, C.}, \bibinfo{author}{Sigalotti, P.},
  \bibinfo{author}{Spampinati, S.}, \bibinfo{author}{Spezzani, C.},
  \bibinfo{author}{Svandrlik, M.}, \bibinfo{author}{Svetina, C.},
  \bibinfo{author}{Trovo, M.}, \bibinfo{author}{Veronese, M.},
  \bibinfo{author}{Zangrando, D.}, \bibinfo{author}{Zangrando, M.},
  \bibinfo{year}{2013}b.
\newblock \bibinfo{title}{Two-stage seeded soft-{X}-ray free-electron laser}.
\newblock \bibinfo{journal}{Nat. Photonics} \bibinfo{volume}{7},
  \bibinfo{pages}{913--918}.
\newblock \DOIprefix\doi{10.1038/nphoton.2013.277}.
%Type = Article
\bibitem[{Alon et~al.(2009)Alon, Streltsov and Cederbaum}]{Alon2009}
\bibinfo{author}{Alon, O.E.}, \bibinfo{author}{Streltsov, A.I.},
  \bibinfo{author}{Cederbaum, L.S.}, \bibinfo{year}{2009}.
\newblock \bibinfo{title}{Many-body theory for systems with particle
  conversion: Extending the multiconfigurational time-dependent {H}artree
  method}.
\newblock \bibinfo{journal}{Phys. Rev. A} \bibinfo{volume}{79},
  \bibinfo{pages}{022503}.
\newblock \DOIprefix\doi{10.1103/PhysRevA.79.022503}.
%Type = Article
\bibitem[{Amann et~al.(2012)Amann, Berg, Blank, Decker, Ding, Emma, Feng,
  Frisch, Fritz, Hastings, Huang, Krzywinski, Lindberg, Loos, Lutman, Nuhn,
  Ratner, Rzepiela, Shu, Shvyd'ko, Spampinati, Stoupin, Terentyev,
  Trakhtenberg, Walz, Welch, Wu, Zholents and Zhu}]{Amann2012}
\bibinfo{author}{Amann, J.}, \bibinfo{author}{Berg, W.},
  \bibinfo{author}{Blank, V.}, \bibinfo{author}{Decker, F.J.},
  \bibinfo{author}{Ding, Y.}, \bibinfo{author}{Emma, P.},
  \bibinfo{author}{Feng, Y.}, \bibinfo{author}{Frisch, J.},
  \bibinfo{author}{Fritz, D.}, \bibinfo{author}{Hastings, J.},
  \bibinfo{author}{Huang, Z.}, \bibinfo{author}{Krzywinski, J.},
  \bibinfo{author}{Lindberg, R.}, \bibinfo{author}{Loos, H.},
  \bibinfo{author}{Lutman, A.}, \bibinfo{author}{Nuhn, H.D.},
  \bibinfo{author}{Ratner, D.}, \bibinfo{author}{Rzepiela, J.},
  \bibinfo{author}{Shu, D.}, \bibinfo{author}{Shvyd'ko, Y.},
  \bibinfo{author}{Spampinati, S.}, \bibinfo{author}{Stoupin, S.},
  \bibinfo{author}{Terentyev, S.}, \bibinfo{author}{Trakhtenberg, E.},
  \bibinfo{author}{Walz, D.}, \bibinfo{author}{Welch, J.}, \bibinfo{author}{Wu,
  J.}, \bibinfo{author}{Zholents, A.}, \bibinfo{author}{Zhu, D.},
  \bibinfo{year}{2012}.
\newblock \bibinfo{title}{Demonstration of self-seeding in a hard-{X}-ray
  free-electron laser}.
\newblock \bibinfo{journal}{Nat. Photonics} \bibinfo{volume}{6},
  \bibinfo{pages}{693--698}.
\newblock \DOIprefix\doi{10.1038/nphoton.2012.180}.
%Type = Book
\bibitem[{Amusia et~al.(2016)Amusia, Semenov and Chernysheva}]{Amusia2016}
\bibinfo{author}{Amusia, M.Y.}, \bibinfo{author}{Semenov, S.K.},
  \bibinfo{author}{Chernysheva, L.V.}, \bibinfo{year}{2016}.
\newblock \bibinfo{title}{{ATOM-M}: algorithms and programs for the study of
  atomic and molecular processes}.
\newblock \bibinfo{publisher}{Nauka}, \bibinfo{address}{St. Petersburg,
  Russia}.
\newblock \bibinfo{note}{In Russian}.
%Type = Article
\bibitem[{Ancilotto et~al.(2017)Ancilotto, Barranco, Coppens, Eloranta,
  Halberstadt, Hernando, Mateo and Pi}]{Ancilotto2017}
\bibinfo{author}{Ancilotto, F.}, \bibinfo{author}{Barranco, M.},
  \bibinfo{author}{Coppens, F.}, \bibinfo{author}{Eloranta, J.},
  \bibinfo{author}{Halberstadt, N.}, \bibinfo{author}{Hernando, A.},
  \bibinfo{author}{Mateo, D.}, \bibinfo{author}{Pi, M.}, \bibinfo{year}{2017}.
\newblock \bibinfo{title}{Density functional theory of doped superfluid liquid
  helium and nanodroplets}.
\newblock \bibinfo{journal}{Int. Rev. Phys. Chem.} \bibinfo{volume}{36},
  \bibinfo{pages}{621--707}.
\newblock \DOIprefix\doi{10.1080/0144235x.2017.1351672}.
%Type = Article
\bibitem[{Arasaki et~al.(2011)Arasaki, Wang, McKoy and Takatsuka}]{Arasaki2011}
\bibinfo{author}{Arasaki, Y.}, \bibinfo{author}{Wang, K.},
  \bibinfo{author}{McKoy, V.}, \bibinfo{author}{Takatsuka, K.},
  \bibinfo{year}{2011}.
\newblock \bibinfo{title}{Monitoring the effect of a control pulse on a conical
  intersection by time-resolved photoelectron spectroscopy}.
\newblock \bibinfo{journal}{Phys. Chem. Chem. Phys.} \bibinfo{volume}{13},
  \bibinfo{pages}{8681--8689}.
\newblock \DOIprefix\doi{10.1039/c0cp02302g}.
%Type = Article
\bibitem[{Artemyev et~al.(2015)Artemyev, M\"uller, Hochstuhl and
  Demekhin}]{Artemyev2015}
\bibinfo{author}{Artemyev, A.N.}, \bibinfo{author}{M\"uller, A.D.},
  \bibinfo{author}{Hochstuhl, D.}, \bibinfo{author}{Demekhin, P.V.},
  \bibinfo{year}{2015}.
\newblock \bibinfo{title}{Photoelectron circular dichroism in the multiphoton
  ionization by short laser pulses. {I}. propagation of single-active-electron
  wave packets in chiral pseudo-potentials}.
\newblock \bibinfo{journal}{J. Chem. Phys.} \bibinfo{volume}{142},
  \bibinfo{pages}{244105}.
\newblock \DOIprefix\doi{10.1063/1.4922690}.
%Type = Article
\bibitem[{Attar et~al.(2017)Attar, Bhattacherjee, Pemmaraju, Schnorr, Closser,
  Prendergast and Leone}]{Attar2017}
\bibinfo{author}{Attar, A.R.}, \bibinfo{author}{Bhattacherjee, A.},
  \bibinfo{author}{Pemmaraju, C.D.}, \bibinfo{author}{Schnorr, K.},
  \bibinfo{author}{Closser, K.D.}, \bibinfo{author}{Prendergast, D.},
  \bibinfo{author}{Leone, S.R.}, \bibinfo{year}{2017}.
\newblock \bibinfo{title}{Femtosecond x-ray spectroscopy of an electrocyclic
  ring-opening reaction}.
\newblock \bibinfo{journal}{Science} \bibinfo{volume}{356},
  \bibinfo{pages}{54--59}.
\newblock \DOIprefix\doi{10.1126/science.aaj2198}.
%Type = Book
\bibitem[{Attwood and Sakdinawat(2016)}]{Attwood2016}
\bibinfo{author}{Attwood, D.}, \bibinfo{author}{Sakdinawat, A.},
  \bibinfo{year}{2016}.
\newblock \bibinfo{title}{X-Rays and Extreme Ultraviolet Radiation}.
\newblock \bibinfo{publisher}{Cambridge University Press}.
\newblock \DOIprefix\doi{10.1017/cbo9781107477629}.
%Type = Article
\bibitem[{Aue et~al.(1976)Aue, Bartholdi and Ernst}]{Aue1976}
\bibinfo{author}{Aue, W.P.}, \bibinfo{author}{Bartholdi, E.},
  \bibinfo{author}{Ernst, R.R.}, \bibinfo{year}{1976}.
\newblock \bibinfo{title}{Two-dimensional spectroscopy. application to nuclear
  magnetic resonance}.
\newblock \bibinfo{journal}{J. Chem. Phys.} \bibinfo{volume}{64},
  \bibinfo{pages}{2229--2246}.
\newblock \DOIprefix\doi{10.1063/1.432450}.
%Type = Article
\bibitem[{Augustin et~al.(2018)Augustin, Schulz, Schmid, Schnorr, Gryzlova,
  Lindenblatt, Meister, Liu, Trost, Fechner, Grum-Grzhimailo, Burkov, Braune,
  Treusch, Gisselbrecht, Schr\"oter, Pfeifer and Moshammer}]{Augustin2018}
\bibinfo{author}{Augustin, S.}, \bibinfo{author}{Schulz, M.},
  \bibinfo{author}{Schmid, G.}, \bibinfo{author}{Schnorr, K.},
  \bibinfo{author}{Gryzlova, E.V.}, \bibinfo{author}{Lindenblatt, H.},
  \bibinfo{author}{Meister, S.}, \bibinfo{author}{Liu, Y.F.},
  \bibinfo{author}{Trost, F.}, \bibinfo{author}{Fechner, L.},
  \bibinfo{author}{Grum-Grzhimailo, A.N.}, \bibinfo{author}{Burkov, S.M.},
  \bibinfo{author}{Braune, M.}, \bibinfo{author}{Treusch, R.},
  \bibinfo{author}{Gisselbrecht, M.}, \bibinfo{author}{Schr\"oter, C.D.},
  \bibinfo{author}{Pfeifer, T.}, \bibinfo{author}{Moshammer, R.},
  \bibinfo{year}{2018}.
\newblock \bibinfo{title}{Signatures of autoionization in the angular electron
  distribution in two-photon double ionization of {Ar}}.
\newblock \bibinfo{journal}{Phys. Rev. A} \bibinfo{volume}{98},
  \bibinfo{pages}{033408}.
\newblock \DOIprefix\doi{10.1103/PhysRevA.98.033408}.
%Type = Article
\bibitem[{Awasthi et~al.(2005)Awasthi, Vanne and Saenz}]{Awasthi2005}
\bibinfo{author}{Awasthi, M.}, \bibinfo{author}{Vanne, Y.V.},
  \bibinfo{author}{Saenz, A.}, \bibinfo{year}{2005}.
\newblock \bibinfo{title}{Non-perturbative solution of the time-dependent
  {S}chr\"odinger equation describing {H}$_2$ in intense short laser pulses}.
\newblock \bibinfo{journal}{J. Phys. B} \bibinfo{volume}{38},
  \bibinfo{pages}{3973--3985}.
\newblock \DOIprefix\doi{10.1088/0953-4075/38/22/005}.
%Type = Article
\bibitem[{Ayvazyan et~al.(2006)Ayvazyan, Baboi, B\"ahr, Balandin, Beutner,
  Brandt, Bohnet, Bolzmann, Brinkmann, Brovko, Carneiro, Casalbuoni,
  Castellano, Castro, Catani, Chiadroni, Choroba, Cianchi, Delsim-Hashemi,
  Di~Pirro, Dohlus, D\"usterer, Edwards, Faatz, Fateev, Feldhaus, Fl\"ottmann,
  Frisch, Fr\"ohlich, Garvey, Gensch, Golubeva, Grabosch, Grigoryan, Grimm,
  Hahn, Han, Hartrott, Honkavaara, H\"uning, Ischebeck, Jaeschke, Jablonka,
  Kammering, Katalev, Keitel, Khodyachykh, Kim, Kocharyan, K\"orfer, Kollewe,
  Kostin, Kr\"amer, Krassilnikov, Kube, Lilje, Limberg, Lipka, L\"ohl, Luong,
  Magne, Menzel, Michelato, Miltchev, Minty, M\"oller, Monaco, M\"uller, Nagl,
  Napoly, Nicolosi, N\"olle, Nu\~nez, Oppelt, Pagani, Paparella, Petersen,
  Petrosyan, Pfl\"uger, Piot, Pl\"onjes, Poletto, Proch, Pugachov, Rehlich,
  Richter, Riemann, Ross, Rossbach, Sachwitz, Saldin, Sandner, Schlarb,
  Schmidt, Schmitz, Schm\"user, Schneider, Schneidmiller, Schreiber, Schreiber,
  Shabunov, Sertore, Setzer, Simrock, Sombrowski, Staykov, Steffen, Stephan,
  Stulle, Sytchev, Thom, Tiedtke, Tischer, Treusch, Trines, Tsakov, Vardanyan,
  Wanzenberg, Weiland, Weise, Wendt, Will, Winter, Wittenburg, Yurkov,
  Zagorodnov, Zambolin and Zapfe}]{Ayvazyan2006}
\bibinfo{author}{Ayvazyan, V.}, \bibinfo{author}{Baboi, N.},
  \bibinfo{author}{B\"ahr, J.}, \bibinfo{author}{Balandin, V.},
  \bibinfo{author}{Beutner, B.}, \bibinfo{author}{Brandt, A.},
  \bibinfo{author}{Bohnet, I.}, \bibinfo{author}{Bolzmann, A.},
  \bibinfo{author}{Brinkmann, R.}, \bibinfo{author}{Brovko, O.I.},
  \bibinfo{author}{Carneiro, J.P.}, \bibinfo{author}{Casalbuoni, S.},
  \bibinfo{author}{Castellano, M.}, \bibinfo{author}{Castro, P.},
  \bibinfo{author}{Catani, L.}, \bibinfo{author}{Chiadroni, E.},
  \bibinfo{author}{Choroba, S.}, \bibinfo{author}{Cianchi, A.},
  \bibinfo{author}{Delsim-Hashemi, H.}, \bibinfo{author}{Di~Pirro, G.},
  \bibinfo{author}{Dohlus, M.}, \bibinfo{author}{D\"usterer, S.},
  \bibinfo{author}{Edwards, H.T.}, \bibinfo{author}{Faatz, B.},
  \bibinfo{author}{Fateev, A.A.}, \bibinfo{author}{Feldhaus, J.},
  \bibinfo{author}{Fl\"ottmann, K.}, \bibinfo{author}{Frisch, J.},
  \bibinfo{author}{Fr\"ohlich, L.}, \bibinfo{author}{Garvey, T.},
  \bibinfo{author}{Gensch, U.}, \bibinfo{author}{Golubeva, N.},
  \bibinfo{author}{Grabosch, H.J.}, \bibinfo{author}{Grigoryan, B.},
  \bibinfo{author}{Grimm, O.}, \bibinfo{author}{Hahn, U.},
  \bibinfo{author}{Han, J.H.}, \bibinfo{author}{Hartrott, M.V.},
  \bibinfo{author}{Honkavaara, K.}, \bibinfo{author}{H\"uning, M.},
  \bibinfo{author}{Ischebeck, R.}, \bibinfo{author}{Jaeschke, E.},
  \bibinfo{author}{Jablonka, M.}, \bibinfo{author}{Kammering, R.},
  \bibinfo{author}{Katalev, V.}, \bibinfo{author}{Keitel, B.},
  \bibinfo{author}{Khodyachykh, S.}, \bibinfo{author}{Kim, Y.},
  \bibinfo{author}{Kocharyan, V.}, \bibinfo{author}{K\"orfer, M.},
  \bibinfo{author}{Kollewe, M.}, \bibinfo{author}{Kostin, D.},
  \bibinfo{author}{Kr\"amer, D.}, \bibinfo{author}{Krassilnikov, M.},
  \bibinfo{author}{Kube, G.}, \bibinfo{author}{Lilje, L.},
  \bibinfo{author}{Limberg, T.}, \bibinfo{author}{Lipka, D.},
  \bibinfo{author}{L\"ohl, F.}, \bibinfo{author}{Luong, M.},
  \bibinfo{author}{Magne, C.}, \bibinfo{author}{Menzel, J.},
  \bibinfo{author}{Michelato, P.}, \bibinfo{author}{Miltchev, V.},
  \bibinfo{author}{Minty, M.}, \bibinfo{author}{M\"oller, W.D.},
  \bibinfo{author}{Monaco, L.}, \bibinfo{author}{M\"uller, W.},
  \bibinfo{author}{Nagl, M.}, \bibinfo{author}{Napoly, O.},
  \bibinfo{author}{Nicolosi, P.}, \bibinfo{author}{N\"olle, D.},
  \bibinfo{author}{Nu\~nez, T.}, \bibinfo{author}{Oppelt, A.},
  \bibinfo{author}{Pagani, C.}, \bibinfo{author}{Paparella, R.},
  \bibinfo{author}{Petersen, B.}, \bibinfo{author}{Petrosyan, B.},
  \bibinfo{author}{Pfl\"uger, J.}, \bibinfo{author}{Piot, P.},
  \bibinfo{author}{Pl\"onjes, E.}, \bibinfo{author}{Poletto, L.},
  \bibinfo{author}{Proch, D.}, \bibinfo{author}{Pugachov, D.},
  \bibinfo{author}{Rehlich, K.}, \bibinfo{author}{Richter, D.},
  \bibinfo{author}{Riemann, S.}, \bibinfo{author}{Ross, M.},
  \bibinfo{author}{Rossbach, J.}, \bibinfo{author}{Sachwitz, M.},
  \bibinfo{author}{Saldin, E.L.}, \bibinfo{author}{Sandner, W.},
  \bibinfo{author}{Schlarb, H.}, \bibinfo{author}{Schmidt, B.},
  \bibinfo{author}{Schmitz, M.}, \bibinfo{author}{Schm\"user, P.},
  \bibinfo{author}{Schneider, J.R.}, \bibinfo{author}{Schneidmiller, E.A.},
  \bibinfo{author}{Schreiber, H.J.}, \bibinfo{author}{Schreiber, S.},
  \bibinfo{author}{Shabunov, A.V.}, \bibinfo{author}{Sertore, D.},
  \bibinfo{author}{Setzer, S.}, \bibinfo{author}{Simrock, S.},
  \bibinfo{author}{Sombrowski, E.}, \bibinfo{author}{Staykov, L.},
  \bibinfo{author}{Steffen, B.}, \bibinfo{author}{Stephan, F.},
  \bibinfo{author}{Stulle, F.}, \bibinfo{author}{Sytchev, K.P.},
  \bibinfo{author}{Thom, H.}, \bibinfo{author}{Tiedtke, K.},
  \bibinfo{author}{Tischer, M.}, \bibinfo{author}{Treusch, R.},
  \bibinfo{author}{Trines, D.}, \bibinfo{author}{Tsakov, I.},
  \bibinfo{author}{Vardanyan, A.}, \bibinfo{author}{Wanzenberg, R.},
  \bibinfo{author}{Weiland, T.}, \bibinfo{author}{Weise, H.},
  \bibinfo{author}{Wendt, M.}, \bibinfo{author}{Will, I.},
  \bibinfo{author}{Winter, A.}, \bibinfo{author}{Wittenburg, K.},
  \bibinfo{author}{Yurkov, M.V.}, \bibinfo{author}{Zagorodnov, I.},
  \bibinfo{author}{Zambolin, P.}, \bibinfo{author}{Zapfe, K.},
  \bibinfo{year}{2006}.
\newblock \bibinfo{title}{First operation of a free-electron laser generating
  {GW} power radiation at 32\,nm wavelength}.
\newblock \bibinfo{journal}{Eur. Phys. J. D} \bibinfo{volume}{37},
  \bibinfo{pages}{297--303}.
\newblock \DOIprefix\doi{10.1140/epjd/e2005-00308-1}.
%Type = Article
\bibitem[{Ayvazyan et~al.(2002)Ayvazyan, Baboi, Schneider, Schneidmiller,
  Schreiber, Schreiber, Sertore, Setzer, Simrock, Sobierajski, Sonntag, Steeg,
  Stephan, Sytchev, Tiedtke, Tonutti, Treusch, Trines, T{\"{u}}rke, Verzilov,
  Wanzenberg, Weiland, Weise, Wendt, Wilhein, Will, Wittenburg, Wolff, Yurkov,
  Zapfe, Bohnet, Brinkmann, Castellano, Castro, Catani, Choroba, Cianchi,
  Dohlus, Edwards, Faatz, Fateev, Feldhaus, Fl{\"{o}}ttmann, Gamp, Garvey,
  Genz, Gerth, Gretchko, Grigoryan, Hahn, Hessler, Honkavaara, H{\"{u}}ning,
  Ischebeck, Jablonka, Kamps, K{\"{o}}rfer, Krassilnikov, Krzywinski, Liepe,
  Liero, Limberg, Loos, Luong, Magne, Menzel, Michelato, Minty, M{\"{u}}ller,
  N{\"{o}}lle, Novokhatski, Pagani, Peters, Pfl{\"{u}}ger, Piot, Plucinski,
  Rehlich, Reyzl, Richter, Rossbach, Saldin, Sandner, Schlarb, Schmidt and
  Schm{\"{u}}ser}]{Ayvazyan2002}
\bibinfo{author}{Ayvazyan, V.}, \bibinfo{author}{Baboi, N.},
  \bibinfo{author}{Schneider, J.R.}, \bibinfo{author}{Schneidmiller, E.A.},
  \bibinfo{author}{Schreiber, H.J.}, \bibinfo{author}{Schreiber, S.},
  \bibinfo{author}{Sertore, D.}, \bibinfo{author}{Setzer, S.},
  \bibinfo{author}{Simrock, S.}, \bibinfo{author}{Sobierajski, R.},
  \bibinfo{author}{Sonntag, B.}, \bibinfo{author}{Steeg, B.},
  \bibinfo{author}{Stephan, F.}, \bibinfo{author}{Sytchev, K.P.},
  \bibinfo{author}{Tiedtke, K.}, \bibinfo{author}{Tonutti, M.},
  \bibinfo{author}{Treusch, R.}, \bibinfo{author}{Trines, D.},
  \bibinfo{author}{T{\"{u}}rke, D.}, \bibinfo{author}{Verzilov, V.},
  \bibinfo{author}{Wanzenberg, R.}, \bibinfo{author}{Weiland, T.},
  \bibinfo{author}{Weise, H.}, \bibinfo{author}{Wendt, M.},
  \bibinfo{author}{Wilhein, T.}, \bibinfo{author}{Will, I.},
  \bibinfo{author}{Wittenburg, K.}, \bibinfo{author}{Wolff, S.},
  \bibinfo{author}{Yurkov, M.V.}, \bibinfo{author}{Zapfe, K.},
  \bibinfo{author}{Bohnet, I.}, \bibinfo{author}{Brinkmann, R.},
  \bibinfo{author}{Castellano, M.}, \bibinfo{author}{Castro, P.},
  \bibinfo{author}{Catani, L.}, \bibinfo{author}{Choroba, S.},
  \bibinfo{author}{Cianchi, A.}, \bibinfo{author}{Dohlus, M.},
  \bibinfo{author}{Edwards, H.T.}, \bibinfo{author}{Faatz, B.},
  \bibinfo{author}{Fateev, A.A.}, \bibinfo{author}{Feldhaus, J.},
  \bibinfo{author}{Fl{\"{o}}ttmann, K.}, \bibinfo{author}{Gamp, A.},
  \bibinfo{author}{Garvey, T.}, \bibinfo{author}{Genz, H.},
  \bibinfo{author}{Gerth, C.}, \bibinfo{author}{Gretchko, V.},
  \bibinfo{author}{Grigoryan, B.}, \bibinfo{author}{Hahn, U.},
  \bibinfo{author}{Hessler, C.}, \bibinfo{author}{Honkavaara, K.},
  \bibinfo{author}{H{\"{u}}ning, M.}, \bibinfo{author}{Ischebeck, R.},
  \bibinfo{author}{Jablonka, M.}, \bibinfo{author}{Kamps, T.},
  \bibinfo{author}{K{\"{o}}rfer, M.}, \bibinfo{author}{Krassilnikov, M.},
  \bibinfo{author}{Krzywinski, J.}, \bibinfo{author}{Liepe, M.},
  \bibinfo{author}{Liero, A.}, \bibinfo{author}{Limberg, T.},
  \bibinfo{author}{Loos, H.}, \bibinfo{author}{Luong, M.},
  \bibinfo{author}{Magne, C.}, \bibinfo{author}{Menzel, J.},
  \bibinfo{author}{Michelato, P.}, \bibinfo{author}{Minty, M.},
  \bibinfo{author}{M{\"{u}}ller, U.C.}, \bibinfo{author}{N{\"{o}}lle, D.},
  \bibinfo{author}{Novokhatski, A.}, \bibinfo{author}{Pagani, C.},
  \bibinfo{author}{Peters, F.}, \bibinfo{author}{Pfl{\"{u}}ger, J.},
  \bibinfo{author}{Piot, P.}, \bibinfo{author}{Plucinski, L.},
  \bibinfo{author}{Rehlich, K.}, \bibinfo{author}{Reyzl, I.},
  \bibinfo{author}{Richter, A.}, \bibinfo{author}{Rossbach, J.},
  \bibinfo{author}{Saldin, E.L.}, \bibinfo{author}{Sandner, W.},
  \bibinfo{author}{Schlarb, H.}, \bibinfo{author}{Schmidt, G.},
  \bibinfo{author}{Schm{\"{u}}ser, P.}, \bibinfo{year}{2002}.
\newblock \bibinfo{title}{A new powerful source for coherent {VUV} radiation:
  Demonstration of exponential growth and saturation at the {TTF} free-electron
  laser}.
\newblock \bibinfo{journal}{Eur. Phys. J. D} \bibinfo{volume}{20},
  \bibinfo{pages}{149--156}.
\newblock \DOIprefix\doi{10.1140/epjd/e2002-00121-4}.
%Type = Article
\bibitem[{Bachau et~al.(2001)Bachau, Cormier, Decleva, Hansen and
  Martin}]{Bachau2001}
\bibinfo{author}{Bachau, H.}, \bibinfo{author}{Cormier, E.},
  \bibinfo{author}{Decleva, P.}, \bibinfo{author}{Hansen, J.E.},
  \bibinfo{author}{Martin, F.}, \bibinfo{year}{2001}.
\newblock \bibinfo{title}{Applications of {B}-splines in atomic and molecular
  physics}.
\newblock \bibinfo{journal}{Rep. Prog. Phys.} \bibinfo{volume}{64},
  \bibinfo{pages}{1815--1942}.
\newblock \DOIprefix\doi{10.1088/0034-4885/64/12/205}.
%Type = Article
\bibitem[{Bahrdt et~al.(1992)Bahrdt, Gaupp, Gudat, Mast, Molter, Peatman,
  Scheer, Schroeter and Wang}]{Bahrdt1992}
\bibinfo{author}{Bahrdt, J.}, \bibinfo{author}{Gaupp, A.},
  \bibinfo{author}{Gudat, W.}, \bibinfo{author}{Mast, M.},
  \bibinfo{author}{Molter, K.}, \bibinfo{author}{Peatman, W.B.},
  \bibinfo{author}{Scheer, M.}, \bibinfo{author}{Schroeter, T.},
  \bibinfo{author}{Wang, C.}, \bibinfo{year}{1992}.
\newblock \bibinfo{title}{Circularly polarized synchrotron radiation from the
  crossed undulator at {BESSY}}.
\newblock \bibinfo{journal}{Rev. Sci. Instrum.} \bibinfo{volume}{63},
  \bibinfo{pages}{339--342}.
\newblock \DOIprefix\doi{10.1063/1.1142750}.
%Type = Book
\bibitem[{Balashov et~al.(2000)Balashov, Grum-Grzhimailo and
  Kabachnik}]{Balashov2000}
\bibinfo{author}{Balashov, V.V.}, \bibinfo{author}{Grum-Grzhimailo, A.N.},
  \bibinfo{author}{Kabachnik, N.M.}, \bibinfo{year}{2000}.
\newblock \bibinfo{title}{Polarization and Correlation Phenomena in Atomic
  Collisions. {A} Practical Theory Course}.
\newblock Physics of Atoms and Molecules, \bibinfo{publisher}{Kluwer
  Academic/Plenum Publishers}, \bibinfo{address}{New York Boston Dordrecht
  London Moscow}.
\newblock \DOIprefix\doi{10.1007/978-1-4757-3228-3}.
%Type = Article
\bibitem[{Bandrauk and Lu(2013)}]{Bandrauk2013}
\bibinfo{author}{Bandrauk, A.D.}, \bibinfo{author}{Lu, H.},
  \bibinfo{year}{2013}.
\newblock \bibinfo{title}{Exponential propagators (integrators) for the
  time-dependent {S}chr\"odinger equation}.
\newblock \bibinfo{journal}{J. Theor. Comp. Chem.} \bibinfo{volume}{12},
  \bibinfo{pages}{1340001}.
\newblock \DOIprefix\doi{10.1142/S0219633613400014}.
%Type = Article
\bibitem[{{Baranova} et~al.(1992){Baranova}, {Beterov}, {Zel'Dovich},
  {Ryabtsev}, {Chudinov} and {Shul'Ginov}}]{Baranova1992}
\bibinfo{author}{{Baranova}, N.B.}, \bibinfo{author}{{Beterov}, I.M.},
  \bibinfo{author}{{Zel'Dovich}, B.Y.}, \bibinfo{author}{{Ryabtsev}, I.I.},
  \bibinfo{author}{{Chudinov}, A.N.}, \bibinfo{author}{{Shul'Ginov}, A.A.},
  \bibinfo{year}{1992}.
\newblock \bibinfo{title}{Observation of an interference of one- and two-photon
  ionization of the sodium 4{s} state}.
\newblock \bibinfo{journal}{JETP Lett.} \bibinfo{volume}{55},
  \bibinfo{pages}{439--444}.
\newblock \URLprefix
  \url{http://www.jetpletters.ac.ru/ps/1275/article_19280.shtml}.
%Type = Book
\bibitem[{Barron(2004)}]{Barron2004}
\bibinfo{author}{Barron, L.D.}, \bibinfo{year}{2004}.
\newblock \bibinfo{title}{Molecular Light Scattering and Optical Activity}.
\newblock \bibinfo{publisher}{Cambridge University Press}.
\newblock \DOIprefix\doi{10.1017/cbo9780511535468}.
%Type = Article
\bibitem[{Bartels et~al.(2000)Bartels, Backus, Zeek, Misoguti, Vdovin,
  Christov, Murnane and Kapteyn}]{Bartels2000}
\bibinfo{author}{Bartels, R.}, \bibinfo{author}{Backus, S.},
  \bibinfo{author}{Zeek, E.}, \bibinfo{author}{Misoguti, L.},
  \bibinfo{author}{Vdovin, G.}, \bibinfo{author}{Christov, I.P.},
  \bibinfo{author}{Murnane, M.M.}, \bibinfo{author}{Kapteyn, H.C.},
  \bibinfo{year}{2000}.
\newblock \bibinfo{title}{Shaped-pulse optimization of coherent emission of
  high-harmonic soft {X}-rays}.
\newblock \bibinfo{journal}{Nature} \bibinfo{volume}{406},
  \bibinfo{pages}{164--166}.
\newblock \DOIprefix\doi{10.1038/35018029}.
%Type = Article
\bibitem[{Bauch et~al.(2014)Bauch, S\o{}rensen and Madsen}]{Bauch2014}
\bibinfo{author}{Bauch, S.}, \bibinfo{author}{S\o{}rensen, L.K.},
  \bibinfo{author}{Madsen, L.B.}, \bibinfo{year}{2014}.
\newblock \bibinfo{title}{Time-dependent generalized-active-space
  configuration-interaction approach to photoionization dynamics of atoms and
  molecules}.
\newblock \bibinfo{journal}{Phys. Rev. A} \bibinfo{volume}{90},
  \bibinfo{pages}{062508}.
\newblock \DOIprefix\doi{10.1103/PhysRevA.90.062508}.
%Type = Article
\bibitem[{Beck et~al.(2000)Beck, Ja\"ckle, Worth and Meyer}]{Beck2000}
\bibinfo{author}{Beck, M.}, \bibinfo{author}{Ja\"ckle, A.},
  \bibinfo{author}{Worth, G.}, \bibinfo{author}{Meyer, H.D.},
  \bibinfo{year}{2000}.
\newblock \bibinfo{title}{The multiconfiguration time-dependent {H}artree
  ({MCTDH}) method: a highly effcient algorithm for propagating wavepackets}.
\newblock \bibinfo{journal}{Phys. Rep.} \bibinfo{volume}{324},
  \bibinfo{pages}{1--105}.
\newblock \DOIprefix\doi{10.1016/S0370-1573(99)00047-2}.
%Type = Article
\bibitem[{Bencivenga et~al.(2015)Bencivenga, Cucini, Capotondi, Battistoni,
  Mincigrucci, Giangrisostomi, Gessini, Manfredda, Nikolov, Pedersoli,
  Principi, Svetina, Parisse, Casolari, Danailov, Kiskinova and
  Masciovecchio}]{Bencivenga2015}
\bibinfo{author}{Bencivenga, F.}, \bibinfo{author}{Cucini, R.},
  \bibinfo{author}{Capotondi, F.}, \bibinfo{author}{Battistoni, A.},
  \bibinfo{author}{Mincigrucci, R.}, \bibinfo{author}{Giangrisostomi, E.},
  \bibinfo{author}{Gessini, A.}, \bibinfo{author}{Manfredda, M.},
  \bibinfo{author}{Nikolov, I.P.}, \bibinfo{author}{Pedersoli, E.},
  \bibinfo{author}{Principi, E.}, \bibinfo{author}{Svetina, C.},
  \bibinfo{author}{Parisse, P.}, \bibinfo{author}{Casolari, F.},
  \bibinfo{author}{Danailov, M.B.}, \bibinfo{author}{Kiskinova, M.},
  \bibinfo{author}{Masciovecchio, C.}, \bibinfo{year}{2015}.
\newblock \bibinfo{title}{Four-wave mixing experiments with extreme ultraviolet
  transient gratings}.
\newblock \bibinfo{journal}{Nature} \bibinfo{volume}{520},
  \bibinfo{pages}{205--208}.
\newblock \DOIprefix\doi{10.1038/nature14341}.
%Type = Article
\bibitem[{Berrah et~al.(2010)Berrah, Bozek, Costello, D{\"u}sterer, Fang,
  Feldhaus, Fukuzawa, Hoener, Jiang, Johnsson, Kennedy, Meyer, Moshammer,
  Radcliffe, Richter, Rouz{\'{e}}e, Rudenko, Sorokin, Tiedtke, Ueda, Ullrich
  and Vrakking}]{Berrah_JModOpt_2010}
\bibinfo{author}{Berrah, N.}, \bibinfo{author}{Bozek, J.},
  \bibinfo{author}{Costello, J.}, \bibinfo{author}{D{\"u}sterer, S.},
  \bibinfo{author}{Fang, L.}, \bibinfo{author}{Feldhaus, J.},
  \bibinfo{author}{Fukuzawa, H.}, \bibinfo{author}{Hoener, M.},
  \bibinfo{author}{Jiang, Y.}, \bibinfo{author}{Johnsson, P.},
  \bibinfo{author}{Kennedy, E.}, \bibinfo{author}{Meyer, M.},
  \bibinfo{author}{Moshammer, R.}, \bibinfo{author}{Radcliffe, P.},
  \bibinfo{author}{Richter, M.}, \bibinfo{author}{Rouz{\'{e}}e, A.},
  \bibinfo{author}{Rudenko, A.}, \bibinfo{author}{Sorokin, A.},
  \bibinfo{author}{Tiedtke, K.}, \bibinfo{author}{Ueda, K.},
  \bibinfo{author}{Ullrich, J.}, \bibinfo{author}{Vrakking, M.},
  \bibinfo{year}{2010}.
\newblock \bibinfo{title}{Non-linear processes in the interaction of atoms and
  molecules with intense {EUV} and {X}-ray fields from {SASE} free electron
  lasers ({FELs})}.
\newblock \bibinfo{journal}{J. Mod. Opt.} \bibinfo{volume}{57},
  \bibinfo{pages}{1015--1040}.
\newblock \DOIprefix\doi{10.1080/09500340.2010.487946}.
%Type = Article
\bibitem[{Berrah et~al.(2011)Berrah, Fang, Murphy, Osipov, Ueda, Kukk, Feifel,
  van~der Meulen, Salen, Schmidt, Thomas, Larsson, Richter, Prince, Bozek,
  Bostedt, Wada, Piancastelli, Tashiro and Ehara}]{Berrah2011}
\bibinfo{author}{Berrah, N.}, \bibinfo{author}{Fang, L.},
  \bibinfo{author}{Murphy, B.}, \bibinfo{author}{Osipov, T.},
  \bibinfo{author}{Ueda, K.}, \bibinfo{author}{Kukk, E.},
  \bibinfo{author}{Feifel, R.}, \bibinfo{author}{van~der Meulen, P.},
  \bibinfo{author}{Salen, P.}, \bibinfo{author}{Schmidt, H.T.},
  \bibinfo{author}{Thomas, R.D.}, \bibinfo{author}{Larsson, M.},
  \bibinfo{author}{Richter, R.}, \bibinfo{author}{Prince, K.C.},
  \bibinfo{author}{Bozek, J.D.}, \bibinfo{author}{Bostedt, C.},
  \bibinfo{author}{Wada, S.}, \bibinfo{author}{Piancastelli, M.N.},
  \bibinfo{author}{Tashiro, M.}, \bibinfo{author}{Ehara, M.},
  \bibinfo{year}{2011}.
\newblock \bibinfo{title}{Double-core-hole spectroscopy for chemical analysis
  with an intense x-ray femtosecond laser}.
\newblock \bibinfo{journal}{Proceedings of the National Academy of Sciences}
  \bibinfo{volume}{108}, \bibinfo{pages}{16912--16915}.
\newblock \DOIprefix\doi{10.1073/pnas.1111380108}.
%Type = Article
\bibitem[{Berrah et~al.(2019)Berrah, Sanchez-Gonzalez, Jurek, Obaid, Xiong,
  Squibb, Osipov, Lutman, Fang, Barillot, Bozek, Cryan, Wolf, Rolles, Coffee,
  Schnorr, Augustin, Fukuzawa, Motomura, Niebuhr, Frasinski, Feifel, Schulz,
  Toyota, Son, Ueda, Pfeifer, Marangos and Santra}]{Berrah2019}
\bibinfo{author}{Berrah, N.}, \bibinfo{author}{Sanchez-Gonzalez, A.},
  \bibinfo{author}{Jurek, Z.}, \bibinfo{author}{Obaid, R.},
  \bibinfo{author}{Xiong, H.}, \bibinfo{author}{Squibb, R.J.},
  \bibinfo{author}{Osipov, T.}, \bibinfo{author}{Lutman, A.},
  \bibinfo{author}{Fang, L.}, \bibinfo{author}{Barillot, T.},
  \bibinfo{author}{Bozek, J.D.}, \bibinfo{author}{Cryan, J.},
  \bibinfo{author}{Wolf, T.J.A.}, \bibinfo{author}{Rolles, D.},
  \bibinfo{author}{Coffee, R.}, \bibinfo{author}{Schnorr, K.},
  \bibinfo{author}{Augustin, S.}, \bibinfo{author}{Fukuzawa, H.},
  \bibinfo{author}{Motomura, K.}, \bibinfo{author}{Niebuhr, N.},
  \bibinfo{author}{Frasinski, L.J.}, \bibinfo{author}{Feifel, R.},
  \bibinfo{author}{Schulz, C.P.}, \bibinfo{author}{Toyota, K.},
  \bibinfo{author}{Son, S.K.}, \bibinfo{author}{Ueda, K.},
  \bibinfo{author}{Pfeifer, T.}, \bibinfo{author}{Marangos, J.P.},
  \bibinfo{author}{Santra, R.}, \bibinfo{year}{2019}.
\newblock \bibinfo{title}{Femtosecond-resolved observation of the fragmentation
  of buckminsterfullerene following {X}-ray multiphoton ionization}.
\newblock \bibinfo{journal}{Nat. Phys.} \bibinfo{volume}{15},
  \bibinfo{pages}{1279--1283}.
\newblock \DOIprefix\doi{10.1038/s41567-019-0665-7}.
%Type = Article
\bibitem[{Berrington et~al.(1995)Berrington, Eissner and
  Norrington}]{Berrington1995}
\bibinfo{author}{Berrington, K.A.}, \bibinfo{author}{Eissner, W.B.},
  \bibinfo{author}{Norrington, P.H.}, \bibinfo{year}{1995}.
\newblock \bibinfo{title}{{RMATRX}1: Belfast atomic {R}-matrix codes}.
\newblock \bibinfo{journal}{Comput. Phys. Commun.} \bibinfo{volume}{92},
  \bibinfo{pages}{290--420}.
\newblock \DOIprefix\doi{10.1016/0010-4655(95)00123-8}.
%Type = Article
\bibitem[{Beutler(1935)}]{Beutler1935}
\bibinfo{author}{Beutler, H.}, \bibinfo{year}{1935}.
\newblock \bibinfo{title}{{\"U}ber {A}bsorptionsserien von {A}rgon, {K}rypton
  und {X}enon zu {T}ermen zwischen den beiden {I}onisierungsgrenzen
  $^{2}${P}$_{3}^{2/0}$ und $^{2}${P}$_{1}^{2/0}$}.
\newblock \bibinfo{journal}{Z. Phys.} \bibinfo{volume}{93},
  \bibinfo{pages}{177--196}.
\newblock \DOIprefix\doi{10.1007/bf01365116}.
%Type = Article
\bibitem[{Bhattacherjee et~al.(2017)Bhattacherjee, Pemmaraju, Schnorr, Attar
  and Leone}]{Bhattacherjee2017}
\bibinfo{author}{Bhattacherjee, A.}, \bibinfo{author}{Pemmaraju, C.D.},
  \bibinfo{author}{Schnorr, K.}, \bibinfo{author}{Attar, A.R.},
  \bibinfo{author}{Leone, S.R.}, \bibinfo{year}{2017}.
\newblock \bibinfo{title}{Ultrafast intersystem crossing in acetylacetone via
  femtosecond {X}-ray transient absorption at the carbon {K}-edge}.
\newblock \bibinfo{journal}{J. Am. Chem. Soc.} \bibinfo{volume}{139},
  \bibinfo{pages}{16576--16583}.
\newblock \DOIprefix\doi{10.1021/jacs.7b07532}.
%Type = Book
\bibitem[{Blum(2012)}]{Blum2012}
\bibinfo{author}{Blum, K.}, \bibinfo{year}{2012}.
\newblock \bibinfo{title}{Density Matrix Theory and Applications}.
  volume~\bibinfo{volume}{64} of \textit{\bibinfo{series}{Springer Series on
  Atomic, Optical, and Plasma Physics}}.
\newblock \bibinfo{edition}{Third} ed., \bibinfo{publisher}{Springer-Verlag},
  \bibinfo{address}{Berlin, Heidelberg}.
\newblock \DOIprefix\doi{10.1007/978-3-642-20561-3}.
%Type = Article
\bibitem[{Boll and Foj\'on(2014)}]{Boll2014}
\bibinfo{author}{Boll, D.I.R.}, \bibinfo{author}{Foj\'on, O.A.},
  \bibinfo{year}{2014}.
\newblock \bibinfo{title}{Interferences and asymmetries in laser-assisted
  photoionization of diatomic molecules}.
\newblock \bibinfo{journal}{Phys. Rev. A} \bibinfo{volume}{90},
  \bibinfo{pages}{053414}.
\newblock \DOIprefix\doi{10.1103/PhysRevA.90.053414}.
%Type = Article
\bibitem[{Boll and Foj\'on(2016)}]{Boll2016}
\bibinfo{author}{Boll, D.I.R.}, \bibinfo{author}{Foj\'on, O.A.},
  \bibinfo{year}{2016}.
\newblock \bibinfo{title}{Atomic {RABBITT}-like experiments framed as diatomic
  molecules}.
\newblock \bibinfo{journal}{J. Phys. B} \bibinfo{volume}{49},
  \bibinfo{pages}{185601}.
\newblock \DOIprefix\doi{10.1088/0953-4075/49/18/185601}.
%Type = Article
\bibitem[{Boll et~al.(2013)Boll, Anielski, Bostedt, Bozek, Christensen, Coffee,
  De, Decleva, Epp, Erk, Foucar, Krasniqi, K\"upper, Rouz\'ee, Rudek, Rudenko,
  Schorb, Stapelfeldt, Stener, Stern, Techert, Trippel, Vrakking, Ullrich and
  Rolles}]{Boll2013}
\bibinfo{author}{Boll, R.}, \bibinfo{author}{Anielski, D.},
  \bibinfo{author}{Bostedt, C.}, \bibinfo{author}{Bozek, J.D.},
  \bibinfo{author}{Christensen, L.}, \bibinfo{author}{Coffee, R.},
  \bibinfo{author}{De, S.}, \bibinfo{author}{Decleva, P.},
  \bibinfo{author}{Epp, S.W.}, \bibinfo{author}{Erk, B.},
  \bibinfo{author}{Foucar, L.}, \bibinfo{author}{Krasniqi, F.},
  \bibinfo{author}{K\"upper, J.}, \bibinfo{author}{Rouz\'ee, A.},
  \bibinfo{author}{Rudek, B.}, \bibinfo{author}{Rudenko, A.},
  \bibinfo{author}{Schorb, S.}, \bibinfo{author}{Stapelfeldt, H.},
  \bibinfo{author}{Stener, M.}, \bibinfo{author}{Stern, S.},
  \bibinfo{author}{Techert, S.}, \bibinfo{author}{Trippel, S.},
  \bibinfo{author}{Vrakking, M.J.J.}, \bibinfo{author}{Ullrich, J.},
  \bibinfo{author}{Rolles, D.}, \bibinfo{year}{2013}.
\newblock \bibinfo{title}{Femtosecond photoelectron diffraction on
  laser-aligned molecules: Towards time-resolved imaging of molecular
  structure}.
\newblock \bibinfo{journal}{Phys. Rev. A} \bibinfo{volume}{88},
  \bibinfo{pages}{061402}.
\newblock \DOIprefix\doi{10.1103/PhysRevA.88.061402}.
%Type = Article
\bibitem[{Bonifacio et~al.(1990a)Bonifacio, Casagrande, Cerchioni,
  de~Salvo~Souza, Pierini and Piovella}]{Bonifacio1990a}
\bibinfo{author}{Bonifacio, R.}, \bibinfo{author}{Casagrande, F.},
  \bibinfo{author}{Cerchioni, G.}, \bibinfo{author}{de~Salvo~Souza, L.},
  \bibinfo{author}{Pierini, P.}, \bibinfo{author}{Piovella, N.},
  \bibinfo{year}{1990}a.
\newblock \bibinfo{title}{Physics of the high-gain {FEL} and superradiance}.
\newblock \bibinfo{journal}{Nuovo Cimento} \bibinfo{volume}{13},
  \bibinfo{pages}{1--69}.
\newblock \DOIprefix\doi{10.1007/BF02770850}.
%Type = Article
\bibitem[{Bonifacio et~al.(1994)Bonifacio, De~Salvo, Pierini, Piovella and
  Pellegrini}]{Bonifacio1994}
\bibinfo{author}{Bonifacio, R.}, \bibinfo{author}{De~Salvo, L.},
  \bibinfo{author}{Pierini, P.}, \bibinfo{author}{Piovella, N.},
  \bibinfo{author}{Pellegrini, C.}, \bibinfo{year}{1994}.
\newblock \bibinfo{title}{Spectrum, temporal structure, and fluctuations in a
  high-gain free-electron laser starting from noise}.
\newblock \bibinfo{journal}{Phys. Rev. Lett.} \bibinfo{volume}{73},
  \bibinfo{pages}{70--73}.
\newblock \DOIprefix\doi{10.1103/PhysRevLett.73.70}.
%Type = Article
\bibitem[{Bonifacio et~al.(1990b)Bonifacio, De~Salvo~Souza, Pierini and
  Scharlemann}]{Bonifacio1990}
\bibinfo{author}{Bonifacio, R.}, \bibinfo{author}{De~Salvo~Souza, L.},
  \bibinfo{author}{Pierini, P.}, \bibinfo{author}{Scharlemann, E.T.},
  \bibinfo{year}{1990}b.
\newblock \bibinfo{title}{Generation of {XUV} light by resonant frequency
  tripling in a two-wiggler {FEL} amplifier}.
\newblock \bibinfo{journal}{Nucl. Instrum. Methods Phys. Res. A}
  \bibinfo{volume}{296}, \bibinfo{pages}{787--790}.
\newblock \DOIprefix\doi{10.1016/0168-9002(90)91307-w}.
%Type = Article
\bibitem[{Bostedt et~al.(2016)Bostedt, Boutet, Fritz, Huang, Lee, Lemke,
  Robert, Schlotter, Turner and Williams}]{Bostedt2016}
\bibinfo{author}{Bostedt, C.}, \bibinfo{author}{Boutet, S.},
  \bibinfo{author}{Fritz, D.M.}, \bibinfo{author}{Huang, Z.},
  \bibinfo{author}{Lee, H.J.}, \bibinfo{author}{Lemke, H.T.},
  \bibinfo{author}{Robert, A.}, \bibinfo{author}{Schlotter, W.F.},
  \bibinfo{author}{Turner, J.J.}, \bibinfo{author}{Williams, G.J.},
  \bibinfo{year}{2016}.
\newblock \bibinfo{title}{{L}inac {C}oherent {L}ight {S}ource: The first five
  years}.
\newblock \bibinfo{journal}{Rev. Mod. Phys.} \bibinfo{volume}{88},
  \bibinfo{pages}{015007}.
\newblock \DOIprefix\doi{10.1103/RevModPhys.88.015007}.
%Type = Article
\bibitem[{Bostedt et~al.(2009)Bostedt, Chapman, Costello, L\'opez-Urrutia,
  D\"usterer, Epp, Feldhaus, F\"ohlisch, Meyer, M\"oller, Moshammer, Richter,
  Sokolowski-Tinten, Sorokin, Tiedtke, Ullrich and Wurth}]{Bostedt2009}
\bibinfo{author}{Bostedt, C.}, \bibinfo{author}{Chapman, H.N.},
  \bibinfo{author}{Costello, J.T.}, \bibinfo{author}{L\'opez-Urrutia, J.R.C.},
  \bibinfo{author}{D\"usterer, S.}, \bibinfo{author}{Epp, S.W.},
  \bibinfo{author}{Feldhaus, J.}, \bibinfo{author}{F\"ohlisch, A.},
  \bibinfo{author}{Meyer, M.}, \bibinfo{author}{M\"oller, T.},
  \bibinfo{author}{Moshammer, R.}, \bibinfo{author}{Richter, M.},
  \bibinfo{author}{Sokolowski-Tinten, K.}, \bibinfo{author}{Sorokin, A.},
  \bibinfo{author}{Tiedtke, K.}, \bibinfo{author}{Ullrich, J.},
  \bibinfo{author}{Wurth, W.}, \bibinfo{year}{2009}.
\newblock \bibinfo{title}{Experiments at {FLASH}}.
\newblock \bibinfo{journal}{Nucl. Instrum. Methods Phys. Res. A}
  \bibinfo{volume}{601}, \bibinfo{pages}{108 -- 122}.
\newblock \DOIprefix\doi{10.1016/j.nima.2008.12.202}. \bibinfo{note}{special
  issue in honour of Prof. Kai Siegbahn}.
%Type = Article
\bibitem[{Burke and Burke(1997)}]{Burke1997}
\bibinfo{author}{Burke, P.G.}, \bibinfo{author}{Burke, V.M.},
  \bibinfo{year}{1997}.
\newblock \bibinfo{title}{Time-dependent {R}-matrix theory of multiphoton
  processes}.
\newblock \bibinfo{journal}{J. Phys. B} \bibinfo{volume}{30},
  \bibinfo{pages}{L383--L391}.
\newblock \DOIprefix\doi{10.1088/0953-4075/30/11/002}.
%Type = Article
\bibitem[{Buth et~al.(2018)Buth, Beerwerth, Obaid, Berrah, Cederbaum and
  Fritzsche}]{Buth2018}
\bibinfo{author}{Buth, C.}, \bibinfo{author}{Beerwerth, R.},
  \bibinfo{author}{Obaid, R.}, \bibinfo{author}{Berrah, N.},
  \bibinfo{author}{Cederbaum, L.S.}, \bibinfo{author}{Fritzsche, S.},
  \bibinfo{year}{2018}.
\newblock \bibinfo{title}{Neon in ultrashort and intense x-rays from free
  electron lasers}.
\newblock \bibinfo{journal}{J. Phys. B} \bibinfo{volume}{51},
  \bibinfo{pages}{055602}.
\newblock \DOIprefix\doi{10.1088/1361-6455/aaa39a}.
%Type = Article
\bibitem[{Caillat et~al.(2005)Caillat, Zanghellini, Kitzler, Koch, Kreuzer and
  Scrinzi}]{Caillat2005}
\bibinfo{author}{Caillat, J.}, \bibinfo{author}{Zanghellini, J.},
  \bibinfo{author}{Kitzler, M.}, \bibinfo{author}{Koch, O.},
  \bibinfo{author}{Kreuzer, W.}, \bibinfo{author}{Scrinzi, A.},
  \bibinfo{year}{2005}.
\newblock \bibinfo{title}{Correlated multielectron systems in strong laser
  fields: A multiconfiguration time-dependent {H}artree-{F}ock approach}.
\newblock \bibinfo{journal}{Phys. Rev. A} \bibinfo{volume}{71},
  \bibinfo{pages}{012712}.
\newblock \DOIprefix\doi{10.1103/PhysRevA.71.012712}.
%Type = Article
\bibitem[{Carpeggiani et~al.(2019)Carpeggiani, Gryzlova, Reduzzi, Dubrouil,
  Faccial{\'{a}}, Negro, Ueda, Burkov, Frassetto, Stienkemeier, Ovcharenko,
  Meyer, Plekan, Finetti, Prince, Callegari, Grum-Grzhimailo and
  Sansone}]{Carpeggiani2019}
\bibinfo{author}{Carpeggiani, P.A.}, \bibinfo{author}{Gryzlova, E.V.},
  \bibinfo{author}{Reduzzi, M.}, \bibinfo{author}{Dubrouil, A.},
  \bibinfo{author}{Faccial{\'{a}}, D.}, \bibinfo{author}{Negro, M.},
  \bibinfo{author}{Ueda, K.}, \bibinfo{author}{Burkov, S.M.},
  \bibinfo{author}{Frassetto, F.}, \bibinfo{author}{Stienkemeier, F.},
  \bibinfo{author}{Ovcharenko, Y.}, \bibinfo{author}{Meyer, M.},
  \bibinfo{author}{Plekan, O.}, \bibinfo{author}{Finetti, P.},
  \bibinfo{author}{Prince, K.C.}, \bibinfo{author}{Callegari, C.},
  \bibinfo{author}{Grum-Grzhimailo, A.N.}, \bibinfo{author}{Sansone, G.},
  \bibinfo{year}{2019}.
\newblock \bibinfo{title}{Complete reconstruction of bound and unbound
  electronic wavefunctions in two-photon double ionization}.
\newblock \bibinfo{journal}{Nat. Phys.} \bibinfo{volume}{15},
  \bibinfo{pages}{170--177}.
\newblock \DOIprefix\doi{10.1038/s41567-018-0340-4}.
%Type = Article
\bibitem[{Cederbaum et~al.(1986)Cederbaum, Tarantelli, Sgamellotti and
  Schirmer}]{Cederbaum1986}
\bibinfo{author}{Cederbaum, L.S.}, \bibinfo{author}{Tarantelli, F.},
  \bibinfo{author}{Sgamellotti, A.}, \bibinfo{author}{Schirmer, J.},
  \bibinfo{year}{1986}.
\newblock \bibinfo{title}{On double vacancies in the core}.
\newblock \bibinfo{journal}{J. Chem. Phys.} \bibinfo{volume}{85},
  \bibinfo{pages}{6513--6523}.
\newblock \DOIprefix\doi{10.1063/1.451432}.
%Type = Article
\bibitem[{Cederbaum et~al.(1997)Cederbaum, Zobeley and
  Tarantelli}]{Cederbaum1997}
\bibinfo{author}{Cederbaum, L.S.}, \bibinfo{author}{Zobeley, J.},
  \bibinfo{author}{Tarantelli, F.}, \bibinfo{year}{1997}.
\newblock \bibinfo{title}{Giant intermolecular decay and fragmentation of
  clusters}.
\newblock \bibinfo{journal}{Phys. Rev. Lett.} \bibinfo{volume}{79},
  \bibinfo{pages}{4778--4781}.
\newblock \DOIprefix\doi{10.1103/PhysRevLett.79.4778}.
%Type = Article
\bibitem[{Chen et~al.(2010)Chen, Arpin, Popmintchev, Gerrity, Zhang, Seaberg,
  Popmintchev, Murnane and Kapteyn}]{Chen2010}
\bibinfo{author}{Chen, M.C.}, \bibinfo{author}{Arpin, P.},
  \bibinfo{author}{Popmintchev, T.}, \bibinfo{author}{Gerrity, M.},
  \bibinfo{author}{Zhang, B.}, \bibinfo{author}{Seaberg, M.},
  \bibinfo{author}{Popmintchev, D.}, \bibinfo{author}{Murnane, M.M.},
  \bibinfo{author}{Kapteyn, H.C.}, \bibinfo{year}{2010}.
\newblock \bibinfo{title}{Bright, coherent, ultrafast soft {X}-ray harmonics
  spanning the water window from a tabletop light source}.
\newblock \bibinfo{journal}{Phys. Rev. Lett.} \bibinfo{volume}{105},
  \bibinfo{pages}{173901}.
\newblock \DOIprefix\doi{10.1103/PhysRevLett.105.173901}.
%Type = Article
\bibitem[{Chen et~al.(2015)Chen, Pabst, Karamatskou and Santra}]{Chen2015}
\bibinfo{author}{Chen, Y.J.}, \bibinfo{author}{Pabst, S.},
  \bibinfo{author}{Karamatskou, A.}, \bibinfo{author}{Santra, R.},
  \bibinfo{year}{2015}.
\newblock \bibinfo{title}{Theoretical characterization of the collective
  resonance states underlying the xenon giant dipole resonance}.
\newblock \bibinfo{journal}{Phys. Rev. A} \bibinfo{volume}{91},
  \bibinfo{pages}{032503}.
\newblock \DOIprefix\doi{10.1103/PhysRevA.91.032503}.
%Type = Article
\bibitem[{Chen et~al.(2019)Chen, Zhang, Wang, Zatsarinny, Bartschat, Morishita
  and Lin}]{Chen2019}
\bibinfo{author}{Chen, Z.}, \bibinfo{author}{Zhang, L.}, \bibinfo{author}{Wang,
  Y.}, \bibinfo{author}{Zatsarinny, O.}, \bibinfo{author}{Bartschat, K.},
  \bibinfo{author}{Morishita, T.}, \bibinfo{author}{Lin, C.D.},
  \bibinfo{year}{2019}.
\newblock \bibinfo{title}{Pulse-duration dependence of the double-to-single
  ionization ratio of {Ne} by intense 780-nm and 800-nm laser fields:
  Comparison of simulations with experiments}.
\newblock \bibinfo{journal}{Phys. Rev. A} \bibinfo{volume}{99},
  \bibinfo{pages}{043408}.
\newblock \DOIprefix\doi{10.1103/PhysRevA.99.043408}.
%Type = Article
\bibitem[{Cho et~al.(2018)Cho, Rouxel, Kowalewski, Saurabh, Lee and
  Mukamel}]{Cho2018}
\bibinfo{author}{Cho, D.}, \bibinfo{author}{Rouxel, J.R.},
  \bibinfo{author}{Kowalewski, M.}, \bibinfo{author}{Saurabh, P.},
  \bibinfo{author}{Lee, J.Y.}, \bibinfo{author}{Mukamel, S.},
  \bibinfo{year}{2018}.
\newblock \bibinfo{title}{Phase cycling {RT}-{TDDFT} simulation protocol for
  nonlinear {XUV} and x-ray molecular spectroscopy}.
\newblock \bibinfo{journal}{J. Phys. Chem. Lett.} \bibinfo{volume}{9},
  \bibinfo{pages}{1072--1078}.
\newblock \DOIprefix\doi{10.1021/acs.jpclett.8b00061}.
%Type = Article
\bibitem[{Cirelli et~al.(2018)Cirelli, Marante, Heuser, Petersson, Gal{\'{a}}n,
  Argenti, Zhong, Busto, Isinger, Nandi, Maclot, Rading, Johnsson,
  Gisselbrecht, Lucchini, Gallmann, Dahlstr\"{o}m, Lindroth, L'Huillier,
  Mart{\'{\i}}n and Keller}]{Cirelli2018}
\bibinfo{author}{Cirelli, C.}, \bibinfo{author}{Marante, C.},
  \bibinfo{author}{Heuser, S.}, \bibinfo{author}{Petersson, C.L.M.},
  \bibinfo{author}{Gal{\'{a}}n, {\'{A}}.J.}, \bibinfo{author}{Argenti, L.},
  \bibinfo{author}{Zhong, S.}, \bibinfo{author}{Busto, D.},
  \bibinfo{author}{Isinger, M.}, \bibinfo{author}{Nandi, S.},
  \bibinfo{author}{Maclot, S.}, \bibinfo{author}{Rading, L.},
  \bibinfo{author}{Johnsson, P.}, \bibinfo{author}{Gisselbrecht, M.},
  \bibinfo{author}{Lucchini, M.}, \bibinfo{author}{Gallmann, L.},
  \bibinfo{author}{Dahlstr\"{o}m, J.M.}, \bibinfo{author}{Lindroth, E.},
  \bibinfo{author}{L'Huillier, A.}, \bibinfo{author}{Mart{\'{\i}}n, F.},
  \bibinfo{author}{Keller, U.}, \bibinfo{year}{2018}.
\newblock \bibinfo{title}{Anisotropic photoemission time delays close to a
  {F}ano resonance}.
\newblock \bibinfo{journal}{Nat. Commun.} \bibinfo{volume}{9},
  \bibinfo{pages}{955}.
\newblock \DOIprefix\doi{10.1038/s41467-018-03009-1}.
%Type = Article
\bibitem[{Codling and Madden(1971)}]{Codling1971}
\bibinfo{author}{Codling, K.}, \bibinfo{author}{Madden, R.P.},
  \bibinfo{year}{1971}.
\newblock \bibinfo{title}{Resonances in the photoionization continuum of {Kr}
  and {Xe}}.
\newblock \bibinfo{journal}{Phys. Rev. A} \bibinfo{volume}{4},
  \bibinfo{pages}{2261--2263}.
\newblock \DOIprefix\doi{10.1103/PhysRevA.4.2261}.
%Type = Article
\bibitem[{Coffee et~al.(2019)Coffee, Cryan, Duris, Helml, Li and
  Marinelli}]{Coffee2019}
\bibinfo{author}{Coffee, R.N.}, \bibinfo{author}{Cryan, J.P.},
  \bibinfo{author}{Duris, J.}, \bibinfo{author}{Helml, W.},
  \bibinfo{author}{Li, S.}, \bibinfo{author}{Marinelli, A.},
  \bibinfo{year}{2019}.
\newblock \bibinfo{title}{Development of ultrafast capabilities for {X}-ray
  free-electron lasers at the linac coherent light source}.
\newblock \bibinfo{journal}{Philos. Trans. R. Soc. A} \bibinfo{volume}{377},
  \bibinfo{pages}{20180386}.
\newblock \DOIprefix\doi{10.1098/rsta.2018.0386}.
%Type = Book
\bibitem[{Cohen-Tannoudji et~al.(2004)Cohen-Tannoudji, Dupont-Roc and
  Grynberg}]{CohenTannoudji2004}
\bibinfo{author}{Cohen-Tannoudji, C.}, \bibinfo{author}{Dupont-Roc, J.},
  \bibinfo{author}{Grynberg, G.}, \bibinfo{year}{2004}.
\newblock \bibinfo{title}{Atom-Photon Interactions}.
\newblock \bibinfo{publisher}{John Wiley \& Sons, Ltd}.
\newblock \DOIprefix\doi{10.1002/9783527617197}.
%Type = Article
\bibitem[{Colgan and Pindzola(2012)}]{Colgan2012}
\bibinfo{author}{Colgan, J.}, \bibinfo{author}{Pindzola, M.S.},
  \bibinfo{year}{2012}.
\newblock \bibinfo{title}{Angular distributions for the complete
  photofragmentation of the {Li} atom}.
\newblock \bibinfo{journal}{Phys. Rev. Lett.} \bibinfo{volume}{108},
  \bibinfo{pages}{053001}.
\newblock \DOIprefix\doi{10.1103/PhysRevLett.108.053001}.
%Type = Article
\bibitem[{Colson(1981)}]{Colson1981}
\bibinfo{author}{Colson, W.}, \bibinfo{year}{1981}.
\newblock \bibinfo{title}{The nonlinear wave equation for higher harmonics in
  free-electron lasers}.
\newblock \bibinfo{journal}{{IEEE} J. Quantum Electron.} \bibinfo{volume}{17},
  \bibinfo{pages}{1417--1427}.
\newblock \DOIprefix\doi{10.1109/jqe.1981.1071273}.
%Type = Article
\bibitem[{Colson(1977)}]{Colson1977}
\bibinfo{author}{Colson, W.B.}, \bibinfo{year}{1977}.
\newblock \bibinfo{title}{One-body electron dynamics in a free electron laser}.
\newblock \bibinfo{journal}{Phys. Lett. A} \bibinfo{volume}{64},
  \bibinfo{pages}{190--192}.
\newblock \DOIprefix\doi{10.1016/0375-9601(77)90712-5}.
%Type = Article
\bibitem[{Colson et~al.(1985)Colson, Dattoli and Ciocci}]{Colson1985}
\bibinfo{author}{Colson, W.B.}, \bibinfo{author}{Dattoli, G.},
  \bibinfo{author}{Ciocci, F.}, \bibinfo{year}{1985}.
\newblock \bibinfo{title}{Angular-gain spectrum of free-electron lasers}.
\newblock \bibinfo{journal}{Phys. Rev. A} \bibinfo{volume}{31},
  \bibinfo{pages}{828--842}.
\newblock \DOIprefix\doi{10.1103/PhysRevA.31.828}.
%Type = Article
\bibitem[{Colson and Sessler(1985)}]{Colson1985a}
\bibinfo{author}{Colson, W.B.}, \bibinfo{author}{Sessler, A.M.},
  \bibinfo{year}{1985}.
\newblock \bibinfo{title}{Free electron lasers}.
\newblock \bibinfo{journal}{Annu. Rev. Nucl. Part. Sci.} \bibinfo{volume}{35},
  \bibinfo{pages}{25--54}.
\newblock \DOIprefix\doi{10.1146/annurev.ns.35.120185.000325}.
%Type = Article
\bibitem[{Cooper and Averbukh(2013)}]{Cooper2013}
\bibinfo{author}{Cooper, B.}, \bibinfo{author}{Averbukh, V.},
  \bibinfo{year}{2013}.
\newblock \bibinfo{title}{Single-photon laser-enabled {A}uger spectroscopy for
  measuring attosecond electron-hole dynamics}.
\newblock \bibinfo{journal}{Phys. Rev. Lett.} \bibinfo{volume}{111},
  \bibinfo{pages}{083004}.
\newblock \DOIprefix\doi{10.1103/PhysRevLett.111.083004}.
%Type = Article
\bibitem[{Cooper et~al.(2014)Cooper, Koloren{\v{c}}, Frasinski, Averbukh and
  Marangos}]{Cooper2014}
\bibinfo{author}{Cooper, B.}, \bibinfo{author}{Koloren{\v{c}}, P.},
  \bibinfo{author}{Frasinski, L.J.}, \bibinfo{author}{Averbukh, V.},
  \bibinfo{author}{Marangos, J.P.}, \bibinfo{year}{2014}.
\newblock \bibinfo{title}{Analysis of a measurement scheme for ultrafast hole
  dynamics by few femtosecond resolution {X}-ray pump-probe {A}uger
  spectroscopy}.
\newblock \bibinfo{journal}{Faraday Discuss.} \bibinfo{volume}{171},
  \bibinfo{pages}{93--111}.
\newblock \DOIprefix\doi{10.1039/c4fd00051j}.
%Type = Article
\bibitem[{Cousin et~al.(2014)Cousin, Silva, Teichmann, Hemmer, Buades and
  Biegert}]{Cousin2014}
\bibinfo{author}{Cousin, S.L.}, \bibinfo{author}{Silva, F.},
  \bibinfo{author}{Teichmann, S.}, \bibinfo{author}{Hemmer, M.},
  \bibinfo{author}{Buades, B.}, \bibinfo{author}{Biegert, J.},
  \bibinfo{year}{2014}.
\newblock \bibinfo{title}{High-flux table-top soft x-ray source driven by
  sub-2-cycle, {CEP} stable, 1.85-$\mu$m 1-{kHz} pulses for carbon {K}-edge
  spectroscopy}.
\newblock \bibinfo{journal}{Opt. Lett.} \bibinfo{volume}{39},
  \bibinfo{pages}{5383}.
\newblock \DOIprefix\doi{10.1364/ol.39.005383}.
%Type = Article
\bibitem[{Dahlstr\"om et~al.(2012)Dahlstr\"om, L'Huillier and
  Maquet}]{Dahlstrom2012}
\bibinfo{author}{Dahlstr\"om, J.M.}, \bibinfo{author}{L'Huillier, A.},
  \bibinfo{author}{Maquet, A.}, \bibinfo{year}{2012}.
\newblock \bibinfo{title}{Introduction to attosecond delays in
  photoionization}.
\newblock \bibinfo{journal}{J. Phys. B} \bibinfo{volume}{45},
  \bibinfo{pages}{183001}.
\newblock \DOIprefix\doi{10.1088/0953-4075/45/18/183001}.
%Type = Inproceedings
\bibitem[{Danailov et~al.(2011)Danailov, Cinquegrana, Demidovich, Ivanov,
  Nikolov and Sigalotti}]{Danailov2011}
\bibinfo{author}{Danailov, M.B.}, \bibinfo{author}{Cinquegrana, P.},
  \bibinfo{author}{Demidovich, A.A.}, \bibinfo{author}{Ivanov, R.},
  \bibinfo{author}{Nikolov, I.}, \bibinfo{author}{Sigalotti, P.},
  \bibinfo{year}{2011}.
\newblock \bibinfo{title}{Design and first experience with the {FERMI} seed
  laser}, in: \bibinfo{booktitle}{Proceedings of the 33rd International Free
  Electron Laser Conference (FEL 2011), Shanghai, China},
  \bibinfo{publisher}{FEL'11/EPS-AG}. p. \bibinfo{pages}{183}.
\newblock \URLprefix
  \url{https://accelconf.web.cern.ch/FEL2011/papers/tuoc4.pdf}.
%Type = Article
\bibitem[{De~Ninno et~al.(2015)De~Ninno, Gauthier, Mahieu, Rebernik~Ribi\v{c},
  Allaria, Cinquegrana, Danailov, Demidovich, Ferrari, Giannessi, Penco,
  Sigalotti and Stupar}]{DeNinno2015}
\bibinfo{author}{De~Ninno, G.}, \bibinfo{author}{Gauthier, D.},
  \bibinfo{author}{Mahieu, B.}, \bibinfo{author}{Rebernik~Ribi\v{c}, P.},
  \bibinfo{author}{Allaria, E.}, \bibinfo{author}{Cinquegrana, P.},
  \bibinfo{author}{Danailov, M.B.}, \bibinfo{author}{Demidovich, A.},
  \bibinfo{author}{Ferrari, E.}, \bibinfo{author}{Giannessi, L.},
  \bibinfo{author}{Penco, G.}, \bibinfo{author}{Sigalotti, P.},
  \bibinfo{author}{Stupar, M.}, \bibinfo{year}{2015}.
\newblock \bibinfo{title}{Single-shot spectro-temporal characterization of
  {XUV} pulses from a seeded free-electron laser}.
\newblock \bibinfo{journal}{Nat. Commun.} \bibinfo{volume}{6},
  \bibinfo{pages}{8075}.
\newblock \DOIprefix\doi{10.1038/ncomms9075}.
%Type = Article
\bibitem[{De~Ninno et~al.(2020)De~Ninno, W\"{a}tzel, Rebernik~Ribi\v{c},
  Allaria, Coreno, Danailov, David, Demidovich, Di~Fraia, Giannessi, Hansen,
  Kru\v{s}i\v{c}, Manfredda, Meyer, Miheli\v{c}, Mirian, Plekan, Ressel,
  R\"{o}sner, Simoncig, Spampinati, Stupar, \v{Z}itnik, Zangrando, Callegari
  and Berakdar}]{DeNinno2020}
\bibinfo{author}{De~Ninno, G.}, \bibinfo{author}{W\"{a}tzel, J.},
  \bibinfo{author}{Rebernik~Ribi\v{c}, P.}, \bibinfo{author}{Allaria, E.},
  \bibinfo{author}{Coreno, M.}, \bibinfo{author}{Danailov, M.B.},
  \bibinfo{author}{David, C.}, \bibinfo{author}{Demidovich, A.},
  \bibinfo{author}{Di~Fraia, M.}, \bibinfo{author}{Giannessi, L.},
  \bibinfo{author}{Hansen, K.}, \bibinfo{author}{Kru\v{s}i\v{c}, {\v S}.},
  \bibinfo{author}{Manfredda, M.}, \bibinfo{author}{Meyer, M.},
  \bibinfo{author}{Miheli\v{c}, A.}, \bibinfo{author}{Mirian, N.},
  \bibinfo{author}{Plekan, O.}, \bibinfo{author}{Ressel, B.},
  \bibinfo{author}{R\"{o}sner, B.}, \bibinfo{author}{Simoncig, A.},
  \bibinfo{author}{Spampinati, S.}, \bibinfo{author}{Stupar, M.},
  \bibinfo{author}{\v{Z}itnik, M.}, \bibinfo{author}{Zangrando, M.},
  \bibinfo{author}{Callegari, C.}, \bibinfo{author}{Berakdar, J.},
  \bibinfo{year}{2020}.
\newblock \bibinfo{title}{Photoelectric effect with a twist}.
\newblock \bibinfo{journal}{Nat. Photon.} \bibinfo{note}{In press}.
%Type = Article
\bibitem[{Decking et~al.(2020)Decking, Abeghyan, Abramian, Abramsky, Aguirre,
  Albrecht, Alou, Altarelli, Altmann, Amyan, Anashin, Apostolov, Appel,
  Auguste, Ayvazyan, Baark, Babies, Baboi, Bak, Balandin, Baldinger, Baranasic,
  Barbanotti, Belikov, Belokurov, Belova, Belyakov, Berry, Bertucci, Beutner,
  Block, Bl\"{o}cher, B\"{o}ckmann, Bohm, B\"{o}hnert, Bondar, Bondarchuk,
  Bonezzi, Borowiec, B\"{o}sch, B\"{o}senberg, Bosotti, B\"{o}spflug,
  Bousonville, Boyd, Bozhko, Brand, Branlard, Briechle, Brinker, Brinker,
  Brinkmann, Brockhauser, Brovko, Br\"{u}ck, Br\"{u}dgam, Butkowski,
  B\"{u}ttner, Calero, Castro-Carballo, Cattalanotto, Charrier, Chen,
  Cherepenko, Cheskidov, Chiodini, Chong, Choroba, Chorowski, Churanov,
  Cichalewski, Clausen, Clement, Clou{\'{e}}, Cobos, Coppola, Cunis, Czuba,
  Czwalinna, D'Almagne, Dammann, Danared, de~Zubiaurre~Wagner, Delfs, Delfs,
  Dietrich, Dietrich, Dohlus, Dommach, Donat, Dong, Doynikov, Dressel, Duda,
  Duda, Eckoldt, Ehsan, Eidam, Eints, Engling, Englisch, Ermakov, Escherich,
  Eschke, Saldin, Faesing, Fallou, Felber, Fenner, Fernandes, Fern{\'{a}}ndez,
  Feuker, Filippakopoulos, Floettmann, Fogel, Fontaine, Franc{\'{e}}s, Martin,
  Freund, Freyermuth, Friedland, Fr\"{o}hlich, Fusetti, Fydrych, Gallas,
  Garc{\'{\i}}a, Garcia-Tabares, Geloni, Gerasimova, Gerth, Ge{\ss}ler,
  Gharibyan, Gloor, G{\l}owinkowski, Goessel, Go{\l}{\k{e}}biewski, Golubeva,
  Grabowski, Graeff, Grebentsov, Grecki, Grevsmuehl, Gross, Grosse-Wortmann,
  Gr\"{u}nert, Grunewald, Grzegory, Feng, Guler, Gusev, Gutierrez, Hagge,
  Hamberg, Hanneken, Harms, Hartl, Hauberg, Hauf, Hauschildt, Hauser, Havlicek,
  Hedqvist, Heidbrook, Hellberg, Henning, Hensler, Hermann, Hidv{\'{e}}gi,
  Hierholzer, Hintz, Hoffmann, Hoffmann, Hoffmann, Holler, H\"{u}ning,
  Ignatenko, Ilchen, Iluk, Iversen, Iversen, Izquierdo, Jachmann, Jardon,
  Jastrow, Jensch, Jensen, Je{\.{z}}abek, Jidda, Jin, Johannson, Jonas, Kaabi,
  Kaefer, Kammering, Kapitza, Karabekyan, Karstensen, Kasprzak, Katalev, Keese,
  Keil, Kholopov, Killenberger, Kitaev, Klimchenko, Klos, Knebel, Koch, Koepke,
  K\"{o}hler, K\"{o}hler, Kohlstrunk, Konopkova, Konstantinov, Kook, Koprek,
  K\"{o}rfer, Korth, Kosarev, Kosi{\'{n}}ski, Kostin, Kot, Kotarba, Kozak,
  Kozak, Kramert, Krasilnikov, Krasnov, Krause, Kravchuk, Krebs, Kretschmer,
  Kreutzkamp, Kr\"{o}plin, Krzysik, Kube, Kuehn, Kujala, Kulikov, Kuzminych,
  Civita, Lacroix, Lamb, Lancetov, Larsson, Pinvidic, Lederer, Lensch, Lenz,
  Leuschner, Levenhagen, Li, Liebing, Lilje, Limberg, Lipka, List, Liu, Liu,
  Lorbeer, Lorkiewicz, Lu, Ludwig, Machau, Maciocha, Madec, Magueur, Maiano,
  Maksimova, Malcher, Maltezopoulos, Mamoshkina, Manschwetus, Marcellini,
  Marinkovic, Martinez, Martirosyan, Maschmann, Maslov, Matheisen, Mavric,
  Mei{\ss}ner, Meissner, Messerschmidt, Meyners, Michalski, Michelato, Mildner,
  Moe, Moglia, Mohr, Mohr, M\"{o}ller, Mommerz, Monaco, Montiel, Moretti,
  Morozov, Morozov, Mross, Mueller, M\"{u}ller, M\"{u}ller, M\"{u}ller,
  Munilla, M\"{u}nnich, Muratov, Napoly, N\"{a}ser, Nefedov, Neumann, Neumann,
  Ngada, Noelle, Obier, Okunev, Oliver, Omet, Oppelt, Ottmar, Oublaid, Pagani,
  Paparella, Paramonov, Peitzmann, Penning, Perus, Peters, Petersen, Petrov,
  Petrov, Pfeiffer, Pfl\"{u}ger, Philipp, Pienaud, Pierini, Pivovarov, Planas,
  P{\l}awski, Pohl, Polinski, Popov, Prat, Prenting, Priebe, Pryschelski,
  Przygoda, Pyata, Racky, Rathjen, Ratuschni, Regnaud-Campderros, Rehlich,
  Reschke, Robson, Roever, Roggli, Rothenburg, Rusi{\'{n}}ski, Rybaniec,
  Sahling, Salmani, Samoylova, Sanzone, Saretzki, Sawlanski, Schaffran,
  Schlarb, Schl\"{o}sser, Schlott, Schmidt, Schmidt-Foehre, Schmitz,
  Schm\"{o}kel, Schnautz, Schneidmiller, Scholz, Sch\"{o}neburg, Schultze,
  Schulz, Schwarz, Sekutowicz, Sellmann, Semenov, Serkez, Sertore, Shehzad,
  Shemarykin, Shi, Sienkiewicz, Sikora, Sikorski, Silenzi, Simon, Singer,
  Singer, Sinn, Sinram, Skvorodnev, Smirnow, Sommer, Sorokin, Stadler, Steckel,
  Steffen, Steinhau-K\"{u}hl, Stephan, Stodulski, Stolper, Sulimov, Susen,
  {\'{S}}wierblewski, Sydlo, Syresin, Sytchev, Szuba, Tesch, Thie, Thiebault,
  Tiedtke, Tischhauser, Tolkiehn, Tomin, Tonisch, Toral, Torbin, Trapp, Treyer,
  Trowitzsch, Trublet, Tschentscher, Ullrich, Vannoni, Varela, Varghese,
  Vashchenko, Vasic, Vazquez-Velez, Verguet, Vilcins-Czvitkovits, Villanueva,
  Visentin, Viti, Vogel, Volobuev, Wagner, Walker, Wamsat, Weddig, Weichert,
  Weise, Wenndorf, Werner, Wichmann, Wiebers, Wiencek, Wilksen, Will,
  Winkelmann, Winkowski, Wittenburg, Witzig, Wlk, Wohlenberg, Wojciechowski,
  Wolff-Fabris, Wrochna, Wrona, Yakopov, Yang, Yang, Yurkov, Zagorodnov,
  Zalden, Zavadtsev, Zavadtsev, Zhirnov, Zhukov, Ziemann, Zolotov, Zolotukhina,
  Zummack and Zybin}]{Decking2020}
\bibinfo{author}{Decking, W.}, \bibinfo{author}{Abeghyan, S.},
  \bibinfo{author}{Abramian, P.}, \bibinfo{author}{Abramsky, A.},
  \bibinfo{author}{Aguirre, A.}, \bibinfo{author}{Albrecht, C.},
  \bibinfo{author}{Alou, P.}, \bibinfo{author}{Altarelli, M.},
  \bibinfo{author}{Altmann, P.}, \bibinfo{author}{Amyan, K.},
  \bibinfo{author}{Anashin, V.}, \bibinfo{author}{Apostolov, E.},
  \bibinfo{author}{Appel, K.}, \bibinfo{author}{Auguste, D.},
  \bibinfo{author}{Ayvazyan, V.}, \bibinfo{author}{Baark, S.},
  \bibinfo{author}{Babies, F.}, \bibinfo{author}{Baboi, N.},
  \bibinfo{author}{Bak, P.}, \bibinfo{author}{Balandin, V.},
  \bibinfo{author}{Baldinger, R.}, \bibinfo{author}{Baranasic, B.},
  \bibinfo{author}{Barbanotti, S.}, \bibinfo{author}{Belikov, O.},
  \bibinfo{author}{Belokurov, V.}, \bibinfo{author}{Belova, L.},
  \bibinfo{author}{Belyakov, V.}, \bibinfo{author}{Berry, S.},
  \bibinfo{author}{Bertucci, M.}, \bibinfo{author}{Beutner, B.},
  \bibinfo{author}{Block, A.}, \bibinfo{author}{Bl\"{o}cher, M.},
  \bibinfo{author}{B\"{o}ckmann, T.}, \bibinfo{author}{Bohm, C.},
  \bibinfo{author}{B\"{o}hnert, M.}, \bibinfo{author}{Bondar, V.},
  \bibinfo{author}{Bondarchuk, E.}, \bibinfo{author}{Bonezzi, M.},
  \bibinfo{author}{Borowiec, P.}, \bibinfo{author}{B\"{o}sch, C.},
  \bibinfo{author}{B\"{o}senberg, U.}, \bibinfo{author}{Bosotti, A.},
  \bibinfo{author}{B\"{o}spflug, R.}, \bibinfo{author}{Bousonville, M.},
  \bibinfo{author}{Boyd, E.}, \bibinfo{author}{Bozhko, Y.},
  \bibinfo{author}{Brand, A.}, \bibinfo{author}{Branlard, J.},
  \bibinfo{author}{Briechle, S.}, \bibinfo{author}{Brinker, F.},
  \bibinfo{author}{Brinker, S.}, \bibinfo{author}{Brinkmann, R.},
  \bibinfo{author}{Brockhauser, S.}, \bibinfo{author}{Brovko, O.},
  \bibinfo{author}{Br\"{u}ck, H.}, \bibinfo{author}{Br\"{u}dgam, A.},
  \bibinfo{author}{Butkowski, L.}, \bibinfo{author}{B\"{u}ttner, T.},
  \bibinfo{author}{Calero, J.}, \bibinfo{author}{Castro-Carballo, E.},
  \bibinfo{author}{Cattalanotto, G.}, \bibinfo{author}{Charrier, J.},
  \bibinfo{author}{Chen, J.}, \bibinfo{author}{Cherepenko, A.},
  \bibinfo{author}{Cheskidov, V.}, \bibinfo{author}{Chiodini, M.},
  \bibinfo{author}{Chong, A.}, \bibinfo{author}{Choroba, S.},
  \bibinfo{author}{Chorowski, M.}, \bibinfo{author}{Churanov, D.},
  \bibinfo{author}{Cichalewski, W.}, \bibinfo{author}{Clausen, M.},
  \bibinfo{author}{Clement, W.}, \bibinfo{author}{Clou{\'{e}}, C.},
  \bibinfo{author}{Cobos, J.A.}, \bibinfo{author}{Coppola, N.},
  \bibinfo{author}{Cunis, S.}, \bibinfo{author}{Czuba, K.},
  \bibinfo{author}{Czwalinna, M.}, \bibinfo{author}{D'Almagne, B.},
  \bibinfo{author}{Dammann, J.}, \bibinfo{author}{Danared, H.},
  \bibinfo{author}{de~Zubiaurre~Wagner, A.}, \bibinfo{author}{Delfs, A.},
  \bibinfo{author}{Delfs, T.}, \bibinfo{author}{Dietrich, F.},
  \bibinfo{author}{Dietrich, T.}, \bibinfo{author}{Dohlus, M.},
  \bibinfo{author}{Dommach, M.}, \bibinfo{author}{Donat, A.},
  \bibinfo{author}{Dong, X.}, \bibinfo{author}{Doynikov, N.},
  \bibinfo{author}{Dressel, M.}, \bibinfo{author}{Duda, M.},
  \bibinfo{author}{Duda, P.}, \bibinfo{author}{Eckoldt, H.},
  \bibinfo{author}{Ehsan, W.}, \bibinfo{author}{Eidam, J.},
  \bibinfo{author}{Eints, F.}, \bibinfo{author}{Engling, C.},
  \bibinfo{author}{Englisch, U.}, \bibinfo{author}{Ermakov, A.},
  \bibinfo{author}{Escherich, K.}, \bibinfo{author}{Eschke, J.},
  \bibinfo{author}{Saldin, E.}, \bibinfo{author}{Faesing, M.},
  \bibinfo{author}{Fallou, A.}, \bibinfo{author}{Felber, M.},
  \bibinfo{author}{Fenner, M.}, \bibinfo{author}{Fernandes, B.},
  \bibinfo{author}{Fern{\'{a}}ndez, J.M.}, \bibinfo{author}{Feuker, S.},
  \bibinfo{author}{Filippakopoulos, K.}, \bibinfo{author}{Floettmann, K.},
  \bibinfo{author}{Fogel, V.}, \bibinfo{author}{Fontaine, M.},
  \bibinfo{author}{Franc{\'{e}}s, A.}, \bibinfo{author}{Martin, I.F.},
  \bibinfo{author}{Freund, W.}, \bibinfo{author}{Freyermuth, T.},
  \bibinfo{author}{Friedland, M.}, \bibinfo{author}{Fr\"{o}hlich, L.},
  \bibinfo{author}{Fusetti, M.}, \bibinfo{author}{Fydrych, J.},
  \bibinfo{author}{Gallas, A.}, \bibinfo{author}{Garc{\'{\i}}a, O.},
  \bibinfo{author}{Garcia-Tabares, L.}, \bibinfo{author}{Geloni, G.},
  \bibinfo{author}{Gerasimova, N.}, \bibinfo{author}{Gerth, C.},
  \bibinfo{author}{Ge{\ss}ler, P.}, \bibinfo{author}{Gharibyan, V.},
  \bibinfo{author}{Gloor, M.}, \bibinfo{author}{G{\l}owinkowski, J.},
  \bibinfo{author}{Goessel, A.}, \bibinfo{author}{Go{\l}{\k{e}}biewski, Z.},
  \bibinfo{author}{Golubeva, N.}, \bibinfo{author}{Grabowski, W.},
  \bibinfo{author}{Graeff, W.}, \bibinfo{author}{Grebentsov, A.},
  \bibinfo{author}{Grecki, M.}, \bibinfo{author}{Grevsmuehl, T.},
  \bibinfo{author}{Gross, M.}, \bibinfo{author}{Grosse-Wortmann, U.},
  \bibinfo{author}{Gr\"{u}nert, J.}, \bibinfo{author}{Grunewald, S.},
  \bibinfo{author}{Grzegory, P.}, \bibinfo{author}{Feng, G.},
  \bibinfo{author}{Guler, H.}, \bibinfo{author}{Gusev, G.},
  \bibinfo{author}{Gutierrez, J.L.}, \bibinfo{author}{Hagge, L.},
  \bibinfo{author}{Hamberg, M.}, \bibinfo{author}{Hanneken, R.},
  \bibinfo{author}{Harms, E.}, \bibinfo{author}{Hartl, I.},
  \bibinfo{author}{Hauberg, A.}, \bibinfo{author}{Hauf, S.},
  \bibinfo{author}{Hauschildt, J.}, \bibinfo{author}{Hauser, J.},
  \bibinfo{author}{Havlicek, J.}, \bibinfo{author}{Hedqvist, A.},
  \bibinfo{author}{Heidbrook, N.}, \bibinfo{author}{Hellberg, F.},
  \bibinfo{author}{Henning, D.}, \bibinfo{author}{Hensler, O.},
  \bibinfo{author}{Hermann, T.}, \bibinfo{author}{Hidv{\'{e}}gi, A.},
  \bibinfo{author}{Hierholzer, M.}, \bibinfo{author}{Hintz, H.},
  \bibinfo{author}{Hoffmann, F.}, \bibinfo{author}{Hoffmann, M.},
  \bibinfo{author}{Hoffmann, M.}, \bibinfo{author}{Holler, Y.},
  \bibinfo{author}{H\"{u}ning, M.}, \bibinfo{author}{Ignatenko, A.},
  \bibinfo{author}{Ilchen, M.}, \bibinfo{author}{Iluk, A.},
  \bibinfo{author}{Iversen, J.}, \bibinfo{author}{Iversen, J.},
  \bibinfo{author}{Izquierdo, M.}, \bibinfo{author}{Jachmann, L.},
  \bibinfo{author}{Jardon, N.}, \bibinfo{author}{Jastrow, U.},
  \bibinfo{author}{Jensch, K.}, \bibinfo{author}{Jensen, J.},
  \bibinfo{author}{Je{\.{z}}abek, M.}, \bibinfo{author}{Jidda, M.},
  \bibinfo{author}{Jin, H.}, \bibinfo{author}{Johannson, N.},
  \bibinfo{author}{Jonas, R.}, \bibinfo{author}{Kaabi, W.},
  \bibinfo{author}{Kaefer, D.}, \bibinfo{author}{Kammering, R.},
  \bibinfo{author}{Kapitza, H.}, \bibinfo{author}{Karabekyan, S.},
  \bibinfo{author}{Karstensen, S.}, \bibinfo{author}{Kasprzak, K.},
  \bibinfo{author}{Katalev, V.}, \bibinfo{author}{Keese, D.},
  \bibinfo{author}{Keil, B.}, \bibinfo{author}{Kholopov, M.},
  \bibinfo{author}{Killenberger, M.}, \bibinfo{author}{Kitaev, B.},
  \bibinfo{author}{Klimchenko, Y.}, \bibinfo{author}{Klos, R.},
  \bibinfo{author}{Knebel, L.}, \bibinfo{author}{Koch, A.},
  \bibinfo{author}{Koepke, M.}, \bibinfo{author}{K\"{o}hler, S.},
  \bibinfo{author}{K\"{o}hler, W.}, \bibinfo{author}{Kohlstrunk, N.},
  \bibinfo{author}{Konopkova, Z.}, \bibinfo{author}{Konstantinov, A.},
  \bibinfo{author}{Kook, W.}, \bibinfo{author}{Koprek, W.},
  \bibinfo{author}{K\"{o}rfer, M.}, \bibinfo{author}{Korth, O.},
  \bibinfo{author}{Kosarev, A.}, \bibinfo{author}{Kosi{\'{n}}ski, K.},
  \bibinfo{author}{Kostin, D.}, \bibinfo{author}{Kot, Y.},
  \bibinfo{author}{Kotarba, A.}, \bibinfo{author}{Kozak, T.},
  \bibinfo{author}{Kozak, V.}, \bibinfo{author}{Kramert, R.},
  \bibinfo{author}{Krasilnikov, M.}, \bibinfo{author}{Krasnov, A.},
  \bibinfo{author}{Krause, B.}, \bibinfo{author}{Kravchuk, L.},
  \bibinfo{author}{Krebs, O.}, \bibinfo{author}{Kretschmer, R.},
  \bibinfo{author}{Kreutzkamp, J.}, \bibinfo{author}{Kr\"{o}plin, O.},
  \bibinfo{author}{Krzysik, K.}, \bibinfo{author}{Kube, G.},
  \bibinfo{author}{Kuehn, H.}, \bibinfo{author}{Kujala, N.},
  \bibinfo{author}{Kulikov, V.}, \bibinfo{author}{Kuzminych, V.},
  \bibinfo{author}{Civita, D.L.}, \bibinfo{author}{Lacroix, M.},
  \bibinfo{author}{Lamb, T.}, \bibinfo{author}{Lancetov, A.},
  \bibinfo{author}{Larsson, M.}, \bibinfo{author}{Pinvidic, D.L.},
  \bibinfo{author}{Lederer, S.}, \bibinfo{author}{Lensch, T.},
  \bibinfo{author}{Lenz, D.}, \bibinfo{author}{Leuschner, A.},
  \bibinfo{author}{Levenhagen, F.}, \bibinfo{author}{Li, Y.},
  \bibinfo{author}{Liebing, J.}, \bibinfo{author}{Lilje, L.},
  \bibinfo{author}{Limberg, T.}, \bibinfo{author}{Lipka, D.},
  \bibinfo{author}{List, B.}, \bibinfo{author}{Liu, J.}, \bibinfo{author}{Liu,
  S.}, \bibinfo{author}{Lorbeer, B.}, \bibinfo{author}{Lorkiewicz, J.},
  \bibinfo{author}{Lu, H.H.}, \bibinfo{author}{Ludwig, F.},
  \bibinfo{author}{Machau, K.}, \bibinfo{author}{Maciocha, W.},
  \bibinfo{author}{Madec, C.}, \bibinfo{author}{Magueur, C.},
  \bibinfo{author}{Maiano, C.}, \bibinfo{author}{Maksimova, I.},
  \bibinfo{author}{Malcher, K.}, \bibinfo{author}{Maltezopoulos, T.},
  \bibinfo{author}{Mamoshkina, E.}, \bibinfo{author}{Manschwetus, B.},
  \bibinfo{author}{Marcellini, F.}, \bibinfo{author}{Marinkovic, G.},
  \bibinfo{author}{Martinez, T.}, \bibinfo{author}{Martirosyan, H.},
  \bibinfo{author}{Maschmann, W.}, \bibinfo{author}{Maslov, M.},
  \bibinfo{author}{Matheisen, A.}, \bibinfo{author}{Mavric, U.},
  \bibinfo{author}{Mei{\ss}ner, J.}, \bibinfo{author}{Meissner, K.},
  \bibinfo{author}{Messerschmidt, M.}, \bibinfo{author}{Meyners, N.},
  \bibinfo{author}{Michalski, G.}, \bibinfo{author}{Michelato, P.},
  \bibinfo{author}{Mildner, N.}, \bibinfo{author}{Moe, M.},
  \bibinfo{author}{Moglia, F.}, \bibinfo{author}{Mohr, C.},
  \bibinfo{author}{Mohr, S.}, \bibinfo{author}{M\"{o}ller, W.},
  \bibinfo{author}{Mommerz, M.}, \bibinfo{author}{Monaco, L.},
  \bibinfo{author}{Montiel, C.}, \bibinfo{author}{Moretti, M.},
  \bibinfo{author}{Morozov, I.}, \bibinfo{author}{Morozov, P.},
  \bibinfo{author}{Mross, D.}, \bibinfo{author}{Mueller, J.},
  \bibinfo{author}{M\"{u}ller, C.}, \bibinfo{author}{M\"{u}ller, J.},
  \bibinfo{author}{M\"{u}ller, K.}, \bibinfo{author}{Munilla, J.},
  \bibinfo{author}{M\"{u}nnich, A.}, \bibinfo{author}{Muratov, V.},
  \bibinfo{author}{Napoly, O.}, \bibinfo{author}{N\"{a}ser, B.},
  \bibinfo{author}{Nefedov, N.}, \bibinfo{author}{Neumann, R.},
  \bibinfo{author}{Neumann, R.}, \bibinfo{author}{Ngada, N.},
  \bibinfo{author}{Noelle, D.}, \bibinfo{author}{Obier, F.},
  \bibinfo{author}{Okunev, I.}, \bibinfo{author}{Oliver, J.A.},
  \bibinfo{author}{Omet, M.}, \bibinfo{author}{Oppelt, A.},
  \bibinfo{author}{Ottmar, A.}, \bibinfo{author}{Oublaid, M.},
  \bibinfo{author}{Pagani, C.}, \bibinfo{author}{Paparella, R.},
  \bibinfo{author}{Paramonov, V.}, \bibinfo{author}{Peitzmann, C.},
  \bibinfo{author}{Penning, J.}, \bibinfo{author}{Perus, A.},
  \bibinfo{author}{Peters, F.}, \bibinfo{author}{Petersen, B.},
  \bibinfo{author}{Petrov, A.}, \bibinfo{author}{Petrov, I.},
  \bibinfo{author}{Pfeiffer, S.}, \bibinfo{author}{Pfl\"{u}ger, J.},
  \bibinfo{author}{Philipp, S.}, \bibinfo{author}{Pienaud, Y.},
  \bibinfo{author}{Pierini, P.}, \bibinfo{author}{Pivovarov, S.},
  \bibinfo{author}{Planas, M.}, \bibinfo{author}{P{\l}awski, E.},
  \bibinfo{author}{Pohl, M.}, \bibinfo{author}{Polinski, J.},
  \bibinfo{author}{Popov, V.}, \bibinfo{author}{Prat, S.},
  \bibinfo{author}{Prenting, J.}, \bibinfo{author}{Priebe, G.},
  \bibinfo{author}{Pryschelski, H.}, \bibinfo{author}{Przygoda, K.},
  \bibinfo{author}{Pyata, E.}, \bibinfo{author}{Racky, B.},
  \bibinfo{author}{Rathjen, A.}, \bibinfo{author}{Ratuschni, W.},
  \bibinfo{author}{Regnaud-Campderros, S.}, \bibinfo{author}{Rehlich, K.},
  \bibinfo{author}{Reschke, D.}, \bibinfo{author}{Robson, C.},
  \bibinfo{author}{Roever, J.}, \bibinfo{author}{Roggli, M.},
  \bibinfo{author}{Rothenburg, J.}, \bibinfo{author}{Rusi{\'{n}}ski, E.},
  \bibinfo{author}{Rybaniec, R.}, \bibinfo{author}{Sahling, H.},
  \bibinfo{author}{Salmani, M.}, \bibinfo{author}{Samoylova, L.},
  \bibinfo{author}{Sanzone, D.}, \bibinfo{author}{Saretzki, F.},
  \bibinfo{author}{Sawlanski, O.}, \bibinfo{author}{Schaffran, J.},
  \bibinfo{author}{Schlarb, H.}, \bibinfo{author}{Schl\"{o}sser, M.},
  \bibinfo{author}{Schlott, V.}, \bibinfo{author}{Schmidt, C.},
  \bibinfo{author}{Schmidt-Foehre, F.}, \bibinfo{author}{Schmitz, M.},
  \bibinfo{author}{Schm\"{o}kel, M.}, \bibinfo{author}{Schnautz, T.},
  \bibinfo{author}{Schneidmiller, E.}, \bibinfo{author}{Scholz, M.},
  \bibinfo{author}{Sch\"{o}neburg, B.}, \bibinfo{author}{Schultze, J.},
  \bibinfo{author}{Schulz, C.}, \bibinfo{author}{Schwarz, A.},
  \bibinfo{author}{Sekutowicz, J.}, \bibinfo{author}{Sellmann, D.},
  \bibinfo{author}{Semenov, E.}, \bibinfo{author}{Serkez, S.},
  \bibinfo{author}{Sertore, D.}, \bibinfo{author}{Shehzad, N.},
  \bibinfo{author}{Shemarykin, P.}, \bibinfo{author}{Shi, L.},
  \bibinfo{author}{Sienkiewicz, M.}, \bibinfo{author}{Sikora, D.},
  \bibinfo{author}{Sikorski, M.}, \bibinfo{author}{Silenzi, A.},
  \bibinfo{author}{Simon, C.}, \bibinfo{author}{Singer, W.},
  \bibinfo{author}{Singer, X.}, \bibinfo{author}{Sinn, H.},
  \bibinfo{author}{Sinram, K.}, \bibinfo{author}{Skvorodnev, N.},
  \bibinfo{author}{Smirnow, P.}, \bibinfo{author}{Sommer, T.},
  \bibinfo{author}{Sorokin, A.}, \bibinfo{author}{Stadler, M.},
  \bibinfo{author}{Steckel, M.}, \bibinfo{author}{Steffen, B.},
  \bibinfo{author}{Steinhau-K\"{u}hl, N.}, \bibinfo{author}{Stephan, F.},
  \bibinfo{author}{Stodulski, M.}, \bibinfo{author}{Stolper, M.},
  \bibinfo{author}{Sulimov, A.}, \bibinfo{author}{Susen, R.},
  \bibinfo{author}{{\'{S}}wierblewski, J.}, \bibinfo{author}{Sydlo, C.},
  \bibinfo{author}{Syresin, E.}, \bibinfo{author}{Sytchev, V.},
  \bibinfo{author}{Szuba, J.}, \bibinfo{author}{Tesch, N.},
  \bibinfo{author}{Thie, J.}, \bibinfo{author}{Thiebault, A.},
  \bibinfo{author}{Tiedtke, K.}, \bibinfo{author}{Tischhauser, D.},
  \bibinfo{author}{Tolkiehn, J.}, \bibinfo{author}{Tomin, S.},
  \bibinfo{author}{Tonisch, F.}, \bibinfo{author}{Toral, F.},
  \bibinfo{author}{Torbin, I.}, \bibinfo{author}{Trapp, A.},
  \bibinfo{author}{Treyer, D.}, \bibinfo{author}{Trowitzsch, G.},
  \bibinfo{author}{Trublet, T.}, \bibinfo{author}{Tschentscher, T.},
  \bibinfo{author}{Ullrich, F.}, \bibinfo{author}{Vannoni, M.},
  \bibinfo{author}{Varela, P.}, \bibinfo{author}{Varghese, G.},
  \bibinfo{author}{Vashchenko, G.}, \bibinfo{author}{Vasic, M.},
  \bibinfo{author}{Vazquez-Velez, C.}, \bibinfo{author}{Verguet, A.},
  \bibinfo{author}{Vilcins-Czvitkovits, S.}, \bibinfo{author}{Villanueva, R.},
  \bibinfo{author}{Visentin, B.}, \bibinfo{author}{Viti, M.},
  \bibinfo{author}{Vogel, E.}, \bibinfo{author}{Volobuev, E.},
  \bibinfo{author}{Wagner, R.}, \bibinfo{author}{Walker, N.},
  \bibinfo{author}{Wamsat, T.}, \bibinfo{author}{Weddig, H.},
  \bibinfo{author}{Weichert, G.}, \bibinfo{author}{Weise, H.},
  \bibinfo{author}{Wenndorf, R.}, \bibinfo{author}{Werner, M.},
  \bibinfo{author}{Wichmann, R.}, \bibinfo{author}{Wiebers, C.},
  \bibinfo{author}{Wiencek, M.}, \bibinfo{author}{Wilksen, T.},
  \bibinfo{author}{Will, I.}, \bibinfo{author}{Winkelmann, L.},
  \bibinfo{author}{Winkowski, M.}, \bibinfo{author}{Wittenburg, K.},
  \bibinfo{author}{Witzig, A.}, \bibinfo{author}{Wlk, P.},
  \bibinfo{author}{Wohlenberg, T.}, \bibinfo{author}{Wojciechowski, M.},
  \bibinfo{author}{Wolff-Fabris, F.}, \bibinfo{author}{Wrochna, G.},
  \bibinfo{author}{Wrona, K.}, \bibinfo{author}{Yakopov, M.},
  \bibinfo{author}{Yang, B.}, \bibinfo{author}{Yang, F.},
  \bibinfo{author}{Yurkov, M.}, \bibinfo{author}{Zagorodnov, I.},
  \bibinfo{author}{Zalden, P.}, \bibinfo{author}{Zavadtsev, A.},
  \bibinfo{author}{Zavadtsev, D.}, \bibinfo{author}{Zhirnov, A.},
  \bibinfo{author}{Zhukov, A.}, \bibinfo{author}{Ziemann, V.},
  \bibinfo{author}{Zolotov, A.}, \bibinfo{author}{Zolotukhina, N.},
  \bibinfo{author}{Zummack, F.}, \bibinfo{author}{Zybin, D.},
  \bibinfo{year}{2020}.
\newblock \bibinfo{title}{A {MHz}-repetition-rate hard x-ray free-electron
  laser driven by a superconducting linear accelerator}.
\newblock \bibinfo{journal}{Nature Photonics} \bibinfo{volume}{14},
  \bibinfo{pages}{391--397}.
\newblock \DOIprefix\doi{10.1038/s41566-020-0607-z}.
%Type = Book
\bibitem[{Delone and Krainov(2000)}]{Delone2000}
\bibinfo{author}{Delone, N.B.}, \bibinfo{author}{Krainov, V.P.},
  \bibinfo{year}{2000}.
\newblock \bibinfo{title}{Multiphoton Processes in Atoms}.
\newblock \bibinfo{publisher}{Springer}, \bibinfo{address}{Berlin Heidelberg}.
\newblock \DOIprefix\doi{10.1007/978-3-642-57208-1}.
%Type = Article
\bibitem[{Demekhin et~al.(2011a)Demekhin, Ehresmann and
  Sukhorukov}]{Demekhin2011a}
\bibinfo{author}{Demekhin, P.V.}, \bibinfo{author}{Ehresmann, A.},
  \bibinfo{author}{Sukhorukov, V.}, \bibinfo{year}{2011}a.
\newblock \bibinfo{title}{Single center method: A computational tool for
  ionization and electronic excitation studies of molecules}.
\newblock \bibinfo{journal}{J. Chem. Phys.} \bibinfo{volume}{134},
  \bibinfo{pages}{024113}.
\newblock \DOIprefix\doi{10.1063/1.3526026}.
%Type = Article
\bibitem[{Demekhin et~al.(2013)Demekhin, Gokhberg, Jabbari, Kopelke, Kuleff and
  Cederbaum}]{Demekhin2013}
\bibinfo{author}{Demekhin, P.V.}, \bibinfo{author}{Gokhberg, K.},
  \bibinfo{author}{Jabbari, G.}, \bibinfo{author}{Kopelke, S.},
  \bibinfo{author}{Kuleff, A.I.}, \bibinfo{author}{Cederbaum, L.S.},
  \bibinfo{year}{2013}.
\newblock \bibinfo{title}{Overcoming blockade in producing doubly excited
  dimers by a single intense pulse and their decay}.
\newblock \bibinfo{journal}{J. Phys. B} \bibinfo{volume}{46},
  \bibinfo{pages}{021001}.
\newblock \DOIprefix\doi{10.1088/0953-4075/46/2/021001}.
%Type = Article
\bibitem[{Demekhin et~al.(2011b)Demekhin, Stoychev, Kuleff and
  Cederbaum}]{Demekhin2011}
\bibinfo{author}{Demekhin, P.V.}, \bibinfo{author}{Stoychev, S.D.},
  \bibinfo{author}{Kuleff, A.I.}, \bibinfo{author}{Cederbaum, L.S.},
  \bibinfo{year}{2011}b.
\newblock \bibinfo{title}{Exploring interatomic coulombic decay by free
  electron lasers}.
\newblock \bibinfo{journal}{Phys. Rev. Lett.} \bibinfo{volume}{107},
  \bibinfo{pages}{273002}.
\newblock \DOIprefix\doi{10.1103/PhysRevLett.107.273002}.
%Type = Incollection
\bibitem[{Devons and Goldfarb(1957)}]{Devons1957}
\bibinfo{author}{Devons, S.}, \bibinfo{author}{Goldfarb, L.J.B.},
  \bibinfo{year}{1957}.
\newblock \bibinfo{title}{Angular correlations}, in: \bibinfo{editor}{Fl\"ugge,
  S.} (Ed.), \bibinfo{booktitle}{Kernreaktionen {III} / Nuclear Reactions
  {III}}. \bibinfo{publisher}{Springer-Verlag}, \bibinfo{address}{Berlin
  Heidelberg}. volume~\bibinfo{volume}{42} of \textit{\bibinfo{series}{Handbuch
  der Physik}}. chapter~\bibinfo{chapter}{5}, pp. \bibinfo{pages}{362--554}.
\newblock \DOIprefix\doi{10.1007/978-3-642-45878-1_5}.
%Type = Article
\bibitem[{Di~Fraia et~al.(2019)Di~Fraia, Plekan, Callegari, Prince, Giannessi,
  Allaria, Badano, De~Ninno, Trov\`o, Diviacco, Gauthier, Mirian, Penco,
  Rebernik~Ribi\v{c}, Spampinati, Spezzani, Gaio, Orimo, Tugs, Sato, Ishikawa,
  Carpeggiani, Csizmadia, F\"ule, Sansone, Kumar~Maroju, D'Elia, Mazza, Meyer,
  Gryzlova, Grum-Grzhimailo, You and Ueda}]{DiFraia2019}
\bibinfo{author}{Di~Fraia, M.}, \bibinfo{author}{Plekan, O.},
  \bibinfo{author}{Callegari, C.}, \bibinfo{author}{Prince, K.C.},
  \bibinfo{author}{Giannessi, L.}, \bibinfo{author}{Allaria, E.},
  \bibinfo{author}{Badano, L.}, \bibinfo{author}{De~Ninno, G.},
  \bibinfo{author}{Trov\`o, M.}, \bibinfo{author}{Diviacco, B.},
  \bibinfo{author}{Gauthier, D.}, \bibinfo{author}{Mirian, N.},
  \bibinfo{author}{Penco, G.}, \bibinfo{author}{Rebernik~Ribi\v{c}, P.},
  \bibinfo{author}{Spampinati, S.}, \bibinfo{author}{Spezzani, C.},
  \bibinfo{author}{Gaio, G.}, \bibinfo{author}{Orimo, Y.},
  \bibinfo{author}{Tugs, O.}, \bibinfo{author}{Sato, T.},
  \bibinfo{author}{Ishikawa, K.L.}, \bibinfo{author}{Carpeggiani, P.A.},
  \bibinfo{author}{Csizmadia, T.}, \bibinfo{author}{F\"ule, M.},
  \bibinfo{author}{Sansone, G.}, \bibinfo{author}{Kumar~Maroju, P.},
  \bibinfo{author}{D'Elia, A.}, \bibinfo{author}{Mazza, T.},
  \bibinfo{author}{Meyer, M.}, \bibinfo{author}{Gryzlova, E.V.},
  \bibinfo{author}{Grum-Grzhimailo, A.N.}, \bibinfo{author}{You, D.},
  \bibinfo{author}{Ueda, K.}, \bibinfo{year}{2019}.
\newblock \bibinfo{title}{Complete characterization of phase and amplitude of
  bichromatic extreme ultraviolet light}.
\newblock \bibinfo{journal}{Phys. Rev. Lett.} \bibinfo{volume}{123},
  \bibinfo{pages}{213904}.
\newblock \DOIprefix\doi{10.1103/PhysRevLett.123.213904}.
%Type = Article
\bibitem[{Ding et~al.(2014)Ding, Xiong, Fan, Hickstein, Popmintchev, Zhang,
  Walls, Murnane and Kapteyn}]{Ding2014}
\bibinfo{author}{Ding, C.}, \bibinfo{author}{Xiong, W.}, \bibinfo{author}{Fan,
  T.}, \bibinfo{author}{Hickstein, D.D.}, \bibinfo{author}{Popmintchev, T.},
  \bibinfo{author}{Zhang, X.}, \bibinfo{author}{Walls, M.},
  \bibinfo{author}{Murnane, M.M.}, \bibinfo{author}{Kapteyn, H.C.},
  \bibinfo{year}{2014}.
\newblock \bibinfo{title}{High flux coherent super-continuum soft {X}-ray
  source driven by a single-stage, 10 {mJ}, {Ti}:sapphire amplifier-pumped
  {OPA}}.
\newblock \bibinfo{journal}{Opt. Express} \bibinfo{volume}{22},
  \bibinfo{pages}{6194}.
\newblock \DOIprefix\doi{10.1364/oe.22.006194}.
%Type = Article
\bibitem[{Ding et~al.(2019)Ding, Rebholz, Aufleger, Hartmann, Meyer, Stoo\ss{},
  Magunia, Wachs, Birk, Mi, Borisova, da~Costa~Castanheira, Rupprecht, Loh,
  Attar, Gaumnitz, Roling, Butz, Zacharias, D\"usterer, Treusch, Cavaletto, Ott
  and Pfeifer}]{Ding2019}
\bibinfo{author}{Ding, T.}, \bibinfo{author}{Rebholz, M.},
  \bibinfo{author}{Aufleger, L.}, \bibinfo{author}{Hartmann, M.},
  \bibinfo{author}{Meyer, K.}, \bibinfo{author}{Stoo\ss{}, V.},
  \bibinfo{author}{Magunia, A.}, \bibinfo{author}{Wachs, D.},
  \bibinfo{author}{Birk, P.}, \bibinfo{author}{Mi, Y.},
  \bibinfo{author}{Borisova, G.D.}, \bibinfo{author}{da~Costa~Castanheira, C.},
  \bibinfo{author}{Rupprecht, P.}, \bibinfo{author}{Loh, Z.H.},
  \bibinfo{author}{Attar, A.R.}, \bibinfo{author}{Gaumnitz, T.},
  \bibinfo{author}{Roling, S.}, \bibinfo{author}{Butz, M.},
  \bibinfo{author}{Zacharias, H.}, \bibinfo{author}{D\"usterer, S.},
  \bibinfo{author}{Treusch, R.}, \bibinfo{author}{Cavaletto, S.M.},
  \bibinfo{author}{Ott, C.}, \bibinfo{author}{Pfeifer, T.},
  \bibinfo{year}{2019}.
\newblock \bibinfo{title}{Nonlinear coherence effects in transient-absorption
  ion spectroscopy with stochastic extreme-ultraviolet free-electron laser
  pulses}.
\newblock \bibinfo{journal}{Phys. Rev. Lett.} \bibinfo{volume}{123},
  \bibinfo{pages}{103001}.
\newblock \DOIprefix\doi{10.1103/PhysRevLett.123.103001}.
%Type = Inproceedings
\bibitem[{Diviacco et~al.(2011)Diviacco, Bracco, Millo and
  Musardo}]{Diviacco2011}
\bibinfo{author}{Diviacco, B.}, \bibinfo{author}{Bracco, R.},
  \bibinfo{author}{Millo, D.}, \bibinfo{author}{Musardo, M.M.},
  \bibinfo{year}{2011}.
\newblock \bibinfo{title}{Phase shifters for the {FERMI@E}lettra undulators},
  in: \bibinfo{editor}{Petit-Jean-Genaz, C.} (Ed.),
  \bibinfo{booktitle}{Proceedings of the 2nd International Conference on
  Particle Accelerators (IPAC 2011), San Sebasti\'{a}n, Spain},
  \bibinfo{publisher}{IPAC'11 EPS-AG}, \bibinfo{address}{Geneva, Switzerland}.
  p. \bibinfo{pages}{3278}.
\newblock \URLprefix
  \url{https://accelconf.web.cern.ch/AccelConf/IPAC2011/papers/thpc164.pdf}.
%Type = Article
\bibitem[{Domke et~al.(1995)Domke, Schulz, Remmers, Guti\'errez, Kaindl and
  Wintgen}]{Domke1995}
\bibinfo{author}{Domke, M.}, \bibinfo{author}{Schulz, K.},
  \bibinfo{author}{Remmers, G.}, \bibinfo{author}{Guti\'errez, A.},
  \bibinfo{author}{Kaindl, G.}, \bibinfo{author}{Wintgen, D.},
  \bibinfo{year}{1995}.
\newblock \bibinfo{title}{Interferences in photoexcited double-excitation
  series of {He}}.
\newblock \bibinfo{journal}{Phys. Rev. A} \bibinfo{volume}{51},
  \bibinfo{pages}{R4309--R4312}.
\newblock \DOIprefix\doi{10.1103/PhysRevA.51.R4309}.
%Type = Article
\bibitem[{Domke et~al.(1996)Domke, Schulz, Remmers, Kaindl and
  Wintgen}]{Domke1996}
\bibinfo{author}{Domke, M.}, \bibinfo{author}{Schulz, K.},
  \bibinfo{author}{Remmers, G.}, \bibinfo{author}{Kaindl, G.},
  \bibinfo{author}{Wintgen, D.}, \bibinfo{year}{1996}.
\newblock \bibinfo{title}{High-resolution study of $^{1}{P}^{o}$
  double-excitation states in helium}.
\newblock \bibinfo{journal}{Phys. Rev. A} \bibinfo{volume}{53},
  \bibinfo{pages}{1424--1438}.
\newblock \DOIprefix\doi{10.1103/PhysRevA.53.1424}.
%Type = Article
\bibitem[{Dorfman et~al.(2016)Dorfman, Zhang and Mukamel}]{Dorfman2016}
\bibinfo{author}{Dorfman, K.E.}, \bibinfo{author}{Zhang, Y.},
  \bibinfo{author}{Mukamel, S.}, \bibinfo{year}{2016}.
\newblock \bibinfo{title}{Coherent control of long-range photoinduced electron
  transfer by stimulated {X}-ray {R}aman processes}.
\newblock \bibinfo{journal}{Proc. Natl. Acad. Sci. U.S.A.}
  \bibinfo{volume}{113}, \bibinfo{pages}{10001--10006}.
\newblock \DOIprefix\doi{10.1073/pnas.1610729113}.
%Type = Article
\bibitem[{Douguet et~al.(2016)Douguet, Grum-Grzhimailo, Gryzlova,
  Staroselskaya, Venzke and Bartschat}]{Douguet2016}
\bibinfo{author}{Douguet, N.}, \bibinfo{author}{Grum-Grzhimailo, A.N.},
  \bibinfo{author}{Gryzlova, E.V.}, \bibinfo{author}{Staroselskaya, E.I.},
  \bibinfo{author}{Venzke, J.}, \bibinfo{author}{Bartschat, K.},
  \bibinfo{year}{2016}.
\newblock \bibinfo{title}{Photoelectron angular distributions in bichromatic
  atomic ionization induced by circularly polarized {VUV} femtosecond pulses}.
\newblock \bibinfo{journal}{Phys. Rev. A} \bibinfo{volume}{93},
  \bibinfo{pages}{033402}.
\newblock \DOIprefix\doi{10.1103/PhysRevA.93.033402}.
%Type = Article
\bibitem[{Douguet et~al.(2017)Douguet, Gryzlova, Staroselskaya, Bartschat and
  Grum-Grzhimailo}]{Douguet2017}
\bibinfo{author}{Douguet, N.}, \bibinfo{author}{Gryzlova, E.V.},
  \bibinfo{author}{Staroselskaya, E.I.}, \bibinfo{author}{Bartschat, K.},
  \bibinfo{author}{Grum-Grzhimailo, A.N.}, \bibinfo{year}{2017}.
\newblock \bibinfo{title}{Photoelectron angular distribution in two-pathway
  ionization of neon with femtosecond {XUV} pulses}.
\newblock \bibinfo{journal}{Eur. Phys. J. D} \bibinfo{volume}{71},
  \bibinfo{pages}{105}.
\newblock \DOIprefix\doi{10.1140/epjd/e2017-70695-7}.
%Type = Article
\bibitem[{Driver et~al.(2020)Driver, Li, Champenois, Duris, Ratner, Lane,
  Rosenberger, Al-Haddad, Averbukh, Barnard, Berrah, Bostedt, Bucksbaum,
  Coffee, DiMauro, Fang, Garratt, Gatton, Guo, Hartmann, Haxton, Helml, Huang,
  LaForge, Kamalov, Kling, Knurr, Lin, Lutman, MacArthur, Marangos, Nantel,
  Natan, Obaid, O'Neal, Shivaram, Schori, Walter, Wang, Wolf, Marinelli and
  Cryan}]{Driver2020}
\bibinfo{author}{Driver, T.}, \bibinfo{author}{Li, S.},
  \bibinfo{author}{Champenois, E.G.}, \bibinfo{author}{Duris, J.},
  \bibinfo{author}{Ratner, D.}, \bibinfo{author}{Lane, T.J.},
  \bibinfo{author}{Rosenberger, P.}, \bibinfo{author}{Al-Haddad, A.},
  \bibinfo{author}{Averbukh, V.}, \bibinfo{author}{Barnard, T.},
  \bibinfo{author}{Berrah, N.}, \bibinfo{author}{Bostedt, C.},
  \bibinfo{author}{Bucksbaum, P.H.}, \bibinfo{author}{Coffee, R.},
  \bibinfo{author}{DiMauro, L.F.}, \bibinfo{author}{Fang, L.},
  \bibinfo{author}{Garratt, D.}, \bibinfo{author}{Gatton, A.},
  \bibinfo{author}{Guo, Z.}, \bibinfo{author}{Hartmann, G.},
  \bibinfo{author}{Haxton, D.}, \bibinfo{author}{Helml, W.},
  \bibinfo{author}{Huang, Z.}, \bibinfo{author}{LaForge, A.},
  \bibinfo{author}{Kamalov, A.}, \bibinfo{author}{Kling, M.F.},
  \bibinfo{author}{Knurr, J.}, \bibinfo{author}{Lin, M.F.},
  \bibinfo{author}{Lutman, A.A.}, \bibinfo{author}{MacArthur, J.P.},
  \bibinfo{author}{Marangos, J.P.}, \bibinfo{author}{Nantel, M.},
  \bibinfo{author}{Natan, A.}, \bibinfo{author}{Obaid, R.},
  \bibinfo{author}{O'Neal, J.T.}, \bibinfo{author}{Shivaram, N.H.},
  \bibinfo{author}{Schori, A.}, \bibinfo{author}{Walter, P.},
  \bibinfo{author}{Wang, A.L.}, \bibinfo{author}{Wolf, T.J.A.},
  \bibinfo{author}{Marinelli, A.}, \bibinfo{author}{Cryan, J.P.},
  \bibinfo{year}{2020}.
\newblock \bibinfo{title}{Attosecond transient absorption spooktroscopy: a
  ghost imaging approach to ultrafast absorption spectroscopy}.
\newblock \bibinfo{journal}{Phys. Chem. Chem. Phys.} \bibinfo{volume}{22},
  \bibinfo{pages}{2704--2712}.
\newblock \DOIprefix\doi{10.1039/c9cp03951a}.
%Type = Article
\bibitem[{Dubrouil et~al.(2015)Dubrouil, Reduzzi, Devetta, Feng, Hummert,
  Finetti, Plekan, Grazioli, Di~Fraia, Lyamayev, La~Forge, Katzy, Stienkemeier,
  Ovcharenko, Coreno, Berrah, Motomura, Mondal, Ueda, Prince, Callegari,
  Kuleff, Demekhin and Sansone}]{Dubrouil2015}
\bibinfo{author}{Dubrouil, A.}, \bibinfo{author}{Reduzzi, M.},
  \bibinfo{author}{Devetta, M.}, \bibinfo{author}{Feng, C.},
  \bibinfo{author}{Hummert, J.}, \bibinfo{author}{Finetti, P.},
  \bibinfo{author}{Plekan, O.}, \bibinfo{author}{Grazioli, C.},
  \bibinfo{author}{Di~Fraia, M.}, \bibinfo{author}{Lyamayev, V.},
  \bibinfo{author}{La~Forge, A.}, \bibinfo{author}{Katzy, R.},
  \bibinfo{author}{Stienkemeier, F.}, \bibinfo{author}{Ovcharenko, Y.},
  \bibinfo{author}{Coreno, M.}, \bibinfo{author}{Berrah, N.},
  \bibinfo{author}{Motomura, K.}, \bibinfo{author}{Mondal, S.},
  \bibinfo{author}{Ueda, K.}, \bibinfo{author}{Prince, K.C.},
  \bibinfo{author}{Callegari, C.}, \bibinfo{author}{Kuleff, A.I.},
  \bibinfo{author}{Demekhin, P.V.}, \bibinfo{author}{Sansone, G.},
  \bibinfo{year}{2015}.
\newblock \bibinfo{title}{Two-photon resonant excitation of interatomic
  coulombic decay in neon dimers}.
\newblock \bibinfo{journal}{J. Phys. B} \bibinfo{volume}{48},
  \bibinfo{pages}{204005}.
\newblock \DOIprefix\doi{10.1088/0953-4075/48/20/204005}.
%Type = Article
\bibitem[{Duris et~al.(2020)Duris, Li, Driver, Champenois, MacArthur, Lutman,
  Zhang, Rosenberger, Aldrich, Coffee, Coslovich, Decker, Glownia, Hartmann,
  Helml, Kamalov, Knurr, Krzywinski, Lin, Marangos, Nantel, Natan, O'Neal,
  Shivaram, Walter, Wang, Welch, Wolf, Xu, Kling, Bucksbaum, Zholents, Huang,
  Cryan and Marinelli}]{Duris2020}
\bibinfo{author}{Duris, J.}, \bibinfo{author}{Li, S.}, \bibinfo{author}{Driver,
  T.}, \bibinfo{author}{Champenois, E.G.}, \bibinfo{author}{MacArthur, J.P.},
  \bibinfo{author}{Lutman, A.A.}, \bibinfo{author}{Zhang, Z.},
  \bibinfo{author}{Rosenberger, P.}, \bibinfo{author}{Aldrich, J.W.},
  \bibinfo{author}{Coffee, R.}, \bibinfo{author}{Coslovich, G.},
  \bibinfo{author}{Decker, F.J.}, \bibinfo{author}{Glownia, J.M.},
  \bibinfo{author}{Hartmann, G.}, \bibinfo{author}{Helml, W.},
  \bibinfo{author}{Kamalov, A.}, \bibinfo{author}{Knurr, J.},
  \bibinfo{author}{Krzywinski, J.}, \bibinfo{author}{Lin, M.F.},
  \bibinfo{author}{Marangos, J.P.}, \bibinfo{author}{Nantel, M.},
  \bibinfo{author}{Natan, A.}, \bibinfo{author}{O'Neal, J.T.},
  \bibinfo{author}{Shivaram, N.}, \bibinfo{author}{Walter, P.},
  \bibinfo{author}{Wang, A.L.}, \bibinfo{author}{Welch, J.J.},
  \bibinfo{author}{Wolf, T.J.A.}, \bibinfo{author}{Xu, J.Z.},
  \bibinfo{author}{Kling, M.F.}, \bibinfo{author}{Bucksbaum, P.H.},
  \bibinfo{author}{Zholents, A.}, \bibinfo{author}{Huang, Z.},
  \bibinfo{author}{Cryan, J.P.}, \bibinfo{author}{Marinelli, A.},
  \bibinfo{year}{2020}.
\newblock \bibinfo{title}{Tunable isolated attosecond {X}-ray pulses with
  gigawatt peak power from a free-electron laser}.
\newblock \bibinfo{journal}{Nat. Photonics} \bibinfo{volume}{14},
  \bibinfo{pages}{30--36}.
\newblock \DOIprefix\doi{10.1038/s41566-019-0549-5}.
%Type = Article
\bibitem[{Einstein(1905)}]{Einstein1905}
\bibinfo{author}{Einstein, A.}, \bibinfo{year}{1905}.
\newblock \bibinfo{title}{\"{U}ber einen die {E}rzeugung und {V}erwandlung des
  {L}ichtes betreffenden heuristischen {G}esichtspunkt}.
\newblock \bibinfo{journal}{Ann. Phys.} \bibinfo{volume}{322},
  \bibinfo{pages}{132--148}.
\newblock \DOIprefix\doi{10.1002/andp.19053220607}.
%Type = Phdthesis
\bibitem[{Eisenbud(1948)}]{Eisenbud1948}
\bibinfo{author}{Eisenbud, L.}, \bibinfo{year}{1948}.
\newblock \bibinfo{title}{Formal properties of nuclear collisions}.
\newblock Ph.D. thesis. Princeton University. \bibinfo{address}{Princeton, NJ,
  U.S.A.}
%Type = Article
\bibitem[{Emma et~al.(2017)Emma, Lutman, Guetg, Krzywinski, Marinelli, Wu and
  Pellegrini}]{Emma2017}
\bibinfo{author}{Emma, C.}, \bibinfo{author}{Lutman, A.},
  \bibinfo{author}{Guetg, M.W.}, \bibinfo{author}{Krzywinski, J.},
  \bibinfo{author}{Marinelli, A.}, \bibinfo{author}{Wu, J.},
  \bibinfo{author}{Pellegrini, C.}, \bibinfo{year}{2017}.
\newblock \bibinfo{title}{Experimental demonstration of fresh bunch
  self-seeding in an {X}-ray free electron laser}.
\newblock \bibinfo{journal}{Appl. Phys. Lett.} \bibinfo{volume}{110},
  \bibinfo{pages}{154101}.
\newblock \DOIprefix\doi{10.1063/1.4980092}.
%Type = Article
\bibitem[{Emma et~al.(2010)Emma, Akre, Arthur, Bionta, Bostedt, Bozek,
  Brachmann, Bucksbaum, Coffee, Decker, Ding, Dowell, Edstrom, Fisher, Frisch,
  Gilevich, Hastings, Hays, Hering, Huang, Iverson, Loos, Messerschmidt,
  Miahnahri, Moeller, Nuhn, Pile, Ratner, Rzepiela, Schultz, Smith, Stefan,
  Tompkins, Turner, Welch, White, Wu, Yocky and Galayda}]{Emma2010}
\bibinfo{author}{Emma, P.}, \bibinfo{author}{Akre, R.},
  \bibinfo{author}{Arthur, J.}, \bibinfo{author}{Bionta, R.},
  \bibinfo{author}{Bostedt, C.}, \bibinfo{author}{Bozek, J.},
  \bibinfo{author}{Brachmann, A.}, \bibinfo{author}{Bucksbaum, P.},
  \bibinfo{author}{Coffee, R.}, \bibinfo{author}{Decker, F.J.},
  \bibinfo{author}{Ding, Y.}, \bibinfo{author}{Dowell, D.},
  \bibinfo{author}{Edstrom, S.}, \bibinfo{author}{Fisher, A.},
  \bibinfo{author}{Frisch, J.}, \bibinfo{author}{Gilevich, S.},
  \bibinfo{author}{Hastings, J.}, \bibinfo{author}{Hays, G.},
  \bibinfo{author}{Hering, P.}, \bibinfo{author}{Huang, Z.},
  \bibinfo{author}{Iverson, R.}, \bibinfo{author}{Loos, H.},
  \bibinfo{author}{Messerschmidt, M.}, \bibinfo{author}{Miahnahri, A.},
  \bibinfo{author}{Moeller, S.}, \bibinfo{author}{Nuhn, H.D.},
  \bibinfo{author}{Pile, G.}, \bibinfo{author}{Ratner, D.},
  \bibinfo{author}{Rzepiela, J.}, \bibinfo{author}{Schultz, D.},
  \bibinfo{author}{Smith, T.}, \bibinfo{author}{Stefan, P.},
  \bibinfo{author}{Tompkins, H.}, \bibinfo{author}{Turner, J.},
  \bibinfo{author}{Welch, J.}, \bibinfo{author}{White, W.},
  \bibinfo{author}{Wu, J.}, \bibinfo{author}{Yocky, G.},
  \bibinfo{author}{Galayda, J.}, \bibinfo{year}{2010}.
\newblock \bibinfo{title}{First lasing and operation of an
  {\aa}ngstrom-wavelength free-electron laser}.
\newblock \bibinfo{journal}{Nat. Photonics} \bibinfo{volume}{4},
  \bibinfo{pages}{641--647}.
\newblock \DOIprefix\doi{10.1038/nphoton.2010.176}.
%Type = Article
\bibitem[{Engin et~al.(2019)Engin, Gonz\'alez-V\'azquez, Maliyar,
  Milosavljevi\'c, Ono, Nandi, Iablonskyi, Kooser, Bozek, Decleva, Kukk, Ueda
  and Mart\'in}]{Engin2019}
\bibinfo{author}{Engin, S.}, \bibinfo{author}{Gonz\'alez-V\'azquez, J.},
  \bibinfo{author}{Maliyar, G.G.}, \bibinfo{author}{Milosavljevi\'c, A.R.},
  \bibinfo{author}{Ono, T.}, \bibinfo{author}{Nandi, S.},
  \bibinfo{author}{Iablonskyi, D.}, \bibinfo{author}{Kooser, K.},
  \bibinfo{author}{Bozek, J.D.}, \bibinfo{author}{Decleva, P.},
  \bibinfo{author}{Kukk, E.}, \bibinfo{author}{Ueda, K.},
  \bibinfo{author}{Mart\'in, F.}, \bibinfo{year}{2019}.
\newblock \bibinfo{title}{Full-dimensional theoretical description of
  vibrationally resolved valence-shell photoionization of {H}$_2${O}}.
\newblock \bibinfo{journal}{Struct. Dyn.} \bibinfo{volume}{6},
  \bibinfo{pages}{054101}.
\newblock \DOIprefix\doi{10.1063/1.5106431}.
%Type = Article
\bibitem[{Erk et~al.(2014)Erk, Boll, Trippel, Anielski, Foucar, Rudek, Epp,
  Coffee, Carron, Schorb, Ferguson, Swiggers, Bozek, Simon, Marchenko, Kupper,
  Schlichting, Ullrich, Bostedt, Rolles and Rudenko}]{Erk2014}
\bibinfo{author}{Erk, B.}, \bibinfo{author}{Boll, R.},
  \bibinfo{author}{Trippel, S.}, \bibinfo{author}{Anielski, D.},
  \bibinfo{author}{Foucar, L.}, \bibinfo{author}{Rudek, B.},
  \bibinfo{author}{Epp, S.W.}, \bibinfo{author}{Coffee, R.},
  \bibinfo{author}{Carron, S.}, \bibinfo{author}{Schorb, S.},
  \bibinfo{author}{Ferguson, K.R.}, \bibinfo{author}{Swiggers, M.},
  \bibinfo{author}{Bozek, J.D.}, \bibinfo{author}{Simon, M.},
  \bibinfo{author}{Marchenko, T.}, \bibinfo{author}{Kupper, J.},
  \bibinfo{author}{Schlichting, I.}, \bibinfo{author}{Ullrich, J.},
  \bibinfo{author}{Bostedt, C.}, \bibinfo{author}{Rolles, D.},
  \bibinfo{author}{Rudenko, A.}, \bibinfo{year}{2014}.
\newblock \bibinfo{title}{Imaging charge transfer in iodomethane upon {x}-ray
  photoabsorption}.
\newblock \bibinfo{journal}{Science} \bibinfo{volume}{345},
  \bibinfo{pages}{288--291}.
\newblock \DOIprefix\doi{10.1126/science.1253607}.
%Type = Book
\bibitem[{Ernst et~al.(1990)Ernst, Wokaun and Bodenhausen}]{Ernst1990}
\bibinfo{author}{Ernst, R.R.}, \bibinfo{author}{Wokaun, A.},
  \bibinfo{author}{Bodenhausen, G.}, \bibinfo{year}{1990}.
\newblock \bibinfo{title}{Principles of Nuclear Magnetic Resonance in One and
  Two Dimensions}.
\newblock \bibinfo{publisher}{Oxford University Press},
  \bibinfo{address}{Oxford, UK}.
%Type = Article
\bibitem[{Faatz et~al.(2017)Faatz, Braune, Hensler, Honkavaara, Kammering,
  Kuhlmann, Ploenjes, Roensch-Schulenburg, Schneidmiller, Schreiber, Tiedtke,
  Tischer, Treusch, Vogt, Wurth, Yurkov and Zemella}]{Faatz2017}
\bibinfo{author}{Faatz, B.}, \bibinfo{author}{Braune, M.},
  \bibinfo{author}{Hensler, O.}, \bibinfo{author}{Honkavaara, K.},
  \bibinfo{author}{Kammering, R.}, \bibinfo{author}{Kuhlmann, M.},
  \bibinfo{author}{Ploenjes, E.}, \bibinfo{author}{Roensch-Schulenburg, J.},
  \bibinfo{author}{Schneidmiller, E.}, \bibinfo{author}{Schreiber, S.},
  \bibinfo{author}{Tiedtke, K.}, \bibinfo{author}{Tischer, M.},
  \bibinfo{author}{Treusch, R.}, \bibinfo{author}{Vogt, M.},
  \bibinfo{author}{Wurth, W.}, \bibinfo{author}{Yurkov, M.},
  \bibinfo{author}{Zemella, J.}, \bibinfo{year}{2017}.
\newblock \bibinfo{title}{The {FLASH} facility: Advanced options for {FLASH}2
  and future perspectives}.
\newblock \bibinfo{journal}{Appl. Sci.} \bibinfo{volume}{7},
  \bibinfo{pages}{1114}.
\newblock \DOIprefix\doi{10.3390/app7111114}.
%Type = Article
\bibitem[{Faisal(1973)}]{Faisal1973}
\bibinfo{author}{Faisal, F.H.M.}, \bibinfo{year}{1973}.
\newblock \bibinfo{title}{Multiple absorption of laser photons by atoms}.
\newblock \bibinfo{journal}{J. Phys. B} \bibinfo{volume}{6},
  \bibinfo{pages}{L89--L92}.
\newblock \DOIprefix\doi{10.1088/0022-3700/6/4/011}.
%Type = Book
\bibitem[{Faisal(1987)}]{Faisal1987}
\bibinfo{author}{Faisal, F.H.M.}, \bibinfo{year}{1987}.
\newblock \bibinfo{title}{Theory of Multiphoton Processes}.
\newblock \bibinfo{publisher}{Plenum Press}, \bibinfo{address}{New York}.
\newblock \DOIprefix\doi{10.1007/978-1-4899-1977-9}.
%Type = Article
\bibitem[{Fano(1935)}]{Fano1935}
\bibinfo{author}{Fano, U.}, \bibinfo{year}{1935}.
\newblock \bibinfo{title}{Sullo spettro di assorbimento dei gas nobili presso
  il limite dello spettro d'arco}.
\newblock \bibinfo{journal}{Il Nuovo Cimento} \bibinfo{volume}{12},
  \bibinfo{pages}{154--161}.
\newblock \DOIprefix\doi{10.1007/bf02958288}. \bibinfo{note}{{I}n Italian;
  translated into English as \cite{Fano2005}}.
%Type = Article
\bibitem[{Fano(1961)}]{Fano1961}
\bibinfo{author}{Fano, U.}, \bibinfo{year}{1961}.
\newblock \bibinfo{title}{Effects of configuration interaction on intensities
  and phase shifts}.
\newblock \bibinfo{journal}{Phys. Rev.} \bibinfo{volume}{124},
  \bibinfo{pages}{1866--1878}.
\newblock \DOIprefix\doi{10.1103/PhysRev.124.1866}.
%Type = Article
\bibitem[{Fano et~al.(2005)Fano, Pupillo, Zannoni and Clark}]{Fano2005}
\bibinfo{author}{Fano, U.}, \bibinfo{author}{Pupillo, G.},
  \bibinfo{author}{Zannoni, A.}, \bibinfo{author}{Clark, C.},
  \bibinfo{year}{2005}.
\newblock \bibinfo{title}{On the absorption spectrum of noble gases at the arc
  spectrum limit}.
\newblock \bibinfo{journal}{J. Res. Nat. Inst. Stand. Technol.}
  \bibinfo{volume}{110}, \bibinfo{pages}{583}.
\newblock \DOIprefix\doi{10.6028/jres.110.083}. \bibinfo{note}{{E}nglish
  translation of \cite{Fano1935}}.
%Type = Book
\bibitem[{Fedorov(1998)}]{Fedorov1998}
\bibinfo{author}{Fedorov, M.V.}, \bibinfo{year}{1998}.
\newblock \bibinfo{title}{Atomic and Free Electrons in a Strong Light Field}.
\newblock \bibinfo{publisher}{{W}orld {S}cientific}.
\newblock \DOIprefix\doi{10.1142/3320}.
%Type = Article
\bibitem[{Feist et~al.(2009)Feist, Nagele, Pazourek, Persson, Schneider,
  Collins and Burgd{\"o}rfer}]{Burgdorfer2009}
\bibinfo{author}{Feist, J.}, \bibinfo{author}{Nagele, S.},
  \bibinfo{author}{Pazourek, R.}, \bibinfo{author}{Persson, E.},
  \bibinfo{author}{Schneider, B.I.}, \bibinfo{author}{Collins, L.A.},
  \bibinfo{author}{Burgd{\"o}rfer, J.}, \bibinfo{year}{2009}.
\newblock \bibinfo{title}{Probing electron correlation via attosecond xuv
  pulses in the two-photon double ionization of helium}.
\newblock \bibinfo{journal}{Phys. Rev. Lett.} \bibinfo{volume}{103},
  \bibinfo{pages}{063002}.
\newblock \DOIprefix\doi{10.1103/PhysRevLett.103.063002}.
%Type = Article
\bibitem[{Feist et~al.(2014)Feist, Zatsarinny, Nagele, Pazourek, Burgd\"orfer,
  Guan, Bartschat and Schneider}]{Feist2014}
\bibinfo{author}{Feist, J.}, \bibinfo{author}{Zatsarinny, O.},
  \bibinfo{author}{Nagele, S.}, \bibinfo{author}{Pazourek, R.},
  \bibinfo{author}{Burgd\"orfer, J.}, \bibinfo{author}{Guan, X.},
  \bibinfo{author}{Bartschat, K.}, \bibinfo{author}{Schneider, B.I.},
  \bibinfo{year}{2014}.
\newblock \bibinfo{title}{Time delays for attosecond streaking in
  photoionization of neon}.
\newblock \bibinfo{journal}{Phys. Rev. A} \bibinfo{volume}{89},
  \bibinfo{pages}{033417}.
\newblock \DOIprefix\doi{10.1103/PhysRevA.89.033417}.
%Type = Article
\bibitem[{Feldhaus et~al.(2005)Feldhaus, Arthur and Hastings}]{Feldhaus2005}
\bibinfo{author}{Feldhaus, J.}, \bibinfo{author}{Arthur, J.},
  \bibinfo{author}{Hastings, J.B.}, \bibinfo{year}{2005}.
\newblock \bibinfo{title}{{X}-ray free-electron lasers}.
\newblock \bibinfo{journal}{J. Phys. B} \bibinfo{volume}{38},
  \bibinfo{pages}{S799--S819}.
\newblock \DOIprefix\doi{10.1088/0953-4075/38/9/023}.
%Type = Article
\bibitem[{Feldhaus et~al.(2013)Feldhaus, Krikunova, Meyer, M\"{o}ller,
  Moshammer, Rudenko, Tschentscher and Ullrich}]{Feldhaus2013}
\bibinfo{author}{Feldhaus, J.}, \bibinfo{author}{Krikunova, M.},
  \bibinfo{author}{Meyer, M.}, \bibinfo{author}{M\"{o}ller, T.},
  \bibinfo{author}{Moshammer, R.}, \bibinfo{author}{Rudenko, A.},
  \bibinfo{author}{Tschentscher, T.}, \bibinfo{author}{Ullrich, J.},
  \bibinfo{year}{2013}.
\newblock \bibinfo{title}{{AMO} science at the {FLASH} and european {XFEL}
  free-electron laser facilities}.
\newblock \bibinfo{journal}{J. Phys. B} \bibinfo{volume}{46},
  \bibinfo{pages}{164002}.
\newblock \DOIprefix\doi{10.1088/0953-4075/46/16/164002}.
%Type = Article
\bibitem[{Feldhaus et~al.(1997)Feldhaus, Saldin, Schneider, Schneidmiller and
  Yurkov}]{Feldhaus1997}
\bibinfo{author}{Feldhaus, J.}, \bibinfo{author}{Saldin, E.L.},
  \bibinfo{author}{Schneider, J.R.}, \bibinfo{author}{Schneidmiller, E.A.},
  \bibinfo{author}{Yurkov, M.V.}, \bibinfo{year}{1997}.
\newblock \bibinfo{title}{Possible application of {X}-ray optical elements for
  reducing the spectral bandwidth of an {X}-ray {SASE} {FEL}}.
\newblock \bibinfo{journal}{Opt. Commun.} \bibinfo{volume}{140},
  \bibinfo{pages}{341--352}.
\newblock \DOIprefix\doi{10.1016/s0030-4018(97)00163-6}.
%Type = Article
\bibitem[{Feng and Deng(2018)}]{Feng2018}
\bibinfo{author}{Feng, C.}, \bibinfo{author}{Deng, H.X.}, \bibinfo{year}{2018}.
\newblock \bibinfo{title}{Review of fully coherent free-electron lasers}.
\newblock \bibinfo{journal}{Nucl. Sci. Tech.} \bibinfo{volume}{29}.
\newblock \DOIprefix\doi{10.1007/s41365-018-0490-1}.
%Type = Article
\bibitem[{Feng et~al.(2020)Feng, Heilmann, Bock, Ehrentraut, Witting, Yu,
  Stiel, Eisebitt and Schn\"{u}rer}]{Feng2020}
\bibinfo{author}{Feng, T.}, \bibinfo{author}{Heilmann, A.},
  \bibinfo{author}{Bock, M.}, \bibinfo{author}{Ehrentraut, L.},
  \bibinfo{author}{Witting, T.}, \bibinfo{author}{Yu, H.},
  \bibinfo{author}{Stiel, H.}, \bibinfo{author}{Eisebitt, S.},
  \bibinfo{author}{Schn\"{u}rer, M.}, \bibinfo{year}{2020}.
\newblock \bibinfo{title}{27 {W} 2.1$\mathrm{\mu}$m {OPCPA} system for coherent
  soft {X}-ray generation operating at 10 {kHz}}.
\newblock \bibinfo{journal}{Opt. Express} \bibinfo{volume}{28},
  \bibinfo{pages}{8724}.
\newblock \DOIprefix\doi{10.1364/oe.386588}.
%Type = Book
\bibitem[{Ferguson(1965)}]{Ferguson1965}
\bibinfo{author}{Ferguson, A.J.}, \bibinfo{year}{1965}.
\newblock \bibinfo{title}{Angular correlation method in gamma ray
  spectroscopy}.
\newblock \bibinfo{publisher}{North Holland}, \bibinfo{address}{Amsterdam}.
%Type = Article
\bibitem[{Ferguson et~al.(2016)Ferguson, Bucher, Gorkhover, Boutet, Fukuzawa,
  Koglin, Kumagai, Lutman, Marinelli, Messerschmidt, Nagaya, Turner, Ueda,
  Williams, Bucksbaum and Bostedt}]{Ferguson2016}
\bibinfo{author}{Ferguson, K.R.}, \bibinfo{author}{Bucher, M.},
  \bibinfo{author}{Gorkhover, T.}, \bibinfo{author}{Boutet, S.},
  \bibinfo{author}{Fukuzawa, H.}, \bibinfo{author}{Koglin, J.E.},
  \bibinfo{author}{Kumagai, Y.}, \bibinfo{author}{Lutman, A.},
  \bibinfo{author}{Marinelli, A.}, \bibinfo{author}{Messerschmidt, M.},
  \bibinfo{author}{Nagaya, K.}, \bibinfo{author}{Turner, J.},
  \bibinfo{author}{Ueda, K.}, \bibinfo{author}{Williams, G.J.},
  \bibinfo{author}{Bucksbaum, P.H.}, \bibinfo{author}{Bostedt, C.},
  \bibinfo{year}{2016}.
\newblock \bibinfo{title}{Transient lattice contraction in the solid-to-plasma
  transition}.
\newblock \bibinfo{journal}{Sci. Adv.} \bibinfo{volume}{2},
  \bibinfo{pages}{e1500837}.
\newblock \DOIprefix\doi{10.1126/sciadv.1500837}.
%Type = Article
\bibitem[{Ferrari et~al.(2019)Ferrari, Roussel, Buck, Callegari, Cucini,
  De~Ninno, Diviacco, Gauthier, Giannessi, Glaser, Hartmann, Penco, Scholz,
  Seltmann, Shevchuk, Viefhaus, Zangrando and Allaria}]{Ferrari2019}
\bibinfo{author}{Ferrari, E.}, \bibinfo{author}{Roussel, E.},
  \bibinfo{author}{Buck, J.}, \bibinfo{author}{Callegari, C.},
  \bibinfo{author}{Cucini, R.}, \bibinfo{author}{De~Ninno, G.},
  \bibinfo{author}{Diviacco, B.}, \bibinfo{author}{Gauthier, D.},
  \bibinfo{author}{Giannessi, L.}, \bibinfo{author}{Glaser, L.},
  \bibinfo{author}{Hartmann, G.}, \bibinfo{author}{Penco, G.},
  \bibinfo{author}{Scholz, F.}, \bibinfo{author}{Seltmann, J.},
  \bibinfo{author}{Shevchuk, I.}, \bibinfo{author}{Viefhaus, J.},
  \bibinfo{author}{Zangrando, M.}, \bibinfo{author}{Allaria, E.M.},
  \bibinfo{year}{2019}.
\newblock \bibinfo{title}{Free electron laser polarization control with
  interfering crossed polarized fields}.
\newblock \bibinfo{journal}{Phys. Rev. Accel. Beams} \bibinfo{volume}{22},
  \bibinfo{pages}{080701}.
\newblock \DOIprefix\doi{10.1103/PhysRevAccelBeams.22.080701}.
%Type = Article
\bibitem[{Ferrari et~al.(2016)Ferrari, Spezzani, Fortuna, Delaunay, Vidal,
  Nikolov, Cinquegrana, Diviacco, Gauthier, Penco, Ribi{\v{c}}, Roussel,
  Trov{\`{o}}, Moussy, Pincelli, Lounis, Manfredda, Pedersoli, Capotondi,
  Svetina, Mahne, Zangrando, Raimondi, Demidovich, Giannessi, De~Ninno,
  Danailov, Allaria and Sacchi}]{Ferrari2016}
\bibinfo{author}{Ferrari, E.}, \bibinfo{author}{Spezzani, C.},
  \bibinfo{author}{Fortuna, F.}, \bibinfo{author}{Delaunay, R.},
  \bibinfo{author}{Vidal, F.}, \bibinfo{author}{Nikolov, I.},
  \bibinfo{author}{Cinquegrana, P.}, \bibinfo{author}{Diviacco, B.},
  \bibinfo{author}{Gauthier, D.}, \bibinfo{author}{Penco, G.},
  \bibinfo{author}{Ribi{\v{c}}, P.R.}, \bibinfo{author}{Roussel, E.},
  \bibinfo{author}{Trov{\`{o}}, M.}, \bibinfo{author}{Moussy, J.B.},
  \bibinfo{author}{Pincelli, T.}, \bibinfo{author}{Lounis, L.},
  \bibinfo{author}{Manfredda, M.}, \bibinfo{author}{Pedersoli, E.},
  \bibinfo{author}{Capotondi, F.}, \bibinfo{author}{Svetina, C.},
  \bibinfo{author}{Mahne, N.}, \bibinfo{author}{Zangrando, M.},
  \bibinfo{author}{Raimondi, L.}, \bibinfo{author}{Demidovich, A.},
  \bibinfo{author}{Giannessi, L.}, \bibinfo{author}{De~Ninno, G.},
  \bibinfo{author}{Danailov, M.B.}, \bibinfo{author}{Allaria, E.},
  \bibinfo{author}{Sacchi, M.}, \bibinfo{year}{2016}.
\newblock \bibinfo{title}{Widely tunable two-colour seeded free-electron laser
  source for resonant-pump resonant-probe magnetic scattering}.
\newblock \bibinfo{journal}{Nat. Commun.} \bibinfo{volume}{7},
  \bibinfo{pages}{10343}.
\newblock \DOIprefix\doi{10.1038/ncomms10343}.
%Type = Article
\bibitem[{Feyer et~al.(2018)Feyer, Prince, Coreno, Melandri, Maris,
  Evangelisti, Caminati, Giuliano, Kjaergaard and Carravetta}]{Feyer2018}
\bibinfo{author}{Feyer, V.}, \bibinfo{author}{Prince, K.C.},
  \bibinfo{author}{Coreno, M.}, \bibinfo{author}{Melandri, S.},
  \bibinfo{author}{Maris, A.}, \bibinfo{author}{Evangelisti, L.},
  \bibinfo{author}{Caminati, W.}, \bibinfo{author}{Giuliano, B.M.},
  \bibinfo{author}{Kjaergaard, H.G.}, \bibinfo{author}{Carravetta, V.},
  \bibinfo{year}{2018}.
\newblock \bibinfo{title}{Quantum effects for a proton in a low-barrier,
  double-well potential: Core level photoemission spectroscopy of
  acetylacetone}.
\newblock \bibinfo{journal}{J. Phys. Chem. Lett.} \bibinfo{volume}{9},
  \bibinfo{pages}{521--526}.
\newblock \DOIprefix\doi{10.1021/acs.jpclett.7b03175}.
%Type = Article
\bibitem[{Finetti et~al.(2017)Finetti, H\"oppner, Allaria, Callegari,
  Capotondi, Cinquegrana, Coreno, Cucini, Danailov, Demidovich, De~Ninno,
  Di~Fraia, Feifel, Ferrari, Fr\"ohlich, Gauthier, Golz, Grazioli, Kai, Kurdi,
  Mahne, Manfredda, Medvedev, Nikolov, Pedersoli, Penco, Plekan, Prandolini,
  Prince, Raimondi, Rebernik, Riedel, Roussel, Sigalotti, Squibb, Stojanovic,
  Stranges, Svetina, Tanikawa, Teubner, Tkachenko, Toleikis, Zangrando, Ziaja,
  Tavella and Giannessi}]{Finetti2017}
\bibinfo{author}{Finetti, P.}, \bibinfo{author}{H\"oppner, H.},
  \bibinfo{author}{Allaria, E.}, \bibinfo{author}{Callegari, C.},
  \bibinfo{author}{Capotondi, F.}, \bibinfo{author}{Cinquegrana, P.},
  \bibinfo{author}{Coreno, M.}, \bibinfo{author}{Cucini, R.},
  \bibinfo{author}{Danailov, M.B.}, \bibinfo{author}{Demidovich, A.},
  \bibinfo{author}{De~Ninno, G.}, \bibinfo{author}{Di~Fraia, M.},
  \bibinfo{author}{Feifel, R.}, \bibinfo{author}{Ferrari, E.},
  \bibinfo{author}{Fr\"ohlich, L.}, \bibinfo{author}{Gauthier, D.},
  \bibinfo{author}{Golz, T.}, \bibinfo{author}{Grazioli, C.},
  \bibinfo{author}{Kai, Y.}, \bibinfo{author}{Kurdi, G.},
  \bibinfo{author}{Mahne, N.}, \bibinfo{author}{Manfredda, M.},
  \bibinfo{author}{Medvedev, N.}, \bibinfo{author}{Nikolov, I.P.},
  \bibinfo{author}{Pedersoli, E.}, \bibinfo{author}{Penco, G.},
  \bibinfo{author}{Plekan, O.}, \bibinfo{author}{Prandolini, M.J.},
  \bibinfo{author}{Prince, K.C.}, \bibinfo{author}{Raimondi, L.},
  \bibinfo{author}{Rebernik, P.}, \bibinfo{author}{Riedel, R.},
  \bibinfo{author}{Roussel, E.}, \bibinfo{author}{Sigalotti, P.},
  \bibinfo{author}{Squibb, R.}, \bibinfo{author}{Stojanovic, N.},
  \bibinfo{author}{Stranges, S.}, \bibinfo{author}{Svetina, C.},
  \bibinfo{author}{Tanikawa, T.}, \bibinfo{author}{Teubner, U.},
  \bibinfo{author}{Tkachenko, V.}, \bibinfo{author}{Toleikis, S.},
  \bibinfo{author}{Zangrando, M.}, \bibinfo{author}{Ziaja, B.},
  \bibinfo{author}{Tavella, F.}, \bibinfo{author}{Giannessi, L.},
  \bibinfo{year}{2017}.
\newblock \bibinfo{title}{Pulse duration of seeded free-electron lasers}.
\newblock \bibinfo{journal}{Phys. Rev. X} \bibinfo{volume}{7},
  \bibinfo{pages}{021043}.
\newblock \DOIprefix\doi{10.1103/PhysRevX.7.021043}.
%Type = Article
\bibitem[{Foumouo et~al.(2006)Foumouo, Kamta, Edah and Piraux}]{Foumouo2006}
\bibinfo{author}{Foumouo, E.}, \bibinfo{author}{Kamta, G.L.},
  \bibinfo{author}{Edah, G.}, \bibinfo{author}{Piraux, B.},
  \bibinfo{year}{2006}.
\newblock \bibinfo{title}{Theory of multiphoton single and double ionization of
  two-electron atomic systems driven by short-wavelength electric fields: An
  \emph{ab initio} treatment}.
\newblock \bibinfo{journal}{Phys. Rev. A} \bibinfo{volume}{74},
  \bibinfo{pages}{063409}.
\newblock \DOIprefix\doi{10.1103/PhysRevA.74.063409}.
%Type = Article
\bibitem[{Frasinski(2016)}]{Frasinski2016}
\bibinfo{author}{Frasinski, L.J.}, \bibinfo{year}{2016}.
\newblock \bibinfo{title}{Covariance mapping techniques}.
\newblock \bibinfo{journal}{J. Phys. B: At., Mol. Opt. Phys.}
  \bibinfo{volume}{49}, \bibinfo{pages}{152004}.
\newblock \DOIprefix\doi{10.1088/0953-4075/49/15/152004}.
%Type = Article
\bibitem[{Frasinski et~al.(1989)Frasinski, Codling and
  Hatherly}]{Frasinski1989}
\bibinfo{author}{Frasinski, L.J.}, \bibinfo{author}{Codling, K.},
  \bibinfo{author}{Hatherly, P.A.}, \bibinfo{year}{1989}.
\newblock \bibinfo{title}{Covariance mapping: A correlation method applied to
  multiphoton multiple ionization}.
\newblock \bibinfo{journal}{Science} \bibinfo{volume}{246},
  \bibinfo{pages}{1029--1031}.
\newblock \DOIprefix\doi{10.1126/science.246.4933.1029}.
%Type = Article
\bibitem[{Frasinski et~al.(2013)Frasinski, Zhaunerchyk, Mucke, Squibb, Siano,
  Eland, Linusson, v.d. Meulen, Sal\'en, Thomas, Larsson, Foucar, Ullrich,
  Motomura, Mondal, Ueda, Osipov, Fang, Murphy, Berrah, Bostedt, Bozek, Schorb,
  Messerschmidt, Glownia, Cryan, Coffee, Takahashi, Wada, Piancastelli,
  Richter, Prince and Feifel}]{Frasinski2013}
\bibinfo{author}{Frasinski, L.J.}, \bibinfo{author}{Zhaunerchyk, V.},
  \bibinfo{author}{Mucke, M.}, \bibinfo{author}{Squibb, R.J.},
  \bibinfo{author}{Siano, M.}, \bibinfo{author}{Eland, J.H.D.},
  \bibinfo{author}{Linusson, P.}, \bibinfo{author}{v.d. Meulen, P.},
  \bibinfo{author}{Sal\'en, P.}, \bibinfo{author}{Thomas, R.D.},
  \bibinfo{author}{Larsson, M.}, \bibinfo{author}{Foucar, L.},
  \bibinfo{author}{Ullrich, J.}, \bibinfo{author}{Motomura, K.},
  \bibinfo{author}{Mondal, S.}, \bibinfo{author}{Ueda, K.},
  \bibinfo{author}{Osipov, T.}, \bibinfo{author}{Fang, L.},
  \bibinfo{author}{Murphy, B.F.}, \bibinfo{author}{Berrah, N.},
  \bibinfo{author}{Bostedt, C.}, \bibinfo{author}{Bozek, J.D.},
  \bibinfo{author}{Schorb, S.}, \bibinfo{author}{Messerschmidt, M.},
  \bibinfo{author}{Glownia, J.M.}, \bibinfo{author}{Cryan, J.P.},
  \bibinfo{author}{Coffee, R.N.}, \bibinfo{author}{Takahashi, O.},
  \bibinfo{author}{Wada, S.}, \bibinfo{author}{Piancastelli, M.N.},
  \bibinfo{author}{Richter, R.}, \bibinfo{author}{Prince, K.C.},
  \bibinfo{author}{Feifel, R.}, \bibinfo{year}{2013}.
\newblock \bibinfo{title}{Dynamics of hollow atom formation in intense x-ray
  pulses probed by partial covariance mapping}.
\newblock \bibinfo{journal}{Phys. Rev. Lett.} \bibinfo{volume}{111},
  \bibinfo{pages}{073002}.
\newblock \DOIprefix\doi{10.1103/PhysRevLett.111.073002}.
%Type = Article
\bibitem[{Friedman et~al.(1988)Friedman, Gover, Kurizki, Ruschin and
  Yariv}]{Friedman1988}
\bibinfo{author}{Friedman, A.}, \bibinfo{author}{Gover, A.},
  \bibinfo{author}{Kurizki, G.}, \bibinfo{author}{Ruschin, S.},
  \bibinfo{author}{Yariv, A.}, \bibinfo{year}{1988}.
\newblock \bibinfo{title}{Spontaneous and stimulated emission from quasifree
  electrons}.
\newblock \bibinfo{journal}{Rev. Mod. Phys.} \bibinfo{volume}{60},
  \bibinfo{pages}{471--535}.
\newblock \DOIprefix\doi{10.1103/RevModPhys.60.471}.
%Type = Article
\bibitem[{Fritzsche(2001)}]{Fritzsche2001}
\bibinfo{author}{Fritzsche, S.}, \bibinfo{year}{2001}.
\newblock \bibinfo{title}{Utilities for the \textsc{Ratip} package}.
\newblock \bibinfo{journal}{Comput. Phys. Commun.} \bibinfo{volume}{141},
  \bibinfo{pages}{163--174}.
\newblock \DOIprefix\doi{10.1016/s0010-4655(01)00400-3}.
%Type = Article
\bibitem[{Fritzsche(2012)}]{Fritzsche2012}
\bibinfo{author}{Fritzsche, S.}, \bibinfo{year}{2012}.
\newblock \bibinfo{title}{The \textsc{Ratip} program for relativistic
  calculations of atomic transition, ionization and recombination properties}.
\newblock \bibinfo{journal}{Comput. Phys. Commun.} \bibinfo{volume}{183},
  \bibinfo{pages}{1525--1559}.
\newblock \DOIprefix\doi{10.1016/j.cpc.2012.02.016}.
%Type = Article
\bibitem[{Fritzsche et~al.(2008)Fritzsche, Grum-Grzhimailo, Gryzlova and
  Kabachnik}]{Fritzsche2008}
\bibinfo{author}{Fritzsche, S.}, \bibinfo{author}{Grum-Grzhimailo, A.N.},
  \bibinfo{author}{Gryzlova, E.V.}, \bibinfo{author}{Kabachnik, N.M.},
  \bibinfo{year}{2008}.
\newblock \bibinfo{title}{Angular distributions and angular correlations in
  sequential two-photon double ionization of atoms}.
\newblock \bibinfo{journal}{J. Phys. B} \bibinfo{volume}{41},
  \bibinfo{pages}{165601}.
\newblock \DOIprefix\doi{10.1088/0953-4075/41/16/165601}.
%Type = Article
\bibitem[{Fritzsche et~al.(2009)Fritzsche, Grum-Grzhimailo, Gryzlova and
  Kabachnik}]{Fritzsche2009}
\bibinfo{author}{Fritzsche, S.}, \bibinfo{author}{Grum-Grzhimailo, A.N.},
  \bibinfo{author}{Gryzlova, E.V.}, \bibinfo{author}{Kabachnik, N.M.},
  \bibinfo{year}{2009}.
\newblock \bibinfo{title}{Sequential two-photon double ionization of {Kr}
  atoms}.
\newblock \bibinfo{journal}{J. Phys. B} \bibinfo{volume}{42},
  \bibinfo{pages}{145602}.
\newblock \DOIprefix\doi{10.1088/0953-4075/42/14/145602}.
%Type = Article
\bibitem[{Fritzsche et~al.(2011)Fritzsche, Grum-Grzhimailo, Gryzlova and
  Kabachnik}]{Fritzsche2011}
\bibinfo{author}{Fritzsche, S.}, \bibinfo{author}{Grum-Grzhimailo, A.N.},
  \bibinfo{author}{Gryzlova, E.V.}, \bibinfo{author}{Kabachnik, N.M.},
  \bibinfo{year}{2011}.
\newblock \bibinfo{title}{Sequential two-photon double ionization of the 4d
  shell in xenon}.
\newblock \bibinfo{journal}{J. Phys. B} \bibinfo{volume}{44},
  \bibinfo{pages}{175602}.
\newblock \DOIprefix\doi{10.1088/0953-4075/44/17/175602}.
%Type = Book
\bibitem[{Froese-Fischer et~al.(1997)Froese-Fischer, Brage and
  J\"onsson}]{Froese-Fischer1997}
\bibinfo{author}{Froese-Fischer, C.}, \bibinfo{author}{Brage, T.},
  \bibinfo{author}{J\"onsson, P.}, \bibinfo{year}{1997}.
\newblock \bibinfo{title}{Computational Atomic Structure: An {MCHF} Approach}.
\newblock \bibinfo{publisher}{CRC Press}.
\newblock \DOIprefix\doi{10.1201/9781315139982}.
%Type = Article
\bibitem[{Fukuzawa et~al.(2010)Fukuzawa, Gryzlova, Motomura, Yamada, Ueda,
  Grum-Grzhimailo, Strakhova, Nagaya, Sugishima, Mizoguchi, Iwayama, Yao,
  Saito, Piseri, Mazza, Devetta, Coreno, Nagasono, Tono, Yabashi, Ishikawa,
  Ohashi, Kimura, Togashi and Senba}]{Fukuzawa2010}
\bibinfo{author}{Fukuzawa, H.}, \bibinfo{author}{Gryzlova, E.V.},
  \bibinfo{author}{Motomura, K.}, \bibinfo{author}{Yamada, A.},
  \bibinfo{author}{Ueda, K.}, \bibinfo{author}{Grum-Grzhimailo, A.N.},
  \bibinfo{author}{Strakhova, S.I.}, \bibinfo{author}{Nagaya, K.},
  \bibinfo{author}{Sugishima, A.}, \bibinfo{author}{Mizoguchi, Y.},
  \bibinfo{author}{Iwayama, H.}, \bibinfo{author}{Yao, M.},
  \bibinfo{author}{Saito, N.}, \bibinfo{author}{Piseri, P.},
  \bibinfo{author}{Mazza, T.}, \bibinfo{author}{Devetta, M.},
  \bibinfo{author}{Coreno, M.}, \bibinfo{author}{Nagasono, M.},
  \bibinfo{author}{Tono, K.}, \bibinfo{author}{Yabashi, M.},
  \bibinfo{author}{Ishikawa, T.}, \bibinfo{author}{Ohashi, H.},
  \bibinfo{author}{Kimura, H.}, \bibinfo{author}{Togashi, T.},
  \bibinfo{author}{Senba, Y.}, \bibinfo{year}{2010}.
\newblock \bibinfo{title}{Photoelectron spectroscopy of sequential three-photon
  double ionization of {Ar} irradiated by {EUV} free-electron laser pulses}.
\newblock \bibinfo{journal}{J. Phys. B} \bibinfo{volume}{43},
  \bibinfo{pages}{111001}.
\newblock \DOIprefix\doi{10.1088/0953-4075/43/11/111001}.
%Type = Article
\bibitem[{Fukuzawa et~al.(2013)Fukuzawa, Son, Motomura, Mondal, Nagaya, Wada,
  Liu, Feifel, Tachibana, Ito, Kimura, Sakai, Matsunami, Hayashita, Kajikawa,
  Johnsson, Siano, Kukk, Rudek, Erk, Foucar, Robert, Miron, Tono, Inubushi,
  Hatsui, Yabashi, Yao, Santra and Ueda}]{FukuzawaPRL2013}
\bibinfo{author}{Fukuzawa, H.}, \bibinfo{author}{Son, S.K.},
  \bibinfo{author}{Motomura, K.}, \bibinfo{author}{Mondal, S.},
  \bibinfo{author}{Nagaya, K.}, \bibinfo{author}{Wada, S.},
  \bibinfo{author}{Liu, X.J.}, \bibinfo{author}{Feifel, R.},
  \bibinfo{author}{Tachibana, T.}, \bibinfo{author}{Ito, Y.},
  \bibinfo{author}{Kimura, M.}, \bibinfo{author}{Sakai, T.},
  \bibinfo{author}{Matsunami, K.}, \bibinfo{author}{Hayashita, H.},
  \bibinfo{author}{Kajikawa, J.}, \bibinfo{author}{Johnsson, P.},
  \bibinfo{author}{Siano, M.}, \bibinfo{author}{Kukk, E.},
  \bibinfo{author}{Rudek, B.}, \bibinfo{author}{Erk, B.},
  \bibinfo{author}{Foucar, L.}, \bibinfo{author}{Robert, E.},
  \bibinfo{author}{Miron, C.}, \bibinfo{author}{Tono, K.},
  \bibinfo{author}{Inubushi, Y.}, \bibinfo{author}{Hatsui, T.},
  \bibinfo{author}{Yabashi, M.}, \bibinfo{author}{Yao, M.},
  \bibinfo{author}{Santra, R.}, \bibinfo{author}{Ueda, K.},
  \bibinfo{year}{2013}.
\newblock \bibinfo{title}{Deep inner-shell multiphoton ionization by intense
  {X}-ray free-electron laser pulses}.
\newblock \bibinfo{journal}{Phys. Rev. Lett.} \bibinfo{volume}{110},
  \bibinfo{pages}{173005}.
\newblock \DOIprefix\doi{10.1103/PhysRevLett.110.173005}.
%Type = Article
\bibitem[{Fukuzawa et~al.(2019)Fukuzawa, Takanashi, Kukk, Motomura, Wada,
  Nagaya, Ito, Nishiyama, Nicolas, Kumagai, Iablonskyi, Mondal, Tachibana, You,
  Yamada, Sakakibara, Asa, Sato, Sakai, Matsunami, Umemoto, Kariyazono,
  Kajimoto, Sotome, Johnsson, Sch\"{o}ffler, Kastirke, Kooser, Liu, Asavei,
  Neagu, Molodtsov, Ochiai, Kanno, Yamazaki, Owada, Ogawa, Katayama, Togashi,
  Tono, Yabashi, Ghosh, Gokhberg, Cederbaum, Kuleff, Fukumura, Kishimoto,
  Rudenko, Miron, Kono and Ueda}]{Fukuzawa2019}
\bibinfo{author}{Fukuzawa, H.}, \bibinfo{author}{Takanashi, T.},
  \bibinfo{author}{Kukk, E.}, \bibinfo{author}{Motomura, K.},
  \bibinfo{author}{Wada, S.}, \bibinfo{author}{Nagaya, K.},
  \bibinfo{author}{Ito, Y.}, \bibinfo{author}{Nishiyama, T.},
  \bibinfo{author}{Nicolas, C.}, \bibinfo{author}{Kumagai, Y.},
  \bibinfo{author}{Iablonskyi, D.}, \bibinfo{author}{Mondal, S.},
  \bibinfo{author}{Tachibana, T.}, \bibinfo{author}{You, D.},
  \bibinfo{author}{Yamada, S.}, \bibinfo{author}{Sakakibara, Y.},
  \bibinfo{author}{Asa, K.}, \bibinfo{author}{Sato, Y.},
  \bibinfo{author}{Sakai, T.}, \bibinfo{author}{Matsunami, K.},
  \bibinfo{author}{Umemoto, T.}, \bibinfo{author}{Kariyazono, K.},
  \bibinfo{author}{Kajimoto, S.}, \bibinfo{author}{Sotome, H.},
  \bibinfo{author}{Johnsson, P.}, \bibinfo{author}{Sch\"{o}ffler, M.S.},
  \bibinfo{author}{Kastirke, G.}, \bibinfo{author}{Kooser, K.},
  \bibinfo{author}{Liu, X.J.}, \bibinfo{author}{Asavei, T.},
  \bibinfo{author}{Neagu, L.}, \bibinfo{author}{Molodtsov, S.},
  \bibinfo{author}{Ochiai, K.}, \bibinfo{author}{Kanno, M.},
  \bibinfo{author}{Yamazaki, K.}, \bibinfo{author}{Owada, S.},
  \bibinfo{author}{Ogawa, K.}, \bibinfo{author}{Katayama, T.},
  \bibinfo{author}{Togashi, T.}, \bibinfo{author}{Tono, K.},
  \bibinfo{author}{Yabashi, M.}, \bibinfo{author}{Ghosh, A.},
  \bibinfo{author}{Gokhberg, K.}, \bibinfo{author}{Cederbaum, L.S.},
  \bibinfo{author}{Kuleff, A.I.}, \bibinfo{author}{Fukumura, H.},
  \bibinfo{author}{Kishimoto, N.}, \bibinfo{author}{Rudenko, A.},
  \bibinfo{author}{Miron, C.}, \bibinfo{author}{Kono, H.},
  \bibinfo{author}{Ueda, K.}, \bibinfo{year}{2019}.
\newblock \bibinfo{title}{Real-time observation of {X}-ray-induced
  intramolecular and interatomic electronic decay in {CH}$_2${I}$_2$}.
\newblock \bibinfo{journal}{Nat. Commun.} \bibinfo{volume}{10},
  \bibinfo{pages}{2186}.
\newblock \DOIprefix\doi{10.1038/s41467-019-10060-z}.
%Type = Article
\bibitem[{Fukuzawa and Ueda(2020)}]{Fukuzawa2020}
\bibinfo{author}{Fukuzawa, H.}, \bibinfo{author}{Ueda, K.},
  \bibinfo{year}{2020}.
\newblock \bibinfo{title}{{X}-ray induced ultrafast dynamics in atoms,
  molecules and clusters: experimental studies at an {X}-ray free-electron
  laser facility {SACLA} and modelling}.
\newblock \bibinfo{journal}{Advances in Physics: X}
  \DOIprefix\doi{10.1080/23746149.2020.1785327}. \bibinfo{note}{in press}.
%Type = Article
\bibitem[{Gao and Starace(1989)}]{Gao1989}
\bibinfo{author}{Gao, B.}, \bibinfo{author}{Starace, A.F.},
  \bibinfo{year}{1989}.
\newblock \bibinfo{title}{Variational principle for high-order perturbations
  with application to multiphoton processes for the {H} atom}.
\newblock \bibinfo{journal}{Phys. Rev. A} \bibinfo{volume}{39},
  \bibinfo{pages}{4550--4560}.
\newblock \DOIprefix\doi{10.1103/PhysRevA.39.4550}.
%Type = Article
\bibitem[{Gauthier et~al.(2016a)Gauthier, Allaria, Coreno, Cudin, Dacasa,
  Danailov, Demidovich, di~Mitri, Diviacco, Ferrari, Finetti, Frassetto,
  Garzella, K{\"u}nzel, Leroux, Mahieu, Mahne, Meyer, Mazza, Miotti, Penco,
  Raimondi, Rebernik~Ribi\v{c}, Richter, Roussel, Schulz, Sturari, Svetina,
  Trov{\`o}, Walker, Zangrando, Callegari, Fajardo, Poletto, Zeitoun, Giannessi
  and de~Ninno}]{Gauthier2016}
\bibinfo{author}{Gauthier, D.}, \bibinfo{author}{Allaria, E.},
  \bibinfo{author}{Coreno, M.}, \bibinfo{author}{Cudin, I.},
  \bibinfo{author}{Dacasa, H.}, \bibinfo{author}{Danailov, M.B.},
  \bibinfo{author}{Demidovich, A.}, \bibinfo{author}{di~Mitri, S.},
  \bibinfo{author}{Diviacco, B.}, \bibinfo{author}{Ferrari, E.},
  \bibinfo{author}{Finetti, P.}, \bibinfo{author}{Frassetto, F.},
  \bibinfo{author}{Garzella, D.}, \bibinfo{author}{K{\"u}nzel, S.},
  \bibinfo{author}{Leroux, V.}, \bibinfo{author}{Mahieu, B.},
  \bibinfo{author}{Mahne, N.}, \bibinfo{author}{Meyer, M.},
  \bibinfo{author}{Mazza, T.}, \bibinfo{author}{Miotti, P.},
  \bibinfo{author}{Penco, G.}, \bibinfo{author}{Raimondi, L.},
  \bibinfo{author}{Rebernik~Ribi\v{c}, P.}, \bibinfo{author}{Richter, R.},
  \bibinfo{author}{Roussel, E.}, \bibinfo{author}{Schulz, S.},
  \bibinfo{author}{Sturari, L.}, \bibinfo{author}{Svetina, C.},
  \bibinfo{author}{Trov{\`o}, M.}, \bibinfo{author}{Walker, P.A.},
  \bibinfo{author}{Zangrando, M.}, \bibinfo{author}{Callegari, C.},
  \bibinfo{author}{Fajardo, M.}, \bibinfo{author}{Poletto, L.},
  \bibinfo{author}{Zeitoun, P.}, \bibinfo{author}{Giannessi, L.},
  \bibinfo{author}{de~Ninno, G.}, \bibinfo{year}{2016}a.
\newblock \bibinfo{title}{Chirped pulse amplification in an extreme-ultraviolet
  free-electron laser}.
\newblock \bibinfo{journal}{Nat. Commun.} \bibinfo{volume}{7},
  \bibinfo{pages}{13688}.
\newblock \DOIprefix\doi{10.1038/ncomms13688}.
%Type = Article
\bibitem[{Gauthier et~al.(2016b)Gauthier, Rebernik~Ribi\v{c}, De~Ninno,
  Allaria, Cinquegrana, Danailov, Demidovich, Ferrari and
  Giannessi}]{Gauthier2016a}
\bibinfo{author}{Gauthier, D.}, \bibinfo{author}{Rebernik~Ribi\v{c}, P.},
  \bibinfo{author}{De~Ninno, G.}, \bibinfo{author}{Allaria, E.},
  \bibinfo{author}{Cinquegrana, P.}, \bibinfo{author}{Danailov, M.B.},
  \bibinfo{author}{Demidovich, A.}, \bibinfo{author}{Ferrari, E.},
  \bibinfo{author}{Giannessi, L.}, \bibinfo{year}{2016}b.
\newblock \bibinfo{title}{Generation of phase-locked pulses from a seeded
  free-electron laser}.
\newblock \bibinfo{journal}{Phys. Rev. Lett.} \bibinfo{volume}{116},
  \bibinfo{pages}{024801}.
\newblock \DOIprefix\doi{10.1103/PhysRevLett.116.024801}.
%Type = Article
\bibitem[{Gauthier et~al.(2015)Gauthier, Rebernik~Ribi\v{c}, De~Ninno, Allaria,
  Cinquegrana, Danailov, Demidovich, Ferrari, Giannessi, Mahieu and
  Penco}]{Gauthier2015}
\bibinfo{author}{Gauthier, D.}, \bibinfo{author}{Rebernik~Ribi\v{c}, P.},
  \bibinfo{author}{De~Ninno, G.}, \bibinfo{author}{Allaria, E.},
  \bibinfo{author}{Cinquegrana, P.}, \bibinfo{author}{Danailov, M.B.},
  \bibinfo{author}{Demidovich, A.}, \bibinfo{author}{Ferrari, E.},
  \bibinfo{author}{Giannessi, L.}, \bibinfo{author}{Mahieu, B.},
  \bibinfo{author}{Penco, G.}, \bibinfo{year}{2015}.
\newblock \bibinfo{title}{Spectrotemporal shaping of seeded free-electron laser
  pulses}.
\newblock \bibinfo{journal}{Phys. Rev. Lett.} \bibinfo{volume}{115},
  \bibinfo{pages}{114801}.
\newblock \DOIprefix\doi{10.1103/PhysRevLett.115.114801}.
%Type = Incollection
\bibitem[{Geloni(2016)}]{Geloni2016}
\bibinfo{author}{Geloni, G.}, \bibinfo{year}{2016}.
\newblock \bibinfo{title}{Self-seeded free-electron lasers}, in:
  \bibinfo{editor}{Jaeschke, E.J.}, \bibinfo{editor}{Khan, S.},
  \bibinfo{editor}{Schneider, J.R.}, \bibinfo{editor}{Hastings, J.B.} (Eds.),
  \bibinfo{booktitle}{Synchrotron Light Sources and Free-Electron Lasers}.
  \bibinfo{publisher}{Springer International Publishing}, pp.
  \bibinfo{pages}{161--193}.
\newblock \DOIprefix\doi{10.1007/978-3-319-14394-1_4}.
%Type = Article
\bibitem[{Geloni et~al.(2011)Geloni, Kocharyan and Saldin}]{Geloni2011}
\bibinfo{author}{Geloni, G.}, \bibinfo{author}{Kocharyan, V.},
  \bibinfo{author}{Saldin, E.}, \bibinfo{year}{2011}.
\newblock \bibinfo{title}{A novel self-seeding scheme for hard {X}-ray {FELs}}.
\newblock \bibinfo{journal}{J. Mod. Opt.} \bibinfo{volume}{58},
  \bibinfo{pages}{1391--1403}.
\newblock \DOIprefix\doi{10.1080/09500340.2011.586473}.
%Type = Article
\bibitem[{Giannessi et~al.(2018)Giannessi, Allaria, Prince, Callegari, Sansone,
  Ueda, Morishita, Liu, Grum-Grzhimailo, Gryzlova, Douguet and
  Bartschat}]{Giannessi2018}
\bibinfo{author}{Giannessi, L.}, \bibinfo{author}{Allaria, E.},
  \bibinfo{author}{Prince, K.C.}, \bibinfo{author}{Callegari, C.},
  \bibinfo{author}{Sansone, G.}, \bibinfo{author}{Ueda, K.},
  \bibinfo{author}{Morishita, T.}, \bibinfo{author}{Liu, C.N.},
  \bibinfo{author}{Grum-Grzhimailo, A.N.}, \bibinfo{author}{Gryzlova, E.V.},
  \bibinfo{author}{Douguet, N.}, \bibinfo{author}{Bartschat, K.},
  \bibinfo{year}{2018}.
\newblock \bibinfo{title}{Coherent control schemes for the photoionization of
  neon and helium in the {E}xtreme {U}ltraviolet spectral region}.
\newblock \bibinfo{journal}{Sci. Rep.} \bibinfo{volume}{8},
  \bibinfo{pages}{7774}.
\newblock \DOIprefix\doi{10.1038/s41598-018-25833-7}.
%Type = Article
\bibitem[{Giannessi et~al.(2013)Giannessi, Bellaveglia, Chiadroni, Cianchi,
  Couprie, Del~Franco, Di~Pirro, Ferrario, Gatti, Labat, Marcus, Mostacci,
  Petralia, Petrillo, Quattromini, Rau, Spampinati and
  Surrenti}]{Giannessi_PRL_2013}
\bibinfo{author}{Giannessi, L.}, \bibinfo{author}{Bellaveglia, M.},
  \bibinfo{author}{Chiadroni, E.}, \bibinfo{author}{Cianchi, A.},
  \bibinfo{author}{Couprie, M.E.}, \bibinfo{author}{Del~Franco, M.},
  \bibinfo{author}{Di~Pirro, G.}, \bibinfo{author}{Ferrario, M.},
  \bibinfo{author}{Gatti, G.}, \bibinfo{author}{Labat, M.},
  \bibinfo{author}{Marcus, G.}, \bibinfo{author}{Mostacci, A.},
  \bibinfo{author}{Petralia, A.}, \bibinfo{author}{Petrillo, V.},
  \bibinfo{author}{Quattromini, M.}, \bibinfo{author}{Rau, J.V.},
  \bibinfo{author}{Spampinati, S.}, \bibinfo{author}{Surrenti, V.},
  \bibinfo{year}{2013}.
\newblock \bibinfo{title}{Superradiant cascade in a seeded free-electron
  laser}.
\newblock \bibinfo{journal}{Phys. Rev. Lett.} \bibinfo{volume}{110},
  \bibinfo{pages}{044801}.
\newblock \DOIprefix\doi{10.1103/PhysRevLett.110.044801}.
%Type = Article
\bibitem[{Ginzburg(1947)}]{Ginzburg1947}
\bibinfo{author}{Ginzburg, V.}, \bibinfo{year}{1947}.
\newblock \bibinfo{title}{On the radiation of microradiowaves and their
  absorbtion in the air}.
\newblock \bibinfo{journal}{Izv. Akad. Nauk SSSR, Ser. Fiz.}
  \bibinfo{volume}{11}, \bibinfo{pages}{165}.
%Type = Article
\bibitem[{Glauber(1963)}]{Glauber1963}
\bibinfo{author}{Glauber, R.J.}, \bibinfo{year}{1963}.
\newblock \bibinfo{title}{The quantum theory of optical coherence}.
\newblock \bibinfo{journal}{Phys. Rev.} \bibinfo{volume}{130},
  \bibinfo{pages}{2529--2539}.
\newblock \DOIprefix\doi{10.1103/PhysRev.130.2529}.
%Type = Article
\bibitem[{Goetz et~al.(2016)Goetz, Karamatskou, Santra and Koch}]{Goetz2016}
\bibinfo{author}{Goetz, R.E.}, \bibinfo{author}{Karamatskou, A.},
  \bibinfo{author}{Santra, R.}, \bibinfo{author}{Koch, C.P.},
  \bibinfo{year}{2016}.
\newblock \bibinfo{title}{Quantum optimal control of photoelectron spectra and
  angular distributions}.
\newblock \bibinfo{journal}{Phys. Rev. A} \bibinfo{volume}{93},
  \bibinfo{pages}{013413}.
\newblock \DOIprefix\doi{10.1103/PhysRevA.93.013413}.
%Type = Article
\bibitem[{Gordon et~al.(1954)Gordon, Zeiger and Townes}]{Gordon1954}
\bibinfo{author}{Gordon, J.P.}, \bibinfo{author}{Zeiger, H.J.},
  \bibinfo{author}{Townes, C.H.}, \bibinfo{year}{1954}.
\newblock \bibinfo{title}{Molecular microwave oscillator and new hyperfine
  structure in the microwave spectrum of {NH}$_{3}$}.
\newblock \bibinfo{journal}{Phys. Rev.} \bibinfo{volume}{95},
  \bibinfo{pages}{282--284}.
\newblock \DOIprefix\doi{10.1103/PhysRev.95.282}.
%Type = Article
\bibitem[{Gordon et~al.(1955)Gordon, Zeiger and Townes}]{Gordon1955}
\bibinfo{author}{Gordon, J.P.}, \bibinfo{author}{Zeiger, H.J.},
  \bibinfo{author}{Townes, C.H.}, \bibinfo{year}{1955}.
\newblock \bibinfo{title}{The maser---new type of microwave amplifier,
  frequency standard, and spectrometer}.
\newblock \bibinfo{journal}{Phys. Rev.} \bibinfo{volume}{99},
  \bibinfo{pages}{1264--1274}.
\newblock \DOIprefix\doi{10.1103/PhysRev.99.1264}.
%Type = Article
\bibitem[{Gordon and Rice(1997)}]{Gordon1997}
\bibinfo{author}{Gordon, R.J.}, \bibinfo{author}{Rice, S.A.},
  \bibinfo{year}{1997}.
\newblock \bibinfo{title}{Active control of the dynamics of atoms and
  molecules}.
\newblock \bibinfo{journal}{Annu. Rev. Phys. Chem.} \bibinfo{volume}{48},
  \bibinfo{pages}{601--641}.
\newblock \DOIprefix\doi{10.1146/annurev.physchem.48.1.601}.
%Type = Article
\bibitem[{Goreslavski et~al.(2004)Goreslavski, Paulus, Popruzhenko and
  Shvetsov-Shilovski}]{Goreslavski2004}
\bibinfo{author}{Goreslavski, S.P.}, \bibinfo{author}{Paulus, G.G.},
  \bibinfo{author}{Popruzhenko, S.V.}, \bibinfo{author}{Shvetsov-Shilovski,
  N.I.}, \bibinfo{year}{2004}.
\newblock \bibinfo{title}{Coulomb asymmetry in above-threshold ionization}.
\newblock \bibinfo{journal}{Phys. Rev. Lett.} \bibinfo{volume}{93},
  \bibinfo{pages}{233002}.
\newblock \DOIprefix\doi{10.1103/PhysRevLett.93.233002}.
%Type = Article
\bibitem[{Gorkhover et~al.(2016)Gorkhover, Schorb, Coffee, Adolph, Foucar,
  Rupp, Aquila, Bozek, Epp, Erk, Gumprecht, Holmegaard, Hartmann, Hartmann,
  Hauser, Holl, H\"omke, Johnsson, Kimmel, K\"uhnel, Messerschmidt, Reich,
  Rouz{\'{e}}e, Rudek, Schmidt, Schulz, Soltau, Stern, Weidenspointner, White,
  K\"upper, Str\"uder, Schlichting, Ullrich, Rolles, Rudenko, M\"oller and
  Bostedt}]{Gorkhover2016}
\bibinfo{author}{Gorkhover, T.}, \bibinfo{author}{Schorb, S.},
  \bibinfo{author}{Coffee, R.}, \bibinfo{author}{Adolph, M.},
  \bibinfo{author}{Foucar, L.}, \bibinfo{author}{Rupp, D.},
  \bibinfo{author}{Aquila, A.}, \bibinfo{author}{Bozek, J.D.},
  \bibinfo{author}{Epp, S.W.}, \bibinfo{author}{Erk, B.},
  \bibinfo{author}{Gumprecht, L.}, \bibinfo{author}{Holmegaard, L.},
  \bibinfo{author}{Hartmann, A.}, \bibinfo{author}{Hartmann, R.},
  \bibinfo{author}{Hauser, G.}, \bibinfo{author}{Holl, P.},
  \bibinfo{author}{H\"omke, A.}, \bibinfo{author}{Johnsson, P.},
  \bibinfo{author}{Kimmel, N.}, \bibinfo{author}{K\"uhnel, K.},
  \bibinfo{author}{Messerschmidt, M.}, \bibinfo{author}{Reich, C.},
  \bibinfo{author}{Rouz{\'{e}}e, A.}, \bibinfo{author}{Rudek, B.},
  \bibinfo{author}{Schmidt, C.}, \bibinfo{author}{Schulz, J.},
  \bibinfo{author}{Soltau, H.}, \bibinfo{author}{Stern, S.},
  \bibinfo{author}{Weidenspointner, G.}, \bibinfo{author}{White, B.},
  \bibinfo{author}{K\"upper, J.}, \bibinfo{author}{Str\"uder, L.},
  \bibinfo{author}{Schlichting, I.}, \bibinfo{author}{Ullrich, J.},
  \bibinfo{author}{Rolles, D.}, \bibinfo{author}{Rudenko, A.},
  \bibinfo{author}{M\"oller, T.}, \bibinfo{author}{Bostedt, C.},
  \bibinfo{year}{2016}.
\newblock \bibinfo{title}{Femtosecond and nanometre visualization of structural
  dynamics in superheated nanoparticles}.
\newblock \bibinfo{journal}{Nat. Photonics} \bibinfo{volume}{10},
  \bibinfo{pages}{93--97}.
\newblock \DOIprefix\doi{10.1038/nphoton.2015.264}.
%Type = Article
\bibitem[{Gorobtsov et~al.(2018)Gorobtsov, Mercurio, Capotondi, Skopintsev,
  Lazarev, Zaluzhnyy, Danailov, Dell'Angela, Manfredda, Pedersoli, Giannessi,
  Kiskinova, Prince, Wurth and Vartanyants}]{Gorobtsov2018}
\bibinfo{author}{Gorobtsov, O.Y.}, \bibinfo{author}{Mercurio, G.},
  \bibinfo{author}{Capotondi, F.}, \bibinfo{author}{Skopintsev, P.},
  \bibinfo{author}{Lazarev, S.}, \bibinfo{author}{Zaluzhnyy, I.A.},
  \bibinfo{author}{Danailov, M.B.}, \bibinfo{author}{Dell'Angela, M.},
  \bibinfo{author}{Manfredda, M.}, \bibinfo{author}{Pedersoli, E.},
  \bibinfo{author}{Giannessi, L.}, \bibinfo{author}{Kiskinova, M.},
  \bibinfo{author}{Prince, K.C.}, \bibinfo{author}{Wurth, W.},
  \bibinfo{author}{Vartanyants, I.A.}, \bibinfo{year}{2018}.
\newblock \bibinfo{title}{Seeded {X}-ray free-electron laser generating
  radiation with laser statistical properties}.
\newblock \bibinfo{journal}{Nat. Commun.} \bibinfo{volume}{9},
  \bibinfo{pages}{4498}.
\newblock \DOIprefix\doi{10.1038/s41467-018-06743-8}.
%Type = Article
\bibitem[{Goswami(2003)}]{Goswami2003}
\bibinfo{author}{Goswami, D.}, \bibinfo{year}{2003}.
\newblock \bibinfo{title}{Optical pulse shaping approaches to coherent
  control}.
\newblock \bibinfo{journal}{Phys. Rep.} \bibinfo{volume}{374},
  \bibinfo{pages}{385--481}.
\newblock \DOIprefix\doi{10.1016/s0370-1573(02)00480-5}.
%Type = Article
\bibitem[{Goulielmakis(2004)}]{Goulielmakis2004}
\bibinfo{author}{Goulielmakis, E.}, \bibinfo{year}{2004}.
\newblock \bibinfo{title}{Direct measurement of light waves}.
\newblock \bibinfo{journal}{Science} \bibinfo{volume}{305},
  \bibinfo{pages}{1267--1269}.
\newblock \DOIprefix\doi{10.1126/science.1100866}.
%Type = Article
\bibitem[{Greene and Zare(1982)}]{Greene_ARPC_1982}
\bibinfo{author}{Greene, C.H.}, \bibinfo{author}{Zare, R.N.},
  \bibinfo{year}{1982}.
\newblock \bibinfo{title}{Photofragment alignment and orientation}.
\newblock \bibinfo{journal}{Annu. Rev. Phys. Chem.} \bibinfo{volume}{33},
  \bibinfo{pages}{119--150}.
\newblock \DOIprefix\doi{10.1146/annurev.pc.33.100182.001003}.
%Type = Article
\bibitem[{Greenman et~al.(2010)Greenman, Ho, Pabst, Kamarchik, Mazziotti and
  Santra}]{Greenman2010}
\bibinfo{author}{Greenman, L.}, \bibinfo{author}{Ho, P.J.},
  \bibinfo{author}{Pabst, S.}, \bibinfo{author}{Kamarchik, E.},
  \bibinfo{author}{Mazziotti, D.A.}, \bibinfo{author}{Santra, R.},
  \bibinfo{year}{2010}.
\newblock \bibinfo{title}{Implementation of the time-dependent
  configuration-interaction singles method for atomic strong-field processes}.
\newblock \bibinfo{journal}{Phys. Rev. A} \bibinfo{volume}{82},
  \bibinfo{pages}{023406}.
\newblock \DOIprefix\doi{10.1103/PhysRevA.82.023406}.
%Type = Article
\bibitem[{Grum-Grzhimailo et~al.(2019)Grum-Grzhimailo, Douguet, Meyer and
  Bartschat}]{Grum2019}
\bibinfo{author}{Grum-Grzhimailo, A.N.}, \bibinfo{author}{Douguet, N.},
  \bibinfo{author}{Meyer, M.}, \bibinfo{author}{Bartschat, K.},
  \bibinfo{year}{2019}.
\newblock \bibinfo{title}{Two-color {XUV} plus near-{IR} multiphoton
  near-threshold ionization of the helium ion by circularly polarized light in
  the vicinity of the {$3p$} resonance}.
\newblock \bibinfo{journal}{Phys. Rev. A} \bibinfo{volume}{100},
  \bibinfo{pages}{033404}.
\newblock \DOIprefix\doi{10.1103/PhysRevA.100.033404}.
%Type = Article
\bibitem[{Grum-Grzhimailo and Gryzlova(2014)}]{Grum2014}
\bibinfo{author}{Grum-Grzhimailo, A.N.}, \bibinfo{author}{Gryzlova, E.V.},
  \bibinfo{year}{2014}.
\newblock \bibinfo{title}{Nondipole effects in the angular distribution of
  photoelectrons in two-photon two-color above-threshold atomic ionization}.
\newblock \bibinfo{journal}{Phys. Rev. A} \bibinfo{volume}{89},
  \bibinfo{pages}{043424}.
\newblock \DOIprefix\doi{10.1103/PhysRevA.89.043424}.
%Type = Article
\bibitem[{Grum-Grzhimailo et~al.(2016)Grum-Grzhimailo, Gryzlova, Fritzsche and
  Kabachnik}]{Grum2016}
\bibinfo{author}{Grum-Grzhimailo, A.N.}, \bibinfo{author}{Gryzlova, E.V.},
  \bibinfo{author}{Fritzsche, S.}, \bibinfo{author}{Kabachnik, N.M.},
  \bibinfo{year}{2016}.
\newblock \bibinfo{title}{Photoelectron angular distributions and correlations
  in sequential double and triple atomic ionization by free electron lasers}.
\newblock \bibinfo{journal}{J. Mod. Opt.} \bibinfo{volume}{60},
  \bibinfo{pages}{334--357}.
\newblock \DOIprefix\doi{10.1080/09500340.2015.1047805}.
%Type = Article
\bibitem[{Grum-Grzhimailo et~al.(2015a)Grum-Grzhimailo, Gryzlova, Kuzmina,
  Chetverkina and Strakhova}]{Grum2015a}
\bibinfo{author}{Grum-Grzhimailo, A.N.}, \bibinfo{author}{Gryzlova, E.V.},
  \bibinfo{author}{Kuzmina, E.I.}, \bibinfo{author}{Chetverkina, A.S.},
  \bibinfo{author}{Strakhova, S.I.}, \bibinfo{year}{2015}a.
\newblock \bibinfo{title}{Two-color above-threshold and two-photon sequential
  double ionization beyond the dipole approximation}.
\newblock \bibinfo{journal}{J. Phys. Conf. Ser.} \bibinfo{volume}{601},
  \bibinfo{pages}{012012}.
\newblock \DOIprefix\doi{10.1088/1742-6596/601/1/012012}.
%Type = Article
\bibitem[{Grum-Grzhimailo et~al.(2012)Grum-Grzhimailo, Gryzlova and
  Meyer}]{Grum2012}
\bibinfo{author}{Grum-Grzhimailo, A.N.}, \bibinfo{author}{Gryzlova, E.V.},
  \bibinfo{author}{Meyer, M.}, \bibinfo{year}{2012}.
\newblock \bibinfo{title}{Non-dipole effects in the angular distribution of
  photoelectrons in sequential two-photon atomic double ionization}.
\newblock \bibinfo{journal}{J. Phys. B} \bibinfo{volume}{45},
  \bibinfo{pages}{215602}.
\newblock \DOIprefix\doi{10.1088/0953-4075/45/21/215602}.
%Type = Article
\bibitem[{Grum-Grzhimailo et~al.(2015b)Grum-Grzhimailo, Gryzlova,
  Staroselskaya, Venzke and Bartschat}]{Grum2015}
\bibinfo{author}{Grum-Grzhimailo, A.N.}, \bibinfo{author}{Gryzlova, E.V.},
  \bibinfo{author}{Staroselskaya, E.I.}, \bibinfo{author}{Venzke, J.},
  \bibinfo{author}{Bartschat, K.}, \bibinfo{year}{2015}b.
\newblock \bibinfo{title}{Interfering one-photon and two-photon ionization by
  femtosecond {VUV} pulses in the region of an intermediate resonance}.
\newblock \bibinfo{journal}{Phys. Rev. A} \bibinfo{volume}{91},
  \bibinfo{pages}{063418}.
\newblock \DOIprefix\doi{10.1103/PhysRevA.91.063418}.
%Type = Article
\bibitem[{Grum-Grzhimailo and Meyer(2009)}]{GrumGrzhimailo2009}
\bibinfo{author}{Grum-Grzhimailo, A.N.}, \bibinfo{author}{Meyer, M.},
  \bibinfo{year}{2009}.
\newblock \bibinfo{title}{Magnetic dichroism in atomic core level
  photoemission}.
\newblock \bibinfo{journal}{Eur. Phys. J ST} \bibinfo{volume}{169},
  \bibinfo{pages}{43--50}.
\newblock \DOIprefix\doi{10.1140/epjst/e2009-00971-2}.
%Type = Article
\bibitem[{Gruson et~al.(2016)Gruson, Barreau, Jim{\'{e}}nez-Galan, Risoud,
  Caillat, Maquet, Carr{\'{e}}, Lepetit, Hergott, Ruchon, Argenti, Ta\"{\i}eb,
  Mart{\'{\i}}n and Sali{\`{e}}res}]{Gruson2016}
\bibinfo{author}{Gruson, V.}, \bibinfo{author}{Barreau, L.},
  \bibinfo{author}{Jim{\'{e}}nez-Galan, {\'{A}}.}, \bibinfo{author}{Risoud,
  F.}, \bibinfo{author}{Caillat, J.}, \bibinfo{author}{Maquet, A.},
  \bibinfo{author}{Carr{\'{e}}, B.}, \bibinfo{author}{Lepetit, F.},
  \bibinfo{author}{Hergott, J.F.}, \bibinfo{author}{Ruchon, T.},
  \bibinfo{author}{Argenti, L.}, \bibinfo{author}{Ta\"{\i}eb, R.},
  \bibinfo{author}{Mart{\'{\i}}n, F.}, \bibinfo{author}{Sali{\`{e}}res, P.},
  \bibinfo{year}{2016}.
\newblock \bibinfo{title}{Attosecond dynamics through a {F}ano resonance:
  Monitoring the birth of a photoelectron}.
\newblock \bibinfo{journal}{Science} \bibinfo{volume}{354},
  \bibinfo{pages}{734--738}.
\newblock \DOIprefix\doi{10.1126/science.aah5188}.
%Type = Article
\bibitem[{Gryzlova et~al.(2010)Gryzlova, Grum-Grzhimailo, Fritzsche and
  Kabachnik}]{Gryzlova2010}
\bibinfo{author}{Gryzlova, E.V.}, \bibinfo{author}{Grum-Grzhimailo, A.N.},
  \bibinfo{author}{Fritzsche, S.}, \bibinfo{author}{Kabachnik, N.M.},
  \bibinfo{year}{2010}.
\newblock \bibinfo{title}{Angular correlations between two electrons emitted in
  the sequential two-photon double ionization of atoms}.
\newblock \bibinfo{journal}{J. Phys. B} \bibinfo{volume}{43},
  \bibinfo{pages}{225602}.
\newblock \DOIprefix\doi{10.1088/0953-4075/43/22/225602}.
%Type = Article
\bibitem[{Gryzlova et~al.(2012)Gryzlova, Grum-Grzhimailo, Kabachnik and
  Fritzsche}]{Gryzlova2012}
\bibinfo{author}{Gryzlova, E.V.}, \bibinfo{author}{Grum-Grzhimailo, A.N.},
  \bibinfo{author}{Kabachnik, N.M.}, \bibinfo{author}{Fritzsche, S.},
  \bibinfo{year}{2012}.
\newblock \bibinfo{title}{Angular distributions and correlations in sequential
  three-photon triple atomic ionization}.
\newblock \bibinfo{journal}{J. Phys. Conf. Ser.} \bibinfo{volume}{388},
  \bibinfo{pages}{012031}.
\newblock \DOIprefix\doi{10.1088/1742-6596/388/1/012031}.
%Type = Article
\bibitem[{Gryzlova et~al.(2019a)Gryzlova, Grum-Grzhimailo, Kiselev and
  Burkov}]{Gryzlova2019a}
\bibinfo{author}{Gryzlova, E.V.}, \bibinfo{author}{Grum-Grzhimailo, A.N.},
  \bibinfo{author}{Kiselev, M.D.}, \bibinfo{author}{Burkov, S.M.},
  \bibinfo{year}{2019}a.
\newblock \bibinfo{title}{Two-photon sequential double ionization of argon in
  the region of {R}ydberg autoionizing states of {Ar}$^+$}.
\newblock \bibinfo{journal}{Eur. Phys. J. D} \bibinfo{volume}{73},
  \bibinfo{pages}{93}.
\newblock \DOIprefix\doi{10.1140/epjd/e2019-90678-x}.
%Type = Article
\bibitem[{Gryzlova et~al.(2014)Gryzlova, Grum-Grzhimailo, Kuzmina and
  Strakhova}]{Gryzlova2014}
\bibinfo{author}{Gryzlova, E.V.}, \bibinfo{author}{Grum-Grzhimailo, A.N.},
  \bibinfo{author}{Kuzmina, E.I.}, \bibinfo{author}{Strakhova, S.I.},
  \bibinfo{year}{2014}.
\newblock \bibinfo{title}{Sequential two-photon double ionization of noble
  gases by circularly polarized {XUV} radiation}.
\newblock \bibinfo{journal}{J. Phys. B} \bibinfo{volume}{47},
  \bibinfo{pages}{195601}.
\newblock \DOIprefix\doi{10.1088/0953-4075/47/19/195601}.
%Type = Article
\bibitem[{Gryzlova et~al.(2018)Gryzlova, Grum-Grzhimailo, Staroselskaya,
  Douguet and Bartschat}]{Gryzlova2018}
\bibinfo{author}{Gryzlova, E.V.}, \bibinfo{author}{Grum-Grzhimailo, A.N.},
  \bibinfo{author}{Staroselskaya, E.I.}, \bibinfo{author}{Douguet, N.},
  \bibinfo{author}{Bartschat, K.}, \bibinfo{year}{2018}.
\newblock \bibinfo{title}{Quantum coherent control of the photoelectron angular
  distribution in bichromatic-field ionization of atomic neon}.
\newblock \bibinfo{journal}{Phys. Rev. A} \bibinfo{volume}{97},
  \bibinfo{pages}{013420}.
\newblock \DOIprefix\doi{10.1103/PhysRevA.97.013420}.
%Type = Article
\bibitem[{Gryzlova et~al.(2015)Gryzlova, Grum-Grzhimailo, Staroselskaya and
  Strakhova}]{Gryzlova2015}
\bibinfo{author}{Gryzlova, E.V.}, \bibinfo{author}{Grum-Grzhimailo, A.N.},
  \bibinfo{author}{Staroselskaya, E.I.}, \bibinfo{author}{Strakhova, S.I.},
  \bibinfo{year}{2015}.
\newblock \bibinfo{title}{Similarity between the angular distributions of the
  first- and second-step electrons in sequential two-photon atomic double
  ionization}.
\newblock \bibinfo{journal}{J. Electron Spectrosc. Relat. Phenom.}
  \bibinfo{volume}{204}, \bibinfo{pages}{277--283}.
\newblock \DOIprefix\doi{10.1016/j.elspec.2015.08.016}.
%Type = Article
\bibitem[{Gryzlova et~al.(2011)Gryzlova, Ma, Fukuzawa, Motomura, Yamada, Ueda,
  Grum-Grzhimailo, Kabachnik, Strakhova, Rouz\'ee, Hundermark, Vrakking,
  Johnsson, Nagaya, Yase, Mizoguchi, Yao, Nagasono, Tono, Togashi, Senba,
  Ohashi, Yabashi and Ishikawa}]{Gryzlova2011}
\bibinfo{author}{Gryzlova, E.V.}, \bibinfo{author}{Ma, R.},
  \bibinfo{author}{Fukuzawa, H.}, \bibinfo{author}{Motomura, K.},
  \bibinfo{author}{Yamada, A.}, \bibinfo{author}{Ueda, K.},
  \bibinfo{author}{Grum-Grzhimailo, A.N.}, \bibinfo{author}{Kabachnik, N.M.},
  \bibinfo{author}{Strakhova, S.I.}, \bibinfo{author}{Rouz\'ee, A.},
  \bibinfo{author}{Hundermark, A.}, \bibinfo{author}{Vrakking, M.J.J.},
  \bibinfo{author}{Johnsson, P.}, \bibinfo{author}{Nagaya, K.},
  \bibinfo{author}{Yase, S.}, \bibinfo{author}{Mizoguchi, Y.},
  \bibinfo{author}{Yao, M.}, \bibinfo{author}{Nagasono, M.},
  \bibinfo{author}{Tono, K.}, \bibinfo{author}{Togashi, T.},
  \bibinfo{author}{Senba, Y.}, \bibinfo{author}{Ohashi, H.},
  \bibinfo{author}{Yabashi, M.}, \bibinfo{author}{Ishikawa, T.},
  \bibinfo{year}{2011}.
\newblock \bibinfo{title}{Doubly resonant three-photon double ionization of
  {Ar} atoms induced by an {EUV} free-electron laser}.
\newblock \bibinfo{journal}{Phys. Rev. A} \bibinfo{volume}{84},
  \bibinfo{pages}{063405}.
\newblock \DOIprefix\doi{10.1103/PhysRevA.84.063405}.
%Type = Article
\bibitem[{Gryzlova et~al.(2019b)Gryzlova, Popova, Grum-Grzhimailo,
  Staroselskaya, Douguet and Bartschat}]{Gryzlova2019}
\bibinfo{author}{Gryzlova, E.V.}, \bibinfo{author}{Popova, M.M.},
  \bibinfo{author}{Grum-Grzhimailo, A.N.}, \bibinfo{author}{Staroselskaya,
  E.I.}, \bibinfo{author}{Douguet, N.}, \bibinfo{author}{Bartschat, K.},
  \bibinfo{year}{2019}b.
\newblock \bibinfo{title}{Coherent control of the photoelectron angular
  distribution in ionization of neon by a circularly polarized bichromatic
  field in the resonance region}.
\newblock \bibinfo{journal}{Phys. Rev. A} \bibinfo{volume}{100},
  \bibinfo{pages}{063417}.
\newblock \DOIprefix\doi{10.1103/PhysRevA.100.063417}.
%Type = Article
\bibitem[{Guan et~al.(2011)Guan, Bartschat and Schneider}]{Guan2011}
\bibinfo{author}{Guan, X.}, \bibinfo{author}{Bartschat, K.},
  \bibinfo{author}{Schneider, B.I.}, \bibinfo{year}{2011}.
\newblock \bibinfo{title}{Breakup of the aligned {H}$_2$ molecule by xuv laser
  pulses: A time-dependent treatment in prolate spheroidal coordinates}.
\newblock \bibinfo{journal}{Phys. Rev. A} \bibinfo{volume}{83},
  \bibinfo{pages}{043403}.
\newblock \DOIprefix\doi{10.1103/PhysRevA.83.043403}.
%Type = Article
\bibitem[{Guan et~al.(2008)Guan, Noble, Zatsarinny, Bartschat and
  Schneider}]{Guan2008}
\bibinfo{author}{Guan, X.}, \bibinfo{author}{Noble, C.J.},
  \bibinfo{author}{Zatsarinny, O.}, \bibinfo{author}{Bartschat, K.},
  \bibinfo{author}{Schneider, B.I.}, \bibinfo{year}{2008}.
\newblock \bibinfo{title}{Time-dependent {R}-matrix calculations for
  multiphoton ionization of argon atoms in strong laser pulses}.
\newblock \bibinfo{journal}{Phys. Rev. A} \bibinfo{volume}{78},
  \bibinfo{pages}{053402}.
\newblock \DOIprefix\doi{10.1103/PhysRevA.78.053402}.
%Type = Article
\bibitem[{Guan et~al.(2007)Guan, Zatsarinny, Bartschat, Schneider, Feist and
  Noble}]{Guan2007}
\bibinfo{author}{Guan, X.}, \bibinfo{author}{Zatsarinny, O.},
  \bibinfo{author}{Bartschat, K.}, \bibinfo{author}{Schneider, B.I.},
  \bibinfo{author}{Feist, J.}, \bibinfo{author}{Noble, C.J.},
  \bibinfo{year}{2007}.
\newblock \bibinfo{title}{General approach to few-cycle intense laser
  interactions with complex atoms}.
\newblock \bibinfo{journal}{Phys. Rev. A} \bibinfo{volume}{76},
  \bibinfo{pages}{053411}.
\newblock \DOIprefix\doi{10.1103/PhysRevA.76.053411}.
%Type = Incollection
\bibitem[{Haake(1973)}]{Haake1973}
\bibinfo{author}{Haake, F.}, \bibinfo{year}{1973}.
\newblock \bibinfo{title}{Statistical treatment of open systems by generalized
  master equations}, in: \bibinfo{editor}{H\"ohler, G.} (Ed.),
  \bibinfo{booktitle}{Quantum Statistics in Optical and Solid-State Physics}.
  \bibinfo{publisher}{Springer-Verlag}. volume~\bibinfo{volume}{66} of
  \textit{\bibinfo{series}{Springer Tracts in Modern Physics}}, pp.
  \bibinfo{pages}{98--168}.
\newblock \DOIprefix\doi{10.1007/978-3-662-40468-3_2}.
%Type = Article
\bibitem[{Haber et~al.(2017)Haber, Kong, Strohm, Willing, Gollwitzer, Bocklage,
  R\"uffer, P{\'{a}}lffy and R\"ohlsberger}]{Haber2017}
\bibinfo{author}{Haber, J.}, \bibinfo{author}{Kong, X.},
  \bibinfo{author}{Strohm, C.}, \bibinfo{author}{Willing, S.},
  \bibinfo{author}{Gollwitzer, J.}, \bibinfo{author}{Bocklage, L.},
  \bibinfo{author}{R\"uffer, R.}, \bibinfo{author}{P{\'{a}}lffy, A.},
  \bibinfo{author}{R\"ohlsberger, R.}, \bibinfo{year}{2017}.
\newblock \bibinfo{title}{Rabi oscillations of {X}-ray radiation between two
  nuclear ensembles}.
\newblock \bibinfo{journal}{Nat. Photonics} \bibinfo{volume}{11},
  \bibinfo{pages}{720--725}.
\newblock \DOIprefix\doi{10.1038/s41566-017-0013-3}.
%Type = Article
\bibitem[{Haber et~al.(2009)Haber, Doughty and Leone}]{Haber2009}
\bibinfo{author}{Haber, L.H.}, \bibinfo{author}{Doughty, B.},
  \bibinfo{author}{Leone, S.R.}, \bibinfo{year}{2009}.
\newblock \bibinfo{title}{Continuum phase shifts and partial cross sections for
  photoionization from excited states of atomic helium measured by high-order
  harmonic optical pump-probe velocity map imaging}.
\newblock \bibinfo{journal}{Phys. Rev. A} \bibinfo{volume}{79},
  \bibinfo{pages}{031401}.
\newblock \DOIprefix\doi{10.1103/PhysRevA.79.031401}.
%Type = Article
\bibitem[{Haber et~al.(2010)Haber, Doughty and Leone}]{Haber2010}
\bibinfo{author}{Haber, L.H.}, \bibinfo{author}{Doughty, B.},
  \bibinfo{author}{Leone, S.R.}, \bibinfo{year}{2010}.
\newblock \bibinfo{title}{Time-resolved photoelectron angular distributions and
  cross-section ratios of two-colour two-photon above threshold ionization of
  helium}.
\newblock \bibinfo{journal}{Molecular Physics} \bibinfo{volume}{108},
  \bibinfo{pages}{1241--1251}.
\newblock \DOIprefix\doi{10.1080/00268976.2010.483133}.
%Type = Article
\bibitem[{Halavanau et~al.(2019)Halavanau, Decker, Emma, Sheppard and
  Pellegrini}]{Halavanau2019}
\bibinfo{author}{Halavanau, A.}, \bibinfo{author}{Decker, F.J.},
  \bibinfo{author}{Emma, C.}, \bibinfo{author}{Sheppard, J.},
  \bibinfo{author}{Pellegrini, C.}, \bibinfo{year}{2019}.
\newblock \bibinfo{title}{Very high brightness and power {LCLS}-{II} hard
  {X}-ray pulses}.
\newblock \bibinfo{journal}{J. Synchrotron Radiat.} \bibinfo{volume}{26},
  \bibinfo{pages}{635--646}.
\newblock \DOIprefix\doi{10.1107/s1600577519002492}.
%Type = Article
\bibitem[{Han et~al.(2019)Han, Zhang and E}]{Han2019}
\bibinfo{author}{Han, J.}, \bibinfo{author}{Zhang, L.}, \bibinfo{author}{E,
  W.}, \bibinfo{year}{2019}.
\newblock \bibinfo{title}{Solving many-electron {S}chr\"odinger equation using
  deep neural networks}.
\newblock \bibinfo{journal}{J. Comp. Phys.} \bibinfo{volume}{399},
  \bibinfo{pages}{108929}.
\newblock \DOIprefix\doi{10.1016/j.jcp.2019.108929}.
%Type = Article
\bibitem[{Hanbury~Brown and Twiss(1956a)}]{HBT1956}
\bibinfo{author}{Hanbury~Brown, R.}, \bibinfo{author}{Twiss, R.Q.},
  \bibinfo{year}{1956}a.
\newblock \bibinfo{title}{Correlation between photons in two coherent beams of
  light}.
\newblock \bibinfo{journal}{Nature} \bibinfo{volume}{177},
  \bibinfo{pages}{27--29}.
\newblock \DOIprefix\doi{10.1038/177027a0}.
%Type = Article
\bibitem[{Hanbury~Brown and Twiss(1956b)}]{HBT1956a}
\bibinfo{author}{Hanbury~Brown, R.}, \bibinfo{author}{Twiss, R.Q.},
  \bibinfo{year}{1956}b.
\newblock \bibinfo{title}{A test of a new type of stellar interferometer on
  {Si}rius}.
\newblock \bibinfo{journal}{Nature} \bibinfo{volume}{178},
  \bibinfo{pages}{1046--1048}.
\newblock \DOIprefix\doi{10.1038/1781046a0}.
%Type = Article
\bibitem[{Hanson et~al.(1997)Hanson, Zhang and Lambropoulos}]{Hanson1997}
\bibinfo{author}{Hanson, L.G.}, \bibinfo{author}{Zhang, J.},
  \bibinfo{author}{Lambropoulos, P.}, \bibinfo{year}{1997}.
\newblock \bibinfo{title}{Manifestations of atomic and core resonances in
  photoelectron energy spectra}.
\newblock \bibinfo{journal}{Phys. Rev. A} \bibinfo{volume}{55},
  \bibinfo{pages}{2232--2244}.
\newblock \DOIprefix\doi{10.1103/PhysRevA.55.2232}.
%Type = Article
\bibitem[{Harries et~al.(2018)Harries, Iwayama, Kuma, Iizawa, Suzuki, Azuma,
  Inoue, Owada, Togashi, Tono, Yabashi and Shigemasa}]{Harries2018}
\bibinfo{author}{Harries, J.R.}, \bibinfo{author}{Iwayama, H.},
  \bibinfo{author}{Kuma, S.}, \bibinfo{author}{Iizawa, M.},
  \bibinfo{author}{Suzuki, N.}, \bibinfo{author}{Azuma, Y.},
  \bibinfo{author}{Inoue, I.}, \bibinfo{author}{Owada, S.},
  \bibinfo{author}{Togashi, T.}, \bibinfo{author}{Tono, K.},
  \bibinfo{author}{Yabashi, M.}, \bibinfo{author}{Shigemasa, E.},
  \bibinfo{year}{2018}.
\newblock \bibinfo{title}{Superfluorescence, free-induction decay, and
  four-wave mixing: Propagation of free-electron laser pulses through a dense
  sample of helium ions}.
\newblock \bibinfo{journal}{Phys. Rev. Lett.} \bibinfo{volume}{121},
  \bibinfo{pages}{263201}.
\newblock \DOIprefix\doi{10.1103/PhysRevLett.121.263201}.
%Type = Article
\bibitem[{van~der Hart et~al.(2007)van~der Hart, Lysaght and
  Burke}]{vanderHart2007}
\bibinfo{author}{van~der Hart, H.W.}, \bibinfo{author}{Lysaght, M.A.},
  \bibinfo{author}{Burke, P.G.}, \bibinfo{year}{2007}.
\newblock \bibinfo{title}{Time-dependent multielectron dynamics of {A}r in
  intense short laser pulses}.
\newblock \bibinfo{journal}{Phys. Rev. A} \bibinfo{volume}{76},
  \bibinfo{pages}{043405}.
\newblock \DOIprefix\doi{10.1103/PhysRevA.76.043405}.
%Type = Article
\bibitem[{van~der Hart et~al.(2008)van~der Hart, Lysaght and
  Burke}]{vanderHart2008}
\bibinfo{author}{van~der Hart, H.W.}, \bibinfo{author}{Lysaght, M.A.},
  \bibinfo{author}{Burke, P.G.}, \bibinfo{year}{2008}.
\newblock \bibinfo{title}{Momentum distributions of electrons ejected during
  ultrashort laser interactions with multielectron atoms described using the
  {R}-matrix basis sets}.
\newblock \bibinfo{journal}{Phys. Rev. A} \bibinfo{volume}{77},
  \bibinfo{pages}{065401}.
\newblock \DOIprefix\doi{10.1103/PhysRevA.77.065401}.
%Type = Article
\bibitem[{Haxton et~al.(2011)Haxton, Lawler and McCurdy}]{Haxton2011}
\bibinfo{author}{Haxton, D.J.}, \bibinfo{author}{Lawler, K.V.},
  \bibinfo{author}{McCurdy, C.W.}, \bibinfo{year}{2011}.
\newblock \bibinfo{title}{Multiconfiguration time-dependent {H}artree-{F}ock
  treatment of electronic and nuclear dynamics in diatomic molecules}.
\newblock \bibinfo{journal}{Phys. Rev. A} \bibinfo{volume}{83},
  \bibinfo{pages}{063416}.
\newblock \DOIprefix\doi{10.1103/PhysRevA.83.063416}.
%Type = Article
\bibitem[{Haxton and McCurdy(2015)}]{Haxton2015}
\bibinfo{author}{Haxton, D.J.}, \bibinfo{author}{McCurdy, C.W.},
  \bibinfo{year}{2015}.
\newblock \bibinfo{title}{Two methods for restricted configuration spaces
  within the multiconfiguration time-dependent {H}artree-{F}ock method}.
\newblock \bibinfo{journal}{Phys. Rev. A} \bibinfo{volume}{91},
  \bibinfo{pages}{012509}.
\newblock \DOIprefix\doi{10.1103/PhysRevA.91.012509}.
%Type = Article
\bibitem[{Hellmann et~al.(2012)Hellmann, Sohrt, Beye, Rohwer, Sorgenfrei,
  Marczynski-B\"uhlow, Kall\"ane, Redlin, Hennies, Bauer, F\"ohlisch, Kipp,
  Wurth and Rossnagel}]{Hellmann2012}
\bibinfo{author}{Hellmann, S.}, \bibinfo{author}{Sohrt, C.},
  \bibinfo{author}{Beye, M.}, \bibinfo{author}{Rohwer, T.},
  \bibinfo{author}{Sorgenfrei, F.}, \bibinfo{author}{Marczynski-B\"uhlow, M.},
  \bibinfo{author}{Kall\"ane, M.}, \bibinfo{author}{Redlin, H.},
  \bibinfo{author}{Hennies, F.}, \bibinfo{author}{Bauer, M.},
  \bibinfo{author}{F\"ohlisch, A.}, \bibinfo{author}{Kipp, L.},
  \bibinfo{author}{Wurth, W.}, \bibinfo{author}{Rossnagel, K.},
  \bibinfo{year}{2012}.
\newblock \bibinfo{title}{Time-resolved x-ray photoelectron spectroscopy at
  {FLASH}}.
\newblock \bibinfo{journal}{New J. Phys.} \bibinfo{volume}{14},
  \bibinfo{pages}{013062}.
\newblock \DOIprefix\doi{10.1088/1367-2630/14/1/013062}.
%Type = Article
\bibitem[{Helml et~al.(2017)Helml, Grgura{\v{s}}, Jurani{\'{c}}, D{\"u}sterer,
  Mazza, Maier, Hartmann, Ilchen, Hartmann, Patthey, Callegari, Costello,
  Meyer, Coffee, Cavalieri and Kienberger}]{Helml2017}
\bibinfo{author}{Helml, W.}, \bibinfo{author}{Grgura{\v{s}}, I.},
  \bibinfo{author}{Jurani{\'{c}}, P.}, \bibinfo{author}{D{\"u}sterer, S.},
  \bibinfo{author}{Mazza, T.}, \bibinfo{author}{Maier, A.},
  \bibinfo{author}{Hartmann, N.}, \bibinfo{author}{Ilchen, M.},
  \bibinfo{author}{Hartmann, G.}, \bibinfo{author}{Patthey, L.},
  \bibinfo{author}{Callegari, C.}, \bibinfo{author}{Costello, J.},
  \bibinfo{author}{Meyer, M.}, \bibinfo{author}{Coffee, R.},
  \bibinfo{author}{Cavalieri, A.}, \bibinfo{author}{Kienberger, R.},
  \bibinfo{year}{2017}.
\newblock \bibinfo{title}{Ultrashort free-electron laser {X}-ray pulses}.
\newblock \bibinfo{journal}{Applied Sciences} \bibinfo{volume}{7},
  \bibinfo{pages}{915}.
\newblock \DOIprefix\doi{10.3390/app7090915}.
%Type = Article
\bibitem[{Helml et~al.(2014)Helml, Maier, Schweinberger, Grgura{\v{s}},
  Radcliffe, Doumy, Roedig, Gagnon, Messerschmidt, Schorb, Bostedt, Gr{\"u}ner,
  DiMauro, Cubaynes, Bozek, Tschentscher, Costello, Meyer, Coffee,
  D{\"u}sterer, Cavalieri and Kienberger}]{Helml2014}
\bibinfo{author}{Helml, W.}, \bibinfo{author}{Maier, A.R.},
  \bibinfo{author}{Schweinberger, W.}, \bibinfo{author}{Grgura{\v{s}}, I.},
  \bibinfo{author}{Radcliffe, P.}, \bibinfo{author}{Doumy, G.},
  \bibinfo{author}{Roedig, C.}, \bibinfo{author}{Gagnon, J.},
  \bibinfo{author}{Messerschmidt, M.}, \bibinfo{author}{Schorb, S.},
  \bibinfo{author}{Bostedt, C.}, \bibinfo{author}{Gr{\"u}ner, F.},
  \bibinfo{author}{DiMauro, L.F.}, \bibinfo{author}{Cubaynes, D.},
  \bibinfo{author}{Bozek, J.D.}, \bibinfo{author}{Tschentscher, T.},
  \bibinfo{author}{Costello, J.T.}, \bibinfo{author}{Meyer, M.},
  \bibinfo{author}{Coffee, R.}, \bibinfo{author}{D{\"u}sterer, S.},
  \bibinfo{author}{Cavalieri, A.L.}, \bibinfo{author}{Kienberger, R.},
  \bibinfo{year}{2014}.
\newblock \bibinfo{title}{Measuring the temporal structure of few-femtosecond
  free-electron laser {X}-ray pulses directly in the time domain}.
\newblock \bibinfo{journal}{Nature Photonics} \bibinfo{volume}{8},
  \bibinfo{pages}{950--957}.
\newblock \DOIprefix\doi{10.1038/nphoton.2014.278}.
%Type = Article
\bibitem[{Hemsing et~al.(2020)Hemsing, Halavanau and Zhang}]{Hemsing2020}
\bibinfo{author}{Hemsing, E.}, \bibinfo{author}{Halavanau, A.},
  \bibinfo{author}{Zhang, Z.}, \bibinfo{year}{2020}.
\newblock \bibinfo{title}{Statistical theory of a self-seeded free electron
  laser with noise pedestal growth}.
\newblock \bibinfo{journal}{Phys. Rev. Accel. Beams} \bibinfo{volume}{23},
  \bibinfo{pages}{010701}.
\newblock \DOIprefix\doi{10.1103/PhysRevAccelBeams.23.010701}.
%Type = Article
\bibitem[{Hemsing et~al.(2019)Hemsing, Marcus, Fawley, Schoenlein, Coffee,
  Dakovski, Hastings, Huang, Ratner, Raubenheimer and Penn}]{Hemsing2019}
\bibinfo{author}{Hemsing, E.}, \bibinfo{author}{Marcus, G.},
  \bibinfo{author}{Fawley, W.M.}, \bibinfo{author}{Schoenlein, R.W.},
  \bibinfo{author}{Coffee, R.}, \bibinfo{author}{Dakovski, G.},
  \bibinfo{author}{Hastings, J.}, \bibinfo{author}{Huang, Z.},
  \bibinfo{author}{Ratner, D.}, \bibinfo{author}{Raubenheimer, T.},
  \bibinfo{author}{Penn, G.}, \bibinfo{year}{2019}.
\newblock \bibinfo{title}{Soft {x}-ray seeding studies for the {SLAC} {L}inac
  {C}oherent {L}ight {S}ource {II}}.
\newblock \bibinfo{journal}{Phys. Rev. Accel. Beams} \bibinfo{volume}{22},
  \bibinfo{pages}{110701}.
\newblock \DOIprefix\doi{10.1103/PhysRevAccelBeams.22.110701}.
%Type = Article
\bibitem[{Heyl et~al.(2016)Heyl, Arnold, Couairon and L'Huillier}]{Heyl2016}
\bibinfo{author}{Heyl, C.M.}, \bibinfo{author}{Arnold, C.L.},
  \bibinfo{author}{Couairon, A.}, \bibinfo{author}{L'Huillier, A.},
  \bibinfo{year}{2016}.
\newblock \bibinfo{title}{Introduction to macroscopic power scaling principles
  for high-order harmonic generation}.
\newblock \bibinfo{journal}{J. Phys. B} \bibinfo{volume}{50},
  \bibinfo{pages}{013001}.
\newblock \DOIprefix\doi{10.1088/1361-6455/50/1/013001}.
%Type = Article
\bibitem[{Hikosaka et~al.(2010)Hikosaka, Fushitani, Matsuda, Tseng, Hishikawa,
  Shigemasa, Nagasono, Tono, Togashi, Ohashi, Kimura, Senba, Yabashi and
  Ishikawa}]{Hikosaka2010}
\bibinfo{author}{Hikosaka, Y.}, \bibinfo{author}{Fushitani, M.},
  \bibinfo{author}{Matsuda, A.}, \bibinfo{author}{Tseng, C.M.},
  \bibinfo{author}{Hishikawa, A.}, \bibinfo{author}{Shigemasa, E.},
  \bibinfo{author}{Nagasono, M.}, \bibinfo{author}{Tono, K.},
  \bibinfo{author}{Togashi, T.}, \bibinfo{author}{Ohashi, H.},
  \bibinfo{author}{Kimura, H.}, \bibinfo{author}{Senba, Y.},
  \bibinfo{author}{Yabashi, M.}, \bibinfo{author}{Ishikawa, T.},
  \bibinfo{year}{2010}.
\newblock \bibinfo{title}{Multiphoton double ionization of {Ar} in intense
  extreme ultraviolet laser fields studied by shot-by-shot photoelectron
  spectroscopy}.
\newblock \bibinfo{journal}{Phys. Rev. Lett.} \bibinfo{volume}{105},
  \bibinfo{pages}{133001}.
\newblock \DOIprefix\doi{10.1103/PhysRevLett.105.133001}.
%Type = Article
\bibitem[{Hikosaka et~al.(2019)Hikosaka, Kaneyasu, Fujimoto, Iwayama and
  Katoh}]{Hikosaka2019}
\bibinfo{author}{Hikosaka, Y.}, \bibinfo{author}{Kaneyasu, T.},
  \bibinfo{author}{Fujimoto, M.}, \bibinfo{author}{Iwayama, H.},
  \bibinfo{author}{Katoh, M.}, \bibinfo{year}{2019}.
\newblock \bibinfo{title}{Coherent control in the extreme ultraviolet and
  attosecond regime by synchrotron radiation}.
\newblock \bibinfo{journal}{Nat. Commun.} \bibinfo{volume}{10},
  \bibinfo{pages}{4988}.
\newblock \DOIprefix\doi{10.1038/s41467-019-12978-w}.
%Type = Article
\bibitem[{Hishikawa et~al.(2011)Hishikawa, Fushitani, Hikosaka, Matsuda, Liu,
  Morishita, Shigemasa, Nagasono, Tono, Togashi, Ohashi, Kimura, Senba, Yabashi
  and Ishikawa}]{Hishikawa2011}
\bibinfo{author}{Hishikawa, A.}, \bibinfo{author}{Fushitani, M.},
  \bibinfo{author}{Hikosaka, Y.}, \bibinfo{author}{Matsuda, A.},
  \bibinfo{author}{Liu, C.N.}, \bibinfo{author}{Morishita, T.},
  \bibinfo{author}{Shigemasa, E.}, \bibinfo{author}{Nagasono, M.},
  \bibinfo{author}{Tono, K.}, \bibinfo{author}{Togashi, T.},
  \bibinfo{author}{Ohashi, H.}, \bibinfo{author}{Kimura, H.},
  \bibinfo{author}{Senba, Y.}, \bibinfo{author}{Yabashi, M.},
  \bibinfo{author}{Ishikawa, T.}, \bibinfo{year}{2011}.
\newblock \bibinfo{title}{Enhanced nonlinear double excitation of {He} in
  intense extreme ultraviolet laser fields}.
\newblock \bibinfo{journal}{Phys. Rev. Lett.} \bibinfo{volume}{107},
  \bibinfo{pages}{243003}.
\newblock \DOIprefix\doi{10.1103/PhysRevLett.107.243003}.
%Type = Article
\bibitem[{Ho et~al.(2014)Ho, Bostedt, Schorb and Young}]{HoPRL2014}
\bibinfo{author}{Ho, P.J.}, \bibinfo{author}{Bostedt, C.},
  \bibinfo{author}{Schorb, S.}, \bibinfo{author}{Young, L.},
  \bibinfo{year}{2014}.
\newblock \bibinfo{title}{Theoretical tracking of resonance-enhanced multiple
  ionization pathways in {X}-ray free-electron laser pulses}.
\newblock \bibinfo{journal}{Phys. Rev. Lett.} \bibinfo{volume}{113},
  \bibinfo{pages}{253001}.
\newblock \DOIprefix\doi{10.1103/PhysRevLett.113.253001}.
%Type = Article
\bibitem[{Ho et~al.(2015)Ho, Kanter and Young}]{Ho2015}
\bibinfo{author}{Ho, P.J.}, \bibinfo{author}{Kanter, E.P.},
  \bibinfo{author}{Young, L.}, \bibinfo{year}{2015}.
\newblock \bibinfo{title}{Resonance-mediated atomic ionization dynamics induced
  by ultraintense {x}-ray pulses}.
\newblock \bibinfo{journal}{Phys. Rev. A} \bibinfo{volume}{92},
  \bibinfo{pages}{063430}.
\newblock \DOIprefix\doi{10.1103/PhysRevA.92.063430}.
%Type = Article
\bibitem[{Hochstuhl and Bonitz(2012)}]{Hochstuhl2012}
\bibinfo{author}{Hochstuhl, D.}, \bibinfo{author}{Bonitz, M.},
  \bibinfo{year}{2012}.
\newblock \bibinfo{title}{Time-dependent restricted-active-space
  configuration-interaction method for the photoionization of many-electron
  atoms}.
\newblock \bibinfo{journal}{Phys. Rev. A} \bibinfo{volume}{86},
  \bibinfo{pages}{053424}.
\newblock \DOIprefix\doi{10.1103/PhysRevA.86.053424}.
%Type = Article
\bibitem[{Hofbrucker et~al.(2018)Hofbrucker, Volotka and
  Fritzsche}]{Hofbrucker2018}
\bibinfo{author}{Hofbrucker, J.}, \bibinfo{author}{Volotka, A.V.},
  \bibinfo{author}{Fritzsche, S.}, \bibinfo{year}{2018}.
\newblock \bibinfo{title}{Maximum elliptical dichroism in atomic two-photon
  ionization}.
\newblock \bibinfo{journal}{Phys. Rev. Lett.} \bibinfo{volume}{121},
  \bibinfo{pages}{053401}.
\newblock \DOIprefix\doi{10.1103/PhysRevLett.121.053401}.
%Type = Article
\bibitem[{Hofstetter et~al.(2011)Hofstetter, Schultze, Fie{\ss}, Dennhardt,
  Guggenmos, Gagnon, Yakovlev, Goulielmakis, Kienberger, Gullikson, Krausz and
  Kleineberg}]{Hofstetter2011}
\bibinfo{author}{Hofstetter, M.}, \bibinfo{author}{Schultze, M.},
  \bibinfo{author}{Fie{\ss}, M.}, \bibinfo{author}{Dennhardt, B.},
  \bibinfo{author}{Guggenmos, A.}, \bibinfo{author}{Gagnon, J.},
  \bibinfo{author}{Yakovlev, V.S.}, \bibinfo{author}{Goulielmakis, E.},
  \bibinfo{author}{Kienberger, R.}, \bibinfo{author}{Gullikson, E.M.},
  \bibinfo{author}{Krausz, F.}, \bibinfo{author}{Kleineberg, U.},
  \bibinfo{year}{2011}.
\newblock \bibinfo{title}{Attosecond dispersion control by extreme ultraviolet
  multilayer mirrors}.
\newblock \bibinfo{journal}{Opt. Express} \bibinfo{volume}{19},
  \bibinfo{pages}{1767--1776}.
\newblock \DOIprefix\doi{10.1364/OE.19.001767}.
%Type = Article
\bibitem[{Hogle et~al.(2015)Hogle, Tong, Martin, Murnane, Kapteyn and
  Ranitovic}]{Hogle2015}
\bibinfo{author}{Hogle, C.W.}, \bibinfo{author}{Tong, X.M.},
  \bibinfo{author}{Martin, L.}, \bibinfo{author}{Murnane, M.M.},
  \bibinfo{author}{Kapteyn, H.C.}, \bibinfo{author}{Ranitovic, P.},
  \bibinfo{year}{2015}.
\newblock \bibinfo{title}{Attosecond coherent control of single and double
  photoionization in argon}.
\newblock \bibinfo{journal}{Phys. Rev. Lett.} \bibinfo{volume}{115},
  \bibinfo{pages}{173004}.
\newblock \DOIprefix\doi{10.1103/PhysRevLett.115.173004}.
%Type = Article
\bibitem[{Holzmeier et~al.(2018)Holzmeier, Bello, Herv\'e, Achner, Baumann,
  Meyer, Finetti, Di~Fraia, Gauthier, Roussel, Plekan, Richter, Prince,
  Callegari, Bachau, Palacios, Mart\'{\i}n and Dowek}]{Holzmeier2018}
\bibinfo{author}{Holzmeier, F.}, \bibinfo{author}{Bello, R.Y.},
  \bibinfo{author}{Herv\'e, M.}, \bibinfo{author}{Achner, A.},
  \bibinfo{author}{Baumann, T.M.}, \bibinfo{author}{Meyer, M.},
  \bibinfo{author}{Finetti, P.}, \bibinfo{author}{Di~Fraia, M.},
  \bibinfo{author}{Gauthier, D.}, \bibinfo{author}{Roussel, E.},
  \bibinfo{author}{Plekan, O.}, \bibinfo{author}{Richter, R.},
  \bibinfo{author}{Prince, K.C.}, \bibinfo{author}{Callegari, C.},
  \bibinfo{author}{Bachau, H.}, \bibinfo{author}{Palacios, A.},
  \bibinfo{author}{Mart\'{\i}n, F.}, \bibinfo{author}{Dowek, D.},
  \bibinfo{year}{2018}.
\newblock \bibinfo{title}{Control of {H}$_{2}$ dissociative ionization in the
  nonlinear regime using vacuum ultraviolet free-electron laser pulses}.
\newblock \bibinfo{journal}{Phys. Rev. Lett.} \bibinfo{volume}{121},
  \bibinfo{pages}{103002}.
\newblock \DOIprefix\doi{10.1103/PhysRevLett.121.103002}.
%Type = Article
\bibitem[{Hong et~al.(2012)Hong, Lai, Gkortsas, Huang, Moses, Granados,
  Bhardwaj and K\"{a}rtner}]{Hong2012}
\bibinfo{author}{Hong, K.H.}, \bibinfo{author}{Lai, C.J.},
  \bibinfo{author}{Gkortsas, V.M.}, \bibinfo{author}{Huang, S.W.},
  \bibinfo{author}{Moses, J.}, \bibinfo{author}{Granados, E.},
  \bibinfo{author}{Bhardwaj, S.}, \bibinfo{author}{K\"{a}rtner, F.X.},
  \bibinfo{year}{2012}.
\newblock \bibinfo{title}{High-order harmonic generation in {Xe}, {Kr}, and
  {Ar} driven by a 2.1-$\mu$m source: High-order harmonic spectroscopy under
  macroscopic effects}.
\newblock \bibinfo{journal}{Phys. Rev. A} \bibinfo{volume}{86},
  \bibinfo{pages}{043412}.
\newblock \DOIprefix\doi{10.1103/physreva.86.043412}.
%Type = Article
\bibitem[{Hong et~al.(2014)Hong, Lai, Siqueira, Krogen, Moses, Chang, Stein,
  Zapata and K\"{a}rtner}]{Hong2014}
\bibinfo{author}{Hong, K.H.}, \bibinfo{author}{Lai, C.J.},
  \bibinfo{author}{Siqueira, J.P.}, \bibinfo{author}{Krogen, P.},
  \bibinfo{author}{Moses, J.}, \bibinfo{author}{Chang, C.L.},
  \bibinfo{author}{Stein, G.J.}, \bibinfo{author}{Zapata, L.E.},
  \bibinfo{author}{K\"{a}rtner, F.X.}, \bibinfo{year}{2014}.
\newblock \bibinfo{title}{Multi-{mJ}, {kHz}, 2.1$\mu$m optical parametric
  chirped-pulse amplifier and high-flux soft x-ray high-harmonic generation}.
\newblock \bibinfo{journal}{Opt. Lett.} \bibinfo{volume}{39},
  \bibinfo{pages}{3145}.
\newblock \DOIprefix\doi{10.1364/ol.39.003145}.
%Type = Article
\bibitem[{Hutchinson et~al.(2013)Hutchinson, Lysaght and van~der
  Hart}]{Hutchinson2013}
\bibinfo{author}{Hutchinson, S.}, \bibinfo{author}{Lysaght, M.A.},
  \bibinfo{author}{van~der Hart, H.W.}, \bibinfo{year}{2013}.
\newblock \bibinfo{title}{Anisotropy parameters in two-color two-photon
  above-threshold ionization}.
\newblock \bibinfo{journal}{Phys. Rev. A} \bibinfo{volume}{88},
  \bibinfo{pages}{023424}.
\newblock \DOIprefix\doi{10.1103/PhysRevA.88.023424}.
%Type = Article
\bibitem[{Iablonskyi et~al.(2016)Iablonskyi, Nagaya, Fukuzawa, Motomura,
  Kumagai, Mondal, Tachibana, Takanashi, Nishiyama, Matsunami, Johnsson,
  Piseri, Sansone, Dubrouil, Reduzzi, Carpeggiani, Vozzi, Devetta, Negro,
  Calegari, Trabattoni, Castrovilli, Faccial{\'{a}}, Ovcharenko, M{\"o}ller,
  Mudrich, Stienkemeier, Coreno, Alagia, Sch{\"u}tte, Berrah, Kuleff, Jabbari,
  Callegari, Plekan, Finetti, Spezzani, Ferrari, Allaria, Penco, Serpico,
  De~Ninno, Nikolov, Diviacco, Di~Mitri, Giannessi, Prince and
  Ueda}]{Iablonskyi2016}
\bibinfo{author}{Iablonskyi, D.}, \bibinfo{author}{Nagaya, K.},
  \bibinfo{author}{Fukuzawa, H.}, \bibinfo{author}{Motomura, K.},
  \bibinfo{author}{Kumagai, Y.}, \bibinfo{author}{Mondal, S.},
  \bibinfo{author}{Tachibana, T.}, \bibinfo{author}{Takanashi, T.},
  \bibinfo{author}{Nishiyama, T.}, \bibinfo{author}{Matsunami, K.},
  \bibinfo{author}{Johnsson, P.}, \bibinfo{author}{Piseri, P.},
  \bibinfo{author}{Sansone, G.}, \bibinfo{author}{Dubrouil, A.},
  \bibinfo{author}{Reduzzi, M.}, \bibinfo{author}{Carpeggiani, P.},
  \bibinfo{author}{Vozzi, C.}, \bibinfo{author}{Devetta, M.},
  \bibinfo{author}{Negro, M.}, \bibinfo{author}{Calegari, F.},
  \bibinfo{author}{Trabattoni, A.}, \bibinfo{author}{Castrovilli, M.C.},
  \bibinfo{author}{Faccial{\'{a}}, D.}, \bibinfo{author}{Ovcharenko, Y.},
  \bibinfo{author}{M{\"o}ller, T.}, \bibinfo{author}{Mudrich, M.},
  \bibinfo{author}{Stienkemeier, F.}, \bibinfo{author}{Coreno, M.},
  \bibinfo{author}{Alagia, M.}, \bibinfo{author}{Sch{\"u}tte, B.},
  \bibinfo{author}{Berrah, N.}, \bibinfo{author}{Kuleff, A.I.},
  \bibinfo{author}{Jabbari, G.}, \bibinfo{author}{Callegari, C.},
  \bibinfo{author}{Plekan, O.}, \bibinfo{author}{Finetti, P.},
  \bibinfo{author}{Spezzani, C.}, \bibinfo{author}{Ferrari, E.},
  \bibinfo{author}{Allaria, E.}, \bibinfo{author}{Penco, G.},
  \bibinfo{author}{Serpico, C.}, \bibinfo{author}{De~Ninno, G.},
  \bibinfo{author}{Nikolov, I.}, \bibinfo{author}{Diviacco, B.},
  \bibinfo{author}{Di~Mitri, S.}, \bibinfo{author}{Giannessi, L.},
  \bibinfo{author}{Prince, K.C.}, \bibinfo{author}{Ueda, K.},
  \bibinfo{year}{2016}.
\newblock \bibinfo{title}{Slow interatomic coulombic decay of multiply excited
  neon clusters}.
\newblock \bibinfo{journal}{Phys. Rev. Lett.} \bibinfo{volume}{117},
  \bibinfo{pages}{276806}.
\newblock \DOIprefix\doi{10.1103/PhysRevLett.117.276806}.
%Type = Article
\bibitem[{Iablonskyi et~al.(2017)Iablonskyi, Ueda, Ishikawa, Kheifets,
  Carpeggiani, Reduzzi, Ahmadi, Comby, Sansone, Csizmadia, Kuehn, Ovcharenko,
  Mazza, Meyer, Fischer, Callegari, Plekan, Finetti, Allaria, Ferrari, Roussel,
  Gauthier, Giannessi and Prince}]{Iablonskyi2017}
\bibinfo{author}{Iablonskyi, D.}, \bibinfo{author}{Ueda, K.},
  \bibinfo{author}{Ishikawa, K.L.}, \bibinfo{author}{Kheifets, A.S.},
  \bibinfo{author}{Carpeggiani, P.}, \bibinfo{author}{Reduzzi, M.},
  \bibinfo{author}{Ahmadi, H.}, \bibinfo{author}{Comby, A.},
  \bibinfo{author}{Sansone, G.}, \bibinfo{author}{Csizmadia, T.},
  \bibinfo{author}{Kuehn, S.}, \bibinfo{author}{Ovcharenko, E.},
  \bibinfo{author}{Mazza, T.}, \bibinfo{author}{Meyer, M.},
  \bibinfo{author}{Fischer, A.}, \bibinfo{author}{Callegari, C.},
  \bibinfo{author}{Plekan, O.}, \bibinfo{author}{Finetti, P.},
  \bibinfo{author}{Allaria, E.}, \bibinfo{author}{Ferrari, E.},
  \bibinfo{author}{Roussel, E.}, \bibinfo{author}{Gauthier, D.},
  \bibinfo{author}{Giannessi, L.}, \bibinfo{author}{Prince, K.C.},
  \bibinfo{year}{2017}.
\newblock \bibinfo{title}{Observation and control of laser-enabled {A}uger
  decay}.
\newblock \bibinfo{journal}{Phys. Rev. Lett.} \bibinfo{volume}{119},
  \bibinfo{pages}{073203}.
\newblock \DOIprefix\doi{10.1103/PhysRevLett.119.073203}.
%Type = Article
\bibitem[{Ilchen et~al.(2017)Ilchen, Douguet, Mazza, Rafipoor, Callegari,
  Finetti, Plekan, Prince, Demidovich, Grazioli, Avaldi, Bolognesi, Coreno,
  Di~Fraia, Devetta, Ovcharenko, D\"usterer, Ueda, Bartschat, Grum-Grzhimailo,
  Bozhevolnov, Kazansky, Kabachnik and Meyer}]{Ilchen2017}
\bibinfo{author}{Ilchen, M.}, \bibinfo{author}{Douguet, N.},
  \bibinfo{author}{Mazza, T.}, \bibinfo{author}{Rafipoor, A.J.},
  \bibinfo{author}{Callegari, C.}, \bibinfo{author}{Finetti, P.},
  \bibinfo{author}{Plekan, O.}, \bibinfo{author}{Prince, K.C.},
  \bibinfo{author}{Demidovich, A.}, \bibinfo{author}{Grazioli, C.},
  \bibinfo{author}{Avaldi, L.}, \bibinfo{author}{Bolognesi, P.},
  \bibinfo{author}{Coreno, M.}, \bibinfo{author}{Di~Fraia, M.},
  \bibinfo{author}{Devetta, M.}, \bibinfo{author}{Ovcharenko, Y.},
  \bibinfo{author}{D\"usterer, S.}, \bibinfo{author}{Ueda, K.},
  \bibinfo{author}{Bartschat, K.}, \bibinfo{author}{Grum-Grzhimailo, A.N.},
  \bibinfo{author}{Bozhevolnov, A.V.}, \bibinfo{author}{Kazansky, A.K.},
  \bibinfo{author}{Kabachnik, N.M.}, \bibinfo{author}{Meyer, M.},
  \bibinfo{year}{2017}.
\newblock \bibinfo{title}{Circular dichroism in multiphoton ionization of
  resonantly excited {He}$^{+}$ ions}.
\newblock \bibinfo{journal}{Phys. Rev. Lett.} \bibinfo{volume}{118},
  \bibinfo{pages}{013002}.
\newblock \DOIprefix\doi{10.1103/PhysRevLett.118.013002}.
%Type = Article
\bibitem[{Ilchen et~al.(2018)Ilchen, Hartmann, Gryzlova, Achner, Allaria,
  Beckmann, Braune, Buck, Callegari, Coffee, Cucini, Danailov, De~Fanis,
  Demidovich, Ferrari, Finetti, Glaser, Knie, Lindahl, Plekan, Mahne, Mazza,
  Raimondi, Roussel, Seltmann, Shevchuk, Svetina, Walter, Zangrando, Viefhaus,
  Grum-Grzhimailo and Meyer}]{Ilchen2018}
\bibinfo{author}{Ilchen, M.}, \bibinfo{author}{Hartmann, G.},
  \bibinfo{author}{Gryzlova, E.V.}, \bibinfo{author}{Achner, A.},
  \bibinfo{author}{Allaria, E.}, \bibinfo{author}{Beckmann, A.},
  \bibinfo{author}{Braune, M.}, \bibinfo{author}{Buck, J.},
  \bibinfo{author}{Callegari, C.}, \bibinfo{author}{Coffee, R.N.},
  \bibinfo{author}{Cucini, R.}, \bibinfo{author}{Danailov, M.},
  \bibinfo{author}{De~Fanis, A.}, \bibinfo{author}{Demidovich, A.},
  \bibinfo{author}{Ferrari, E.}, \bibinfo{author}{Finetti, P.},
  \bibinfo{author}{Glaser, L.}, \bibinfo{author}{Knie, A.},
  \bibinfo{author}{Lindahl, A.O.}, \bibinfo{author}{Plekan, O.},
  \bibinfo{author}{Mahne, N.}, \bibinfo{author}{Mazza, T.},
  \bibinfo{author}{Raimondi, L.}, \bibinfo{author}{Roussel, E.},
  \bibinfo{author}{Seltmann, F.S.J.}, \bibinfo{author}{Shevchuk, I.},
  \bibinfo{author}{Svetina, C.}, \bibinfo{author}{Walter, P.},
  \bibinfo{author}{Zangrando, M.}, \bibinfo{author}{Viefhaus, J.},
  \bibinfo{author}{Grum-Grzhimailo, A.N.}, \bibinfo{author}{Meyer, M.},
  \bibinfo{year}{2018}.
\newblock \bibinfo{title}{Symmetry breakdown of electron emission in extreme
  ultraviolet photoionization of argon}.
\newblock \bibinfo{journal}{Nat. Commun.} \bibinfo{volume}{8},
  \bibinfo{pages}{4659}.
\newblock \DOIprefix\doi{10.1038/s41467-018-07152-7}.
%Type = Article
\bibitem[{Inhester et~al.(2016)Inhester, Hanasaki, Hao, Son and
  Santra}]{Inhester2016}
\bibinfo{author}{Inhester, L.}, \bibinfo{author}{Hanasaki, K.},
  \bibinfo{author}{Hao, Y.}, \bibinfo{author}{Son, S.K.},
  \bibinfo{author}{Santra, R.}, \bibinfo{year}{2016}.
\newblock \bibinfo{title}{X-ray multiphoton ionization dynamics of a water
  molecule irradiated by an x-ray free-electron laser pulse}.
\newblock \bibinfo{journal}{Phys. Rev. A} \bibinfo{volume}{94},
  \bibinfo{pages}{023422}.
\newblock \DOIprefix\doi{10.1103/PhysRevA.94.023422}.
%Type = Article
\bibitem[{Inoue et~al.(2019)Inoue, Osaka, Hara, Tanaka, Inagaki, Fukui, Goto,
  Inubushi, Kimura, Kinjo, Ohashi, Togawa, Tono, Yamaga, Tanaka, Ishikawa and
  Yabashi}]{Inoue2019}
\bibinfo{author}{Inoue, I.}, \bibinfo{author}{Osaka, T.},
  \bibinfo{author}{Hara, T.}, \bibinfo{author}{Tanaka, T.},
  \bibinfo{author}{Inagaki, T.}, \bibinfo{author}{Fukui, T.},
  \bibinfo{author}{Goto, S.}, \bibinfo{author}{Inubushi, Y.},
  \bibinfo{author}{Kimura, H.}, \bibinfo{author}{Kinjo, R.},
  \bibinfo{author}{Ohashi, H.}, \bibinfo{author}{Togawa, K.},
  \bibinfo{author}{Tono, K.}, \bibinfo{author}{Yamaga, M.},
  \bibinfo{author}{Tanaka, H.}, \bibinfo{author}{Ishikawa, T.},
  \bibinfo{author}{Yabashi, M.}, \bibinfo{year}{2019}.
\newblock \bibinfo{title}{Generation of narrow-band {X}-ray free-electron laser
  via reflection self-seeding}.
\newblock \bibinfo{journal}{Nat. Photonics} \bibinfo{volume}{13},
  \bibinfo{pages}{319--322}.
\newblock \DOIprefix\doi{10.1038/s41566-019-0365-y}.
%Type = Article
\bibitem[{Ishikawa et~al.(2010)Ishikawa, Kawazura and Ueda}]{Ishikawa2010}
\bibinfo{author}{Ishikawa, K.L.}, \bibinfo{author}{Kawazura, Y.},
  \bibinfo{author}{Ueda, K.}, \bibinfo{year}{2010}.
\newblock \bibinfo{title}{Two-photon ionization of atoms by ultrashort laser
  pulses}.
\newblock \bibinfo{journal}{J. Mod. Opt.} \bibinfo{volume}{57},
  \bibinfo{pages}{999--1007}.
\newblock \DOIprefix\doi{10.1080/09500340903511703}.
%Type = Article
\bibitem[{Ishikawa et~al.(2014)Ishikawa, Kazansky, Kabachnik and
  Ueda}]{Ishikawa2014}
\bibinfo{author}{Ishikawa, K.L.}, \bibinfo{author}{Kazansky, A.K.},
  \bibinfo{author}{Kabachnik, N.M.}, \bibinfo{author}{Ueda, K.},
  \bibinfo{year}{2014}.
\newblock \bibinfo{title}{Theoretical study of pulse delay effects in the
  photoelectron angular distribution of near-threshold {EUV} + {IR} two-photon
  ionization of atoms}.
\newblock \bibinfo{journal}{Phys. Rev. A} \bibinfo{volume}{90},
  \bibinfo{pages}{023408}.
\newblock \DOIprefix\doi{10.1103/PhysRevA.90.023408}.
%Type = Article
\bibitem[{Ishikawa and Midorikawa(2005)}]{Ishikawa2005}
\bibinfo{author}{Ishikawa, K.L.}, \bibinfo{author}{Midorikawa, K.},
  \bibinfo{year}{2005}.
\newblock \bibinfo{title}{Above-threshold double ionization of helium with
  attosecond intense soft {x}-ray pulses}.
\newblock \bibinfo{journal}{Phys. Rev. A} \bibinfo{volume}{72},
  \bibinfo{pages}{013407}.
\newblock \DOIprefix\doi{10.1103/PhysRevA.72.013407}.
%Type = Article
\bibitem[{Ishikawa and Sato(2015)}]{Ishikawa2015}
\bibinfo{author}{Ishikawa, K.L.}, \bibinfo{author}{Sato, T.},
  \bibinfo{year}{2015}.
\newblock \bibinfo{title}{A review on \emph{ab initio} approaches for
  multielectron dynamics}.
\newblock \bibinfo{journal}{IEEE J. Sel. Top. Quantum Electron.}
  \bibinfo{volume}{21}, \bibinfo{pages}{8700916}.
\newblock \DOIprefix\doi{10.1109/JSTQE.2015.2438827}.
%Type = Article
\bibitem[{Ishikawa and Ueda(2012)}]{Ishikawa2012a}
\bibinfo{author}{Ishikawa, K.L.}, \bibinfo{author}{Ueda, K.},
  \bibinfo{year}{2012}.
\newblock \bibinfo{title}{Competition of resonant and nonresonant paths in
  resonance-enhanced two-photon single ionization of {He} by an ultrashort
  extreme-ultraviolet pulse}.
\newblock \bibinfo{journal}{Phys. Rev. Lett.} \bibinfo{volume}{108},
  \bibinfo{pages}{033003}.
\newblock \DOIprefix\doi{10.1103/PhysRevLett.108.033003}.
%Type = Article
\bibitem[{Ishikawa and Ueda(2013)}]{Ishikawa2013}
\bibinfo{author}{Ishikawa, K.L.}, \bibinfo{author}{Ueda, K.},
  \bibinfo{year}{2013}.
\newblock \bibinfo{title}{Photoelectron angular distribution and phase in
  two-photon single ionization of {H} and {He} by a femtosecond and attosecond
  extreme-ultraviolet pulse}.
\newblock \bibinfo{journal}{Applied Sciences} \bibinfo{volume}{3},
  \bibinfo{pages}{189--213}.
\newblock \DOIprefix\doi{10.3390/app3010189}.
%Type = Article
\bibitem[{Ishikawa et~al.(2012)Ishikawa, Aoyagi, Asaka, Asano, Azumi, Bizen,
  Ego, Fukami, Fukui, Furukawa, Goto, Hanaki, Hara, Hasegawa, Hatsui,
  Higashiya, Hirono, Hosoda, Ishii, Inagaki, Inubushi, Itoga, Joti, Kago,
  Kameshima, Kimura, Kirihara, Kiyomichi, Kobayashi, Kondo, Kudo, Maesaka,
  Mar{\'{e}}chal, Masuda, Matsubara, Matsumoto, Matsushita, Matsui, Nagasono,
  Nariyama, Ohashi, Ohata, Ohshima, Ono, Otake, Saji, Sakurai, Sato, Sawada,
  Seike, Shirasawa, Sugimoto, Suzuki, Takahashi, Takebe, Takeshita, Tamasaku,
  Tanaka, Tanaka, Tanaka, Togashi, Togawa, Tokuhisa, Tomizawa, Tono, Wu,
  Yabashi, Yamaga, Yamashita, Yanagida, Zhang, Shintake, Kitamura and
  Kumagai}]{Ishikawa2012}
\bibinfo{author}{Ishikawa, T.}, \bibinfo{author}{Aoyagi, H.},
  \bibinfo{author}{Asaka, T.}, \bibinfo{author}{Asano, Y.},
  \bibinfo{author}{Azumi, N.}, \bibinfo{author}{Bizen, T.},
  \bibinfo{author}{Ego, H.}, \bibinfo{author}{Fukami, K.},
  \bibinfo{author}{Fukui, T.}, \bibinfo{author}{Furukawa, Y.},
  \bibinfo{author}{Goto, S.}, \bibinfo{author}{Hanaki, H.},
  \bibinfo{author}{Hara, T.}, \bibinfo{author}{Hasegawa, T.},
  \bibinfo{author}{Hatsui, T.}, \bibinfo{author}{Higashiya, A.},
  \bibinfo{author}{Hirono, T.}, \bibinfo{author}{Hosoda, N.},
  \bibinfo{author}{Ishii, M.}, \bibinfo{author}{Inagaki, T.},
  \bibinfo{author}{Inubushi, Y.}, \bibinfo{author}{Itoga, T.},
  \bibinfo{author}{Joti, Y.}, \bibinfo{author}{Kago, M.},
  \bibinfo{author}{Kameshima, T.}, \bibinfo{author}{Kimura, H.},
  \bibinfo{author}{Kirihara, Y.}, \bibinfo{author}{Kiyomichi, A.},
  \bibinfo{author}{Kobayashi, T.}, \bibinfo{author}{Kondo, C.},
  \bibinfo{author}{Kudo, T.}, \bibinfo{author}{Maesaka, H.},
  \bibinfo{author}{Mar{\'{e}}chal, X.M.}, \bibinfo{author}{Masuda, T.},
  \bibinfo{author}{Matsubara, S.}, \bibinfo{author}{Matsumoto, T.},
  \bibinfo{author}{Matsushita, T.}, \bibinfo{author}{Matsui, S.},
  \bibinfo{author}{Nagasono, M.}, \bibinfo{author}{Nariyama, N.},
  \bibinfo{author}{Ohashi, H.}, \bibinfo{author}{Ohata, T.},
  \bibinfo{author}{Ohshima, T.}, \bibinfo{author}{Ono, S.},
  \bibinfo{author}{Otake, Y.}, \bibinfo{author}{Saji, C.},
  \bibinfo{author}{Sakurai, T.}, \bibinfo{author}{Sato, T.},
  \bibinfo{author}{Sawada, K.}, \bibinfo{author}{Seike, T.},
  \bibinfo{author}{Shirasawa, K.}, \bibinfo{author}{Sugimoto, T.},
  \bibinfo{author}{Suzuki, S.}, \bibinfo{author}{Takahashi, S.},
  \bibinfo{author}{Takebe, H.}, \bibinfo{author}{Takeshita, K.},
  \bibinfo{author}{Tamasaku, K.}, \bibinfo{author}{Tanaka, H.},
  \bibinfo{author}{Tanaka, R.}, \bibinfo{author}{Tanaka, T.},
  \bibinfo{author}{Togashi, T.}, \bibinfo{author}{Togawa, K.},
  \bibinfo{author}{Tokuhisa, A.}, \bibinfo{author}{Tomizawa, H.},
  \bibinfo{author}{Tono, K.}, \bibinfo{author}{Wu, S.},
  \bibinfo{author}{Yabashi, M.}, \bibinfo{author}{Yamaga, M.},
  \bibinfo{author}{Yamashita, A.}, \bibinfo{author}{Yanagida, K.},
  \bibinfo{author}{Zhang, C.}, \bibinfo{author}{Shintake, T.},
  \bibinfo{author}{Kitamura, H.}, \bibinfo{author}{Kumagai, N.},
  \bibinfo{year}{2012}.
\newblock \bibinfo{title}{A compact x-ray free-electron laser emitting in the
  sub-{\aa}ngstr{\"o}m region}.
\newblock \bibinfo{journal}{Nat. Photonics} \bibinfo{volume}{6},
  \bibinfo{pages}{540--544}.
\newblock \DOIprefix\doi{10.1038/nphoton.2012.141}.
%Type = Article
\bibitem[{Ivanov et~al.(2005)Ivanov, Spanner and Smirnova}]{Ivanov2005}
\bibinfo{author}{Ivanov, M.Y.}, \bibinfo{author}{Spanner, M.},
  \bibinfo{author}{Smirnova, O.}, \bibinfo{year}{2005}.
\newblock \bibinfo{title}{Anatomy of strong field ionization}.
\newblock \bibinfo{journal}{J. Mod. Opt.} \bibinfo{volume}{52},
  \bibinfo{pages}{165--184}.
\newblock \DOIprefix\doi{10.1080/0950034042000275360}.
%Type = Article
\bibitem[{Jahnke(2015)}]{Jahnke2015}
\bibinfo{author}{Jahnke, T.}, \bibinfo{year}{2015}.
\newblock \bibinfo{title}{Interatomic and intermolecular {C}oulombic decay: the
  coming of age story}.
\newblock \bibinfo{journal}{J. Phys. B} \bibinfo{volume}{48},
  \bibinfo{pages}{082001}.
\newblock \DOIprefix\doi{10.1088/0953-4075/48/8/082001}.
%Type = Article
\bibitem[{Jiang et~al.(2010)Jiang, Rudenko, Herrwerth, Foucar, Kurka, K\"uhnel,
  Lezius, Kling, van Tilborg, Belkacem, Ueda, D\"usterer, Treusch, Schr\"oter,
  Moshammer and Ullrich}]{Jiang2010}
\bibinfo{author}{Jiang, Y.H.}, \bibinfo{author}{Rudenko, A.},
  \bibinfo{author}{Herrwerth, O.}, \bibinfo{author}{Foucar, L.},
  \bibinfo{author}{Kurka, M.}, \bibinfo{author}{K\"uhnel, K.U.},
  \bibinfo{author}{Lezius, M.}, \bibinfo{author}{Kling, M.F.},
  \bibinfo{author}{van Tilborg, J.}, \bibinfo{author}{Belkacem, A.},
  \bibinfo{author}{Ueda, K.}, \bibinfo{author}{D\"usterer, S.},
  \bibinfo{author}{Treusch, R.}, \bibinfo{author}{Schr\"oter, C.D.},
  \bibinfo{author}{Moshammer, R.}, \bibinfo{author}{Ullrich, J.},
  \bibinfo{year}{2010}.
\newblock \bibinfo{title}{Ultrafast extreme ultraviolet induced isomerization
  of acetylene cations}.
\newblock \bibinfo{journal}{Phys. Rev. Lett.} \bibinfo{volume}{105},
  \bibinfo{pages}{263002}.
\newblock \DOIprefix\doi{10.1103/PhysRevLett.105.263002}.
%Type = Article
\bibitem[{Jim\'enez~Gal\'an et~al.(2013)Jim\'enez~Gal\'an, Argenti and
  Mart\'{\i}n}]{Galan2013}
\bibinfo{author}{Jim\'enez~Gal\'an, A.}, \bibinfo{author}{Argenti, L.},
  \bibinfo{author}{Mart\'{\i}n, F.}, \bibinfo{year}{2013}.
\newblock \bibinfo{title}{The soft-photon approximation in
  infrared-laser-assisted atomic ionization by extreme-ultraviolet
  attosecond-pulse trains}.
\newblock \bibinfo{journal}{New J. Phys.} \bibinfo{volume}{15},
  \bibinfo{pages}{113009}.
\newblock \DOIprefix\doi{10.1088/1367-2630/15/11/113009}.
%Type = Book
\bibitem[{Joachain et~al.(2012)Joachain, Kylstra and Potvliege}]{Joachain2012}
\bibinfo{author}{Joachain, C.J.}, \bibinfo{author}{Kylstra, N.J.},
  \bibinfo{author}{Potvliege, R.M.}, \bibinfo{year}{2012}.
\newblock \bibinfo{title}{Atoms in Intense Laser Fields}.
\newblock \bibinfo{publisher}{Cambridge University Press}.
\newblock \DOIprefix\doi{10.1017/cbo9780511993459}.
%Type = Article
\bibitem[{Jonas(2003)}]{Jonas2003}
\bibinfo{author}{Jonas, D.M.}, \bibinfo{year}{2003}.
\newblock \bibinfo{title}{Two-dimensional femtosecond spectroscopy}.
\newblock \bibinfo{journal}{Annu. Rev. Phys. Chem.} \bibinfo{volume}{54},
  \bibinfo{pages}{425--463}.
\newblock \DOIprefix\doi{10.1146/annurev.physchem.54.011002.103907}.
%Type = Article
\bibitem[{Kaldun et~al.(2016)Kaldun, Bl\"{a}ttermann, Stoo{\ss}, Donsa, Wei,
  Pazourek, Nagele, Ott, Lin, Burgd\"{o}rfer and Pfeifer}]{Kaldun2016}
\bibinfo{author}{Kaldun, A.}, \bibinfo{author}{Bl\"{a}ttermann, A.},
  \bibinfo{author}{Stoo{\ss}, V.}, \bibinfo{author}{Donsa, S.},
  \bibinfo{author}{Wei, H.}, \bibinfo{author}{Pazourek, R.},
  \bibinfo{author}{Nagele, S.}, \bibinfo{author}{Ott, C.},
  \bibinfo{author}{Lin, C.D.}, \bibinfo{author}{Burgd\"{o}rfer, J.},
  \bibinfo{author}{Pfeifer, T.}, \bibinfo{year}{2016}.
\newblock \bibinfo{title}{Observing the ultrafast buildup of a {F}ano resonance
  in the time domain}.
\newblock \bibinfo{journal}{Science} \bibinfo{volume}{354},
  \bibinfo{pages}{738--741}.
\newblock \DOIprefix\doi{10.1126/science.aah6972}.
%Type = Article
\bibitem[{Kaneyasu et~al.(2019)Kaneyasu, Hikosaka, Fujimoto, Iwayama and
  Katoh}]{Kaneyasu2019}
\bibinfo{author}{Kaneyasu, T.}, \bibinfo{author}{Hikosaka, Y.},
  \bibinfo{author}{Fujimoto, M.}, \bibinfo{author}{Iwayama, H.},
  \bibinfo{author}{Katoh, M.}, \bibinfo{year}{2019}.
\newblock \bibinfo{title}{Controlling the orbital alignment in atoms using
  cross-circularly polarized extreme ultraviolet wave packets}.
\newblock \bibinfo{journal}{Phys. Rev. Lett.} \bibinfo{volume}{123},
  \bibinfo{pages}{233401}.
\newblock \DOIprefix\doi{10.1103/PhysRevLett.123.233401}.
%Type = Article
\bibitem[{Kaneyasu et~al.(2017)Kaneyasu, Hikosaka, Fujimoto, Konomi, Katoh,
  Iwayama and Shigemasa}]{Kaneyasu2017}
\bibinfo{author}{Kaneyasu, T.}, \bibinfo{author}{Hikosaka, Y.},
  \bibinfo{author}{Fujimoto, M.}, \bibinfo{author}{Konomi, T.},
  \bibinfo{author}{Katoh, M.}, \bibinfo{author}{Iwayama, H.},
  \bibinfo{author}{Shigemasa, E.}, \bibinfo{year}{2017}.
\newblock \bibinfo{title}{Limitations in photoionization of helium by an
  extreme ultraviolet optical vortex}.
\newblock \bibinfo{journal}{Phys. Rev. A} \bibinfo{volume}{95},
  \bibinfo{pages}{023413}.
\newblock \DOIprefix\doi{10.1103/PhysRevA.95.023413}.
%Type = Article
\bibitem[{Kang et~al.(2017)Kang, Min, Heo, Kim, Yang, Kim, Nam, Baek, Choi,
  Mun, Park, Suh, Shin, Hu, Hong, Jung, Kim, Kim, Na, Park, Park, Han, Jung,
  Jeong, Lee, Lee, Lee, Lee, Oh, Suh, Parc, Park, Kim, Jung, Kim, Lee, Lee,
  Sung, Mok, Yang, Lee, Shin, Kim, Kim, Lee, Park, Kim, Park, Eom, Rah, Kim,
  Nam, Park, Park, Kim, Kwon, Park, Kim, Hyun, Kim, Kim, Hwang, Kim, Lim, Yu,
  Kim, Kang, Kim, Kim, Lee, Lee, Park, Koo, Kim and Ko}]{Kang2017}
\bibinfo{author}{Kang, H.S.}, \bibinfo{author}{Min, C.K.},
  \bibinfo{author}{Heo, H.}, \bibinfo{author}{Kim, C.}, \bibinfo{author}{Yang,
  H.}, \bibinfo{author}{Kim, G.}, \bibinfo{author}{Nam, I.},
  \bibinfo{author}{Baek, S.Y.}, \bibinfo{author}{Choi, H.J.},
  \bibinfo{author}{Mun, G.}, \bibinfo{author}{Park, B.R.},
  \bibinfo{author}{Suh, Y.J.}, \bibinfo{author}{Shin, D.C.},
  \bibinfo{author}{Hu, J.}, \bibinfo{author}{Hong, J.}, \bibinfo{author}{Jung,
  S.}, \bibinfo{author}{Kim, S.H.}, \bibinfo{author}{Kim, K.},
  \bibinfo{author}{Na, D.}, \bibinfo{author}{Park, S.S.},
  \bibinfo{author}{Park, Y.J.}, \bibinfo{author}{Han, J.H.},
  \bibinfo{author}{Jung, Y.G.}, \bibinfo{author}{Jeong, S.H.},
  \bibinfo{author}{Lee, H.G.}, \bibinfo{author}{Lee, S.}, \bibinfo{author}{Lee,
  S.}, \bibinfo{author}{Lee, W.W.}, \bibinfo{author}{Oh, B.},
  \bibinfo{author}{Suh, H.S.}, \bibinfo{author}{Parc, Y.W.},
  \bibinfo{author}{Park, S.J.}, \bibinfo{author}{Kim, M.H.},
  \bibinfo{author}{Jung, N.S.}, \bibinfo{author}{Kim, Y.C.},
  \bibinfo{author}{Lee, M.S.}, \bibinfo{author}{Lee, B.H.},
  \bibinfo{author}{Sung, C.W.}, \bibinfo{author}{Mok, I.S.},
  \bibinfo{author}{Yang, J.M.}, \bibinfo{author}{Lee, C.S.},
  \bibinfo{author}{Shin, H.}, \bibinfo{author}{Kim, J.H.},
  \bibinfo{author}{Kim, Y.}, \bibinfo{author}{Lee, J.H.},
  \bibinfo{author}{Park, S.Y.}, \bibinfo{author}{Kim, J.},
  \bibinfo{author}{Park, J.}, \bibinfo{author}{Eom, I.}, \bibinfo{author}{Rah,
  S.}, \bibinfo{author}{Kim, S.}, \bibinfo{author}{Nam, K.H.},
  \bibinfo{author}{Park, J.}, \bibinfo{author}{Park, J.}, \bibinfo{author}{Kim,
  S.}, \bibinfo{author}{Kwon, S.}, \bibinfo{author}{Park, S.H.},
  \bibinfo{author}{Kim, K.S.}, \bibinfo{author}{Hyun, H.},
  \bibinfo{author}{Kim, S.N.}, \bibinfo{author}{Kim, S.},
  \bibinfo{author}{Hwang, S.m.}, \bibinfo{author}{Kim, M.J.},
  \bibinfo{author}{Lim, C.y.}, \bibinfo{author}{Yu, C.J.},
  \bibinfo{author}{Kim, B.S.}, \bibinfo{author}{Kang, T.H.},
  \bibinfo{author}{Kim, K.W.}, \bibinfo{author}{Kim, S.H.},
  \bibinfo{author}{Lee, H.S.}, \bibinfo{author}{Lee, H.S.},
  \bibinfo{author}{Park, K.H.}, \bibinfo{author}{Koo, T.Y.},
  \bibinfo{author}{Kim, D.E.}, \bibinfo{author}{Ko, I.S.},
  \bibinfo{year}{2017}.
\newblock \bibinfo{title}{Hard {X}-ray free-electron laser with
  femtosecond-scale timing jitter}.
\newblock \bibinfo{journal}{Nat. Photonics} \bibinfo{volume}{11},
  \bibinfo{pages}{708--713}.
\newblock \DOIprefix\doi{10.1038/s41566-017-0029-8}.
%Type = Article
\bibitem[{Karamatskou et~al.(2014)Karamatskou, Pabst, Chen and
  Santra}]{Karamatskou2014}
\bibinfo{author}{Karamatskou, A.}, \bibinfo{author}{Pabst, S.},
  \bibinfo{author}{Chen, Y.J.}, \bibinfo{author}{Santra, R.},
  \bibinfo{year}{2014}.
\newblock \bibinfo{title}{Calculation of photoelectron spectra within the
  time-dependent configuration-interaction singles scheme}.
\newblock \bibinfo{journal}{Phys. Rev. A} \bibinfo{volume}{89},
  \bibinfo{pages}{033415}.
\newblock \DOIprefix\doi{10.1103/PhysRevA.89.033415}.
%Type = Article
\bibitem[{Karnakov et~al.(2015)Karnakov, Mur, Popruzhenko and
  Popov}]{Karnakov2015}
\bibinfo{author}{Karnakov, B.M.}, \bibinfo{author}{Mur, V.D.},
  \bibinfo{author}{Popruzhenko, S.V.}, \bibinfo{author}{Popov, V.S.},
  \bibinfo{year}{2015}.
\newblock \bibinfo{title}{Current progress in developing the nonlinear
  ionization theory of atoms and ions}.
\newblock \bibinfo{journal}{Physics-Uspekhi} \bibinfo{volume}{58},
  \bibinfo{pages}{3--32}.
\newblock \DOIprefix\doi{10.3367/ufne.0185.201501b.0003}.
%Type = Article
\bibitem[{Kastirke et~al.(2020)}]{Kastirke2020}
\bibinfo{author}{Kastirke, G.}, et~al., \bibinfo{year}{2020}.
\newblock \bibinfo{title}{Photoelectron diffraction imaging of a molecular
  breakup using an {X}-ray free-electron laser}.
\newblock \bibinfo{journal}{Phys. Rev. X} \bibinfo{volume}{10},
  \bibinfo{pages}{021052}.
\newblock \URLprefix
  \url{https://journals.aps.org/prx/accepted/3e079K1fY5616308f3c282754576ab2113a3a587f}.
%Type = Article
\bibitem[{Kato and Kono(2004)}]{Kato2004}
\bibinfo{author}{Kato, T.}, \bibinfo{author}{Kono, H.}, \bibinfo{year}{2004}.
\newblock \bibinfo{title}{Time-dependent multiconfiguration theory for
  electronic dynamics of molecules in an intense laser field}.
\newblock \bibinfo{journal}{Chem. Phys. Lett.} \bibinfo{volume}{392},
  \bibinfo{pages}{533--540}.
\newblock \DOIprefix\doi{10.1016/j.cplett.2004.05.106}.
%Type = Article
\bibitem[{Kazansky et~al.(2011)Kazansky, Grigorieva and
  Kabachnik}]{Kazansky2011}
\bibinfo{author}{Kazansky, A.K.}, \bibinfo{author}{Grigorieva, A.V.},
  \bibinfo{author}{Kabachnik, N.M.}, \bibinfo{year}{2011}.
\newblock \bibinfo{title}{Circular dichroism in laser-assisted short-pulse
  photoionization}.
\newblock \bibinfo{journal}{Phys. Rev. Lett.} \bibinfo{volume}{107},
  \bibinfo{pages}{253002}.
\newblock \DOIprefix\doi{10.1103/PhysRevLett.107.253002}.
%Type = Article
\bibitem[{Kazansky and Kabachnik(2006)}]{Kazansky2006}
\bibinfo{author}{Kazansky, A.K.}, \bibinfo{author}{Kabachnik, N.M.},
  \bibinfo{year}{2006}.
\newblock \bibinfo{title}{Calculations of the double differential cross section
  for attosecond laser-assisted photoionization of atoms}.
\newblock \bibinfo{journal}{J. Phys. B} \bibinfo{volume}{39},
  \bibinfo{pages}{5173--5186}.
\newblock \DOIprefix\doi{10.1088/0953-4075/39/24/014}.
%Type = Article
\bibitem[{Kazansky and Kabachnik(2007)}]{Kazansky2007}
\bibinfo{author}{Kazansky, A.K.}, \bibinfo{author}{Kabachnik, N.M.},
  \bibinfo{year}{2007}.
\newblock \bibinfo{title}{Theoretical description of atomic photoionization by
  an attosecond {XUV} pulse in a strong laser field: effects of rescattering
  and orbital polarization}.
\newblock \bibinfo{journal}{J. Phys. B} \bibinfo{volume}{40},
  \bibinfo{pages}{2163--2177}.
\newblock \DOIprefix\doi{10.1088/0953-4075/40/11/017}.
%Type = Article
\bibitem[{Kazansky et~al.(2010)Kazansky, Sazhina and Kabachnik}]{Kazansky2010}
\bibinfo{author}{Kazansky, A.K.}, \bibinfo{author}{Sazhina, I.P.},
  \bibinfo{author}{Kabachnik, N.M.}, \bibinfo{year}{2010}.
\newblock \bibinfo{title}{Angle-resolved electron spectra in short-pulse
  two-color {XUV}$+${IR} photoionization of atoms}.
\newblock \bibinfo{journal}{Phys. Rev. A} \bibinfo{volume}{82},
  \bibinfo{pages}{033420}.
\newblock \DOIprefix\doi{10.1103/PhysRevA.82.033420}.
%Type = Article
\bibitem[{Keldysh(1965)}]{Keldysh1965_en}
\bibinfo{author}{Keldysh, L.V.}, \bibinfo{year}{1965}.
\newblock \bibinfo{title}{Ionization in the field of a strong electromagnetic
  wave}.
\newblock \bibinfo{journal}{Sov. Phys. JETP} \bibinfo{volume}{20},
  \bibinfo{pages}{1307--1314}.
\newblock \URLprefix
  \url{http://www.jetp.ac.ru/cgi-bin/e/index/e/20/5/p1307?a=list}.
%Type = Article
\bibitem[{Kheifets(2007)}]{Kheifets2007}
\bibinfo{author}{Kheifets, A.S.}, \bibinfo{year}{2007}.
\newblock \bibinfo{title}{Sequential two-photon double ionization of noble gas
  atoms}.
\newblock \bibinfo{journal}{J. Phys. B} \bibinfo{volume}{40},
  \bibinfo{pages}{F313--F318}.
\newblock \DOIprefix\doi{10.1088/0953-4075/40/22/F02}.
%Type = Article
\bibitem[{Khokhlova et~al.(2019)Khokhlova, Cooper, Ueda, Prince, Koloren\v{c},
  Ivanov and Averbukh}]{Khokhlova2019}
\bibinfo{author}{Khokhlova, M.A.}, \bibinfo{author}{Cooper, B.},
  \bibinfo{author}{Ueda, K.}, \bibinfo{author}{Prince, K.C.},
  \bibinfo{author}{Koloren\v{c}, P.}, \bibinfo{author}{Ivanov, M.Y.},
  \bibinfo{author}{Averbukh, V.}, \bibinfo{year}{2019}.
\newblock \bibinfo{title}{Molecular {A}uger interferometry}.
\newblock \bibinfo{journal}{Phys. Rev. Lett.} \bibinfo{volume}{122},
  \bibinfo{pages}{233001}.
\newblock \DOIprefix\doi{10.1103/PhysRevLett.122.233001}.
%Type = Book
\bibitem[{Kim et~al.(2017)Kim, Huang and Lindberg}]{Kim2017}
\bibinfo{author}{Kim, K.J.}, \bibinfo{author}{Huang, Z.},
  \bibinfo{author}{Lindberg, R.}, \bibinfo{year}{2017}.
\newblock \bibinfo{title}{Synchrotron Radiation and Free-Electron Lasers}.
\newblock \bibinfo{publisher}{Cambridge University Press}.
\newblock \DOIprefix\doi{10.1017/9781316677377}.
%Type = Article
\bibitem[{Kleine et~al.(2019)Kleine, Ekimova, Goldsztejn, Raabe, Str\"uber,
  Ludwig, Yarlagadda, Eisebitt, Vrakking, Elsaesser, Nibbering and
  Rouz{\'{e}}e}]{Kleine2019}
\bibinfo{author}{Kleine, C.}, \bibinfo{author}{Ekimova, M.},
  \bibinfo{author}{Goldsztejn, G.}, \bibinfo{author}{Raabe, S.},
  \bibinfo{author}{Str\"uber, C.}, \bibinfo{author}{Ludwig, J.},
  \bibinfo{author}{Yarlagadda, S.}, \bibinfo{author}{Eisebitt, S.},
  \bibinfo{author}{Vrakking, M.J.J.}, \bibinfo{author}{Elsaesser, T.},
  \bibinfo{author}{Nibbering, E.T.J.}, \bibinfo{author}{Rouz{\'{e}}e, A.},
  \bibinfo{year}{2019}.
\newblock \bibinfo{title}{Soft {X}-ray absorption spectroscopy of aqueous
  solutions using a table-top femtosecond soft {X}-ray source}.
\newblock \bibinfo{journal}{J. Phys. Chem. Lett.} \bibinfo{volume}{10},
  \bibinfo{pages}{52--58}.
\newblock \DOIprefix\doi{10.1021/acs.jpclett.8b03420}.
%Type = Article
\bibitem[{Kondratenko and Saldin(1979)}]{Kondratenko1979}
\bibinfo{author}{Kondratenko, A.M.}, \bibinfo{author}{Saldin, E.L.},
  \bibinfo{year}{1979}.
\newblock \bibinfo{title}{Generation of coherent radiation by relativistic
  electron beam in an ondulator}.
\newblock \bibinfo{journal}{Dokl. Akad. Nauk SSSR} \bibinfo{volume}{249},
  \bibinfo{pages}{843--847}.
\newblock \URLprefix \url{http://mi.mathnet.ru/dan43196}.
%Type = Article
\bibitem[{Kondratenko and Saldin(1980)}]{Kondratenko1980}
\bibinfo{author}{Kondratenko, A.M.}, \bibinfo{author}{Saldin, E.L.},
  \bibinfo{year}{1980}.
\newblock \bibinfo{title}{Generation of coherent radiation by a relativistic
  electron beam in an ondulator}.
\newblock \bibinfo{journal}{Part. Accel.} \bibinfo{volume}{10},
  \bibinfo{pages}{207--216}.
\newblock \URLprefix \url{https://cds.cern.ch/record/1107977}.
%Type = Article
\bibitem[{Kornilov et~al.(2013)Kornilov, Eckstein, Rosenblatt, Schulz,
  Motomura, Rouz{\'{e}}e, Klei, Foucar, Siano, L\"ubcke, Schapper, Johnsson,
  Holland, Schlath\"olter, Marchenko, D\"usterer, Ueda, Vrakking and
  Frasinski}]{Kornilov2013}
\bibinfo{author}{Kornilov, O.}, \bibinfo{author}{Eckstein, M.},
  \bibinfo{author}{Rosenblatt, M.}, \bibinfo{author}{Schulz, C.P.},
  \bibinfo{author}{Motomura, K.}, \bibinfo{author}{Rouz{\'{e}}e, A.},
  \bibinfo{author}{Klei, J.}, \bibinfo{author}{Foucar, L.},
  \bibinfo{author}{Siano, M.}, \bibinfo{author}{L\"ubcke, A.},
  \bibinfo{author}{Schapper, F.}, \bibinfo{author}{Johnsson, P.},
  \bibinfo{author}{Holland, D.M.P.}, \bibinfo{author}{Schlath\"olter, T.},
  \bibinfo{author}{Marchenko, T.}, \bibinfo{author}{D\"usterer, S.},
  \bibinfo{author}{Ueda, K.}, \bibinfo{author}{Vrakking, M.J.J.},
  \bibinfo{author}{Frasinski, L.J.}, \bibinfo{year}{2013}.
\newblock \bibinfo{title}{Coulomb explosion of diatomic molecules in intense
  {XUV} fields mapped by partial covariance}.
\newblock \bibinfo{journal}{J. Phys. B} \bibinfo{volume}{46},
  \bibinfo{pages}{164028}.
\newblock \DOIprefix\doi{10.1088/0953-4075/46/16/164028}.
%Type = Article
\bibitem[{Kosloff et~al.(1997)Kosloff, Rice, Gaspard, Tersigni and
  Tannor}]{Kosloff1989}
\bibinfo{author}{Kosloff, R.}, \bibinfo{author}{Rice, S.A.},
  \bibinfo{author}{Gaspard, P.}, \bibinfo{author}{Tersigni, S.},
  \bibinfo{author}{Tannor, D.J.}, \bibinfo{year}{1997}.
\newblock \bibinfo{title}{Wavepacket dancing: Achieving chemical sensitivity by
  shaping light pulses}.
\newblock \bibinfo{journal}{Chem. Phys.} \bibinfo{volume}{48},
  \bibinfo{pages}{601--641}.
\newblock \DOIprefix\doi{10.1016/0301-0104(89)90012-8}.
%Type = Article
\bibitem[{Kotur et~al.(2016)Kotur, Gu{\'{e}}not, Jim{\'{e}}nez-Gal{\'{a}}n,
  Kroon, Larsen, Louisy, Bengtsson, Miranda, Mauritsson, Arnold, Canton,
  Gisselbrecht, Carette, Dahlstr\"{o}m, Lindroth, Maquet, Argenti,
  Mart{\'{\i}}n and L'Huillier}]{Kotur2016}
\bibinfo{author}{Kotur, M.}, \bibinfo{author}{Gu{\'{e}}not, D.},
  \bibinfo{author}{Jim{\'{e}}nez-Gal{\'{a}}n, {\'{A}}.},
  \bibinfo{author}{Kroon, D.}, \bibinfo{author}{Larsen, E.W.},
  \bibinfo{author}{Louisy, M.}, \bibinfo{author}{Bengtsson, S.},
  \bibinfo{author}{Miranda, M.}, \bibinfo{author}{Mauritsson, J.},
  \bibinfo{author}{Arnold, C.L.}, \bibinfo{author}{Canton, S.E.},
  \bibinfo{author}{Gisselbrecht, M.}, \bibinfo{author}{Carette, T.},
  \bibinfo{author}{Dahlstr\"{o}m, J.M.}, \bibinfo{author}{Lindroth, E.},
  \bibinfo{author}{Maquet, A.}, \bibinfo{author}{Argenti, L.},
  \bibinfo{author}{Mart{\'{\i}}n, F.}, \bibinfo{author}{L'Huillier, A.},
  \bibinfo{year}{2016}.
\newblock \bibinfo{title}{Spectral phase measurement of a {F}ano resonance
  using tunable attosecond pulses}.
\newblock \bibinfo{journal}{Nat. Commun.} \bibinfo{volume}{7},
  \bibinfo{pages}{10566}.
\newblock \DOIprefix\doi{10.1038/ncomms10566}.
%Type = Article
\bibitem[{Kowalewski et~al.(2015)Kowalewski, Bennett, Dorfman and
  Mukamel}]{Kowalewski2015}
\bibinfo{author}{Kowalewski, M.}, \bibinfo{author}{Bennett, K.},
  \bibinfo{author}{Dorfman, K.E.}, \bibinfo{author}{Mukamel, S.},
  \bibinfo{year}{2015}.
\newblock \bibinfo{title}{Catching conical intersections in the act: Monitoring
  transient electronic coherences by attosecond stimulated x-ray {R}aman
  signals}.
\newblock \bibinfo{journal}{Phys. Rev. Lett.} \bibinfo{volume}{115},
  \bibinfo{pages}{193003}.
\newblock \DOIprefix\doi{10.1103/PhysRevLett.115.193003}.
%Type = Article
\bibitem[{Kowalewski et~al.(2017)Kowalewski, Fingerhut, Dorfman, Bennett and
  Mukamel}]{Kowalewski2017}
\bibinfo{author}{Kowalewski, M.}, \bibinfo{author}{Fingerhut, B.P.},
  \bibinfo{author}{Dorfman, K.E.}, \bibinfo{author}{Bennett, K.},
  \bibinfo{author}{Mukamel, S.}, \bibinfo{year}{2017}.
\newblock \bibinfo{title}{Simulating coherent multidimensional spectroscopy of
  nonadiabatic molecular processes: From the infrared to the {X}-ray regime}.
\newblock \bibinfo{journal}{Chem. Rev.} \bibinfo{volume}{117},
  \bibinfo{pages}{12165--12226}.
\newblock \DOIprefix\doi{10.1021/acs.chemrev.7b00081}.
%Type = Article
\bibitem[{Krasniqi et~al.(2010)Krasniqi, Najjari, Str\"uder, Rolles, Voitkiv
  and Ullrich}]{Krasniqi2010}
\bibinfo{author}{Krasniqi, F.}, \bibinfo{author}{Najjari, B.},
  \bibinfo{author}{Str\"uder, L.}, \bibinfo{author}{Rolles, D.},
  \bibinfo{author}{Voitkiv, A.}, \bibinfo{author}{Ullrich, J.},
  \bibinfo{year}{2010}.
\newblock \bibinfo{title}{Imaging molecules from within: Ultrafast
  angstr\"om-scale structure determination of molecules via photoelectron
  holography using free-electron lasers}.
\newblock \bibinfo{journal}{Phys. Rev. A} \bibinfo{volume}{81},
  \bibinfo{pages}{033411}.
\newblock \DOIprefix\doi{10.1103/PhysRevA.81.033411}.
%Type = Article
\bibitem[{Krausz and Ivanov(2009)}]{Krausz2009}
\bibinfo{author}{Krausz, F.}, \bibinfo{author}{Ivanov, M.},
  \bibinfo{year}{2009}.
\newblock \bibinfo{title}{Attosecond physics}.
\newblock \bibinfo{journal}{Rev. Mod. Phys.} \bibinfo{volume}{81},
  \bibinfo{pages}{163--234}.
\newblock \DOIprefix\doi{10.1103/RevModPhys.81.163}.
%Type = Article
\bibitem[{Kroll et~al.(2016)Kroll, Kern, Kubin, Ratner, Gul, Fuller, L\"ochel,
  Krzywinski, Lutman, Ding, Dakovski, Moeller, Turner, Alonso-Mori, Nordlund,
  Rehanek, Weniger, Firsov, Brzhezinskaya, Chatterjee, Lassalle-Kaiser, Sierra,
  Laksmono, Hill, Borovik, Erko, F\"ohlisch, Mitzner, Yachandra, Yano, Wernet
  and Bergmann}]{kroll2016}
\bibinfo{author}{Kroll, T.}, \bibinfo{author}{Kern, J.},
  \bibinfo{author}{Kubin, M.}, \bibinfo{author}{Ratner, D.},
  \bibinfo{author}{Gul, S.}, \bibinfo{author}{Fuller, F.D.},
  \bibinfo{author}{L\"ochel, H.}, \bibinfo{author}{Krzywinski, J.},
  \bibinfo{author}{Lutman, A.}, \bibinfo{author}{Ding, Y.},
  \bibinfo{author}{Dakovski, G.L.}, \bibinfo{author}{Moeller, S.},
  \bibinfo{author}{Turner, J.J.}, \bibinfo{author}{Alonso-Mori, R.},
  \bibinfo{author}{Nordlund, D.L.}, \bibinfo{author}{Rehanek, J.},
  \bibinfo{author}{Weniger, C.}, \bibinfo{author}{Firsov, A.},
  \bibinfo{author}{Brzhezinskaya, M.}, \bibinfo{author}{Chatterjee, R.},
  \bibinfo{author}{Lassalle-Kaiser, B.}, \bibinfo{author}{Sierra, R.G.},
  \bibinfo{author}{Laksmono, H.}, \bibinfo{author}{Hill, E.},
  \bibinfo{author}{Borovik, A.}, \bibinfo{author}{Erko, A.},
  \bibinfo{author}{F\"ohlisch, A.}, \bibinfo{author}{Mitzner, R.},
  \bibinfo{author}{Yachandra, V.K.}, \bibinfo{author}{Yano, J.},
  \bibinfo{author}{Wernet, P.}, \bibinfo{author}{Bergmann, U.},
  \bibinfo{year}{2016}.
\newblock \bibinfo{title}{{X}-ray absorption spectroscopy using a self-seeded
  soft {X}-ray free-electron laser}.
\newblock \bibinfo{journal}{Opt. Express} \bibinfo{volume}{24},
  \bibinfo{pages}{22469}.
\newblock \DOIprefix\doi{10.1364/oe.24.022469}.
%Type = Article
\bibitem[{Kulander(1987)}]{Kulander1987}
\bibinfo{author}{Kulander, K.C.}, \bibinfo{year}{1987}.
\newblock \bibinfo{title}{Multiphoton ionization of hydrogen: A time-dependent
  theory}.
\newblock \bibinfo{journal}{Phys. Rev. A} \bibinfo{volume}{35},
  \bibinfo{pages}{445--447}.
\newblock \DOIprefix\doi{10.1103/PhysRevA.35.445}.
%Type = Article
\bibitem[{Kuleff et~al.(2010)Kuleff, Gokhberg, Kopelke and
  Cederbaum}]{Kuleff2010}
\bibinfo{author}{Kuleff, A.I.}, \bibinfo{author}{Gokhberg, K.},
  \bibinfo{author}{Kopelke, S.}, \bibinfo{author}{Cederbaum, L.S.},
  \bibinfo{year}{2010}.
\newblock \bibinfo{title}{Ultrafast interatomic electronic decay in multiply
  excited clusters}.
\newblock \bibinfo{journal}{Phys. Rev. Lett.} \bibinfo{volume}{105},
  \bibinfo{pages}{043004}.
\newblock \DOIprefix\doi{10.1103/PhysRevLett.105.043004}.
%Type = Article
\bibitem[{Kuleff et~al.(2016)Kuleff, Kryzhevoi, Pernpointner and
  Cederbaum}]{Kuleff2016}
\bibinfo{author}{Kuleff, A.I.}, \bibinfo{author}{Kryzhevoi, N.V.},
  \bibinfo{author}{Pernpointner, M.}, \bibinfo{author}{Cederbaum, L.S.},
  \bibinfo{year}{2016}.
\newblock \bibinfo{title}{Core ionization initiates subfemtosecond charge
  migration in the valence shell of molecules}.
\newblock \bibinfo{journal}{Phys. Rev. Lett.} \bibinfo{volume}{117},
  \bibinfo{pages}{093002}.
\newblock \DOIprefix\doi{10.1103/PhysRevLett.117.093002}.
%Type = Article
\bibitem[{Kumagai et~al.(2018)Kumagai, Fukuzawa, Motomura, Iablonskyi, Nagaya,
  Wada, Ito, Takanashi, Sakakibara, You, Nishiyama, Asa, Sato, Umemoto,
  Kariyazono, Kukk, Kooser, Nicolas, Miron, Asavei, Neagu, Sch\"offler,
  Kastirke, Liu, Owada, Katayama, Togashi, Tono, Yabashi, Golubev, Gokhberg,
  Cederbaum, Kuleff and Ueda}]{Kumagai2018}
\bibinfo{author}{Kumagai, Y.}, \bibinfo{author}{Fukuzawa, H.},
  \bibinfo{author}{Motomura, K.}, \bibinfo{author}{Iablonskyi, D.},
  \bibinfo{author}{Nagaya, K.}, \bibinfo{author}{Wada, S.i.},
  \bibinfo{author}{Ito, Y.}, \bibinfo{author}{Takanashi, T.},
  \bibinfo{author}{Sakakibara, Y.}, \bibinfo{author}{You, D.},
  \bibinfo{author}{Nishiyama, T.}, \bibinfo{author}{Asa, K.},
  \bibinfo{author}{Sato, Y.}, \bibinfo{author}{Umemoto, T.},
  \bibinfo{author}{Kariyazono, K.}, \bibinfo{author}{Kukk, E.},
  \bibinfo{author}{Kooser, K.}, \bibinfo{author}{Nicolas, C.},
  \bibinfo{author}{Miron, C.}, \bibinfo{author}{Asavei, T.},
  \bibinfo{author}{Neagu, L.}, \bibinfo{author}{Sch\"offler, M.S.},
  \bibinfo{author}{Kastirke, G.}, \bibinfo{author}{Liu, X.j.},
  \bibinfo{author}{Owada, S.}, \bibinfo{author}{Katayama, T.},
  \bibinfo{author}{Togashi, T.}, \bibinfo{author}{Tono, K.},
  \bibinfo{author}{Yabashi, M.}, \bibinfo{author}{Golubev, N.V.},
  \bibinfo{author}{Gokhberg, K.}, \bibinfo{author}{Cederbaum, L.S.},
  \bibinfo{author}{Kuleff, A.I.}, \bibinfo{author}{Ueda, K.},
  \bibinfo{year}{2018}.
\newblock \bibinfo{title}{Following the birth of a nanoplasma produced by an
  ultrashort hard-x-ray laser in xenon clusters}.
\newblock \bibinfo{journal}{Phys. Rev. X} \bibinfo{volume}{8},
  \bibinfo{pages}{031034}.
\newblock \DOIprefix\doi{10.1103/PhysRevX.8.031034}.
%Type = Article
\bibitem[{Kurka et~al.(2009)Kurka, Rudenko, Foucar, K{\"u}hnel, Jiang, Ergler,
  Havermeier, Smolarski, Sch{\"o}ssler, Cole, Sch{\"o}ffler, D{\"o}rner,
  Gensch, D{\"u}sterer, Treusch, Fritzsche, Grum-Grzhimailo, Gryzlova,
  Kabachnik, Schr{\"o}ter, Moshammer and Ullrich}]{Kurka2009}
\bibinfo{author}{Kurka, M.}, \bibinfo{author}{Rudenko, A.},
  \bibinfo{author}{Foucar, L.}, \bibinfo{author}{K{\"u}hnel, K.U.},
  \bibinfo{author}{Jiang, Y.H.}, \bibinfo{author}{Ergler, T.},
  \bibinfo{author}{Havermeier, T.}, \bibinfo{author}{Smolarski, M.},
  \bibinfo{author}{Sch{\"o}ssler, S.}, \bibinfo{author}{Cole, K.},
  \bibinfo{author}{Sch{\"o}ffler, M.}, \bibinfo{author}{D{\"o}rner, R.},
  \bibinfo{author}{Gensch, M.}, \bibinfo{author}{D{\"u}sterer, S.},
  \bibinfo{author}{Treusch, R.}, \bibinfo{author}{Fritzsche, S.},
  \bibinfo{author}{Grum-Grzhimailo, A.N.}, \bibinfo{author}{Gryzlova, E.V.},
  \bibinfo{author}{Kabachnik, N.M.}, \bibinfo{author}{Schr{\"o}ter, C.D.},
  \bibinfo{author}{Moshammer, R.}, \bibinfo{author}{Ullrich, J.},
  \bibinfo{year}{2009}.
\newblock \bibinfo{title}{Two-photon double ionization of {Ne} by free-electron
  laser radiation: a kinematically complete experiment}.
\newblock \bibinfo{journal}{J. Phys. B} \bibinfo{volume}{42},
  \bibinfo{pages}{141002}.
\newblock \DOIprefix\doi{10.1088/0953-4075/42/14/141002}.
%Type = Article
\bibitem[{Kvaal(2012)}]{Kvaal2012}
\bibinfo{author}{Kvaal, S.}, \bibinfo{year}{2012}.
\newblock \bibinfo{title}{Ab initio quantum dynamics using coupled-cluster}.
\newblock \bibinfo{journal}{J. Chem. Phys.} \bibinfo{volume}{136},
  \bibinfo{pages}{194109}.
\newblock \DOIprefix\doi{10.1063/1.4718427}.
%Type = Article
\bibitem[{Lablanquie et~al.(2011)Lablanquie, Grozdanov, \v{Z}itnik, Carniato,
  Selles, Andric, Palaudoux, Penent, Iwayama, Shigemasa, Hikosaka, Soejima,
  Nakano, Suzuki and Ito}]{Lablanquie2011}
\bibinfo{author}{Lablanquie, P.}, \bibinfo{author}{Grozdanov, T.P.},
  \bibinfo{author}{\v{Z}itnik, M.}, \bibinfo{author}{Carniato, S.},
  \bibinfo{author}{Selles, P.}, \bibinfo{author}{Andric, L.},
  \bibinfo{author}{Palaudoux, J.}, \bibinfo{author}{Penent, F.},
  \bibinfo{author}{Iwayama, H.}, \bibinfo{author}{Shigemasa, E.},
  \bibinfo{author}{Hikosaka, Y.}, \bibinfo{author}{Soejima, K.},
  \bibinfo{author}{Nakano, M.}, \bibinfo{author}{Suzuki, I.H.},
  \bibinfo{author}{Ito, K.}, \bibinfo{year}{2011}.
\newblock \bibinfo{title}{Evidence of single-photon two-site core double
  ionization of {C}$_{2}${H}$_{2}$ molecules}.
\newblock \bibinfo{journal}{Phys. Rev. Lett.} \bibinfo{volume}{107},
  \bibinfo{pages}{193004}.
\newblock \DOIprefix\doi{10.1103/PhysRevLett.107.193004}.
%Type = Article
\bibitem[{LaForge et~al.(2014)LaForge, Drabbels, Brauer, Coreno, Devetta,
  Di~Fraia, Finetti, Grazioli, Katzy, Lyamayev, Mazza, Mudrich, O'Keeffe,
  Ovcharenko, Piseri, Plekan, Prince, Richter, Stranges, Callegari, M{\"o}ller
  and Stienkemeier}]{LaForge2014}
\bibinfo{author}{LaForge, A.C.}, \bibinfo{author}{Drabbels, M.},
  \bibinfo{author}{Brauer, N.B.}, \bibinfo{author}{Coreno, M.},
  \bibinfo{author}{Devetta, M.}, \bibinfo{author}{Di~Fraia, M.},
  \bibinfo{author}{Finetti, P.}, \bibinfo{author}{Grazioli, C.},
  \bibinfo{author}{Katzy, R.}, \bibinfo{author}{Lyamayev, V.},
  \bibinfo{author}{Mazza, T.}, \bibinfo{author}{Mudrich, M.},
  \bibinfo{author}{O'Keeffe, P.}, \bibinfo{author}{Ovcharenko, Y.},
  \bibinfo{author}{Piseri, P.}, \bibinfo{author}{Plekan, O.},
  \bibinfo{author}{Prince, K.C.}, \bibinfo{author}{Richter, R.},
  \bibinfo{author}{Stranges, S.}, \bibinfo{author}{Callegari, C.},
  \bibinfo{author}{M{\"o}ller, T.}, \bibinfo{author}{Stienkemeier, F.},
  \bibinfo{year}{2014}.
\newblock \bibinfo{title}{Collective autoionization in multiply-excited
  systems: A novel ionization process observed in helium nanodroplets}.
\newblock \bibinfo{journal}{Sci. Rep.} \bibinfo{volume}{4},
  \bibinfo{pages}{3621}.
\newblock \DOIprefix\doi{10.1038/srep03621}.
%Type = Article
\bibitem[{Lam et~al.(2018)Lam, Raj, Pascal, Pemmaraju, Foglia, Simoncig,
  Fabris, Miotti, Hull, Rizzuto, Smith, Mincigrucci, Masciovecchio, Gessini,
  Allaria, De~Ninno, Diviacco, Roussel, Spampinati, Penco, Di~Mitri, Trov\`o,
  Danailov, Christensen, Sokaras, Weng, Coreno, Poletto, Drisdell, Prendergast,
  Giannessi, Principi, Nordlund, Saykally and Schwartz}]{Lam2018}
\bibinfo{author}{Lam, R.K.}, \bibinfo{author}{Raj, S.L.},
  \bibinfo{author}{Pascal, T.A.}, \bibinfo{author}{Pemmaraju, C.D.},
  \bibinfo{author}{Foglia, L.}, \bibinfo{author}{Simoncig, A.},
  \bibinfo{author}{Fabris, N.}, \bibinfo{author}{Miotti, P.},
  \bibinfo{author}{Hull, C.J.}, \bibinfo{author}{Rizzuto, A.M.},
  \bibinfo{author}{Smith, J.W.}, \bibinfo{author}{Mincigrucci, R.},
  \bibinfo{author}{Masciovecchio, C.}, \bibinfo{author}{Gessini, A.},
  \bibinfo{author}{Allaria, E.}, \bibinfo{author}{De~Ninno, G.},
  \bibinfo{author}{Diviacco, B.}, \bibinfo{author}{Roussel, E.},
  \bibinfo{author}{Spampinati, S.}, \bibinfo{author}{Penco, G.},
  \bibinfo{author}{Di~Mitri, S.}, \bibinfo{author}{Trov\`o, M.},
  \bibinfo{author}{Danailov, M.}, \bibinfo{author}{Christensen, S.T.},
  \bibinfo{author}{Sokaras, D.}, \bibinfo{author}{Weng, T.C.},
  \bibinfo{author}{Coreno, M.}, \bibinfo{author}{Poletto, L.},
  \bibinfo{author}{Drisdell, W.S.}, \bibinfo{author}{Prendergast, D.},
  \bibinfo{author}{Giannessi, L.}, \bibinfo{author}{Principi, E.},
  \bibinfo{author}{Nordlund, D.}, \bibinfo{author}{Saykally, R.J.},
  \bibinfo{author}{Schwartz, C.P.}, \bibinfo{year}{2018}.
\newblock \bibinfo{title}{Soft {X}-ray second harmonic generation as an
  interfacial probe}.
\newblock \bibinfo{journal}{Phys. Rev. Lett.} \bibinfo{volume}{120},
  \bibinfo{pages}{023901}.
\newblock \DOIprefix\doi{10.1103/PhysRevLett.120.023901}.
%Type = Article
\bibitem[{Lambropoulos et~al.(1998)Lambropoulos, Maragakis and
  Zhang}]{Lambropoulos1998}
\bibinfo{author}{Lambropoulos, P.}, \bibinfo{author}{Maragakis, P.},
  \bibinfo{author}{Zhang, J.}, \bibinfo{year}{1998}.
\newblock \bibinfo{title}{Two-electron atoms in strong fields}.
\newblock \bibinfo{journal}{Phys. Rep.} \bibinfo{volume}{305},
  \bibinfo{pages}{203--293}.
\newblock \DOIprefix\doi{10.1016/s0370-1573(98)00027-1}.
%Type = Article
\bibitem[{Lambropoulos and Nikolopoulos(2013)}]{Lambropoulos2013}
\bibinfo{author}{Lambropoulos, P.}, \bibinfo{author}{Nikolopoulos, G.M.},
  \bibinfo{year}{2013}.
\newblock \bibinfo{title}{Multiple ionization under strong {XUV} to {X}-ray
  radiation}.
\newblock \bibinfo{journal}{Eur. Phys. J. Special Topics}
  \bibinfo{volume}{222}, \bibinfo{pages}{2067--2084}.
\newblock \DOIprefix\doi{10.1140/epjst/e2013-01987-7}.
%Type = Book
\bibitem[{Lambropoulos and Petrosyan(2007)}]{Lambropoulos2007}
\bibinfo{author}{Lambropoulos, P.}, \bibinfo{author}{Petrosyan, D.},
  \bibinfo{year}{2007}.
\newblock \bibinfo{title}{Fundamentals of Quantum Optics and Quantum
  Information}.
\newblock \bibinfo{publisher}{Springer}, \bibinfo{address}{Berlin Heidelberg}.
\newblock \DOIprefix\doi{10.1007/978-3-540-34572-5}.
%Type = Article
\bibitem[{Leitner et~al.(2018)Leitner, Josefsson, Mazza, Miedema, Schr\"oder,
  Beye, Kunnus, Schreck, D\"usterer, F\"ohlisch, Meyer, Odelius and
  Wernet}]{Leitner2018}
\bibinfo{author}{Leitner, T.}, \bibinfo{author}{Josefsson, I.},
  \bibinfo{author}{Mazza, T.}, \bibinfo{author}{Miedema, P.S.},
  \bibinfo{author}{Schr\"oder, H.}, \bibinfo{author}{Beye, M.},
  \bibinfo{author}{Kunnus, K.}, \bibinfo{author}{Schreck, S.},
  \bibinfo{author}{D\"usterer, S.}, \bibinfo{author}{F\"ohlisch, A.},
  \bibinfo{author}{Meyer, M.}, \bibinfo{author}{Odelius, M.},
  \bibinfo{author}{Wernet, P.}, \bibinfo{year}{2018}.
\newblock \bibinfo{title}{Time-resolved electron spectroscopy for chemical
  analysis of photodissociation: Photoelectron spectra of {Fe}({CO})$_5$,
  {Fe}({CO})$_4$, and {Fe}({CO})$_3$}.
\newblock \bibinfo{journal}{J. Chem. Phys.} \bibinfo{volume}{149},
  \bibinfo{pages}{044307}.
\newblock \DOIprefix\doi{10.1063/1.5035149}.
%Type = Article
\bibitem[{Lewenstein et~al.(1994)Lewenstein, Balcou, Ivanov, L'Huillier and
  Corkum}]{Lewenstein1994}
\bibinfo{author}{Lewenstein, M.}, \bibinfo{author}{Balcou, P.},
  \bibinfo{author}{Ivanov, M.Y.}, \bibinfo{author}{L'Huillier, A.},
  \bibinfo{author}{Corkum, P.B.}, \bibinfo{year}{1994}.
\newblock \bibinfo{title}{Theory of high-harmonic generation by low-frequency
  laser fields}.
\newblock \bibinfo{journal}{Phys. Rev. A} \bibinfo{volume}{49},
  \bibinfo{pages}{2117--2132}.
\newblock \DOIprefix\doi{10.1103/PhysRevA.49.2117}.
%Type = Article
\bibitem[{Li et~al.(2018)Li, Guo, Coffee, Hegazy, Huang, Natan, Osipov, Ray,
  Marinelli and Cryan}]{Li_OE_2018}
\bibinfo{author}{Li, S.}, \bibinfo{author}{Guo, Z.}, \bibinfo{author}{Coffee,
  R.N.}, \bibinfo{author}{Hegazy, K.}, \bibinfo{author}{Huang, Z.},
  \bibinfo{author}{Natan, A.}, \bibinfo{author}{Osipov, T.},
  \bibinfo{author}{Ray, D.}, \bibinfo{author}{Marinelli, A.},
  \bibinfo{author}{Cryan, J.P.}, \bibinfo{year}{2018}.
\newblock \bibinfo{title}{Characterizing isolated attosecond pulses with
  angular streaking}.
\newblock \bibinfo{journal}{Opt. Express} \bibinfo{volume}{26},
  \bibinfo{pages}{4531--4547}.
\newblock \DOIprefix\doi{10.1364/OE.26.004531}.
%Type = Article
\bibitem[{Li et~al.(2014)Li, El-Amine~Madjet, Vendrell and Santra}]{Li2014}
\bibinfo{author}{Li, Z.}, \bibinfo{author}{El-Amine~Madjet, M.},
  \bibinfo{author}{Vendrell, O.}, \bibinfo{author}{Santra, R.},
  \bibinfo{year}{2014}.
\newblock \bibinfo{title}{Core-level transient absorption spectroscopy as a
  probe of electron hole relaxation in photoionized
  \ensuremath{{\mathrm{H}}^{+}({\mathrm{H}}_{2}{\mathrm{O}})_{n}}}.
\newblock \bibinfo{journal}{Faraday Discuss.} \bibinfo{volume}{171},
  \bibinfo{pages}{457--470}.
\newblock \DOIprefix\doi{10.1039/c4fd00078a}.
%Type = Article
\bibitem[{Liekhus-Schmaltz et~al.(2015)Liekhus-Schmaltz, Tenney, Osipov,
  Sanchez-Gonzalez, Berrah, Boll, Bomme, Bostedt, Bozek, Carron, Coffee, Devin,
  Erk, Ferguson, Field, Foucar, Frasinski, Glownia, G\"uhr, Kamalov,
  Krzywinski, Li, Marangos, Martinez, McFarland, Miyabe, Murphy, Natan, Rolles,
  Rudenko, Siano, Simpson, Spector, Swiggers, Walke, Wang, Weber, Bucksbaum and
  Petrovic}]{Liekhus-Schmaltz2015}
\bibinfo{author}{Liekhus-Schmaltz, C.E.}, \bibinfo{author}{Tenney, I.},
  \bibinfo{author}{Osipov, T.}, \bibinfo{author}{Sanchez-Gonzalez, A.},
  \bibinfo{author}{Berrah, N.}, \bibinfo{author}{Boll, R.},
  \bibinfo{author}{Bomme, C.}, \bibinfo{author}{Bostedt, C.},
  \bibinfo{author}{Bozek, J.D.}, \bibinfo{author}{Carron, S.},
  \bibinfo{author}{Coffee, R.}, \bibinfo{author}{Devin, J.},
  \bibinfo{author}{Erk, B.}, \bibinfo{author}{Ferguson, K.R.},
  \bibinfo{author}{Field, R.W.}, \bibinfo{author}{Foucar, L.},
  \bibinfo{author}{Frasinski, L.J.}, \bibinfo{author}{Glownia, J.M.},
  \bibinfo{author}{G\"uhr, M.}, \bibinfo{author}{Kamalov, A.},
  \bibinfo{author}{Krzywinski, J.}, \bibinfo{author}{Li, H.},
  \bibinfo{author}{Marangos, J.P.}, \bibinfo{author}{Martinez, T.J.},
  \bibinfo{author}{McFarland, B.K.}, \bibinfo{author}{Miyabe, S.},
  \bibinfo{author}{Murphy, B.}, \bibinfo{author}{Natan, A.},
  \bibinfo{author}{Rolles, D.}, \bibinfo{author}{Rudenko, A.},
  \bibinfo{author}{Siano, M.}, \bibinfo{author}{Simpson, E.R.},
  \bibinfo{author}{Spector, L.}, \bibinfo{author}{Swiggers, M.},
  \bibinfo{author}{Walke, D.}, \bibinfo{author}{Wang, S.},
  \bibinfo{author}{Weber, T.}, \bibinfo{author}{Bucksbaum, P.H.},
  \bibinfo{author}{Petrovic, V.S.}, \bibinfo{year}{2015}.
\newblock \bibinfo{title}{Ultrafast isomerization initiated by {X}-ray core
  ionization}.
\newblock \bibinfo{journal}{Nat. Commun.} \bibinfo{volume}{6},
  \bibinfo{pages}{8199}.
\newblock \DOIprefix\doi{10.1038/ncomms9199}.
%Type = Article
\bibitem[{Liu et~al.(2016)Liu, Berrah, Cederbaum, Cryan, Glownia, Schafer and
  Buth}]{Liu2016}
\bibinfo{author}{Liu, J.C.}, \bibinfo{author}{Berrah, N.},
  \bibinfo{author}{Cederbaum, L.S.}, \bibinfo{author}{Cryan, J.P.},
  \bibinfo{author}{Glownia, J.M.}, \bibinfo{author}{Schafer, K.J.},
  \bibinfo{author}{Buth, C.}, \bibinfo{year}{2016}.
\newblock \bibinfo{title}{Rate equations for nitrogen molecules in ultrashort
  and intense {x}-ray pulses}.
\newblock \bibinfo{journal}{J. Phys. B} \bibinfo{volume}{49},
  \bibinfo{pages}{075602}.
\newblock \DOIprefix\doi{10.1088/0953-4075/49/7/075602}.
%Type = Article
\bibitem[{Liu et~al.(2019)Liu, Miron, \AA{}gren, Polyutov and
  Gel'mukhanov}]{Liu2019}
\bibinfo{author}{Liu, J.C.}, \bibinfo{author}{Miron, C.},
  \bibinfo{author}{\AA{}gren, H.}, \bibinfo{author}{Polyutov, S.},
  \bibinfo{author}{Gel'mukhanov, F.}, \bibinfo{year}{2019}.
\newblock \bibinfo{title}{Resonant x-ray second-harmonic generation in atomic
  gases}.
\newblock \bibinfo{journal}{Phys. Rev. A} \bibinfo{volume}{100},
  \bibinfo{pages}{063403}.
\newblock \DOIprefix\doi{10.1103/PhysRevA.100.063403}.
%Type = Article
\bibitem[{L\'opez-Martens et~al.(2005)L\'opez-Martens, Varj\'u, Johnsson,
  Mauritsson, Mairesse, Sali\`eres, Gaarde, Schafer, Persson, Svanberg,
  Wahlstr\"om and L'Huillier}]{Martens2005}
\bibinfo{author}{L\'opez-Martens, R.}, \bibinfo{author}{Varj\'u, K.},
  \bibinfo{author}{Johnsson, P.}, \bibinfo{author}{Mauritsson, J.},
  \bibinfo{author}{Mairesse, Y.}, \bibinfo{author}{Sali\`eres, P.},
  \bibinfo{author}{Gaarde, M.B.}, \bibinfo{author}{Schafer, K.J.},
  \bibinfo{author}{Persson, A.}, \bibinfo{author}{Svanberg, S.},
  \bibinfo{author}{Wahlstr\"om, C.G.}, \bibinfo{author}{L'Huillier, A.},
  \bibinfo{year}{2005}.
\newblock \bibinfo{title}{Amplitude and phase control of attosecond light
  pulses}.
\newblock \bibinfo{journal}{Phys. Rev. Lett.} \bibinfo{volume}{94},
  \bibinfo{pages}{033001}.
\newblock \DOIprefix\doi{10.1103/PhysRevLett.94.033001}.
%Type = Article
\bibitem[{Lorenz et~al.(2012)Lorenz, Kabachnik, Weckert and
  Vartanyants}]{Lorenz2012}
\bibinfo{author}{Lorenz, U.}, \bibinfo{author}{Kabachnik, N.M.},
  \bibinfo{author}{Weckert, E.}, \bibinfo{author}{Vartanyants, I.A.},
  \bibinfo{year}{2012}.
\newblock \bibinfo{title}{Impact of ultrafast electronic damage in
  single-particle x-ray imaging experiments}.
\newblock \bibinfo{journal}{Phys. Rev. E} \bibinfo{volume}{86},
  \bibinfo{pages}{051911}.
\newblock \DOIprefix\doi{10.1103/PhysRevE.86.051911}.
%Type = Article
\bibitem[{Lucchese and McKoy(1983)}]{Lucchese1983}
\bibinfo{author}{Lucchese, R.R.}, \bibinfo{author}{McKoy, V.},
  \bibinfo{year}{1983}.
\newblock \bibinfo{title}{Pad\'e-approximant corrections to general variational
  expressions of scattering theory: Application to $5\ensuremath{\sigma}$
  photoionization of carbon monoxide}.
\newblock \bibinfo{journal}{Phys. Rev. A} \bibinfo{volume}{28},
  \bibinfo{pages}{1382--1394}.
\newblock \DOIprefix\doi{10.1103/PhysRevA.28.1382}.
%Type = Article
\bibitem[{Lunin et~al.(2015)Lunin, Grum-Grzhimailo, Gryzlova, Sinitsyn,
  Petrova, Lunina, Balabaev, Tereshkina, Stepanov and Krupyanskii}]{Lunin2015}
\bibinfo{author}{Lunin, V.Y.}, \bibinfo{author}{Grum-Grzhimailo, A.N.},
  \bibinfo{author}{Gryzlova, E.V.}, \bibinfo{author}{Sinitsyn, D.O.},
  \bibinfo{author}{Petrova, T.E.}, \bibinfo{author}{Lunina, N.L.},
  \bibinfo{author}{Balabaev, N.K.}, \bibinfo{author}{Tereshkina, K.B.},
  \bibinfo{author}{Stepanov, A.S.}, \bibinfo{author}{Krupyanskii, Y.F.},
  \bibinfo{year}{2015}.
\newblock \bibinfo{title}{Efficient calculation of diffracted intensities in
  the case of nonstationary scattering by biological macromolecules under
  {XFEL} pulses}.
\newblock \bibinfo{journal}{Acta Cryst. D} \bibinfo{volume}{71},
  \bibinfo{pages}{293--303}.
\newblock \DOIprefix\doi{10.1107/S1399004714025450}.
%Type = Article
\bibitem[{Lysaght et~al.(2008)Lysaght, Burke and van~der Hart}]{Lysaght2008}
\bibinfo{author}{Lysaght, M.A.}, \bibinfo{author}{Burke, P.G.},
  \bibinfo{author}{van~der Hart, H.W.}, \bibinfo{year}{2008}.
\newblock \bibinfo{title}{Ultrafast laser-driven excitation dynamics in {Ne}:
  An \emph{ab initio} time-dependent {R}-matrix approach}.
\newblock \bibinfo{journal}{Phys. Rev. Lett.} \bibinfo{volume}{101},
  \bibinfo{pages}{253001}.
\newblock \DOIprefix\doi{10.1103/PhysRevLett.101.253001}.
%Type = Article
\bibitem[{Lysaght et~al.(2009)Lysaght, van~der Hart and Burke}]{Lysaght2009}
\bibinfo{author}{Lysaght, M.A.}, \bibinfo{author}{van~der Hart, H.W.},
  \bibinfo{author}{Burke, P.G.}, \bibinfo{year}{2009}.
\newblock \bibinfo{title}{Time-dependent {R}-matrix theory for ultrafast atomic
  processes}.
\newblock \bibinfo{journal}{Phys. Rev. A} \bibinfo{volume}{79},
  \bibinfo{pages}{053411}.
\newblock \DOIprefix\doi{10.1103/PhysRevA.79.053411}.
%Type = Article
\bibitem[{Ma et~al.(2013)Ma, Motomura, Ishikawa, Mondal, Fukuzawa, Yamada,
  Ueda, Nagaya, Yase, Mizoguchi, Yao, Rouze, Hundermark, Vrakking, Johnsson,
  Nagasono, Tono, Togashi, Senba, Ohashi, Yabashi and
  Ishikawa}]{Ma_JPhysB_2013}
\bibinfo{author}{Ma, R.}, \bibinfo{author}{Motomura, K.},
  \bibinfo{author}{Ishikawa, K.L.}, \bibinfo{author}{Mondal, S.},
  \bibinfo{author}{Fukuzawa, H.}, \bibinfo{author}{Yamada, A.},
  \bibinfo{author}{Ueda, K.}, \bibinfo{author}{Nagaya, K.},
  \bibinfo{author}{Yase, S.}, \bibinfo{author}{Mizoguchi, Y.},
  \bibinfo{author}{Yao, M.}, \bibinfo{author}{Rouze, A.},
  \bibinfo{author}{Hundermark, A.}, \bibinfo{author}{Vrakking, M.J.J.},
  \bibinfo{author}{Johnsson, P.}, \bibinfo{author}{Nagasono, M.},
  \bibinfo{author}{Tono, K.}, \bibinfo{author}{Togashi, T.},
  \bibinfo{author}{Senba, Y.}, \bibinfo{author}{Ohashi, H.},
  \bibinfo{author}{Yabashi, M.}, \bibinfo{author}{Ishikawa, T.},
  \bibinfo{year}{2013}.
\newblock \bibinfo{title}{Photoelectron angular distributions for the
  two-photon ionization of helium by ultrashort extreme ultraviolet
  free-electron laser pulses}.
\newblock \bibinfo{journal}{J. Phys. B} \bibinfo{volume}{46},
  \bibinfo{pages}{164018}.
\newblock \DOIprefix\doi{10.1088/0953-4075/46/16/164018}.
%Type = Article
\bibitem[{Machado and Masili(2004)}]{Machado2004}
\bibinfo{author}{Machado, A.M.}, \bibinfo{author}{Masili, M.},
  \bibinfo{year}{2004}.
\newblock \bibinfo{title}{Variationally stable calculations for molecular
  systems: Polarizabilities and two-photon ionization cross section for the
  hydrogen molecule}.
\newblock \bibinfo{journal}{J. Chem. Phys.} \bibinfo{volume}{120},
  \bibinfo{pages}{7505--7511}.
\newblock \DOIprefix\doi{10.1063/1.1687677}.
%Type = Article
\bibitem[{Madden and Codling(1963)}]{Madden1963}
\bibinfo{author}{Madden, R.P.}, \bibinfo{author}{Codling, K.},
  \bibinfo{year}{1963}.
\newblock \bibinfo{title}{New autoionizing atomic energy levels in {He}, {Ne},
  and {Ar}}.
\newblock \bibinfo{journal}{Phys. Rev. Lett.} \bibinfo{volume}{10},
  \bibinfo{pages}{516--518}.
\newblock \DOIprefix\doi{10.1103/PhysRevLett.10.516}.
%Type = Article
\bibitem[{Madden et~al.(1969)Madden, Ederer and Codling}]{Madden1969}
\bibinfo{author}{Madden, R.P.}, \bibinfo{author}{Ederer, D.L.},
  \bibinfo{author}{Codling, K.}, \bibinfo{year}{1969}.
\newblock \bibinfo{title}{Resonances in the photo-ionization continuum of {Ar}
  \textsc{i} (20--150 {eV})}.
\newblock \bibinfo{journal}{Phys. Rev.} \bibinfo{volume}{177},
  \bibinfo{pages}{136--151}.
\newblock \DOIprefix\doi{10.1103/PhysRev.177.136}.
%Type = Article
\bibitem[{Madey(1971)}]{Madey1971}
\bibinfo{author}{Madey, J.M.J.}, \bibinfo{year}{1971}.
\newblock \bibinfo{title}{Stimulated emission of {B}remsstrahlung in a periodic
  magnetic field}.
\newblock \bibinfo{journal}{J. Appl. Phys.} \bibinfo{volume}{42},
  \bibinfo{pages}{1906--1913}.
\newblock \DOIprefix\doi{10.1063/1.1660466}.
%Type = Article
\bibitem[{Madey(2014)}]{Madey2014}
\bibinfo{author}{Madey, J.M.J.}, \bibinfo{year}{2014}.
\newblock \bibinfo{title}{Wilson prize article: From vacuum tubes to lasers and
  back again}.
\newblock \bibinfo{journal}{Phys. Rev. ST Accel. Beams} \bibinfo{volume}{17},
  \bibinfo{pages}{074901}.
\newblock \DOIprefix\doi{10.1103/PhysRevSTAB.17.074901}.
%Type = Article
\bibitem[{Madey et~al.(1973)Madey, Schwettman and Fairbank}]{Madey1973}
\bibinfo{author}{Madey, J.M.J.}, \bibinfo{author}{Schwettman, H.A.},
  \bibinfo{author}{Fairbank, W.M.}, \bibinfo{year}{1973}.
\newblock \bibinfo{title}{A free electron laser}.
\newblock \bibinfo{journal}{{IEEE} Trans. Nucl. Sci.} \bibinfo{volume}{20},
  \bibinfo{pages}{980--983}.
\newblock \DOIprefix\doi{10.1109/tns.1973.4327304}.
%Type = Article
\bibitem[{Maeda et~al.(1993)Maeda, Ueda and Ito}]{Maeda1993}
\bibinfo{author}{Maeda, K.}, \bibinfo{author}{Ueda, K.}, \bibinfo{author}{Ito,
  K.}, \bibinfo{year}{1993}.
\newblock \bibinfo{title}{High-resolution measurement for photoabsorption cross
  sections in the autoionization regions of {Ar}, {Kr} and {Xe}}.
\newblock \bibinfo{journal}{J. Phys. B} \bibinfo{volume}{26},
  \bibinfo{pages}{1541--1555}.
\newblock \DOIprefix\doi{10.1088/0953-4075/26/9/003}.
%Type = Article
\bibitem[{Maiman(1960)}]{Maiman1960}
\bibinfo{author}{Maiman, T.H.}, \bibinfo{year}{1960}.
\newblock \bibinfo{title}{Stimulated optical radiation in ruby}.
\newblock \bibinfo{journal}{Nature} \bibinfo{volume}{187},
  \bibinfo{pages}{493--494}.
\newblock \DOIprefix\doi{10.1038/187493a0}.
%Type = Article
\bibitem[{Makris et~al.(2009)Makris, Lambropoulos and Miheli\v{c}}]{Makris2009}
\bibinfo{author}{Makris, M.G.}, \bibinfo{author}{Lambropoulos, P.},
  \bibinfo{author}{Miheli\v{c}, A.}, \bibinfo{year}{2009}.
\newblock \bibinfo{title}{Theory of multiphoton multielectron ionization of
  xenon under strong 93-{eV} radiation}.
\newblock \bibinfo{journal}{Phys. Rev. Lett.} \bibinfo{volume}{102},
  \bibinfo{pages}{033002}.
\newblock \DOIprefix\doi{10.1103/PhysRevLett.102.033002}.
%Type = Article
\bibitem[{Manakov et~al.(1986)Manakov, Ovsiannikov and Rapoport}]{Manakov1986}
\bibinfo{author}{Manakov, N.L.}, \bibinfo{author}{Ovsiannikov, V.D.},
  \bibinfo{author}{Rapoport, L.P.}, \bibinfo{year}{1986}.
\newblock \bibinfo{title}{Atoms in a laser field}.
\newblock \bibinfo{journal}{Phys. Rep.} \bibinfo{volume}{141},
  \bibinfo{pages}{320--433}.
\newblock \DOIprefix\doi{10.1016/s0370-1573(86)80001-1}.
%Type = Article
\bibitem[{Maquet et~al.(2004)Maquet, Caillat and Ta\"\i{}eb}]{Maquet2014}
\bibinfo{author}{Maquet, A.}, \bibinfo{author}{Caillat, J.},
  \bibinfo{author}{Ta\"\i{}eb, R.}, \bibinfo{year}{2004}.
\newblock \bibinfo{title}{Attosecond delays in photoionization: time and
  quantum mechanics}.
\newblock \bibinfo{journal}{J. Phys. B} \bibinfo{volume}{47},
  \bibinfo{pages}{204004}.
\newblock \DOIprefix\doi{10.1088/0953-4075/47/20/204004}.
%Type = Article
\bibitem[{Maquet and Ta\"\i{}eb(2007)}]{Maquet2007}
\bibinfo{author}{Maquet, A.}, \bibinfo{author}{Ta\"\i{}eb, R.},
  \bibinfo{year}{2007}.
\newblock \bibinfo{title}{Two-colour {IR}+{XUV} spectroscopies: the
  ``soft-photon approximation''}.
\newblock \bibinfo{journal}{J. Mod. Opt.} \bibinfo{volume}{54},
  \bibinfo{pages}{1847--1857}.
\newblock \DOIprefix\doi{10.1080/09500340701306751}.
%Type = Article
\bibitem[{Maquet et~al.(1998)Maquet, V\'eniard and Marian}]{Maquet1998}
\bibinfo{author}{Maquet, A.}, \bibinfo{author}{V\'eniard, V.},
  \bibinfo{author}{Marian, T.A.}, \bibinfo{year}{1998}.
\newblock \bibinfo{title}{The {C}oulomb {G}reen's function and multiphoton
  calculations}.
\newblock \bibinfo{journal}{J. Phys. B} \bibinfo{volume}{31},
  \bibinfo{pages}{3743--3764}.
\newblock \DOIprefix\doi{10.1088/0953-4075/31/17/004}.
%Type = Article
\bibitem[{Marcus et~al.(2019)Marcus, Fawley, Bohler, Ding, Feng, Hemsing,
  Huang, Krzywinski, Lutman and Ratner}]{Marcus2019}
\bibinfo{author}{Marcus, G.}, \bibinfo{author}{Fawley, W.M.},
  \bibinfo{author}{Bohler, D.}, \bibinfo{author}{Ding, Y.},
  \bibinfo{author}{Feng, Y.}, \bibinfo{author}{Hemsing, E.},
  \bibinfo{author}{Huang, Z.}, \bibinfo{author}{Krzywinski, J.},
  \bibinfo{author}{Lutman, A.}, \bibinfo{author}{Ratner, D.},
  \bibinfo{year}{2019}.
\newblock \bibinfo{title}{Experimental observations of seed growth and
  accompanying pedestal contamination in a self-seeded, soft {x}-ray
  free-electron laser}.
\newblock \bibinfo{journal}{Phys. Rev. Accel. Beams} \bibinfo{volume}{22},
  \bibinfo{pages}{080702}.
\newblock \DOIprefix\doi{10.1103/PhysRevAccelBeams.22.080702}.
%Type = Article
\bibitem[{Maroju et~al.(2020)Maroju, Grazioli, Di~Fraia, Moioli, Ertel, Ahmadi,
  Plekan, Finetti, Allaria, Giannessi, De~Ninno, Spezzani, Penco, Spampinati,
  Demidovich, Danailov, Borghes, Kourousias, Sanches Dos~Reis, Billé, Lutman,
  Squibb, Feifel, Carpeggiani, Reduzzi, Mazza, Meyer, Bengtsson, Ibrakovic,
  Simpson, Mauritsson, Csizmadia, Dumergue, K{\"u}hn, Nandiga~Gopalakrishna,
  You, Ueda, Labeye, B\ae{}kh\o{}j, Schafer, Gryzlova, Grum-Grzhimailo, Prince,
  Callegari and Sansone}]{Maroju2020}
\bibinfo{author}{Maroju, P.K.}, \bibinfo{author}{Grazioli, C.},
  \bibinfo{author}{Di~Fraia, M.}, \bibinfo{author}{Moioli, M.},
  \bibinfo{author}{Ertel, D.}, \bibinfo{author}{Ahmadi, H.},
  \bibinfo{author}{Plekan, O.}, \bibinfo{author}{Finetti, P.},
  \bibinfo{author}{Allaria, E.}, \bibinfo{author}{Giannessi, L.},
  \bibinfo{author}{De~Ninno, G.}, \bibinfo{author}{Spezzani, C.},
  \bibinfo{author}{Penco, G.}, \bibinfo{author}{Spampinati, S.},
  \bibinfo{author}{Demidovich, A.}, \bibinfo{author}{Danailov, M.B.},
  \bibinfo{author}{Borghes, R.}, \bibinfo{author}{Kourousias, G.},
  \bibinfo{author}{Sanches Dos~Reis, C.E.}, \bibinfo{author}{Billé, F.},
  \bibinfo{author}{Lutman, A.A.}, \bibinfo{author}{Squibb, R.J.},
  \bibinfo{author}{Feifel, R.}, \bibinfo{author}{Carpeggiani, P.},
  \bibinfo{author}{Reduzzi, M.}, \bibinfo{author}{Mazza, T.},
  \bibinfo{author}{Meyer, M.}, \bibinfo{author}{Bengtsson, S.},
  \bibinfo{author}{Ibrakovic, N.}, \bibinfo{author}{Simpson, E.R.},
  \bibinfo{author}{Mauritsson, J.}, \bibinfo{author}{Csizmadia, T.},
  \bibinfo{author}{Dumergue, M.}, \bibinfo{author}{K{\"u}hn, S.},
  \bibinfo{author}{Nandiga~Gopalakrishna, H.}, \bibinfo{author}{You, D.},
  \bibinfo{author}{Ueda, K.}, \bibinfo{author}{Labeye, M.},
  \bibinfo{author}{B\ae{}kh\o{}j, J.E.}, \bibinfo{author}{Schafer, K.J.},
  \bibinfo{author}{Gryzlova, E.V.}, \bibinfo{author}{Grum-Grzhimailo, A.N.},
  \bibinfo{author}{Prince, K.C.}, \bibinfo{author}{Callegari, C.},
  \bibinfo{author}{Sansone, G.}, \bibinfo{year}{2020}.
\newblock \bibinfo{title}{Attosecond pulse shaping using a seeded free-electron
  laser}.
\newblock \bibinfo{journal}{Nature} \bibinfo{volume}{578},
  \bibinfo{pages}{386--391}.
\newblock \DOIprefix\doi{10.1038/s41586-020-2005-6}.
%Type = Article
\bibitem[{Mart\'{\i}n(1999)}]{Martin1999}
\bibinfo{author}{Mart\'{\i}n, F.}, \bibinfo{year}{1999}.
\newblock \bibinfo{title}{Ionization and dissociation using {B}-splines:
  photoionization of the hydrogen molecule}.
\newblock \bibinfo{journal}{J. Phys. B} \bibinfo{volume}{32},
  \bibinfo{pages}{R197--R231}.
\newblock \DOIprefix\doi{10.1088/0953-4075/32/16/201}.
%Type = Article
\bibitem[{Masuda et~al.(2019)Masuda, Yoshimi, Fujieda, Fujimoto, Haba, Hara,
  Hiraki, Kaino, Kasamatsu, Kitao, Konashi, Miyamoto, Okai, Okubo, Sasao, Seto,
  Schumm, Shigekawa, Suzuki, Stellmer, Tamasaku, Uetake, Watanabe, Watanabe,
  Yasuda, Yamaguchi, Yoda, Yokokita, Yoshimura and Yoshimura}]{Masuda2019}
\bibinfo{author}{Masuda, T.}, \bibinfo{author}{Yoshimi, A.},
  \bibinfo{author}{Fujieda, A.}, \bibinfo{author}{Fujimoto, H.},
  \bibinfo{author}{Haba, H.}, \bibinfo{author}{Hara, H.},
  \bibinfo{author}{Hiraki, T.}, \bibinfo{author}{Kaino, H.},
  \bibinfo{author}{Kasamatsu, Y.}, \bibinfo{author}{Kitao, S.},
  \bibinfo{author}{Konashi, K.}, \bibinfo{author}{Miyamoto, Y.},
  \bibinfo{author}{Okai, K.}, \bibinfo{author}{Okubo, S.},
  \bibinfo{author}{Sasao, N.}, \bibinfo{author}{Seto, M.},
  \bibinfo{author}{Schumm, T.}, \bibinfo{author}{Shigekawa, Y.},
  \bibinfo{author}{Suzuki, K.}, \bibinfo{author}{Stellmer, S.},
  \bibinfo{author}{Tamasaku, K.}, \bibinfo{author}{Uetake, S.},
  \bibinfo{author}{Watanabe, M.}, \bibinfo{author}{Watanabe, T.},
  \bibinfo{author}{Yasuda, Y.}, \bibinfo{author}{Yamaguchi, A.},
  \bibinfo{author}{Yoda, Y.}, \bibinfo{author}{Yokokita, T.},
  \bibinfo{author}{Yoshimura, M.}, \bibinfo{author}{Yoshimura, K.},
  \bibinfo{year}{2019}.
\newblock \bibinfo{title}{X-ray pumping of the \ensuremath{^{229}}{Th} nuclear
  clock isomer}.
\newblock \bibinfo{journal}{Nature} \bibinfo{volume}{573},
  \bibinfo{pages}{238--242}.
\newblock \DOIprefix\doi{10.1038/s41586-019-1542-3}.
%Type = Article
\bibitem[{Matthews et~al.(1985)Matthews, Hagelstein, Rosen, Eckart, Ceglio,
  Hazi, Medecki, MacGowan, Trebes, Whitten, Campbell, Hatcher, Hawryluk,
  Kauffman, Pleasance, Rambach, Scofield, Stone and Weaver}]{Matthews1985}
\bibinfo{author}{Matthews, D.L.}, \bibinfo{author}{Hagelstein, P.L.},
  \bibinfo{author}{Rosen, M.D.}, \bibinfo{author}{Eckart, M.J.},
  \bibinfo{author}{Ceglio, N.M.}, \bibinfo{author}{Hazi, A.U.},
  \bibinfo{author}{Medecki, H.}, \bibinfo{author}{MacGowan, B.J.},
  \bibinfo{author}{Trebes, J.E.}, \bibinfo{author}{Whitten, B.L.},
  \bibinfo{author}{Campbell, E.M.}, \bibinfo{author}{Hatcher, C.W.},
  \bibinfo{author}{Hawryluk, A.M.}, \bibinfo{author}{Kauffman, R.L.},
  \bibinfo{author}{Pleasance, L.D.}, \bibinfo{author}{Rambach, G.},
  \bibinfo{author}{Scofield, J.H.}, \bibinfo{author}{Stone, G.},
  \bibinfo{author}{Weaver, T.A.}, \bibinfo{year}{1985}.
\newblock \bibinfo{title}{Demonstration of a soft {X}-ray amplifier}.
\newblock \bibinfo{journal}{Phys. Rev. Lett.} \bibinfo{volume}{54},
  \bibinfo{pages}{110--113}.
\newblock \DOIprefix\doi{10.1103/physrevlett.54.110}.
%Type = Article
\bibitem[{Mazza et~al.(2015a)Mazza, Gryzlova, Grum-Grzhimailo, Kazansky,
  Kabachnik and Meyer}]{Mazza2015}
\bibinfo{author}{Mazza, T.}, \bibinfo{author}{Gryzlova, E.V.},
  \bibinfo{author}{Grum-Grzhimailo, A.N.}, \bibinfo{author}{Kazansky, A.K.},
  \bibinfo{author}{Kabachnik, N.M.}, \bibinfo{author}{Meyer, M.},
  \bibinfo{year}{2015}a.
\newblock \bibinfo{title}{Dichroism in the photoionisation of atoms at {XUV}
  free-electron lasers}.
\newblock \bibinfo{journal}{J. Electr. Spectr. Rel. Phenom.}
  \bibinfo{volume}{204}, \bibinfo{pages}{313--321}.
\newblock \DOIprefix\doi{10.1016/j.elspec.2015.08.011}.
%Type = Article
\bibitem[{Mazza et~al.(2014)Mazza, Ilchen, Rafipoor, Callegari, Finetti,
  Plekan, Prince, Richter, Danailov, Demidovich, De~Ninno, Grazioli, Ivanov,
  Mahne, Raimondi, Svetina, Avaldi, Bolognesi, Coreno, O'Keeffe, Di~Fraia,
  Devetta, Ovcharenko, M\"{o}ller, Lyamayev, Stienkemeier, D\"{u}sterer, Ueda,
  Costello, Kazansky, Kabachnik and Meyer}]{Mazza2014}
\bibinfo{author}{Mazza, T.}, \bibinfo{author}{Ilchen, M.},
  \bibinfo{author}{Rafipoor, A.J.}, \bibinfo{author}{Callegari, C.},
  \bibinfo{author}{Finetti, P.}, \bibinfo{author}{Plekan, O.},
  \bibinfo{author}{Prince, K.C.}, \bibinfo{author}{Richter, R.},
  \bibinfo{author}{Danailov, M.B.}, \bibinfo{author}{Demidovich, A.},
  \bibinfo{author}{De~Ninno, G.}, \bibinfo{author}{Grazioli, C.},
  \bibinfo{author}{Ivanov, R.}, \bibinfo{author}{Mahne, N.},
  \bibinfo{author}{Raimondi, L.}, \bibinfo{author}{Svetina, C.},
  \bibinfo{author}{Avaldi, L.}, \bibinfo{author}{Bolognesi, P.},
  \bibinfo{author}{Coreno, M.}, \bibinfo{author}{O'Keeffe, P.},
  \bibinfo{author}{Di~Fraia, M.}, \bibinfo{author}{Devetta, M.},
  \bibinfo{author}{Ovcharenko, Y.}, \bibinfo{author}{M\"{o}ller, T.},
  \bibinfo{author}{Lyamayev, V.}, \bibinfo{author}{Stienkemeier, F.},
  \bibinfo{author}{D\"{u}sterer, S.}, \bibinfo{author}{Ueda, K.},
  \bibinfo{author}{Costello, J.T.}, \bibinfo{author}{Kazansky, A.K.},
  \bibinfo{author}{Kabachnik, N.M.}, \bibinfo{author}{Meyer, M.},
  \bibinfo{year}{2014}.
\newblock \bibinfo{title}{Determining the polarization state of an extreme
  ultraviolet free-electron laser beam using atomic circular dichroism}.
\newblock \bibinfo{journal}{Nat. Commun.} \bibinfo{volume}{5},
  \bibinfo{pages}{3648}.
\newblock \DOIprefix\doi{10.1038/ncomms4648}.
%Type = Article
\bibitem[{Mazza et~al.(2016)Mazza, Ilchen, Rafipoor, Callegari, Finetti,
  Plekan, Prince, Richter, Demidovich, Grazioli, Avaldi, Bolognesi, Coreno,
  O'Keeffe, Di~Fraia, Devetta, Ovcharenko, Lyamayev, D\"usterer, Ueda,
  Costello, Gryzlova, Strakhova, Grum-Grzhimailo, Bozhevolnov, Kazansky,
  Kabachnik and Meyer}]{Mazza2016}
\bibinfo{author}{Mazza, T.}, \bibinfo{author}{Ilchen, M.},
  \bibinfo{author}{Rafipoor, A.J.}, \bibinfo{author}{Callegari, C.},
  \bibinfo{author}{Finetti, P.}, \bibinfo{author}{Plekan, O.},
  \bibinfo{author}{Prince, K.C.}, \bibinfo{author}{Richter, R.},
  \bibinfo{author}{Demidovich, A.}, \bibinfo{author}{Grazioli, C.},
  \bibinfo{author}{Avaldi, L.}, \bibinfo{author}{Bolognesi, P.},
  \bibinfo{author}{Coreno, M.}, \bibinfo{author}{O'Keeffe, P.},
  \bibinfo{author}{Di~Fraia, M.}, \bibinfo{author}{Devetta, M.},
  \bibinfo{author}{Ovcharenko, Y.}, \bibinfo{author}{Lyamayev, V.},
  \bibinfo{author}{D\"usterer, S.}, \bibinfo{author}{Ueda, K.},
  \bibinfo{author}{Costello, J.T.}, \bibinfo{author}{Gryzlova, E.V.},
  \bibinfo{author}{Strakhova, S.I.}, \bibinfo{author}{Grum-Grzhimailo, A.N.},
  \bibinfo{author}{Bozhevolnov, A.V.}, \bibinfo{author}{Kazansky, A.K.},
  \bibinfo{author}{Kabachnik, N.M.}, \bibinfo{author}{Meyer, M.},
  \bibinfo{year}{2016}.
\newblock \bibinfo{title}{Angular distribution and circular dichroism in the
  two-colour {XUV} + {NIR} above-threshold ionization of helium}.
\newblock \bibinfo{journal}{J. Mod. Opt.} \bibinfo{volume}{63},
  \bibinfo{pages}{367--382}.
\newblock \DOIprefix\doi{10.1080/09500340.2015.1119897}.
%Type = Article
\bibitem[{Mazza et~al.(2015b)Mazza, Karamatskou, Ilchen, Bakhtiarzadeh,
  Rafipoor, O'Keeffe, Kelly, Walsh, Costello, Meyer and Santra}]{Mazza2015a}
\bibinfo{author}{Mazza, T.}, \bibinfo{author}{Karamatskou, A.},
  \bibinfo{author}{Ilchen, M.}, \bibinfo{author}{Bakhtiarzadeh, S.},
  \bibinfo{author}{Rafipoor, A.J.}, \bibinfo{author}{O'Keeffe, P.},
  \bibinfo{author}{Kelly, T.J.}, \bibinfo{author}{Walsh, N.},
  \bibinfo{author}{Costello, J.T.}, \bibinfo{author}{Meyer, M.},
  \bibinfo{author}{Santra, R.}, \bibinfo{year}{2015}b.
\newblock \bibinfo{title}{Sensitivity of nonlinear photoionization to resonance
  substructure in collective excitation}.
\newblock \bibinfo{journal}{Nat. Commun.} \bibinfo{volume}{6},
  \bibinfo{pages}{6799}.
\newblock \DOIprefix\doi{10.1038/ncomms7799}.
%Type = Article
\bibitem[{Meister et~al.(2020)Meister, Lindenblatt, Trost, Schnorr, Augustin,
  Braune, Treusch, Pfeifer and Moshammer}]{Meister2020}
\bibinfo{author}{Meister, S.}, \bibinfo{author}{Lindenblatt, H.},
  \bibinfo{author}{Trost, F.}, \bibinfo{author}{Schnorr, K.},
  \bibinfo{author}{Augustin, S.}, \bibinfo{author}{Braune, M.},
  \bibinfo{author}{Treusch, R.}, \bibinfo{author}{Pfeifer, T.},
  \bibinfo{author}{Moshammer, R.}, \bibinfo{year}{2020}.
\newblock \bibinfo{title}{Atomic, molecular and cluster science with the
  reaction microscope endstation at {FLASH}2}.
\newblock \bibinfo{journal}{Appl. Sci.} \bibinfo{volume}{10},
  \bibinfo{pages}{2953}.
\newblock \DOIprefix\doi{10.3390/app10082953}.
%Type = Article
\bibitem[{Mercouris et~al.(1994)Mercouris, Komninos, Dionissopoulou and
  Nicolaides}]{Mercouris1994}
\bibinfo{author}{Mercouris, T.}, \bibinfo{author}{Komninos, Y.},
  \bibinfo{author}{Dionissopoulou, S.}, \bibinfo{author}{Nicolaides, C.A.},
  \bibinfo{year}{1994}.
\newblock \bibinfo{title}{Computation of strong-field multiphoton processes in
  polyelectronic atoms: State-specific method and applications to {H} and
  {Li}\ensuremath{^-}}.
\newblock \bibinfo{journal}{Phys. Rev. A} \bibinfo{volume}{50},
  \bibinfo{pages}{4109--4121}.
\newblock \DOIprefix\doi{10.1103/PhysRevA.50.4109}.
%Type = Incollection
\bibitem[{Mercouris et~al.(2010)Mercouris, Komninos and
  Nicolaides}]{Mercouris2010}
\bibinfo{author}{Mercouris, T.}, \bibinfo{author}{Komninos, Y.},
  \bibinfo{author}{Nicolaides, C.A.}, \bibinfo{year}{2010}.
\newblock \bibinfo{title}{The state-specific expansion approach to the solution
  of the polyelectronic time-dependent {S}chr{\"o}dinger equation for atoms and
  molecules in unstable states}, in: \bibinfo{editor}{Nicolaides, C.A.},
  \bibinfo{editor}{Br{\"a}ndas, E.} (Eds.), \bibinfo{booktitle}{Unstable States
  in the Continuous Spectra, Part {I}: Analysis, Concepts, Methods, and
  Results}. \bibinfo{publisher}{Academic Press}. volume~\bibinfo{volume}{60} of
  \textit{\bibinfo{series}{Advances in Quantum Chemistry}}.
  chapter~\bibinfo{chapter}{6}, pp. \bibinfo{pages}{333--405}.
\newblock \DOIprefix\doi{10.1016/s0065-3276(10)60006-8}.
%Type = Article
\bibitem[{Mercouris et~al.(2016)Mercouris, Komninos and
  Nicolaides}]{Mercouris2016}
\bibinfo{author}{Mercouris, T.}, \bibinfo{author}{Komninos, Y.},
  \bibinfo{author}{Nicolaides, C.A.}, \bibinfo{year}{2016}.
\newblock \bibinfo{title}{{EUV} two-photon-ionization cross sections of helium
  from the solution of the time-dependent {S}chr{\"o}dinger equation, and
  comparison with measurements using free-electron lasers}.
\newblock \bibinfo{journal}{Phys. Rev. A} \bibinfo{volume}{94},
  \bibinfo{pages}{063406}.
\newblock \DOIprefix\doi{10.1103/PhysRevA.94.063406}.
%Type = Article
\bibitem[{Meyer et~al.(2010)Meyer, Costello, D\"usterer, Li and
  Radcliffe}]{Meyer2010}
\bibinfo{author}{Meyer, M.}, \bibinfo{author}{Costello, J.T.},
  \bibinfo{author}{D\"usterer, S.}, \bibinfo{author}{Li, W.B.},
  \bibinfo{author}{Radcliffe, P.}, \bibinfo{year}{2010}.
\newblock \bibinfo{title}{Two-colour experiments in the gas phase}.
\newblock \bibinfo{journal}{J. Phys. B} \bibinfo{volume}{43},
  \bibinfo{pages}{194006}.
\newblock \DOIprefix\doi{10.1088/0953-4075/43/19/194006}.
%Type = Article
\bibitem[{Meyer et~al.(2006)Meyer, Cubaynes, O'Keeffe, Luna, Yeates, Kennedy,
  Costello, Orr, Ta\"{\i}eb, Maquet, D\"usterer, Radcliffe, Redlin, Azima,
  Pl\"onjes and Feldhaus}]{Meyer2006}
\bibinfo{author}{Meyer, M.}, \bibinfo{author}{Cubaynes, D.},
  \bibinfo{author}{O'Keeffe, P.}, \bibinfo{author}{Luna, H.},
  \bibinfo{author}{Yeates, P.}, \bibinfo{author}{Kennedy, E.T.},
  \bibinfo{author}{Costello, J.T.}, \bibinfo{author}{Orr, P.},
  \bibinfo{author}{Ta\"{\i}eb, R.}, \bibinfo{author}{Maquet, A.},
  \bibinfo{author}{D\"usterer, S.}, \bibinfo{author}{Radcliffe, P.},
  \bibinfo{author}{Redlin, H.}, \bibinfo{author}{Azima, A.},
  \bibinfo{author}{Pl\"onjes, E.}, \bibinfo{author}{Feldhaus, J.},
  \bibinfo{year}{2006}.
\newblock \bibinfo{title}{Two-color photoionization in xuv free-electron and
  visible laser fields}.
\newblock \bibinfo{journal}{Phys. Rev. A} \bibinfo{volume}{74},
  \bibinfo{pages}{011401}.
\newblock \DOIprefix\doi{10.1103/PhysRevA.74.011401}.
%Type = Article
\bibitem[{Meyer et~al.(2012)Meyer, Radcliffe, Tschentscher, Costello,
  Cavalieri, Grguras, Maier, Kienberger, Bozek, Bostedt, Schorb, Coffee,
  Messerschmidt, Roedig, Sistrunk, Di~Mauro, Doumy, Ueda, Wada, D\"usterer,
  Kazansky and Kabachnik}]{Meyer2012}
\bibinfo{author}{Meyer, M.}, \bibinfo{author}{Radcliffe, P.},
  \bibinfo{author}{Tschentscher, T.}, \bibinfo{author}{Costello, J.T.},
  \bibinfo{author}{Cavalieri, A.L.}, \bibinfo{author}{Grguras, I.},
  \bibinfo{author}{Maier, A.R.}, \bibinfo{author}{Kienberger, R.},
  \bibinfo{author}{Bozek, J.}, \bibinfo{author}{Bostedt, C.},
  \bibinfo{author}{Schorb, S.}, \bibinfo{author}{Coffee, R.},
  \bibinfo{author}{Messerschmidt, M.}, \bibinfo{author}{Roedig, C.},
  \bibinfo{author}{Sistrunk, E.}, \bibinfo{author}{Di~Mauro, L.F.},
  \bibinfo{author}{Doumy, G.}, \bibinfo{author}{Ueda, K.},
  \bibinfo{author}{Wada, S.}, \bibinfo{author}{D\"usterer, S.},
  \bibinfo{author}{Kazansky, A.K.}, \bibinfo{author}{Kabachnik, N.M.},
  \bibinfo{year}{2012}.
\newblock \bibinfo{title}{Angle-resolved electron spectroscopy of
  laser-assisted {A}uger decay induced by a few-femtosecond {X}-ray pulse}.
\newblock \bibinfo{journal}{Phys. Rev. Lett.} \bibinfo{volume}{108},
  \bibinfo{pages}{063007}.
\newblock \DOIprefix\doi{10.1103/PhysRevLett.108.063007}.
%Type = Article
\bibitem[{Miao et~al.(2015)Miao, Ishikawa, Robinson and Murnane}]{Miao2015}
\bibinfo{author}{Miao, J.}, \bibinfo{author}{Ishikawa, T.},
  \bibinfo{author}{Robinson, I.K.}, \bibinfo{author}{Murnane, M.M.},
  \bibinfo{year}{2015}.
\newblock \bibinfo{title}{Beyond crystallography: Diffractive imaging using
  coherent {x}-ray light sources}.
\newblock \bibinfo{journal}{Science} \bibinfo{volume}{348},
  \bibinfo{pages}{530--535}.
\newblock \DOIprefix\doi{10.1126/science.aaa1394}.
%Type = Article
\bibitem[{Middleton and Nikolopoulos(2012)}]{Middleton2012}
\bibinfo{author}{Middleton, D.}, \bibinfo{author}{Nikolopoulos, L.},
  \bibinfo{year}{2012}.
\newblock \bibinfo{title}{Effects of autoionising states on the single and
  double ionisation yields of neon with soft {X}-ray fields}.
\newblock \bibinfo{journal}{J. Mod. Optics} \bibinfo{volume}{59},
  \bibinfo{pages}{1653--1663}.
\newblock \DOIprefix\doi{10.1080/09500340.2012.737481}.
%Type = Article
\bibitem[{Milne et~al.(2017)Milne, Schietinger, Aiba, Alarcon, Alex, Anghel,
  Arsov, Beard, Beaud, Bettoni, Bopp, Brands, Br{\"o}nnimann, Brunnenkant,
  Calvi, Citterio, Craievich, Csatari~Divall, D\"{a}llenbach, D'Amico, Dax,
  Deng, Dietrich, Dinapoli, Divall, Dordevic, Ebner, Erny, Fitze, Flechsig,
  Follath, Frei, G\"{a}rtner, Ganter, Garvey, Geng, Gorgisyan, Gough, Hauff,
  Hauri, Hiller, Humar, Hunziker, Ingold, Ischebeck, Janousch, Jurani{\'{c}},
  Jurcevic, Kaiser, Kalantari, Kalt, Keil, Kittel, Knopp, Koprek, Lemke,
  Lippuner, Llorente~Sancho, L{\"o}hl, Lopez-Cuenca, M\"{a}rki, Marcellini,
  Marinkovic, Martiel, Menzel, Mozzanica, Nass, Orlandi, Ozkan~Loch, Panepucci,
  Paraliev, Patterson, Pedrini, Pedrozzi, Pollet, Pradervand, Prat, Radi,
  Raguin, Redford, Rehanek, R{\'{e}}hault, Reiche, Ringele, Rittmann, Rivkin,
  Romann, Ruat, Ruder, Sala, Schebacher, Schilcher, Schlott, Schmidt, Schmitt,
  Shi, Stadler, Stingelin, Sturzenegger, Szlachetko, Thattil, Treyer, Trisorio,
  Tron, Vetter, Vicario, Voulot, Wang, Zamofing, Zellweger, Zennaro, Zimoch,
  Abela, Patthey and Braun}]{Milne2017}
\bibinfo{author}{Milne, C.}, \bibinfo{author}{Schietinger, T.},
  \bibinfo{author}{Aiba, M.}, \bibinfo{author}{Alarcon, A.},
  \bibinfo{author}{Alex, J.}, \bibinfo{author}{Anghel, A.},
  \bibinfo{author}{Arsov, V.}, \bibinfo{author}{Beard, C.},
  \bibinfo{author}{Beaud, P.}, \bibinfo{author}{Bettoni, S.},
  \bibinfo{author}{Bopp, M.}, \bibinfo{author}{Brands, H.},
  \bibinfo{author}{Br{\"o}nnimann, M.}, \bibinfo{author}{Brunnenkant, I.},
  \bibinfo{author}{Calvi, M.}, \bibinfo{author}{Citterio, A.},
  \bibinfo{author}{Craievich, P.}, \bibinfo{author}{Csatari~Divall, M.},
  \bibinfo{author}{D\"{a}llenbach, M.}, \bibinfo{author}{D'Amico, M.},
  \bibinfo{author}{Dax, A.}, \bibinfo{author}{Deng, Y.},
  \bibinfo{author}{Dietrich, A.}, \bibinfo{author}{Dinapoli, R.},
  \bibinfo{author}{Divall, E.}, \bibinfo{author}{Dordevic, S.},
  \bibinfo{author}{Ebner, S.}, \bibinfo{author}{Erny, C.},
  \bibinfo{author}{Fitze, H.}, \bibinfo{author}{Flechsig, U.},
  \bibinfo{author}{Follath, R.}, \bibinfo{author}{Frei, F.},
  \bibinfo{author}{G\"{a}rtner, F.}, \bibinfo{author}{Ganter, R.},
  \bibinfo{author}{Garvey, T.}, \bibinfo{author}{Geng, Z.},
  \bibinfo{author}{Gorgisyan, I.}, \bibinfo{author}{Gough, C.},
  \bibinfo{author}{Hauff, A.}, \bibinfo{author}{Hauri, C.},
  \bibinfo{author}{Hiller, N.}, \bibinfo{author}{Humar, T.},
  \bibinfo{author}{Hunziker, S.}, \bibinfo{author}{Ingold, G.},
  \bibinfo{author}{Ischebeck, R.}, \bibinfo{author}{Janousch, M.},
  \bibinfo{author}{Jurani{\'{c}}, P.}, \bibinfo{author}{Jurcevic, M.},
  \bibinfo{author}{Kaiser, M.}, \bibinfo{author}{Kalantari, B.},
  \bibinfo{author}{Kalt, R.}, \bibinfo{author}{Keil, B.},
  \bibinfo{author}{Kittel, C.}, \bibinfo{author}{Knopp, G.},
  \bibinfo{author}{Koprek, W.}, \bibinfo{author}{Lemke, H.},
  \bibinfo{author}{Lippuner, T.}, \bibinfo{author}{Llorente~Sancho, D.},
  \bibinfo{author}{L{\"o}hl, F.}, \bibinfo{author}{Lopez-Cuenca, C.},
  \bibinfo{author}{M\"{a}rki, F.}, \bibinfo{author}{Marcellini, F.},
  \bibinfo{author}{Marinkovic, G.}, \bibinfo{author}{Martiel, I.},
  \bibinfo{author}{Menzel, R.}, \bibinfo{author}{Mozzanica, A.},
  \bibinfo{author}{Nass, K.}, \bibinfo{author}{Orlandi, G.},
  \bibinfo{author}{Ozkan~Loch, C.}, \bibinfo{author}{Panepucci, E.},
  \bibinfo{author}{Paraliev, M.}, \bibinfo{author}{Patterson, B.},
  \bibinfo{author}{Pedrini, B.}, \bibinfo{author}{Pedrozzi, M.},
  \bibinfo{author}{Pollet, P.}, \bibinfo{author}{Pradervand, C.},
  \bibinfo{author}{Prat, E.}, \bibinfo{author}{Radi, P.},
  \bibinfo{author}{Raguin, J.Y.}, \bibinfo{author}{Redford, S.},
  \bibinfo{author}{Rehanek, J.}, \bibinfo{author}{R{\'{e}}hault, J.},
  \bibinfo{author}{Reiche, S.}, \bibinfo{author}{Ringele, M.},
  \bibinfo{author}{Rittmann, J.}, \bibinfo{author}{Rivkin, L.},
  \bibinfo{author}{Romann, A.}, \bibinfo{author}{Ruat, M.},
  \bibinfo{author}{Ruder, C.}, \bibinfo{author}{Sala, L.},
  \bibinfo{author}{Schebacher, L.}, \bibinfo{author}{Schilcher, T.},
  \bibinfo{author}{Schlott, V.}, \bibinfo{author}{Schmidt, T.},
  \bibinfo{author}{Schmitt, B.}, \bibinfo{author}{Shi, X.},
  \bibinfo{author}{Stadler, M.}, \bibinfo{author}{Stingelin, L.},
  \bibinfo{author}{Sturzenegger, W.}, \bibinfo{author}{Szlachetko, J.},
  \bibinfo{author}{Thattil, D.}, \bibinfo{author}{Treyer, D.},
  \bibinfo{author}{Trisorio, A.}, \bibinfo{author}{Tron, W.},
  \bibinfo{author}{Vetter, S.}, \bibinfo{author}{Vicario, C.},
  \bibinfo{author}{Voulot, D.}, \bibinfo{author}{Wang, M.},
  \bibinfo{author}{Zamofing, T.}, \bibinfo{author}{Zellweger, C.},
  \bibinfo{author}{Zennaro, R.}, \bibinfo{author}{Zimoch, E.},
  \bibinfo{author}{Abela, R.}, \bibinfo{author}{Patthey, L.},
  \bibinfo{author}{Braun, H.H.}, \bibinfo{year}{2017}.
\newblock \bibinfo{title}{{SwissFEL}: The swiss {X}-ray free electron laser}.
\newblock \bibinfo{journal}{Appl. Sci.} \bibinfo{volume}{7},
  \bibinfo{pages}{720}.
\newblock \DOIprefix\doi{10.3390/app7070720}.
%Type = Article
\bibitem[{Milton et~al.(2001)Milton, Gluskin, Arnold, Benson, Berg, Biedron,
  Borland, Chae, Dejus, Den~Hartog, Deriy, Erdmann, Eidelman, Hahne, Huang,
  Kim, Lewellen, Li, Lumpkin, Makarov, Moog, Nassiri, Sajaev, Soliday, Tieman,
  Trakhtenberg, Travish, Vasserman, Vinokurov, Wang, Wiemerslage and
  Yang}]{Milton2001}
\bibinfo{author}{Milton, S.V.}, \bibinfo{author}{Gluskin, E.},
  \bibinfo{author}{Arnold, N.D.}, \bibinfo{author}{Benson, C.},
  \bibinfo{author}{Berg, W.}, \bibinfo{author}{Biedron, S.G.},
  \bibinfo{author}{Borland, M.}, \bibinfo{author}{Chae, Y.C.},
  \bibinfo{author}{Dejus, R.J.}, \bibinfo{author}{Den~Hartog, P.K.},
  \bibinfo{author}{Deriy, B.}, \bibinfo{author}{Erdmann, M.},
  \bibinfo{author}{Eidelman, Y.I.}, \bibinfo{author}{Hahne, M.W.},
  \bibinfo{author}{Huang, Z.}, \bibinfo{author}{Kim, K.J.},
  \bibinfo{author}{Lewellen, J.W.}, \bibinfo{author}{Li, Y.},
  \bibinfo{author}{Lumpkin, A.H.}, \bibinfo{author}{Makarov, O.},
  \bibinfo{author}{Moog, E.R.}, \bibinfo{author}{Nassiri, A.},
  \bibinfo{author}{Sajaev, V.}, \bibinfo{author}{Soliday, R.},
  \bibinfo{author}{Tieman, B.J.}, \bibinfo{author}{Trakhtenberg, E.M.},
  \bibinfo{author}{Travish, G.}, \bibinfo{author}{Vasserman, I.B.},
  \bibinfo{author}{Vinokurov, N.A.}, \bibinfo{author}{Wang, X.J.},
  \bibinfo{author}{Wiemerslage, G.}, \bibinfo{author}{Yang, B.X.},
  \bibinfo{year}{2001}.
\newblock \bibinfo{title}{Exponential gain and saturation of a self-amplified
  spontaneous emission free-electron laser}.
\newblock \bibinfo{journal}{Science} \bibinfo{volume}{292},
  \bibinfo{pages}{2037--2041}.
\newblock \DOIprefix\doi{10.1126/science.1059955}.
%Type = Article
\bibitem[{Milton et~al.(2000)Milton, Gluskin, Biedron, Dejus, Den~Hartog,
  Galayda, Kim, Lewellen, Moog, Sajaev, Sereno, Travish, Vinokurov, Arnold,
  Benson, Berg, Biggs, Borland, Carwardine, Chae, Decker, Deriy, Erdmann,
  Friedsam, Gold, Grelick, Hahne, Harkay, Huang, Lessner, Lill, Lumpkin,
  Makarov, Markovich, Meyer, Nassiri, Noonan, Pasky, Pile, Smith, Soliday,
  Tieman, Trakhtenberg, Trento, Vasserman, Walters, Wang, Wiemerslage, Xu and
  Yang}]{Milton2000}
\bibinfo{author}{Milton, S.V.}, \bibinfo{author}{Gluskin, E.},
  \bibinfo{author}{Biedron, S.G.}, \bibinfo{author}{Dejus, R.J.},
  \bibinfo{author}{Den~Hartog, P.K.}, \bibinfo{author}{Galayda, J.N.},
  \bibinfo{author}{Kim, K.J.}, \bibinfo{author}{Lewellen, J.W.},
  \bibinfo{author}{Moog, E.R.}, \bibinfo{author}{Sajaev, V.},
  \bibinfo{author}{Sereno, N.S.}, \bibinfo{author}{Travish, G.},
  \bibinfo{author}{Vinokurov, N.A.}, \bibinfo{author}{Arnold, N.D.},
  \bibinfo{author}{Benson, C.}, \bibinfo{author}{Berg, W.},
  \bibinfo{author}{Biggs, J.A.}, \bibinfo{author}{Borland, M.},
  \bibinfo{author}{Carwardine, J.A.}, \bibinfo{author}{Chae, Y.C.},
  \bibinfo{author}{Decker, G.}, \bibinfo{author}{Deriy, B.N.},
  \bibinfo{author}{Erdmann, M.J.}, \bibinfo{author}{Friedsam, H.},
  \bibinfo{author}{Gold, C.}, \bibinfo{author}{Grelick, A.E.},
  \bibinfo{author}{Hahne, M.W.}, \bibinfo{author}{Harkay, K.C.},
  \bibinfo{author}{Huang, Z.}, \bibinfo{author}{Lessner, E.S.},
  \bibinfo{author}{Lill, R.M.}, \bibinfo{author}{Lumpkin, A.H.},
  \bibinfo{author}{Makarov, O.A.}, \bibinfo{author}{Markovich, G.M.},
  \bibinfo{author}{Meyer, D.}, \bibinfo{author}{Nassiri, A.},
  \bibinfo{author}{Noonan, J.R.}, \bibinfo{author}{Pasky, S.J.},
  \bibinfo{author}{Pile, G.}, \bibinfo{author}{Smith, T.L.},
  \bibinfo{author}{Soliday, R.}, \bibinfo{author}{Tieman, B.J.},
  \bibinfo{author}{Trakhtenberg, E.M.}, \bibinfo{author}{Trento, G.F.},
  \bibinfo{author}{Vasserman, I.B.}, \bibinfo{author}{Walters, D.R.},
  \bibinfo{author}{Wang, X.J.}, \bibinfo{author}{Wiemerslage, G.},
  \bibinfo{author}{Xu, S.}, \bibinfo{author}{Yang, B.X.}, \bibinfo{year}{2000}.
\newblock \bibinfo{title}{Observation of self-amplified spontaneous emission
  and exponential growth at 530 nm}.
\newblock \bibinfo{journal}{Phys. Rev. Lett.} \bibinfo{volume}{85},
  \bibinfo{pages}{988--991}.
\newblock \DOIprefix\doi{10.1103/PhysRevLett.85.988}.
%Type = Article
\bibitem[{Min et~al.(2019)Min, Nam, Yang, Kim, Shim, Ko, Cho, Heo, Oh, Suh,
  Kim, Na, Kim, Kim, Chun, Lee, Kim, Kim, Eom, Kim, Koo, Rah, Shvyd'ko, Shu,
  Kim, Terentyev, Blank and Kang}]{Min2019}
\bibinfo{author}{Min, C.K.}, \bibinfo{author}{Nam, I.}, \bibinfo{author}{Yang,
  H.}, \bibinfo{author}{Kim, G.}, \bibinfo{author}{Shim, C.H.},
  \bibinfo{author}{Ko, J.H.}, \bibinfo{author}{Cho, M.H.},
  \bibinfo{author}{Heo, H.}, \bibinfo{author}{Oh, B.}, \bibinfo{author}{Suh,
  Y.J.}, \bibinfo{author}{Kim, M.J.}, \bibinfo{author}{Na, D.},
  \bibinfo{author}{Kim, C.}, \bibinfo{author}{Kim, Y.}, \bibinfo{author}{Chun,
  S.H.}, \bibinfo{author}{Lee, J.H.}, \bibinfo{author}{Kim, J.},
  \bibinfo{author}{Kim, S.}, \bibinfo{author}{Eom, I.}, \bibinfo{author}{Kim,
  S.N.}, \bibinfo{author}{Koo, T.Y.}, \bibinfo{author}{Rah, S.},
  \bibinfo{author}{Shvyd'ko, Y.}, \bibinfo{author}{Shu, D.},
  \bibinfo{author}{Kim, K.J.}, \bibinfo{author}{Terentyev, S.},
  \bibinfo{author}{Blank, V.}, \bibinfo{author}{Kang, H.S.},
  \bibinfo{year}{2019}.
\newblock \bibinfo{title}{Hard x-ray self-seeding commissioning at
  {PAL}-{XFEL}}.
\newblock \bibinfo{journal}{J. Synchrotron Radiat.} \bibinfo{volume}{26},
  \bibinfo{pages}{1101--1109}.
\newblock \DOIprefix\doi{10.1107/s1600577519005460}.
%Type = Article
\bibitem[{Minemoto et~al.(2016)Minemoto, Teramoto, Akagi, Fujikawa, Majima,
  Nakajima, Niki, Owada, Sakai, Togashi, Tono, Tsuru, Wada, Yabashi, Yoshida
  and Yagishita}]{Minemoto2016}
\bibinfo{author}{Minemoto, S.}, \bibinfo{author}{Teramoto, T.},
  \bibinfo{author}{Akagi, H.}, \bibinfo{author}{Fujikawa, T.},
  \bibinfo{author}{Majima, T.}, \bibinfo{author}{Nakajima, K.},
  \bibinfo{author}{Niki, K.}, \bibinfo{author}{Owada, S.},
  \bibinfo{author}{Sakai, H.}, \bibinfo{author}{Togashi, T.},
  \bibinfo{author}{Tono, K.}, \bibinfo{author}{Tsuru, S.},
  \bibinfo{author}{Wada, K.}, \bibinfo{author}{Yabashi, M.},
  \bibinfo{author}{Yoshida, S.}, \bibinfo{author}{Yagishita, A.},
  \bibinfo{year}{2016}.
\newblock \bibinfo{title}{Structure determination of molecules in an alignment
  laser field by femtosecond photoelectron diffraction using an {X}-ray
  free-electron laser}.
\newblock \bibinfo{journal}{Sci. Rep.} \bibinfo{volume}{6},
  \bibinfo{pages}{38654}.
\newblock \DOIprefix\doi{10.1038/srep38654}.
%Type = Article
\bibitem[{Miyagi and Madsen(2013)}]{Miyagi2013}
\bibinfo{author}{Miyagi, H.}, \bibinfo{author}{Madsen, L.B.},
  \bibinfo{year}{2013}.
\newblock \bibinfo{title}{Time-dependent restricted-active-space
  self-consistent-field theory for laser-driven many-electron dynamics}.
\newblock \bibinfo{journal}{Phys. Rev. A} \bibinfo{volume}{87},
  \bibinfo{pages}{062511}.
\newblock \DOIprefix\doi{10.1103/PhysRevA.87.062511}.
%Type = Article
\bibitem[{Miyauchi et~al.(2011)Miyauchi, Adachi, Yagishita, Sako, Koike, Sato,
  Iwasaki, Okino, Yamanouchi, Midorikawa, Yamakawa, Kannari, Nakano, Nagasono,
  Tono, Yabashi, Ishikawa, Togashi, Ohashi, Kimura and Senba}]{Miyauchi2011}
\bibinfo{author}{Miyauchi, N.}, \bibinfo{author}{Adachi, J.},
  \bibinfo{author}{Yagishita, A.}, \bibinfo{author}{Sako, T.},
  \bibinfo{author}{Koike, F.}, \bibinfo{author}{Sato, T.},
  \bibinfo{author}{Iwasaki, A.}, \bibinfo{author}{Okino, T.},
  \bibinfo{author}{Yamanouchi, K.}, \bibinfo{author}{Midorikawa, K.},
  \bibinfo{author}{Yamakawa, K.}, \bibinfo{author}{Kannari, F.},
  \bibinfo{author}{Nakano, H.}, \bibinfo{author}{Nagasono, M.},
  \bibinfo{author}{Tono, K.}, \bibinfo{author}{Yabashi, M.},
  \bibinfo{author}{Ishikawa, T.}, \bibinfo{author}{Togashi, T.},
  \bibinfo{author}{Ohashi, H.}, \bibinfo{author}{Kimura, H.},
  \bibinfo{author}{Senba, Y.}, \bibinfo{year}{2011}.
\newblock \bibinfo{title}{Three-photon double ionization of {Ar} studied by
  photoelectron spectroscopy using an extreme ultraviolet free-electron laser:
  manifestation of resonance states of an intermediate {Ar}$^{+}$ ion}.
\newblock \bibinfo{journal}{J. Phys. B} \bibinfo{volume}{44},
  \bibinfo{pages}{071001}.
\newblock \DOIprefix\doi{10.1088/0953-4075/44/7/071001}.
%Type = Article
\bibitem[{Moler and Van~Loan(2003)}]{Moler2003}
\bibinfo{author}{Moler, C.}, \bibinfo{author}{Van~Loan, C.},
  \bibinfo{year}{2003}.
\newblock \bibinfo{title}{Nineteen dubious ways to compute the exponential of a
  matrix, twenty five years later}.
\newblock \bibinfo{journal}{SIAM Review} \bibinfo{volume}{45},
  \bibinfo{pages}{3--49}.
\newblock \DOIprefix\doi{10.1137/S00361445024180}.
%Type = Article
\bibitem[{Mondal et~al.(2014)Mondal, Fukuzawa, Motomura, Tachibana, Nagaya,
  Sakai, Matsunami, Yase, Yao, Wada, Hayashita, Saito, Callegari, Prince,
  Miron, Nagasono, Togashi, Yabashi, Ishikawa, Kazansky, Kabachnik and
  Ueda}]{Mondal2014}
\bibinfo{author}{Mondal, S.}, \bibinfo{author}{Fukuzawa, H.},
  \bibinfo{author}{Motomura, K.}, \bibinfo{author}{Tachibana, T.},
  \bibinfo{author}{Nagaya, K.}, \bibinfo{author}{Sakai, T.},
  \bibinfo{author}{Matsunami, K.}, \bibinfo{author}{Yase, S.},
  \bibinfo{author}{Yao, M.}, \bibinfo{author}{Wada, S.},
  \bibinfo{author}{Hayashita, H.}, \bibinfo{author}{Saito, N.},
  \bibinfo{author}{Callegari, C.}, \bibinfo{author}{Prince, K.C.},
  \bibinfo{author}{Miron, C.}, \bibinfo{author}{Nagasono, M.},
  \bibinfo{author}{Togashi, T.}, \bibinfo{author}{Yabashi, M.},
  \bibinfo{author}{Ishikawa, K.L.}, \bibinfo{author}{Kazansky, A.K.},
  \bibinfo{author}{Kabachnik, N.M.}, \bibinfo{author}{Ueda, K.},
  \bibinfo{year}{2014}.
\newblock \bibinfo{title}{Pulse-delay effects in the angular distribution of
  near-threshold {EUV} + {IR} two-photon ionization of {Ne}}.
\newblock \bibinfo{journal}{Phys. Rev. A} \bibinfo{volume}{89},
  \bibinfo{pages}{013415}.
\newblock \DOIprefix\doi{10.1103/PhysRevA.89.013415}.
%Type = Article
\bibitem[{Mondal et~al.(2013)Mondal, Fukuzawa, Motomura, Tachibana, Nagaya,
  Sakai, Matsunami, Yase, Yao, Wada, Hayashita, Saito, Callegari, Prince,
  O'Keeffe, Bolognesi, Avaldi, Miron, Nagasono, Togashi, Yabashi, Ishikawa,
  Sazhina, Kazansky, Kabachnik and Ueda}]{Mondal2013}
\bibinfo{author}{Mondal, S.}, \bibinfo{author}{Fukuzawa, H.},
  \bibinfo{author}{Motomura, K.}, \bibinfo{author}{Tachibana, T.},
  \bibinfo{author}{Nagaya, K.}, \bibinfo{author}{Sakai, T.},
  \bibinfo{author}{Matsunami, K.}, \bibinfo{author}{Yase, S.},
  \bibinfo{author}{Yao, M.}, \bibinfo{author}{Wada, S.},
  \bibinfo{author}{Hayashita, H.}, \bibinfo{author}{Saito, N.},
  \bibinfo{author}{Callegari, C.}, \bibinfo{author}{Prince, K.C.},
  \bibinfo{author}{O'Keeffe, P.}, \bibinfo{author}{Bolognesi, P.},
  \bibinfo{author}{Avaldi, L.}, \bibinfo{author}{Miron, C.},
  \bibinfo{author}{Nagasono, M.}, \bibinfo{author}{Togashi, T.},
  \bibinfo{author}{Yabashi, M.}, \bibinfo{author}{Ishikawa, K.L.},
  \bibinfo{author}{Sazhina, I.P.}, \bibinfo{author}{Kazansky, A.K.},
  \bibinfo{author}{Kabachnik, N.M.}, \bibinfo{author}{Ueda, K.},
  \bibinfo{year}{2013}.
\newblock \bibinfo{title}{Photoelectron angular distributions in infrared
  one-photon and two-photon ionization of {FEL}-pumped {R}ydberg states of
  helium}.
\newblock \bibinfo{journal}{J. Phys. B} \bibinfo{volume}{46},
  \bibinfo{pages}{205601}.
\newblock \DOIprefix\doi{10.1088/0953-4075/46/20/205601}.
%Type = Article
\bibitem[{Moore et~al.(2011)Moore, Lysaght, Nikolopoulos, Parker, van~der Hart
  and Taylor}]{Moore2011}
\bibinfo{author}{Moore, L.R.}, \bibinfo{author}{Lysaght, M.A.},
  \bibinfo{author}{Nikolopoulos, L.A.A.}, \bibinfo{author}{Parker, J.S.},
  \bibinfo{author}{van~der Hart, H.W.}, \bibinfo{author}{Taylor, K.T.},
  \bibinfo{year}{2011}.
\newblock \bibinfo{title}{The {RMT} method for many-electron atomic systems in
  intense short-pulse laser light}.
\newblock \bibinfo{journal}{J. Mod. Opt.} \bibinfo{volume}{58},
  \bibinfo{pages}{1132--1140}.
\newblock \DOIprefix\doi{10.1080/09500340.2011.559315}.
%Type = Article
\bibitem[{Motomura et~al.(2009)Motomura, Fukuzawa, Foucar, Liu, Pr{\"u}mper,
  Ueda, Saito, Iwayama, Nagaya, Murakami, Yao, Belkacem, Nagasono, Higashiya,
  Yabashi, Ishikawa, Ohashi and Kimura}]{Motomura2009}
\bibinfo{author}{Motomura, K.}, \bibinfo{author}{Fukuzawa, H.},
  \bibinfo{author}{Foucar, L.}, \bibinfo{author}{Liu, X.J.},
  \bibinfo{author}{Pr{\"u}mper, G.}, \bibinfo{author}{Ueda, K.},
  \bibinfo{author}{Saito, N.}, \bibinfo{author}{Iwayama, H.},
  \bibinfo{author}{Nagaya, K.}, \bibinfo{author}{Murakami, H.},
  \bibinfo{author}{Yao, M.}, \bibinfo{author}{Belkacem, A.},
  \bibinfo{author}{Nagasono, M.}, \bibinfo{author}{Higashiya, A.},
  \bibinfo{author}{Yabashi, M.}, \bibinfo{author}{Ishikawa, T.},
  \bibinfo{author}{Ohashi, H.}, \bibinfo{author}{Kimura, H.},
  \bibinfo{year}{2009}.
\newblock \bibinfo{title}{Multiple ionization of atomic argon irradiated by
  {EUV} free-electron laser pulses at 62 nm: evidence of sequential electron
  strip}.
\newblock \bibinfo{journal}{J. Phys. B} \bibinfo{volume}{42},
  \bibinfo{pages}{221003}.
\newblock \DOIprefix\doi{10.1088/0953-4075/42/22/221003}.
%Type = Article
\bibitem[{Motz(1951)}]{Motz1951}
\bibinfo{author}{Motz, H.}, \bibinfo{year}{1951}.
\newblock \bibinfo{title}{Applications of the radiation from fast electron
  beams}.
\newblock \bibinfo{journal}{J. Appl. Phys.} \bibinfo{volume}{22},
  \bibinfo{pages}{527--535}.
\newblock \DOIprefix\doi{10.1063/1.1700002}.
%Type = Article
\bibitem[{Motz et~al.(1953)Motz, Thon and Whitehurst}]{Motz1953}
\bibinfo{author}{Motz, H.}, \bibinfo{author}{Thon, W.},
  \bibinfo{author}{Whitehurst, R.N.}, \bibinfo{year}{1953}.
\newblock \bibinfo{title}{Experiments on radiation by fast electron beams}.
\newblock \bibinfo{journal}{J. Appl. Phys.} \bibinfo{volume}{24},
  \bibinfo{pages}{826--833}.
\newblock \DOIprefix\doi{10.1063/1.1721389}.
%Type = Article
\bibitem[{Mourou(2019)}]{Mourou2019}
\bibinfo{author}{Mourou, G.}, \bibinfo{year}{2019}.
\newblock \bibinfo{title}{Nobel lecture: Extreme light physics and
  application}.
\newblock \bibinfo{journal}{Rev. Mod. Phys.} \bibinfo{volume}{91},
  \bibinfo{pages}{030501}.
\newblock \DOIprefix\doi{10.1103/RevModPhys.91.030501}.
%Type = Article
\bibitem[{Mudrich et~al.(2020)Mudrich, LaForge, Ciavardini, O'Keeffe,
  Callegari, Coreno, Demidovich, Devetta, Di~Fraia, Drabbels, Finetti, Gessner,
  Grazioli, Hernando, Neumark, Ovcharenko, Piseri, Plekan, Prince, Richter,
  Ziemkiewicz, M{\"o}ller, Eloranta, Pi, Barranco and
  Stienkemeier}]{Mudrich2020}
\bibinfo{author}{Mudrich, M.}, \bibinfo{author}{LaForge, A.C.},
  \bibinfo{author}{Ciavardini, A.}, \bibinfo{author}{O'Keeffe, P.},
  \bibinfo{author}{Callegari, C.}, \bibinfo{author}{Coreno, M.},
  \bibinfo{author}{Demidovich, A.}, \bibinfo{author}{Devetta, M.},
  \bibinfo{author}{Di~Fraia, M.}, \bibinfo{author}{Drabbels, M.},
  \bibinfo{author}{Finetti, P.}, \bibinfo{author}{Gessner, O.},
  \bibinfo{author}{Grazioli, C.}, \bibinfo{author}{Hernando, A.},
  \bibinfo{author}{Neumark, D.M.}, \bibinfo{author}{Ovcharenko, Y.},
  \bibinfo{author}{Piseri, P.}, \bibinfo{author}{Plekan, O.},
  \bibinfo{author}{Prince, K.C.}, \bibinfo{author}{Richter, R.},
  \bibinfo{author}{Ziemkiewicz, M.P.}, \bibinfo{author}{M{\"o}ller, T.},
  \bibinfo{author}{Eloranta, J.}, \bibinfo{author}{Pi, M.},
  \bibinfo{author}{Barranco, M.}, \bibinfo{author}{Stienkemeier, F.},
  \bibinfo{year}{2020}.
\newblock \bibinfo{title}{Ultrafast relaxation of photoexcited superfluid {He}
  nanodroplets}.
\newblock \bibinfo{journal}{Nat. Commun.} \bibinfo{volume}{11},
  \bibinfo{pages}{112}.
\newblock \DOIprefix\doi{10.1038/s41467-019-13681-6}.
%Type = Article
\bibitem[{Mukamel et~al.(2009)Mukamel, Abramavicius, Yang, Zhuang, Schweigert
  and Voronine}]{Mukamel2009}
\bibinfo{author}{Mukamel, S.}, \bibinfo{author}{Abramavicius, D.},
  \bibinfo{author}{Yang, L.}, \bibinfo{author}{Zhuang, W.},
  \bibinfo{author}{Schweigert, I.V.}, \bibinfo{author}{Voronine, D.V.},
  \bibinfo{year}{2009}.
\newblock \bibinfo{title}{Coherent multidimensional optical probes for electron
  correlations and exciton dynamics: From {NMR} to x-rays}.
\newblock \bibinfo{journal}{Acc. Chem. Res.} \bibinfo{volume}{42},
  \bibinfo{pages}{553--562}.
\newblock \DOIprefix\doi{10.1021/ar800258z}.
%Type = Article
\bibitem[{M{\"{u}}ller et~al.(2018)M{\"{u}}ller, Artemyev and
  Demekhin}]{Muller2018}
\bibinfo{author}{M{\"{u}}ller, A.D.}, \bibinfo{author}{Artemyev, A.N.},
  \bibinfo{author}{Demekhin, P.V.}, \bibinfo{year}{2018}.
\newblock \bibinfo{title}{Photoelectron circular dichroism in the multiphoton
  ionization by short laser pulses. {II}.~{T}hree- and four-photon ionization
  of fenchone and camphor}.
\newblock \bibinfo{journal}{J. Chem. Phys.} \bibinfo{volume}{148},
  \bibinfo{pages}{214307}.
\newblock \DOIprefix\doi{10.1063/1.5032295}.
%Type = Article
\bibitem[{Muller(2002)}]{Muller2002}
\bibinfo{author}{Muller, H.G.}, \bibinfo{year}{2002}.
\newblock \bibinfo{title}{Reconstruction of attosecond harmonic beating by
  interference of two-photon transitions}.
\newblock \bibinfo{journal}{Appl. Phys. B} \bibinfo{volume}{74},
  \bibinfo{pages}{s17--s21}.
\newblock \DOIprefix\doi{10.1007/s00340-002-0894-8}.
%Type = Article
\bibitem[{Murphy and Pellegrini(1985)}]{Murphy1985}
\bibinfo{author}{Murphy, J.B.}, \bibinfo{author}{Pellegrini, C.},
  \bibinfo{year}{1985}.
\newblock \bibinfo{title}{Free electron lasers for the {XUV} spectral region}.
\newblock \bibinfo{journal}{Nucl. Instrum. Methods Phys. Res. A}
  \bibinfo{volume}{237}, \bibinfo{pages}{159--167}.
\newblock \DOIprefix\doi{10.1016/0168-9002(85)90344-4}.
%Type = Article
\bibitem[{Nagasono et~al.(2011)Nagasono, Harries, Iwayama, Togashi, Tono,
  Yabashi, Senba, Ohashi, Ishikawa and Shigemasa}]{Nagasono2011}
\bibinfo{author}{Nagasono, M.}, \bibinfo{author}{Harries, J.R.},
  \bibinfo{author}{Iwayama, H.}, \bibinfo{author}{Togashi, T.},
  \bibinfo{author}{Tono, K.}, \bibinfo{author}{Yabashi, M.},
  \bibinfo{author}{Senba, Y.}, \bibinfo{author}{Ohashi, H.},
  \bibinfo{author}{Ishikawa, T.}, \bibinfo{author}{Shigemasa, E.},
  \bibinfo{year}{2011}.
\newblock \bibinfo{title}{Observation of free-electron-laser-induced collective
  spontaneous emission (superfluorescence)}.
\newblock \bibinfo{journal}{Phys. Rev. Lett.} \bibinfo{volume}{107},
  \bibinfo{pages}{193603}.
\newblock \DOIprefix\doi{10.1103/PhysRevLett.107.193603}.
%Type = Article
\bibitem[{Nagaya et~al.(2016)Nagaya, Iablonskyi, Golubev, Matsunami, Fukuzawa,
  Motomura, Nishiyama, Sakai, Tachibana, Mondal, Wada, Prince, Callegari,
  Miron, Saito, Yabashi, Demekhin, Cederbaum, Kuleff, Yao and
  Ueda}]{Nagaya2016}
\bibinfo{author}{Nagaya, K.}, \bibinfo{author}{Iablonskyi, D.},
  \bibinfo{author}{Golubev, N.V.}, \bibinfo{author}{Matsunami, K.},
  \bibinfo{author}{Fukuzawa, H.}, \bibinfo{author}{Motomura, K.},
  \bibinfo{author}{Nishiyama, T.}, \bibinfo{author}{Sakai, T.},
  \bibinfo{author}{Tachibana, T.}, \bibinfo{author}{Mondal, S.},
  \bibinfo{author}{Wada, S.}, \bibinfo{author}{Prince, K.C.},
  \bibinfo{author}{Callegari, C.}, \bibinfo{author}{Miron, C.},
  \bibinfo{author}{Saito, N.}, \bibinfo{author}{Yabashi, M.},
  \bibinfo{author}{Demekhin, P.V.}, \bibinfo{author}{Cederbaum, L.S.},
  \bibinfo{author}{Kuleff, A.I.}, \bibinfo{author}{Yao, M.},
  \bibinfo{author}{Ueda, K.}, \bibinfo{year}{2016}.
\newblock \bibinfo{title}{Interatomic {C}oulombic decay cascades in multiply
  excited neon clusters}.
\newblock \bibinfo{journal}{Nat. Commun.} \bibinfo{volume}{7},
  \bibinfo{pages}{13477}.
\newblock \DOIprefix\doi{10.1038/ncomms13477}.
%Type = Article
\bibitem[{Nakajima and Nikolopoulos(2002)}]{Nakajima2002}
\bibinfo{author}{Nakajima, T.}, \bibinfo{author}{Nikolopoulos, L.A.A.},
  \bibinfo{year}{2002}.
\newblock \bibinfo{title}{Use of helium double ionization for autocorrelation
  of an xuv pulse}.
\newblock \bibinfo{journal}{Phys. Rev. A} \bibinfo{volume}{66},
  \bibinfo{pages}{041402R}.
\newblock \DOIprefix\doi{10.1103/PhysRevA.66.041402}.
%Type = Article
\bibitem[{Nayak et~al.(2018)Nayak, Orfanos, Makos, Dumergue, K\"{u}hn,
  Skantzakis, Bodi, Varju, Kalpouzos, Banks, Emmanouilidou, Charalambidis and
  Tzallas}]{Nayak2018}
\bibinfo{author}{Nayak, A.}, \bibinfo{author}{Orfanos, I.},
  \bibinfo{author}{Makos, I.}, \bibinfo{author}{Dumergue, M.},
  \bibinfo{author}{K\"{u}hn, S.}, \bibinfo{author}{Skantzakis, E.},
  \bibinfo{author}{Bodi, B.}, \bibinfo{author}{Varju, K.},
  \bibinfo{author}{Kalpouzos, C.}, \bibinfo{author}{Banks, H.I.B.},
  \bibinfo{author}{Emmanouilidou, A.}, \bibinfo{author}{Charalambidis, D.},
  \bibinfo{author}{Tzallas, P.}, \bibinfo{year}{2018}.
\newblock \bibinfo{title}{Multiple ionization of argon via multi-{XUV}-photon
  absorption induced by 20-{GW} high-order harmonic laser pulses}.
\newblock \bibinfo{journal}{Phys. Rev. A} \bibinfo{volume}{98},
  \bibinfo{pages}{023426}.
\newblock \DOIprefix\doi{10.1103/physreva.98.023426}.
%Type = Article
\bibitem[{Nest(2009)}]{Nest2009}
\bibinfo{author}{Nest, M.}, \bibinfo{year}{2009}.
\newblock \bibinfo{title}{The multi-configuration electron-nuclear dynamics
  method}.
\newblock \bibinfo{journal}{Chem. Phys. Lett.} \bibinfo{volume}{472},
  \bibinfo{pages}{171--174}.
\newblock \DOIprefix\doi{10.1016/j.cplett.2009.03.013}.
%Type = Article
\bibitem[{Nikolopoulos and Lambropoulos(2015)}]{Nikolopoulos2015}
\bibinfo{author}{Nikolopoulos, G.M.}, \bibinfo{author}{Lambropoulos, P.},
  \bibinfo{year}{2015}.
\newblock \bibinfo{title}{Resonantly enhanced multiphoton ionization under
  {XUV} {FEL} radiation: a case study of the role of harmonics}.
\newblock \bibinfo{journal}{J. Phys. B: At. Mol. Opt. Phys.}
  \bibinfo{volume}{48}, \bibinfo{pages}{244006}.
\newblock \DOIprefix\doi{10.1088/0953-4075/48/24/244006}.
%Type = Article
\bibitem[{Nikolopoulos(2013)}]{Nikolopoulos2013}
\bibinfo{author}{Nikolopoulos, L.A.A.}, \bibinfo{year}{2013}.
\newblock \bibinfo{title}{Time-dependent theory of angular correlations in
  sequential double ionization}.
\newblock \bibinfo{journal}{Phys. Rev. Lett.} \bibinfo{volume}{111},
  \bibinfo{pages}{093001}.
\newblock \DOIprefix\doi{10.1103/PhysRevLett.111.093001}.
%Type = Article
\bibitem[{Nikolopoulos et~al.(2011)Nikolopoulos, Kelly and
  Costello}]{Nikolopoulos2011}
\bibinfo{author}{Nikolopoulos, L.A.A.}, \bibinfo{author}{Kelly, T.J.},
  \bibinfo{author}{Costello, J.T.}, \bibinfo{year}{2011}.
\newblock \bibinfo{title}{Theory of ac {S}tark splitting in core-resonant
  {A}uger decay in strong x-ray fields}.
\newblock \bibinfo{journal}{Phys. Rev. A} \bibinfo{volume}{84},
  \bibinfo{pages}{063419}.
\newblock \DOIprefix\doi{10.1103/PhysRevA.84.063419}.
%Type = Article
\bibitem[{Nikolopoulos et~al.(2008)Nikolopoulos, Parker and
  Taylor}]{Nikolopoulos2008}
\bibinfo{author}{Nikolopoulos, L.A.A.}, \bibinfo{author}{Parker, J.S.},
  \bibinfo{author}{Taylor, K.T.}, \bibinfo{year}{2008}.
\newblock \bibinfo{title}{Combined {R}-matrix eigenstate basis set and
  finite-difference propagation method for the time-dependent {S}chr{\"o}dinger
  equation: The one-electron case}.
\newblock \bibinfo{journal}{Phys. Rev. A} \bibinfo{volume}{78},
  \bibinfo{pages}{063420}.
\newblock \DOIprefix\doi{10.1103/PhysRevA.78.063420}.
%Type = Article
\bibitem[{Nishiyama et~al.(2019)Nishiyama, Kumagai, Niozu, Fukuzawa, Motomura,
  Bucher, Ito, Takanashi, Asa, Sato, You, Li, Ono, Kukk, Miron, Neagu,
  Callegari, Di~Fraia, Rossi, Galli, Pincelli, Colombo, Kameshima, Joti,
  Hatsui, Owada, Katayama, Togashi, Tono, Yabashi, Matsuda, Bostedt, Nagaya and
  Ueda}]{Nishiyama2019}
\bibinfo{author}{Nishiyama, T.}, \bibinfo{author}{Kumagai, Y.},
  \bibinfo{author}{Niozu, A.}, \bibinfo{author}{Fukuzawa, H.},
  \bibinfo{author}{Motomura, K.}, \bibinfo{author}{Bucher, M.},
  \bibinfo{author}{Ito, Y.}, \bibinfo{author}{Takanashi, T.},
  \bibinfo{author}{Asa, K.}, \bibinfo{author}{Sato, Y.}, \bibinfo{author}{You,
  D.}, \bibinfo{author}{Li, Y.}, \bibinfo{author}{Ono, T.},
  \bibinfo{author}{Kukk, E.}, \bibinfo{author}{Miron, C.},
  \bibinfo{author}{Neagu, L.}, \bibinfo{author}{Callegari, C.},
  \bibinfo{author}{Di~Fraia, M.}, \bibinfo{author}{Rossi, G.},
  \bibinfo{author}{Galli, D.E.}, \bibinfo{author}{Pincelli, T.},
  \bibinfo{author}{Colombo, A.}, \bibinfo{author}{Kameshima, T.},
  \bibinfo{author}{Joti, Y.}, \bibinfo{author}{Hatsui, T.},
  \bibinfo{author}{Owada, S.}, \bibinfo{author}{Katayama, T.},
  \bibinfo{author}{Togashi, T.}, \bibinfo{author}{Tono, K.},
  \bibinfo{author}{Yabashi, M.}, \bibinfo{author}{Matsuda, K.},
  \bibinfo{author}{Bostedt, C.}, \bibinfo{author}{Nagaya, K.},
  \bibinfo{author}{Ueda, K.}, \bibinfo{year}{2019}.
\newblock \bibinfo{title}{Ultrafast structural dynamics of nanoparticles in
  intense laser fields}.
\newblock \bibinfo{journal}{Phys. Rev. Lett.} \bibinfo{volume}{123},
  \bibinfo{pages}{123201}.
\newblock \DOIprefix\doi{10.1103/PhysRevLett.123.123201}.
%Type = Misc
\bibitem[{nuClock()}]{nuclock}
nuClock, \bibinfo{year}{2015}.
\newblock \bibinfo{title}{The project {nuC}lock}.
\newblock \URLprefix \url{www.nuclock.eu}. \bibinfo{note}{{A}ccessed May 26,
  2020}.
%Type = Article
\bibitem[{{\'{O}}~Broin and Nikolopoulos(2015)}]{Broin2015}
\bibinfo{author}{{\'{O}}~Broin, C.}, \bibinfo{author}{Nikolopoulos, L.A.A.},
  \bibinfo{year}{2015}.
\newblock \bibinfo{title}{R-matrix-incorporating-time method for {H}$_2^+$ in
  short and intense laser fields}.
\newblock \bibinfo{journal}{Phys. Rev. A} \bibinfo{volume}{92},
  \bibinfo{pages}{063428}.
\newblock \DOIprefix\doi{10.1103/PhysRevA.92.063428}.
%Type = Article
\bibitem[{{\'{O}}~Broin and Nikolopoulos(2017)}]{Broin2017}
\bibinfo{author}{{\'{O}}~Broin, C.}, \bibinfo{author}{Nikolopoulos, L.A.A.},
  \bibinfo{year}{2017}.
\newblock \bibinfo{title}{R-matrix-incorporating-time theory of one-electron
  atomic and molecular systems in intense laser fields}.
\newblock \bibinfo{journal}{J. Phys. B} \bibinfo{volume}{50},
  \bibinfo{pages}{033001}.
\newblock \DOIprefix\doi{10.1088/1361-6455/aa51a5}.
%Type = Article
\bibitem[{Oberli et~al.(2019)Oberli, Gonz{\'{a}}lez-V{\'{a}}zquez,
  Rodr{\'{\i}}guez-Perell{\'{o}}, Sodupe, Mart{\'{\i}}n and
  Pic{\'{o}}n}]{Oberli2019}
\bibinfo{author}{Oberli, S.}, \bibinfo{author}{Gonz{\'{a}}lez-V{\'{a}}zquez,
  J.}, \bibinfo{author}{Rodr{\'{\i}}guez-Perell{\'{o}}, E.},
  \bibinfo{author}{Sodupe, M.}, \bibinfo{author}{Mart{\'{\i}}n, F.},
  \bibinfo{author}{Pic{\'{o}}n, A.}, \bibinfo{year}{2019}.
\newblock \bibinfo{title}{Site-selective-induced isomerization of formamide}.
\newblock \bibinfo{journal}{Phys. Chem. Chem. Phys.} \bibinfo{volume}{21},
  \bibinfo{pages}{25626--25634}.
\newblock \DOIprefix\doi{10.1039/c9cp04441h}.
%Type = Article
\bibitem[{Ohmori et~al.(2006)Ohmori, Katsuki, Chiba, Honda, Hagihara, Fujiwara,
  Sato and Ueda}]{Ohmori2006}
\bibinfo{author}{Ohmori, K.}, \bibinfo{author}{Katsuki, H.},
  \bibinfo{author}{Chiba, H.}, \bibinfo{author}{Honda, M.},
  \bibinfo{author}{Hagihara, Y.}, \bibinfo{author}{Fujiwara, K.},
  \bibinfo{author}{Sato, Y.}, \bibinfo{author}{Ueda, K.}, \bibinfo{year}{2006}.
\newblock \bibinfo{title}{Real-time observation of phase-controlled molecular
  wave-packet interference}.
\newblock \bibinfo{journal}{Phys. Rev. Lett.} \bibinfo{volume}{96},
  \bibinfo{pages}{093002}.
\newblock \DOIprefix\doi{10.1103/PhysRevLett.96.093002}.
%Type = Article
\bibitem[{{O'Keeffe} et~al.(2012){O'Keeffe}, {Feyer}, {Bolognesi}, {Coreno},
  {Callegari}, {Cautero}, {Moise}, {Prince}, {Richter}, {Sergo}, {Alagia}, {de
  Simone}, {Kivim{\"a}ki}, {Devetta}, {Mazza}, {Piseri}, {Lyamayev}, {Katzy},
  {Stienkemeier}, {Ovcharenko}, {M{\"o}ller} and {Avaldi}}]{OKeeffe2012}
\bibinfo{author}{{O'Keeffe}, P.}, \bibinfo{author}{{Feyer}, V.},
  \bibinfo{author}{{Bolognesi}, P.}, \bibinfo{author}{{Coreno}, M.},
  \bibinfo{author}{{Callegari}, C.}, \bibinfo{author}{{Cautero}, G.},
  \bibinfo{author}{{Moise}, A.}, \bibinfo{author}{{Prince}, K.C.},
  \bibinfo{author}{{Richter}, R.}, \bibinfo{author}{{Sergo}, R.},
  \bibinfo{author}{{Alagia}, M.}, \bibinfo{author}{{de Simone}, M.},
  \bibinfo{author}{{Kivim{\"a}ki}, A.}, \bibinfo{author}{{Devetta}, M.},
  \bibinfo{author}{{Mazza}, T.}, \bibinfo{author}{{Piseri}, P.},
  \bibinfo{author}{{Lyamayev}, V.}, \bibinfo{author}{{Katzy}, R.},
  \bibinfo{author}{{Stienkemeier}, F.}, \bibinfo{author}{{Ovcharenko}, Y.},
  \bibinfo{author}{{M{\"o}ller}, T.}, \bibinfo{author}{{Avaldi}, L.},
  \bibinfo{year}{2012}.
\newblock \bibinfo{title}{{A velocity map imaging apparatus for gas phase
  studies at {FERMI@E}lettra}}.
\newblock \bibinfo{journal}{Nucl. Instrum. Methods Phys. Res. B}
  \bibinfo{volume}{284}, \bibinfo{pages}{69--73}.
\newblock \DOIprefix\doi{10.1016/j.nimb.2011.07.020}.
%Type = Article
\bibitem[{O'Keeffe et~al.(2013)O'Keeffe, Miheli{\v{c}}, Bolognesi,
  {\v{Z}}itnik, Moise, Richter and Avaldi}]{OKeeffe2013}
\bibinfo{author}{O'Keeffe, P.}, \bibinfo{author}{Miheli{\v{c}}, A.},
  \bibinfo{author}{Bolognesi, P.}, \bibinfo{author}{{\v{Z}}itnik, M.},
  \bibinfo{author}{Moise, A.}, \bibinfo{author}{Richter, R.},
  \bibinfo{author}{Avaldi, L.}, \bibinfo{year}{2013}.
\newblock \bibinfo{title}{Near-threshold photoelectron angular distributions
  from two-photon resonant photoionization of {He}}.
\newblock \bibinfo{journal}{New J. Phys.} \bibinfo{volume}{15},
  \bibinfo{pages}{013023}.
\newblock \DOIprefix\doi{10.1088/1367-2630/15/1/013023}.
%Type = Article
\bibitem[{Omiste and Madsen(2019)}]{Omiste2019}
\bibinfo{author}{Omiste, J.J.}, \bibinfo{author}{Madsen, L.B.},
  \bibinfo{year}{2019}.
\newblock \bibinfo{title}{Effects of core space and excitation levels on
  ground-state correlation and photoionization dynamics of {Be} and {Ne}}.
\newblock \bibinfo{journal}{J. Chem. Phys.} \bibinfo{volume}{150},
  \bibinfo{pages}{084305}.
\newblock \DOIprefix\doi{10.1063/1.5082940}.
%Type = Article
\bibitem[{Orimo et~al.(2019)Orimo, Sato and Ishikawa}]{Orimo2019}
\bibinfo{author}{Orimo, Y.}, \bibinfo{author}{Sato, T.},
  \bibinfo{author}{Ishikawa, K.L.}, \bibinfo{year}{2019}.
\newblock \bibinfo{title}{Application of the time-dependent surface flux method
  to the time-dependent multiconfiguration self-consistent-field method}.
\newblock \bibinfo{journal}{Phys. Rev. A} \bibinfo{volume}{100},
  \bibinfo{pages}{013419}.
\newblock \DOIprefix\doi{10.1103/PhysRevA.100.013419}.
%Type = Article
\bibitem[{Orimo et~al.(2018)Orimo, Sato, Scrinzi and Ishikawa}]{Orimo2018}
\bibinfo{author}{Orimo, Y.}, \bibinfo{author}{Sato, T.},
  \bibinfo{author}{Scrinzi, A.}, \bibinfo{author}{Ishikawa, K.L.},
  \bibinfo{year}{2018}.
\newblock \bibinfo{title}{Implementation of the infinite-range exterior complex
  scaling to the time-dependent complete-active-space self-consistent-field
  method}.
\newblock \bibinfo{journal}{Phys. Rev. A} \bibinfo{volume}{97},
  \bibinfo{pages}{023423}.
\newblock \DOIprefix\doi{10.1103/PhysRevA.97.023423}.
%Type = Article
\bibitem[{Ott et~al.(2014)Ott, Kaldun, Argenti, Raith, Meyer, Laux, Zhang,
  Bl\"attermann, Hagstotz, Ding, Heck, Madro{\~{n}}ero, Mart{\'{\i}}n and
  Pfeifer}]{Ott2014}
\bibinfo{author}{Ott, C.}, \bibinfo{author}{Kaldun, A.},
  \bibinfo{author}{Argenti, L.}, \bibinfo{author}{Raith, P.},
  \bibinfo{author}{Meyer, K.}, \bibinfo{author}{Laux, M.},
  \bibinfo{author}{Zhang, Y.}, \bibinfo{author}{Bl\"attermann, A.},
  \bibinfo{author}{Hagstotz, S.}, \bibinfo{author}{Ding, T.},
  \bibinfo{author}{Heck, R.}, \bibinfo{author}{Madro{\~{n}}ero, J.},
  \bibinfo{author}{Mart{\'{\i}}n, F.}, \bibinfo{author}{Pfeifer, T.},
  \bibinfo{year}{2014}.
\newblock \bibinfo{title}{Reconstruction and control of a time-dependent
  two-electron wave packet}.
\newblock \bibinfo{journal}{Nature} \bibinfo{volume}{516},
  \bibinfo{pages}{374--378}.
\newblock \DOIprefix\doi{10.1038/nature14026}.
%Type = Article
\bibitem[{Ott et~al.(2013)Ott, Kaldun, Raith, Meyer, Laux, Evers, Keitel,
  Greene and Pfeifer}]{Ott2013}
\bibinfo{author}{Ott, C.}, \bibinfo{author}{Kaldun, A.},
  \bibinfo{author}{Raith, P.}, \bibinfo{author}{Meyer, K.},
  \bibinfo{author}{Laux, M.}, \bibinfo{author}{Evers, J.},
  \bibinfo{author}{Keitel, C.H.}, \bibinfo{author}{Greene, C.H.},
  \bibinfo{author}{Pfeifer, T.}, \bibinfo{year}{2013}.
\newblock \bibinfo{title}{{L}orentz meets {F}ano in spectral line shapes: A
  universal phase and its laser control}.
\newblock \bibinfo{journal}{Science} \bibinfo{volume}{340},
  \bibinfo{pages}{716--720}.
\newblock \DOIprefix\doi{10.1126/science.1234407}.
%Type = Article
\bibitem[{Ovcharenko et~al.(2020)Ovcharenko, LaForge, Langbehn, Plekan, Cucini,
  Finetti, O'Keeffe, Iablonskyi, Nishiyama, Ueda, Piseri, DiFraia, Richter,
  Coreno, Callegari, Prince, Stienkemeier, Moeller and
  Mudrich}]{Ovcharenko2020}
\bibinfo{author}{Ovcharenko, Y.}, \bibinfo{author}{LaForge, A.},
  \bibinfo{author}{Langbehn, B.}, \bibinfo{author}{Plekan, O.},
  \bibinfo{author}{Cucini, R.}, \bibinfo{author}{Finetti, P.},
  \bibinfo{author}{O'Keeffe, P.}, \bibinfo{author}{Iablonskyi, D.},
  \bibinfo{author}{Nishiyama, T.}, \bibinfo{author}{Ueda, K.},
  \bibinfo{author}{Piseri, P.}, \bibinfo{author}{DiFraia, M.},
  \bibinfo{author}{Richter, R.}, \bibinfo{author}{Coreno, M.},
  \bibinfo{author}{Callegari, C.}, \bibinfo{author}{Prince, K.C.},
  \bibinfo{author}{Stienkemeier, F.}, \bibinfo{author}{Moeller, T.},
  \bibinfo{author}{Mudrich, M.}, \bibinfo{year}{2020}.
\newblock \bibinfo{title}{Autoionization dynamics of {He} nanodroplets
  resonantly excited by intense {XUV} laser pulses}.
\newblock \bibinfo{journal}{New J. Phys.}
  \DOIprefix\doi{10.1088/1367-2630/ab9554}. \bibinfo{note}{accepted
  manuscript}.
%Type = Article
\bibitem[{Ovcharenko et~al.(2014)Ovcharenko, Lyamayev, Katzy, Devetta, LaForge,
  O'Keeffe, Plekan, Finetti, Di~Fraia, Mudrich, Krikunova, Piseri, Coreno,
  Brauer, Mazza, Stranges, Grazioli, Richter, Prince, Drabbels, Callegari,
  Stienkemeier and M\"oller}]{Ovcharenko2014}
\bibinfo{author}{Ovcharenko, Y.}, \bibinfo{author}{Lyamayev, V.},
  \bibinfo{author}{Katzy, R.}, \bibinfo{author}{Devetta, M.},
  \bibinfo{author}{LaForge, A.}, \bibinfo{author}{O'Keeffe, P.},
  \bibinfo{author}{Plekan, O.}, \bibinfo{author}{Finetti, P.},
  \bibinfo{author}{Di~Fraia, M.}, \bibinfo{author}{Mudrich, M.},
  \bibinfo{author}{Krikunova, M.}, \bibinfo{author}{Piseri, P.},
  \bibinfo{author}{Coreno, M.}, \bibinfo{author}{Brauer, N.B.},
  \bibinfo{author}{Mazza, T.}, \bibinfo{author}{Stranges, S.},
  \bibinfo{author}{Grazioli, C.}, \bibinfo{author}{Richter, R.},
  \bibinfo{author}{Prince, K.C.}, \bibinfo{author}{Drabbels, M.},
  \bibinfo{author}{Callegari, C.}, \bibinfo{author}{Stienkemeier, F.},
  \bibinfo{author}{M\"oller, T.}, \bibinfo{year}{2014}.
\newblock \bibinfo{title}{Novel collective autoionization process observed in
  electron spectra of {He} clusters}.
\newblock \bibinfo{journal}{Phys. Rev. Lett.} \bibinfo{volume}{112},
  \bibinfo{pages}{073401}.
\newblock \DOIprefix\doi{10.1103/PhysRevLett.112.073401}.
%Type = Article
\bibitem[{Owada et~al.(2018)Owada, Togawa, Inagaki, Hara, Tanaka, Joti, Koyama,
  Nakajima, Ohashi, Senba, Togashi, Tono, Yamaga, Yumoto, Yabashi, Tanaka and
  Ishikawa}]{Owada2018}
\bibinfo{author}{Owada, S.}, \bibinfo{author}{Togawa, K.},
  \bibinfo{author}{Inagaki, T.}, \bibinfo{author}{Hara, T.},
  \bibinfo{author}{Tanaka, T.}, \bibinfo{author}{Joti, Y.},
  \bibinfo{author}{Koyama, T.}, \bibinfo{author}{Nakajima, K.},
  \bibinfo{author}{Ohashi, H.}, \bibinfo{author}{Senba, Y.},
  \bibinfo{author}{Togashi, T.}, \bibinfo{author}{Tono, K.},
  \bibinfo{author}{Yamaga, M.}, \bibinfo{author}{Yumoto, H.},
  \bibinfo{author}{Yabashi, M.}, \bibinfo{author}{Tanaka, H.},
  \bibinfo{author}{Ishikawa, T.}, \bibinfo{year}{2018}.
\newblock \bibinfo{title}{A soft x-ray free-electron laser beamline at {SACLA}:
  the light source, photon beamline and experimental station}.
\newblock \bibinfo{journal}{J. Synchrotron Radiat.} \bibinfo{volume}{25},
  \bibinfo{pages}{282--288}.
\newblock \DOIprefix\doi{10.1107/s1600577517015685}.
%Type = Article
\bibitem[{Palacios et~al.(2006)Palacios, Bachau and Mart\'{\i}n}]{Palacios2006}
\bibinfo{author}{Palacios, A.}, \bibinfo{author}{Bachau, H.},
  \bibinfo{author}{Mart\'{\i}n, F.}, \bibinfo{year}{2006}.
\newblock \bibinfo{title}{Enhancement and control of {H}$_{2}$ dissociative
  ionization by femtosecond {VUV} laser pulses}.
\newblock \bibinfo{journal}{Phys. Rev. Lett.} \bibinfo{volume}{96},
  \bibinfo{pages}{143001}.
\newblock \DOIprefix\doi{10.1103/PhysRevLett.96.143001}.
%Type = Article
\bibitem[{Palacios and Mart\'in(2020)}]{Palacios2020}
\bibinfo{author}{Palacios, A.}, \bibinfo{author}{Mart\'in, F.},
  \bibinfo{year}{2020}.
\newblock \bibinfo{title}{The quantum chemistry of attosecond molecular
  science}.
\newblock \bibinfo{journal}{WIREs Comput. Mol. Sci.} \bibinfo{volume}{10},
  \bibinfo{pages}{e1430}.
\newblock \DOIprefix\doi{10.1002/wcms.1430}.
%Type = Article
\bibitem[{Palacios et~al.(2015)Palacios, Sanz-Vicario and
  Mart\'in}]{Palacios2015}
\bibinfo{author}{Palacios, A.}, \bibinfo{author}{Sanz-Vicario, J.L.},
  \bibinfo{author}{Mart\'in, F.}, \bibinfo{year}{2015}.
\newblock \bibinfo{title}{Theoretical methods for attosecond electron and
  nuclear dynamics: Applications to the {H}$_2$ molecule}.
\newblock \bibinfo{journal}{J. Phys. B} \bibinfo{volume}{48},
  \bibinfo{pages}{242001}.
\newblock \DOIprefix\doi{10.1088/0953-4075/48/24/242001}.
%Type = Article
\bibitem[{Parker et~al.(2001)Parker, Moore, Meharg, Dundas and
  Taylor}]{Parker2001}
\bibinfo{author}{Parker, J.S.}, \bibinfo{author}{Moore, L.R.},
  \bibinfo{author}{Meharg, K.J.}, \bibinfo{author}{Dundas, D.},
  \bibinfo{author}{Taylor, K.T.}, \bibinfo{year}{2001}.
\newblock \bibinfo{title}{Double-electron above threshold ionization of
  helium}.
\newblock \bibinfo{journal}{J. Phys. B} \bibinfo{volume}{34},
  \bibinfo{pages}{L69--L78}.
\newblock \DOIprefix\doi{10.1088/0953-4075/34/3/103}.
%Type = Article
\bibitem[{Pathak et~al.(2020a)Pathak, Sato and Ishikawa}]{Pathak2020}
\bibinfo{author}{Pathak, H.}, \bibinfo{author}{Sato, T.},
  \bibinfo{author}{Ishikawa, K.L.}, \bibinfo{year}{2020}a.
\newblock \bibinfo{title}{Time-dependent optimized coupled-cluster method for
  multielectron dynamics. {II}.~{A} coupled electron-pair approximation}.
\newblock \bibinfo{journal}{J. Chem. Phys.} \bibinfo{volume}{152},
  \bibinfo{pages}{124115}.
\newblock \DOIprefix\doi{10.1063/1.5143747}.
%Type = Article
\bibitem[{Pathak et~al.(2020b)Pathak, Ibele, Boll, Callegari, Demidovich, Erk,
  Feifel, Forbes, Di~Fraia, Giannessi, Hansen, Holland, Ingle, Mason, Plekan,
  Prince, Rouz\'{e}e, Squibb, Tross, Ashfold, Curchod and Rolles}]{Pathak2020b}
\bibinfo{author}{Pathak, S.}, \bibinfo{author}{Ibele, L.M.},
  \bibinfo{author}{Boll, R.}, \bibinfo{author}{Callegari, C.},
  \bibinfo{author}{Demidovich, A.}, \bibinfo{author}{Erk, B.},
  \bibinfo{author}{Feifel, R.}, \bibinfo{author}{Forbes, R.},
  \bibinfo{author}{Di~Fraia, M.}, \bibinfo{author}{Giannessi, L.},
  \bibinfo{author}{Hansen, C.S.}, \bibinfo{author}{Holland, D.M.P.},
  \bibinfo{author}{Ingle, R.A.}, \bibinfo{author}{Mason, R.},
  \bibinfo{author}{Plekan, O.}, \bibinfo{author}{Prince, K.C.},
  \bibinfo{author}{Rouz\'{e}e, A.}, \bibinfo{author}{Squibb, R.J.},
  \bibinfo{author}{Tross, J.}, \bibinfo{author}{Ashfold, M.N.R.},
  \bibinfo{author}{Curchod, B.F.E.}, \bibinfo{author}{Rolles, D.},
  \bibinfo{year}{2020}b.
\newblock \bibinfo{title}{Tracking the ultraviolet photochemistry of
  thiophenone during and beyond the initial ultrafast ring opening}.
\newblock \bibinfo{journal}{Nat. Chem.} \bibinfo{volume}{in press}.
\newblock \href{http://arxiv.org/abs/1912.00531}{\tt arXiv:1912.00531}.
%Type = Article
\bibitem[{{Paul} et~al.(2001){Paul}, {Toma}, {Breger}, {Mullot}, {Aug{\'e}},
  {Balcou}, {Muller} and {Agostini}}]{Paul2001}
\bibinfo{author}{{Paul}, P.M.}, \bibinfo{author}{{Toma}, E.S.},
  \bibinfo{author}{{Breger}, P.}, \bibinfo{author}{{Mullot}, G.},
  \bibinfo{author}{{Aug{\'e}}, F.}, \bibinfo{author}{{Balcou}, P.},
  \bibinfo{author}{{Muller}, H.G.}, \bibinfo{author}{{Agostini}, P.},
  \bibinfo{year}{2001}.
\newblock \bibinfo{title}{Observation of a train of attosecond pulses from high
  harmonic generation}.
\newblock \bibinfo{journal}{Science} \bibinfo{volume}{292},
  \bibinfo{pages}{1689--1692}.
\newblock \DOIprefix\doi{10.1126/science.1059413}.
%Type = Article
\bibitem[{Pazourek et~al.(2015)Pazourek, Nagele and
  Burgd\"orfer}]{Pazourek2015}
\bibinfo{author}{Pazourek, R.}, \bibinfo{author}{Nagele, S.},
  \bibinfo{author}{Burgd\"orfer, J.}, \bibinfo{year}{2015}.
\newblock \bibinfo{title}{Attosecond chronoscopy of photoemission}.
\newblock \bibinfo{journal}{Rev. Mod. Phys.} \bibinfo{volume}{87},
  \bibinfo{pages}{765--802}.
\newblock \DOIprefix\doi{10.1103/RevModPhys.87.765}.
%Type = Article
\bibitem[{Penco et~al.(2015)Penco, Allaria, De~Ninno, Ferrari and
  Giannessi}]{Penco2015}
\bibinfo{author}{Penco, G.}, \bibinfo{author}{Allaria, E.},
  \bibinfo{author}{De~Ninno, G.}, \bibinfo{author}{Ferrari, E.},
  \bibinfo{author}{Giannessi, L.}, \bibinfo{year}{2015}.
\newblock \bibinfo{title}{Experimental demonstration of enhanced self-amplified
  spontaneous emission by an optical klystron}.
\newblock \bibinfo{journal}{Phys. Rev. Lett.} \bibinfo{volume}{114},
  \bibinfo{pages}{013901}.
\newblock \DOIprefix\doi{10.1103/PhysRevLett.114.013901}.
%Type = Article
\bibitem[{Phillips(1960)}]{Phillips1960}
\bibinfo{author}{Phillips, R.}, \bibinfo{year}{1960}.
\newblock \bibinfo{title}{The {U}bitron, a high-power traveling-wave tube based
  on a periodic beam interaction in unloaded waveguide}.
\newblock \bibinfo{journal}{{IRE} Trans. Electron Devices} \bibinfo{volume}{7},
  \bibinfo{pages}{231--241}.
\newblock \DOIprefix\doi{10.1109/t-ed.1960.14687}.
%Type = Article
\bibitem[{Phillips(1988)}]{Phillips1988}
\bibinfo{author}{Phillips, R.M.}, \bibinfo{year}{1988}.
\newblock \bibinfo{title}{History of the ubitron}.
\newblock \bibinfo{journal}{Nuclear Instrum. Methods Phys. Research A}
  \bibinfo{volume}{272}, \bibinfo{pages}{1--9}.
\newblock \DOIprefix\doi{10.1016/0168-9002(88)90185-4}.
%Type = Article
\bibitem[{Picard et~al.(2004)Picard, Manschwetus, G\`el\`eoc, B\"ottcher,
  Casagrande, Lin, Ruchon, Carr\`e, Hergott, Lepetit, Ta\"\i{}eb, Maquet and
  Huetz}]{Picard2014}
\bibinfo{author}{Picard, Y.J.}, \bibinfo{author}{Manschwetus, B.},
  \bibinfo{author}{G\`el\`eoc, M.}, \bibinfo{author}{B\"ottcher, M.},
  \bibinfo{author}{Casagrande, E.M.S.}, \bibinfo{author}{Lin, N.},
  \bibinfo{author}{Ruchon, T.}, \bibinfo{author}{Carr\`e, B.},
  \bibinfo{author}{Hergott, J.F.}, \bibinfo{author}{Lepetit, F.},
  \bibinfo{author}{Ta\"\i{}eb, R.}, \bibinfo{author}{Maquet, A.},
  \bibinfo{author}{Huetz, A.}, \bibinfo{year}{2004}.
\newblock \bibinfo{title}{Attosecond evolution of energy- and angle-resolved
  photoemission spectra in two-color ({XUV} + {IR}) ionization of rare gases}.
\newblock \bibinfo{journal}{Phys. Rev. A} \bibinfo{volume}{89},
  \bibinfo{pages}{031401(R)}.
\newblock \DOIprefix\doi{10.1103/PhysRevA.89.031401}.
%Type = Article
\bibitem[{Pindzola et~al.(2019)Pindzola, Abdel-Naby and Colgan}]{Pindzola2019}
\bibinfo{author}{Pindzola, M.S.}, \bibinfo{author}{Abdel-Naby, S.A.},
  \bibinfo{author}{Colgan, J.P.}, \bibinfo{year}{2019}.
\newblock \bibinfo{title}{Triple autoionization of atomic ions}.
\newblock \bibinfo{journal}{J. Phys. B} \bibinfo{volume}{52},
  \bibinfo{pages}{095201}.
\newblock \DOIprefix\doi{10.1088/1361-6455/aafa37}.
%Type = Article
\bibitem[{Pl\'esiat et~al.(2012)Pl\'esiat, Decleva and Mart\'in}]{Plesiat2012}
\bibinfo{author}{Pl\'esiat, E.}, \bibinfo{author}{Decleva, P.},
  \bibinfo{author}{Mart\'in, F.}, \bibinfo{year}{2012}.
\newblock \bibinfo{title}{Vibrationally resolved photoelectron angular
  distributions from randomly oriented and fixed-in-space {N}$_2$ and {CO}
  molecules}.
\newblock \bibinfo{journal}{J. Phys. B: At. Mol. Opt. Phys.}
  \bibinfo{volume}{45}, \bibinfo{pages}{194008}.
\newblock \DOIprefix\doi{10.1088/0953-4075/45/19/194008}.
%Type = Article
\bibitem[{Prat et~al.(2020)Prat, Dijkstal, Ferrari and Reiche}]{Prat2020}
\bibinfo{author}{Prat, E.}, \bibinfo{author}{Dijkstal, P.},
  \bibinfo{author}{Ferrari, E.}, \bibinfo{author}{Reiche, S.},
  \bibinfo{year}{2020}.
\newblock \bibinfo{title}{Demonstration of large bandwidth hard {X}-ray
  free-electron laser pulses at {S}wiss{FEL}}.
\newblock \bibinfo{journal}{Phys. Rev. Lett.} \bibinfo{volume}{124},
  \bibinfo{pages}{074801}.
\newblock \DOIprefix\doi{10.1103/PhysRevLett.124.074801}.
%Type = Article
\bibitem[{Prat and Reiche(2018)}]{Prat2018}
\bibinfo{author}{Prat, E.}, \bibinfo{author}{Reiche, S.}, \bibinfo{year}{2018}.
\newblock \bibinfo{title}{Compact coherence enhancement by subharmonic
  self-seeding in {X}-ray free-electron laser facilities}.
\newblock \bibinfo{journal}{J. Synchrotron Radiat.} \bibinfo{volume}{25},
  \bibinfo{pages}{329--335}.
\newblock \DOIprefix\doi{10.1107/s1600577518000395}.
%Type = Article
\bibitem[{Prat and Reiche(2019)}]{Prat2019}
\bibinfo{author}{Prat, E.}, \bibinfo{author}{Reiche, S.}, \bibinfo{year}{2019}.
\newblock \bibinfo{title}{A simple and compact scheme to enhance the brightness
  of self-amplified spontaneous emission free-electron-lasers}.
\newblock \bibinfo{journal}{J. Synchrotron Radiat.} \bibinfo{volume}{26},
  \bibinfo{pages}{1085--1091}.
\newblock \DOIprefix\doi{10.1107/s1600577519005435}.
%Type = Article
\bibitem[{Prince et~al.(2016)Prince, Allaria, Callegari, Cucini, De~Ninno,
  Di~Mitri, Diviacco, Ferrari, Finetti, Gauthier, Giannessi, Mahne, Penco,
  Plekan, Raimondi, Rebernik, Roussel, Svetina, Trov{\`{o}}, Zangrando, Negro,
  Carpeggiani, Reduzzi, Sansone, Grum-Grzhimailo, Gryzlova, Strakhova,
  Bartschat, Douguet, Venzke, Iablonskyi, Kumagai, Takanashi, Ueda, Fischer,
  Coreno, Stienkemeier, Ovcharenko, Mazza and Meyer}]{Prince2016}
\bibinfo{author}{Prince, K.C.}, \bibinfo{author}{Allaria, E.},
  \bibinfo{author}{Callegari, C.}, \bibinfo{author}{Cucini, R.},
  \bibinfo{author}{De~Ninno, G.}, \bibinfo{author}{Di~Mitri, S.},
  \bibinfo{author}{Diviacco, B.}, \bibinfo{author}{Ferrari, E.},
  \bibinfo{author}{Finetti, P.}, \bibinfo{author}{Gauthier, D.},
  \bibinfo{author}{Giannessi, L.}, \bibinfo{author}{Mahne, N.},
  \bibinfo{author}{Penco, G.}, \bibinfo{author}{Plekan, O.},
  \bibinfo{author}{Raimondi, L.}, \bibinfo{author}{Rebernik, P.},
  \bibinfo{author}{Roussel, E.}, \bibinfo{author}{Svetina, C.},
  \bibinfo{author}{Trov{\`{o}}, M.}, \bibinfo{author}{Zangrando, M.},
  \bibinfo{author}{Negro, M.}, \bibinfo{author}{Carpeggiani, P.},
  \bibinfo{author}{Reduzzi, M.}, \bibinfo{author}{Sansone, G.},
  \bibinfo{author}{Grum-Grzhimailo, A.N.}, \bibinfo{author}{Gryzlova, E.V.},
  \bibinfo{author}{Strakhova, S.I.}, \bibinfo{author}{Bartschat, K.},
  \bibinfo{author}{Douguet, N.}, \bibinfo{author}{Venzke, J.},
  \bibinfo{author}{Iablonskyi, D.}, \bibinfo{author}{Kumagai, Y.},
  \bibinfo{author}{Takanashi, T.}, \bibinfo{author}{Ueda, K.},
  \bibinfo{author}{Fischer, A.}, \bibinfo{author}{Coreno, M.},
  \bibinfo{author}{Stienkemeier, F.}, \bibinfo{author}{Ovcharenko, Y.},
  \bibinfo{author}{Mazza, T.}, \bibinfo{author}{Meyer, M.},
  \bibinfo{year}{2016}.
\newblock \bibinfo{title}{Coherent control with a short-wavelength
  free-electron laser}.
\newblock \bibinfo{journal}{Nat. Photonics} \bibinfo{volume}{10},
  \bibinfo{pages}{176--179}.
\newblock \DOIprefix\doi{10.1038/nphoton.2016.13}.
%Type = Article
\bibitem[{Qu{\'{e}}r{\'{e}} et~al.(2005)Qu{\'{e}}r{\'{e}}, Mairesse and
  Itatani}]{Quere2005}
\bibinfo{author}{Qu{\'{e}}r{\'{e}}, F.}, \bibinfo{author}{Mairesse, Y.},
  \bibinfo{author}{Itatani, J.}, \bibinfo{year}{2005}.
\newblock \bibinfo{title}{Temporal characterization of attosecond {XUV}
  fields}.
\newblock \bibinfo{journal}{J. Mod. Opt.} \bibinfo{volume}{52},
  \bibinfo{pages}{339--360}.
\newblock \DOIprefix\doi{10.1080/09500340412331307942}.
%Type = Article
\bibitem[{Ranitovic et~al.(2011)Ranitovic, Tong, Hogle, Zhou, Liu, Toshima,
  Murnane and Kapteyn}]{Ranitovic2011}
\bibinfo{author}{Ranitovic, P.}, \bibinfo{author}{Tong, X.M.},
  \bibinfo{author}{Hogle, C.W.}, \bibinfo{author}{Zhou, X.},
  \bibinfo{author}{Liu, Y.}, \bibinfo{author}{Toshima, N.},
  \bibinfo{author}{Murnane, M.M.}, \bibinfo{author}{Kapteyn, H.C.},
  \bibinfo{year}{2011}.
\newblock \bibinfo{title}{Laser-enabled {A}uger decay in rare-gas atoms}.
\newblock \bibinfo{journal}{Phys. Rev. Lett.} \bibinfo{volume}{106},
  \bibinfo{pages}{053002}.
\newblock \DOIprefix\doi{10.1103/PhysRevLett.106.053002}.
%Type = Article
\bibitem[{Ratner et~al.(2015)Ratner, Abela, Amann, Behrens, Bohler, Bouchard,
  Bostedt, Boyes, Chow, Cocco, Decker, Ding, Eckman, Emma, Fairley, Feng,
  Field, Flechsig, Gassner, Hastings, Heimann, Huang, Kelez, Krzywinski, Loos,
  Lutman, Marinelli, Marcus, Maxwell, Montanez, Moeller, Morton, Nuhn, Rodes,
  Schlotter, Serkez, Stevens, Turner, Walz, Welch and Wu}]{Ratner2015}
\bibinfo{author}{Ratner, D.}, \bibinfo{author}{Abela, R.},
  \bibinfo{author}{Amann, J.}, \bibinfo{author}{Behrens, C.},
  \bibinfo{author}{Bohler, D.}, \bibinfo{author}{Bouchard, G.},
  \bibinfo{author}{Bostedt, C.}, \bibinfo{author}{Boyes, M.},
  \bibinfo{author}{Chow, K.}, \bibinfo{author}{Cocco, D.},
  \bibinfo{author}{Decker, F.J.}, \bibinfo{author}{Ding, Y.},
  \bibinfo{author}{Eckman, C.}, \bibinfo{author}{Emma, P.},
  \bibinfo{author}{Fairley, D.}, \bibinfo{author}{Feng, Y.},
  \bibinfo{author}{Field, C.}, \bibinfo{author}{Flechsig, U.},
  \bibinfo{author}{Gassner, G.}, \bibinfo{author}{Hastings, J.},
  \bibinfo{author}{Heimann, P.}, \bibinfo{author}{Huang, Z.},
  \bibinfo{author}{Kelez, N.}, \bibinfo{author}{Krzywinski, J.},
  \bibinfo{author}{Loos, H.}, \bibinfo{author}{Lutman, A.},
  \bibinfo{author}{Marinelli, A.}, \bibinfo{author}{Marcus, G.},
  \bibinfo{author}{Maxwell, T.}, \bibinfo{author}{Montanez, P.},
  \bibinfo{author}{Moeller, S.}, \bibinfo{author}{Morton, D.},
  \bibinfo{author}{Nuhn, H.D.}, \bibinfo{author}{Rodes, N.},
  \bibinfo{author}{Schlotter, W.}, \bibinfo{author}{Serkez, S.},
  \bibinfo{author}{Stevens, T.}, \bibinfo{author}{Turner, J.},
  \bibinfo{author}{Walz, D.}, \bibinfo{author}{Welch, J.}, \bibinfo{author}{Wu,
  J.}, \bibinfo{year}{2015}.
\newblock \bibinfo{title}{Experimental demonstration of a soft {X}-ray
  self-seeded free-electron laser}.
\newblock \bibinfo{journal}{Phys. Rev. Lett.} \bibinfo{volume}{114},
  \bibinfo{pages}{054801}.
\newblock \DOIprefix\doi{10.1103/PhysRevLett.114.054801}.
%Type = Inproceedings
\bibitem[{Raubenheimer(2018)}]{Raubenheimer2018}
\bibinfo{author}{Raubenheimer, T.}, \bibinfo{year}{2018}.
\newblock \bibinfo{title}{The {LCLS-II-HE}, a high energy upgrade of the
  {LCLS-II}}, in: \bibinfo{booktitle}{Proc. 60th ICFA Advanced Beam Dynamics
  Workshop (FLS'18), Shanghai, China, 5-9 March 2018},
  \bibinfo{publisher}{JACoW Publishing}, \bibinfo{address}{Geneva,
  Switzerland}. pp. \bibinfo{pages}{6--11}.
\newblock \DOIprefix\doi{10.18429/JACoW-FLS2018-MOP1WA02}.
%Type = Article
\bibitem[{Rebernik~Ribi{\v{c}} et~al.(2019)Rebernik~Ribi{\v{c}}, Abrami,
  Badano, Bossi, Braun, Bruchon, Capotondi, Castronovo, Cautero, Cinquegrana,
  Coreno, Couprie, Cudin, Danailov, De~Ninno, Demidovich, Di~Mitri, Diviacco,
  Fawley, Feng, Ferianis, Ferrari, Foglia, Frassetto, Gaio, Garzella, Ghaith,
  Giacuzzo, Giannessi, Grattoni, Grulja, Hemsing, Iazzourene, Kurdi, Lonza,
  Mahne, Malvestuto, Manfredda, Masciovecchio, Miotti, Mirian, Nikolov, Penco,
  Penn, Poletto, Pop, Prat, Principi, Raimondi, Reiche, Roussel, Sauro,
  Scafuri, Sigalotti, Spampinati, Spezzani, Sturari, Svandrlik, Tanikawa,
  Trov{\'{o}}, Veronese, Vivoda, Xiang, Zaccaria, Zangrando, Zangrando and
  Allaria}]{RebernikRibic_NatPhot_2019}
\bibinfo{author}{Rebernik~Ribi{\v{c}}, P.}, \bibinfo{author}{Abrami, A.},
  \bibinfo{author}{Badano, L.}, \bibinfo{author}{Bossi, M.},
  \bibinfo{author}{Braun, H.H.}, \bibinfo{author}{Bruchon, N.},
  \bibinfo{author}{Capotondi, F.}, \bibinfo{author}{Castronovo, D.},
  \bibinfo{author}{Cautero, M.}, \bibinfo{author}{Cinquegrana, P.},
  \bibinfo{author}{Coreno, M.}, \bibinfo{author}{Couprie, M.E.},
  \bibinfo{author}{Cudin, I.}, \bibinfo{author}{Danailov, M.B.},
  \bibinfo{author}{De~Ninno, G.}, \bibinfo{author}{Demidovich, A.},
  \bibinfo{author}{Di~Mitri, S.}, \bibinfo{author}{Diviacco, B.},
  \bibinfo{author}{Fawley, W.M.}, \bibinfo{author}{Feng, C.},
  \bibinfo{author}{Ferianis, M.}, \bibinfo{author}{Ferrari, E.},
  \bibinfo{author}{Foglia, L.}, \bibinfo{author}{Frassetto, F.},
  \bibinfo{author}{Gaio, G.}, \bibinfo{author}{Garzella, D.},
  \bibinfo{author}{Ghaith, A.}, \bibinfo{author}{Giacuzzo, F.},
  \bibinfo{author}{Giannessi, L.}, \bibinfo{author}{Grattoni, V.},
  \bibinfo{author}{Grulja, S.}, \bibinfo{author}{Hemsing, E.},
  \bibinfo{author}{Iazzourene, F.}, \bibinfo{author}{Kurdi, G.},
  \bibinfo{author}{Lonza, M.}, \bibinfo{author}{Mahne, N.},
  \bibinfo{author}{Malvestuto, M.}, \bibinfo{author}{Manfredda, M.},
  \bibinfo{author}{Masciovecchio, C.}, \bibinfo{author}{Miotti, P.},
  \bibinfo{author}{Mirian, N.S.}, \bibinfo{author}{Nikolov, I.P.},
  \bibinfo{author}{Penco, G.M.}, \bibinfo{author}{Penn, G.},
  \bibinfo{author}{Poletto, L.}, \bibinfo{author}{Pop, M.},
  \bibinfo{author}{Prat, E.}, \bibinfo{author}{Principi, E.},
  \bibinfo{author}{Raimondi, L.}, \bibinfo{author}{Reiche, S.},
  \bibinfo{author}{Roussel, E.}, \bibinfo{author}{Sauro, R.},
  \bibinfo{author}{Scafuri, C.}, \bibinfo{author}{Sigalotti, P.},
  \bibinfo{author}{Spampinati, S.}, \bibinfo{author}{Spezzani, C.},
  \bibinfo{author}{Sturari, L.}, \bibinfo{author}{Svandrlik, M.},
  \bibinfo{author}{Tanikawa, T.}, \bibinfo{author}{Trov{\'{o}}, M.},
  \bibinfo{author}{Veronese, M.}, \bibinfo{author}{Vivoda, D.},
  \bibinfo{author}{Xiang, D.}, \bibinfo{author}{Zaccaria, M.},
  \bibinfo{author}{Zangrando, D.}, \bibinfo{author}{Zangrando, M.},
  \bibinfo{author}{Allaria, E.M.}, \bibinfo{year}{2019}.
\newblock \bibinfo{title}{Coherent soft {X}-ray pulses from an echo-enabled
  harmonic generation free-electron laser}.
\newblock \bibinfo{journal}{Nat. Photonics} \bibinfo{volume}{13},
  \bibinfo{pages}{555--561}.
\newblock \DOIprefix\doi{10.1038/s41566-019-0427-1}.
%Type = Article
\bibitem[{Reiss(1980)}]{Reiss1980}
\bibinfo{author}{Reiss, H.R.}, \bibinfo{year}{1980}.
\newblock \bibinfo{title}{Effect of an intense electromagnetic field on a
  weakly bound system}.
\newblock \bibinfo{journal}{Phys. Rev. A} \bibinfo{volume}{22},
  \bibinfo{pages}{1786--1813}.
\newblock \DOIprefix\doi{10.1103/PhysRevA.22.1786}.
%Type = Article
\bibitem[{Rice(1992)}]{Rice1992}
\bibinfo{author}{Rice, S.A.}, \bibinfo{year}{1992}.
\newblock \bibinfo{title}{New ideas for guiding the evolution of a quantum
  system}.
\newblock \bibinfo{journal}{Science} \bibinfo{volume}{258},
  \bibinfo{pages}{412}.
\newblock \DOIprefix\doi{10.1126/science.258.5081.412}.
%Type = Article
\bibitem[{Richardson et~al.(2010)Richardson, Costello, Cubaynes, D\"usterer,
  Feldhaus, van~der Hart, Jurani\'{c}, Li, Meyer, Richter, Sorokin and
  Tiedke}]{Richardson2010}
\bibinfo{author}{Richardson, V.}, \bibinfo{author}{Costello, J.T.},
  \bibinfo{author}{Cubaynes, D.}, \bibinfo{author}{D\"usterer, S.},
  \bibinfo{author}{Feldhaus, J.}, \bibinfo{author}{van~der Hart, H.W.},
  \bibinfo{author}{Jurani\'{c}, P.}, \bibinfo{author}{Li, W.B.},
  \bibinfo{author}{Meyer, M.}, \bibinfo{author}{Richter, M.},
  \bibinfo{author}{Sorokin, A.A.}, \bibinfo{author}{Tiedke, K.},
  \bibinfo{year}{2010}.
\newblock \bibinfo{title}{Two-photon inner-shell ionization in the extreme
  ultraviolet}.
\newblock \bibinfo{journal}{Phys. Rev. Lett.} \bibinfo{volume}{105},
  \bibinfo{pages}{013001}.
\newblock \DOIprefix\doi{10.1103/PhysRevLett.105.013001}.
%Type = Book
\bibitem[{Roberson and Sprangle(1989a)}]{Roberson1989Book}
\bibinfo{author}{Roberson, C.W.}, \bibinfo{author}{Sprangle, P.},
  \bibinfo{year}{1989}a.
\newblock \bibinfo{title}{A Review of Free Electron Lasers}.
\newblock \bibinfo{publisher}{Naval Research Laboratory (U.S.)}.
\newblock \bibinfo{note}{Also published as \cite{Roberson1989}}.
%Type = Article
\bibitem[{Roberson and Sprangle(1989b)}]{Roberson1989}
\bibinfo{author}{Roberson, C.W.}, \bibinfo{author}{Sprangle, P.},
  \bibinfo{year}{1989}b.
\newblock \bibinfo{title}{A review of free-electron lasers}.
\newblock \bibinfo{journal}{Phys. Fluids B: Plasma Phys.} \bibinfo{volume}{1},
  \bibinfo{pages}{3--42}.
\newblock \DOIprefix\doi{10.1063/1.859102}.
%Type = Article
\bibitem[{Rohringer et~al.(2012)Rohringer, Ryan, London, Purvis, Albert, Dunn,
  Bozek, Bostedt, Graf, Hill, Hau-Riege and Rocca}]{Rohringer2012}
\bibinfo{author}{Rohringer, N.}, \bibinfo{author}{Ryan, D.},
  \bibinfo{author}{London, R.A.}, \bibinfo{author}{Purvis, M.},
  \bibinfo{author}{Albert, F.}, \bibinfo{author}{Dunn, J.},
  \bibinfo{author}{Bozek, J.D.}, \bibinfo{author}{Bostedt, C.},
  \bibinfo{author}{Graf, A.}, \bibinfo{author}{Hill, R.},
  \bibinfo{author}{Hau-Riege, S.P.}, \bibinfo{author}{Rocca, J.J.},
  \bibinfo{year}{2012}.
\newblock \bibinfo{title}{Atomic inner-shell {X}-ray laser at 1.46 nanometres
  pumped by an {X}-ray free-electron laser}.
\newblock \bibinfo{journal}{Nature} \bibinfo{volume}{481},
  \bibinfo{pages}{488--491}.
\newblock \DOIprefix\doi{10.1038/nature10721}.
%Type = Article
\bibitem[{Rohringer and Santra(2012)}]{Rohringer2012a}
\bibinfo{author}{Rohringer, N.}, \bibinfo{author}{Santra, R.},
  \bibinfo{year}{2012}.
\newblock \bibinfo{title}{Strongly driven resonant {A}uger effect treated by an
  open-quantum-system approach}.
\newblock \bibinfo{journal}{Phys. Rev. A} \bibinfo{volume}{86},
  \bibinfo{pages}{043434}.
\newblock \DOIprefix\doi{10.1103/PhysRevA.86.043434}.
%Type = Article
\bibitem[{Roussel et~al.(2015)Roussel, Ferrari, Allaria, Penco, Di~Mitri,
  Veronese, Danailov, Gauthier and Giannessi}]{Roussel2015}
\bibinfo{author}{Roussel, E.}, \bibinfo{author}{Ferrari, E.},
  \bibinfo{author}{Allaria, E.}, \bibinfo{author}{Penco, G.},
  \bibinfo{author}{Di~Mitri, S.}, \bibinfo{author}{Veronese, M.},
  \bibinfo{author}{Danailov, M.}, \bibinfo{author}{Gauthier, D.},
  \bibinfo{author}{Giannessi, L.}, \bibinfo{year}{2015}.
\newblock \bibinfo{title}{Multicolor high-gain free-electron laser driven by
  seeded microbunching instability}.
\newblock \bibinfo{journal}{Phys. Rev. Lett.} \bibinfo{volume}{115},
  \bibinfo{pages}{214801}.
\newblock \DOIprefix\doi{10.1103/PhysRevLett.115.214801}.
%Type = Article
\bibitem[{Rouxel et~al.(2017)Rouxel, Kowalewski and Mukamel}]{Rouxel2017}
\bibinfo{author}{Rouxel, J.R.}, \bibinfo{author}{Kowalewski, M.},
  \bibinfo{author}{Mukamel, S.}, \bibinfo{year}{2017}.
\newblock \bibinfo{title}{Photoinduced molecular chirality probed by ultrafast
  resonant {X}-ray spectroscopy}.
\newblock \bibinfo{journal}{Struct. Dyn.} \bibinfo{volume}{4},
  \bibinfo{pages}{044006}.
\newblock \DOIprefix\doi{10.1063/1.4974260}.
%Type = Article
\bibitem[{Rouz\'ee et~al.(2011)Rouz\'ee, Johnsson, Gryzlova, Fukuzawa, Yamada,
  Siu, Huismans, Louis, Bijkerk, Holland, Grum-Grzhimailo, Kabachnik, Vrakking
  and Ueda}]{Rouzee2011}
\bibinfo{author}{Rouz\'ee, A.}, \bibinfo{author}{Johnsson, P.},
  \bibinfo{author}{Gryzlova, E.V.}, \bibinfo{author}{Fukuzawa, H.},
  \bibinfo{author}{Yamada, A.}, \bibinfo{author}{Siu, W.},
  \bibinfo{author}{Huismans, Y.}, \bibinfo{author}{Louis, E.},
  \bibinfo{author}{Bijkerk, F.}, \bibinfo{author}{Holland, D.M.P.},
  \bibinfo{author}{Grum-Grzhimailo, A.N.}, \bibinfo{author}{Kabachnik, N.M.},
  \bibinfo{author}{Vrakking, M.J.J.}, \bibinfo{author}{Ueda, K.},
  \bibinfo{year}{2011}.
\newblock \bibinfo{title}{Angle-resolved photoelectron spectroscopy of
  sequential three-photon triple ionization of neon at 90.5 {eV} photon
  energy}.
\newblock \bibinfo{journal}{Phys. Rev. A} \bibinfo{volume}{83},
  \bibinfo{pages}{031401}.
\newblock \DOIprefix\doi{10.1103/PhysRevA.83.031401}.
%Type = Article
\bibitem[{Rouz{\'{e}}e et~al.(2013)Rouz{\'{e}}e, Johnsson, Rading, Hundertmark,
  Siu, Huismans, D\"usterer, Redlin, Tavella, Stojanovic, Al-Shemmary,
  L{\'{e}}pine, Holland, Schlatholter, Hoekstra, Fukuzawa, Ueda and
  Vrakking}]{Rouzee2013}
\bibinfo{author}{Rouz{\'{e}}e, A.}, \bibinfo{author}{Johnsson, P.},
  \bibinfo{author}{Rading, L.}, \bibinfo{author}{Hundertmark, A.},
  \bibinfo{author}{Siu, W.}, \bibinfo{author}{Huismans, Y.},
  \bibinfo{author}{D\"usterer, S.}, \bibinfo{author}{Redlin, H.},
  \bibinfo{author}{Tavella, F.}, \bibinfo{author}{Stojanovic, N.},
  \bibinfo{author}{Al-Shemmary, A.}, \bibinfo{author}{L{\'{e}}pine, F.},
  \bibinfo{author}{Holland, D.M.P.}, \bibinfo{author}{Schlatholter, T.},
  \bibinfo{author}{Hoekstra, R.}, \bibinfo{author}{Fukuzawa, H.},
  \bibinfo{author}{Ueda, K.}, \bibinfo{author}{Vrakking, M.J.J.},
  \bibinfo{year}{2013}.
\newblock \bibinfo{title}{Towards imaging of ultrafast molecular dynamics using
  {FELs}}.
\newblock \bibinfo{journal}{J. Phys. B} \bibinfo{volume}{46},
  \bibinfo{pages}{164029}.
\newblock \DOIprefix\doi{10.1088/0953-4075/46/16/164029}.
%Type = Article
\bibitem[{Rudawski et~al.(2013)Rudawski, Heyl, Brizuela, Schwenke, Persson,
  Mansten, Rakowski, Rading, Campi, Kim, Johnsson and
  L'Huillier}]{Rudawski2013}
\bibinfo{author}{Rudawski, P.}, \bibinfo{author}{Heyl, C.M.},
  \bibinfo{author}{Brizuela, F.}, \bibinfo{author}{Schwenke, J.},
  \bibinfo{author}{Persson, A.}, \bibinfo{author}{Mansten, E.},
  \bibinfo{author}{Rakowski, R.}, \bibinfo{author}{Rading, L.},
  \bibinfo{author}{Campi, F.}, \bibinfo{author}{Kim, B.},
  \bibinfo{author}{Johnsson, P.}, \bibinfo{author}{L'Huillier, A.},
  \bibinfo{year}{2013}.
\newblock \bibinfo{title}{A high-flux high-order harmonic source}.
\newblock \bibinfo{journal}{Rev. Sci. Instrum.} \bibinfo{volume}{84},
  \bibinfo{pages}{073103}.
\newblock \DOIprefix\doi{10.1063/1.4812266}.
%Type = Article
\bibitem[{Rudek et~al.(2012)Rudek, Son, Foucar, Epp, Erk, Hartmann, Adolph,
  Andritschke, Aquila, Berrah, Bostedt, Bozek, Coppola, Filsinger, Gorke,
  Gorkhover, Graafsma, Gumprecht, Hartmann, Hauser, Herrmann, Hirsemann, Holl,
  H{\"o}mke, Journel, Kaiser, Kimmel, Krasniqi, K{\"u}hnel, Matysek,
  Messerschmidt, Miesner, M{\"o}ller, Moshammer, Nagaya, Nilsson, Potdevin,
  Pietschner, Reich, Rupp, Schaller, Schlichting, Schmidt, Schopper, Schorb,
  Schr{\"o}ter, Schulz, Simon, Soltau, Str{\"u}der, Ueda, Weidenspointner,
  Santra, Ullrich, Rudenko and Rolles}]{Rudek2012}
\bibinfo{author}{Rudek, B.}, \bibinfo{author}{Son, S.K.},
  \bibinfo{author}{Foucar, L.}, \bibinfo{author}{Epp, S.W.},
  \bibinfo{author}{Erk, B.}, \bibinfo{author}{Hartmann, R.},
  \bibinfo{author}{Adolph, M.}, \bibinfo{author}{Andritschke, R.},
  \bibinfo{author}{Aquila, A.}, \bibinfo{author}{Berrah, N.},
  \bibinfo{author}{Bostedt, C.}, \bibinfo{author}{Bozek, J.},
  \bibinfo{author}{Coppola, N.}, \bibinfo{author}{Filsinger, F.},
  \bibinfo{author}{Gorke, H.}, \bibinfo{author}{Gorkhover, T.},
  \bibinfo{author}{Graafsma, H.}, \bibinfo{author}{Gumprecht, L.},
  \bibinfo{author}{Hartmann, A.}, \bibinfo{author}{Hauser, G.},
  \bibinfo{author}{Herrmann, S.}, \bibinfo{author}{Hirsemann, H.},
  \bibinfo{author}{Holl, P.}, \bibinfo{author}{H{\"o}mke, A.},
  \bibinfo{author}{Journel, L.}, \bibinfo{author}{Kaiser, C.},
  \bibinfo{author}{Kimmel, N.}, \bibinfo{author}{Krasniqi, F.},
  \bibinfo{author}{K{\"u}hnel, K.U.}, \bibinfo{author}{Matysek, M.},
  \bibinfo{author}{Messerschmidt, M.}, \bibinfo{author}{Miesner, D.},
  \bibinfo{author}{M{\"o}ller, T.}, \bibinfo{author}{Moshammer, R.},
  \bibinfo{author}{Nagaya, K.}, \bibinfo{author}{Nilsson, B.},
  \bibinfo{author}{Potdevin, G.}, \bibinfo{author}{Pietschner, D.},
  \bibinfo{author}{Reich, C.}, \bibinfo{author}{Rupp, D.},
  \bibinfo{author}{Schaller, G.}, \bibinfo{author}{Schlichting, I.},
  \bibinfo{author}{Schmidt, C.}, \bibinfo{author}{Schopper, F.},
  \bibinfo{author}{Schorb, S.}, \bibinfo{author}{Schr{\"o}ter, C.D.},
  \bibinfo{author}{Schulz, J.}, \bibinfo{author}{Simon, M.},
  \bibinfo{author}{Soltau, H.}, \bibinfo{author}{Str{\"u}der, L.},
  \bibinfo{author}{Ueda, K.}, \bibinfo{author}{Weidenspointner, G.},
  \bibinfo{author}{Santra, R.}, \bibinfo{author}{Ullrich, J.},
  \bibinfo{author}{Rudenko, A.}, \bibinfo{author}{Rolles, D.},
  \bibinfo{year}{2012}.
\newblock \bibinfo{title}{Ultra-efficient ionization of heavy atoms by intense
  {X}-ray free-electron laser pulses}.
\newblock \bibinfo{journal}{Nat. Photonics} \bibinfo{volume}{6},
  \bibinfo{pages}{858--865}.
\newblock \DOIprefix\doi{10.1038/nphoton.2012.261}.
%Type = Article
\bibitem[{Rudek et~al.(2018)Rudek, Toyota, Foucar, Erk, Boll, Bomme, Correa,
  Carron, Boutet, Williams, Ferguson, Alonso-Mori, Koglin, Gorkhover, Bucher,
  Lehmann, Kr{\"a}ssig, Southworth, Young, Bostedt, Ueda, Marchenko, Simon,
  Jurek, Santra, Rudenko, Son and Rolles}]{Rudek2018}
\bibinfo{author}{Rudek, B.}, \bibinfo{author}{Toyota, K.},
  \bibinfo{author}{Foucar, L.}, \bibinfo{author}{Erk, B.},
  \bibinfo{author}{Boll, R.}, \bibinfo{author}{Bomme, C.},
  \bibinfo{author}{Correa, J.}, \bibinfo{author}{Carron, S.},
  \bibinfo{author}{Boutet, S.}, \bibinfo{author}{Williams, G.J.},
  \bibinfo{author}{Ferguson, K.R.}, \bibinfo{author}{Alonso-Mori, R.},
  \bibinfo{author}{Koglin, J.E.}, \bibinfo{author}{Gorkhover, T.},
  \bibinfo{author}{Bucher, M.}, \bibinfo{author}{Lehmann, C.S.},
  \bibinfo{author}{Kr{\"a}ssig, B.}, \bibinfo{author}{Southworth, S.H.},
  \bibinfo{author}{Young, L.}, \bibinfo{author}{Bostedt, C.},
  \bibinfo{author}{Ueda, K.}, \bibinfo{author}{Marchenko, T.},
  \bibinfo{author}{Simon, M.}, \bibinfo{author}{Jurek, Z.},
  \bibinfo{author}{Santra, R.}, \bibinfo{author}{Rudenko, A.},
  \bibinfo{author}{Son, S.K.}, \bibinfo{author}{Rolles, D.},
  \bibinfo{year}{2018}.
\newblock \bibinfo{title}{Relativistic and resonant effects in the ionization
  of heavy atoms by ultra-intense hard {X}-rays}.
\newblock \bibinfo{journal}{Nat. Commun.} \bibinfo{volume}{9},
  \bibinfo{pages}{4200}.
\newblock \DOIprefix\doi{10.1038/s41467-018-06745-6}.
%Type = Article
\bibitem[{Rudenko et~al.(2008)Rudenko, Foucar, Kurka, Ergler, K\"uhnel, Jiang,
  Voitkiv, Najjari, Kheifets, L\"udemann, Havermeier, Smolarski, Sch\"ossler,
  Cole, Sch\"offler, D\"orner, D\"usterer, Li, Keitel, Treusch, Gensch,
  Schr\"oter, Moshammer and Ullrich}]{Rudenko2008}
\bibinfo{author}{Rudenko, A.}, \bibinfo{author}{Foucar, L.},
  \bibinfo{author}{Kurka, M.}, \bibinfo{author}{Ergler, T.},
  \bibinfo{author}{K\"uhnel, K.U.}, \bibinfo{author}{Jiang, Y.H.},
  \bibinfo{author}{Voitkiv, A.}, \bibinfo{author}{Najjari, B.},
  \bibinfo{author}{Kheifets, A.}, \bibinfo{author}{L\"udemann, S.},
  \bibinfo{author}{Havermeier, T.}, \bibinfo{author}{Smolarski, M.},
  \bibinfo{author}{Sch\"ossler, S.}, \bibinfo{author}{Cole, K.},
  \bibinfo{author}{Sch\"offler, M.}, \bibinfo{author}{D\"orner, R.},
  \bibinfo{author}{D\"usterer, S.}, \bibinfo{author}{Li, W.},
  \bibinfo{author}{Keitel, B.}, \bibinfo{author}{Treusch, R.},
  \bibinfo{author}{Gensch, M.}, \bibinfo{author}{Schr\"oter, C.D.},
  \bibinfo{author}{Moshammer, R.}, \bibinfo{author}{Ullrich, J.},
  \bibinfo{year}{2008}.
\newblock \bibinfo{title}{Recoil-ion momentum distributions for two-photon
  double ionization of {He} and {Ne} by 44 {eV} free-electron laser radiation}.
\newblock \bibinfo{journal}{Phys. Rev. Lett.} \bibinfo{volume}{101},
  \bibinfo{pages}{073003}.
\newblock \DOIprefix\doi{10.1103/PhysRevLett.101.073003}.
%Type = Article
\bibitem[{Saldin et~al.(2008)Saldin, Schneidmiller and Yurkov}]{Saldin2008}
\bibinfo{author}{Saldin, E.}, \bibinfo{author}{Schneidmiller, E.},
  \bibinfo{author}{Yurkov, M.}, \bibinfo{year}{2008}.
\newblock \bibinfo{title}{Coherence properties of the radiation from {X}-ray
  free electron laser}.
\newblock \bibinfo{journal}{Opt. Commun.} \bibinfo{volume}{281},
  \bibinfo{pages}{1179--1188}.
\newblock \DOIprefix\doi{10.1016/j.optcom.2007.10.044}.
%Type = Article
\bibitem[{Sal\'en et~al.(2012)Sal\'en, van~der Meulen, Schmidt, Thomas,
  Larsson, Feifel, Piancastelli, Fang, Murphy, Osipov, Berrah, Kukk, Ueda,
  Bozek, Bostedt, Wada, Richter, Feyer and Prince}]{Salen2012}
\bibinfo{author}{Sal\'en, P.}, \bibinfo{author}{van~der Meulen, P.},
  \bibinfo{author}{Schmidt, H.T.}, \bibinfo{author}{Thomas, R.D.},
  \bibinfo{author}{Larsson, M.}, \bibinfo{author}{Feifel, R.},
  \bibinfo{author}{Piancastelli, M.N.}, \bibinfo{author}{Fang, L.},
  \bibinfo{author}{Murphy, B.}, \bibinfo{author}{Osipov, T.},
  \bibinfo{author}{Berrah, N.}, \bibinfo{author}{Kukk, E.},
  \bibinfo{author}{Ueda, K.}, \bibinfo{author}{Bozek, J.D.},
  \bibinfo{author}{Bostedt, C.}, \bibinfo{author}{Wada, S.},
  \bibinfo{author}{Richter, R.}, \bibinfo{author}{Feyer, V.},
  \bibinfo{author}{Prince, K.C.}, \bibinfo{year}{2012}.
\newblock \bibinfo{title}{Experimental verification of the chemical sensitivity
  of two-site double core-hole states formed by an {X}-ray free-electron
  laser}.
\newblock \bibinfo{journal}{Phys. Rev. Lett.} \bibinfo{volume}{108},
  \bibinfo{pages}{153003}.
\newblock \DOIprefix\doi{10.1103/PhysRevLett.108.153003}.
%Type = Article
\bibitem[{Santra et~al.(2009)Santra, Kryzhevoi and Cederbaum}]{Santra2009}
\bibinfo{author}{Santra, R.}, \bibinfo{author}{Kryzhevoi, N.V.},
  \bibinfo{author}{Cederbaum, L.S.}, \bibinfo{year}{2009}.
\newblock \bibinfo{title}{X-ray two-photon photoelectron spectroscopy: A
  theoretical study of inner-shell spectra of the organic para-aminophenol
  molecule}.
\newblock \bibinfo{journal}{Phys. Rev. Lett.} \bibinfo{volume}{103},
  \bibinfo{pages}{013002}.
\newblock \DOIprefix\doi{10.1103/PhysRevLett.103.013002}.
%Type = Article
\bibitem[{Sasaki(1994)}]{Sasaki1994}
\bibinfo{author}{Sasaki, S.}, \bibinfo{year}{1994}.
\newblock \bibinfo{title}{Analyses for a planar variably-polarizing undulator}.
\newblock \bibinfo{journal}{Nucl. Instrum. Methods Phys. Res. A}
  \bibinfo{volume}{347}, \bibinfo{pages}{83--86}.
\newblock \DOIprefix\doi{10.1016/0168-9002(94)91859-7}.
%Type = Article
\bibitem[{{Sassoli de Bianchi}(2012)}]{SassolideBianchi2012}
\bibinfo{author}{{Sassoli de Bianchi}, M.}, \bibinfo{year}{2012}.
\newblock \bibinfo{title}{Time-delay of classical and quantum scattering
  processes: a conceptual overview and a general definition}.
\newblock \bibinfo{journal}{Open Phys.} \bibinfo{volume}{10},
  \bibinfo{pages}{282--319}.
\newblock \DOIprefix\doi{10.2478/s11534-011-0105-5}.
%Type = Article
\bibitem[{Sato and Ishikawa(2013)}]{Sato2013}
\bibinfo{author}{Sato, T.}, \bibinfo{author}{Ishikawa, K.L.},
  \bibinfo{year}{2013}.
\newblock \bibinfo{title}{Time-dependent complete-active-space
  self-consistent-field method for multielectron dynamics in intense laser
  fields}.
\newblock \bibinfo{journal}{Phys. Rev. A} \bibinfo{volume}{88},
  \bibinfo{pages}{023402}.
\newblock \DOIprefix\doi{10.1103/PhysRevA.88.023402}.
%Type = Article
\bibitem[{Sato and Ishikawa(2015)}]{Sato2015}
\bibinfo{author}{Sato, T.}, \bibinfo{author}{Ishikawa, K.L.},
  \bibinfo{year}{2015}.
\newblock \bibinfo{title}{Time-dependent multiconfiguration
  self-consistent-field method based on the occupation-restricted
  multiple-active-space model for multielectron dynamics in intense laser
  fields}.
\newblock \bibinfo{journal}{Phys. Rev. A} \bibinfo{volume}{91},
  \bibinfo{pages}{023417}.
\newblock \DOIprefix\doi{10.1103/PhysRevA.91.023417}.
%Type = Article
\bibitem[{Sato et~al.(2016)Sato, Ishikawa, B\v{r}ezinov\'a, Lackner, Nagele and
  Burgd\"orfer}]{Sato2016}
\bibinfo{author}{Sato, T.}, \bibinfo{author}{Ishikawa, K.L.},
  \bibinfo{author}{B\v{r}ezinov\'a, I.}, \bibinfo{author}{Lackner, F.},
  \bibinfo{author}{Nagele, S.}, \bibinfo{author}{Burgd\"orfer, J.},
  \bibinfo{year}{2016}.
\newblock \bibinfo{title}{Time-dependent complete-active-space
  self-consistent-field method for atoms: Application to high-order harmonic
  generation}.
\newblock \bibinfo{journal}{Phys. Rev. A} \bibinfo{volume}{94},
  \bibinfo{pages}{023405}.
\newblock \DOIprefix\doi{10.1103/PhysRevA.94.023405}.
%Type = Incollection
\bibitem[{Sato et~al.(2018a)Sato, Orimo, Teramura, Tugs and
  Ishikawa}]{Sato2018}
\bibinfo{author}{Sato, T.}, \bibinfo{author}{Orimo, Y.},
  \bibinfo{author}{Teramura, T.}, \bibinfo{author}{Tugs, O.},
  \bibinfo{author}{Ishikawa, K.L.}, \bibinfo{year}{2018}a.
\newblock \bibinfo{title}{Time-dependent complete-active-space
  self-consistent-field method for ultrafast intense laser science}, in:
  \bibinfo{editor}{Yamanouchi, K.}, \bibinfo{editor}{Martin, P.},
  \bibinfo{editor}{Sentis, M.}, \bibinfo{editor}{Ruxin, L.},
  \bibinfo{editor}{Normand, D.} (Eds.), \bibinfo{booktitle}{Progress in
  Ultrafast Intense Laser Science {XIV}}. \bibinfo{publisher}{Springer
  International Publishing}. volume \bibinfo{volume}{118} of
  \textit{\bibinfo{series}{Springer Series in Chemical Physics}}, pp.
  \bibinfo{pages}{143--171}.
\newblock \DOIprefix\doi{10.1007/978-3-030-03786-4_8}.
%Type = Article
\bibitem[{Sato et~al.(2018b)Sato, Pathak, Orimo and Ishikawa}]{Sato2018b}
\bibinfo{author}{Sato, T.}, \bibinfo{author}{Pathak, H.},
  \bibinfo{author}{Orimo, Y.}, \bibinfo{author}{Ishikawa, K.L.},
  \bibinfo{year}{2018}b.
\newblock \bibinfo{title}{Communication: {T}ime-dependent optimized
  coupled-cluster method for multielectron dynamics}.
\newblock \bibinfo{journal}{J. Chem. Phys.} \bibinfo{volume}{148},
  \bibinfo{pages}{051101}.
\newblock \DOIprefix\doi{10.1063/1.5020633}.
%Type = Article
\bibitem[{Sato et~al.(2018c)Sato, Teramura and Ishikawa}]{Sato2018a}
\bibinfo{author}{Sato, T.}, \bibinfo{author}{Teramura, T.},
  \bibinfo{author}{Ishikawa, K.L.}, \bibinfo{year}{2018}c.
\newblock \bibinfo{title}{Gauge-invariant formulation of time-dependent
  configuration interaction singles method}.
\newblock \bibinfo{journal}{Appl. Sci.} \bibinfo{volume}{8},
  \bibinfo{pages}{433}.
\newblock \DOIprefix\doi{10.3390/app8030433}.
%Type = Article
\bibitem[{Schawlow and Townes(1958)}]{Schawlow1958}
\bibinfo{author}{Schawlow, A.L.}, \bibinfo{author}{Townes, C.H.},
  \bibinfo{year}{1958}.
\newblock \bibinfo{title}{Infrared and optical masers}.
\newblock \bibinfo{journal}{Phys. Rev.} \bibinfo{volume}{112},
  \bibinfo{pages}{1940--1949}.
\newblock \DOIprefix\doi{10.1103/PhysRev.112.1940}.
%Type = Article
\bibitem[{Schins et~al.(1994)Schins, Breger, Agostini, Constantinescu, Muller,
  Grillon, Antonetti and Mysyrowicz}]{Schins1994}
\bibinfo{author}{Schins, J.M.}, \bibinfo{author}{Breger, P.},
  \bibinfo{author}{Agostini, P.}, \bibinfo{author}{Constantinescu, R.C.},
  \bibinfo{author}{Muller, H.G.}, \bibinfo{author}{Grillon, G.},
  \bibinfo{author}{Antonetti, A.}, \bibinfo{author}{Mysyrowicz, A.},
  \bibinfo{year}{1994}.
\newblock \bibinfo{title}{Observation of laser-assisted {A}uger decay in
  argon}.
\newblock \bibinfo{journal}{Phys. Rev. Lett.} \bibinfo{volume}{73},
  \bibinfo{pages}{2180--2183}.
\newblock \DOIprefix\doi{10.1103/PhysRevLett.73.2180}.
%Type = Article
\bibitem[{Schmiegelow et~al.(2016)Schmiegelow, Schulz, Kaufmann, Ruster,
  Poschinger and Schmidt-Kaler}]{Schmiegelow2016}
\bibinfo{author}{Schmiegelow, C.T.}, \bibinfo{author}{Schulz, J.},
  \bibinfo{author}{Kaufmann, H.}, \bibinfo{author}{Ruster, T.},
  \bibinfo{author}{Poschinger, U.G.}, \bibinfo{author}{Schmidt-Kaler, F.},
  \bibinfo{year}{2016}.
\newblock \bibinfo{title}{Transfer of optical orbital angular momentum to a
  bound electron}.
\newblock \bibinfo{journal}{Nature Communications} \bibinfo{volume}{7},
  \bibinfo{pages}{12998}.
\newblock \DOIprefix\doi{10.1038/ncomms12998}.
%Type = Article
\bibitem[{Schnorr et~al.(2013)Schnorr, Senftleben, Kurka, Rudenko, Foucar,
  Schmid, Broska, Pfeifer, Meyer, Anielski, Boll, Rolles, K\"ubel, Kling,
  Jiang, Mondal, Tachibana, Ueda, Marchenko, Simon, Brenner, Treusch, Scheit,
  Averbukh, Ullrich, Schr\"oter and Moshammer}]{Schnorr2013}
\bibinfo{author}{Schnorr, K.}, \bibinfo{author}{Senftleben, A.},
  \bibinfo{author}{Kurka, M.}, \bibinfo{author}{Rudenko, A.},
  \bibinfo{author}{Foucar, L.}, \bibinfo{author}{Schmid, G.},
  \bibinfo{author}{Broska, A.}, \bibinfo{author}{Pfeifer, T.},
  \bibinfo{author}{Meyer, K.}, \bibinfo{author}{Anielski, D.},
  \bibinfo{author}{Boll, R.}, \bibinfo{author}{Rolles, D.},
  \bibinfo{author}{K\"ubel, M.}, \bibinfo{author}{Kling, M.F.},
  \bibinfo{author}{Jiang, Y.H.}, \bibinfo{author}{Mondal, S.},
  \bibinfo{author}{Tachibana, T.}, \bibinfo{author}{Ueda, K.},
  \bibinfo{author}{Marchenko, T.}, \bibinfo{author}{Simon, M.},
  \bibinfo{author}{Brenner, G.}, \bibinfo{author}{Treusch, R.},
  \bibinfo{author}{Scheit, S.}, \bibinfo{author}{Averbukh, V.},
  \bibinfo{author}{Ullrich, J.}, \bibinfo{author}{Schr\"oter, C.D.},
  \bibinfo{author}{Moshammer, R.}, \bibinfo{year}{2013}.
\newblock \bibinfo{title}{Time-resolved measurement of interatomic coulombic
  decay in {Ne}$_{2}$}.
\newblock \bibinfo{journal}{Phys. Rev. Lett.} \bibinfo{volume}{111},
  \bibinfo{pages}{093402}.
\newblock \DOIprefix\doi{10.1103/PhysRevLett.111.093402}.
%Type = Article
\bibitem[{Schoenlein et~al.(2019)Schoenlein, Elsaesser, Holldack, Huang,
  Kapteyn, Murnane and Woerner}]{Schoenlein2019}
\bibinfo{author}{Schoenlein, R.}, \bibinfo{author}{Elsaesser, T.},
  \bibinfo{author}{Holldack, K.}, \bibinfo{author}{Huang, Z.},
  \bibinfo{author}{Kapteyn, H.}, \bibinfo{author}{Murnane, M.},
  \bibinfo{author}{Woerner, M.}, \bibinfo{year}{2019}.
\newblock \bibinfo{title}{Recent advances in ultrafast {X}-ray sources}.
\newblock \bibinfo{journal}{Phil. Trans. R. Soc. A} \bibinfo{volume}{377},
  \bibinfo{pages}{20180384}.
\newblock \DOIprefix\doi{10.1098/rsta.2018.0384}.
%Type = Article
\bibitem[{Schultze et~al.(2010)Schultze, Fie{\ss}, Karpowicz, Gagnon, Korbman,
  Hofstetter, Neppl, Cavalieri, Komninos, Mercouris, Nicolaides, Pazourek,
  Nagele, Feist, Burgd{\"o}rfer, Azzeer, Ernstorfer, Kienberger, Kleineberg,
  Goulielmakis, Krausz and Yakovlev}]{Schultze2010}
\bibinfo{author}{Schultze, M.}, \bibinfo{author}{Fie{\ss}, M.},
  \bibinfo{author}{Karpowicz, N.}, \bibinfo{author}{Gagnon, J.},
  \bibinfo{author}{Korbman, M.}, \bibinfo{author}{Hofstetter, M.},
  \bibinfo{author}{Neppl, S.}, \bibinfo{author}{Cavalieri, A.L.},
  \bibinfo{author}{Komninos, Y.}, \bibinfo{author}{Mercouris, T.},
  \bibinfo{author}{Nicolaides, C.A.}, \bibinfo{author}{Pazourek, R.},
  \bibinfo{author}{Nagele, S.}, \bibinfo{author}{Feist, J.},
  \bibinfo{author}{Burgd{\"o}rfer, J.}, \bibinfo{author}{Azzeer, A.M.},
  \bibinfo{author}{Ernstorfer, R.}, \bibinfo{author}{Kienberger, R.},
  \bibinfo{author}{Kleineberg, U.}, \bibinfo{author}{Goulielmakis, E.},
  \bibinfo{author}{Krausz, F.}, \bibinfo{author}{Yakovlev, V.S.},
  \bibinfo{year}{2010}.
\newblock \bibinfo{title}{Delay in photoemission}.
\newblock \bibinfo{journal}{Science} \bibinfo{volume}{328},
  \bibinfo{pages}{1658--1662}.
\newblock \DOIprefix\doi{10.1126/science.1189401}.
%Type = Article
\bibitem[{Sch{\"{u}}tte et~al.(2015)Sch{\"{u}}tte, Arbeiter, Fennel, Jabbari,
  Kuleff, Vrakking and Rouz{\'{e}}e}]{Schuette2015}
\bibinfo{author}{Sch{\"{u}}tte, B.}, \bibinfo{author}{Arbeiter, M.},
  \bibinfo{author}{Fennel, T.}, \bibinfo{author}{Jabbari, G.},
  \bibinfo{author}{Kuleff, A.}, \bibinfo{author}{Vrakking, M.},
  \bibinfo{author}{Rouz{\'{e}}e, A.}, \bibinfo{year}{2015}.
\newblock \bibinfo{title}{Observation of correlated electronic decay in
  expanding clusters triggered by near-infrared fields}.
\newblock \bibinfo{journal}{Nat. Commun.} \bibinfo{volume}{6},
  \bibinfo{pages}{8596}.
\newblock \DOIprefix\doi{10.1038/ncomms9596}.
%Type = Article
\bibitem[{Seddon et~al.(2017)Seddon, Clarke, Dunning, Masciovecchio, Milne,
  Parmigiani, Rugg, Spence, Thompson, Ueda, Vinko, Wark and Wurth}]{Seddon2017}
\bibinfo{author}{Seddon, E.A.}, \bibinfo{author}{Clarke, J.A.},
  \bibinfo{author}{Dunning, D.J.}, \bibinfo{author}{Masciovecchio, C.},
  \bibinfo{author}{Milne, C.J.}, \bibinfo{author}{Parmigiani, F.},
  \bibinfo{author}{Rugg, D.}, \bibinfo{author}{Spence, J.C.H.},
  \bibinfo{author}{Thompson, N.R.}, \bibinfo{author}{Ueda, K.},
  \bibinfo{author}{Vinko, S.M.}, \bibinfo{author}{Wark, J.S.},
  \bibinfo{author}{Wurth, W.}, \bibinfo{year}{2017}.
\newblock \bibinfo{title}{Short-wavelength free-electron laser sources and
  science: a review}.
\newblock \bibinfo{journal}{Rep. Prog. Phys.} \bibinfo{volume}{80},
  \bibinfo{pages}{115901}.
\newblock \DOIprefix\doi{10.1088/1361-6633/aa7cca}.
%Type = Article
\bibitem[{Semenov et~al.(2000)Semenov, Cherepkov, Fecher and
  Sch\"onhense}]{Semenov2000}
\bibinfo{author}{Semenov, S.K.}, \bibinfo{author}{Cherepkov, N.A.},
  \bibinfo{author}{Fecher, G.H.}, \bibinfo{author}{Sch\"onhense, G.},
  \bibinfo{year}{2000}.
\newblock \bibinfo{title}{Generalization of the atomic
  random-phase-approximation method for diatomic molecules: N$_2$
  photoionization cross-section calculations}.
\newblock \bibinfo{journal}{Phys. Rev. A} \bibinfo{volume}{61},
  \bibinfo{pages}{032704}.
\newblock \DOIprefix\doi{10.1103/PhysRevA.61.032704}.
%Type = Article
\bibitem[{Seres et~al.(2006)Seres, Seres and Spielmann}]{Seres2006}
\bibinfo{author}{Seres, E.}, \bibinfo{author}{Seres, J.},
  \bibinfo{author}{Spielmann, C.}, \bibinfo{year}{2006}.
\newblock \bibinfo{title}{X-ray absorption spectroscopy in the {keV} range with
  laser generated high harmonic radiation}.
\newblock \bibinfo{journal}{Appl. Phys. Lett.} \bibinfo{volume}{89},
  \bibinfo{pages}{181919}.
\newblock \DOIprefix\doi{10.1063/1.2364126}.
%Type = Article
\bibitem[{Serkez et~al.(2020)Serkez, Decking, Froehlich, Gerasimova,
  Gr\"{u}nert, Guetg, Huttula, Karabekyan, Koch, Kocharyan, Kot, Kukk, Laksman,
  Lytaev, Maltezopoulos, Mazza, Meyer, Saldin, Schneidmiller, Scholz, Tomin,
  Vannoni, Wohlenberg, Yurkov, Zagorodnov and Geloni}]{Serkez2020}
\bibinfo{author}{Serkez, S.}, \bibinfo{author}{Decking, W.},
  \bibinfo{author}{Froehlich, L.}, \bibinfo{author}{Gerasimova, N.},
  \bibinfo{author}{Gr\"{u}nert, J.}, \bibinfo{author}{Guetg, M.},
  \bibinfo{author}{Huttula, M.}, \bibinfo{author}{Karabekyan, S.},
  \bibinfo{author}{Koch, A.}, \bibinfo{author}{Kocharyan, V.},
  \bibinfo{author}{Kot, Y.}, \bibinfo{author}{Kukk, E.},
  \bibinfo{author}{Laksman, J.}, \bibinfo{author}{Lytaev, P.},
  \bibinfo{author}{Maltezopoulos, T.}, \bibinfo{author}{Mazza, T.},
  \bibinfo{author}{Meyer, M.}, \bibinfo{author}{Saldin, E.},
  \bibinfo{author}{Schneidmiller, E.}, \bibinfo{author}{Scholz, M.},
  \bibinfo{author}{Tomin, S.}, \bibinfo{author}{Vannoni, M.},
  \bibinfo{author}{Wohlenberg, T.}, \bibinfo{author}{Yurkov, M.},
  \bibinfo{author}{Zagorodnov, I.}, \bibinfo{author}{Geloni, G.},
  \bibinfo{year}{2020}.
\newblock \bibinfo{title}{Opportunities for two-color experiments in the soft
  {X}-ray regime at the european {XFEL}}.
\newblock \bibinfo{journal}{Appl. Sci.} \bibinfo{volume}{10},
  \bibinfo{pages}{2728}.
\newblock \DOIprefix\doi{10.3390/app10082728}.
%Type = Article
\bibitem[{Serkez et~al.(2018)Serkez, Geloni, Tomin, Feng, Gryzlova,
  Grum-Grzhimailo and Meyer}]{Serkez2018}
\bibinfo{author}{Serkez, S.}, \bibinfo{author}{Geloni, G.},
  \bibinfo{author}{Tomin, S.}, \bibinfo{author}{Feng, G.},
  \bibinfo{author}{Gryzlova, E.V.}, \bibinfo{author}{Grum-Grzhimailo, A.N.},
  \bibinfo{author}{Meyer, M.}, \bibinfo{year}{2018}.
\newblock \bibinfo{title}{Overview of options for generating high-brightness
  attosecond x-ray pulses at free-electron lasers and applications at the
  {E}uropean {XFEL}}.
\newblock \bibinfo{journal}{J. Opt.} \bibinfo{volume}{20},
  \bibinfo{pages}{024005}.
\newblock \DOIprefix\doi{10.1088/2040-8986/aa9f4f}.
%Type = Article
\bibitem[{Shintake et~al.(2008)Shintake, Tanaka, Hara, Tanaka, Togawa, Yabashi,
  Otake, Asano, Bizen, Fukui, Goto, Higashiya, Hirono, Hosoda, Inagaki, Inoue,
  Ishii, Kim, Kimura, Kitamura, Kobayashi, Maesaka, Masuda, Matsui, Matsushita,
  Mar{\'{e}}chal, Nagasono, Ohashi, Ohata, Ohshima, Onoe, Shirasawa, Takagi,
  Takahashi, Takeuchi, Tamasaku, Tanaka, Tanaka, Tanikawa, Togashi, Wu,
  Yamashita, Yanagida, Zhang, Kitamura and Ishikawa}]{Shintake2008}
\bibinfo{author}{Shintake, T.}, \bibinfo{author}{Tanaka, H.},
  \bibinfo{author}{Hara, T.}, \bibinfo{author}{Tanaka, T.},
  \bibinfo{author}{Togawa, K.}, \bibinfo{author}{Yabashi, M.},
  \bibinfo{author}{Otake, Y.}, \bibinfo{author}{Asano, Y.},
  \bibinfo{author}{Bizen, T.}, \bibinfo{author}{Fukui, T.},
  \bibinfo{author}{Goto, S.}, \bibinfo{author}{Higashiya, A.},
  \bibinfo{author}{Hirono, T.}, \bibinfo{author}{Hosoda, N.},
  \bibinfo{author}{Inagaki, T.}, \bibinfo{author}{Inoue, S.},
  \bibinfo{author}{Ishii, M.}, \bibinfo{author}{Kim, Y.},
  \bibinfo{author}{Kimura, H.}, \bibinfo{author}{Kitamura, M.},
  \bibinfo{author}{Kobayashi, T.}, \bibinfo{author}{Maesaka, H.},
  \bibinfo{author}{Masuda, T.}, \bibinfo{author}{Matsui, S.},
  \bibinfo{author}{Matsushita, T.}, \bibinfo{author}{Mar{\'{e}}chal, X.},
  \bibinfo{author}{Nagasono, M.}, \bibinfo{author}{Ohashi, H.},
  \bibinfo{author}{Ohata, T.}, \bibinfo{author}{Ohshima, T.},
  \bibinfo{author}{Onoe, K.}, \bibinfo{author}{Shirasawa, K.},
  \bibinfo{author}{Takagi, T.}, \bibinfo{author}{Takahashi, S.},
  \bibinfo{author}{Takeuchi, M.}, \bibinfo{author}{Tamasaku, K.},
  \bibinfo{author}{Tanaka, R.}, \bibinfo{author}{Tanaka, Y.},
  \bibinfo{author}{Tanikawa, T.}, \bibinfo{author}{Togashi, T.},
  \bibinfo{author}{Wu, S.}, \bibinfo{author}{Yamashita, A.},
  \bibinfo{author}{Yanagida, K.}, \bibinfo{author}{Zhang, C.},
  \bibinfo{author}{Kitamura, H.}, \bibinfo{author}{Ishikawa, T.},
  \bibinfo{year}{2008}.
\newblock \bibinfo{title}{A compact free-electron laser for generating coherent
  radiation in the extreme ultraviolet region}.
\newblock \bibinfo{journal}{Nat. Photonics} \bibinfo{volume}{2},
  \bibinfo{pages}{555--559}.
\newblock \DOIprefix\doi{10.1038/nphoton.2008.134}.
%Type = Book
\bibitem[{Siegbahn(1971)}]{Siegbahn1971}
\bibinfo{author}{Siegbahn, K.}, \bibinfo{year}{1971}.
\newblock \bibinfo{title}{ESCA applied to free molecules}.
\newblock \bibinfo{publisher}{North-Holland Publishing Company},
  \bibinfo{address}{Amsterdam}.
%Type = Article
\bibitem[{Smith(1960)}]{Smith1960}
\bibinfo{author}{Smith, F.T.}, \bibinfo{year}{1960}.
\newblock \bibinfo{title}{Lifetime matrix in collision theory}.
\newblock \bibinfo{journal}{Phys. Rev.} \bibinfo{volume}{118},
  \bibinfo{pages}{349--356}.
\newblock \DOIprefix\doi{10.1103/PhysRev.118.349}.
%Type = Article
\bibitem[{Smyth et~al.(1998)Smyth, Parker and Taylor}]{Smyth1998}
\bibinfo{author}{Smyth, E.S.}, \bibinfo{author}{Parker, J.S.},
  \bibinfo{author}{Taylor, K.}, \bibinfo{year}{1998}.
\newblock \bibinfo{title}{Numerical integration of the time-dependent
  {S}chr\"odinger equation for laser-driven helium}.
\newblock \bibinfo{journal}{Comput. Phys. Commun.} \bibinfo{volume}{114},
  \bibinfo{pages}{1--14}.
\newblock \DOIprefix\doi{10.1016/S0010-4655(98)00083-6}.
%Type = Article
\bibitem[{Sommerfeld(1916a)}]{Sommerfeld1916}
\bibinfo{author}{Sommerfeld, A.}, \bibinfo{year}{1916}a.
\newblock \bibinfo{title}{Zur {Q}uantentheorie der {S}pektrallinien}.
\newblock \bibinfo{journal}{Ann. Phys.} \bibinfo{volume}{356},
  \bibinfo{pages}{1--94}.
\newblock \DOIprefix\doi{10.1002/andp.19163561702}.
%Type = Article
\bibitem[{Sommerfeld(1916b)}]{Sommerfeld1916a}
\bibinfo{author}{Sommerfeld, A.}, \bibinfo{year}{1916}b.
\newblock \bibinfo{title}{Zur {Q}uantentheorie der {S}pektrallinien}.
\newblock \bibinfo{journal}{Ann. Phys.} \bibinfo{volume}{356},
  \bibinfo{pages}{125--167}.
\newblock \DOIprefix\doi{10.1002/andp.19163561802}.
%Type = Article
\bibitem[{Son and Santra(2011)}]{Son2011}
\bibinfo{author}{Son, S.K.}, \bibinfo{author}{Santra, R.},
  \bibinfo{year}{2011}.
\newblock \bibinfo{title}{Impact of hollow-atom formation on coherent {x}-ray
  scattering at high intensity}.
\newblock \bibinfo{journal}{Phys. Rev. A} \bibinfo{volume}{83},
  \bibinfo{pages}{033402}.
\newblock \DOIprefix\doi{10.1103/PhysRevA.83.033402}.
%Type = Article
\bibitem[{Son and Santra(2012)}]{Son2012}
\bibinfo{author}{Son, S.K.}, \bibinfo{author}{Santra, R.},
  \bibinfo{year}{2012}.
\newblock \bibinfo{title}{{M}onte {C}arlo calculation of ion, electron, and
  photon spectra of xenon atoms in x-ray free-electron laser pulses}.
\newblock \bibinfo{journal}{Phys. Rev. A} \bibinfo{volume}{85},
  \bibinfo{pages}{063415}.
\newblock \DOIprefix\doi{10.1103/PhysRevA.85.063415}.
%Type = Article
\bibitem[{Sorensen et~al.(1994)Sorensen, \AA{}berg, Tulkki, Rachlew-K\"allne,
  Sundstr\"om and Kirm}]{Sorensen1994}
\bibinfo{author}{Sorensen, S.L.}, \bibinfo{author}{\AA{}berg, T.},
  \bibinfo{author}{Tulkki, J.}, \bibinfo{author}{Rachlew-K\"allne, E.},
  \bibinfo{author}{Sundstr\"om, G.}, \bibinfo{author}{Kirm, M.},
  \bibinfo{year}{1994}.
\newblock \bibinfo{title}{Argon {3s} autoionization resonances}.
\newblock \bibinfo{journal}{Phys. Rev. A} \bibinfo{volume}{50},
  \bibinfo{pages}{1218--1230}.
\newblock \DOIprefix\doi{10.1103/PhysRevA.50.1218}.
%Type = Article
\bibitem[{Sorokin et~al.(2007)Sorokin, Wellh\"ofer, Bobashev, Tiedtke and
  Richter}]{Sorokin2007}
\bibinfo{author}{Sorokin, A.A.}, \bibinfo{author}{Wellh\"ofer, M.},
  \bibinfo{author}{Bobashev, S.V.}, \bibinfo{author}{Tiedtke, K.},
  \bibinfo{author}{Richter, M.}, \bibinfo{year}{2007}.
\newblock \bibinfo{title}{X-ray-laser interaction with matter and the role of
  multiphoton ionization: Free-electron-laser studies on neon and helium}.
\newblock \bibinfo{journal}{Phys. Rev. A} \bibinfo{volume}{75},
  \bibinfo{pages}{051402}.
\newblock \DOIprefix\doi{10.1103/PhysRevA.75.051402}.
%Type = Article
\bibitem[{Squibb et~al.(2018)Squibb, Sapunar, Ponzi, Richter, Kivim\"{a}ki,
  Plekan, Finetti, Sisourat, Zhaunerchyk, Marchenko, Journel, Guillemin,
  Cucini, Coreno, Grazioli, Di~Fraia, Callegari, Prince, Decleva, Simon, Eland,
  Do{\v{s}}li{\'{c}}, Feifel and Piancastelli}]{Squibb2018}
\bibinfo{author}{Squibb, R.J.}, \bibinfo{author}{Sapunar, M.},
  \bibinfo{author}{Ponzi, A.}, \bibinfo{author}{Richter, R.},
  \bibinfo{author}{Kivim\"{a}ki, A.}, \bibinfo{author}{Plekan, O.},
  \bibinfo{author}{Finetti, P.}, \bibinfo{author}{Sisourat, N.},
  \bibinfo{author}{Zhaunerchyk, V.}, \bibinfo{author}{Marchenko, T.},
  \bibinfo{author}{Journel, L.}, \bibinfo{author}{Guillemin, R.},
  \bibinfo{author}{Cucini, R.}, \bibinfo{author}{Coreno, M.},
  \bibinfo{author}{Grazioli, C.}, \bibinfo{author}{Di~Fraia, M.},
  \bibinfo{author}{Callegari, C.}, \bibinfo{author}{Prince, K.C.},
  \bibinfo{author}{Decleva, P.}, \bibinfo{author}{Simon, M.},
  \bibinfo{author}{Eland, J.H.D.}, \bibinfo{author}{Do{\v{s}}li{\'{c}}, N.},
  \bibinfo{author}{Feifel, R.}, \bibinfo{author}{Piancastelli, M.N.},
  \bibinfo{year}{2018}.
\newblock \bibinfo{title}{Acetylacetone photodynamics at a seeded free-electron
  laser}.
\newblock \bibinfo{journal}{Nat. Commun.} \bibinfo{volume}{9},
  \bibinfo{pages}{63}.
\newblock \DOIprefix\doi{10.1038/s41467-017-02478-0}.
%Type = Book
\bibitem[{Starke(2000)}]{Starke2000}
\bibinfo{author}{Starke, K.}, \bibinfo{year}{2000}.
\newblock \bibinfo{title}{Magnetic Dichroism in Core-Level Photoemission}.
  volume \bibinfo{volume}{159} of \textit{\bibinfo{series}{Springer Tracts in
  Modern Physics}}.
\newblock \bibinfo{publisher}{Springer}, \bibinfo{address}{Berlin Heidelberg}.
\newblock \DOIprefix\doi{10.1007/bfb0109607}.
%Type = Article
\bibitem[{Stener et~al.(2007)Stener, Toffoli, Fronzoni and
  Decleva}]{Stener2007}
\bibinfo{author}{Stener, M.}, \bibinfo{author}{Toffoli, D.},
  \bibinfo{author}{Fronzoni, G.}, \bibinfo{author}{Decleva, P.},
  \bibinfo{year}{2007}.
\newblock \bibinfo{title}{Recent advances in molecular photoionization by
  density functional theory based approaches}.
\newblock \bibinfo{journal}{Theor. Chem. Acc.} \bibinfo{volume}{117},
  \bibinfo{pages}{943--956}.
\newblock \DOIprefix\doi{10.1007/s00214-006-0212-3}.
%Type = Article
\bibitem[{Stratmann and Lucchese(1995)}]{Stratmann1995}
\bibinfo{author}{Stratmann, R.E.}, \bibinfo{author}{Lucchese, R.R.},
  \bibinfo{year}{1995}.
\newblock \bibinfo{title}{A graphical unitary group approach to study multiplet
  specific multichannel electron correlation effects in the photoionization of
  {O}$_2$}.
\newblock \bibinfo{journal}{J. Chem. Phys.} \bibinfo{volume}{102},
  \bibinfo{pages}{8493--8505}.
\newblock \DOIprefix\doi{10.1063/1.468841}.
%Type = Article
\bibitem[{Strickland(2019)}]{Strickland2019}
\bibinfo{author}{Strickland, D.}, \bibinfo{year}{2019}.
\newblock \bibinfo{title}{Nobel lecture: Generating high-intensity ultrashort
  optical pulses}.
\newblock \bibinfo{journal}{Rev. Mod. Phys.} \bibinfo{volume}{91},
  \bibinfo{pages}{030502}.
\newblock \DOIprefix\doi{10.1103/RevModPhys.91.030502}.
%Type = Article
\bibitem[{Suckewer and Jaegl{\'{e}}(2009)}]{Suckewer2009}
\bibinfo{author}{Suckewer, S.}, \bibinfo{author}{Jaegl{\'{e}}, P.},
  \bibinfo{year}{2009}.
\newblock \bibinfo{title}{X-ray laser: past, present, and future}.
\newblock \bibinfo{journal}{Laser Phys. Lett.} \bibinfo{volume}{6},
  \bibinfo{pages}{411--436}.
\newblock \DOIprefix\doi{10.1002/lapl.200910023}.
%Type = Article
\bibitem[{Suckewer et~al.(1985)Suckewer, Skinner, Milchberg, Keane and
  Voorhees}]{Suckewer1985}
\bibinfo{author}{Suckewer, S.}, \bibinfo{author}{Skinner, C.H.},
  \bibinfo{author}{Milchberg, H.}, \bibinfo{author}{Keane, C.},
  \bibinfo{author}{Voorhees, D.}, \bibinfo{year}{1985}.
\newblock \bibinfo{title}{Amplification of stimulated soft {x}-ray emission in
  a confined plasma column}.
\newblock \bibinfo{journal}{Phys. Rev. Lett.} \bibinfo{volume}{55},
  \bibinfo{pages}{1753--1756}.
\newblock \DOIprefix\doi{10.1103/physrevlett.55.1753}.
%Type = Article
\bibitem[{Sun et~al.(2010)Sun, Rinkevicius, Wang, Carniato, Simon, Ta\"{\i}eb
  and Gel'mukhanov}]{Sun2010}
\bibinfo{author}{Sun, Y.P.}, \bibinfo{author}{Rinkevicius, Z.},
  \bibinfo{author}{Wang, C.K.}, \bibinfo{author}{Carniato, S.},
  \bibinfo{author}{Simon, M.}, \bibinfo{author}{Ta\"{\i}eb, R.},
  \bibinfo{author}{Gel'mukhanov, F.}, \bibinfo{year}{2010}.
\newblock \bibinfo{title}{Two-photon-induced x-ray emission in neon atoms}.
\newblock \bibinfo{journal}{Phys. Rev. A} \bibinfo{volume}{82},
  \bibinfo{pages}{043430}.
\newblock \DOIprefix\doi{10.1103/PhysRevA.82.043430}.
%Type = Article
\bibitem[{Suzuki(2006)}]{Suzuki2006}
\bibinfo{author}{Suzuki, T.}, \bibinfo{year}{2006}.
\newblock \bibinfo{title}{Femtosecond time-resolved photoelectron imaging}.
\newblock \bibinfo{journal}{Annu. Rev. Phys. Chem.} \bibinfo{volume}{57},
  \bibinfo{pages}{555--592}.
\newblock \DOIprefix\doi{10.1146/annurev.physchem.57.032905.104601}.
%Type = Article
\bibitem[{Takahashi et~al.(2002)Takahashi, Nabekawa and
  Midorikawa}]{Takahashi2002}
\bibinfo{author}{Takahashi, E.}, \bibinfo{author}{Nabekawa, Y.},
  \bibinfo{author}{Midorikawa, K.}, \bibinfo{year}{2002}.
\newblock \bibinfo{title}{Generation of 10-$\mu${J} coherent
  extreme-ultraviolet light by use of high-order harmonics}.
\newblock \bibinfo{journal}{Opt. Lett.} \bibinfo{volume}{27},
  \bibinfo{pages}{1920}.
\newblock \DOIprefix\doi{10.1364/ol.27.001920}.
%Type = Article
\bibitem[{Takahashi et~al.(2011)Takahashi, Tashiro, Ehara, Yamasaki and
  Ueda}]{Takahashi2011}
\bibinfo{author}{Takahashi, O.}, \bibinfo{author}{Tashiro, M.},
  \bibinfo{author}{Ehara, M.}, \bibinfo{author}{Yamasaki, K.},
  \bibinfo{author}{Ueda, K.}, \bibinfo{year}{2011}.
\newblock \bibinfo{title}{Theoretical molecular double-core-hole spectroscopy
  of nucleobases}.
\newblock \bibinfo{journal}{J. Phys. Chem.~A} \bibinfo{volume}{115},
  \bibinfo{pages}{12070--12082}.
\newblock \DOIprefix\doi{10.1021/jp205923m}.
%Type = Article
\bibitem[{Takanashi et~al.(2017)Takanashi, Golubev, Callegari, Fukuzawa,
  Motomura, Iablonskyi, Kumagai, Mondal, Tachibana, Nagaya, Nishiyama,
  Matsunami, Johnsson, Piseri, Sansone, Dubrouil, Reduzzi, Carpeggiani, Vozzi,
  Devetta, Negro, Faccial\`a, Calegari, Trabattoni, Castrovilli, Ovcharenko,
  Mudrich, Stienkemeier, Coreno, Alagia, Sch\"utte, Berrah, Plekan, Finetti,
  Spezzani, Ferrari, Allaria, Penco, Serpico, De~Ninno, Diviacco, Di~Mitri,
  Giannessi, Jabbari, Prince, Cederbaum, Demekhin, Kuleff and
  Ueda}]{Takanashi2017}
\bibinfo{author}{Takanashi, T.}, \bibinfo{author}{Golubev, N.V.},
  \bibinfo{author}{Callegari, C.}, \bibinfo{author}{Fukuzawa, H.},
  \bibinfo{author}{Motomura, K.}, \bibinfo{author}{Iablonskyi, D.},
  \bibinfo{author}{Kumagai, Y.}, \bibinfo{author}{Mondal, S.},
  \bibinfo{author}{Tachibana, T.}, \bibinfo{author}{Nagaya, K.},
  \bibinfo{author}{Nishiyama, T.}, \bibinfo{author}{Matsunami, K.},
  \bibinfo{author}{Johnsson, P.}, \bibinfo{author}{Piseri, P.},
  \bibinfo{author}{Sansone, G.}, \bibinfo{author}{Dubrouil, A.},
  \bibinfo{author}{Reduzzi, M.}, \bibinfo{author}{Carpeggiani, P.},
  \bibinfo{author}{Vozzi, C.}, \bibinfo{author}{Devetta, M.},
  \bibinfo{author}{Negro, M.}, \bibinfo{author}{Faccial\`a, D.},
  \bibinfo{author}{Calegari, F.}, \bibinfo{author}{Trabattoni, A.},
  \bibinfo{author}{Castrovilli, M.C.}, \bibinfo{author}{Ovcharenko, Y.},
  \bibinfo{author}{Mudrich, M.}, \bibinfo{author}{Stienkemeier, F.},
  \bibinfo{author}{Coreno, M.}, \bibinfo{author}{Alagia, M.},
  \bibinfo{author}{Sch\"utte, B.}, \bibinfo{author}{Berrah, N.},
  \bibinfo{author}{Plekan, O.}, \bibinfo{author}{Finetti, P.},
  \bibinfo{author}{Spezzani, C.}, \bibinfo{author}{Ferrari, E.},
  \bibinfo{author}{Allaria, E.}, \bibinfo{author}{Penco, G.},
  \bibinfo{author}{Serpico, C.}, \bibinfo{author}{De~Ninno, G.},
  \bibinfo{author}{Diviacco, B.}, \bibinfo{author}{Di~Mitri, S.},
  \bibinfo{author}{Giannessi, L.}, \bibinfo{author}{Jabbari, G.},
  \bibinfo{author}{Prince, K.C.}, \bibinfo{author}{Cederbaum, L.S.},
  \bibinfo{author}{Demekhin, P.V.}, \bibinfo{author}{Kuleff, A.I.},
  \bibinfo{author}{Ueda, K.}, \bibinfo{year}{2017}.
\newblock \bibinfo{title}{Time-resolved measurement of interatomic coulombic
  decay induced by two-photon double excitation of {Ne}\ensuremath{_{2}}}.
\newblock \bibinfo{journal}{Phys. Rev. Lett.} \bibinfo{volume}{118},
  \bibinfo{pages}{033202}.
\newblock \DOIprefix\doi{10.1103/PhysRevLett.118.033202}.
%Type = Article
\bibitem[{Tan(2008)}]{Tan2008}
\bibinfo{author}{Tan, H.S.}, \bibinfo{year}{2008}.
\newblock \bibinfo{title}{Theory and phase-cycling scheme selection principles
  of collinear phase coherent multi-dimensional optical spectroscopy}.
\newblock \bibinfo{journal}{J. Chem. Phys.} \bibinfo{volume}{129},
  \bibinfo{pages}{124501}.
\newblock \DOIprefix\doi{10.1063/1.2978381}.
%Type = Article
\bibitem[{Tanaka and Mukamel(2002)}]{Tanaka2002}
\bibinfo{author}{Tanaka, S.}, \bibinfo{author}{Mukamel, S.},
  \bibinfo{year}{2002}.
\newblock \bibinfo{title}{X-ray four-wave mixing in molecules}.
\newblock \bibinfo{journal}{J. Chem. Phys.} \bibinfo{volume}{116},
  \bibinfo{pages}{1877--1891}.
\newblock \DOIprefix\doi{10.1063/1.1429950}.
%Type = Book
\bibitem[{Tannor(2007)}]{Tannor2007}
\bibinfo{author}{Tannor, D.J.}, \bibinfo{year}{2007}.
\newblock \bibinfo{title}{Introduction to Quantum Mechanics. A Time-Dependent
  Perspective}.
\newblock \bibinfo{publisher}{University science Books},
  \bibinfo{address}{Sausalito, California}.
%Type = Article
\bibitem[{Tashiro et~al.(2010a)Tashiro, Ehara, Fukuzawa, Ueda, Buth, Kryzhevoi
  and Cederbaum}]{Tashiro2010}
\bibinfo{author}{Tashiro, M.}, \bibinfo{author}{Ehara, M.},
  \bibinfo{author}{Fukuzawa, H.}, \bibinfo{author}{Ueda, K.},
  \bibinfo{author}{Buth, C.}, \bibinfo{author}{Kryzhevoi, N.V.},
  \bibinfo{author}{Cederbaum, L.S.}, \bibinfo{year}{2010}a.
\newblock \bibinfo{title}{Molecular double core hole electron spectroscopy for
  chemical analysis}.
\newblock \bibinfo{journal}{J. Chem. Phys.} \bibinfo{volume}{132},
  \bibinfo{pages}{184302}.
\newblock \DOIprefix\doi{10.1063/1.3408251}.
%Type = Article
\bibitem[{Tashiro et~al.(2010b)Tashiro, Ehara and Ueda}]{Tashiro2010a}
\bibinfo{author}{Tashiro, M.}, \bibinfo{author}{Ehara, M.},
  \bibinfo{author}{Ueda, K.}, \bibinfo{year}{2010}b.
\newblock \bibinfo{title}{Double core-hole electron spectroscopy for open-shell
  molecules: Theoretical perspective}.
\newblock \bibinfo{journal}{Chem. Phys. Lett.} \bibinfo{volume}{496},
  \bibinfo{pages}{217--222}.
\newblock \DOIprefix\doi{10.1016/j.cplett.2010.07.046}.
%Type = Article
\bibitem[{Teichmann et~al.(2016)Teichmann, Silva, Cousin, Hemmer and
  Biegert}]{Teichmann2016}
\bibinfo{author}{Teichmann, S.M.}, \bibinfo{author}{Silva, F.},
  \bibinfo{author}{Cousin, S.L.}, \bibinfo{author}{Hemmer, M.},
  \bibinfo{author}{Biegert, J.}, \bibinfo{year}{2016}.
\newblock \bibinfo{title}{0.5-{keV} {S}oft {X}-ray attosecond continua}.
\newblock \bibinfo{journal}{Nat. Commun.} \bibinfo{volume}{7},
  \bibinfo{pages}{11493}.
\newblock \DOIprefix\doi{10.1038/ncomms11493}.
%Type = Article
\bibitem[{Tilley et~al.(2015)Tilley, Karamatskou and Santra}]{Tilley2015}
\bibinfo{author}{Tilley, M.}, \bibinfo{author}{Karamatskou, A.},
  \bibinfo{author}{Santra, R.}, \bibinfo{year}{2015}.
\newblock \bibinfo{title}{Wave-packet propagation based calculation of
  above-threshold ionization in the {x}-ray regime}.
\newblock \bibinfo{journal}{J. Phys. B} \bibinfo{volume}{48},
  \bibinfo{pages}{124001}.
\newblock \DOIprefix\doi{10.1088/0953-4075/48/12/124001}.
%Type = Article
\bibitem[{Toffoli et~al.(2002)Toffoli, Stener, Fronzoni and
  Decleva}]{Toffoli2002}
\bibinfo{author}{Toffoli, D.}, \bibinfo{author}{Stener, M.},
  \bibinfo{author}{Fronzoni, G.}, \bibinfo{author}{Decleva, P.},
  \bibinfo{year}{2002}.
\newblock \bibinfo{title}{Convergence of the multicenter {B}-spline {DFT}
  approach for the continuum}.
\newblock \bibinfo{journal}{Chem. Phys.} \bibinfo{volume}{276},
  \bibinfo{pages}{25--43}.
\newblock \DOIprefix\doi{10.1016/S0301-0104(01)00549-3}.
%Type = Article
\bibitem[{Tong et~al.(2011)Tong, Ranitovic, Hogle, Murnane, Kapteyn and
  Toshima}]{Tong2011}
\bibinfo{author}{Tong, X.M.}, \bibinfo{author}{Ranitovic, P.},
  \bibinfo{author}{Hogle, C.W.}, \bibinfo{author}{Murnane, M.M.},
  \bibinfo{author}{Kapteyn, H.C.}, \bibinfo{author}{Toshima, N.},
  \bibinfo{year}{2011}.
\newblock \bibinfo{title}{Theory and experiment on laser-enabled inner-valence
  {A}uger decay of rare-gas atoms}.
\newblock \bibinfo{journal}{Phys. Rev. A} \bibinfo{volume}{84},
  \bibinfo{pages}{013405}.
\newblock \DOIprefix\doi{10.1103/PhysRevA.84.013405}.
%Type = Article
\bibitem[{Tschentscher et~al.(2017)Tschentscher, Bressler, Gr{\"u}nert, Madsen,
  Mancuso, Meyer, Scherz, Sinn and Zastrau}]{Tschentscher2017}
\bibinfo{author}{Tschentscher, T.}, \bibinfo{author}{Bressler, C.},
  \bibinfo{author}{Gr{\"u}nert, J.}, \bibinfo{author}{Madsen, A.},
  \bibinfo{author}{Mancuso, A.}, \bibinfo{author}{Meyer, M.},
  \bibinfo{author}{Scherz, A.}, \bibinfo{author}{Sinn, H.},
  \bibinfo{author}{Zastrau, U.}, \bibinfo{year}{2017}.
\newblock \bibinfo{title}{Photon beam transport and scientific instruments at
  the {E}uropean {XFEL}}.
\newblock \bibinfo{journal}{Applied Sciences} \bibinfo{volume}{7},
  \bibinfo{pages}{592}.
\newblock \DOIprefix\doi{10.3390/app7060592}.
%Type = Article
\bibitem[{Ullrich et~al.(2012)Ullrich, Rudenko and Moshammer}]{Ullrich2012}
\bibinfo{author}{Ullrich, J.}, \bibinfo{author}{Rudenko, A.},
  \bibinfo{author}{Moshammer, R.}, \bibinfo{year}{2012}.
\newblock \bibinfo{title}{Free-electron lasers: New avenues in molecular
  physics and photochemistry}.
\newblock \bibinfo{journal}{Annu. Rev. Phys. Chem.} \bibinfo{volume}{63},
  \bibinfo{pages}{635--660}.
\newblock \DOIprefix\doi{10.1146/annurev-physchem-032511-143720}.
%Type = Article
\bibitem[{Usenko et~al.(2017)Usenko, Przystawik, Jakob, Lazzarino, Brenner,
  Toleikis, Haunhorst, Kip and Laarmann}]{Usenko2017}
\bibinfo{author}{Usenko, S.}, \bibinfo{author}{Przystawik, A.},
  \bibinfo{author}{Jakob, M.A.}, \bibinfo{author}{Lazzarino, L.L.},
  \bibinfo{author}{Brenner, G.}, \bibinfo{author}{Toleikis, S.},
  \bibinfo{author}{Haunhorst, C.}, \bibinfo{author}{Kip, D.},
  \bibinfo{author}{Laarmann, T.}, \bibinfo{year}{2017}.
\newblock \bibinfo{title}{Attosecond interferometry with self-amplified
  spontaneous emission of a free-electron laser}.
\newblock \bibinfo{journal}{Nature Communications} \bibinfo{volume}{8},
  \bibinfo{pages}{15626}.
\newblock \DOIprefix\doi{10.1038/ncomms15626}.
%Type = Book
\bibitem[{Vrakking and Lepine(2019)}]{Vrakking2018}
\bibinfo{editor}{Vrakking, M.J.J.}, \bibinfo{editor}{Lepine, F.} (Eds.),
  \bibinfo{year}{2019}.
\newblock \bibinfo{title}{Attosecond Molecular Dynamics}.
\newblock Theoretical and Computational Chemistry Series,
  \bibinfo{publisher}{The Royal Society of Chemistry}.
\newblock \DOIprefix\doi{10.1039/9781788012669}.
%Type = Article
\bibitem[{Wang et~al.(2010)Wang, Chini, Chen, Zhang, He, Cheng, Wu, Thumm and
  Chang}]{PRL-Chang-2010}
\bibinfo{author}{Wang, H.}, \bibinfo{author}{Chini, M.}, \bibinfo{author}{Chen,
  S.}, \bibinfo{author}{Zhang, C.H.}, \bibinfo{author}{He, F.},
  \bibinfo{author}{Cheng, Y.}, \bibinfo{author}{Wu, Y.},
  \bibinfo{author}{Thumm, U.}, \bibinfo{author}{Chang, Z.},
  \bibinfo{year}{2010}.
\newblock \bibinfo{title}{Attosecond time-resolved autoionization of argon}.
\newblock \bibinfo{journal}{Phys. Rev. Lett.} \bibinfo{volume}{105},
  \bibinfo{pages}{143002}.
\newblock \DOIprefix\doi{10.1103/PhysRevLett.105.143002}.
%Type = Article
\bibitem[{Wang et~al.(2008)Wang, Granados, Pedaci, Alessi, Luther, Berrill and
  Rocca}]{Wang2008}
\bibinfo{author}{Wang, Y.}, \bibinfo{author}{Granados, E.},
  \bibinfo{author}{Pedaci, F.}, \bibinfo{author}{Alessi, D.},
  \bibinfo{author}{Luther, B.}, \bibinfo{author}{Berrill, M.},
  \bibinfo{author}{Rocca, J.J.}, \bibinfo{year}{2008}.
\newblock \bibinfo{title}{Phase-coherent, injection-seeded, table-top
  soft-{X}-ray lasers at 18.9~nm and 13.9~nm}.
\newblock \bibinfo{journal}{Nat. Photonics} \bibinfo{volume}{2},
  \bibinfo{pages}{94--98}.
\newblock \DOIprefix\doi{10.1038/nphoton.2007.280}.
%Type = Article
\bibitem[{Wang et~al.(2018)Wang, Guo, Li, Zhao, Yin, Ren, Li, Wu, Weidman,
  Chang, Jager, Kaplan, Geneaux, Ott, Neumark and Leone}]{Wang2018}
\bibinfo{author}{Wang, Y.}, \bibinfo{author}{Guo, T.}, \bibinfo{author}{Li,
  J.}, \bibinfo{author}{Zhao, J.}, \bibinfo{author}{Yin, Y.},
  \bibinfo{author}{Ren, X.}, \bibinfo{author}{Li, J.}, \bibinfo{author}{Wu,
  Y.}, \bibinfo{author}{Weidman, M.}, \bibinfo{author}{Chang, Z.},
  \bibinfo{author}{Jager, M.F.}, \bibinfo{author}{Kaplan, C.J.},
  \bibinfo{author}{Geneaux, R.}, \bibinfo{author}{Ott, C.},
  \bibinfo{author}{Neumark, D.M.}, \bibinfo{author}{Leone, S.R.},
  \bibinfo{year}{2018}.
\newblock \bibinfo{title}{Enhanced high-order harmonic generation driven by a
  wavefront corrected high-energy laser}.
\newblock \bibinfo{journal}{J. Phys. B} \bibinfo{volume}{51},
  \bibinfo{pages}{134005}.
\newblock \DOIprefix\doi{10.1088/1361-6455/aac59e}.
%Type = Article
\bibitem[{Wang and Elliott(2001)}]{Wang2001}
\bibinfo{author}{Wang, Z.M.}, \bibinfo{author}{Elliott, D.S.},
  \bibinfo{year}{2001}.
\newblock \bibinfo{title}{Determination of the phase difference between even
  and odd continuum wave functions in atoms through quantum interference
  measurements}.
\newblock \bibinfo{journal}{Phys. Rev. Lett.} \bibinfo{volume}{87},
  \bibinfo{pages}{173001}.
\newblock \DOIprefix\doi{10.1103/PhysRevLett.87.173001}.
%Type = Article
\bibitem[{Wigner(1955)}]{Wigner1955}
\bibinfo{author}{Wigner, E.P.}, \bibinfo{year}{1955}.
\newblock \bibinfo{title}{Lower limit for the energy derivative of the
  scattering phase shift}.
\newblock \bibinfo{journal}{Phys. Rev.} \bibinfo{volume}{98},
  \bibinfo{pages}{145}.
\newblock \DOIprefix\doi{10.1103/PhysRev.98.145}.
%Type = Article
\bibitem[{Wituschek et~al.(2020)Wituschek, Bruder, Allaria, Bangert, Binz,
  Callegari, Cerullo, Cinquegrana, Gianessi, Danailov, Demidovich, Di~Fraia,
  Drabbels, Feifel, Laarmann, Michiels, Mirian, Mudrich, Nikolov, O'Shea,
  Penco, Piseri, Plekan, Prince, Przystawik, Ribi{\v{c}}, Sansone, Sigalotti,
  Spampinati, Spezzani, Squibb, Stranges, Uhl and Stienkemeier}]{Wituschek2020}
\bibinfo{author}{Wituschek, A.}, \bibinfo{author}{Bruder, L.},
  \bibinfo{author}{Allaria, E.}, \bibinfo{author}{Bangert, U.},
  \bibinfo{author}{Binz, M.}, \bibinfo{author}{Callegari, C.},
  \bibinfo{author}{Cerullo, G.}, \bibinfo{author}{Cinquegrana, P.},
  \bibinfo{author}{Gianessi, L.}, \bibinfo{author}{Danailov, M.},
  \bibinfo{author}{Demidovich, A.}, \bibinfo{author}{Di~Fraia, M.},
  \bibinfo{author}{Drabbels, M.}, \bibinfo{author}{Feifel, R.},
  \bibinfo{author}{Laarmann, T.}, \bibinfo{author}{Michiels, R.},
  \bibinfo{author}{Mirian, N.S.}, \bibinfo{author}{Mudrich, M.},
  \bibinfo{author}{Nikolov, I.}, \bibinfo{author}{O'Shea, F.H.},
  \bibinfo{author}{Penco, G.}, \bibinfo{author}{Piseri, P.},
  \bibinfo{author}{Plekan, O.}, \bibinfo{author}{Prince, K.C.},
  \bibinfo{author}{Przystawik, A.}, \bibinfo{author}{Ribi{\v{c}}, P.R.},
  \bibinfo{author}{Sansone, G.}, \bibinfo{author}{Sigalotti, P.},
  \bibinfo{author}{Spampinati, S.}, \bibinfo{author}{Spezzani, C.},
  \bibinfo{author}{Squibb, R.J.}, \bibinfo{author}{Stranges, S.},
  \bibinfo{author}{Uhl, D.}, \bibinfo{author}{Stienkemeier, F.},
  \bibinfo{year}{2020}.
\newblock \bibinfo{title}{Tracking attosecond electronic coherences using
  phase-manipulated extreme ultraviolet pulses}.
\newblock \bibinfo{journal}{Nat. Commun.} \bibinfo{volume}{11},
  \bibinfo{pages}{883}.
\newblock \DOIprefix\doi{10.1038/s41467-020-14721-2}.
%Type = Article
\bibitem[{Wolf et~al.(2017a)Wolf, Holzmeier, Wagner, Berrah, Bostedt, Bozek,
  Bucksbaum, Coffee, Cryan, Farrell, Feifel, Martinez, McFarland, Mucke, Nandi,
  Tarantelli, Fischer and G{\"u}hr}]{Wolf2017a}
\bibinfo{author}{Wolf, T.}, \bibinfo{author}{Holzmeier, F.},
  \bibinfo{author}{Wagner, I.}, \bibinfo{author}{Berrah, N.},
  \bibinfo{author}{Bostedt, C.}, \bibinfo{author}{Bozek, J.},
  \bibinfo{author}{Bucksbaum, P.}, \bibinfo{author}{Coffee, R.},
  \bibinfo{author}{Cryan, J.}, \bibinfo{author}{Farrell, J.},
  \bibinfo{author}{Feifel, R.}, \bibinfo{author}{Martinez, T.},
  \bibinfo{author}{McFarland, B.}, \bibinfo{author}{Mucke, M.},
  \bibinfo{author}{Nandi, S.}, \bibinfo{author}{Tarantelli, F.},
  \bibinfo{author}{Fischer, I.}, \bibinfo{author}{G{\"u}hr, M.},
  \bibinfo{year}{2017}a.
\newblock \bibinfo{title}{Observing femtosecond fragmentation using ultrafast
  {X}-ray-induced {A}uger spectra}.
\newblock \bibinfo{journal}{Appl. Sci.} \bibinfo{volume}{7},
  \bibinfo{pages}{681}.
\newblock \DOIprefix\doi{10.3390/app7070681}.
%Type = Article
\bibitem[{Wolf et~al.(2017b)Wolf, Myhre, Cryan, Coriani, Squibb, Battistoni,
  Berrah, Bostedt, Bucksbaum, Coslovich, Feifel, Gaffney, Grilj, Martinez,
  Miyabe, Moeller, Mucke, Natan, Obaid, Osipov, Plekan, Wang, Koch and
  G{\"u}hr}]{Wolf2017}
\bibinfo{author}{Wolf, T.J.A.}, \bibinfo{author}{Myhre, R.H.},
  \bibinfo{author}{Cryan, J.P.}, \bibinfo{author}{Coriani, S.},
  \bibinfo{author}{Squibb, R.J.}, \bibinfo{author}{Battistoni, A.},
  \bibinfo{author}{Berrah, N.}, \bibinfo{author}{Bostedt, C.},
  \bibinfo{author}{Bucksbaum, P.}, \bibinfo{author}{Coslovich, G.},
  \bibinfo{author}{Feifel, R.}, \bibinfo{author}{Gaffney, K.J.},
  \bibinfo{author}{Grilj, J.}, \bibinfo{author}{Martinez, T.J.},
  \bibinfo{author}{Miyabe, S.}, \bibinfo{author}{Moeller, S.P.},
  \bibinfo{author}{Mucke, M.}, \bibinfo{author}{Natan, A.},
  \bibinfo{author}{Obaid, R.}, \bibinfo{author}{Osipov, T.},
  \bibinfo{author}{Plekan, O.}, \bibinfo{author}{Wang, S.},
  \bibinfo{author}{Koch, H.}, \bibinfo{author}{G{\"u}hr, M.},
  \bibinfo{year}{2017}b.
\newblock \bibinfo{title}{Probing ultrafast $\pi\pi^*/n\pi^*$ internal
  conversion in organic chromophores via {K}-edge resonant absorption}.
\newblock \bibinfo{journal}{Nat. Commun.} \bibinfo{volume}{8},
  \bibinfo{pages}{29}.
\newblock \DOIprefix\doi{10.1038/s41467-017-00069-7}.
%Type = Article
\bibitem[{Wolkow(1935)}]{Volkov1935}
\bibinfo{author}{Wolkow, D.V.}, \bibinfo{year}{1935}.
\newblock \bibinfo{title}{{\"U}ber eine {K}lasse von {L}\"osungen der
  {D}iracschen {G}leichung}.
\newblock \bibinfo{journal}{Z. Phys} \bibinfo{volume}{94},
  \bibinfo{pages}{250--260}.
\newblock \DOIprefix\doi{10.1007/BF01331022}.
%Type = Article
\bibitem[{Wragg et~al.(2015)Wragg, Parker and van~der Hart}]{Wragg2015}
\bibinfo{author}{Wragg, J.}, \bibinfo{author}{Parker, J.S.},
  \bibinfo{author}{van~der Hart, H.W.}, \bibinfo{year}{2015}.
\newblock \bibinfo{title}{Double ionization in {R}-matrix theory using a
  two-electron outer region}.
\newblock \bibinfo{journal}{Phys. Rev. A} \bibinfo{volume}{92},
  \bibinfo{pages}{022504}.
\newblock \DOIprefix\doi{10.1103/PhysRevA.92.022504}.
%Type = Article
\bibitem[{Yabashi et~al.(2013)Yabashi, Tanaka, Tanaka, Tomizawa, Togashi,
  Nagasono, Ishikawa, Harries, Hikosaka, Hishikawa, Nagaya, Saito, Shigemasa,
  Yamanouchi and Ueda}]{Yabashi2013}
\bibinfo{author}{Yabashi, M.}, \bibinfo{author}{Tanaka, H.},
  \bibinfo{author}{Tanaka, T.}, \bibinfo{author}{Tomizawa, H.},
  \bibinfo{author}{Togashi, T.}, \bibinfo{author}{Nagasono, M.},
  \bibinfo{author}{Ishikawa, T.}, \bibinfo{author}{Harries, J.R.},
  \bibinfo{author}{Hikosaka, Y.}, \bibinfo{author}{Hishikawa, A.},
  \bibinfo{author}{Nagaya, K.}, \bibinfo{author}{Saito, N.},
  \bibinfo{author}{Shigemasa, E.}, \bibinfo{author}{Yamanouchi, K.},
  \bibinfo{author}{Ueda, K.}, \bibinfo{year}{2013}.
\newblock \bibinfo{title}{Compact {XFEL} and {AMO} sciences: {SACLA} and
  {SCSS}}.
\newblock \bibinfo{journal}{J. Phys. B} \bibinfo{volume}{46},
  \bibinfo{pages}{164001}.
\newblock \DOIprefix\doi{10.1088/0953-4075/46/16/164001}.
%Type = Article
\bibitem[{Yamazaki and Elliott(2007)}]{Yamazaki2007}
\bibinfo{author}{Yamazaki, R.}, \bibinfo{author}{Elliott, D.S.},
  \bibinfo{year}{2007}.
\newblock \bibinfo{title}{Observation of the phase lag in the asymmetric
  photoelectron angular distributions of atomic barium}.
\newblock \bibinfo{journal}{Phys. Rev. Lett.} \bibinfo{volume}{98},
  \bibinfo{pages}{053001}.
\newblock \DOIprefix\doi{10.1103/PhysRevLett.98.053001}.
%Type = Article
\bibitem[{Yan et~al.(2019)Yan, Mueller, Ahmed, Hao, Huang, Li, Litvinenko, Liu,
  Mikhailov, Popov, Sikora, Vinokurov and Wu}]{Yan2019}
\bibinfo{author}{Yan, J.}, \bibinfo{author}{Mueller, J.M.},
  \bibinfo{author}{Ahmed, M.W.}, \bibinfo{author}{Hao, H.},
  \bibinfo{author}{Huang, S.}, \bibinfo{author}{Li, J.},
  \bibinfo{author}{Litvinenko, V.N.}, \bibinfo{author}{Liu, P.},
  \bibinfo{author}{Mikhailov, S.F.}, \bibinfo{author}{Popov, V.G.},
  \bibinfo{author}{Sikora, M.H.}, \bibinfo{author}{Vinokurov, N.A.},
  \bibinfo{author}{Wu, Y.K.}, \bibinfo{year}{2019}.
\newblock \bibinfo{title}{Precision control of gamma-ray polarization using a
  crossed helical undulator free-electron laser}.
\newblock \bibinfo{journal}{Nat. Photonics} \bibinfo{volume}{13},
  \bibinfo{pages}{629--635}.
\newblock \DOIprefix\doi{10.1038/s41566-019-0467-6}.
%Type = Article
\bibitem[{Yang et~al.(2020)Yang, Mirian and Giannessi}]{Xi2020}
\bibinfo{author}{Yang, X.}, \bibinfo{author}{Mirian, N.},
  \bibinfo{author}{Giannessi, L.}, \bibinfo{year}{2020}.
\newblock \bibinfo{title}{Postsaturation dynamics and superluminal propagation
  of a superradiant spike in a free-electron laser amplifier}.
\newblock \bibinfo{journal}{Phys. Rev. Accel. Beams} \bibinfo{volume}{23},
  \bibinfo{pages}{010703}.
\newblock \DOIprefix\doi{10.1103/PhysRevAccelBeams.23.010703}.
%Type = Article
\bibitem[{Ye et~al.(2019)Ye, Rouxel, Asban, R\"osner and Mukamel}]{Le2019}
\bibinfo{author}{Ye, L.}, \bibinfo{author}{Rouxel, J.R.},
  \bibinfo{author}{Asban, S.}, \bibinfo{author}{R\"osner, B.},
  \bibinfo{author}{Mukamel, S.}, \bibinfo{year}{2019}.
\newblock \bibinfo{title}{Probing molecular chirality by
  orbital-angular-momentum-carrying x-ray pulses}.
\newblock \bibinfo{journal}{J. Chem. Theory Comput.} \bibinfo{volume}{15},
  \bibinfo{pages}{4180--4186}.
\newblock \DOIprefix\doi{10.1021/acs.jctc.9b00346}.
%Type = Article
\bibitem[{Yin et~al.(1992)Yin, Chen, Elliott and Smith}]{Yin1992}
\bibinfo{author}{Yin, Y.Y.}, \bibinfo{author}{Chen, C.},
  \bibinfo{author}{Elliott, D.S.}, \bibinfo{author}{Smith, A.V.},
  \bibinfo{year}{1992}.
\newblock \bibinfo{title}{Asymmetric photoelectron angular distributions from
  interfering photoionization processes}.
\newblock \bibinfo{journal}{Phys. Rev. Lett.} \bibinfo{volume}{69},
  \bibinfo{pages}{2353--2356}.
\newblock \DOIprefix\doi{10.1103/PhysRevLett.69.2353}.
%Type = Article
\bibitem[{Yoneda et~al.(2015)Yoneda, Inubushi, Nagamine, Michine, Ohashi,
  Yumoto, Yamauchi, Mimura, Kitamura, Katayama, Ishikawa and
  Yabashi}]{Yoneda2015}
\bibinfo{author}{Yoneda, H.}, \bibinfo{author}{Inubushi, Y.},
  \bibinfo{author}{Nagamine, K.}, \bibinfo{author}{Michine, Y.},
  \bibinfo{author}{Ohashi, H.}, \bibinfo{author}{Yumoto, H.},
  \bibinfo{author}{Yamauchi, K.}, \bibinfo{author}{Mimura, H.},
  \bibinfo{author}{Kitamura, H.}, \bibinfo{author}{Katayama, T.},
  \bibinfo{author}{Ishikawa, T.}, \bibinfo{author}{Yabashi, M.},
  \bibinfo{year}{2015}.
\newblock \bibinfo{title}{Atomic inner-shell laser at 1.5 {\aa}ngstr\"om
  wavelength pumped by an {X}-ray free-electron laser}.
\newblock \bibinfo{journal}{Nature} \bibinfo{volume}{524},
  \bibinfo{pages}{446--449}.
\newblock \DOIprefix\doi{10.1038/nature14894}.
%Type = Article
\bibitem[{Yong et~al.(2019)Yong, Qinming, Jiayue, Guanglei, Lei, Hongli, Kai,
  Zhenxing, Zhigang, Zhichao, Yuhuan, Dongxu, Guorong, Weiqing, Xueming, Chao,
  Si, Zhen, Duan, Jie, Xiaoqing, Taihe, Lie, Wenyan, Shaopeng, Junqiang, Lin,
  Chengcheng, Hao, Huan, Guanghua, Haijun, Jianguo, Maomao, Wei, Longwei,
  Fubin, Guanghong, Shengwang, Yiyong, Sen, Fei, Zhiqiang, Xiaoxuan, Yongfang,
  Yonghua, Zhihao, Ruiping, Dazhang, Meng, Haixiao, Bin, Guoqiang, Luyang,
  Yingbing, Shanchuan, Xiaobin, Qiaogen, Bo, Qiang, Ming, Guoping, Yongbin,
  Lixin, Dong and Zhentang}]{Yong2019}
\bibinfo{author}{Yong, Y.}, \bibinfo{author}{Qinming, L.},
  \bibinfo{author}{Jiayue, Y.}, \bibinfo{author}{Guanglei, W.},
  \bibinfo{author}{Lei, S.}, \bibinfo{author}{Hongli, D.},
  \bibinfo{author}{Kai, T.}, \bibinfo{author}{Zhenxing, T.},
  \bibinfo{author}{Zhigang, H.}, \bibinfo{author}{Zhichao, C.},
  \bibinfo{author}{Yuhuan, T.}, \bibinfo{author}{Dongxu, D.},
  \bibinfo{author}{Guorong, W.}, \bibinfo{author}{Weiqing, Z.},
  \bibinfo{author}{Xueming, Y.}, \bibinfo{author}{Chao, F.},
  \bibinfo{author}{Si, C.}, \bibinfo{author}{Zhen, W.}, \bibinfo{author}{Duan,
  G.}, \bibinfo{author}{Jie, C.}, \bibinfo{author}{Xiaoqing, L.},
  \bibinfo{author}{Taihe, L.}, \bibinfo{author}{Lie, F.},
  \bibinfo{author}{Wenyan, Z.}, \bibinfo{author}{Shaopeng, Z.},
  \bibinfo{author}{Junqiang, Z.}, \bibinfo{author}{Lin, L.},
  \bibinfo{author}{Chengcheng, X.}, \bibinfo{author}{Hao, L.},
  \bibinfo{author}{Huan, Z.}, \bibinfo{author}{Guanghua, C.},
  \bibinfo{author}{Haijun, Z.}, \bibinfo{author}{Jianguo, D.},
  \bibinfo{author}{Maomao, H.}, \bibinfo{author}{Wei, Z.},
  \bibinfo{author}{Longwei, L.}, \bibinfo{author}{Fubin, Y.},
  \bibinfo{author}{Guanghong, W.}, \bibinfo{author}{Shengwang, X.},
  \bibinfo{author}{Yiyong, H.X.}, \bibinfo{author}{Sen, S.},
  \bibinfo{author}{Fei, G.}, \bibinfo{author}{Zhiqiang, J.},
  \bibinfo{author}{Xiaoxuan, Z.}, \bibinfo{author}{Yongfang, L.},
  \bibinfo{author}{Yonghua, W.}, \bibinfo{author}{Zhihao, C.},
  \bibinfo{author}{Ruiping, W.}, \bibinfo{author}{Dazhang, H.},
  \bibinfo{author}{Meng, Z.}, \bibinfo{author}{Haixiao, D.},
  \bibinfo{author}{Bin, L.}, \bibinfo{author}{Guoqiang, L.},
  \bibinfo{author}{Luyang, Y.}, \bibinfo{author}{Yingbing, Y.},
  \bibinfo{author}{Shanchuan, Y.}, \bibinfo{author}{Xiaobin, X.},
  \bibinfo{author}{Qiaogen, Z.}, \bibinfo{author}{Bo, L.},
  \bibinfo{author}{Qiang, G.}, \bibinfo{author}{Ming, G.},
  \bibinfo{author}{Guoping, F.}, \bibinfo{author}{Yongbin, L.},
  \bibinfo{author}{Lixin, Y.}, \bibinfo{author}{Dong, W.},
  \bibinfo{author}{Zhentang, Z.}, \bibinfo{year}{2019}.
\newblock \bibinfo{title}{Dalian extreme ultraviolet coherent light source}.
\newblock \bibinfo{journal}{Chinese J. Lasers} \bibinfo{volume}{46},
  \bibinfo{pages}{0100005}.
\newblock \DOIprefix\doi{10.3788/cjl201946.0100005}. \bibinfo{note}{in
  Chinese}.
%Type = Article
\bibitem[{You et~al.(2020a)You, Fukuzawa, Luo, Saito, Berholts, Gaumnitz,
  Huttula, Johnsson, Kishimoto, Myllynen, Nemer, Niozu, Patanen, Pelimanni,
  Takanashi, Wada, Yokono, Owada, Tono, Yabashi, Nagaya, Kukk and
  Ueda}]{You2020}
\bibinfo{author}{You, D.}, \bibinfo{author}{Fukuzawa, H.},
  \bibinfo{author}{Luo, Y.}, \bibinfo{author}{Saito, S.},
  \bibinfo{author}{Berholts, M.}, \bibinfo{author}{Gaumnitz, T.},
  \bibinfo{author}{Huttula, M.}, \bibinfo{author}{Johnsson, P.},
  \bibinfo{author}{Kishimoto, N.}, \bibinfo{author}{Myllynen, H.},
  \bibinfo{author}{Nemer, A.}, \bibinfo{author}{Niozu, A.},
  \bibinfo{author}{Patanen, M.}, \bibinfo{author}{Pelimanni, E.},
  \bibinfo{author}{Takanashi, T.}, \bibinfo{author}{Wada, S.},
  \bibinfo{author}{Yokono, N.}, \bibinfo{author}{Owada, S.},
  \bibinfo{author}{Tono, K.}, \bibinfo{author}{Yabashi, M.},
  \bibinfo{author}{Nagaya, K.}, \bibinfo{author}{Kukk, E.},
  \bibinfo{author}{Ueda, K.}, \bibinfo{year}{2020}a.
\newblock \bibinfo{title}{Multi-particle momentum correlations extracted using
  covariance methods on multiple-ionization of diiodomethane molecules by
  soft-{X}-ray free-electron laser pulses}.
\newblock \bibinfo{journal}{Phys. Chem. Chem. Phys.} \bibinfo{volume}{22},
  \bibinfo{pages}{2648--2659}.
\newblock \DOIprefix\doi{10.1039/c9cp03638e}.
%Type = Misc
\bibitem[{You et~al.(2020b)You, Ueda, Gryzlova, Grum-Grzhimailo, Popova,
  Staroselskaya, Tugs, Orimo, Sato, Ishikawa, Carpeggiani, Csizmadia, F{\"u}le,
  Sansone, Maroju, D'Elia, Mazza, Meyer, Callegari, Di~Fraia, Plekan, Richter,
  Giannessi, Allaria, De~Ninno, Trov{\`o}, Badano, Diviacco, Gauthier, Mirian,
  Penco, Rebernik~Ribi{\v{c}}, Spampinati, Spezzani, Gaio and
  Prince}]{You2020b}
\bibinfo{author}{You, D.}, \bibinfo{author}{Ueda, K.},
  \bibinfo{author}{Gryzlova, E.V.}, \bibinfo{author}{Grum-Grzhimailo, A.N.},
  \bibinfo{author}{Popova, M.M.}, \bibinfo{author}{Staroselskaya, E.I.},
  \bibinfo{author}{Tugs, O.}, \bibinfo{author}{Orimo, Y.},
  \bibinfo{author}{Sato, T.}, \bibinfo{author}{Ishikawa, K.L.},
  \bibinfo{author}{Carpeggiani, P.A.}, \bibinfo{author}{Csizmadia, T.},
  \bibinfo{author}{F{\"u}le, M.}, \bibinfo{author}{Sansone, G.},
  \bibinfo{author}{Maroju, P.K.}, \bibinfo{author}{D'Elia, A.},
  \bibinfo{author}{Mazza, T.}, \bibinfo{author}{Meyer, M.},
  \bibinfo{author}{Callegari, C.}, \bibinfo{author}{Di~Fraia, M.},
  \bibinfo{author}{Plekan, O.}, \bibinfo{author}{Richter, R.},
  \bibinfo{author}{Giannessi, L.}, \bibinfo{author}{Allaria, E.},
  \bibinfo{author}{De~Ninno, G.}, \bibinfo{author}{Trov{\`o}, M.},
  \bibinfo{author}{Badano, L.}, \bibinfo{author}{Diviacco, B.},
  \bibinfo{author}{Gauthier, D.}, \bibinfo{author}{Mirian, N.},
  \bibinfo{author}{Penco, G.}, \bibinfo{author}{Rebernik~Ribi{\v{c}}, P.},
  \bibinfo{author}{Spampinati, S.}, \bibinfo{author}{Spezzani, C.},
  \bibinfo{author}{Gaio, G.}, \bibinfo{author}{Prince, K.C.},
  \bibinfo{year}{2020}b.
\newblock \bibinfo{title}{A new method for measuring angle-resolved phases in
  photoemission}.
\newblock \bibinfo{note}{Phys. Rev. X, in press}.
%Type = Article
\bibitem[{You et~al.(2019)You, Ueda, Ruberti, Ishikawa, Carpeggiani, Csizmadia,
  Oldal, NG, Sansone, Maroju, Kooser, Callegari, Di~Fraia, Plekan, Giannessi,
  Allaria, De~Ninno, Trov\`{o}, Badano, Diviacco, Gauthier, Mirian, Penco,
  Ribi{\v{c}}, Spampinati, Spezzani, Di~Mitri, Gaio and Prince}]{You2019}
\bibinfo{author}{You, D.}, \bibinfo{author}{Ueda, K.},
  \bibinfo{author}{Ruberti, M.}, \bibinfo{author}{Ishikawa, K.L.},
  \bibinfo{author}{Carpeggiani, P.A.}, \bibinfo{author}{Csizmadia, T.},
  \bibinfo{author}{Oldal, L.G.}, \bibinfo{author}{NG, H.},
  \bibinfo{author}{Sansone, G.}, \bibinfo{author}{Maroju, P.K.},
  \bibinfo{author}{Kooser, K.}, \bibinfo{author}{Callegari, C.},
  \bibinfo{author}{Di~Fraia, M.}, \bibinfo{author}{Plekan, O.},
  \bibinfo{author}{Giannessi, L.}, \bibinfo{author}{Allaria, E.},
  \bibinfo{author}{De~Ninno, G.}, \bibinfo{author}{Trov\`{o}, M.},
  \bibinfo{author}{Badano, L.}, \bibinfo{author}{Diviacco, B.},
  \bibinfo{author}{Gauthier, D.}, \bibinfo{author}{Mirian, N.},
  \bibinfo{author}{Penco, G.}, \bibinfo{author}{Ribi{\v{c}}, P.R.},
  \bibinfo{author}{Spampinati, S.}, \bibinfo{author}{Spezzani, C.},
  \bibinfo{author}{Di~Mitri, S.}, \bibinfo{author}{Gaio, G.},
  \bibinfo{author}{Prince, K.C.}, \bibinfo{year}{2019}.
\newblock \bibinfo{title}{A detailed investigation of single-photon laser
  enabled {A}uger decay in neon}.
\newblock \bibinfo{journal}{New J. Phys.} \bibinfo{volume}{21},
  \bibinfo{pages}{113036}.
\newblock \DOIprefix\doi{10.1088/1367-2630/ab520d}.
%Type = Article
\bibitem[{Young et~al.(2010)Young, Kanter, Kr{\"a}ssig, Li, March, Pratt,
  Santra, Southworth, Rohringer, DiMauro, Doumy, Roedig, Berrah, Fang, Hoener,
  Bucksbaum, Cryan, Ghimire, Glownia, Reis, Bozek, Bostedt and
  Messerschmidt}]{YoungNature2010}
\bibinfo{author}{Young, L.}, \bibinfo{author}{Kanter, E.P.},
  \bibinfo{author}{Kr{\"a}ssig, B.}, \bibinfo{author}{Li, Y.},
  \bibinfo{author}{March, A.M.}, \bibinfo{author}{Pratt, S.T.},
  \bibinfo{author}{Santra, R.}, \bibinfo{author}{Southworth, S.H.},
  \bibinfo{author}{Rohringer, N.}, \bibinfo{author}{DiMauro, L.F.},
  \bibinfo{author}{Doumy, G.}, \bibinfo{author}{Roedig, C.A.},
  \bibinfo{author}{Berrah, N.}, \bibinfo{author}{Fang, L.},
  \bibinfo{author}{Hoener, M.}, \bibinfo{author}{Bucksbaum, P.H.},
  \bibinfo{author}{Cryan, J.P.}, \bibinfo{author}{Ghimire, S.},
  \bibinfo{author}{Glownia, J.M.}, \bibinfo{author}{Reis, D.A.},
  \bibinfo{author}{Bozek, J.D.}, \bibinfo{author}{Bostedt, C.},
  \bibinfo{author}{Messerschmidt, M.}, \bibinfo{year}{2010}.
\newblock \bibinfo{title}{Femtosecond electronic response of atoms to
  ultra-intense {X}-rays}.
\newblock \bibinfo{journal}{Nature} \bibinfo{volume}{466},
  \bibinfo{pages}{56--61}.
\newblock \DOIprefix\doi{10.1038/nature09177}.
%Type = Article
\bibitem[{Yu(1984)}]{Yu1984}
\bibinfo{author}{Yu, L.H.}, \bibinfo{year}{1984}.
\newblock \bibinfo{title}{Optical klystron harmonic generator with electron
  microbunches induced and frozen by lasers as an intense coherent soft {X}-ray
  source}.
\newblock \bibinfo{journal}{Phys. Rev. Lett.} \bibinfo{volume}{53},
  \bibinfo{pages}{254--257}.
\newblock \DOIprefix\doi{10.1103/PhysRevLett.53.254}.
%Type = Article
\bibitem[{Yu(1991)}]{Yu1991}
\bibinfo{author}{Yu, L.H.}, \bibinfo{year}{1991}.
\newblock \bibinfo{title}{Generation of intense {uv} radiation by
  subharmonically seeded single-pass free-electron lasers}.
\newblock \bibinfo{journal}{Phys. Rev. A} \bibinfo{volume}{44},
  \bibinfo{pages}{5178--5193}.
\newblock \DOIprefix\doi{10.1103/PhysRevA.44.5178}.
%Type = Article
\bibitem[{Yu et~al.(2000)Yu, Babzien, Ben-Zvi, Dimauro, Doyuran, Graves,
  Johnson, Krinsky, Malone, Pogorelsky, Skaritka, Rakowsky, Solomon, X.~J.,
  Woodle, Yakimenko, Biedron, Galayda, Gluskin, Jagger, Sajaev and
  Vasserman}]{Yu2000}
\bibinfo{author}{Yu, L.H.}, \bibinfo{author}{Babzien, M.},
  \bibinfo{author}{Ben-Zvi, I.}, \bibinfo{author}{Dimauro, L.F.},
  \bibinfo{author}{Doyuran, A.}, \bibinfo{author}{Graves, W.},
  \bibinfo{author}{Johnson, E.}, \bibinfo{author}{Krinsky, S.},
  \bibinfo{author}{Malone, R.}, \bibinfo{author}{Pogorelsky, I.},
  \bibinfo{author}{Skaritka, J.}, \bibinfo{author}{Rakowsky, G.},
  \bibinfo{author}{Solomon, L.}, \bibinfo{author}{X.~J., W.A.N.G.},
  \bibinfo{author}{Woodle, M.}, \bibinfo{author}{Yakimenko, V.},
  \bibinfo{author}{Biedron, S.G.}, \bibinfo{author}{Galayda, J.N.},
  \bibinfo{author}{Gluskin, E.}, \bibinfo{author}{Jagger, J.},
  \bibinfo{author}{Sajaev, V.}, \bibinfo{author}{Vasserman, I.},
  \bibinfo{year}{2000}.
\newblock \bibinfo{title}{High-gain harmonic-generation free-electron laser}.
\newblock \bibinfo{journal}{Science} \bibinfo{volume}{289},
  \bibinfo{pages}{932--934}.
\newblock \DOIprefix\doi{10.1126/science.289.5481.932}.
%Type = Article
\bibitem[{Yu et~al.(2003)Yu, DiMauro, Doyuran, Graves, Johnson, Heese, Krinsky,
  Loos, Murphy, Rakowsky, Rose, Shaftan, Sheehy, Skaritka, Wang and
  Wu}]{Yu2003}
\bibinfo{author}{Yu, L.H.}, \bibinfo{author}{DiMauro, L.},
  \bibinfo{author}{Doyuran, A.}, \bibinfo{author}{Graves, W.S.},
  \bibinfo{author}{Johnson, E.D.}, \bibinfo{author}{Heese, R.},
  \bibinfo{author}{Krinsky, S.}, \bibinfo{author}{Loos, H.},
  \bibinfo{author}{Murphy, J.B.}, \bibinfo{author}{Rakowsky, G.},
  \bibinfo{author}{Rose, J.}, \bibinfo{author}{Shaftan, T.},
  \bibinfo{author}{Sheehy, B.}, \bibinfo{author}{Skaritka, J.},
  \bibinfo{author}{Wang, X.J.}, \bibinfo{author}{Wu, Z.}, \bibinfo{year}{2003}.
\newblock \bibinfo{title}{First ultraviolet high-gain harmonic-generation
  free-electron laser}.
\newblock \bibinfo{journal}{Phys. Rev. Lett.} \bibinfo{volume}{91},
  \bibinfo{pages}{074801}.
\newblock \DOIprefix\doi{10.1103/PhysRevLett.91.074801}.
%Type = Article
\bibitem[{Yuan and Bandrauk(2019)}]{Yuan2019}
\bibinfo{author}{Yuan, K.J.}, \bibinfo{author}{Bandrauk, A.D.},
  \bibinfo{year}{2019}.
\newblock \bibinfo{title}{Probing attosecond electron coherence in molecular
  charge migration by ultrafast {X}-ray photoelectron imaging}.
\newblock \bibinfo{journal}{Appl. Sci.} \bibinfo{volume}{9},
  \bibinfo{pages}{1941}.
\newblock \DOIprefix\doi{doi:10.3390/app9091941}.
%Type = Article
\bibitem[{Zandee and Bernstein(1979)}]{Zandee1979}
\bibinfo{author}{Zandee, L.}, \bibinfo{author}{Bernstein, R.B.},
  \bibinfo{year}{1979}.
\newblock \bibinfo{title}{Resonance-enhanced multiphoton ionization and
  fragmentation of molecular beams: {NO}, {I$_2$}, benzene, and butadiene}.
\newblock \bibinfo{journal}{J. Chem. Phys.} \bibinfo{volume}{71},
  \bibinfo{pages}{1359--1371}.
\newblock \DOIprefix\doi{10.1063/1.438436}.
%Type = Article
\bibitem[{Zanghellini et~al.(2003)Zanghellini, Kitzler, Fabian, Brabec and
  Scrinzi}]{Zanghellini2003}
\bibinfo{author}{Zanghellini, J.}, \bibinfo{author}{Kitzler, M.},
  \bibinfo{author}{Fabian, C.}, \bibinfo{author}{Brabec, T.},
  \bibinfo{author}{Scrinzi, A.}, \bibinfo{year}{2003}.
\newblock \bibinfo{title}{An {MCTDHF} approach to multielectron dynamics in
  laser fields}.
\newblock \bibinfo{journal}{Laser Phys.} \bibinfo{volume}{13},
  \bibinfo{pages}{1064--1068}.
%Type = Article
\bibitem[{Zatsarinny(2006)}]{Zatsarinny2006}
\bibinfo{author}{Zatsarinny, O.}, \bibinfo{year}{2006}.
\newblock \bibinfo{title}{{BSR}: {B}-spline atomic {R}-matrix codes}.
\newblock \bibinfo{journal}{Comput. Phys. Commun.} \bibinfo{volume}{174},
  \bibinfo{pages}{273--356}.
\newblock \DOIprefix\doi{10.1016/j.cpc.2005.10.006}.
%Type = Article
\bibitem[{Zatsarinny and Bartschat(2013)}]{Zatsarinny2013}
\bibinfo{author}{Zatsarinny, O.}, \bibinfo{author}{Bartschat, K.},
  \bibinfo{year}{2013}.
\newblock \bibinfo{title}{The {B}-spline {R}-matrix method for atomic
  processes: application to atomic structure, electron collisions and
  photoionization}.
\newblock \bibinfo{journal}{J. Phys. B} \bibinfo{volume}{46},
  \bibinfo{pages}{112001}.
\newblock \DOIprefix\doi{10.1088/0953-4075/46/11/112001}.
%Type = Article
\bibitem[{Zewail(2000)}]{Zewail2000}
\bibinfo{author}{Zewail, A.H.}, \bibinfo{year}{2000}.
\newblock \bibinfo{title}{Femtochemistry: Atomic-scale dynamics of the chemical
  bond using ultrafast lasers ({N}obel {L}ecture)}.
\newblock \bibinfo{journal}{Angew. Chem. Int. Ed.} \bibinfo{volume}{39},
  \bibinfo{pages}{2586--2631}.
\newblock
  \DOIprefix\doi{10.1002/1521-3773(20000804)39:15<2586::aid-anie2586>3.0.co;2-o}.
%Type = Article
\bibitem[{Zhang et~al.(2018)Zhang, Zheng, Li, Jia, Li, Xu, Leng, Zeng, Li and
  Xu}]{Zhang2018}
\bibinfo{author}{Zhang, L.}, \bibinfo{author}{Zheng, Y.}, \bibinfo{author}{Li,
  G.}, \bibinfo{author}{Jia, Z.}, \bibinfo{author}{Li, Y.},
  \bibinfo{author}{Xu, Y.}, \bibinfo{author}{Leng, Y.}, \bibinfo{author}{Zeng,
  Z.}, \bibinfo{author}{Li, R.}, \bibinfo{author}{Xu, Z.},
  \bibinfo{year}{2018}.
\newblock \bibinfo{title}{Bright high-order harmonic generation around 30 nm
  using hundred-terawatt-level laser system for seeding full coherent {XFEL}}.
\newblock \bibinfo{journal}{Applied Sciences} \bibinfo{volume}{8},
  \bibinfo{pages}{1446}.
\newblock \DOIprefix\doi{10.3390/app8091446}.
%Type = Article
\bibitem[{Zhang et~al.(2016)Zhang, Kimberg and Rohringer}]{Zhang2016}
\bibinfo{author}{Zhang, S.B.}, \bibinfo{author}{Kimberg, V.},
  \bibinfo{author}{Rohringer, N.}, \bibinfo{year}{2016}.
\newblock \bibinfo{title}{Nonlinear resonant auger spectroscopy in {CO} using
  an {x}-ray pump-control scheme}.
\newblock \bibinfo{journal}{Phys. Rev. A} \bibinfo{volume}{94},
  \bibinfo{pages}{063413}.
\newblock \DOIprefix\doi{10.1103/PhysRevA.94.063413}.
%Type = Article
\bibitem[{Zhao et~al.(2017a)Zhao, Wang, Gu, Yin, Fang, Gu, Leng, Zhou, Liu,
  Tang, Huang, Liu and Jiang}]{Zhao2017}
\bibinfo{author}{Zhao, Z.}, \bibinfo{author}{Wang, D.}, \bibinfo{author}{Gu,
  Q.}, \bibinfo{author}{Yin, L.}, \bibinfo{author}{Fang, G.},
  \bibinfo{author}{Gu, M.}, \bibinfo{author}{Leng, Y.}, \bibinfo{author}{Zhou,
  Q.}, \bibinfo{author}{Liu, B.}, \bibinfo{author}{Tang, C.},
  \bibinfo{author}{Huang, W.}, \bibinfo{author}{Liu, Z.},
  \bibinfo{author}{Jiang, H.}, \bibinfo{year}{2017}a.
\newblock \bibinfo{title}{{SXFEL}: A soft {X}-ray free electron laser in
  {C}hina}.
\newblock \bibinfo{journal}{Synchrotron Radiat. News} \bibinfo{volume}{30},
  \bibinfo{pages}{29--33}.
\newblock \DOIprefix\doi{10.1080/08940886.2017.1386997}.
%Type = Article
\bibitem[{Zhao et~al.(2017b)Zhao, Wang, Gu, Yin, Gu, Leng and Liu}]{Zhao2017a}
\bibinfo{author}{Zhao, Z.}, \bibinfo{author}{Wang, D.}, \bibinfo{author}{Gu,
  Q.}, \bibinfo{author}{Yin, L.}, \bibinfo{author}{Gu, M.},
  \bibinfo{author}{Leng, Y.}, \bibinfo{author}{Liu, B.}, \bibinfo{year}{2017}b.
\newblock \bibinfo{title}{Status of the {SXFEL} facility}.
\newblock \bibinfo{journal}{Applied Sciences} \bibinfo{volume}{7},
  \bibinfo{pages}{607}.
\newblock \DOIprefix\doi{10.3390/app7060607}.
%Type = Article
\bibitem[{Zhao et~al.(2019)Zhao, Wang, Yin, Gu, Fang, Gu, Leng, Zhou, Liu,
  Tang, Huang, Liu, Jiang and Weng}]{Zhao2019}
\bibinfo{author}{Zhao, Z.}, \bibinfo{author}{Wang, D.}, \bibinfo{author}{Yin,
  L.}, \bibinfo{author}{Gu, Q.}, \bibinfo{author}{Fang, G.},
  \bibinfo{author}{Gu, M.}, \bibinfo{author}{Leng, Y.}, \bibinfo{author}{Zhou,
  Q.}, \bibinfo{author}{Liu, B.}, \bibinfo{author}{Tang, C.},
  \bibinfo{author}{Huang, W.}, \bibinfo{author}{Liu, Z.},
  \bibinfo{author}{Jiang, H.}, \bibinfo{author}{Weng, Z.},
  \bibinfo{year}{2019}.
\newblock \bibinfo{title}{Shanghai soft {X}-ray free-electron laser facility}.
\newblock \bibinfo{journal}{Chinese J. Lasers} \bibinfo{volume}{46},
  \bibinfo{pages}{0100004}.
\newblock \DOIprefix\doi{10.3788/cjl201946.0100004}. \bibinfo{note}{in
  chinese}.
%Type = Article
\bibitem[{{\v{Z}}itnik et~al.(2014){\v{Z}}itnik, Miheli\v{c}, Bu\v{c}ar,
  Kav\v{c}i\v{c}, Rubensson, Svanquist, S\"oderstr\"om, Feifel, S\aa{}the,
  Ovcharenko, Lyamayev, Mazza, Meyer, Simon, Journel, L\"uning, Plekan, Coreno,
  Devetta, Di~Fraia, Finetti, Richter, Grazioli, Prince and
  Callegari}]{Zitnik_PRL_2014}
\bibinfo{author}{{\v{Z}}itnik, M.}, \bibinfo{author}{Miheli\v{c}, A.},
  \bibinfo{author}{Bu\v{c}ar, K.}, \bibinfo{author}{Kav\v{c}i\v{c}, M.},
  \bibinfo{author}{Rubensson, J.E.}, \bibinfo{author}{Svanquist, M.},
  \bibinfo{author}{S\"oderstr\"om, J.}, \bibinfo{author}{Feifel, R.},
  \bibinfo{author}{S\aa{}the, C.}, \bibinfo{author}{Ovcharenko, Y.},
  \bibinfo{author}{Lyamayev, V.}, \bibinfo{author}{Mazza, T.},
  \bibinfo{author}{Meyer, M.}, \bibinfo{author}{Simon, M.},
  \bibinfo{author}{Journel, L.}, \bibinfo{author}{L\"uning, J.},
  \bibinfo{author}{Plekan, O.}, \bibinfo{author}{Coreno, M.},
  \bibinfo{author}{Devetta, M.}, \bibinfo{author}{Di~Fraia, M.},
  \bibinfo{author}{Finetti, P.}, \bibinfo{author}{Richter, R.},
  \bibinfo{author}{Grazioli, C.}, \bibinfo{author}{Prince, K.C.},
  \bibinfo{author}{Callegari, C.}, \bibinfo{year}{2014}.
\newblock \bibinfo{title}{High resolution multiphoton spectroscopy by a tunable
  free-electron-laser light}.
\newblock \bibinfo{journal}{Phys. Rev. Lett.} \bibinfo{volume}{113},
  \bibinfo{pages}{193201}.
\newblock \DOIprefix\doi{10.1103/PhysRevLett.113.193201}.
%Type = Article
\bibitem[{Zwanzig(1964)}]{Zwanzig1964}
\bibinfo{author}{Zwanzig, R.}, \bibinfo{year}{1964}.
\newblock \bibinfo{title}{On the identity of three generalized master
  equations}.
\newblock \bibinfo{journal}{Physica} \bibinfo{volume}{30},
  \bibinfo{pages}{1109--1123}.
\newblock \DOIprefix\doi{10.1016/0031-8914(64)90102-8}.

\end{thebibliography}

\end{document}